\documentclass[a4paper,12pt,twoside,fleqn]{report}
\usepackage[a4paper]{geometry}
\usepackage[polutonikogreek,italian,english]{babel}
\usepackage{indentfirst}
\usepackage[T1]{fontenc}
\usepackage[sc]{mathpazo} 
\usepackage{eulervm}
\usepackage[utf8]{inputenc} 
\usepackage{graphicx,xcolor}
\usepackage{epsf}
\usepackage{amsmath} 
\usepackage{amsfonts} 
\usepackage{amssymb} 
\usepackage{xcolor}
\usepackage[colorlinks,
            bookmarksnumbered=true,
            bookmarks=true,
            hyperfootnotes=true,
            backref=page
            ]
           {hyperref}
\hypersetup{
 linktocpage = true,
 linkcolor = blue,
 urlcolor = blue,
 colorlinks = true,
 linkcolor = teal,
 anchorcolor = blue,
 citecolor = teal,
 urlcolor = teal,
 pdftitle = {High-energy resummation in semi-hard processes at the LHC},                
 pdfauthor = {Francesco Giovanni Celiberto},                 
 pdfsubject = {High-energy phenomenology},               
 pdfcreator = {Celiberto Francesco Giovanni FGC},               
 pdfproducer = {Celiberto Francesco Giovanni FGC},
 pdfstartview = {XYZ null null 1.25},
 pdfkeywords = {hep-ph, QCD, BFKL, LHC, semi-hard processes, inclusive processes, jets, jet phsysics, hadrons, hadron physics, multi-jet production, phenomenology, physics, hep-th ,theoretical physics, theory}, 
           }
\usepackage{backrefx}
  \renewcommand{\backrefpagesname}{Cited on page~}
  \renewcommand{\backrefpagesnames}{Cited on pages~}
\usepackage{doi}
\usepackage{booktabs}
\usepackage{siunitx}
\usepackage{graphics}
\DeclareGraphicsExtensions{.jpg}
\usepackage[format=hang,font=small]{caption}
\usepackage{subfig}
\usepackage{appendix}
\usepackage[format=hang,font=small]{caption}
\usepackage{subfig}
\usepackage{axodraw4j}
\usepackage{caption}[2004/07/16]
\usepackage{ragged2e}
\usepackage{float}
\usepackage{slashed} 
\usepackage{pstricks} 
\usepackage{footmisc} 
\usepackage{cite}
\usepackage{listings} 
\usepackage{fancyhdr} 
\newcounter{appcnt}
\newcounter{tmp}

\newcommand{à}{\`a}
\newcommand{è}{\`e}
\newcommand{é}{\'e}
\newcommand{ì}{\`i}
\newcommand{ò}{\`o}
\newcommand{ù}{\`u}

\newcommand{\greek}[1]{\begin{otherlanguage*}{polutonikogreek}#1\end{otherlanguage*}}

\newcommand{\beq}{\begin{equation}}
\newcommand{\eeq}{\end{equation}}
\def\beeq{\begin{eqnarray}}
\def\eeeq{\end{eqnarray}}
\def\bq{\begin{equation}}
\def\eq{\end{equation}}
\def\ba{\begin{eqnarray}}
\def\ea{\end{eqnarray}}

\newcommand{\ti}{\textit}

\newcommand{\stringa}{\ttfamily\lstinline}
\def\cod#1{{\stringa!#1!}}

\newcommand{\as}{\alpha_s}
\newcommand{\asq}{\alpha_s^2}
\newcommand{\asb}{\bar{\alpha}_s}

\newcommand{\mur}{\mu_R}
\newcommand{\om}{\omega}

\newcommand{\bea}{\begin{eqnarray}}
\newcommand{\eea}{\end{eqnarray}}
\newcommand{\real}{{\sf I}\kern-.12em{\sf R}}
\newcommand{\comp}{{\sf I}\kern-.50em{\sf C}}
\newcommand{\unity}{{\sf I}\kern-.54em{\sf 1}}

\newcommand{\stjp}{{\hat\sigma^{3-{\rm jet}}}_{r,s}}
\newcommand{\sfjp}{{\hat\sigma^{4-{\rm jet}}}_{r,s}}

\newcommand{\phiaj}{\Delta\phi_{\widehat{AJ}}}
\newcommand{\phijb}{\Delta\phi_{\widehat{JB}}}

\def\ap#1#2#3   {{\em Ann. Phys. (NY)} {\bf#1} (#2) #3}
\def\apj#1#2#3  {{\em Astrophys. J.} {\bf#1} (#2) #3}
\def\apjl#1#2#3 {{\em Astrophys. J. Lett.} {\bf#1} (#2) #3}
\def\app#1#2#3  {{\em Acta. Phys. Pol.} {\bf#1} (#2) #3}
\def\ar#1#2#3   {{\em Ann. Rev. Nucl. Part. Sci.} {\bf#1} (#2) #3}
\def\cpc#1#2#3  {{\em Computer Phys. Comm.} {\bf#1} (#2) #3}
\def\err#1#2#3  {{\it Erratum} {\bf#1} (#2) #3}
\def\ib#1#2#3   {{\it ibid.} {\bf#1} (#2) #3}
\def\jmp#1#2#3  {{\em J. Math. Phys.} {\bf#1} (#2) #3}
\def\ijmp#1#2#3 {{\em Int. J. Mod. Phys.} {\bf#1} (#2) #3}
\def\jetp#1#2#3 {{\em JETP Lett.} {\bf#1} (#2) #3}
\def\jpg#1#2#3  {{\em J. Phys. G.} {\bf#1} (#2) #3}
\def\mpl#1#2#3  {{\em Mod. Phys. Lett.} {\bf#1} (#2) #3}
\def\nat#1#2#3  {{\em Nature (London)} {\bf#1} (#2) #3}
\def\nc#1#2#3   {{\em Nuovo Cim.} {\bf#1} (#2) #3}
\def\nim#1#2#3  {{\em Nucl. Instr. Meth.} {\bf#1} (#2) #3}
\def\np#1#2#3   {{\em Nucl. Phys.} {\bf#1} (#2) #3}
\def\pcps#1#2#3 {{\em Proc. Cam. Phil. Soc.} {\bf#1} (#2) #3}
\def\pl#1#2#3   {{\em Phys. Lett.} {\bf#1} (#2) #3}
\def\prep#1#2#3 {{\em Phys. Rep.} {\bf#1} (#2) #3}
\def\prev#1#2#3 {{\em Phys. Rev.} {\bf#1} (#2) #3}
\def\prl#1#2#3  {{\em Phys. Rev. Lett.} {\bf#1} (#2) #3}
\def\prs#1#2#3  {{\em Proc. Roy. Soc.} {\bf#1} (#2) #3}
\def\ptp#1#2#3  {{\em Prog. Th. Phys.} {\bf#1} (#2) #3}
\def\ps#1#2#3   {{\em Physica Scripta} {\bf#1} (#2) #3}
\def\rmp#1#2#3  {{\em Rev. Mod. Phys.} {\bf#1} (#2) #3}
\def\rpp#1#2#3  {{\em Rep. Prog. Phys.} {\bf#1} (#2) #3}
\def\sjnp#1#2#3 {{\em Sov. J. Nucl. Phys.} {\bf#1} (#2) #3}
\def\spj#1#2#3  {{\em Sov. Phys. JEPT} {\bf#1} (#2) #3}
\def\spu#1#2#3  {{\em Sov. Phys.-Usp.} {\bf#1} (#2) #3}
\def\zp#1#2#3   {{\em Zeit. Phys.} {\bf#1} (#2) #3}
\def\ap#1#2#3{Ann.\ Phys.\ (NY) #1 (19#3) #2}
\def\ar#1#2#3{Ann.\ Rev.\ Nucl.\ Part.\ Sci.\ #1 (19#3) #2}

\def\cpc#1#2#3{Computer Phys.\ Comm.\ #1 (19#3) #2}
\def\ib#1#2#3{ibid.\ #1 (19#3) #2}
\def\np#1#2#3{Nucl.\ Phys.\ B#1 (19#3) #2}
\def\pl#1#2#3{Phys.\ Lett.\ #1B (19#3) #2}

\def\prep#1#2#3{Phys.\ Rep.\ #1 (19#3) #2}
\def\prl#1#2#3{Phys.\ Rev.\ Lett.\ #1 (19#3) #2}
\def\rmp#1#2#3{Rev.\ Mod.\ Phys.\ #1 (19#3) #2}

\def\zp#1#2#3{Zeit.\ Phys.\ C#1 (19#3) #2}

\begin{document}

\setcounter{secnumdepth}{4}
\setcounter{tocdepth}{4}

\begin{titlepage}
 \thispagestyle{plain}
 \newgeometry{left=0cm,right=0cm,top=0cm,bottom=0cm}
 \begin{figure}[H]
  \centering
  \includegraphics[scale=0.99]{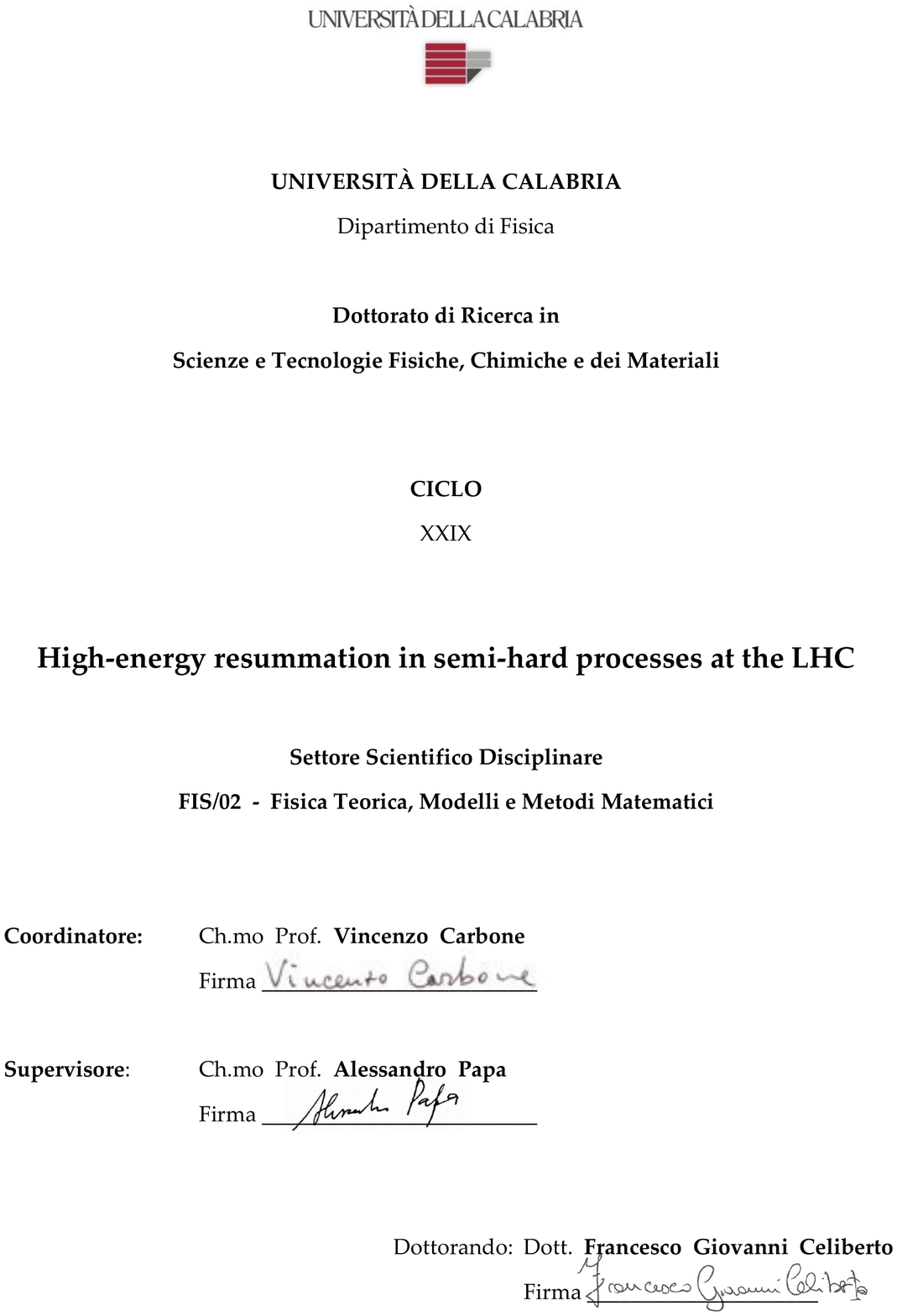}
 \end{figure} 
 \restoregeometry
\end{titlepage}

\input epsf
\pagenumbering{roman}
\setcounter{page}{0}

\begin{flushright}
 \thispagestyle{empty}
 \null\vspace{\stretch{1.5}}
 {\textit{Alla mia famiglia 
          \\ 
          A zio Pinuccio $\,$}}
 \vspace{\stretch{1}}\null
\end{flushright}

\newpage
\begingroup
 \hypersetup{linktoc = all, 
             linkcolor = black
             }
 \pdfbookmark[1]{Contents}{toc}
 \tableofcontents
 \newpage
 \pdfbookmark[1]{List of Tables}{tables}
 \listoftables
 \newpage
 \pdfbookmark[1]{List of Figures}{figures}
 \listoffigures
\endgroup
\newpage

\baselineskip=22pt

\pagenumbering{arabic}

\chapter*{Abstract}
\label{abstract}
\addcontentsline{toc}{chapter}{Abstract}
\markboth{ABSTRACT}{}
\markright{ABSTRACT}{}
Semi-hard processes in the large center-of-mass energy limit offer us an exclusive chance to test the dynamics behind strong interactions in kinematical sectors so far
unexplored, the high luminosity and the record energies 
of the LHC 
providing us with a richness of useful data. 
In the Regge limit, $s\gg |t|$,  
fixed-order calculations in perturbative QCD based on \emph{collinear factorisation}
miss the effect of large energy logarithms, which are so large
to compensate the small QCD coupling $\alpha_s$ and must therefore be accounted for to all perturbative orders. 
The BFKL approach represents the most powerful tool to perform the resummation to all orders of these large logarithms 
both in the LLA, which means inclusion of all terms proportional to $(\alpha_s\ln(s))^n$,
and NLA, which means inclusion of all terms proportional to $\alpha_s(\alpha_s\ln(s))^n$. The inclusive hadroproduction of forward jets with high transverse momenta separated by a large rapidity gap at the LHC, the so-called Mueller--Navelet jets, has been one of the most studied reactions so far. Interesting observables associated to this process are the azimuthal correlation momenta, showing a very
good agreement with experimental data at the LHC. 
However, new BFKL-sensitive observables should be considered in the context of the LHC physics program. 
With the aim the to further and deeply probe the
dynamics of QCD in the Regge limit, we give phenomenological predictions for four distinct semi-hard process. 
On one hand, we continue the analysis of reactions with two objects identified in the final state $(i)$ by addressing open problems in the Mueller--Navelet sector and $(ii)$ by studying the inclusive dihadron production in the full NLA BKFL accuracy. 
Hadrons can be detected at the LHC at much smaller values of the transverse momentum than jets, allowing us to explore an additional kinematical range, complementary to the one studied typical of Mueller--Navelet jets. Furthermore, this process permits to constrain not only the parton distribution functions for the initial proton, but also the parton fragmentation functions describing the detected hadron in the final state.
On the other hand, we show how inclusive multi-jet production processes allow us to define new, generalised and suitable BFKL observables, where transverse momenta and rapidities of the tagged jets, well separated in rapidity from each other, appear in new combinations. 
We give the first phenomenological predictions for the inclusive three-jet production, encoding the effects of higher-order BFKL corrections. Then, making use of the same formalism, we present the first complete BFKL analysis for the four-jet production. 

\newpage

\begin{otherlanguage*}{italian}

\chapter*{Sintesi in lingua italiana}
\label{sintesi}
\addcontentsline{toc}{chapter}{Sintesi in lingua italiana}
\markboth{SINTESI IN LINGUA ITALIANA}{}
\markright{SINTESI IN LINGUA ITALIANA}{}

Per quanto una teoria fisica possa apparire complessa e formalmente ardua l'origine della sua eleganza risiede quasi sempre in un'idea semplice 
e concreta. 
Il Modello Standard (MS) delle particelle elementari, solidamente edificato sull'esistenza di costituenti fondamentali di natura 
fermionica che interagiscono tra loro attraverso lo scambio di bosoni vettori intermedi, è tra gli esempi più significativi. 
All'interno del MS, la Cromodinamica Quantistica (QCD) è la teoria che descrive le \emph{in\-te\-ra\-zio\-ni forti} tra i quark, particelle costituenti di natura fermionica, e i gluoni, bosoni mediatori dell'interazione stessa. 

Nel limite di alte energie nel centro di massa $\sqrt s$, lo studio dei processi \emph{semiduri} (ovvero quei processi caratterizzati da scale dure molto maggiori della scala della QCD $\Lambda_{\rm QCD}$ ma, al contempo, notevolmente inferiori rispetto a $\sqrt s$) permette senza dubbio di effettuare prove stringenti della dinamica delle interazioni forti in regimi cinematici ad ora inesplorati. Nel limite di Regge ($s \gg |t|$, con $t$ la variabile di Mandelstam rappresentante il quadrato della quantità di momento trasferito), le predizioni teoriche di QCD perturbativa ad ordine fissato, basate sulla fattorizzazione collineare, non possono tener conto dell'effeto non trascurabile dei logaritmi in energia, il cui contributo è tale da compensare quello della costante d'accoppiamento della QCD $\alpha_s$ e, per tale ragione, deve essere tenuto in conto a tutti gli ordini dello sviluppo perturbativo. L'approccio Balitsky--Fadin--Kuraev--Lipatov (BFKL) rappresenta di certo lo strumento più potente in grado di risommare a tutti gli ordini il contributo di tali logaritmi, sia in approssimazione logaritmica dominante (LLA), ossia risommazione di tutti i termini proporzionali a $(\alpha_s\ln(s))^n$, sia in quella sottodominante (NLA), ossia risommazione dei fattori del tipo $\alpha_s(\alpha_s\ln(s))^n$. 

Il processo di produzione inclusiva ``in avanti'' di jet con alto momento trasverso e separati da un grande intervallo di rapidità, meglio noto come produzione di jet di \emph{Mueller--Navelet}, è, ad oggi, tra le reazioni più studiate. La ragione della sua popolarità in ambito scientifico risiede soprattutto nell'aver fornito la possibilità di definire i \emph{momenti di correlazione azimutale}, osservabili \emph{infrared-safe} le cui predizioni teoriche sono in buon accordo con i dati sperimentali ottenuti al Large Hadron Collider (LHC). 
\`E tuttavia necessario che nuove osservabili, sensibili alla dinamica BFKL, vengano considerate nell'ambito della fenomenologia di LHC. 

Perseguendo lo scopo di approfondire ed estendere la conoscenza della dinamica delle interazioni forti nel limite di Regge, si propone lo studio di quattro distinti processi semiduri. 

Nella prima parte dell'analisi fenomenologica presentata, ci si propone di continuare l'indagine di processi caratterizzati da due oggetti identificati nello stato finale, proseguendo lo studio dei problemi aperti nel processo di produzione di jet di Mueller--Navelet e, nello stesso tempo, affiancando ad esso 
quello della produzione inclusiva di una coppia adrone-antiadrone (\emph{dihadron} system) carico leggero del tipo $\pi^{\pm}, K^{\pm}, p,\bar p$, entrambi caratterizzati da alto momento trasverso e fortemente separati in rapidità. La possibilità di rivelare gli adroni ad LHC a valori del momento trasverso di gran lunga inferiori rispetto ai jet consente di esplorare un settore cinematico complementare a quello studiato attraverso il canale di Mueller--Navelet. La produzione di adroni offre, inoltre, la possibilità di investigare simultaneamente il comportamento di oggetti non perturbativi, quali le funzioni di distribuzione partonica (PDF) del protone nello stato iniziale e le funzioni di frammentazione (FF) caratterizzanti l'adrone rivelato nello stato finale.

Nella seconda parte della tesi, si pone e si evidenzia come lo studio della produzione di più jet nello stato finale 
(\emph{multi-jet} production) fornisca la possibilità di generalizzare le osservabili definite nel caso di processi con due oggetti nello stato finale, costruendone delle nuove, maggiormente sensibili alla dinamica BFKL a causa della loro dipendenza da momenti trasversi e rapidità dei jet rivelati nelle regioni centrali dei rivelatori. 
\`E presentata la prima analisi fenomenologica sulla produzione di tre jet, tenendo conto degli effetti dovuti all'inclusione delle correzioni d'ordine superiore in risommazione BFKL. 
Infine, facendo uso dello stesso formalismo, viene presentato il primo studio completo sulla produzione di quattro jet.

\end{otherlanguage*}

\renewcommand{\theequation}
             {\arabic{chapter}.\arabic{equation}}
\chapter{Introduction}
\label{chap:intro}

Insofar as a physical theory may appear complex and formally
arduous, the origin of its elegance lies almost always 
on a simple and concrete idea.
The Standard Model (SM) of elementary particles, solidly built up
on the existence of fermionic fundamental constituents, 
their mutual interaction being mediated via the exchange of intermediate vector bosons, represents one of the most significant examples. 
Inside the SM, Quantum Chromodynamics (QCD) is the theory of strong interactions, describing how fermionc quarks and bosonic gluons, the elementary constituents of hadrons~\footnote{From the ancient-greek word \greek{<adr\'os}, which means 'strong'.}, such as the proton and the neutron, interact with each other. 
What makes QCD a challenging sector surrounded by a broad and constant interest in its phenomenology, is the duality between non-perturbative and perturbative aspects 
which comes from the coexistence of two peculiar and concurrent properties, 
as \emph{confinement} and \emph{asymptotic freedom}. 
The striking feature of confinement is the increasing of the strong coupling 
$\alpha_s$ with distance. This means that hadrons are described by bound states of quarks and gluons, which cannot be described at the hand of any perturbative calculation. Conversely, the short-distance regime is ruled by asymptotic freedom, 
so that quarks and gluons behave as quasi-free particles, making it possible to use perturbative approaches.

The high luminosity and the record energies 
of the Large Hadron Collider (LHC) 
provide us with a wealth of useful data. 
A peerless opportunity to test strong interactions in this so far
unexplored kinematical configuration of large center-of-mass energy is given by the study of \emph{semi-hard} processes, \emph{i.e.} hard processes in the kinematical region where the center-of-mass energy squared $s$ is substantially larger than one or more hard scales $Q^2_i$ (large squared transverse momenta, large squared quark masses and/or $t$), $s\gg Q^2_i$, which satisfy in turn $Q^2_i \gg \Lambda_{\rm QCD}^2$, with $\Lambda_{\rm QCD}$ the QCD scale. 
In the kinematical regime~\footnote{Here $t$ represents the $t$-channel Mandelstam variable~\cite{Mandelstam:1958xc}.}, known as \emph{Regge limit} 
(see also Section~\ref{sec:bfkl-regge}), $s\gg |t|$,  
fixed-order calculations in perturbative QCD 
based on \emph{collinear factorisation}~\footnote{It is worth to remember that the factorisation theorem allows to write QCD cross sections as the convolution of a  hard process-dependent cross section with universal
parton distribution functions (PDFs) which are described by the Dokshitzer--Gribov--Lipatov--Altarelli--Parisi (DGLAP) evolution equation~\cite{DGLAP,DGLAP_2,DGLAP_3,DGLAP_4,DGLAP_5}.}
miss the effect of large energy logarithms, entering the perturbative series
with a power increasing with the order and thus compensating the smallness
of the coupling $\alpha_s$. The Balitsky--Fadin--Kuraev--Lipatov (BFKL)
approach~\cite{BFKL,BFKL_2,BFKL_3,BFKL_4} serves as the most powerful tool 
to perform the all-order resummation of these large energy logarithms 
both in the leading approximation (LLA), 
which means inclusion of all terms proportional to $(\alpha_s\ln(s))^n$,
and the next-to-leading approximation (NLA),
which means inclusion of all terms proportional to $\alpha_s(\alpha_s\ln(s))^n$.
In the BFKL formalism, it is possible to express the cross section of
an LHC process falling in the domain of perturbative QCD 
as the convolution between two impact factors, which describe 
the transition from each colliding proton to the respective 
final-state object, and a process-independent Green's function.
The BFKL Green's function obeys an integral equation, whose
kernel is known at the next-to-leading order (NLO) both for forward
scattering ({\it i.e.} for $t=0$ and colour singlet in the
$t$-channel)~\cite{Fadin:1998py,Ciafaloni:1998gs} 
and for any fixed (not growing with energy)
momentum transfer $t$ and any possible two-gluon colour state in the
$t$-channel~\cite{Fadin:1998jv,Fadin:2000kx,FF05,FF05_2}.

The too low $\sqrt{s}$, bringing to small rapidity intervals
among the tagged objects in the final state, 
had been so far the weakness point of the search for BFKL effects. 
Furthermore, too inclusive observables were considered. 
A striking example is the growth of the hadron structure functions 
at small Bjorken-$x$ values in Deep Inelastic Scattering (DIS).
Although NLA BFKL predictions for the structure function $F_{2,L}$ 
have shown a good agreement with the HERA 
data~\cite{Hentschinski:2012kr,Hentschinski:2013id}, 
also other approaches can fit these data. 
The LHC record energy, together with the good resolution 
in azimuthal angles of the particle detectors, can address these issues: 
on one side larger rapidity intervals in the final state are reachable, 
allowing us to study a kinematical regime where 
it is possible to disentangle the BFKL dynamics from other resummations;
on the other side, there is enough statistics 
to define and investigate more exclusive observables, 
which can, in principle, be only described by the BFKL framework.

With this aim, the production of two jets featuring transverse momenta much
larger than $\Lambda^2_{\rm QCD}$ and well separated in rapidity, 
known as \emph{Mueller--Navelet jets}, was proposed~\cite{Mueller:1986ey} 
as a tool to investigate semi-hard parton scatterings at a hadron collider. 
This reaction represents a unique venue where two main resummations, collinear
and BFKL ones, play their role at the same time in the context of perturbative
QCD. On one hand, the rapidity ranges in the final state are large enough 
to let the NLA BFKL resummation of the energy logarithms come into play. 
The process-dependent part of the information needed to build up 
the cross section is encoded in the impact factors
(the so-called ``jet vertices''), which are known 
up to NLO
~\cite{Fadin:1999de,Fadin:1999df,Ciafaloni:1998kx,Ciafaloni:1998hu,Bartels:2001ge,Bartels:2002yj,Caporale:2011cc,Ivanov:2012ms,Colferai:2015zfa}.
On the other hand, the jet vertex can be expressed, within collinear
factorisation at the leading twist, as the convolution of the PDF of the colliding proton, obeying the standard
DGLAP evolution, with the hard process describing the transition
from the parton emitted by the proton to the forward jet in the final state.

A large number of numerical analyses~\cite{Colferai:2010wu,Angioni:2011wj,Caporale:2012ih,Ducloue:2013wmi,Ducloue:2013bva,Caporale:2013uva,Ducloue:2014koa,Caporale:2014gpa,Ducloue:2015jba,Caporale:2015uva,Chachamis:2015crx} has appeared so far, devoted to NLA BFKL
predictions for the Mueller--Navelet jet production process. 
All these studies are involved in calculating cross sections 
and azimuthal angle correlations~\cite{DelDuca:1993mn,Stirling:1994he} 
between the two measured jets, {\it i.e.} average values of $\cos{(n \phi)}$, 
where $n$ is an integer and $\phi$ is the angle in the azimuthal plane
between the direction of one jet and the direction opposite to the other jet,
and ratios of two such cosines~\cite{Vera:2006un,Vera:2007kn}. 
Recently~\cite{Khachatryan:2016udy}, 
the CMS Collaboration presented the first measurements of the azimuthal 
correlation of the Mueller--Navelet jets at $\sqrt{s}=7$~{TeV} at the LHC.  
Further experimental studies of the Mueller--Navelet jets 
at higher LHC energies and larger rapidity intervals, including also the
effects of using {\it asymmetric} cuts for the jet transverse momenta, are
expected. 

In order to reveal the dynamical mechanisms behind partonic interactions
in the Regge limit, new observables, sensitive to the BFKL dynamics and
more exclusive than the Mueller--Navelet ones, need to be proposed and
considered in the next LHC analyses. 

A first step in this direction is the study of another reaction, less inclusive than Mueller--Navelet jets although sharing with it the theoretical framework, {\it i.e.} 
the inclusive detection of two charged light hadrons (a \emph{dihadron} system) $\pi^{\pm}, K^{\pm}, p,\bar p$ having high transverse momenta and well separated in rapidity.
Since the key ingredient beyond the NLA BFKL Green's function, {\it i.e.}
the process-dependent vertex describing the production of an identified hadron, was obtained
with NLA in~\cite{hadrons}, it is possible to study this process in the NLA BFKL approach. 
After the renormalisation of the QCD coupling 
and the ensuing removal of the ultraviolet divergences, soft and virtual 
infrared divergences cancel each other, whereas the  surviving  infrared 
collinear ones are compensated by the collinear counterterms related to 
the renormalisation of PDFs for the initial proton and 
parton fragmentation functions (FFs) describing the detected hadron in the 
final state within collinear factorisation. 
All the theoretical criteria are thus met to give
infrared-safe NLA predictions, thus making of this process an additional 
clear channel to test the BFKL dynamics at the LHC. The fact that hadrons 
can be detected at the LHC at much smaller values of the transverse
momentum than jets, allows to explore a kinematical range outside the
reach of the Mueller--Navelet channel, so that the reaction 
can be considered complementary to Mueller--Navelet jet production. Furthermore, it represents the best context to simultaneously constrain both PDFs and FFs.

The second advance towards further and deeply probing BFKL dynamics is the study of inclusive \emph{multi-jet} production processes where, besides two external jets typical of Mueller--Navelet reactions, the tagging of further jets in more central regions of the detectors and with a relative separation in rapidity from each other is demanded. This allows for the study of even more differential distributions in the transverse momenta, azimuthal angles and rapidities of the central jets, by generalising the two-jet azimuthal correlations $R_{nm} \equiv \cos{(n \phi)}/\cos{(m \phi)}$ to new, suitable BFKL observables sensitive to the azimuthal configurations of the tagged extra particles.

Aware of the importance to pursue the phenomenological paths traced above, we work toward the goal of giving testable predictions on the QCD semi-hard sector by proposing the study of four distinct processes. 
First, we continue the study of Mueller--Navelet jets by addressing issues that still wait to be answered, as the comparison of BFKL with NLO fixed-order perturbative approaches~\cite{Celiberto:2015yba} and a 13 TeV analysis, together with the study of the effect of imposing dynamic constraints in the central rapidity region~\cite{Celiberto:2016ygs}. 
Second, we will give the first phenomenological results for cross sections and azimuthal correlations in the inclusive dihadron production. 
Third, we will show how the inclusive three-jet production process allows to define in a very natural and elegant way new, generalised and suitable BFKL observables. 
Finally, we will investigate the inclusive four-jet production, extending the BFKL formalism defined and used in the three-jet case.

This thesis is organised as follows. In Chapter~\ref{chap:bfkl} a brief overview of the BFKL approach is given, while phenomenological predictions at LHC energies for the considered semi-hard processes are shown in the next four Chapters. 
In particular, Mueller--Navelet jets and the inclusive dihadron production are discussed in Chapters~\ref{chap:mn-jets} and~\ref{chap:dihadron}, respectively, showing the lastest results at full NLA accuracy, together with a study on the effect of using different values for the renormalisation and factorisation scales. 
In Chapter~\ref{chap:3j} the first complete analysis of the inclusive three-jet production process is presented, including the effects of higher-order BFKL corrections. The four-jet production process is investigated in Chapter~\ref{chap:4j}, giving the first results at LLA accuracy. In each Chapter devoted to phenomenology a related summary Section is provided, while the general Conclusions, together with Outlook, are drawn in Chapter~\ref{chap:conclusions}.

\renewcommand{\theequation}
             {\arabic{chapter}.\arabic{equation}}
\chapter{The BFKL resummation}
\label{chap:bfkl}

\section{The Regge theory}
\label{sec:bfkl-regge}

In 1959 the Italian physicist T.~Regge~\cite{Regge:1959mz} found that, when considering solutions of the Schr{\"o}\-din\-ger
equation for non-relativistic potential scattering, it can be advantageous to regard the angular
momentum, $l$, as a complex variable. He showed that, for a wide class of
potentials, the only singularities of the scattering amplitude in the complex $l$-plane are poles, called \emph{Regge poles}~\cite{Collins,Barone-Predazzi} after him. If these poles appear in coincidence with
integer values of $l$, they correspond to bound states or resonances and turn out to be important for the analytic properties of the amplitudes. They occur at the values given by the relation
\begin{equation}
\label{regge_trajectory}
 l = \alpha(k) \; ,
\end{equation}
where $\alpha(k)$ is a function of the energy, known as \emph{Regge trajectory} or \emph{Reggeon}. Each class of
bound states or resonances is related to a single trajectory like~(\ref{regge_trajectory}). The energies of these states are obtained from Eq.~(\ref{regge_trajectory}), giving physical (integer) values to the angular momentum $l$.
The extension of the Regge's approach to high-energy particle physics was formerly due to Chew and Frautschi~\cite{Chew:1961ev} and Gribov~\cite{Gribov:1961fr}, but many other physicists gave their contribution to the theory and its applications. 
Using the general properties of the $S$-matrix, the relativistic partial wave amplitude $\mathcal{A}_l(t)$ can be analytically continued to complex $l$ values in a unique way. 
The resulting function, $A(l,t)$, shows simple poles at 
\begin{equation}
\label{regge_poles}
 l = \alpha(t) \; .
\end{equation}
Each pole contributes to the scattering amplitude with a term which asymptotically behaves
(\emph{i.e.} for $s \to +\infty$ and for fixed $t$) as
\begin{equation}
\label{regge_asymptotic}
 A(s,t) \sim s^{\alpha(t)} \; ,
\end{equation}
where $s$ and $-t$ are the square of the center-of-mass energy and of the momentum
transfer, respectively.
The leading singularity in the
$t$-channel is the one with the largest real part, and rules the asymptotic behaviour of the scattering amplitude in the $s$-channel.
The triumph of Regge theory in its simplest form, \emph{i.e.} the fact that a large class of processes is accurately described by such simple predictions as Eq.~(\ref{regge_asymptotic}), was simply surprising.

\subsection{The Pomeron}
\label{sub:bfkl-pomeron}

Regge theory belongs to the class of the so-called $t$-channel models. They describe hadronic processes in terms of the exchange of ``some objects'' in the $t$-channel.
In the Regge theory a Reggeon plays the same role as an exchanged virtual particle in a tree-level perturbative process, with the important difference that the Reggeon represents a whole class of resonances, instead of a single particle.
In the limit of large $s$, a hadronic process is governed by the exchange of one or more Reggeons in the $t$-channel.
The exchange of Reggeons instead of particles gives rise to scattering amplitudes of the type of Eq.~(\ref{regge_asymptotic}).
Using the \emph{optical theorem}~\cite{optical_theorem_Newton} together with Eq.~(\ref{regge_asymptotic}), we can write the Regge total cross section:
\begin{equation}
\label{regge_cs}
 \sigma_{\rm TOT} \simeq 
 \frac{{\rm Im} \, A(s,t=0)}{s} \simeq
 s^{\alpha(0)-1} \; .
\end{equation}
We know from experiments that hadronic total cross sections, as a function of~$s$, are rather flat around $\sqrt{s} \simeq (10 \div 20)$ GeV and rise slowly as $\sqrt s$ increases.
If the considered process is described by the exchange of a single Regge pole, then it follows that the intercept $\alpha(0)$ of the exchanged Reggeon is greater than 1, leading to the power growth with energy of the cross section in Eq.~(\ref{regge_cs}).
This Reggeon is called \emph{Pomeron}, 
in honour of I.Ya.~Pomeranchuk. 
Particles which would provide the resonances for integer values of $\alpha(t)$ for $t>0$ have not been conclusively identified. 
Natural candidates in QCD are the so-called \emph{glueballs}. The Pomeron trajectory represents the dominant trajectory in elastic and diffractive processes, namely
reactions featuring the exchange of vacuum quantum numbers in the $t$-channel.
The power growth of the cross section violates the Froissart bound~\cite{Froissart:1961ux} and hence the unitarity, which has to be restored through unitarisation techniques (see for instance Ref.~\cite{Forshaw-Ross} and references therein).

\section{Towards the BFKL equation}
\label{sec:bfkl-equation-intro}

The BFKL equation~\cite{BFKL,BFKL_2,BFKL_3,BFKL_4} made the grade when the growth of the $\gamma^*p$ cross section at increasing energy, predicted by Balitsky, Fadin, Kuraev,
and Lipatov, was experimentally confirmed at HERA. 
Therefore this equation
is usually associated with the evolution of the unintegrated gluon distribution.

The PDF evolution with 
$\tau = \ln\left(Q^2/\Lambda_{\rm QCD}\right)$ is determined by
the DGLAP equations~\cite{DGLAP,DGLAP_2,DGLAP_3,DGLAP_4,DGLAP_5}, which allow to resum to all orders collinear logarithms $\ln Q^2$ picked up from the region of small angles between parton momenta.
There is another class of logarithms to be taken into account: soft logarithms which originate from ratios of
parton energies and are present both in PDFs and
in partonic cross sections. At small values 
of the ratio $x = \ln Q^2/s$ soft logarithms
are even larger than collinear ones.

The BFKL approach describes QCD scattering amplitudes in the limit of small $x$, $s \gg |t|$, and $t$ not growing with $s$ (Regge limit).
The evolution equation for the
unintegrated gluon distribution appears in this approach as a particular result
for the imaginary part of the forward scattering amplitude ($t=0$ and vacuum
quantum numbers in the $t$-channel). This approach was developed (and is more suitable) for the description of processes with just one hard scale, such as $\gamma^*\gamma^*$
scattering with both photon virtualities of the same order, where the DGLAP
evolution is not appropriate.
The BFKL approach relies on \emph{gluon Reggeisation}, which can be described as the appearance of a modified propagator in the Feynman gauge, of
the form~\cite{Barone-Predazzi}
\begin{equation}
\label{regge_propagator}
 D_{\mu\nu}(s,q^2)=
 -i\frac{g_{\mu\nu}}{q^2}
  \left(\frac{s}{s_0}\right)^{\alpha_g(q^2)-1} \; ,
\end{equation}
where $\alpha_g(q^2)=1+\epsilon(q^2)$ is the gluon Regge trajectory.

\subsection{Gluon Reggeisation}
\label{sub:bfkl-gluon-Reggeisation}

The Reggeisation of an elementary particle featuring spin $j_0$ and mass $m$ was introduced in Ref.~\cite{Gell-Mann:ep_III} and it means~\cite{IFL} that, in the Regge limit, a factor $s^{j(t)-j_0}$, with $j_0 \equiv j(m^2)$ appears in Born amplitudes with exchange of this particle in the $t$-channel. This phenomenon was discovered originally in QED via the backward Compton scattering~\cite{Gell-Mann:ep_III}. It was called Reggeisation
because just such form of amplitudes is given by the Regge poles (moving poles in the complex $l$-plane~\cite{Regge:1959mz}).
In contrast to QED, where the electron reggeises in perturbation theory~\cite{Gell-Mann:ep_III},
but the photon remains elementary~\cite{Mandelstam:PR137B949:1965}, in QCD the gluon reggeises~\cite{BFKL,BFKL_2,Grisaru:1973vw,Grisaru:1974cf,Lipatov:1976zz} as well as the quark~\cite{Fadin:1976nw,Fadin:1977jr,Bogdan:2002sr,Kotsky:2002aq}. Therefore QCD is the unique theory where all elementary particles reggeise.

Reggeisation represents a key-ingredient for the theoretical description of high-energy processes with fixed momentum transfer. Gluon Reggeisation is particularly important, because cross sections non-decreasing with energy are provided by gluon exchanges, and it determines the form of QCD amplitudes 
in the Regge limit.
The simplest realisation of the gluon Reggeisation happens in the elastic process $A + B \to A^\prime + B^\prime$, 
\begin{figure}[t]
  \centering
    \includegraphics[scale=1.00]{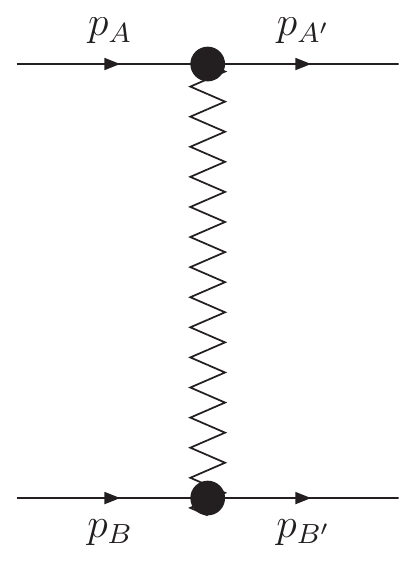}
  \caption[Reggeised gluon exchange ] 
  {Diagrammatical representation of the process 
   $A + B \to A^\prime + B^\prime$ with
   colour octet in the $t$-channel. The waggle line represents the Reggeised gluon exchange.}
\label{fig:regge_gluon}
\end{figure}
where amplitudes with a colour-octet exchange in the $t$-channel and negative signature (see Fig.~\ref{fig:regge_gluon} for a diagrammatical representation) assume
the form
\begin{equation}
\label{octet_amplitude}
 \left(\mathcal{A}_8\right)^{A^\prime B^\prime}_{AB} =
 \Gamma^c_{A^\prime  A}\frac{s}{t} 
 \left[\left(\left(\frac{s}{-t}\right)^{\omega(t)} + 
             \left(\frac{-s}{-t}\right)^{\omega(t)}
 \right)\right]
 \Gamma^c_{B^\prime B} \; ,
\end{equation}
where
\begin{align}
&
s = (p_A + p_B)^2 \; ,
\\ \nonumber &
t = q^2 \; , 
\\ \nonumber &
q = p_A - p_{A^\prime} \; ,
\end{align}
$\omega(t)$ is the gluon trajectory, $c$ is the colour index, and $\Gamma^c_{B^\prime B}$ are the
particle-particle-Reggeon (PPR) vertices which are independent of $s$.
The factorisation given in Eq.~(\ref{octet_amplitude}) represents the analytical structure of the scattering
amplitude, which is quite simple in the elastic case. It is valid both in the LLA and in the NLA. 
In particular, it holds when one of the particles $A^\prime$ and $B^\prime$ is replaced by a jet.
In general the PPR vertex can be written in the form
$\Gamma^C_{P^\prime P}=g_s\left\langle P^\prime|T^c|P \right\rangle\Gamma_{P^\prime P}$,
where $g_s$ is the QCD coupling and $\left\langle P^\prime|T^c|P \right\rangle$ is the matrix element of the
colour-group generator in the fundamental (adjoint) representation for quarks (gluons).
In the LLA this form of amplitude was proved in Refs.~\cite{BFKL,BFKL_2,BFKL_3,BFKL_4}.
In this approximation, the helicity $\lambda_p$ of the scattered particle $P$ is a conserved quantity,
so $\Gamma^{(0)}_{P^\prime P}$
is given by $\delta_{\lambda_{P^\prime}\lambda_P}$ and the Reggeised gluon trajectory is calculated with
1-loop accuracy~\cite{Fadin:1998sh}, having so
\begin{equation}
\label{trajectory_1-loop}
 \omega(t) \simeq \omega^{(1)}(t) = 
 \frac{g_s^2 t}{(2\pi)^{(D-1)}} \frac{N_c}{2} 
 \int \frac{d^{D-2} k_\bot}
           {k_\bot^2 (q-k_\bot)^2_\bot} 
\end{equation}
\[ =
 -\frac{g_s^2 N_c \Gamma(1-\epsilon)}
       {(4\pi)^\frac{D}{2}}
 \frac{\Gamma^2(\epsilon)}
      {\Gamma(2\epsilon)}
 \left(\vec{q}^2\right)^\epsilon
\]
where $t = q^2 \approx q_\bot^2$, $D = 4+2$ is the space-time dimension and $N_c$ is the number
of QCD colours. The $\epsilon$ parameter has been introduced in order to regularise the
infrared divergences, while the integration is done in a $(D-2)$-dimensional space, orthogonal to the momenta of the initial colliding particles $p_A$ and $p_B$.
The gluon Reggeisation determines also the form of inelastic amplitudes in the multi-Regge kinematics
(MRK), namely where all particles are strongly ordered in the rapidity
space with limited transverse momenta and the squared invariant masses $s_{ij} = (k_i + k_j)^2$ of any pair of produced particles $i$ and $j$ are large and increasing with $s$.
This kinematics gives the leading contribution to QCD cross sections.
In the LLA, there are exchanges of vector particles (QCD gluons) in all channels.
In the NLA, as opposed to LLA, MRK is not the solely contributing kinematic configuration.
It can happen then one (and just one) of the produced particles can have a fixed (not increasing
with $s$) invariant mass, \emph{i.e.} components of this pair can have rapidities of
the same order. This is known as quasi-multi-Regge-kinematics (QMRK)~\cite{Fadin:1989kf}.
In the NLA the expression given in Eq.~(\ref{octet_amplitude}) was checked initially 
at the first three perturbative orders~\cite{Fadin:1995dd,Fadin:1995km,Kotsky:1996xm,Fadin:1995xg,Fadin:1996tb}. A rigorous proof of gluon Reggeisation, based on some stringent self-consistency conditions (\emph{bootstrap} conditions ~\cite{Fadin:2006bj,Kozlov:2011zza,Kozlov:2012zza}), was subsequently given with full NLA accuracy.

\section{The amplitude in multi-Regge kinematics}
\label{sec:bfkl-amplitude}

\begin{figure}[t]
  \centering
    \includegraphics[scale=1.00]{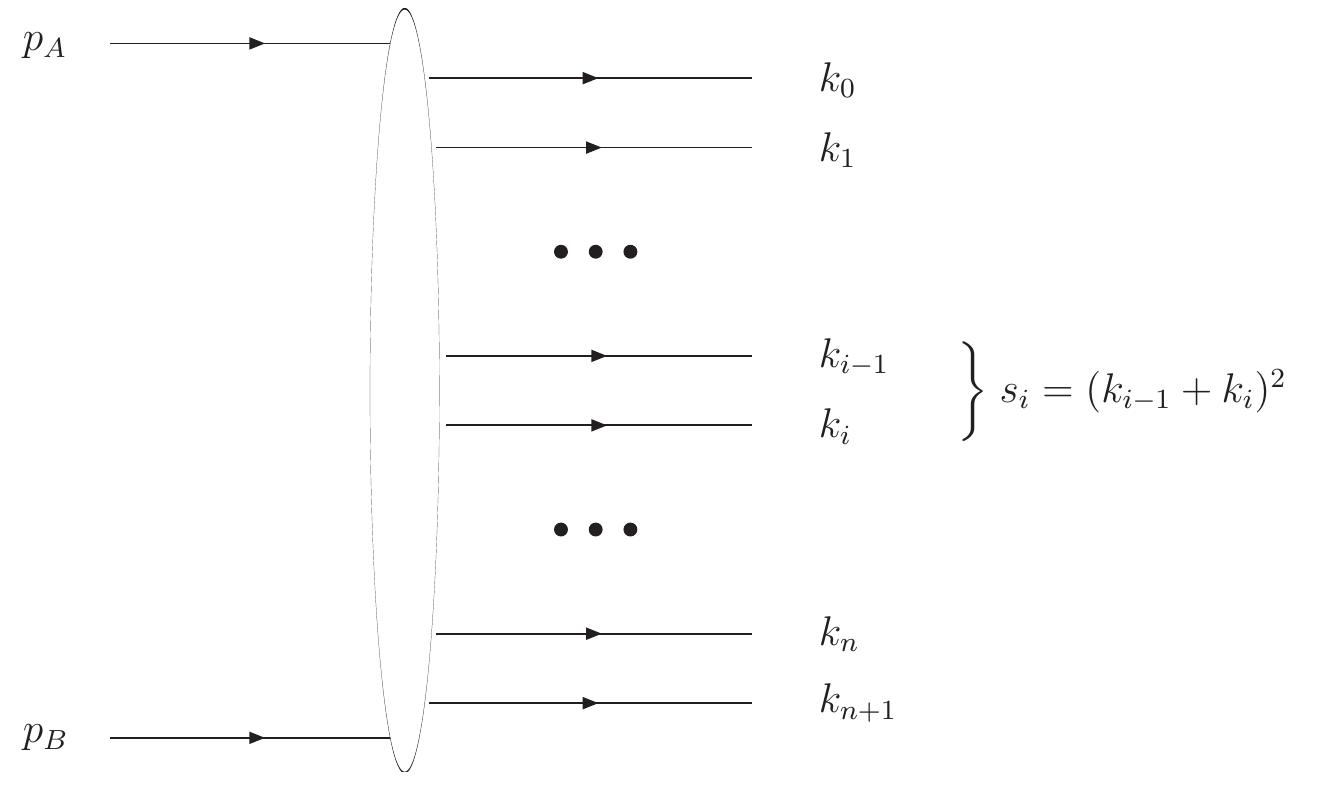}
  \caption[Diagrammatic representation 
           of the $s$-channel unitarity relation]
  {Diagrammatic representation 
   of the $s$-channel unitarity relation.}
\label{fig:s_channel_unitarity}
\end{figure}

The gluon Reggeisation governs amplitudes with colour-octet states and negative signature in the $t$-channel. In the BFKL approach, amplitudes with other quantum numbers
can be obtained by using $s$-channel unitarity relations, where the contribution of order $s$ is given by the MRK. Large logarithms come from the integration over longitudinal momenta of the final-state particles.
In an elastic process $A + B \to A^\prime B^\prime$, according to the Cutkosky rule~\cite{Cutkosky:1960sp} and to the unitarity relation in the $s$-channel, the imaginary
part of the elastic scattering amplitude 
$\mathcal{A}^{A^\prime B^\prime}_{AB}$ can be presented as 
\begin{equation}
\label{s_unitarity_amplitude}
 {\rm Im}_s \mathcal{A}^{A^\prime B^\prime}_{AB}
 = \frac{1}{2} \sum_{n=0}^{\infty} \sum_{\{f\}} 
 \int 
 \left|\mathcal{A}^{\tilde{A} \tilde{B} n}_{AB}\right|^2 
 d\Phi_{\tilde{A} \tilde{B} n} \; ,
\end{equation}
where $\mathcal{A}^{\tilde{A} \tilde{B} n}_{AB}$ is the amplitude for the production of $n+2$ particles 
(see Fig.~\ref{fig:s_channel_unitarity}) with momenta $k_i$, $i = 0,1,\dots,n,n+1$ in the process 
$A + B \to \tilde{A} + \tilde{B} + n$, while $d\Phi_{\tilde{A} \tilde{B} n}$ represents the intermediate phase-space element and $\sum_{\{f\}}$ is over the discrete quantum numbers $\{f\}$ of the intermediate particles.
The initial particle momenta $p_A$ and $p_B$ are assumed to be equal to
$p_A=p_1+(m^2_A/s)p_2$ and 
$p_B=p_2+(m^2_B/s)p_1$, respectively. For any momentum $k_i$ the Sudakov decomposition is satisfied by the relation
\begin{equation}
\label{sudakov_general}
 k_i = \beta_ip_1+\alpha_ip_2+k_{i\bot} \; ,
\end{equation}
where $p_1$ and $p_2$ are light-like vectors and 
$(p_1+p_2)^2=2p_1\cdot p_2=s$,
\begin{equation}
\label{sudakov_alpha-beta}
 \alpha_i \beta_i = 
 \frac{k_i^2-k_{i\bot}^2}{s} =
 \frac{k_i^2+\vec{k}_i^2}{s}\; ,
\end{equation}
with $\vec{k}_{i\bot}$ transverse component with respect to the plane generated by $p_1$ and $p_2$, and $k_{i\bot}^2=-\vec{k}_i^2$.

\begin{figure}[t]
  \centering
    \includegraphics[scale=0.80]{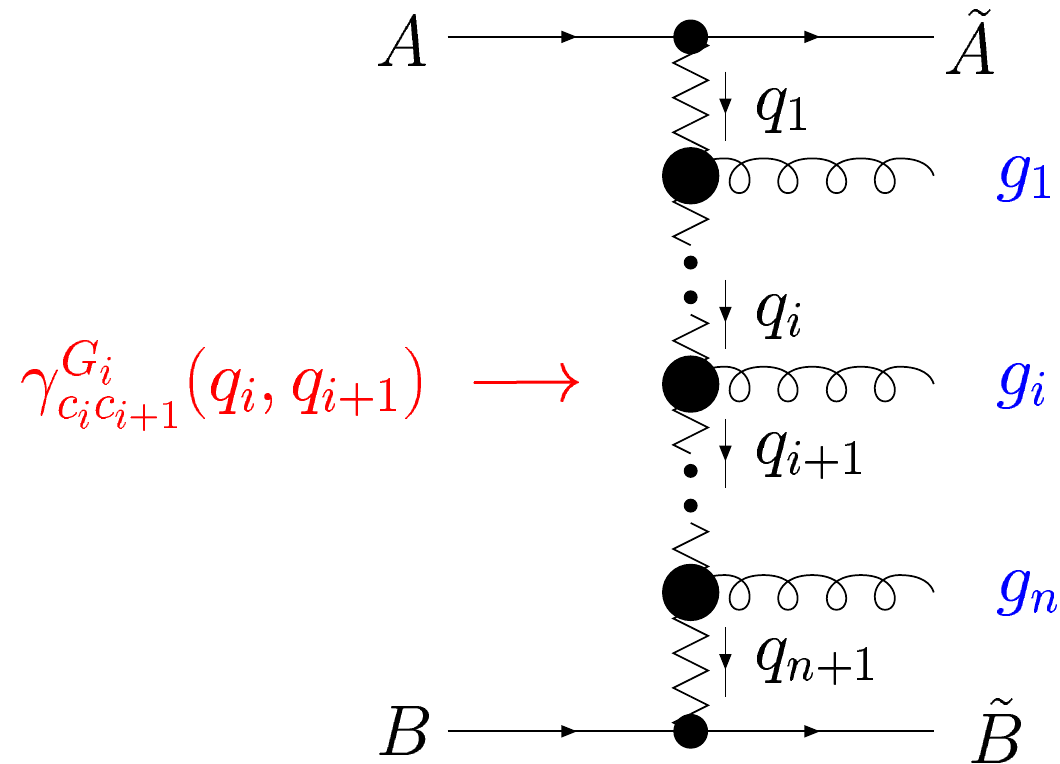}
  \caption[Diagrammatic representation 
           of the amplitude $\mathcal{A}^{\tilde{A} \tilde{B} n}_{AB}$]
  {Diagrammatic representation 
   of the amplitude $\mathcal{A}^{\tilde{A} \tilde{B} n}_{AB}$.}
\label{fig:bfkl_mrk_amplitude}
\end{figure}

The Sudakov decomposition allows us to write the following expression for the phase space:
\begin{equation}
\label{sudakov_ps}
 d\Phi_{\tilde{A} \tilde{B} n} = 
 \frac{2}{s} (2\pi)^D
 \delta\left(1+\frac{m_A^2}{s}-\sum_{i=0}^{n+1}\alpha_i\right)
 \delta\left(1+\frac{m_B^2}{s}-\sum_{i=0}^{n+1}\beta_i\right)
\end{equation}
\[ \times \, 
 \delta^{D-2}\left(\sum_{i=0}^{n+1}k_{i\bot}\right)
 \frac{d\beta_{n+1}}{2\beta_{n+1}}
 \frac{d\alpha_0}{2\alpha_0}
 \prod_{i=0}^{n} \frac{d\beta_i}{2\beta_i}
 \prod_{i=1}^{n+1}\frac{d^{D-2}k_{i\bot}}{(2\pi)^{D-1}} \; , 
\]
where 
$p_{\tilde{A}} = k_0$ ; $p_{\tilde{B}} = k_{n+1}$.
In the unitarity condition (Eq.~(\ref{s_unitarity_amplitude})), the dominant contribution ($\sim \! s$) in the
LLA is given by the region of limited (not growing with $s$) transverse momenta of produced particles.
As we said, large logarithms come from the integration over longitudinal momenta of the produced particles. In particular, we have a logarithm of $s$ for every
particle produced according to MRK. By definition, in this kinematics transverse momenta of the produced
particles are limited and their Sudakov variables $\alpha_i$ and $\beta_i$ are strongly ordered
in the rapidity space, having so
\begin{align}
\label{sudakov_order}
 &
 \alpha_{n+1} \gg \alpha_n \dots \gg \alpha_0 \; , \\ \nonumber &
 \beta_0 \gg \beta_1 \dots \gg \beta_{n+1} \; .
\end{align}
Eqs.~(\ref{sudakov_general}) and (\ref{sudakov_order}) ensure the squared invariant masses of neighbouring particles,
\begin{equation}
\label{sudakov_invariant}
 s_i = (k_{i-1} + k_i)^2 \approx 
 s\beta_{i-1}\alpha_i=\frac{\beta_{i-1}}{\beta_i}
 \left(k_i^2+\vec{k}_i^2\right) \; ,
\end{equation}
to be large with respect to the squared transverse momenta:
\begin{equation}
 s_i \gg \vec{k}_i^2 \sim \left|t_i\right| = q_i^2 \; ,
\end{equation}
with
\begin{equation}
 t_i = q_i^2 \approx q_{i\bot}^2 = - \vec{q}_i^2
\end{equation}
and
\begin{equation}
 \prod_{i=1}^{n+1}s_i =
 s \prod_{i=1}^n \left(k_i^2+\vec{k}_i^2\right) \; .
\end{equation}
In order to obtain the large logarithm from the integration over $\beta_i$ for each
produced particle in the phase space given in Eq.~(\ref{sudakov_ps}), the amplitude in the r.h.s. in Eq.~(\ref{s_unitarity_amplitude})
must not decrease with the growth of the invariant masses. This is true only when there are exchanges of vector particles (gluons) in all channels with momentum transfers 
$q_{i=1,\dots,n+1}$ with
\begin{equation}
 q_i = p_a - \sum_{j=0}^{i-1}k_j = 
 - \left(p_B - \sum_{l=i}^{n+1}k_l\right)  
\end{equation}
\[ \simeq
 \beta_ip_1-\alpha_{i-1}p_2-\sum_{j=0}^{i-1}k_{j\bot} 
\]
and
\begin{equation}
 q_i^2 \simeq q_{i\bot}^2 = - \vec{q}_i^2  
\end{equation}
The dominant amplitudes at every expansion order can be diagrammatically represented
as in Fig.~\ref{fig:bfkl_mrk_amplitude}. Multi-particle amplitudes show a complicated analytical
structure even in MRK (see Refs.~\cite{Bartels:1974tj,Bartels:1974tk,Bartels:1980pe,Fadin:1993wh}). Fortunately, only real parts of these amplitudes are used in the BFKL
approach in NLA as well as in LLA. Considering just the real parts, it is possible to write~\cite{Fadin:1998fv}
\begin{equation}
\label{bfkl_real_amplitude}
 \mathcal{A}^{\tilde{A} \tilde{B} n}_{AB} = 
 2s \Gamma_{\tilde{A}A}^{c_1} 
 \left(\prod_{i=1}^n\gamma_{c_ic_{i+1}}^{P_i}(q_i,q_{i+1})\left(\frac{s_i}{s_R}\right)^{\omega(t_i)}\frac{1}{t_i}\right)
\end{equation}
\[ \times \,
 \frac{1}{t_{n+1}}\left(\frac{s_{n+1}}{s_R}\right)
 ^{\omega(t_{n+1})}
 \Gamma_{\tilde{B}B}^{c_{n+1}}
\]
with $s_R$ being an arbitrary energy scale, irrelevant at LLA. Here $\omega(t)$ and $\Gamma_{P^\prime P}^a$ are
the gluon Regge trajectory and the PPR (see Eq.~(\ref{s_unitarity_amplitude})), while 
$\gamma_{c_ic_{i+1}}^{P_i}$ are the Reggeon-Reggeon-Particle (RRP) vertices, \emph{i.e.} the effective vertices for the
production of particles $P_i$ with momenta $q_i-q_{i+1}$ in collisions of Reggeised gluons with momenta $q_i$ and $-q_{i+1}$ and colour indices $c_i$ and
$c_{i+1}$, respectively.
In the LLA only one gluon can be produced in the RRP
vertex. For this reason, final-state particles are massless. The
Reggeon-Reggeon-Gluon (RRG) vertex takes the form~\cite{BFKL,BFKL_2,BFKL_3,BFKL_4}
\begin{equation}
\label{rrg_vertex}
 \gamma_{c_ic_{i+1}}^{G_i}(q_i,q_{i+1}) = 
 g_sT_{c_ic_{i+1}}^{d_i}e_{\mu}^{\ast}(k_i)
 C^{\mu}(q_{i+1},q_i) \; ,
\end{equation}
where $T_{c_ic_{i+1}}^{d_i}$ are the matrix elements of the $SU(N_c)$ group generators in the adjoint representation, $d_i$ is the colour index of the produced gluon with polarisation vector $e_{\mu}^{\ast}(k_i)$, $k_i=q_i-q_{i+1}$ its momentum and 
\begin{equation}
\label{rrg_current}
 C^{\mu}(q_{i+1},q_i) = 
 - q_i^{\mu} - q_{i+1}^{\mu} + p_1^{\mu} 
 \left(\frac{q_i^2}{k_i \cdot p_1} + 
      2\frac{k_i \cdot p_2}{p_1 \cdot p_2}\right)
\end{equation}
\[ \times \,
 - p_2 ^{\mu} 
  \left(\frac{q_{i+1}^2}{k_i \cdot p_2} + 
      2\frac{k_i \cdot p_1}{p_1 \cdot p_2}\right) \; .
\]
The structure of $C^{\mu}$ given in Eq.~(\ref{rrg_current}) reflects the current conservation property $(k_i)_\mu C^{\mu} = 0$,
which permits to choose an arbitrary gauge for each of the produced gluons. 
%
Let us introduce now the following decomposition: 
\begin{equation}
\label{rrg_decomposition}
 T_{c_ic_{i+1}}^{d_i}
 \left(T_{c_ic_{i+1}}^{d_i}\right)^{\ast} = 
 \sum_R c_R \left\langle c_i c_i^\prime \left| 
             \hat{\mathcal{P}}_R \right|
             c_{i+1} c_{i+1}^\prime \right\rangle         
\end{equation}
where $\hat{\mathcal{P}}_R$ is the projection operator of the two-gluon colour states on the irreducible representation $R$ of the colour group.
For the singlet (vacuum) and antisymmetric octet (gluon) representations one has respectively 
\begin{equation}
 \left\langle c_i c_i^\prime \left| 
             \hat{\mathcal{P}}_0 \right|
             c_{i+1} c_{i+1}^\prime \right\rangle 
 = \frac{\delta_{c_i c_i^\prime}
         \delta_{c_{i+1} c_{i+1}^\prime}}
        {N_c^2 - 1}
\end{equation}
and
\begin{equation}
 \left\langle c_i c_i^\prime \left| 
             \hat{\mathcal{P}}_8 \right|
             c_{i+1} c_{i+1}^\prime \right\rangle 
 = \frac{f_{a c_i c_i^\prime}
         f_{a c_{i+1} c_{i+1}^\prime}}
        {N_c} \; ,
\end{equation}
where $f_{abc}$ are the $(SU N_c)$ structure constants. 
It is possible to prove that
\begin{equation}
 c_0 = N_c \;, \quad\quad 
 c_8 = \frac{N_c}{2} \; .
\end{equation}
Using the decomposition given in Eq.~(\ref{rrg_decomposition}), we can write
\begin{equation}
 \sum_{G_i} \gamma_{c_ic_{i+1}}^{G_i}(q_i,q_{i+1}) 
 \left( 
  \gamma_{c_ic_{i+1}}^{G_i}(q_i,q_{i+1})
 \right)^{\ast} 
\end{equation}
\[ =
 2(2\pi)^{D-1} \sum_R 
 \left\langle c_i c_i^\prime \left| 
             \hat{\mathcal{P}}_R \right|
             c_{i+1} c_{i+1}^\prime \right\rangle
 \mathcal{K}_r^{(R)}(\vec{q}_i,\vec{q}_{i+1};\vec{q})
\]
where the sum is taken over colour and polarisation states of the produced gluon
and $\mathcal{K}_r^{(R)}(\vec{q}_i,\vec{q}_{i+1};\vec{q})$ is the so-called \emph{real part} of the kernel.

\subsection{The BFKL equation}
\label{sub:bfkl-equation}

\begin{figure}[t]
  \centering
    \includegraphics[scale=1.00]{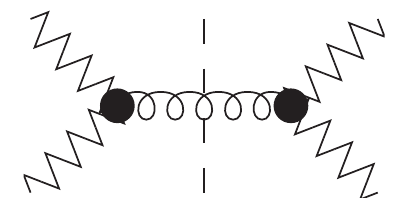}
  \caption[real part of the BFKL kernel at the Born level]
  {Diagrammatic representation of the real part of the BFKL kernel 
   at the Born level.}
\label{fig:born_real_kernel}
\end{figure}

The BFKL equation at LLA is obtained from the amplitude given in 
Eq.~(\ref{bfkl_real_amplitude}), using the unitarity relation 
(see Eq.~(\ref{s_unitarity_amplitude})) for the $s$-channel imaginary part of the elastic amplitude, which,  
according to the decomposition in Eq.~(\ref{rrg_decomposition}) 
can be written as
\begin{equation}
 \mathcal{A}^{A^\prime B^\prime}_{AB} = 
 \sum_R\left(\mathcal{A}_R\right)^{A^\prime B^\prime}_{AB} \; ,
\end{equation}
where $\left(\mathcal{A}_R\right)^{A^\prime B^\prime}_{AB}$ 
is the part of the scattering amplitude corresponding to a definite
irreducible representation $R$ of the colour group in the $t$-channel.
Using the amplitude~(\ref{bfkl_real_amplitude}) in the unitarity relation~(\ref{s_unitarity_amplitude}) for the $s$-channel imaginary
part of the elastic scattering amplitude, one obtain an expression
\begin{figure}[t]
  \centering
    \includegraphics[scale=1.00]{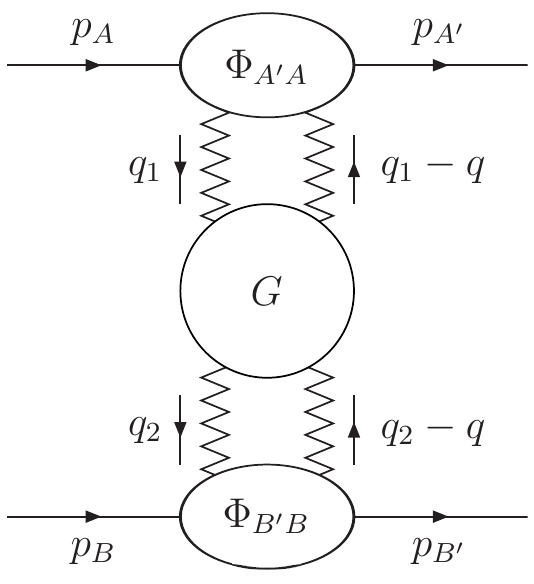}
  \caption[Diagrammatic representation of the BFKL amplitude]
  {Diagrammatic representation of the imaginary part  
   of the BFKL amplitude.}
\label{fig:bfkl_elastic_amplitude}
\end{figure}
which can be factorised~\cite{Fadin:1998sh} in the following way 
(see Fig.~\ref{fig:bfkl_elastic_amplitude}):
\begin{equation}
\label{eq_factorisation}
 {\rm Im} \, \left(\mathcal{A}_R\right)^{A^\prime B^\prime}_{AB} =
 \frac{s}{(2\pi)^{D-2}}
 \int\frac{d^{D-2}q_1}{\vec{q}_1^2 (\vec{q}_1-\vec{q})^2}
 \int\frac{d^{D-2}q_2}{\vec{q}_2^2 (\vec{q}_2-\vec{q})^2}
\end{equation}
\[ \times \,
\sum_{\nu} \Phi_{A^\prime A}^{(R,\nu)}(\vec{q}_1;\vec{q},s_0)
\int_{\delta-i\infty}^{\delta+i\infty}\frac{d\omega}{2\pi i}
\left[\left(\frac{s}{s_0}\right)^\omega G_{\omega}^{(R)}(\vec{q}_1,\vec{q}_2,\vec{q})\right] \Phi_{B^\prime B}^{(R,\nu)}(-\vec{q}_2;-\vec{q},s_0) \; .
\]
Here $\vec{q}_1$ and $\vec{q}_2$ are the transverse momenta of the Reggeised gluons, while $s_0$ is an arbitrary energy scale introduced in order to define
the partial wave expansion of the scattering amplitudes via the (inverse) \emph{Mellin transform} (see Appendix~\hyperlink{app:mellin-link}{A} for further details), while the $\nu$ index identifies the state in the irreducible representation $R$. 
$\Phi_{P^\prime P}^{(R,\nu)}$ are the so-called impact factors, obtained through the convolution of two PPR vertices. 
$G_{\omega}^{(R)}$, defined via a Mellin transform, is the Green's function for scattering of two Reggeised gluons and is universal (it does not depend on the particular process). 
Conversely, the impact factors are specific of the particles on the external lines and can be expressed through the imaginary part of the particle-Reggeon scattering amplitudes, in the form
\begin{equation}
\label{impact_factors_general}
 \Phi_{P^\prime P}^{(R,\nu)}(\vec{q}_R;\vec{q};s_0) =
 \int \frac{ds_{PR}}{2\pi s}
 {\rm Im}  \, 
 \mathcal{A}_{P^\prime P}^{(R,\nu)}(p_P,q_R;\vec{q};s_0)
 \theta(s_{\Lambda}-s_{PR})
\end{equation}
\[
 -\frac{1}{2}\int\frac{d^{D-2}q^\prime}
                      {\vec{q}^{\prime 2} (\vec{q}^\prime  
                                        -\vec{q}^\prime)^2}
 \Phi_{P^\prime P}^{(R,\nu)B}(\vec{q}^\prime,\vec{q})
 \mathcal{K}_r^{(R)B}(\vec{q}^\prime,\vec{q}_R)
 \ln\left(\frac{s_{\Lambda}^2}{(\vec{q}^\prime-\vec{q}_R)s_0}\right) \; ,
\]
where $s_{PR}=(q_P-q_R)^2$ is the squared particle-Reggeon invariant mass and ${\rm Im} \mathcal{A}_{P^\prime}^{(R,\nu)}$ 
is the $s_{PR}$-channel imaginary part of the scattering amplitude of the
particle $P$ with momentum $p_P$ off the Reggeon with momentum $q_R$, while $q$ is the transferred momentum. This definition is valid both in the LLA and in the NLA.
The parameter $s_{\Lambda}$, which plays the role of a cutoff for the $s_{PR}$-integration, is introduced to separate the
contributions from MRK and QMRK and must be
considered in the limit $s_{\Lambda} \to +\infty$. In this way, the second term in the r.h.s. of Eq.~(\ref{impact_factors_general}) works as a counterterm for the large $s_{PR}$.
\begin{figure}[t]
  \centering
    \includegraphics[scale=1.00]{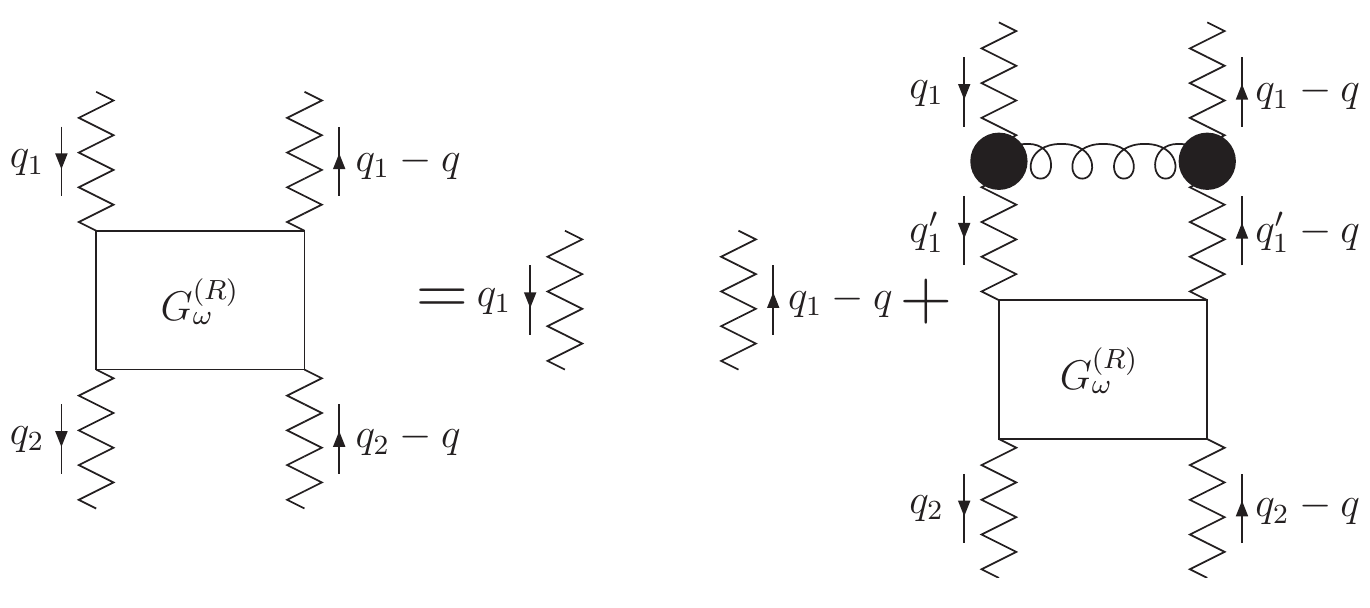}
  \caption[Generalised BFKL integral equation]
  {Diagrammatic representation of the generalised BFKL integral equation.}
\label{fig:kernel_equation}
\end{figure}
The Green's function obeys the following integral equation (Fig.~\ref{fig:kernel_equation}), known as \emph{generalised BFKL equation}:
\begin{equation}
\label{bfkl_kernel_equation}
 \omega G^{(R)}_{\omega}({\vec q}_1, {\vec q}_2; {\vec q}) =
 \vec q_1^{~2}({\vec q}_{1}-{\vec q})^2\delta^{(D-2)}
 ({\vec q}_1- {\vec q}_2)
\end{equation}
\[
 +\int \frac{d^{D-2}q_{1\perp}^{~\prime}}
{\vec q_1^{~\prime 2}({\vec q}_{1}^{~\prime}-{\vec q})^2 }
{\cal K}^{(R)}({\vec q}_1, {\vec q}_1^{~\prime}; {\vec q})
G^{(R)}_{\omega}({\vec q}_1^{~\prime}, {\vec q}_2; {\vec q}) \; ,
\]
where the kernel
\begin{equation}
\label{bfkl_kernel_parts}
{\cal K}^{(R)}({\vec q}_1, {\vec q}_2; {\vec q}) =(\omega(q_{1\perp}^2)+
\omega((q_1-q)_{\perp}^2))\vec q_1^{~2}({\vec q}_{1}-{\vec q})^2
\delta^{(D-2)}({\vec q}_1- {\vec q}_2)
\end{equation}
\[
+{\cal K}^{(R)}_r({\vec q}_1, {\vec q}_2; {\vec q})
\]
consists of two parts: the first one is the so-called \emph{virtual part} and 
is expressed in terms of the gluon Regge trajectory; the second one, known as \emph{real part} (see Fig.~\ref{fig:born_real_kernel}), is related to the real particle production and reads: 
\begin{equation}
\label{bfkl_kernel_real}
{\cal K}_r^{(R)}(\vec q_i,\vec q_{i+1};\vec q )=
-\frac{g^2c_R}{2(2\pi )^{D-1}}
C^{\mu }(q_{i+1},q_i) C_{\mu }(q_{i+1}-q,q_i-q)
\end{equation}
\[
=
\frac{g^2c_R}{(2\pi )^{D-1}}
\left(\frac{{\vec q}_i^2{(\vec q_{i+1} -\vec q)}^2
+{\vec q}_{i+1}^2{(\vec q_{i} -\vec q)}^2}{({\vec q}_i-{\vec q}_{i+1})^2}
-\vec q^{~2} \right) \; .
\]
If $R=0$ (colour-singlet representation) the Eq.~(\ref{bfkl_kernel_equation}) is called 
\emph{BFKL equation}.
The BFKL equation iterative: knowing the kernel
in the Born level, it permits to get all the LLA terms of the Green's function. Similarly, knowing all the NLA corrections to the gluon trajectory and to the real part of the kernel, one can get all the NLA terms of the Green's function.
In order to obtain a full amplitude the impact factors are needed, which depend on the process though and have to be calculated time by time at the requested perturbative accuracy. Furthermore, in most cases impact factors encode non-perturbative objects for real processes, \emph{e.g.} the PDF of the of the parton emitted from the initial state parent hadron and/or the FF describing the detected hadron in the final state within collinear factorisation (in the case of processes with identified particles in the final state). 
Impact factors are known in the NLA just for few processes.

\subsection{The BFKL equation in the NLA}
\label{sub:bfkl-equation-nla}

\begin{figure}[t]
  \centering
    \includegraphics[scale=1.00]{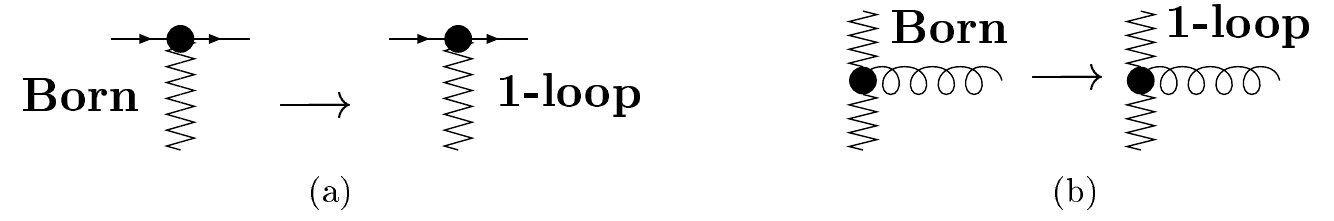}
  \caption[NLA vertex replacements (1)]
  {Diagrammatic representations of replacements 
  $\Gamma_{P^\prime P}^{\mathcal{C}({\rm Born})} \to
 \Gamma_{P^\prime P}^{\mathcal{C}(1-{\rm loop})}$ (a) and $\Gamma_{c_i c_{i+1}}^{\mathcal{G}_i({\rm Born})} \to
 \Gamma_{c_i c_{i+1}}^{\mathcal{G}_i(1-{\rm loop})}$ (b).}
\label{fig:vertex_replacement_1}
\end{figure}

In order to derive the BFKL equation in the NLA, gluon Reggeisation is assumed to be valid to all orders of perturbation theory. As we said, it has been recently 
shown that Reggeisation is fulfilled also in the NLA, through the study of the bootstrap conditions~\cite{Fadin:2006bj,Kozlov:2011zza,Kozlov:2012zza}. 
In the NLA, where all terms of the type
$\alpha_s [\alpha_s ln (s)]^n$ need be collected, the PPR vertex in Eq.~(\ref{octet_amplitude}) assumes the following expression:
\begin{equation}
  \Gamma_{P^\prime P} = 
  \delta_{\lambda P, \lambda P^\prime}\Gamma_{PP}^{(+)} + 
  \delta_{\lambda P^\prime, - \lambda P}\Gamma_{PP}^{(-)} \; .
\end{equation}
In this approximation a term in which the helicity of the scattering particle $P$ is not conserved appears.
To obtain production amplitudes in the NLA it is sufficient to take one of
the vertices or the trajectory in Eq.~(\ref{bfkl_real_amplitude}) in the NLO. In the LLA, the Reggeised
gluon trajectory is needed at 1-loop accuracy and the only contribution to the
real part of the kernel is from the production of one gluon at Born level in the collision of two Reggeons ($K_{RRG}^B$)~\cite{Fadin:1998sh}. In the NLA the gluon trajectory is
taken in the NLO (2-loop accuracy~\cite{Fadin:1995dd,Fadin:1995km,Kotsky:1996xm,Fadin:1995xg,Fadin:1996tb}) and the real part includes
the contributions coming from: 
one-gluon ($K_{RRG}^1$)~\cite{Fadin:1992rh}, 
two-gluon ($K_{RRGG}^B$)~\cite{Fadin:2000kx,Fadin:1996nw,Fadin:1997zv,Kotsky:1998ug},
and quark-antiquark pair ($K_{RRQ\bar{Q}}^B$)~\cite{Fadin:1998jv,Fadin:1997hr,Catani:1990xk,Catani:1990eg} production at Born level~\cite{Fadin:1998sh}.

\begin{figure}[t]
  \centering
    \includegraphics[scale=1.00]{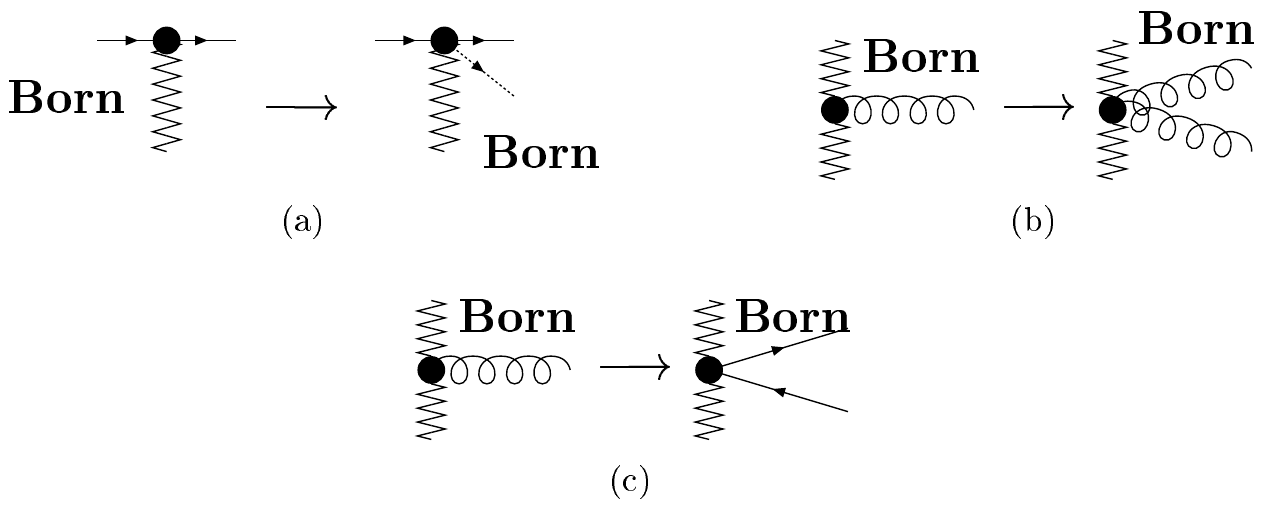}
  \caption[NLA vertex replacements (2)]
  {Diagrammatical representations of replacements 
  $\Gamma_{P^\prime P}^{\mathcal{C}({\rm Born})} \to
 \Gamma_{\{f\} P}^{\mathcal{C}({\rm Born})}$ (a), $ \gamma_{c_i c_{i+1}}^{\mathcal{G}_i({\rm Born})} \to
 \gamma_{c_i c_{i+1}}^{\mathcal{Q\bar{Q}}({\rm Born})}$ (b) 
 and $\gamma_{c_i c_{i+1}}^{\mathcal{G}_i({\rm Born})} \to
 \gamma_{c_i c_{i+1}}^{\mathcal{GG}({\rm Born})}$ (c)
 .}
\label{fig:vertex_replacement_2}
\end{figure}

The first set of corrections is realised by performing,
only in one place, one of the following replacements in the production
amplitude (see Eq.~(\ref{bfkl_real_amplitude})) entering the $s$-channel unitarity relation:
\begin{equation}
 \Gamma_{P^\prime P}^{\mathcal{C}({\rm Born})} \to
 \Gamma_{P^\prime P}^{\mathcal{C}(1-{\rm loop})} \; , \quad\quad
 \gamma_{c_i c_{i+1}}^{\mathcal{G}_i({\rm Born})} \to
 \gamma_{c_i c_{i+1}}^{\mathcal{G}_i(1-{\rm loop})} \; ,
\end{equation}
diagrammatically shown in Fig.~\ref{fig:vertex_replacement_1}.
The second set of corrections consists in allowing the production in the $s$-channel intermediate state of one pair of particles with rapidities of the same
order of magnitude, both in the central or in the fragmentation region (QMRK). This implies one replacement among the following in
the production amplitude:
\begin{equation}
 \Gamma_{P^\prime P}^{\mathcal{C}({\rm Born})} \to
 \Gamma_{\{f\} P}^{\mathcal{C}({\rm Born})} \; , \quad
 \gamma_{c_i c_{i+1}}^{\mathcal{G}_i({\rm Born})} \to
 \gamma_{c_i c_{i+1}}^{\mathcal{Q\bar{Q}}({\rm Born})} \; , \quad
 \gamma_{c_i c_{i+1}}^{\mathcal{G}_i({\rm Born})} \to
 \gamma_{c_i c_{i+1}}^{\mathcal{GG}({\rm Born})} \; .
\end{equation}
Here $\Gamma_{\{f\} P}$ stands for the production of a state containing an extra-particle
in the fragmentation region of the particle $P$ in the scattering off the Reggeon, $\gamma_{c_i c_{i+1}}^{\mathcal{Q\bar{Q}}({\rm Born})}$ and $\gamma_{c_i c_{i+1}}^{\mathcal{GG}({\rm Born})}$ are the effective vertices for the production of a quark-antiquark
pair and of a two-gluon pair, respectively, in the collision of two
Reggeons. This second set of replacements is shown in Fig.~\ref{fig:vertex_replacement_2}.

\subsection{The BFKL cross section}
\label{sub:bfkl-cross-section}

The total cross section 
and many other physical observables are directly 
related to the imaginary part of the forward scattering amplitude 
($\vec{q}=0$) via the optical theorem. The cross section can be expressed by 
\begin{equation}
\label{sigma_s}
 \sigma(s) = \frac{{\rm Im}_s \mathcal{A}_{AB}^{AB}}{s} \; ,
\end{equation}
with ${\rm Im}_s \mathcal{A}_{AB}^{AB}$ given in Eq.~(\ref{eq_factorisation}). 
It is possible to make the following redefinition of the Green's function:
\begin{equation}
\label{ggf_redefinition}
 G_\omega (\vec{q}_1,\vec{q}_2)=
 \frac{ G^{(0)}_{\omega}({\vec q}_{1}, {\vec q}_{2}; 0)}
 {\vec q_{1}^{~2}\vec q_{2}^{~2}} \; ,
\end{equation}
where $\vec{q}_{1,2}$ are two-dimensional vectors and $G^{(0)}_{\omega}({\vec q}_{1}, {\vec q}_{2}; 0)$ is the forward Green's function in the singlet-colour representation which obeys the BFKL equation given in Eq.~(\ref{bfkl_kernel_equation}) with $R=0$. 
This leads to a simplification of the expressions of the BFKL equation and of the BFKL kernel~(\ref{bfkl_kernel_parts}), which now read
\begin{equation}
 \omega \,G_\omega (\vec{q}_1,\vec{q}_2)=\delta
 ^{D-2}(\vec{q}_1-\vec{q}_2)+\int d^{D-2}%
 \widetilde{q}\,\,\,{\cal K}(\vec{q}_1,\vec{\widetilde{q}})
 \,G_\omega (\vec{\widetilde{q}},\vec {q}_2)\,
\label{bfkl_equation_forward}
\end{equation}
and
\begin{equation}
 {\cal K}(\vec{q}_1,\vec{q}_2) =
 \frac{{\cal K}^{(0)}({\vec q}_1, {\vec q}_2; 0)}{{\vec q}_1^{~2}{\vec q}_2^{2}}
  =
 2\,\omega (-\vec q_1^{2})\,\delta^{(D-2)}(\vec{q}_1-\vec{q}_2)
 +{\cal K}_r(\vec{q}_1, \vec{q}_2) \, ,
\label{bfkl_kernel_forward}
\end{equation}
respectively. Here $\omega (-\vec {q}^{~2})$ is the gluon Regge trajectory given in Eq.~(\ref{trajectory_1-loop}). 
Due to scale invariance of the kernel, we can take its eigenfunctions as powers of one of the two squared momenta $q_{1,2}^{~2}$, say $(\vec q_2^{~2})^{\gamma-1}$ with $\gamma$ being a complex number. 
Denoting the corresponding eigenvalues as $\frac{N\alpha_s}{\pi}\chi^B(\gamma)$, we can write:
\begin{equation}
 \int d^{D-2}q_2\,\,{\cal K}(\vec{q}_1,\vec {q}_2)
 (\vec q_2^{~2})^{\gamma-1}=
 \frac{N\alpha_s}{\pi}\chi^B(\gamma)(\vec q_1^{~2})^{\gamma-1} \; ,
\label{eigenfunctions_kernel_averaged}
\end{equation}
with~\cite{BFKL,BFKL_2,BFKL_3,BFKL_4}
\begin{equation}
 \chi^B(\gamma)=2\psi (1)-\psi (\gamma )-\psi (1-\gamma )\,,\,\,\,\psi (\gamma )=\Gamma ^{\prime }(\gamma )/\Gamma (\gamma ) \, .
\label{chi_averaged}
\end{equation}
The set of functions $(\vec q_2^{~2})^{\gamma-1}$ with $\gamma =1/2+i\nu,
\,\,\, -\infty <\nu<\infty $ is complete and represent the eigenfunctions of the LO BFKL kernel averaged on the azimuthal angle between $\vec{q}_1$ and $\vec{q}_2$. Taking its projection onto them in Eq.~(\ref{eq_factorisation}), one can find the following simple expression for the cross section: 
\begin{align}
\label{sigma}
&
\sigma(s)=
\frac{1}{(2\pi)^{2}}\int\frac{d^{2}\vec q_1}{\vec
q_1^{\,\, 2}}\Phi_1(\vec q_1,s_0)\int
\frac{d^{2}\vec q_2}{\vec q_2^{\,\,2}} \Phi_2(-\vec q_2,s_0)
\\ & \nonumber \times \,
\int\limits^{\delta +i\infty}_{\delta
-i\infty}\frac{d\omega}{2\pi i}\left(\frac{s}{s_0}\right)^\omega
G_\omega (\vec q_1, \vec q_2)\, ,
\end{align}
which holds with NLA accuracy. 
All momenta entering this expression are defined on the transverse plane
and are therefore two-dimensional.
$\Phi_{1,2}$ are the NLO impact factors specific of the process.
From this equation it is possible to see that if the Green's function 
$G_\omega$ has a pole at $\omega^\prime$, the cross section at LLA takes the form
\begin{equation}
 \sigma^{\rm (LLA)} \sim \frac{s^{\omega_P^B}}{\sqrt{\ln s}} \; , \quad
 \omega^\prime \equiv \omega_P^B \; ,
\end{equation}
where $\omega_P^B$ is the $(t=0)$-intercept of the Regge trajectory that rules the asymptotic behaviour in $s$ of the amplitude with exchange of the vacuum quantum numbers in the $t$-channel. 
It is equal to $4 N_c (\alpha_s / \pi)\ln 2$, which implies violation of the Froissart bound~\cite{Froissart:1961ux}, 
giving rise to a power-like behaviour of cross section with energy. The BFKL unitarity restoration is an open issue, which goes beyond the scope of this thesis. We mention here some of the solution methods proposed so far: the Balitsky--Kovchegov (BK) scheme~\cite{Balitsky:1995ub,Kovchegov:1999yj}, which generalises the BFKL evolution equation~(\ref{bfkl_kernel_equation}) through the inclusion of non-linear terms that tame the growth of the cross section; the Bartels--Kwiecinski--Praszalowicz (BKP) method~\cite{Bartels:1980pe,Kwiecinski:1980wb}, which introduces composite states of several Reggeised gluons; approaches based on gauge-invariant effective field theories for the Reggeised gluon interactions~\cite{Lipatov:1995pn,Lipatov:1996ts}.

Besides the unitarity issue, there is another important question that should be contemplated, {\it i.e.} whether the characteristic growth with energy of sufficiently inclusive cross sections, which represents the most striking prediction of the BFKL Pomeron, could be observed in actual and forthcoming LHC analyses. 
This possibility will be examined in the course of our study on inclusive dijet production (see Section~\ref{sec:mn-jets-summary}).

As we saw in Section~\ref{sub:bfkl-equation}, the Green's function $G_\omega$ takes care of the universal, energy-dependent part of the amplitude and obeys the BFKL equation~(\ref{bfkl_kernel_equation}).

In this section we derive a general form for the cross section in the so-called $(\nu,n)$-representation (for more details, see Refs.~\cite{Ivanov:2005gn,Ivanov:2006gt}), which will provide us with the starting point of our further analysis. First of all, it is convenient to work in the 
transverse momentum representation, defined by
\beq\label{transv}
\hat{\vec q}\: |\vec q_i\rangle = \vec q_i|\vec q_i\rangle\;, 
\quad\quad
\langle\vec q_1|\vec q_2\rangle =\delta^{(2)}(\vec q_1 - \vec q_2) \;,
\eeq
\[
\langle A|B\rangle =
\langle A|\vec k\rangle\langle\vec k|B\rangle =
\int d^2k A(\vec k)B(\vec k)\;.
\]
In this representation, the total cross section given in Eq.~(\ref{sigma}) takes the simple form
\beq\label{ampl-transv}
\sigma =\frac{1}{(2\pi)^2}
\int_{\delta-i\infty}^{\delta+i\infty}\frac{d\omega}{2\pi i}
\, \left(\frac{s}{s_0}\right)^\omega
\langle\frac{\Phi_1}{\vec q_1^{\,\,2}}|\hat G_\omega|\frac{\Phi_2}{\vec q_2^{\,\,2}}
\rangle \ .
\eeq 

The kernel of the operator $\hat K$ becomes
\beq\label{kernel-op}
K(\vec q_2, \vec q_1) = \langle\vec q_2| \hat K |\vec q_1\rangle
\eeq
and the equation for the Green's function reads
\beq\label{Groper}
\hat 1=(\omega-\hat K)\hat G_\omega\;,
\eeq
its solution being
\beq\label{Groper1}
\hat G_\omega=(\omega-\hat K)^{-1} \, .
\eeq

The kernel is given as an expansion in the strong coupling,
\beq\label{kern}
\hat K=\bar \alpha_s \hat K^0 + \bar \alpha_s^2 \hat K^1\;,
\eeq
where
\beq\label{baral}
{\bar \alpha_s} \equiv \frac{N_c}{\pi} \alpha_s 
\eeq
and $N_c$ is the number of colours. In Eq.~(\ref{kern}) $\hat K^0$ is the
BFKL kernel in the leading order (LO), while $\hat K^1$ represents the NLO correction.

To determine the cross section with NLA accuracy we need an
approximate solution of Eq.~(\ref{Groper1}). With the required accuracy this
solution is
\begin{align}\label{exp}
& \hat G_\omega = (\omega-\bar \alpha_s\hat K^0)^{-1} 
\\ \nonumber & 
+
(\omega-\bar \alpha_s\hat K^0)^{-1}\left(\bar \alpha_s^2 \hat K^1\right)
(\omega-\bar \alpha_s \hat
K^0)^{-1}+ {\cal O}\left[\left(\bar \alpha_s^2 \hat K^1\right)^2\right]
\, .
\end{align}

In Eq.~(\ref{eigenfunctions_kernel_averaged}) we gave the expressions for the eigenfunctions of the LO kernel averaged on the azimuthal angle. 
In the general case the basis of eigenfunctions of the LO kernel,
\begin{align}
\label{KLLA}
&
\hat K^0 |n,\nu\rangle = \chi(n,\nu)|n,\nu\rangle \, ,
\\ & \nonumber
\chi (n,\nu)=2\psi(1)-\psi\left(\frac{n}{2}+\frac{1}{2}+i\nu\right)
-\psi\left(\frac{n}{2}+\frac{1}{2}-i\nu\right)\, ,
\end{align}
is given by the following set of functions:
\beq\label{nuLLA}
\langle\vec q\, |n,\nu\rangle =\frac{1}{\pi\sqrt{2}}
\left(\vec q^{\,\, 2}\right)^{i\nu-\frac{1}{2}}e^{in\phi} \;,
\eeq
which now depend not only on $\nu$, but also on the integer $n$, called \emph{conformal spin}.
Here $\phi$ is the azimuthal angle of the vector $\vec q$ counted from
some fixed direction in the transverse space, $\cos\phi \equiv q_x/|\vec q\,|$.
Then, the orthonormality and completeness conditions take the form
\beq\label{ort}
\langle n',\nu^\prime | n,\nu\rangle =\int \frac{d^2 \vec q}
{2 \pi^2 }\left(\vec q^{\,\, 2}\right)^{i\nu-i\nu^\prime -1}
e^{i(n-n')\phi}=\delta(\nu-\nu^\prime)\, \delta_{nn'}
\eeq
and
\beq\label{comp}
\hat 1 =\sum^{\infty}_{n=-\infty}\int\limits^{\infty}_{-\infty}d\nu \, 
| n,\nu\rangle\langle n,\nu |\ .
\eeq

The action of the full NLO BFKL kernel on these functions may be expressed
as follows:
\begin{align}\label{Konnu}
& \hat K|n,\nu\rangle =
\bar \alpha_s(\mu_R) \chi(n,\nu)|n,\nu\rangle
\\ \nonumber &
+ \bar \alpha_s^2(\mu_R)\left(\chi^{(1)}(n,\nu)
+\frac{\beta_0}{4N_c}\chi(n,\nu)\ln(\mu^2_R)\right)|n,\nu\rangle
\\ \nonumber &
+ \bar
\alpha_s^2(\mu_R)\frac{\beta_0}{4N_c}\chi(n,\nu)
\left(i\frac{\partial}{\partial \nu}\right)|n,\nu\rangle \;,
\end{align}
where $\mu_R$ is the renormalisation scale of the QCD coupling; the first
term represents the action of LO kernel, while the second and the third ones
stand for the diagonal and the non-diagonal parts of the NLO kernel and we
have used
\beq\label{beta00}
\beta_0=\frac{11 N_c}{3}-\frac{2 n_f}{3}\;,
\eeq
where $n_f$ is the number of active quark flavours.

The function $\chi^{(1)}(n,\nu)$, calculated in~Ref.~\cite{Kotikov:2000pm} (see
also Ref.~\cite{Kotikov:2002ab}), is conveniently represented in the form
\beq\label{ch11}
\chi^{(1)}(n,\nu)=-\frac{\beta_0}{8\, N_c}\left(\chi^2(n,\nu)-\frac{10}{3}
\chi(n,\nu)-i\chi^\prime(n,\nu)\right) + {\bar \chi}(n,\nu)\, ,
\eeq
where
\beq\label{chibar}
\bar \chi(n,\nu)\,=\,-\frac{1}{4}\left[\frac{\pi^2-4}{3}\chi(n,\nu)
-6\zeta(3)
-\chi^{\prime\prime}(n,\nu) +\,2\,\phi(n,\nu)
\right.
\eeq
\[
+\,2\,\phi(n,-\nu)
+\frac{\pi^2\sinh(\pi\nu)}{2\,\nu\, \cosh^2(\pi\nu)}
\left(
\left(3+\left(1+\frac{n_f}{N_c^3}\right)\frac{11+12\nu^2}{16(1+\nu^2)}\right)
\delta_{n0}
\right.
\]
\[
\left.\left.
-\left(1+\frac{n_f}{N_c^3}\right)\frac{1+4\nu^2}{32(1+\nu^2)}\delta_{n2}
\right)\right] \, ,
\]
and
\beq\label{phi}
\phi(n,\nu)\,=\,
-\int\limits_0^1dx\,\frac{x^{-1/2+i\nu+n/2}}{1+x}
\left[\frac{1}{2}\left(\psi'\left(\frac{n+1}{2}\right)-\zeta(2)\right) \right.
\eeq
\[
+\mbox{Li}_2(x)+\mbox{Li}_2(-x)
+\ln x \left(\psi(n+1)-\psi(1)+\ln(1+x)+\sum_{k=1}^\infty\frac{(-x)^k}
{k+n}\right)
\]
\[
\left. 
+\sum_{k=1}^\infty\frac{x^k}{(k+n)^2}(1-(-1)^k)\right]
\]
\[
=\sum_{k=0}^\infty\frac{(-1)^{k+1}}{k+(n+1)/2+i\nu}\left[\psi'(k+n+1)
-\psi'(k+1)
\right.
\]
\[
+(-1)^{k+1}(\beta'(k+n+1)+\beta'(k+1))
\]
\[
\left.
-\frac{1}{k+(n+1)/2+i\nu}(\psi(k+n+1)-\psi(k+1))\right] \; ,
\]
\begin{equation}
\beta'(z)=\frac{1}{4}\left[\psi'\left(\frac{z+1}{2}\right)
-\psi'\left(\frac{z}{2}\right)\right] \; ,
\end{equation}
\begin{equation}
\mbox{Li}_2(x)=-\int\limits_0^xdt\,\frac{\ln(1-t)}{t} \; .
\end{equation}
Here and below $\chi^\prime(n,\nu) \equiv d\chi(n,\nu)/d\nu$ and
$\chi^{\prime\prime}(n,\nu) \equiv d^2\chi(n,\nu)/d^2\nu$.

The projection of the impact factors onto the eigenfunctions 
of the LO BFKL 
kernel, {\it i.e.} the transfer to the $(\nu,n)$-representation, 
is done as follows:
\begin{equation}
\frac{\Phi_1(\vec q_1)}{\vec q_1^{\,\, 2}}=\sum^{+\infty}_{n=-\infty}
\int\limits^{+\infty}_{-\infty}
d\nu \, \Phi_1(\nu,n)\langle n,\nu| \vec q_1\rangle\, \; , 
\end{equation}
\[
\frac{\Phi_2(-\vec{q_2})}{\vec q_2^{\,\, 2}}=\sum^{+\infty}_{n=-\infty}
\int\limits^{+\infty}_{-\infty} d\nu \, \Phi_2(\nu,n)
\langle \vec q_2 |n,\nu \rangle \; ,
\]
\beq\label{nu_rep}
\Phi_1(\nu,n)=
\langle\frac{\Phi_1(\vec q_1)}{\vec q_1^{\,\,2}}|n,\nu\rangle \equiv
\int d^2 q_1 \,\frac{\Phi_1(\vec q_1)}{\vec q_1^{\,\, 2}}
\frac{1}{\pi \sqrt{2}} \left(\vec q_1^{\,\, 2}\right)^{i\nu-\frac{1}{2}}
e^{i n \phi_1}\;,
\eeq
\[
\Phi_2(\nu,n)=
\langle n,\nu|\frac{\Phi_2(-\vec q_2)}{\vec q_2^{\,\,2}}\rangle 
\equiv
\int d^2 q_2 \,\frac{\Phi_2(-\vec q_2)}{\vec q_2^{\,\, 2}}
\frac{1}{\pi \sqrt{2}} \left(\vec q_2^{\,\, 2}\right)^{-i\nu-\frac{1}{2}}
e^{-i n \phi_2}\;.
\]
The impact factors can be represented as an expansion in $\alpha_s$,
\beq\label{if}
\Phi_{1,2}(\vec q\,)=\alpha_s(\mu_R)\left[ v_{1,2}(\vec q\, ) 
+ \bar \alpha_s(\mu_R)
v_{1,2}^{(1)}(\vec q\, )\right]
\eeq
and
\beq\label{vertex-exp}
\Phi_{1,2}(n,\nu)=\alpha_s(\mu_R)\left[ c_{1,2}(n,\nu) 
+ \bar \alpha_s(\mu_R)
c_{1,2}^{(1)}(n,\nu) \right]\, .
\eeq

To obtain our representation of the cross section, the matrix 
element of the BFKL Green's function is needed. According to Eq.~(\ref{exp}), one has 
\beq\label{Greens}
\langle n,\nu|\hat G_\omega|n^\prime,\nu^\prime\rangle 
= \delta_{n,n^\prime}\left[
\delta(\nu-\nu^\prime)\left(
\frac{1}{\omega-\bar \alpha_s (\mu_R)\chi(n,\nu)}
\right.\right.
\eeq
\[
\left.
+\frac{1}{(\omega-\bar \alpha_s(\mu_R) 
\chi(n,\nu))^2}
\left(
\bar \alpha_s^2(\mu_R)(\bar \chi(n,\nu)
\right.\right.
\]
\[
\left.\left.
+\frac{\beta_0}{8 N_c}(-\chi^2(n,\nu)+\frac{10}{3}\chi(n,\nu)
+2\chi(n,\nu)
\ln \mu_R^2+i\frac{d}{d\nu}\chi(n,\nu)))
\right)
\right)
\]
\[
\left.
+\frac{\frac{\beta_0}{4 N_c}\bar \alpha_s^2(\mu_R)\chi(n,\nu^\prime)}
{(\omega-\bar \alpha_s(\mu_R) \chi(n,\nu))(\omega-\bar \alpha_s(\mu_R) 
\chi(n,\nu^\prime))}\left(i\frac{d}{d\nu^\prime}
\delta(\nu-\nu^\prime)\right)
\right]\ .
\]

Inserting twice the unity operator, written according to the completeness 
condition given in Eq.~(\ref{comp}), into Eq.~(\ref{ampl-transv}), we get
\beq\label{sigma-f}
\sigma = \frac{1}{(2\pi)^2}
\sum^{\infty}_{n=-\infty}
\int\limits^{\infty}_{-\infty} d\nu\sum^{\infty}_{n^\prime =-\infty} 
\int\limits^{\infty}_{-\infty} d\nu^\prime 
\int_{\delta-i\infty}^{\delta+i\infty}
\frac{d\omega}{2\pi i}
\left(\frac{1}{s_0}\right)^\omega
\eeq
\[
\langle
\frac{\Phi_1}{\vec q_1^{\,\,2}}|n,\nu\rangle\langle n,\nu|\hat 
G_\omega| n^\prime,\nu^\prime\rangle\langle n^\prime,\nu^\prime |
\frac{\Phi_2}{\vec q_2^{\,\,2}}\rangle \ ,
\]
and, after some algebra and integration by parts, finally
\beq\label{sigma-ff}
\sigma = \frac{1}{(2\pi)^2}\sum^{\infty}_{n=-\infty}
\int\limits^{\infty}_{-\infty} d\nu \left(\frac{s}{s_0}\right)^{\bar \alpha_s(\mu_R)
\chi(n,\nu)} 
\eeq
\[
\alpha_s^2(\mu_R) c_1(n,\nu)c_2(n,\nu)\left[1+\bar \alpha_s(\mu_R)\left(\frac{c^{(1)}_1(n,\nu)}{c_1(n,\nu)}
+\frac{c^{(1)}_2(n,\nu)}{c_2(n,\nu)}\right) \right.
\]
\[
+\bar \alpha^2_s(\mu_R)\ln\frac{s}{s_0}\left\{\bar \chi(n,\nu)
+\frac{\beta_0}{8 N_c}\chi(n,\nu)
\right.
\]
\[
\left.\left.
\left(
-\chi(n,\nu)+\frac{10}{3}+2\ln \mu_R^2 +i\frac{d}{d\nu}\ln\frac{c_1(n,\nu)}
{c_2(n,\nu)}\right)\right\}\right] \, .
\]

In order to assess the relative weight of NLA corrections with respect to the LLA contribution, one can confront $\chi^{(1)}(n,\nu)$~(Eq.~(\ref{ch11})) with $\chi(n,\nu)$~(Eq.~(\ref{KLLA})) for $n=0$ and $\nu=0$, \emph{i.e.} for the point in the $(\nu,n)$-space which determines the energy asymptotic behaviour in the LLA case. The result found for the ratio $|\chi^{(1)}(0,0)/\chi(0,0)|$ is large ($\simeq 6.5$)~\cite{Fadin:1998py,Ciafaloni:1998gs}, thus leading to instabilities in the BFKL perturbative expansion, which have to be controlled through some optimisation procedure. In Section~\ref{sec:bfkl_blm} we will discuss one of them, which represent perhaps the most effetive tool to quench the oscillating behaviour of the BFKL series.

\subsubsection{Representation equivalence}
\label{ssub:bfkl_cs_representation}

The expression for the cross section given in Eq.~(\ref{sigma-ff}) is 
valid both in the LLA and in the NLA. However, it is not the only possible one. Actually, several NLA-equivalent expressions can be adopted. One can consider alternative representations aiming at catching some of the unknown next-to-NLA corrections. 
Here we show two examples, which have been used in recent phenomenological analyses (for more details, see Ref.~\cite{Caporale:2014gpa}):
\begin{itemize}
\item the so-called \emph{exponentiated} representation,
\begin{equation}
\label{sigma_repr_exp}
\sigma^{\rm exp} = \frac{1}{(2\pi)^2}\sum^{\infty}_{n=-\infty}
\int\limits^{\infty}_{-\infty} d\nu \,
\alpha_s^2(\mu_R) c_1(n,\nu)c_2(n,\nu)
\end{equation}
\[ \times \,
\left(\frac{s}{s_0}\right)^{\bar \alpha_s(\mu_R)\chi 
+\bar \alpha^2_s(\mu_R)\left\{\bar \chi
+\frac{\beta_0}{8 N_c}\chi
\left(
-\chi+\frac{10}{3}+2\ln \mu_R^2 +i\frac{d}{d\nu}\ln\frac{c_1(n,\nu)}
{c_2(n,\nu)}\right)\right\}} 
\]
\[ \times \,
\left[1+\bar \alpha_s(\mu_R)\left(\frac{c^{(1)}_1(n,\nu)}{c_1(n,\nu)}
+\frac{c^{(1)}_2(n,\nu)}{c_2(n,\nu)}\right) \right] \; ;
\]

\item the exponentiated representation with an extra, irrelevant in the NLA term, given by the product of the NLO corrections of the two impact factors,
\begin{equation}
\label{sigma_repr_cross}
\sigma^{\rm exp \, (c)} = \frac{1}{(2\pi)^2}\sum^{\infty}_{n=-\infty}
\int\limits^{\infty}_{-\infty} d\nu \,
\alpha_s^2(\mu_R) c_1(n,\nu)c_2(n,\nu)
\end{equation}
\[ \times \,
\left(\frac{s}{s_0}\right)^{\bar \alpha_s(\mu_R)\chi 
+\bar \alpha^2_s(\mu_R)\left\{\bar \chi
+\frac{\beta_0}{8 N_c}\chi
\left(
-\chi+\frac{10}{3}+2\ln \mu_R^2 +i\frac{d}{d\nu}\ln\frac{c_1(n,\nu)}
{c_2(n,\nu)}\right)\right\}} 
\]
\[ \times \,
\left[1+\bar \alpha_s(\mu_R)\left(\frac{c^{(1)}_1(n,\nu)}{c_1(n,\nu)}
+\frac{c^{(1)}_2(n,\nu)}{c_2(n,\nu)}\right) \right.
\]
\[
\left. + \, \alpha_s^2\left(\mu_R\right)
\left( \frac{c_1^{\left(1\right)}\left(n,\nu\right)}{c_1\left(n,\nu\right)}
\frac{c_2^{\left(1\right)}\left(n,\nu\right)}{c_2\left(n,\nu\right)}\right)
\right] 
\; .
\]

\end{itemize}

In Eqs.~(\ref{sigma_repr_exp}) and~(\ref{sigma_repr_cross}) 
is the LO kernel eigenvalue given in~(\ref{KLLA}).
In our calculations for Mueller--Navelet jet (see Chapter~\ref{chap:mn-jets}) and dihadron (see Chapter~\ref{chap:mn-jets}) production processes, we will adopt the exponentiated representation (Eq.~(\ref{sigma_repr_exp})).

\section{The BLM optimisation procedure}
\label{sec:bfkl_blm}

It is well known that the BFKL approach is plagued
by large NLA corrections (see the discussion at the end of Section~\ref{sub:bfkl-cross-section}), both in the kernel of the Green's function and in the process-dependent impact factors, as well
as by large uncertainties in the renormalisation scale setting. 
As an example, the NLA BFKL corrections for the $n=0$ conformal spin are with opposite sign with respect to the
LLA results and large in absolute value. 
All that calls for some optimisation procedure of the perturbative series, which can
consist in (i) including some pieces of the (unknown) next-to-NLA corrections,
such as those dictated by renormalisation group, as in \emph{collinear
improvement}~\cite{Caporale:2013uva,Vera:2007kn,Salam:1998tj,Ciafaloni:1998iv,Ciafaloni:1999yw,Ciafaloni:1999au,Ciafaloni:2000cb,Ciafaloni:2002xk,Ciafaloni:2002xf,Ciafaloni:2003ek,Ciafaloni:2003rd,Ciafaloni:2003kd,Vera:2005jt,Caporale:2007vs}, or by energy-momentum 
conservation~\cite{Kwiecinski:1999yx}, or (ii) suppressing the emission of gluons which are close by in rapidity in the BFKL framework (\emph{rapidity veto} approach~\cite{Schmidt:1999mz,Forshaw:1999xm}), and/or (iii) suitably choosing the values 
of the energy and renormalisation scales, which, though arbitrary within the 
NLO, can have a sizeable numerical impact through subleading terms. Common
optimisation methods are those inspired by the \emph{principle of minimum 
sensitivity} (PMS)~\cite{PMS,PMS_2}, the \emph{fast apparent convergence} 
(FAC)~\cite{FAC,FAC_2,FAC_3} and the \emph{Brodsky--Lepage--Mackenzie method} 
(BLM)~\cite{BLM,BLM_2,BLM_3,BLM_4,BLM_5}.

In this Section we present and discuss 
the widely-used BLM approach, 
which relies on the removal
of the renormalisation scale ambiguity by absorbing the non-conformal 
$\beta_0$-terms into the running coupling. It is known that after BLM scale 
setting, the QCD perturbative convergence can be greatly improved due to the 
elimination of renormalon terms in the perturbative QCD series. Moreover, with 
the BLM scale setting, the BFKL Pomeron intercept has a weak dependence on the 
virtuality of the Reggeised gluon~\cite{BLM_4,BLM_5}. 

We provide, as result, an exact implementation of the BLM method, together with two other, approximated ones, 
which were used earlier in the literature of the BLM method for different semi-hard processes 
(see a more detailed discussion in~\cite{Caporale:2015uva}). 

We consider the BLM scale setting for the separate contributions to 
the cross section, specified in Eq.~(\ref{sigma-ff}) by different values of $n$, denoted in the following by ${\cal C}_n$.
The starting point of our considerations is the expression for ${\cal C}_n$
in the $\overline{\rm MS}$ scheme (see Eq.~(\ref{sigma-ff})),
\beq\label{c_n}
{\cal C}_n
=\frac{1}{(2\pi)^2}\int\limits^{\infty}_{-\infty} d\nu 
\left(\frac{s}{s_0}\right)^{\bar \alpha_s(\mu_R)\chi(n,\nu)} \alpha_s^2(\mu_R) 
c_1(n,\nu)c_2(n,\nu)
\eeq
\[ \times \,
\left[1+\bar \alpha_s(\mu_R)\left(\frac{c^{(1)}_1(n,\nu)}{c_1(n,\nu)}
+\frac{c^{(1)}_2(n,\nu)}{c_2(n,\nu)}\right)
\right.
\]
\[
+ \, \bar \alpha^2_s(\mu_R)\ln\frac{s}{s_0}\left\{\bar \chi(n,\nu) 
+\frac{\beta_0}{8 N_c}\chi(n,\nu)\right.
\]
\[ \times \,
\left.\left.
\left(
-\chi(n,\nu)+\frac{10}{3}+2\ln \mu_R^2 +i\frac{d}{d\nu}
\ln\frac{c_1(n,\nu)}{c_2(n,\nu)}\right)\right\}\right] \, .
\]
In the r.h.s. of this expression we have terms $\sim \alpha_s$ originated 
from the NLO corrections to the impact factors, and terms 
$\sim \alpha^2_s\ln(s/s_0)$ coming from NLA corrections to the BFKL kernel. In 
the latter case, the terms proportional to the QCD $\beta$-function are 
explicitly shown. For our further consideration of the BLM scale setting, 
similar contributions have to be separated also from the NLO impact factors.

In fact, the contribution to an NLO impact factor that is proportional to 
$\beta_0$ is universally expressed through the LO impact factor,
\beq\label{beta-if}
v^{(1)}(\vec q\,)=v(\vec q\,)\frac{\beta_0}{4 N_c}\left(\ln\left(\frac{\mu_R^2}
{\vec q\,^2}\right)+\frac{5}{3}\right)+\dots \ ,
\eeq
where the dots stand for the other terms, not proportional to $\beta_0$. This 
statement becomes evident if one considers the part of the strong coupling 
renormalisation proportional to $n_f$ and related to the contributions of 
light quark flavours. Such contribution to the NLO impact factor originates only 
from diagrams with the light quark loop insertion in the Reggeised gluon 
propagator. The results for such contributions can be found, for instance, in   
Eq.~(5.1) of~\cite{Fadin:2001ap}. Tracing there the terms $\sim n_f$ and 
performing the QCD charge renormalisation, one can indeed
confirm~(Eq.~(\ref{beta-if})).

Transforming Eq.~(\ref{beta-if}) to the $\nu$-representation according 
to Eq.~(\ref{nu_rep}), we obtain 
\bea\label{if2}
{\tilde{c}}_1^{\left(1\right)}(\nu, n)&=& \frac{\beta_0}{4 N_c}
\left[+i\frac{d}{d\nu} c_1(\nu,n)+\left(\ln \mu_R^2+\frac{5}{3}\right)
c_1(\nu, n)\right]\ ,
\nonumber
\\
{\tilde{c}}_2^{\left(1\right)}(\nu, n)&=&\frac{\beta_0}{4 N_c}
\left[-i\frac{d}{d\nu} c_2(\nu,n)+\left(\ln \mu_R^2+\frac{5}{3}\right)
c_2(\nu, n)\right] \ ,
\eea
and
\beq\label{}
\frac{{\tilde{c}}_1^{\left(1\right)}}{c_1}+\frac{{\tilde{c}}_2^{\left(1\right)}}{c_2}
=\frac{\beta_0}{4 N_c}\left[i\frac{d}{d\nu}\ln\left(\frac{c_1}{c_2}\right)
+2\left(\ln \mu_R^2+\frac{5}{3}\right)\right] \ .
\eeq

It is convenient to introduce the function $f\left(\nu\right)$, defined
through
\beq
\label{f_nu_general}
i\frac{d}{d\nu}\ln\left(\frac{c_1}{c_2}\right)\equiv 2 \left[f(\nu)
-\ln\left(Q_1 Q_2\right)\right]\ ,
\eeq
that depends on the given process, where $Q_{1,2}$ denote here the hard scales 
which enter the impact factors $c_{1,2}$. 
The specific form of the function $f(\nu)$ depends on the particular process.

Now, we present again our result for the generic observable ${\cal C}_n$, 
showing explicitly all contributions proportional to the QCD $\beta$-function,
{\it i.e.} also those originating from the impact factors:
\beq\label{c_nn}
{\cal C}_n
=\frac{1}{(2\pi)^2}\int\limits^{\infty}_{-\infty} d\nu \left(\frac{s}{s_0}
\right)^{\bar \alpha_s(\mu_R)\chi(n,\nu)} \alpha_s^2(\mu_R) c_1(n,\nu)c_2(n,\nu)
\eeq
\[ \times \,
\left[1+\bar \alpha_s(\mu_R)\left(\frac{\bar c^{(1)}_1(n,\nu)}{c_1(n,\nu)}
+\frac{\bar c^{(1)}_2(n,\nu)}{c_2(n,\nu)}
+\frac{\beta_0}{2 N_c}\left(\frac{5}{3}+\ln \frac{\mu_R^2}{Q_1 Q_2} +f(\nu)
\right)\right)\right.
\]
\[
\left.
+ \, \bar \alpha^2_s(\mu_R)\ln\frac{s}{s_0}\left\{\bar \chi(n,\nu)
+\frac{\beta_0}{4 N_c}\chi(n,\nu)\left(
-\frac{\chi(n,\nu)}{2}+\frac{5}{3}+\ln \frac{\mu_R^2}{Q_1 Q_2} +f(\nu)\right)
\right\}\right] \, ,
\]
where $\bar c^{(1)}_{1,2} \equiv c^{(1)}_{1,2}- \tilde c^{(1)}_{1,2}$. We note that 
the dependence of Eq.~(\ref{c_nn}) on the scale $\mu_R$ is subleading:
performing in Eq.~(\ref{c_nn}) the replacement
\beq\label{alphaSrun}
\alpha_s(\mu_R)=\alpha_s(\mu^\prime_R)\left(1-\bar\alpha_s(\mu^\prime_R)
\frac{\beta_0}{2N_c}\ln\frac{\mu_R}{\mu^\prime_R}\right) \, ,
\eeq
one indeed obtains the same expression as before with the
new scale $\mu_R^\prime$ at the place of the old one $\mu_R$, 
plus some additional contributions which are beyond the NLA accuracy.

As the next step, we perform a finite renormalisation from the 
$\overline{\rm MS}$ to the physical MOM scheme, that means:
\beq
\label{scheme}
\alpha_s^{\overline{\rm MS}}=\alpha_s^{\rm MOM}\left(1+\frac{\alpha_s^{\rm MOM}}{\pi}T 
\right)\;,
\eeq
with 
\beq
\label{T_Tbeta_Tconf}
T=T^{\beta}+T^{\rm conf} \; ,
\eeq
\[
T^{\beta}=-\frac{\beta_0}{2}\left( 1+\frac{2}{3}I \right) \; ,
\]
\[
T^{\rm conf}= \frac{C_A}{8}\left[ \frac{17}{2}I +\frac{3}{2}\left(I-1\right)\xi
+\left( 1-\frac{1}{3}I\right)\xi^2-\frac{1}{6}\xi^3 \right] \; ,
\]
where $C_A \equiv N_c$ is the colour factor associated with gluon emission from a gluon, $I=-2\int_0^1dx\frac{\ln\left(x\right)}{x^2-x+1}\simeq2.3439$ and $\xi$ 
is a gauge parameter, fixed at zero in the following.

Inserting Eq.~(\ref{scheme}) into Eq.~(\ref{c_nn}) and expanding the result, 
we obtain, within NLA accuracy,
\beq\label{c_nnn}
{\cal C}^{\rm MOM}_n =\frac{1}{(2\pi)^2}\int\limits^{\infty}_{-\infty} 
d\nu \left(\frac{s}{s_0}\right)^{\bar \alpha^{\rm MOM}_s(\mu_R)\chi(n,\nu)} 
\left(\alpha^{\rm MOM}_s (\mu_R)\right)^2 
\eeq
\[ \times \,
c_1(n,\nu)c_2(n,\nu)\left[1+\bar \alpha^{\rm MOM}_s(\mu_R)\left\{\frac{\bar c^{(1)}_1(n,\nu)}
{c_1(n,\nu)}+\frac{\bar c^{(1)}_2(n,\nu)}{c_2(n,\nu)}+\frac{2T^{\rm conf}}{N_c}
\right.\right.
\]
\[
\left.
+ \, \frac{\beta_0}{2 N_c}\left(\frac{5}{3}+\ln \frac{\mu_R^2}{Q_1 Q_2} +f(\nu)
-2\left( 1+\frac{2}{3}I \right)\right)
\right\}
\]
\[
+ \, \left(\bar \alpha^{\rm MOM}_s(\mu_R)\right)^2\ln\frac{s}{s_0}
\left\{\bar \chi(n,\nu) +\frac{T^{\rm conf}}{N_c}\chi(n,\nu)\right.
\]
\[
\left.\left.
+ \, \frac{\beta_0}{4 N_c}\chi(n,\nu)\left(
-\frac{\chi(n,\nu)}{2}+\frac{5}{3}+\ln \frac{\mu_R^2}{Q_1 Q_2} +f(\nu)
-2\left(1+\frac{2}{3}I \right)\right)\right\}\right] \, .
\]
The optimal scale $\mu_R^{\rm BLM}$ is the value of $\mu_R$ that makes the 
expression proportional to $\beta_0$ vanish. We thus have
\beq\label{c_nnnbeta}
{\cal C}^{\beta}_n
=\frac{1}{(2\pi)^2}\int\limits^{\infty}_{-\infty} d\nu 
\left(\frac{s}{s_0}\right)^{\bar \alpha^{\rm MOM}_s(\mu^{\rm BLM}_R)\chi(n,\nu)} 
\left(\alpha^{\rm MOM}_s (\mu^{\rm BLM}_R)\right)^3
\eeq
\[ \times \,
c_1(n,\nu)c_2(n,\nu) \frac{\beta_0}{2 N_c} \left[\frac{5}{3}
+\ln \frac{(\mu^{\rm BLM}_R)^2}{Q_1 Q_2} +f(\nu)-2\left( 1+\frac{2}{3}I \right)
\right.
\]
\[
+ \, \bar \alpha^{\rm MOM}_s(\mu^{\rm BLM}_R)\ln\frac{s}{s_0} \: \frac{\chi(n,\nu)}{2}
\]
\[ \times \,
\left.
\left(-\frac{\chi(n,\nu)}{2}+\frac{5}{3}+\ln \frac{(\mu^{\rm BLM}_R)^2}{Q_1 Q_2} 
+f(\nu)-2\left( 1+\frac{2}{3}I \right)\right)\right]=0 \, .
\]
In the r.h.s. of Eq.~(\ref{c_nnnbeta}) we have two groups of contributions. The 
first one originates from the $\beta_0$-dependent part of NLO impact 
factor~(\ref{beta-if}) and also from the expansion of the common 
$\alpha^2_s$ pre-factor in Eq.~(\ref{c_nn}) after expressing it in terms of 
$\alpha_s^{\rm MOM}$.
The other group are the terms proportional to $\bar \alpha_s^{\rm MOM}\ln s/s_0$. 
These contributions are those $\beta_0$-dependent terms that are proportional 
to $\ln s/s_0$ in Eq.~(\ref{c_nn}) and also the one coming from the expansion of  
the $(s/s_0)^{\bar \alpha_s \chi(n,\nu)}$ factor in Eq.~(\ref{c_nn}) after expressing it 
in terms of $\alpha_s^{\rm MOM}$.

The solution of Eq.~(\ref{c_nnnbeta}) gives us the value of BLM scale. Note 
that this solution depends on the energy (on the ratio $s/s_0$).
Such scale setting procedure is a direct application of the original BLM 
approach to semi-hard processes. Finally, our expression for the observable reads
\beq\label{c_BLMmain}
{\cal C}^{\rm BLM}_n
=\frac{1}{(2\pi)^2}\int\limits^{\infty}_{-\infty} d\nu \left(\frac{s}{s_0}
\right)^{\bar \alpha^{\rm MOM}_s(\mu^{\rm BLM}_R)\left[\chi(n,\nu)
+\bar \alpha^{\rm MOM}_s(\mu^{\rm BLM}_R)\left(\bar \chi(n,\nu) +\frac{T^{\rm conf}}
{N_c}\chi(n,\nu)\right)\right]}
\eeq
\[ \times \,
\left(\alpha^{\rm MOM}_s (\mu^{\rm BLM}_R)\right)^2 c_1(n,\nu)c_2(n,\nu)
\]
\[ \times \,
\left[1+\bar \alpha^{\rm MOM}_s(\mu^{\rm BLM}_R)\left\{\frac{\bar c^{(1)}_1(n,\nu)}
{c_1(n,\nu)}+\frac{\bar c^{(1)}_2(n,\nu)}{c_2(n,\nu)}+\frac{2T^{\rm conf}}{N_c}
\right\} \right] \, ,
\]
where we put at the exponent the terms $\sim \bar \alpha_s^{\rm MOM}\ln s/s_0$,
which is allowed within the NLA accuracy 
(see Section~\ref{ssub:bfkl_cs_representation}).

Eq.~(\ref{c_nnnbeta}) can be solved only numerically. 
For this reason, we give also two analytic approximate approaches to the BLM scale setting. We consider the BLM scale as a function of $\nu$ and chose it in order to make vanish either the first or the second ($\sim \bar \alpha_s^{\rm MOM}\ln s/s_0$)  group of terms in the Eq.~(\ref{c_nnnbeta}). 
In these two cases one gets simpler analytical expressions for the BLM scales which do not depend on the
energy. We thus have: 
\begin{itemize}
\item case $(a)$
\beq\label{casea}
\left(\mu_{R, a}^{\rm BLM}\right)^2=Q_1Q_2\ \exp\left[2\left(1+\frac{2}{3}I\right)
-f\left(\nu\right)-\frac{5}{3}\right]\ ,
\eeq
\beq\label{c_BLMa}
{\cal C}^{\rm BLM, a}_n
= \frac{1}{(2\pi)^2}\int\limits^{\infty}_{-\infty} d\nu
\, 
\left(\alpha^{\rm MOM}_s (\mu^{\rm BLM}_{R, a})\right)^2 c_1(n,\nu)c_2(n,\nu)
\eeq
\[ \times \,
\left(\frac{s}{s_0}
\right)^{\bar \alpha^{\rm MOM}_s(\mu^{\rm BLM}_{R, a})\left[\chi+\bar \alpha^{\rm MOM}_s
(\mu^{\rm BLM}_{R, a})\left(\bar \chi +\frac{T^{\rm conf}}{N_c}\chi
-\frac{\beta_0}{8 N_c}\chi^2\right)\right]}
\]
\[ \times \,
\left[1+\bar \alpha^{\rm MOM}_s(\mu^{\rm BLM}_{R, a})
\left\{\frac{\bar c^{(1)}_1(n,\nu)}{c_1(n,\nu)}+\frac{\bar c^{(1)}_2(n,\nu)}
{c_2(n,\nu)}+\frac{2T^{\rm conf}}{N_c}
\right\} \right] \, ,
\]
which corresponds to the removal of the $\beta_0$-dependent terms in the impact factors;
\item case $(b)$
\beq\label{caseb}
\left(\mu_{R, b}^{\rm BLM}\right)^2=Q_1Q_2\ \exp\left[2\left(1+\frac{2}{3}I\right)
-f\left(\nu\right)-\frac{5}{3}+\frac{1}{2}\chi\left(\nu,n\right)\right]\ ,
\eeq
\beq\label{c_BLMb}
{\cal C}^{\rm BLM, b}_n
=\frac{1}{(2\pi)^2}\int\limits^{\infty}_{-\infty} d\nu 
\,
\left(\alpha^{\rm MOM}_s (\mu^{\rm BLM}_{R, b})\right)^2 c_1(n,\nu)c_2(n,\nu)
\eeq
\[ \times \,
\left(\frac{s}{s_0}
\right)^{\bar \alpha^{\rm MOM}_s(\mu^{\rm BLM}_{R, b})\left[\chi+\bar \alpha^{\rm MOM}_s
(\mu^{\rm BLM}_{R, b})\left(\bar \chi +\frac{T^{\rm conf}}{N_c}\chi
\right)\right]}
\]
\[ \times \,
\left[1+\bar \alpha^{\rm MOM}_s(\mu^{\rm BLM}_{R, b})\left\{\frac{\bar c^{(1)}_1
(n,\nu)}{c_1(n,\nu)}+\frac{\bar c^{(1)}_2(n,\nu)}{c_2(n,\nu)}
+\frac{2T^{\rm conf}}{N_c}+\frac{\beta_0}{4 N_c}\chi(n,\nu)
\right\}\right]\, ,
\]
which corresponds to the removal of the $\beta_0$-dependent terms in the BFKL kernel.
\end{itemize}
Note that the two approximated approaches (a) and (b) discussed above and 
given in Eqs. (\ref{c_BLMa}) and~(\ref{c_BLMb}), could be 
applicable only to processes characterised by a real-valued function 
$f(\nu)$. For some processes this is not the case. In particular, the inclusive dihadron production (see Chapter~\ref{chap:dihadron}), is described by a complex-valued 
function, $f^*(\nu)=f(-\nu)$ (Eq.~(\ref{dh-f-nu})).
In such cases one can use only the exact-scale fixing method 
which relies on the numerical solution of Eq.~(\ref{c_nnnbeta}).

\setcounter{appcnt}{0}
\renewcommand{\theequation}{A.\arabic{appcnt}}
\setcounter{tmp}{1}
\clearpage
\hypertarget{app:mellin-link}{}
\chapter*{Appendix~A} 
\vspace{-0.5cm} 
\noindent
{\Huge \bf The Mellin transform}
\label{app:mellin}
\addcontentsline{toc}{chapter}{\numberline{\Alph{tmp}}
 The Mellin transform} 
\markboth{The Mellin transform}{}
\markright{APPENDIX A}{}
\vspace{1.3cm} 
 
R.H.~Mellin was a Finnish mathematician who studied 
under K.~Weierstrass. He is accredited as the developer of the integral transform 
\begin{equation}
\stepcounter{appcnt}
\label{app-mellin:mellin} 
 \textbf{M}\left[f\right](\om) \equiv
 \mathcal{F}(\om) = 
 \int_0^{+\infty} k^{\om-1} f(k) dk \; ,
\end{equation}
known as \emph{Mellin transform}.
Here $f$ is a complex function of the real variable $k$ 
and $\om$ is a complex variable. 
The inverse transform is given by
\begin{equation}
\stepcounter{appcnt}
\label{app-mellin:inverse}
 \textbf{M}^{-1}\left[F\right](k) \equiv
 f(k) = \frac{1}{2 \pi i}
 \int_{c-i\infty}^{c+i\infty} k^{-\om} \mathcal{F}(\om) d\om \; ,
\end{equation}
where the line integral is taken over the line $\om=c$ in the complex-$\om$ plane. 
Conditions under which this inversion is valid are given in the Mellin inversion theorem. In particular, if $\mathcal{F}(\om)$ is analytic in the strip 
$a < {\rm Re} \, \om < b$, and if it tends to zero uniformly as 
${\rm Im} \, \om \to \pm\infty$ for any real value $c \; | \; a < c < b$, 
then we can recover $f(k)$ from $\mathcal{F}(\om)$ via the inverse transform. 
The functions $\mathcal{F}(\om)$ and $f(k)$ are called a Mellin transform pair. 

There is a relation between the Mellin transform $\textbf{M}\left[f\right](\om)$ and the two-sided Laplace transform $\textbf{L}\left[f\right](\om)$. In fact, by letting $k=e^{-x}$, $dk=-e^{-x}dx$, the transform becomes
\begin{equation}
\stepcounter{appcnt}
\label{app-mellin:laplace} 
 \textbf{M}\left[f\right](\om) = 
 \int_{-\infty}^{+\infty} k^{\om x} f(e^{-x}) dx \equiv 
 \textbf{L}\left[f\right](\om) \; .
\end{equation}
Conversely, one can get the two-sided Laplace transform from the Mellin transform by
\begin{equation}
\stepcounter{appcnt}
\label{app-mellin:laplace_reverse}
 \textbf{L}\left[f\right](\om) = 
 \textbf{M}\left[f(-\ln x)\right](\om) \; .
\end{equation}
It is also possible to define the Fourier transform $\textbf{F}\left[f\right](\beta)$ in terms of the Mellin transform and vice versa by setting $\omega = \alpha + 2 \pi i \beta$, with 
$\alpha$ and $\beta$ real, and letting again $k=e^{-x}$:
\begin{equation}
\stepcounter{appcnt}
\label{app-mellin:fourier} 
 \textbf{M}\left[f\right](\om) = 
 \int_{-\infty}^{+\infty} 
  f(e^{-x}) e^{-\alpha x} e^{- 2 \pi i \beta x} dx \equiv 
 \textbf{F}\left[k^{\alpha}f(k)\right](\beta) \; .
\end{equation}

An important example of Mellin transform is the relation between the Riemann function $R(k)$ and the Riemann zeta function $\zeta(\om)$ 
(see Ref.~\cite{Riemann} for further details):
\begin{equation}
\stepcounter{appcnt}
\label{app-mellin:riemann}
 R(k) = \lim_{y \to + \infty} \frac{1}{2 \pi i}
  \int_{2-iy}^{2+iy} \frac{k^{\om}}{\om} \ln(\zeta(\om)) d\om \; ,
\end{equation}
and
\begin{equation}
\stepcounter{appcnt}
\label{app-mellin:riemann_zeta}
 \frac{\ln(\zeta(\om))}{\om} = 
 \int_1^{+\infty} R(k) k^{-\om-1} dk \; .
\end{equation}

 \section*{\Alph{tmp}.1 \;\, Properties of the Mellin transform}
 \label{appsec:mellin-properties}
 \addcontentsline{toc}{section}{\numberline{\Alph{tmp}.1}
  Properties of the Mellin transform} 
 \markboth{\Alph{tmp}.1 \;\, Properties of the Mellin transform}{}
 \markright{APPENDIX A}{}
 
 A list of some general properties of the Mellin transform is given below 
 (to know more, see for instance Refs.~\cite{Mellin_1,Mellin_2,Mellin_3,Mellin_4,Mellin_5,Mellin_6}).
 
 \begin{enumerate}
  \item \textbf{Scaling}
   \begin{equation}
   \stepcounter{appcnt}
    \textbf{M}\left[f(\xi k)\right](\om) = 
    \int_0^{+\infty} f(\xi k) k^{\om-1} d\om 
   \end{equation}
   \[ = 
    \xi^{-\om} \int_0^{+\infty} f(x) x^{\om-1} dx = 
    \xi^{-\om} \mathcal{F}(\om) \; .
   \]
  \item \textbf{Multiplication by $k^{\xi}$}
   \begin{equation}
   \stepcounter{appcnt}
    \textbf{M}\left[k^{\xi} f(k)\right](\om) = 
    \int_0^{+\infty} f(k) k^{\om+\xi-1} d\om =
    \mathcal{F}(\om+\xi) \; .
   \end{equation}
  \item \textbf{Raising the independent variable to a real power}
   \begin{equation}
   \stepcounter{appcnt}
    \textbf{M}\left[f(k^{\xi})\right](\om) = 
    \int_0^{+\infty} f(k^{\xi}) k^{\om-1} d\om 
   \end{equation}
   \[ = 
    \int_0^{+\infty} f(x) x^{\frac{x-1}{\xi}} 
     \left(\frac{1}{\xi}x^{\frac{1}{\xi}-1}\right) dx =
    \xi^{-1} \mathcal{F}\left(\frac{\om}{\xi}\right) \; , \; \xi > 0 \; .
   \]
  \item \textbf{Inverse of the independent variable}
   \begin{equation}
   \stepcounter{appcnt}
    \textbf{M}\left[k^{-1} f(k^{-1})\right](\om) = 
    \mathcal{F}(1-\om) \; .
   \end{equation}
  \item \textbf{Multiplication by $\ln(k)$}
   \begin{equation}
   \stepcounter{appcnt}
    \textbf{M}\left[\ln(k) \, f(k)\right](\om) = 
    \frac{d}{d\om} \mathcal{F}(\om) \; .
   \end{equation}
  \item \textbf{Multiplication by a power of $\ln(k)$}
   \begin{equation}
   \stepcounter{appcnt}
    \textbf{M}\left[(\ln(k))^n \, f(k)\right](\om) = 
    \frac{d^n}{d\om^n} \mathcal{F}(\om) \; .
   \end{equation}
  \item \textbf{Derivative}
   \begin{equation}
   \stepcounter{appcnt}
    \textbf{M}\left[f^{(n)}(k)\right](\om) = 
    (-1)^n \frac{\Gamma(\om)}{\Gamma(\om-n)} \mathcal{F}(\om-n) \; . 
   \end{equation}
  \item \textbf{Derivative multiplied by the independent variable}
   \begin{equation}
   \stepcounter{appcnt}
    \textbf{M}\left[k^n f^{(n)}(k)\right](\om) = 
    (-1)^n \frac{\Gamma(\om+n)}{\Gamma(\om)} \mathcal{F}(\om) \; . 
   \end{equation}
  \item \textbf{Integral}
   \begin{equation}
   \stepcounter{appcnt}
    \textbf{M}\left[\int_0^k f(k^\prime) dk^\prime\right](\om) = 
    -\frac{1}{\om} 
     \textbf{M}\left[\int_0^k f(k^\prime) dk^\prime\right](\om+1) \; . 
   \end{equation}
  \item \textbf{Convolution}
   \begin{equation}
   \stepcounter{appcnt}
    \textbf{M}\left[f(k) g(k)\right](\om) = 
    \frac{1}{2 \pi i} \int_{c-i\infty}^{c+i\infty}
     \mathcal{F}(\om^\prime) \mathcal{G}(\om-\om^\prime) d\om^\prime\; . 
   \end{equation}
  \item \textbf{Multiplicative convolution}
   \begin{equation}
   \stepcounter{appcnt}
    \textbf{M}\left[f \diamond g\right](\om) \equiv 
    \textbf{M}\left[\int_0^{+\infty} 
     f \left(\frac{k}{k^\prime}\right) 
     g(k^\prime) \frac{dk^\prime}{k ^\prime}\right](\om) = 
    \mathcal{F}(\om) \mathcal{G}(\om) \; . 
   \end{equation}
    
 \end{enumerate}

\renewcommand{\theequation}
             {\arabic{chapter}.\arabic{equation}}
\chapter{Mueller--Navelet jets}
\label{chap:mn-jets}

As we anticipated in the Introduction~\ref{chap:intro}, 
Mueller--Navelet jet production has been one of the so far most studied semi-hard processes, having allowed the possibility to define infrared-safe observables whose theoretical predictions (see for instance Refs.~\cite{Ducloue:2013bva,Caporale:2014gpa}) are in a very good agreement with experimental data~\cite{Khachatryan:2016udy}. 

The analysis given in this Chapter, devoted to address the open issues in the Mueller--Navelet sector, is based on 
the work done in Refs.~\cite{Caporale:2014gpa,Celiberto:2015yba,Celiberto:2016ygs} and presented in 
Refs.~\cite{Celiberto:2015mpa,Celiberto:2016vva}.

\begin{figure}[t]
 \centering
  \includegraphics[scale=0.7]{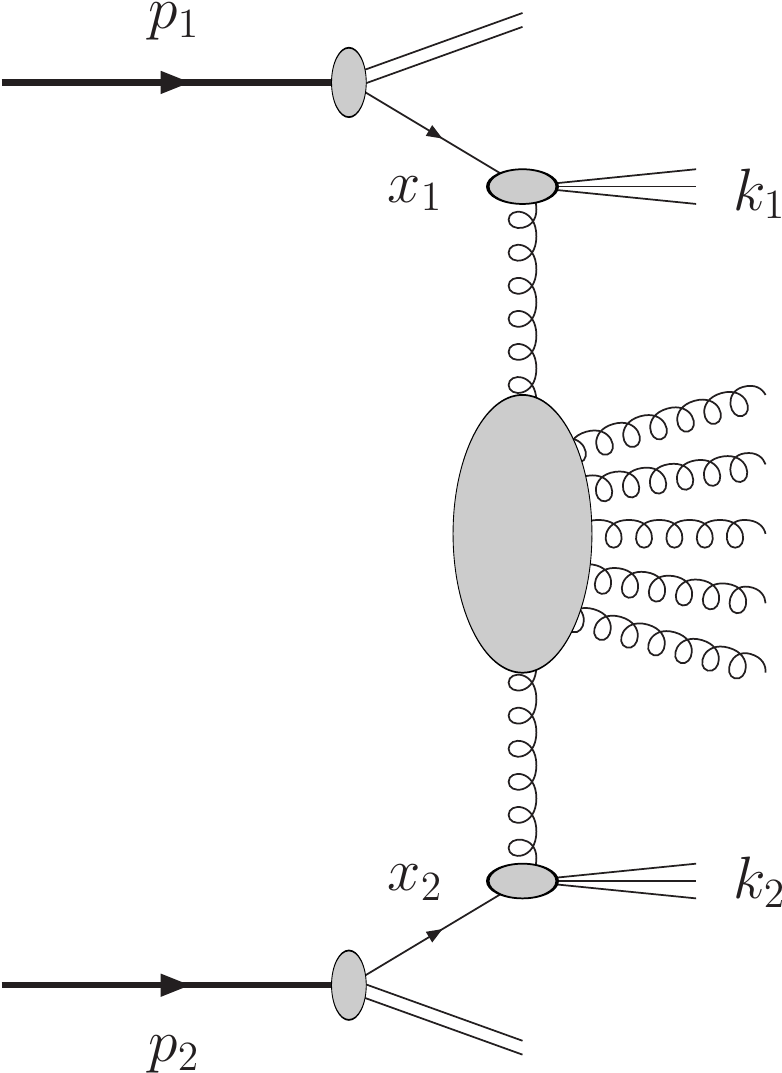}
 \caption[Mueller--Navelet jet production process 
          in multi-Regge kinematics]
 {
 Mueller--Navelet jet production process in multi-Regge kinematics.}
\label{fig:mn-jets}
\end{figure}

 \section{Theoretical framework} 
 \label{sec:mn-jets-theory}
 
 In this Section the BFKL cross section and the azimuthal corrections
 for the Mueller--Navelet jet process are presented.

  \subsection{Inclusive dijet production in proton-proton collisions}
  \label{sub:mn-jets-if}
  
  The reaction under exam is the inclusive production of two jets (a \emph{dijet} system) in proton-proton collisions
  \begin{eqnarray}
  \label{process-mn}
  p(p_1) + p(p_2) \to {\rm jet}(k_{J_1}) + {\rm jet}(k_{J_2})+ X \;,
  \end{eqnarray}
  where the two jets are characterised by high transverse momenta,
  $\vec k_{J_1}^2\sim \vec k_{J_2}^2\gg \Lambda_{\rm QCD}^2$ and large separation
  in rapidity; $p_1$ and $p_2$ are taken as Sudakov vectors (see Eq.~(\ref{sudakov_general})) satisfying
  $p_1^2=p_2^2=0$ and $2\left( p_1 p_2\right)=s$, working at leading twist and neglecting the proton mass and other power suppressed corrections.
  
  At LHC energies, the theoretical description of this reaction lies at the crossing point of two distinct approaches: collinear factorisation and BFKL resummation. On one side, at leading twist the process can be seen as the hard scattering of two partons, each emitted by one
  of the colliding hadrons according to the appropriate PDF, see Fig.~\ref{fig:mn-jets}. Collinear factorisation takes care to systematically resum the logarithms of the hard scale, through the standard DGLAP evolution of the PDFs and the fixed-order radiative corrections to the parton scattering cross section.
  The other resummation mechanism at work, justified by the large center-of-mass energy $\sqrt{s}$ available at the LHC, is the BFKL resummation of energy logarithms, which are so large to compensate the small QCD coupling and must therefore be accounted for to all orders of perturbation.
 
  The expression of the cross section (Eq.~(\ref{sigma-ff})), 
  which takes the form a convolution between two, process-dependent impact factors and a process-independent Green's function, is valid for a fully inclusive process, \emph{i.e.} without any particle/object identified/tagged in the final state. 
  By demanding the tagging of two forward jets, each of them produced in the fragmentation region of the respective parent proton, we are relaxing the inclusiveness condition 
  requested in Eq.~(\ref{sigma-ff}). This has an impact on the form of the process-dependent part of the cross section, namely the forward jet impact factors (also known as jet vertices). 
  \begin{figure}[t]
   \centering
      \includegraphics[scale=0.85]{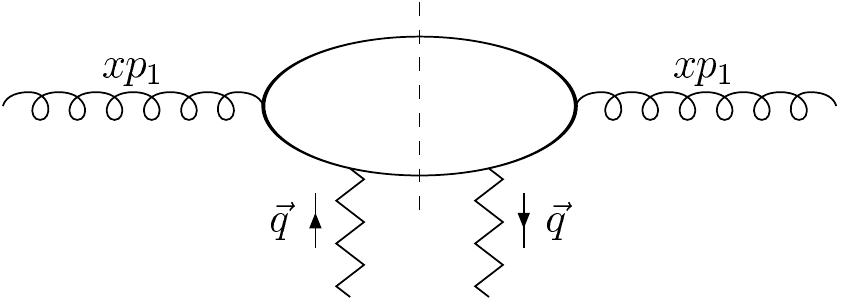}
      \includegraphics[scale=0.85]{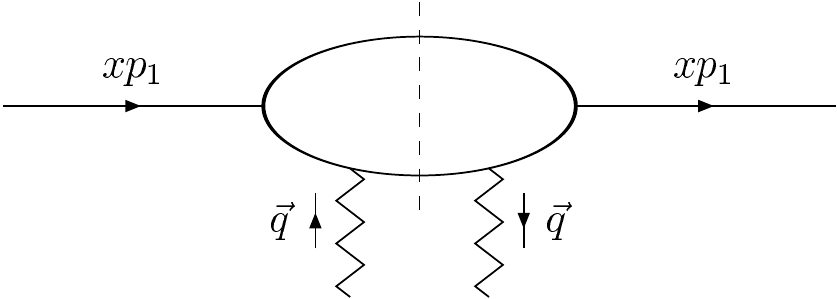}
    \caption[Forward parton impact factor]
    {Schematic view of the forward gluon (left) and the forward quark (right) impact factor. 
    Here $p_1$ is the proton momentum, $x_1$ is the fraction of proton momentum carried by the parton and $\vec q$ is the transverse momentum of the incoming Reggeised gluon.}
   \label{fig:if-parton}
  \end{figure}

\begin{figure}[t]
   \centering
      \includegraphics[scale=0.85]{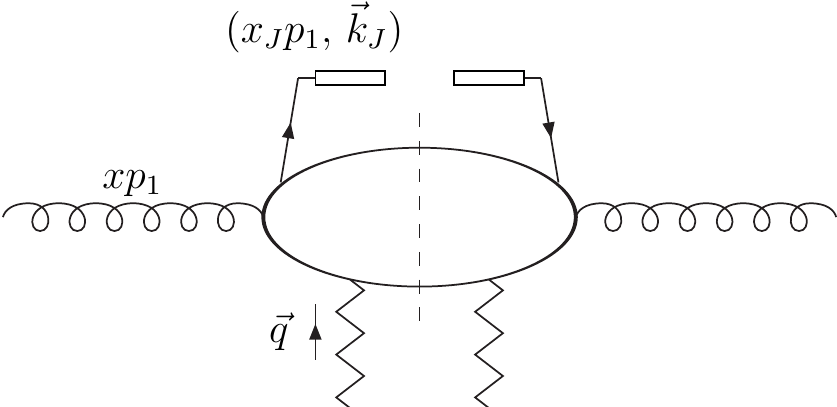}
      \includegraphics[scale=0.85]{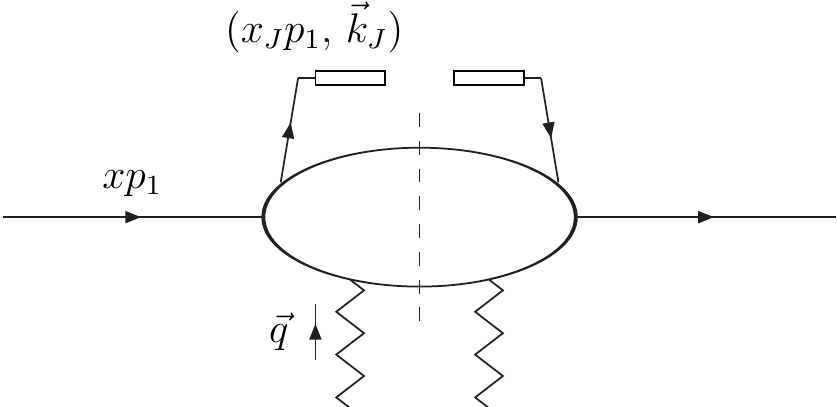}
    \caption[Forward jet impact factor]
    {Schematic view of the vertex for the forward jet production in the case of incoming gluon (left) or quark (right).
    Here $p_1$ is the proton momentum, $x$ is the
    fraction of proton momentum carried by the gluon/quark, $x_J p_1$ is the longitudinal jet momentum, $\vec k_J$ is the transverse jet momentum and $\vec q$ is the transverse momentum of the incoming Reggeised gluon.}
   \label{fig:if-jet}
  \end{figure}

  The starting point is provided by the impact factors for the colliding partons, calculated with NLO accuracy in Refs.~\cite{Fadin:1999de,Fadin:1999df} (see Fig.~\ref{fig:if-parton}). To obtain the impact factor for a tagged forward jet (see Fig.~\ref{fig:if-jet}), the first step is to `open' one of the integrations over the intermediate-state phase space to allow one parton to generate the jet~\footnote{This is achieved by introducing into the phase-space integration a suitably defined function which identifies the jet momentum with the momentum of one parton or with the sum of the two or more parton momenta when the jet is originated from the a multi-parton intermediate state}. Then, according to QCD collinear factorisation, take the convolution with the parent parton PDFs. The jet can be formed by one parton in LO and by one or two partons when the process is considered in NLO. In the simplest case, the jet momentum is identified with the momentum of the parton in the intermediate state $k$ by the following jet function~\cite{Ellis:1989vm}:
  \begin{equation}
  \label{jetF0}
   S_J^{(2)}(\vec k;x)=\delta(x-x_J)\delta^{(2)}(\vec k-\vec k_J)
   \; ,
  \end{equation}
  where $x$ is the fraction of proton momentum carried by the quark, $x_J$ is the longitudinal fraction of the jet momentum and $\vec k_J$ is the transverse jet momentum.
  We get the expression for the jet impact factor, differential with respect to the variables parameterising the jet phase space, at the LO level as
  \begin{equation}
   \label{if-jet-lo}
   \frac{d\Phi^{(0)}_J(\vec q\,)}{dx_J dk_J}=
    2\pi\alpha_s\sqrt{\frac{2C_F}{C_A}}
   \int_0^1 dx
   \left(\frac{C_A}{C_F}f_g(x)+\sum_{a=q, \bar q} f_{a}(x)\right)S_J^{(2)}
   (\vec q;x)\;,
  \end{equation}
  given as the sum of the gluon and all possible quark and antiquark PDF contributions $f_g(x)$, $f_a(x)$. In Eq.~(\ref{if-jet-lo}) $C_F$ is the colour factor associated with gluon emission from a quark, $C_F = (N_c^2-1)/(2N_c)$ 
  The last step to do is to project Eq.~(\ref{if-jet-lo}) onto the eigenfunctions (Eq.~(\ref{nuLLA})) of the LO BFKL kernel (\ref{KLLA}), \emph{i.e.} transfer to the $(\nu,n)$-representation (see Section~\ref{sub:bfkl-cross-section}). The expression for the LO forward jet vertex will be given in Eq.~(\ref{c1}) of Section~\ref{sub:mn-jets-blm}. 
  
  In the NLO case, both the one-loop virtual corrections to the amplitude with one parton state and the terms coming from the two-partons final-state amplitude have to be taken into account. In this last case, when the jet originates from a state of two partons, we need another jet selection function $S_J^{(3)}$, whose explicit form depends on the chosen jet algorithm. 
  We will use the NLO jet vertex calculated in the {\em small-cone} approximation~\cite{Furman:1981kf,Aversa:1988vb}, \emph{i.e.} for small jet-cone aperture in the rapidity-azimuthal angle plane, 
  which allow to get a simple analytic result in the $(\nu,n)$-representation (see Eq.~(\hyperlink{c11-jets}{B.1})).

  \subsection{Dijet cross section and azimuthal correlations}
  \label{sub:mn-jets-cs}
  
  In QCD collinear factorisation the cross section of the process~(\ref{process-mn})
  reads
  \beq
  \frac{d\sigma}{dx_{J_1}dx_{J_2}d^2k_{J_1}d^2k_{J_2}}
  =
  \sum_{r,s=q,{\bar q},g}\int_0^1 dx_1 \int_0^1 dx_2\ f_r\left(x_1,\mu_F\right)
  \ f_s\left(x_2,\mu_F\right)
  \eeq
  \[ =
  \frac{d{\hat\sigma}_{r,s}\left(x_1x_2s,\mu_F\right)}
  {dx_{J_1}dx_{J_2}d^2k_{J_1}d^2k_{J_2}} \; ,
  \]
  where the $r, s$ indices specify the parton types (quarks $q = u, d, s, c, b$;
  antiquarks $\bar q = \bar u, \bar d, \bar s, \bar c, \bar b$; or gluon $g$),
  $f_i\left(x, \mu_F \right)$ denotes the initial proton PDFs; $x_{1,2}$ are
  the longitudinal fractions of the partons involved in the hard subprocess,
  while $x_{J_{1,2}}$ are the jet longitudinal fractions; $\mu_F$ is the
  factorisation scale; $d\hat\sigma_{r,s}\left(x_1x_2s, \mu_F \right)$ is
  the partonic cross section for the production of jets and
  $x_1x_2s \equiv \hat s$ is the squared center-of-mass energy of the
  parton-parton collision subprocess (see Fig.~\ref{fig:mn-jets}).
  
  The cross section of the process can be presented as 
  (see Section~\ref{sub:bfkl-cross-section} for the details of the derivation)
  \beq
  \frac{d\sigma}
  {dy_{J_1}dy_{J_2}\, d|\vec k_{J_1}| \, d|\vec k_{J_2}|
  d\phi_{J_1} d\phi_{J_2}}
  =\frac{1}{(2\pi)^2}\left[{\cal C}_0+\sum_{n=1}^\infty  2\cos (n\phi )\,
  {\cal C}_n\right]\, ,
  \eeq
  where $\phi=\phi_{J_1}-\phi_{J_2}-\pi$, while ${\cal C}_0$ gives the total
  cross section and the other coefficients ${\cal C}_n$ determine the distribution
  of the azimuthal angle of the two jets.
  
  Since the main object of the present analysis is the impact of jet produced
  in the central region on azimuthal coefficients, we will concentrate just on one representation for ${\cal C}_n$, out of the many possible
  NLA-equivalent options (see Section.~\ref{ssub:bfkl_cs_representation} for a discussion).
  In particular, we will use the exponentiated representation
  together with the BLM optimisation method, whose details are given in Section~\ref{sec:bfkl_blm} on scale $\mu_R$ and the factorisation scale $\mu_F$. 
  In our calculation we will use the exact implementation of BLM method, 
  given in Eq.~(\ref{c_BLMmain}), together with the two approximate, semianalytic $(a)$ and $(b)$ cases (Eqs.~(\ref{casea}) and (\ref{caseb}), respectively), in order to keep contact with with 
  previous applications of BLM method where approximate approaches were used.

  \subsection{BLM scale setting} 
  \label{sub:mn-jets-blm}

  In this Section the expressions for the azimuthal coefficients ${\cal C}_n$, using the BLM prescription 
  (see Section~\ref{sec:bfkl_blm}) are given. 
  For the approximated $(a)$ (Eq.~(\ref{c_BLMa})) and $(b)$ (Eq.~(\ref{c_BLMb})) cases, 
  we present also the expressions in the fixed-order DGLAP approach at the NLO, which will be used in the phenomenology Section~\ref{sec:mn-jets-BFKL-vs-DGLAP}. The ${\cal C}_n^{\rm DGLAP}$ coefficients are nothing but the truncation of the respective  BFKL expressions ${\cal C}_n$ up to inclusions of NLO terms.

  Introducing, for the sake of brevity, the definitions
  \begin{equation}
  \label{Y-x_J}
   Y=y_{J_1}-y_{J_2}=\ln\frac{x_{J_1}x_{J_2}s}{|\vec k_{J_1}||\vec k_{J_2}|}\;,
   \;\;\;\;\;
   Y_0=\ln\frac{s_0}{|\vec k_{J_1}||\vec k_{J_2}|}\;,
  \end{equation}
  we will present in what follows 
  the three different expressions for the coefficients ${\cal C}_n$.
  
  $\bullet$ case ``exact'' 
  
  We remember that the BLM optimal scale $\mu_R^{\rm BLM}$ is defined as the value of $\mu_R$ that makes all contributions to the considered observables which are  
  proportional to the QCD $\beta-$ function, $\beta_0$, vanish, such that Eq.~(\ref{c_nnnbeta}) is satisfied.
  After that we have the following expression for our observables:
  \beq\label{blm_exact-jets}
  {\cal C}_n=\frac{x_{J_1}x_{J_2}}{|\vec k_{J_1}||\vec k_{J_2}|}
  \int_{-\infty}^{+\infty}d\nu \ e^{(Y-Y_0) \bar \alpha_s^{\rm MOM}(\mu_R^{\rm BLM})
  \left[\chi+\bar\alpha_s^{\rm MOM}(\mu_R^{\rm BLM})
  \left(\bar\chi+\frac{T^{\rm conf}}{N_c}\chi\right)
  \right]}
  \end{equation}
  \[ \times \,
  (\alpha_s^{\rm MOM}(\mu_R^{\rm BLM}))^2
  c_1(n,\nu,|\vec k_{J_1}|, x_{J_1}) c_2(n,\nu,|\vec k_{J_2}|,x_{J_2})
  \]
  \[ \times \,
  \left[1 + \alpha_s^{\rm MOM}(\mu_R^{\rm BLM})
  \left\{\frac{\bar c_1^{\left(1\right)}(n,\nu,|\vec k_{J_1}|,x_{J_1})}
  {c_1(n,\nu,|\vec k_{J_1}|, x_{J_1})}
  +\frac{\bar c_2^{\left(1\right)}(n,\nu,|\vec k_{J_2}|, x_{J_2})}
  {c_2(n,\nu,|\vec k_{J_2}|, x_{J_2})}+\frac{2T^{\rm conf}}{N_c}\right\}\right] \; .
  \]
  In the above equation, $\alpha_s^{\rm MOM}$ is the QCD coupling 
  in the physical momentum subtraction (MOM) scheme, related to 
  $\alpha_s^{\overline{\rm MS}}$ by the finite renormalisation
  given in Eq.~(\ref{scheme}), while 
  ${\bar \alpha_s^{\rm MOM}} = N_c/\pi \, \alpha_s^{\rm MOM}$ as in Eq.~(\ref{baral}), with $N_c$ the number of colours.
  Then,
  \beq
  \label{c1}
  c_1(n,\nu,|\vec k|,x)=2\sqrt{\frac{C_F}{C_A}}
  (\vec k^{\,2})^{i\nu-1/2}\,\left(\frac{C_A}{C_F}f_g(x,\mu_F)
  +\sum_{a=q,\bar q}f_a(x,\mu_F)\right)
  \eeq
  and
  \beq
  \label{c2}
  c_2(n,\nu,|\vec k|,x)=\biggl[c_1(n,\nu,|\vec k|,x) \biggr]^* \;,
  \eeq
  are the LO jet vertices in the $\nu$-representation (see Eq.~(\ref{if-jet-lo}) for the corresponding expression in the momentum space) and $\chi = \chi(n,\nu)$ is the eigenvalue of the LO BFKL kernel (Eq.~(\ref{KLLA})). 
  Note that, since $c_{1,2}$ do not depend on $\nu$, 
  the $f(\nu)$ function, whose general expression 
  is given in Eq.~(\ref{f_nu_general}), is zero 
  for this process.
  The remaining objects
  are related to the NLO corrections of the BFKL kernel, ($\bar \chi(n,\nu)$,
  given in Eq.~(\ref{chibar})) and of the jet vertices in
  the small-cone approximation ($c_{1,2}^{(1)}(n,\nu,|\vec k_{J_{1,2}}|, x_{J_{1,2}})$,
  given in Eq.~(\hyperlink{c11-jets}{B.1}) of Appendix~\hyperlink{app:jet-nlo-if-link}{B}.
  The functions $\bar c_{1,2}^{(1)}(n,\nu,|\vec k_{J_2}|, x_{J_2})$ are the same as
  $c_{1,2}^{(1)}(n,\nu,|\vec k_{J_{1,2}}|, x_{J_{1,2}})$ with all terms proportional to
  $\beta_0$ removed.
  
  $\bullet$ case $(a)$
  \[
  (\mu_{R,a}^{\rm BLM})^2 = k_{J_1} k_{J_2} \exp\left[2\left(1+\frac{2}{3}I\right)
  -\frac{5}{3}\right]\;,
  \]
  with 
  \beq\label{casea-jets}
  {\cal C}_n^{\rm (a)}= \frac{x_{J_1}x_{J_2}}{|\vec k_{J_1}||\vec k_{J_2}|}
  \int_{-\infty}^{+\infty}d\nu \ e^{(Y-Y_0)
  \left[\bar \alpha_s^{\rm MOM}(\mu_{R,a}^{\rm BLM})\chi
  + (\bar{\alpha}_s^{\rm MOM}(\mu_{R,a}^{\rm BLM}))^2
  \left( \bar \chi+\frac{T^{\rm conf}}{N_c}\chi
  - \frac{\beta_0}{8N_c}\chi^2\right)\right]}
  \eeq
  \[ \times \,
  (\alpha_s^{\rm MOM}(\mu_{R,a}^{\rm BLM}))^2
  c_1(n,\nu,|\vec k_{J_1}|, x_{J_1}) c_2(n,\nu,|\vec k_{J_2}|,x_{J_2})
  \]
  \[ \times \,
  \left[1 + \alpha_s^{\rm MOM}(\mu_{R,a}^{\rm BLM})
  \left\{\frac{\bar c_1^{\left(1\right)}(n,\nu,|\vec k_{J_1}|,x_{J_1})}
  {c_1(n,\nu,|\vec k_{J_1}|, x_{J_1})}
  +\frac{\bar c_2^{\left(1\right)}(n,\nu,|\vec k_{J_2}|, x_{J_2})}
  {c_2(n,\nu,|\vec k_{J_2}|, x_{J_2})}+\frac{2T^{\rm conf}}{N_c}\right\}\right] \; ;
  \]

  \beq
  \label{casea-dglap-jets}
  {\cal C}_n^{\rm DGLAP \, (a)}  = \frac{x_{J_1}x_{J_2}}{|\vec k_{J_1}|
  |\vec k_{J_2}|}\int_{-\infty}^{+\infty}d\nu
  \ \asq 
  c_1(n,\nu,|\vec k_{J_1}|, x_{J_1}) c_2(n,\nu,|\vec k_{J_2}|,x_{J_2})
  \eeq
  \[ \times \,  
  \left[ 1 + \frac{2}{\pi} \as T^{\rm conf}
  + \asb \left( Y - Y_0 \right) \chi \right.
  \]
  \[
  \left.
  + \, \as \left(\frac{\bar c_1^{\left(1\right)}
  (n,\nu,|\vec k_{J_1}|,x_{J_1})}{c_1(n,\nu,|\vec k_{J_1}|, x_{J_1})}
  +\frac{\bar c_2^{\left(1\right)}(n,\nu,|\vec k_{J_2}|, x_{J_2})}
  {c_2(n,\nu,|\vec k_{J_2}|, x_{J_2})}\right)\right] \;,
  \]
  
  $\bullet$ case $(b)$ 
  \[
  (\mu_{R,b}^{\rm BLM})^2 = k_{J_1} k_{J_2} \exp\left[2\left(1+\frac{2}{3}I\right)
  -\frac{5}{3}+\frac{1}{2}\, \chi(n,\nu)\right]\;,
  \]
  with
  \beq\label{caseb-jets}
  {\cal C}_n^{(b)}= \frac{x_{J_1}x_{J_2}}{|\vec k_{J_1}||\vec k_{J_2}|}
  \int_{-\infty}^{+\infty}d\nu \ e^{(Y-Y_0)
  \left[\bar \alpha_s^{\rm MOM}(\mu_{R,b}^{\rm BLM})\chi
  + (\bar{\alpha}_s^{\rm MOM}(\mu_{R,b}^{\rm BLM}))^2
  \left( \bar \chi+\frac{T^{\rm conf}}{N_c}\chi\right)\right]}
  \eeq
  \[ \times \,
  (\alpha_s^{\rm MOM}(\mu_{R,b}^{\rm BLM}))^2
  c_1(n,\nu,|\vec k_{J_1}|, x_{J_1}) c_2(n,\nu,|\vec k_{J_2}|,x_{J_2})
  \]
  \[ \times \,
  \left[1 + \alpha_s^{\rm MOM}(\mu_{R,b}^{\rm BLM})
  \left\{\frac{\bar c_1^{\left(1\right)}(n,\nu,|\vec k_{J_1}|,x_{J_1})}
  {c_1(n,\nu,|\vec k_{J_1}|, x_{J_1})}
  +\frac{\bar c_2^{\left(1\right)}(n,\nu,|\vec k_{J_2}|, x_{J_2})}
  {c_2(n,\nu,|\vec k_{J_2}|, x_{J_2})}+\frac{2T^{\rm conf}}{N_c}
  + \frac{\beta_0}{4N_c}\chi \right\}\right] \; ,
  \]
  
  \begin{equation}
  \label{caseb-dglap-jets}
  {\cal C}_n^{\rm DGLAP \, (b)}  = \frac{x_{J_1}x_{J_2}}{|\vec k_{J_1}|
  |\vec k_{J_2}|}\int_{-\infty}^{+\infty}d\nu
  \ \asq 
  c_1(n,\nu,|\vec k_{J_1}|, x_{J_1}) c_2(n,\nu,|\vec k_{J_2}|,x_{J_2})
  \end{equation}
  \[ \times \,
  \left[ 1 + \alpha_s\left(\mu_R\right)\left(\frac{\beta_0}{4\pi}
  \chi + 2\frac{T^{\rm conf}}{\pi}\right)
  + \asb \left( Y - Y_0 \right) \chi \right.
  \]
  \[
  \left.
  + \, \as \left(\frac{\bar c_1^{\left(1\right)}
  (n,\nu,|\vec k_{J_1}|,x_{J_1})}{c_1(n,\nu,|\vec k_{J_1}|, x_{J_1})}
  +\frac{\bar c_2^{\left(1\right)}(n,\nu,|\vec k_{J_2}|, x_{J_2})}
  {c_2(n,\nu,|\vec k_{J_2}|, x_{J_2})}\right)\right] \;.
  \]
 
  Note that, in the above equations the scale $s_0$ entering $Y_0$ is the 
  artificial energy scale introduced in the BFKL approach to perform the Mellin transform from the $s$-space to the complex angular momentum plane and cancels 
  in the full expression, up to terms beyond the NLA. In our analysis it will 
  always be fixed at the ``natural'' value $Y_0=0$, given by the kinematical of 
  Mueller--Navelet process.
  
  Although the final expressions in Eqs.~(\ref{blm_exact-jets}), (\ref{casea-jets}), 
  (\ref{casea-dglap-jets}), (\ref{caseb-jets}), and  
  (\ref{caseb-dglap-jets}) are given in terms of $\alpha^{\rm MOM}$ in the MOM scheme, it is possible to use analogous 
  expressions in the $\overline{\rm MS}$ scheme. 
  The way to do that 
  is to start from the general expressions, then perform the 
  change of scheme $\overline{\rm MS}$ $\to$ MOM as an intermediate step, and finally, after setting the BLM scales, go back again to the $\overline{\rm MS}$ scheme. 
  From a practical point of view, one obtains the expressions 
  in the $\overline{\rm MS}$ scheme, starting from MOM, by making the change
  \begin{equation}
  \label{Tconf_to_Tbeta}
   T^{\rm conf} \,\to\, - T^{\beta} \; ,
  \end{equation} 
  with $T^{\beta}$ given in Eq.~(\ref{T_Tbeta_Tconf}), 
  in the expressions cited above. 
  In Sections~\ref{sec:mn-jets-theory-vs-experiment} 
  and~\ref{sec:mn-jets-BFKL-vs-DGLAP} 
  we will give predictions for our observables 
  in the $\overline{\rm MS}$ scheme.

  \subsection{Integration over the final-state phase space} 
  \label{sec:mn-jets-phase-space}
  
  In order to match the kinematical cuts used by the CMS collaboration (see for instance Ref.~\cite{Khachatryan:2016udy}), we will
  consider the \emph{integrated coefficients} given by
  \begin{equation}
  \label{Cm_int-mn-jets}
   \int_{y_{J_2}^{\rm min}}^{y_{J_2}^{\rm max}}dy_{J_2}\int_{k_{J_1}^{\rm min}}^{\infty}dk_{J_1}
   \int_{k_{J_2}^{\rm min}}^{\infty}dk_{J_2} \, \delta\left(y_{J_1}-y_{J_2}-Y\right)
  \end{equation}
  \[ \times \,
   {\cal C}_n\left(y_{J_1},y_{J_2},k_{J_1},k_{J_2} \right)
  \]
  and their ratios $R_{nm}\equiv C_n/C_m$. 
  Among them, the ratios of the form $R_{n0}$ have a simple physical interpretation, being the azimuthal
  correlations $\langle \cos(n\phi)\rangle$.
  We will take jet rapidities in the range delimited by $y_{J_1}^{\rm min}=y_{J_2}^{\rm min}=-4.7$ and 
  $y_{J_1}^{\rm max}=y_{J_2}^{\rm max}=4.7$ 
  and study the dependence of the $R_{nm}$ ratios 
  as function of the jet rapidity separation $Y \equiv y_{J_1}-y_{J_2}$.
  Concerning the jet transverse momenta $k_{J_{1,2}}$, differently from most previous analyses, we make several different choices, which include \emph{asymmetric} cuts (see the next three Sections for further details).
  The jet-cone size $R$ entering the NLO-jet vertices is fixed at the value $R=0.5$ and, as anticipated,
  $Y_0=0$. Finally, we will consider two characteristic values for the center-of-mass energy, \emph{i.e.} $\sqrt s=7$ TeV, for which experimental anaylises with \emph{symmetric} configuration for the outgoing jet momenta
  already exist (see Section~\ref{sec:mn-jets-theory-vs-experiment}), and $\sqrt s=13$ TeV.

 \section{Theory versus experiment} 
 \label{sec:mn-jets-theory-vs-experiment}
 
 In this Section we present the analysis of 
 Ref.~\cite{Caporale:2014gpa}, in which predictions 
 for the $R_{10}$, $R_{20}$, $R_{30}$, $R_{21}$ and $R_{32}$ 
 are given and compared with recent CMS data at $7$ TeV~\cite{Khachatryan:2016udy}. 
 BLM scale optimisation, in both variants $(a)$ (Eq.~(\ref{casea-jets})) and $(b)$ (Eq.~(\ref{caseb-jets})) 
 is used, while final calculations are done in the $\overline{\rm MS}$. We remember that the expressions above cited are given 
 in the MOM scheme and it is possible to obtain the analogous 
 ones in the $\overline{\rm MS}$ through the substitution 
 $T^{\rm conf} \to - T^{\beta}$, with $T^{\rm conf}$ and $T^{\beta}$ given in Eq.~(\ref{T_Tbeta_Tconf}). 
 
\begin{figure}[H]
  \centering
    \includegraphics[scale=0.38]{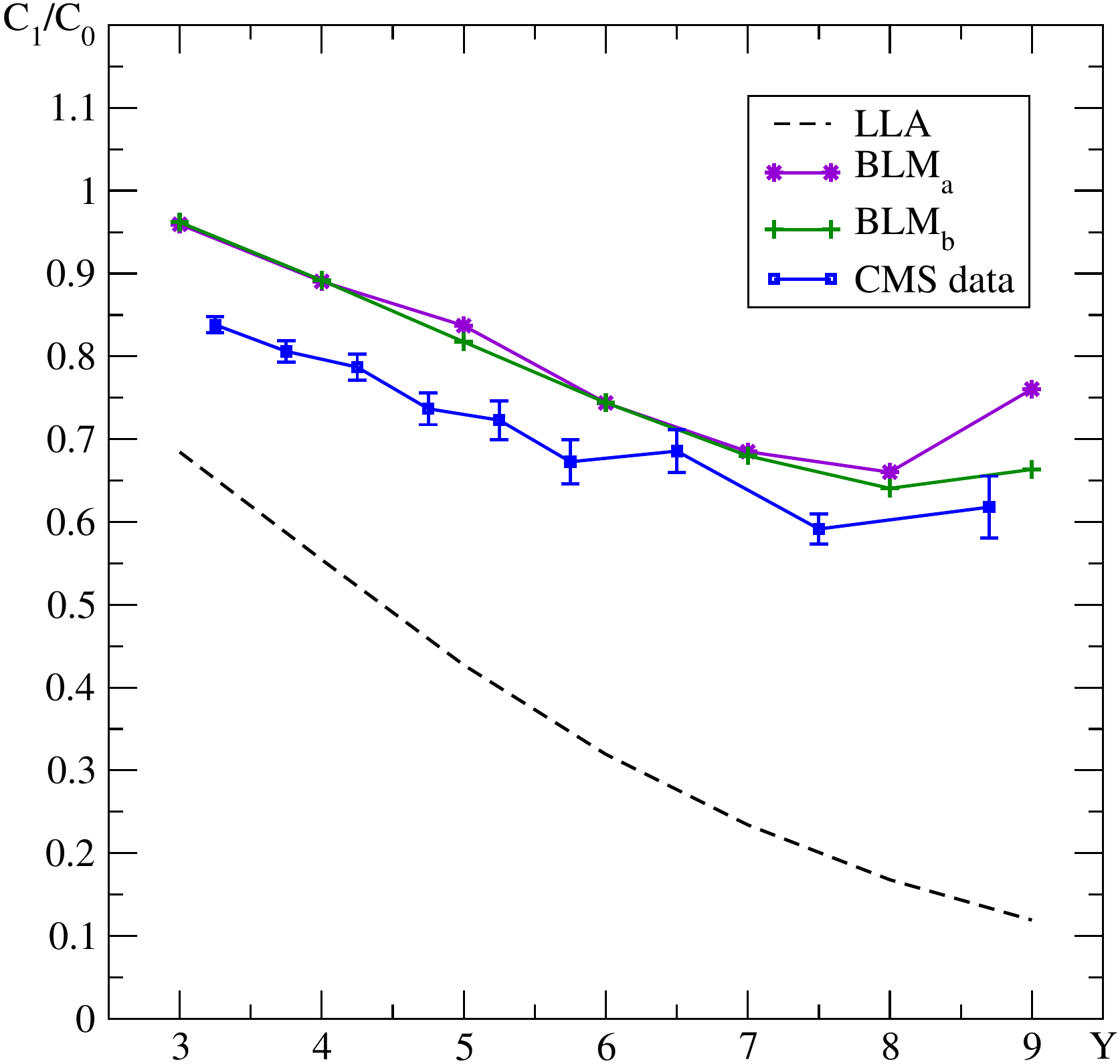}
    \includegraphics[scale=0.38]{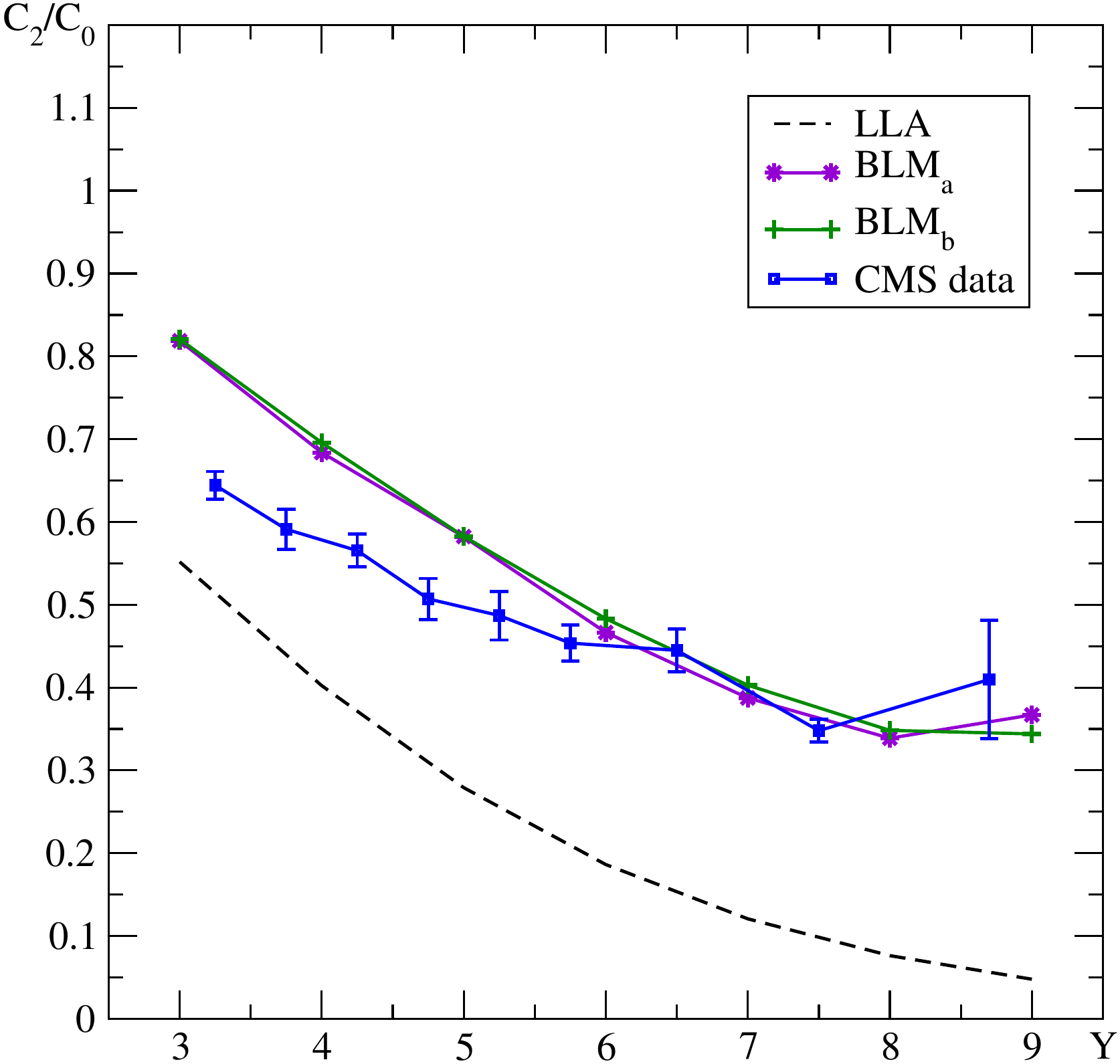}

    \includegraphics[scale=0.38]{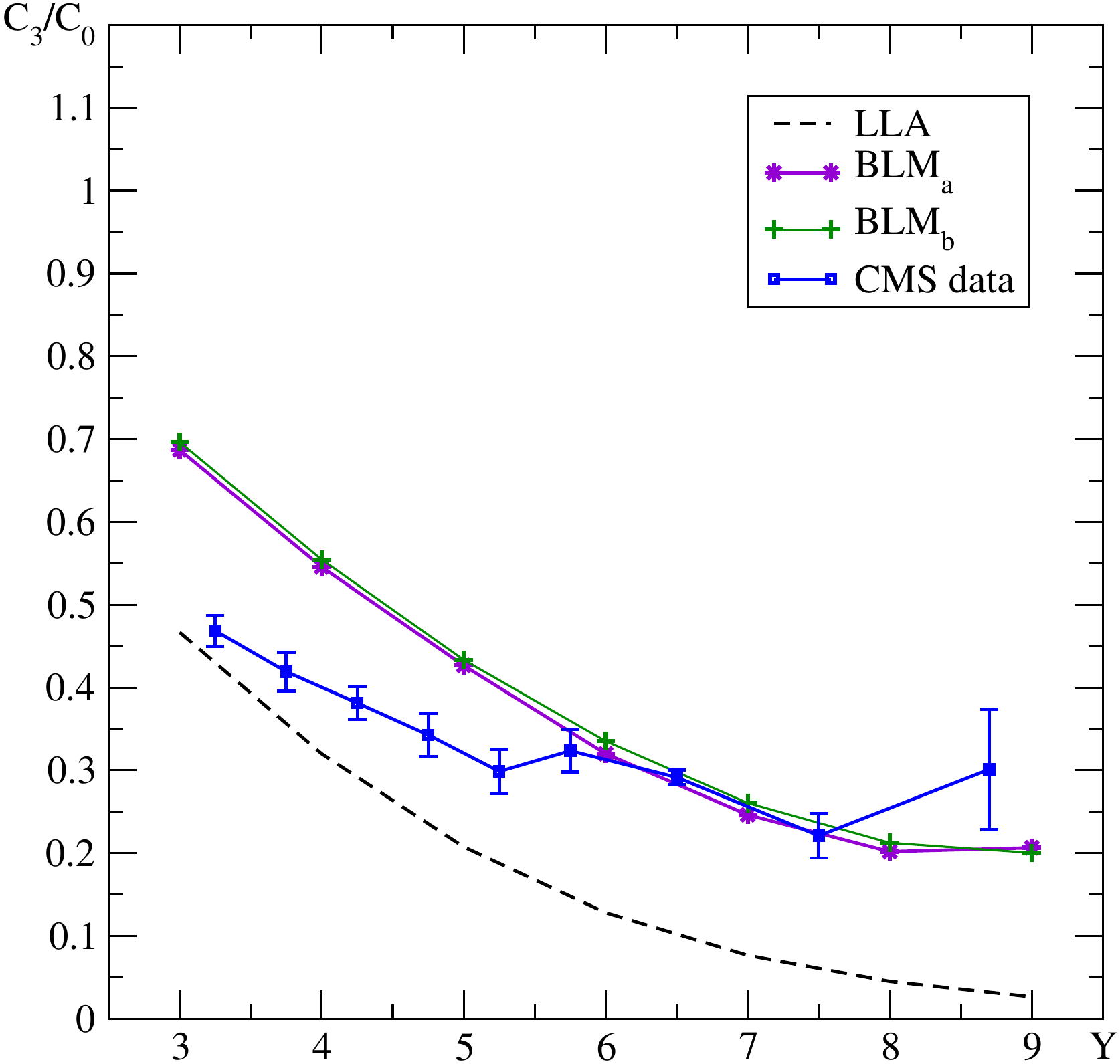}
    \includegraphics[scale=0.38]{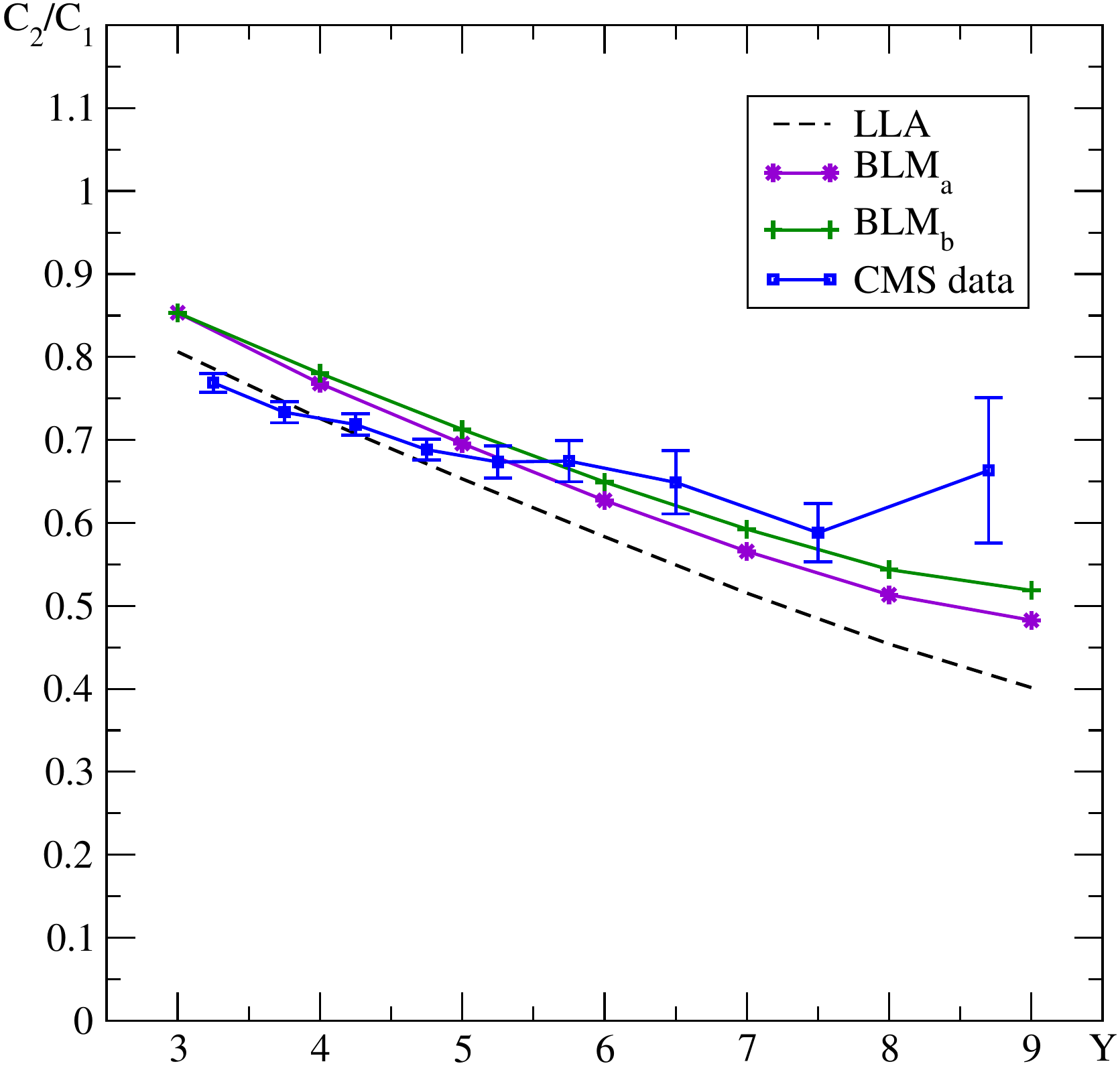}

    \includegraphics[scale=0.38]{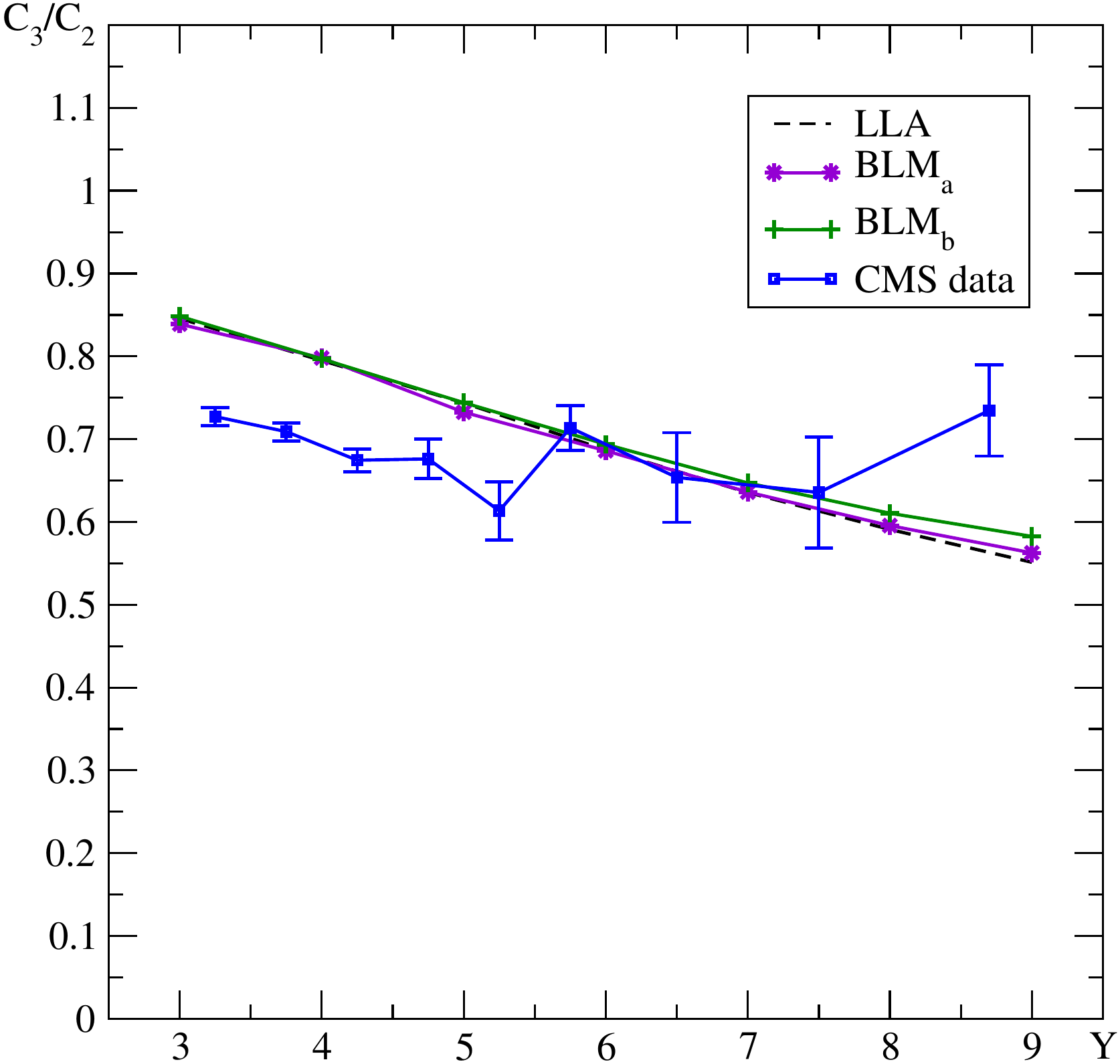}
  \caption[Comparison of dijet BFKL predictions
           with CMS data at 7 TeV]
  {$Y$-dependence of 
   $R_{10}$, $R_{20}$, $R_{30}$, $R_{21}$, and $R_{32}$ 
   at $\sqrt{s} = 7$ TeV~\cite{Khachatryan:2016udy} 
   in the \emph{symmetric} configuration $k_{J_{1,2}}^{\rm min} = 35$ GeV. 
   Results were obtained with the two variants $(a)$ (Eq.~(\ref{casea-jets})) and $(b)$ (Eq.~(\ref{caseb-jets})) 
   of the BLM method. The dashed line gives the LLA BFKL result.}
\label{fig:tve_BLM}
\end{figure}

 Results are reported in Table~\ref{tab:tve_ratios_BLM}
 and in Fig.~\ref{fig:tve_BLM}. 
 We clearly see that the pure LLA calculations (\emph{i.e.} 
 considering just the LO kernel contribution and neglecting the NLO corrections to the impact factors) overestimate the decorrelation by far in all $R_{nm}$ ratios. 
 Introducing NLA BFKL corrections and using the BLM method 
 we can see that, except for the ratio $C_1/C_0$,
 the agreement with experimental data becomes very good, for both variants $(a)$ and $(b)$, at the larger values of $Y$. 
 
 Meanwhile, it would be also useful to address, on the experimental side, some
 possible issues which could be sources of mismatch with the way
 in which Mueller--Navelet jets are defined in theory and that are not
 easy to be revealed in the comparison with theoretical predictions,
 for being the latter affected in their turn by systematic effects of the same
 amount. We list below two of them.
 
 \begin{itemize}
 
 \item The use of \emph{symmetric} cuts in the values of $k_{J_i}^{\rm min}$
 maximises the contribution of the Born term in $C_0$, which is present
 for back-to-back jets (see Fig.~\ref{fig:mn-jets-lla}) only and is expected to be large, therefore making
 less visible the effect of the BFKL resummation in all observables involving
 $C_0$. The use of \emph{asymmetric} cuts can reduce the contribution of the Born
 term and enhance effects with additional undetected hard gluon radiation,
 which makes the visibility of BFKL effect more clear in comparison to the 
 descriptions based on fixed-order DGLAP approach.
 
 \item In data analysis defining the $Y$ value for a given final state with two 
 jets, the rapidity of one of the two jets could be so small, say 
 $|y_{J_i}|\lesssim 2$, that this jet is actually produced in the central 
 region, rather than in one of the two forward regions. The longitudinal 
 momentum fractions of the parent partons that generate a central jet are very 
 small, and one can naturally expect sizable corrections to the vertex of this 
 jet, due to the fact that the collinear factorisation approach used in the
 derivation of the result for jet vertex is not designed for the region of 
 small $x$. 
 
 \end{itemize}
 
 These issues are addressed in Sections~\ref{sec:mn-jets-BFKL-vs-DGLAP} and~\ref{sec:mn-jets-rapidity}, respectively. 
 
 \newpage
 
 \begin{table}[H]
  \centering
  \caption[Dijet BFKL predictions at 7 TeV]
  {$R_{10}$, $R_{20}$, $R_{30}$, $R_{21}$, and $R_{32}$
   with the BLM method, in both variants $(a)$ (Eq.~(\ref{casea-jets})) and $(b)$ (Eq.~(\ref{caseb-jets})).}
  \label{tab:tve_ratios_BLM}
  \begin{tabular}{c|cc|cc|cc|cc|cc}
   \hline\noalign{\smallskip}
   & \multicolumn{2}{c|}{$C_1/C_0$} & \multicolumn{2}{c|}{$C_2/C_0$}
   & \multicolumn{2}{c|}{$C_3/C_0$} & \multicolumn{2}{c|}{$C_2/C_1$}
   & \multicolumn{2}{c}{$C_3/C_2$} \\
   $Y$ & $(a)$ & $(b)$ & $(a)$ & $(b)$ & $(a)$ & $(b)$ & $(a)$ & $(b)$
   & $(a)$ & $(b)$ \\
   \noalign{\smallskip}\hline\noalign{\smallskip}
   3 & 0.960 & 0.962 & 0.819 & 0.821 & 0.687 & 0.696 & 0.853 & 0.853 & 0.839 & 0.848 \\
   4 & 0.890 & 0.892 & 0.684 & 0.696 & 0.548 & 0.555 & 0.768 & 0.780 & 0.798 & 0.797 \\
   5 & 0.837 & 0.818 & 0.582 & 0.587 & 0.427 & 0.434 & 0.696 & 0.713 & 0.733 & 0.744 \\
   6 & 0.744 & 0.744 & 0.447 & 0.483 & 0.320 & 0.335 & 0.627 & 0.649 & 0.686 & 0.694 \\
   7 & 0.685 & 0.680 & 0.387 & 0.403 & 0.246 & 0.261 & 0.566 & 0.593 & 0.636 & 0.647 \\
   8 & 0.660 & 0.641 & 0.339 & 0.348 & 0.202 & 0.213 & 0.513 & 0.544 & 0.596 & 0.611 \\
   9 & 0.760 & 0.663 & 0.367 & 0.344 & 0.207 & 0.201 & 0.483 & 0.519 & 0.563 & 0.583 \\
   \noalign{\smallskip}\hline
  \end{tabular}
 \end{table}
 

 \section{BFKL versus high-energy DGLAP} 
 \label{sec:mn-jets-BFKL-vs-DGLAP}

  \subsection{Motivation}
  \label{sub:mn-jets-DGLAP-motivation}

\begin{figure}[t]
   \centering
    \includegraphics[scale=0.7]{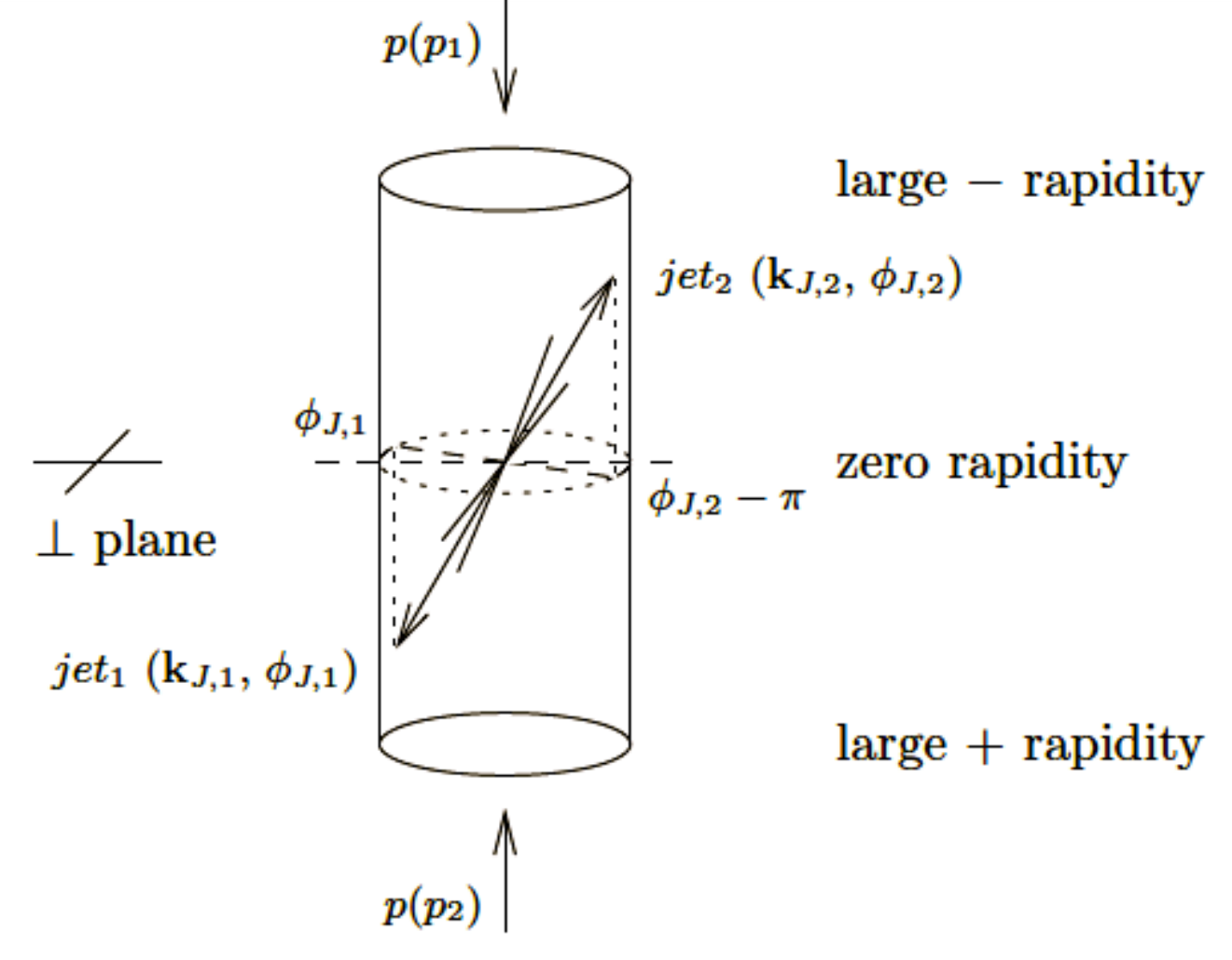}
   \caption[Mueller--Navelet jets at LLA: a back-to-back dijet reaction]
   {Mueller--Navelet jets at LLA: a back-to-back dijet reaction. Picture from Ref.~\cite{Colferai:2010wu}.}
   \label{fig:mn-jets-lla}
  \end{figure}

  As we saw at the end of Section~\ref{sec:mn-jets-theory-vs-experiment}, the effect of Born contribution to the cross section $C_0$, present only for back-to-back jets (see Fig.~\ref{fig:mn-jets-lla}), is maximised when \emph{symmetric}
  cuts in the values of the forward jet transverse
  momenta are used; on the contrary, in the case of \emph{asymmetric} cuts,
  the Born term is suppressed and the effects of the additional undetected 
  hard gluon radiation is enhanced, thus making more visible the BFKL
  resummation, in comparison to descriptions based on the fixed-order DGLAP 
  approach, in all observables involving $C_0$.

  For this purpose, we compare predictions for several azimuthal correlations 
  and their ratios obtained, on one side, by a fixed-order DGLAP calculation
  at the NLO and, on the other side, by BFKL resummation in the NLA. 

  We remember that our implementation of 
  the NLO DGLAP calculation is an approximate one. 
  We just use here 
  NLA BFKL expressions, given in Eqs.~(\ref{casea-dglap-jets}) and~(\ref{caseb-dglap-jets}), for the observables that are truncated to the 
  ${\cal O}\left(\alpha_s^3\right)$ order.
  In this way we take into account the leading power asymptotic of the exact NLO 
  DGLAP prediction and neglect terms that are suppressed by the inverse powers of 
  the energy of the parton-parton collisions. Such approach is legitimate in the 
  region of large $Y$ which we consider here. The exact implementation of NLO 
  DGLAP for Mueller--Navelet jets is important, because it allows to understand 
  better the region of applicability of our approach, but it requires more 
  involved Monte Carlo calculations.
  We use the BLM scheme in both semianalytic $(a)$ (Eq.~(\ref{c_BLMa})) and $(b)$ (Eq.~(\ref{c_BLMb})) cases in 
  order to compare BFKL (Eqs.~(\ref{casea-jets}) and (\ref{caseb-jets}) with DGLAP (Eqs.~(\ref{casea-dglap-jets}) and~(\ref{casea-dglap-jets}) predictions.
  As done in Section~\ref{sec:mn-jets-theory-vs-experiment}, we perform all calculations in the $\overline{\rm MS}$ scheme. 
  We remember that all the four expressions above cited are given in the MOM scheme and it is possible to obtain the analogous ones in the $\overline{\rm MS}$ through the substitution $T^{\rm conf} \to - T^{\beta}$, with $T^{\rm conf}$ and  $T^{\beta}$ given in~(\ref{T_Tbeta_Tconf}).
  
  Another important benefit from the use of \emph{asymmetric} cuts, pointed out 
  in~\cite{Ducloue:2014koa}, is that the effect of violation of the 
  energy-momentum conservation in the NLA is strongly suppressed with respect 
  to what happens in the LLA.

  \subsection{Results and discussion}
  \label{sub:mn-jets-DGLAP-results}
  
  We study the $Y$-dependence of ratios $R_{nm}$ of the integrated coefficients given in Eq.~(\ref{Cm_int-mn-jets}), 
  fixing the center-of-mass energy at $\sqrt s=7$ TeV and
  making two \emph{asymmetric} choices for the jet transverse momenta:
  
  \begin{enumerate}
  \item 
  \label{cut1-jets-DGLAP}
  $
   k_{J_1}^{\rm min} = 35\, \text{GeV}, 
   \quad\quad\quad
   k_{J_2}^{\rm min} = 45\, \text{GeV};
  $
  \item 
  $
   k_{J_1}^{\rm min} = 35\, \text{GeV}, 
   \quad\quad\quad
   k_{J_2}^{\rm min} = 50\, \text{GeV}.
  $
  \end{enumerate}
  
\begin{figure}[H]
  \vspace{-0.50cm}
  \centering
    \includegraphics[scale=0.37]{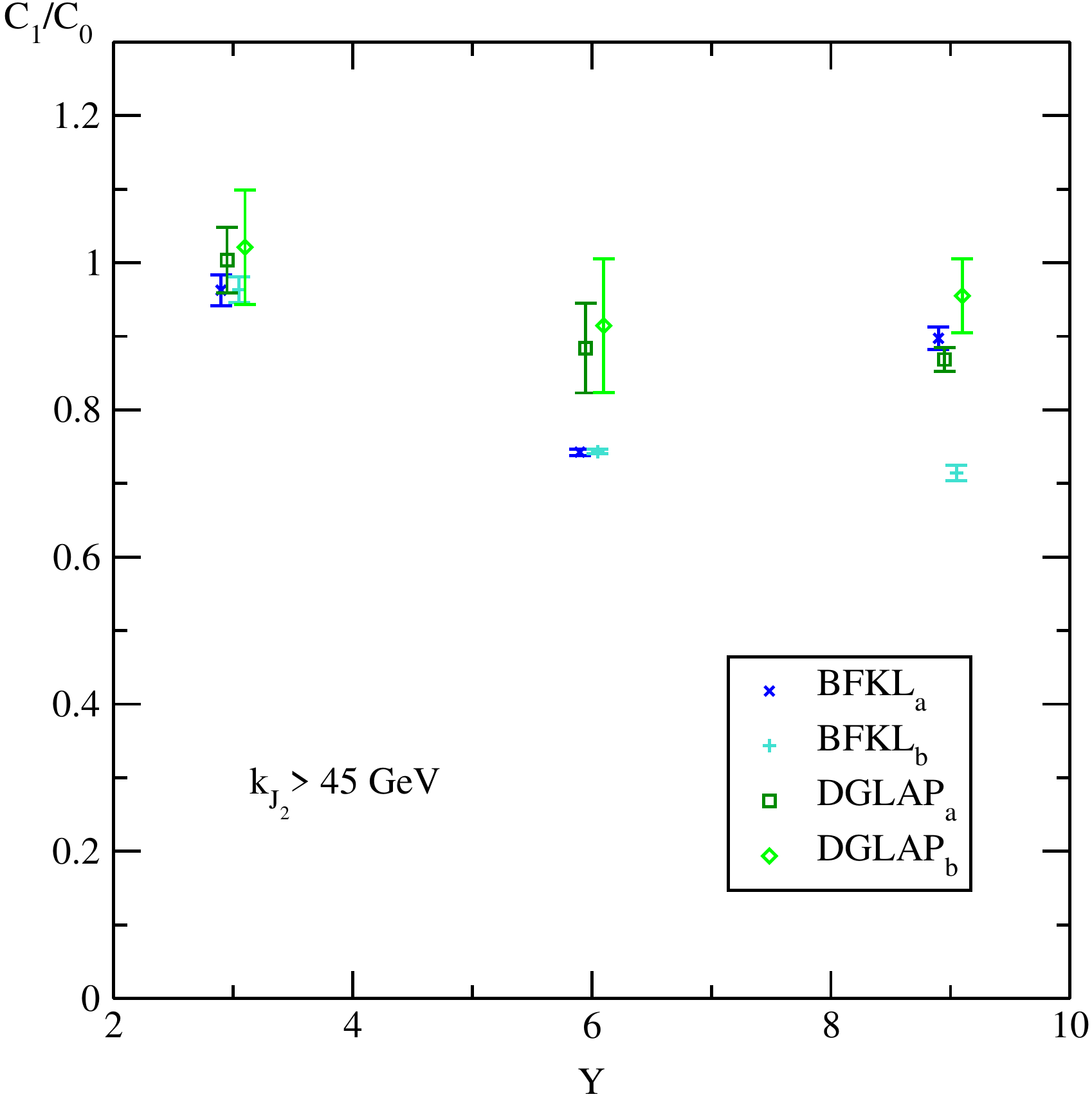}
    \includegraphics[scale=0.37]{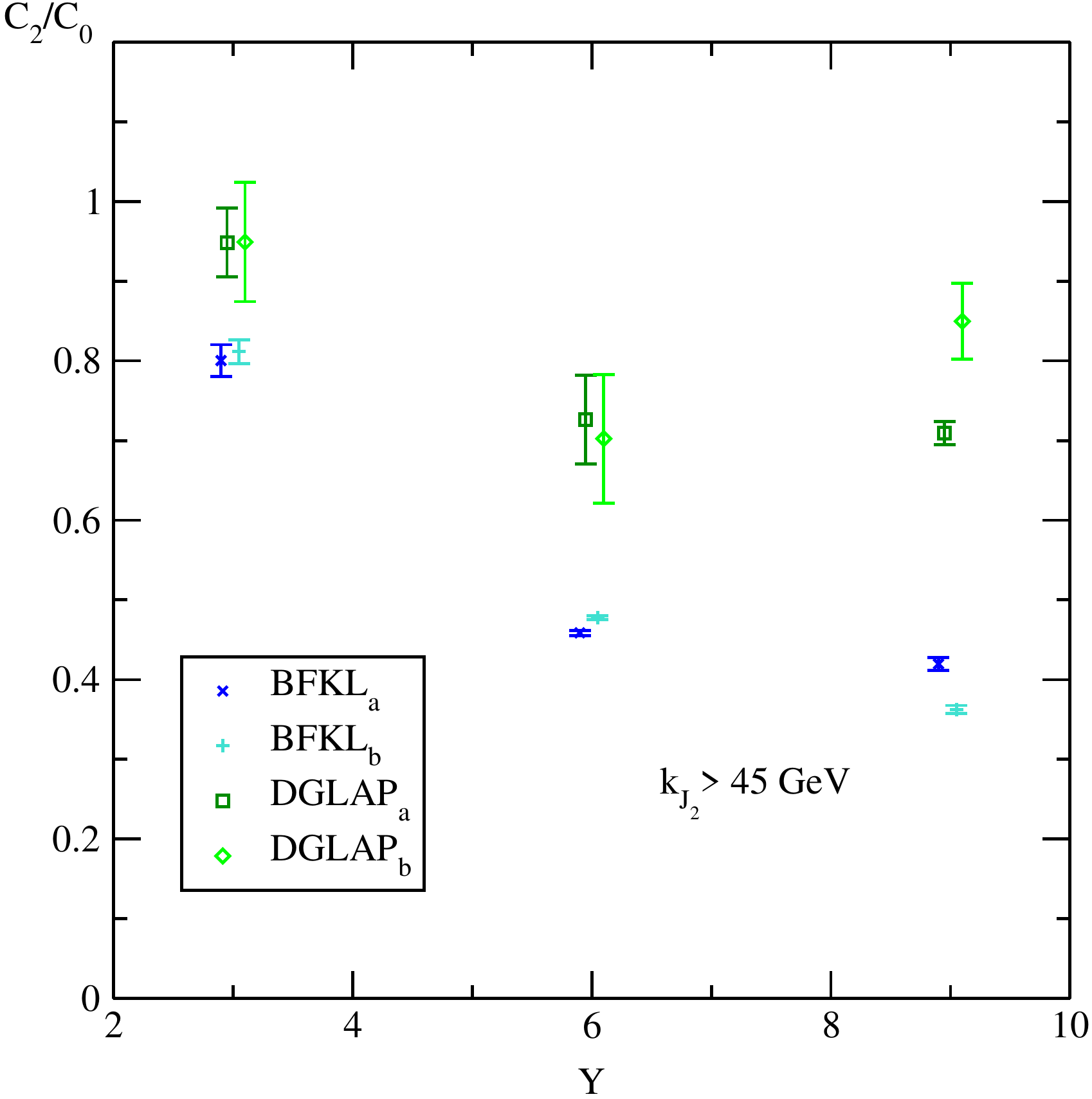}

    \includegraphics[scale=0.37]{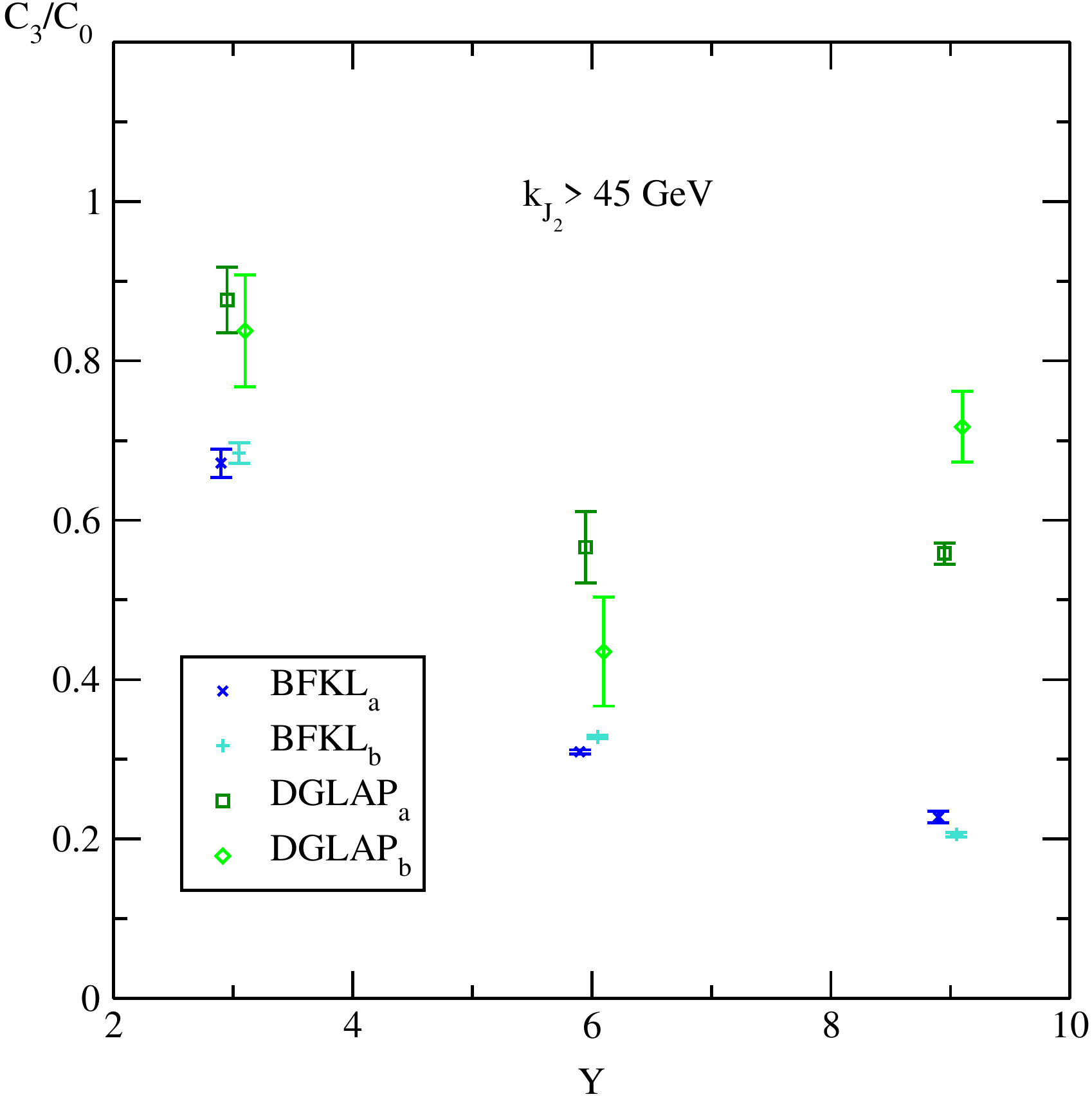}
    \includegraphics[scale=0.37]{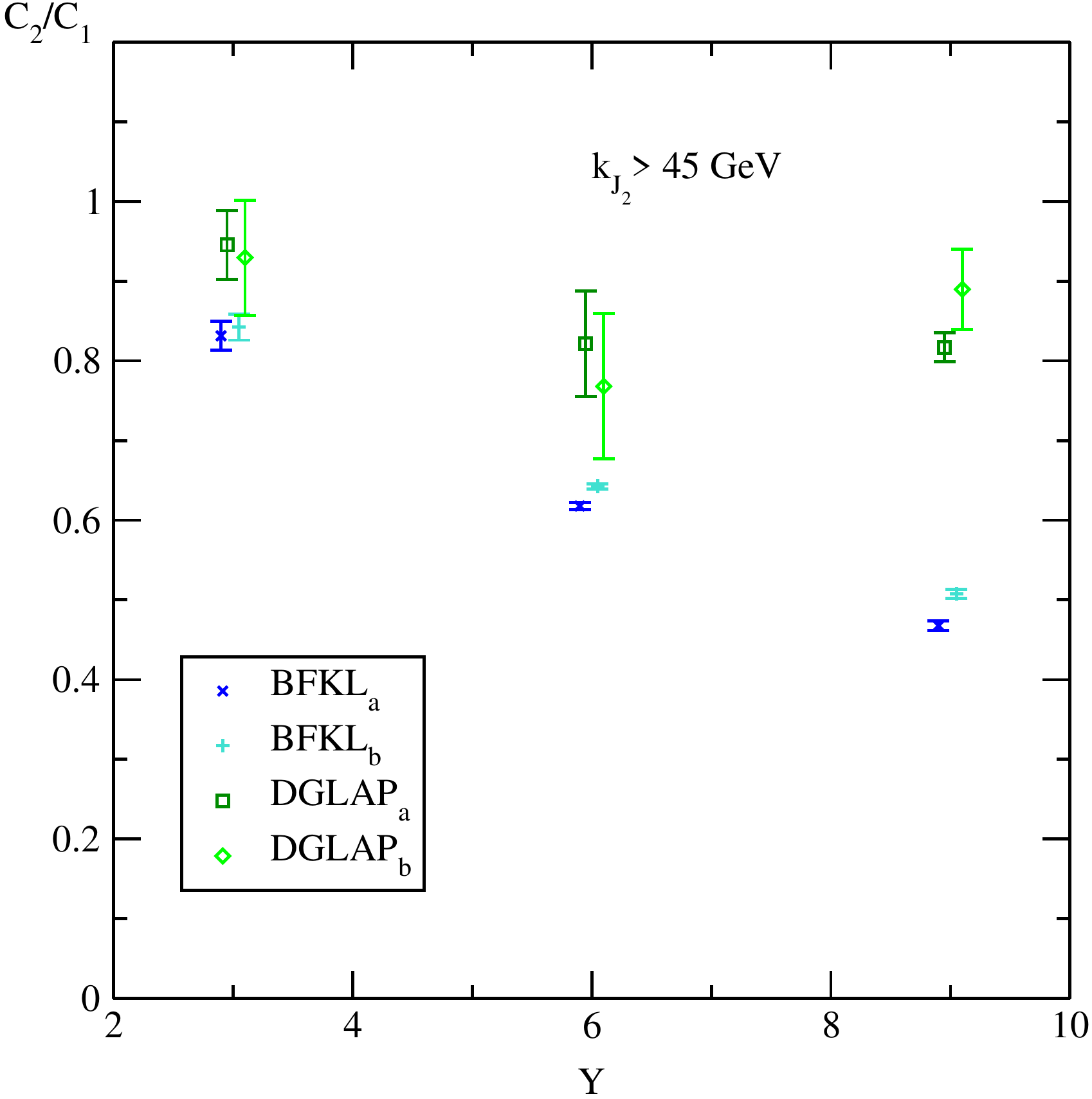}

    \includegraphics[scale=0.37]{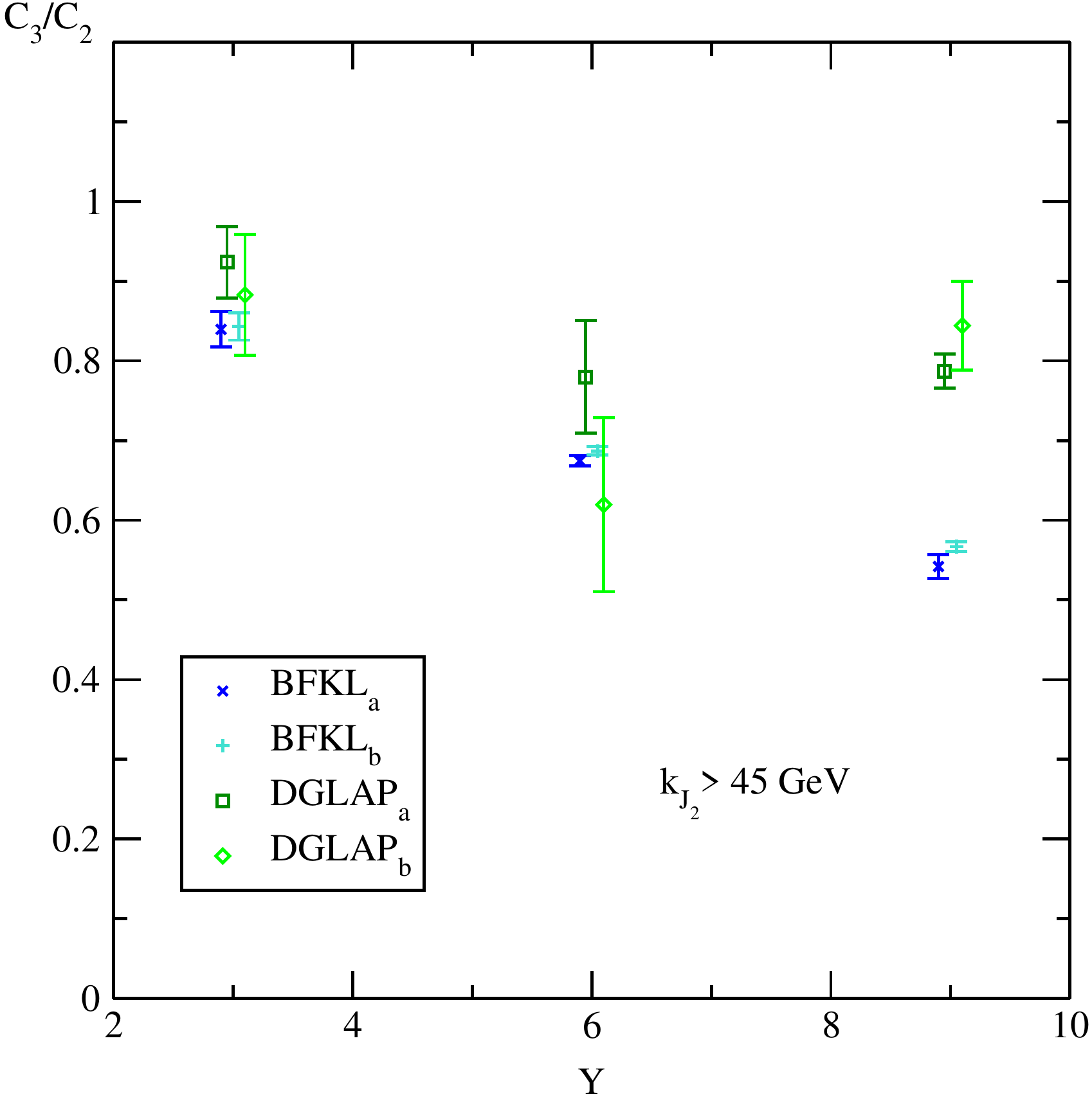}
  \caption[Dijet BFKL and high-energy DGLAP predictions 
           at 7 TeV and for $k_{J_2}^{\rm min}=45$ GeV]
  {$Y$-dependence of several ratios $C_m/C_n$ for $k_{J_1}^{\rm min}=35$ 
   GeV and $k_{J_2}^{\rm min}=45$ GeV, for BFKL and high-energy DGLAP 
   in the two variants $(a)$ (Eq.~(\ref{casea-jets})) and $(b)$ (Eq.~(\ref{caseb-jets})) of the BLM method 
   and for $\sqrt s = 7$ TeV
   (data points have been slightly shifted along 
   the horizontal axis for the sake of readability).}
 \label{plot45}
 \end{figure}

\begin{figure}[H]
  \vspace{-0.50cm}
  \centering
   \includegraphics[scale=0.37]{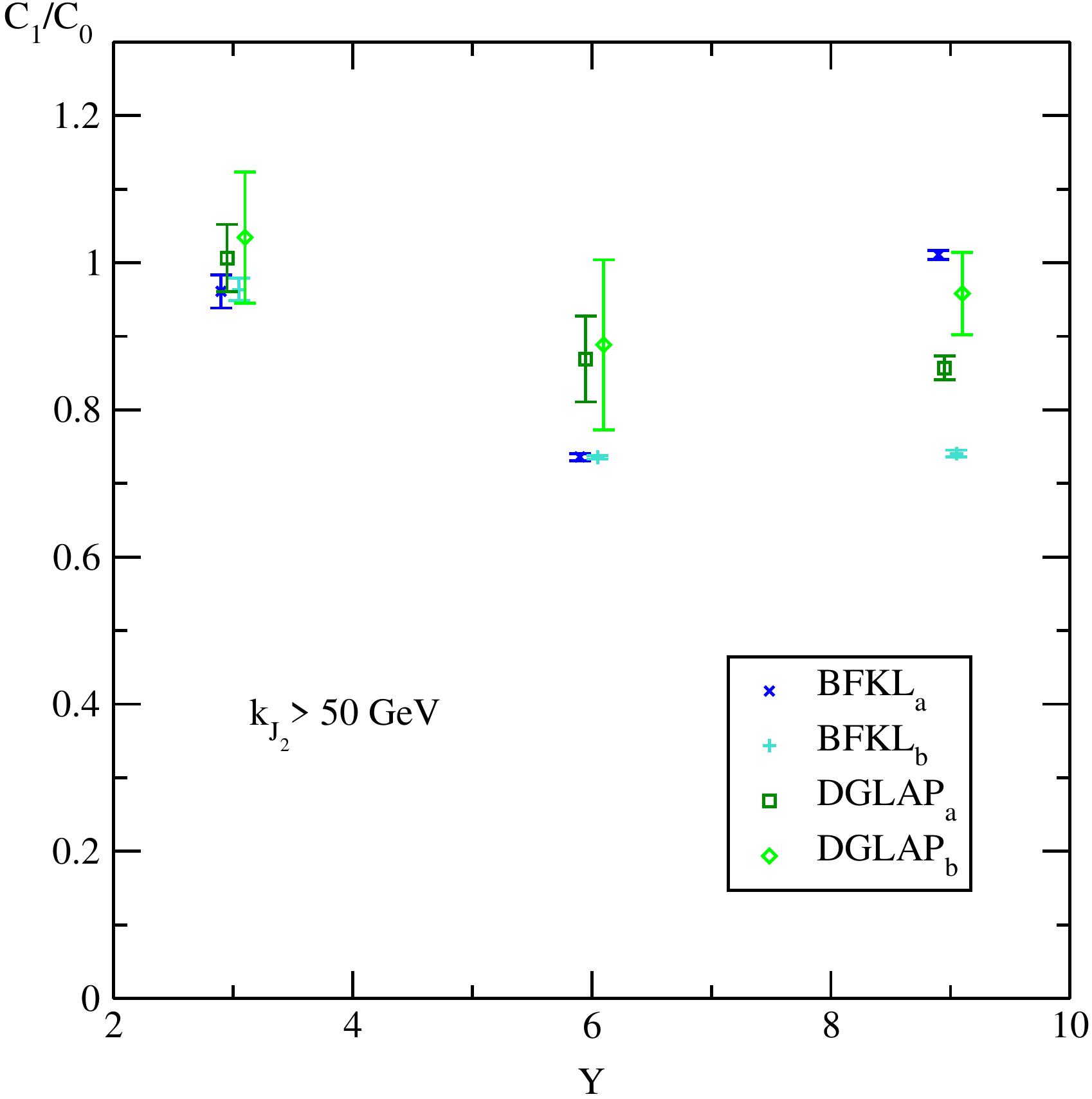}
   \includegraphics[scale=0.37]{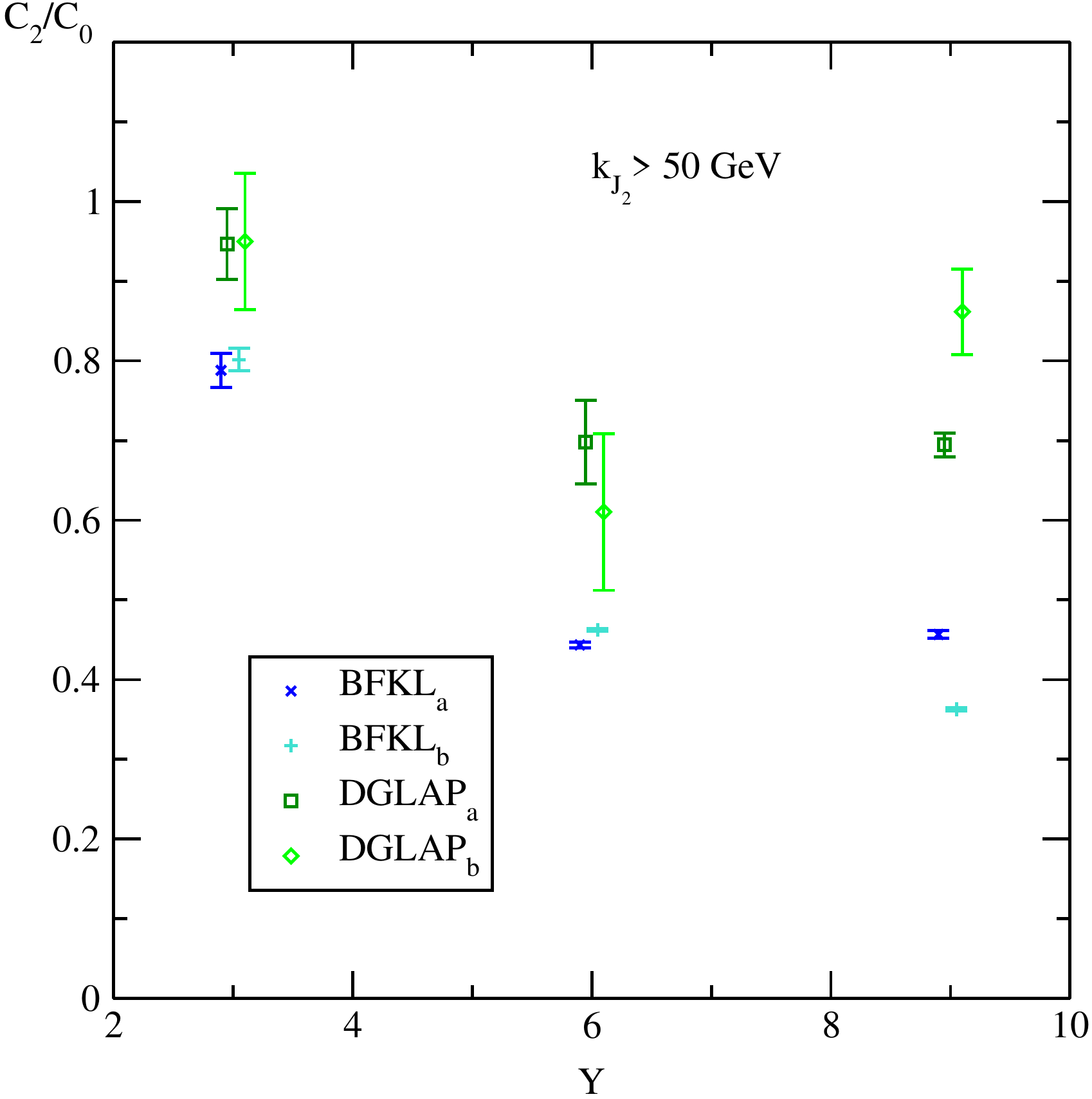}

   \includegraphics[scale=0.37]{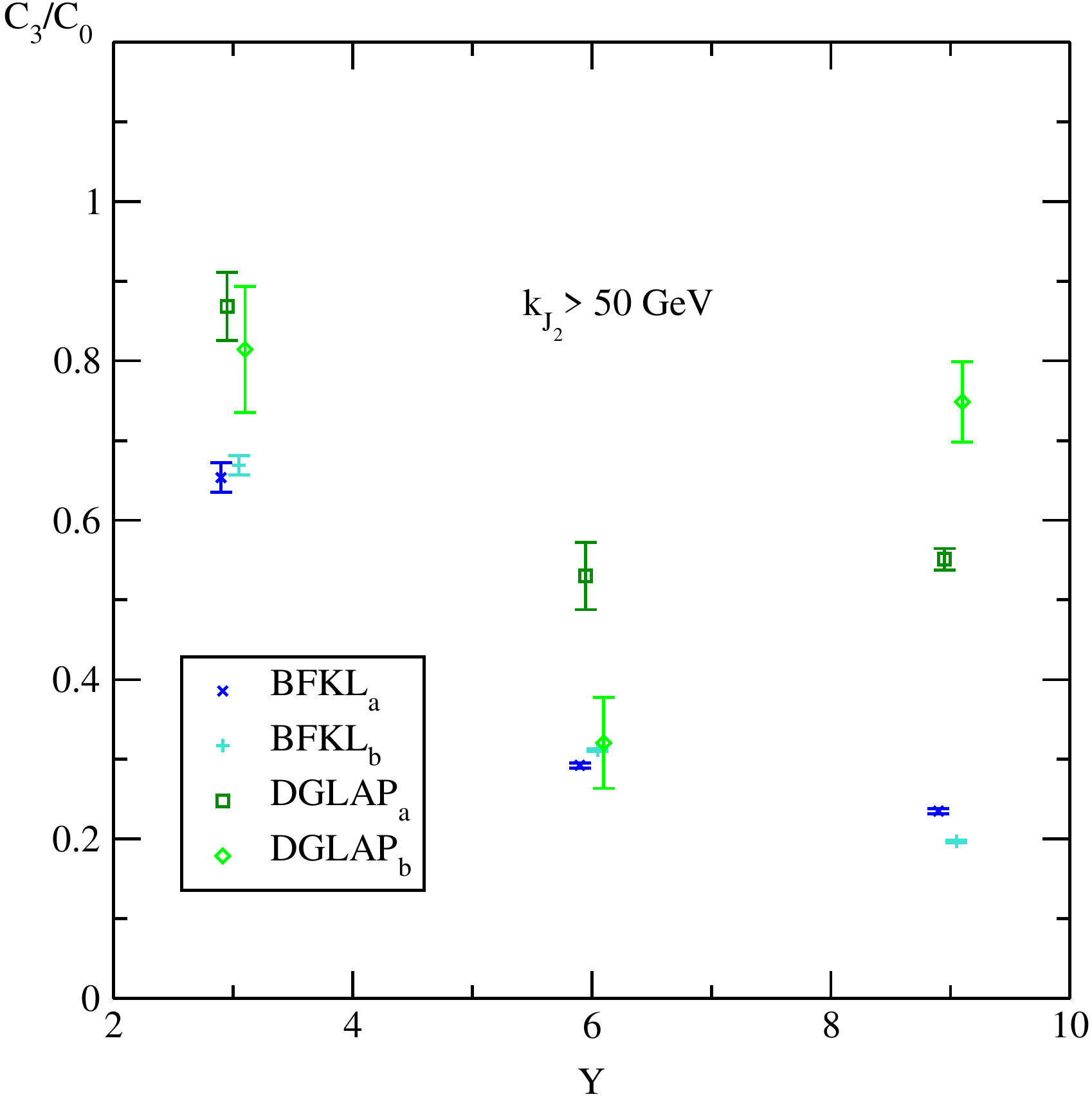}
   \includegraphics[scale=0.37]{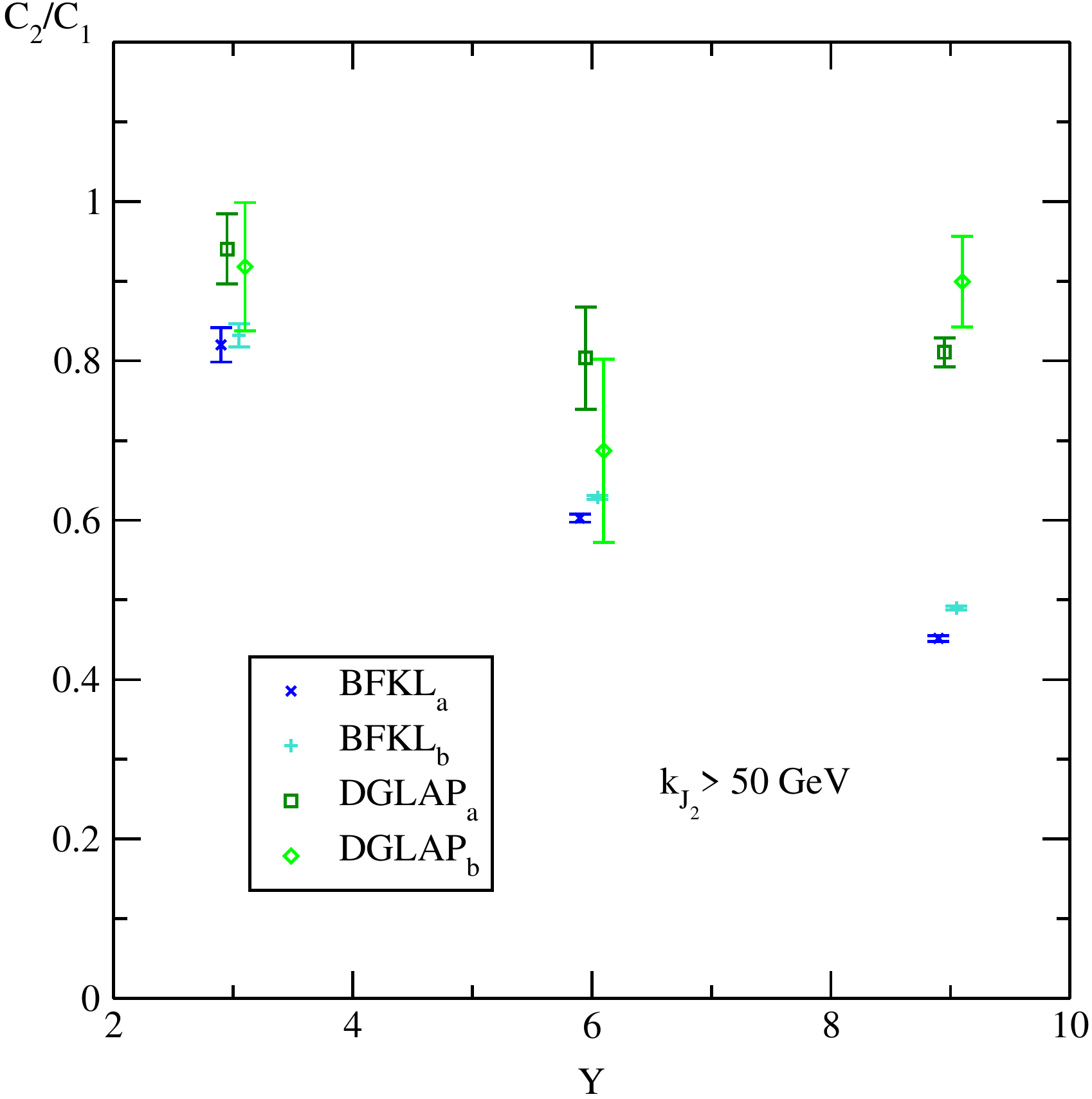}

   \includegraphics[scale=0.37]{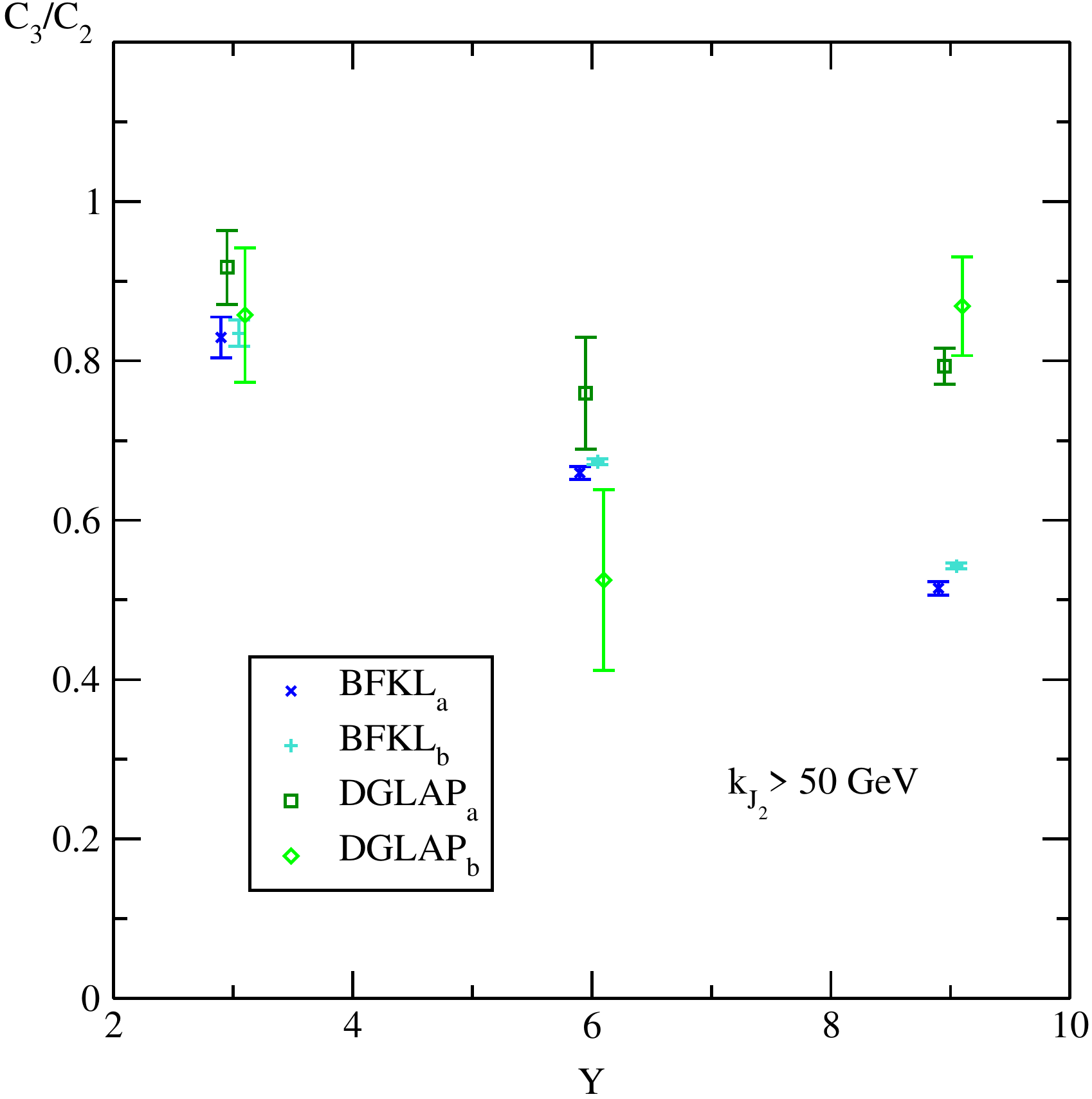}
  \caption[Dijet BFKL and high-energy DGLAP predictions 
           at 7 TeV and for $k_{J_2}^{\rm min}=50$ GeV]
  {$Y$-dependence of several ratios $C_m/C_n$ for $k_{J_1}^{\rm min}=35$ 
   GeV and $k_{J_2}^{\rm min}=50$ GeV, for BFKL and high-energy DGLAP 
   in the two variants $(a)$ (Eq.~(\ref{casea-jets})) and $(b)$ (Eq.~(\ref{caseb-jets})) of the BLM method 
   and for $\sqrt s = 7$ TeV
   (data points have been slightly shifted along 
   the horizontal axis for the sake of readability).}
 \label{plots50}
 \end{figure}

  We summarise our results in Tables~\ref{tab:45} and~\ref{tab:50} and in 
  Figs.~\ref{plot45} and~\ref{plots50}. We can clearly see that, at 
  $Y=9$, BFKL and DGLAP, in both variants $(a)$ (Eq.~(\ref{casea-jets})) and $(b)$ (Eq.~(\ref{caseb-jets})) of the
  BLM setting, give quite different predictions for the all considered ratios 
  except $C_1/C_0$; at $Y=6$ this happens in fewer cases, while at $Y=3$ 
  BFKL and DGLAP cannot be distinguished with given uncertainties. 
  In particular, taking one of the cuts 
  at 35~GeV (as done by the CMS collaboration~\cite{Khachatryan:2016udy}) and the other at 
  45~GeV or 50~GeV, 
  we can clearly see that predictions from BFKL and DGLAP 
  become separate for most azimuthal correlations and ratios between them,
  this effect being more and more visible as the rapidity gap between the 
  jets, $Y$, increases. In other words, in this kinematics the additional 
  undetected parton radiation between the jets which is present in the resummed 
  BFKL series, in comparison to just one undetected parton allowed by the NLO 
  DGLAP approach, makes its difference and leads to more azimuthal angle 
  decorrelation between the jets, in full agreement with the original proposal of 
  Mueller and Navelet.
  
  This result was not unexpected: the use of {\it symmetric} cuts for jet
  transverse momenta maximises the contribution of the Born term, which is present
  for back-to-back jets only and is expected to be large, therefore making
  less visible the effect of the BFKL resummation. This phenomenon
  could be at the origin of the instabilities observed in the NLO fixed-order
  calculations of Refs.~\cite{Andersen:2001kta,Fontannaz:2001nq}.

  One may argue that using \emph{disjoint} intervals for the two jet transverse momenta would be an even cleaner setup. 
  However, since the majority of dijet events are characterised by the lowest possible values for the jet transverse momenta in the selected range, our setup with two different lower cuts but overlapping intervals is  not effectively different from the setup with \emph{disjoint} transverse momenta ranges. 
  Furthermore, independently of the cutoff procedure, there is a non-escapable limitation, namely that the actual energies of partonic subprocesses at the LHC are not much larger than the final-object transverse momenta and, therefore, not too many additional hard parton emissions can occur. This implies that, even after \emph{asymmetric} or \emph{disjoint} configurations in the transverse momentum space, the BFKL and the full NLO DGLAP approaches are not expected to be largely different.

  \newpage 
 

 \begin{table}[H]
 \centering
 \caption[Dijet BFKL vs high-energy DGLAP predictions 
          at 7 TeV and for $k_{J_2}^{\rm min}=45$ GeV]
 {Ratios $C_n/C_m$ at 7 TeV and for $k_{J_1}^{\rm min}=35$ GeV 
  and $k_{J_2}^{\rm min}=45$ GeV.}
 \label{tab:45}
 \begin{tabular}{c|c|llll}
 \toprule
           & $Y$ & BFKL$_{(a)}$  & DGLAP$_{(a)}$  & BFKL$_{(b)}$  & DGLAP$_{(b)}$ \\
 \midrule
           & 3.0 & 0.963(21)   & 1.003(44)   & 0.964(17)   & 1.021(78)   \\
 $C_1/C_0$ & 6.0 & 0.7426(43)  & 0.884(61)   & 0.7433(30)  & 0.914(91)   \\
           & 9.0 & 0.897(15)   & 0.868(16)   & 0.714(10)   & 0.955(50)   \\
 \midrule
           & 3.0 & 0.80(2)     & 0.948(43)   & 0.812(15)   & 0.949(75)   \\
 $C_2/C_0$ & 6.0 & 0.4588(32)  & 0.726(56)   & 0.4777(26)  & 0.702(81)   \\
           & 9.0 & 0.4197(79)  & 0.710(15)   & 0.3627(50)  & 0.850(48)   \\
 \midrule
           & 3.0 & 0.672(18)   & 0.876(41)   & 0.684(13)   & 0.838(70)   \\
 $C_3/C_0$ & 6.0 & 0.3095(26)  & 0.566(45)   & 0.3282(21)  & 0.435(68)   \\
           & 9.0 & 0.2275(72)  & 0.558(13)   & 0.2057(29)  & 0.717(44)   \\
 \midrule
           & 3.0 & 0.831(18)   & 0.945(43)   & 0.842(16)   & 0.929(72)   \\
 $C_2/C_1$ & 6.0 & 0.6178(43)  & 0.821(66)   & 0.6427(34)  & 0.768(91)   \\
           & 9.0 & 0.4677(63)  & 0.817(18)   & 0.5079(56)  & 0.890(51)   \\
 \midrule
           & 3.0 & 0.839(22)   & 0.924(45)   & 0.843(17)   & 0.883(76)   \\
 $C_3/C_2$ & 6.0 & 0.6745(64)  & 0.780(71)   & 0.6869(52)  & 0.62(11)    \\
           & 9.0 & 0.542(15)   & 0.787(21)   & 0.5670(59)  & 0.844(56)   \\
 \bottomrule
 \end{tabular}
 \end{table}

 \begin{table}[H]
 \centering
 \caption[Dijet BFKL vs high-energy DGLAP predictions 
          at 7 TeV and for $k_{J_2}^{\rm min}=50$ GeV]
 {Ratios $C_n/C_m$ at 7 TeV and for $k_{J_1}^{\rm min}=35$ GeV 
  and $k_{J_2}^{\rm min}=50$ GeV.}
 \label{tab:50}
 \begin{tabular}{c|c|llll}
 \toprule
           & $Y$ & BFKL$_{(a)}$  & DGLAP$_{(a)}$ & BFKL$_{(b)}$  & DGLAP$_{(b)}$  \\
 \midrule
           & 3.0 & 0.961(23)   & 1.006(46)   & 0.964(15)   & 1.034(89)   \\
 $C_1/C_0$ & 6.0 & 0.7360(49)  & 0.869(58)   & 0.7357(25)  & 0.89(12)    \\
           & 9.0 & 1.0109(61)  & 0.857(16)   & 0.7406(46)  & 0.958(56)   \\
 \midrule
           & 3.0 & 0.788(21)   & 0.946(44)   & 0.801(14)   & 0.950(85)   \\
 $C_2/C_0$ & 6.0 & 0.4436(37)  & 0.698(53)   & 0.4626(19)  & 0.611(98)   \\
           & 9.0 & 0.4568(50)  & 0.695(15)   & 0.3629(23)  & 0.862(54)   \\
 \midrule
           & 3.0 & 0.653(19)   & 0.868(43)   & 0.669(12)   & 0.814(79)   \\
 $C_3/C_0$ & 6.0 & 0.2925(31)  & 0.530(42)   & 0.3115(15)  & 0.320(57)   \\
           & 9.0 & 0.2351(35)  & 0.551(17)   & 0.1969(17)  & 0.748(50)   \\
 \midrule
           & 3.0 & 0.820(21)   & 0.940(44)   & 0.832(15)   & 0.918(81)   \\
 $C_2/C_1$ & 6.0 & 0.6027(51)  & 0.803(64)   & 0.6288(26)  & 0.69(12)    \\
           & 9.0 & 0.4518(35)  & 0.811(18)   & 0.4900(24)  & 0.899(57)   \\
 \midrule
           & 3.0 & 0.829(26)   & 0.917(46)   & 0.835(17)   & 0.857(85)   \\
 $C_3/C_2$ & 6.0 & 0.6595(82)  & 0.759(70)   & 0.6733(36)  & 0.52(11)    \\
           & 9.0 & 0.5146(85)  & 0.793(23)   & 0.5426(38)  & 0.869(62)   \\
 \bottomrule
 \end{tabular}
 \end{table}



 \newpage
 \section{Central rapidity range exclusion} 
 \label{sec:mn-jets-rapidity}

  \subsection{Motivation}
  \label{sub:mn-jets-rapidity-motivation}
  
  In the last Section we studied the effect of using \emph{asymmetric} cuts for the jet transverse momenta, 
  comparing full NLA BFKL predictions with fixed-order DGLAP calculations in the high-energy limit. 
  Here we want do deal with another issue, which deserves some care and has not been taken
  into consideration both in theoretical and experimental analyses so far.
  As anticipated in the discussion at the end of Section~\ref{sec:mn-jets-theory-vs-experiment},
  in defining the jet rapidity separation $Y$ for a given final state with two jets, the rapidity
  of one of the two jets could be so small, say $|y_{J_i}|\lesssim 2$, that this jet
  is actually produced in the central region, rather than in one of the two
  forward regions. Since the longitudinal momentum fractions of the parent
  partons $x$ that generate such central jet are very small, one can naturally 
  expect sizable corrections to the vertex of this jet, due to the fact that the
  collinear factorisation approach used in the derivation of the result for jet
  vertex could not be accurate enough in our kinematical region, where $x$ values can be as small as, $x \sim 10^{-3}$.
   
  The use of collinear factorisation methods in the case of central jet production
  in our kinematical range deserves some discussion. On one hand, at 
  $x \sim 10^{-3}$ and at scales of the order of the jet transverse momenta which 
  we consider here, $\sim 20\div 40$~GeV, PDFs are well constrained, mainly from 
  DIS HERA data. On the other hand, in this kinematical region PDF 
  parameterisations extracted in next-to-NLO (NNLO) and in NLO approximations start to differ 
  one from the other, which indicates that NNLO effects become essential in the 
  DIS cross sections. The situation with central jet production in proton-proton 
  collisions may be different. Recently, in Ref.~\cite{Currie:2013dwa} results 
  for NNLO corrections to the dijet production originated from the gluonic 
  subprocesses were presented. In the region $|y_{J_{1,2}}|<0.3$ and for jet 
  transverse momenta $\sim 100$~GeV, the account of NNLO effects leads to an
  increase of the cross section by $\sim 25\  $. For our kinematics, 
  featuring smaller jet transverse momenta and ``less inclusive'' coverage of jet 
  rapidities, one could expect even larger NNLO corrections.               
   
  Conceptually, instead of the collinear approach, for jets produced in the 
  central rapidity region (at very small $x$) a promising approach would be to 
  use a \emph{high-energy factorisation} scheme (often also referred as 
  $k_T$-\emph{factorisation})~\cite{Catani:1990eg,Collins:1991ty,Gribov:1984tu,Levin:1991ry} together with the NLO central jet vertex calculated 
  in Ref.~\cite{Bartels:2006hg}~\footnote{For the discussion of different approaches 
  to factorisation for dijet production see, {\it e.g.}, the recent review 
  paper~\cite{Sapeta:2015gee}.}.
   
  We suggest to compare BFKL theory 
  predictions with data in a region where theoretical uncertainties related to 
  other kind of physics are most possibly reduced. Therefore we propose to 
  return to the original Mueller--Navelet idea, to study the inclusive production 
  of two forward jets separated by a large rapidity gap, and to remove from the 
  analysis those regions where jets are produced at central rapidities.   
   
  As a contribution to the assessment of this effect, in this Section we will study 
  the $Y$-dependence of several azimuthal correlations and ratios among them, 
  imposing an additional constraint, that the rapidity of a Mueller--Navelet jet 
  cannot be smaller than a given value. Then we will compare this option with 
  the case when the constraint is absent.

  \subsection{Phase-space constraints}
  \label{sub:mn-jets-rapidity-range}

\begin{figure}[H]
   \centering
   \includegraphics[scale=0.5]{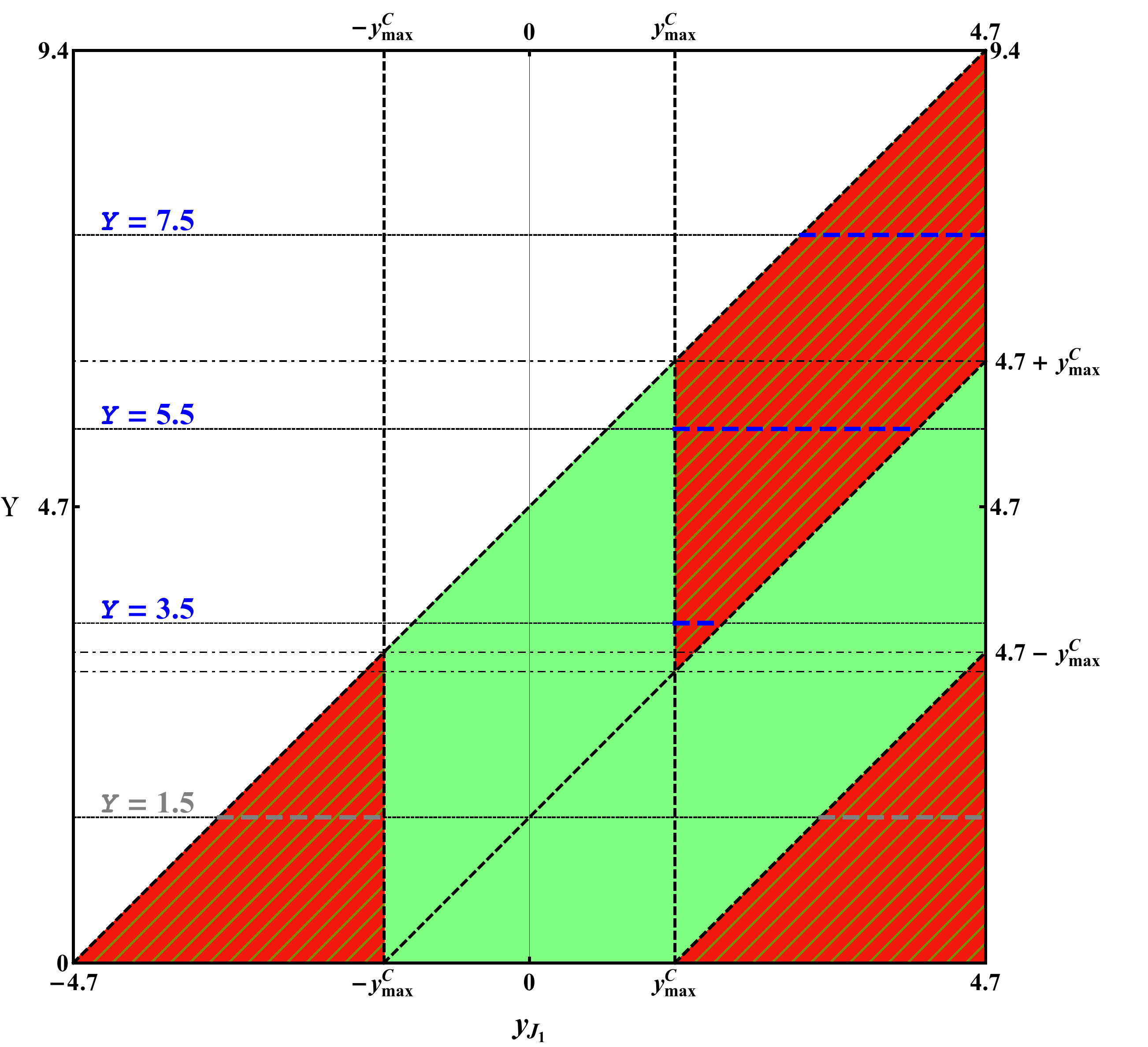}
   \caption[Central rapidity range exclusion in dijet production]
   {Central rapidity range exclusion 
    in the $(y_{J_1},Y)$ phase-subspace 
    for Mueller--Navelet jets. 
    The green triangle represents the area bounded 
    by the kinematical cuts on $y_{J_1}$ and $y_{J_2}$, 
    while the red striped area is the region allowed 
    by imposing the constraint on the central rapidity 
    $y_{J_{1,2}} \ge y^{\rm C}_{\rm max}$. 
    Finally, for any given value of $Y$, 
    the $y_{J_1}$ integration is done 
    over the blue dashed segment.}
   \label{fig:jets-rapidity-range}
  \end{figure}

  In order to exclude events where, for a given $Y$, one of the two jets is produced in the central region, 
  we need to slightly modify the definition of the \emph{integrated coefficients} given in Eq.~(\ref{Cm_int-mn-jets}) in the following way:
  \begin{align}
  \label{Cm_int-mn-jets-rapidity}
  &
  C_n=\int_{y_{J_1}^{\rm min}}^{y_{J_1}^{\rm max}}dy_1
  \int_{y_{J_2}^{\rm min}}^{y_{J_2}^{\rm max}}dy_2\int_{k_{J_1}^{\rm min}}^{\infty}dk_{J_1}
  \int_{k_{J_2}^{\rm min}}^{\infty}dk_{J_2}
  \delta\left(y_{J_1}-y_{J_2}-Y\right)
  \\ \nonumber &  \times \,
  \theta\left(|y_{J_1}| - y^{\rm C}_{\rm max}\right) \theta\left(|y_{J_2}| - y^{\rm C}_{\rm max}\right) 
  {\cal C}_n \left(y_{J_1},y_{J_2},k_{J_1},k_{J_2} \right) \; .
  \end{align}
  In Eq.~(\ref{Cm_int-mn-jets-rapidity}),
  the two step-functions force the exclusion of jets whose rapidity is smaller
  than a cutoff value, given by $y^{\rm C}_{\rm max}$, which delimits the central
  rapidity region (see Fig.~\ref{fig:jets-rapidity-range} for a schematic view). 
  
  As for the values of $y^{\rm C}_{\rm max}$, we will consider three cases:
  $y^{\rm C}_{\rm max}=0$, which means no exclusion from jets in the central
  region, as in all the numerical analyses so far; $y^{\rm C}_{\rm max}=1.5$,
  corresponding to a central region with size equal to about one third of the
  maximum possible rapidity span $Y=9.4$ and $y^{\rm C}_{\rm max}=2.5$, a
  control value, to check the stability of our results.
  
  Concerning the jet transverse momenta, differently from most previous analyses,
  we make the following five choices, which include \emph{asymmetric} cuts:
  \begin{enumerate}
  \item 
  \label{cut1-jets-rapidity}
  $
   k_{J_1}^{\rm \, min} = 20\, \text{GeV}, 
   \quad\quad\quad
   k_{J_2}^{\rm \, min} = 20\, \text{GeV};
  $
  \item 
  \label{cut2-jets-rapidity}
  $
   k_{J_1}^{\rm \, min} = 20\, \text{GeV}, 
   \quad\quad\quad
   k_{J_2}^{\rm \, min} = 30\, \text{GeV};
  $
  \item
  \label{cut3-jets-rapidity}
  $
   k_{J_1}^{\rm \, min} = 20\, \text{GeV}, 
   \quad\quad\quad
   k_{J_2}^{\rm \, min} = 35\, \text{GeV};
  $
  \item
  \label{cut4-jets-rapidity}
  $
   k_{J_1}^{\rm \, min} = 20\, \text{GeV}, 
   \quad\quad\quad
   k_{J_2}^{\rm \, min} = 40\, \text{GeV};
  $
  \item
  \label{cut5-jets-rapidity}
  $
   k_{J_1}^{\rm \, min} = 35\, \text{GeV}, 
   \quad\quad\quad
   k_{J_2}^{\rm \, min} = 35\, \text{GeV};
  $
  \end{enumerate}
  
  We will give the first phenomenological predictions
  for the ratios $R_{nm}\equiv C_n/C_m$ 
  at center-of-mass energy $\sqrt{s} = 13$ TeV, using
  the BLM scheme in its exact implementation (Eq.~(\ref{blm_exact-jets})). 
  We will perform also some calculations with these approximate schemes $(a)$ (Eq.~(\ref{casea-jets})) and $(b)$ (Eq.~(\ref{caseb-jets})), in order to get an idea about the inaccuracy of our predictions coming from such approximate implementations of the BLM scale setting.
  Differently from Sections~\ref{sec:mn-jets-theory-vs-experiment} and~\ref{sec:mn-jets-BFKL-vs-DGLAP}, 
  all calculations are done in the MOM renormalisation scheme.
  
  \subsection{Results and discussion}
  \label{sub:mn-jets-rapidity-results}
  
\begin{figure}[H]
 \centering
 
    \includegraphics[scale=0.38]{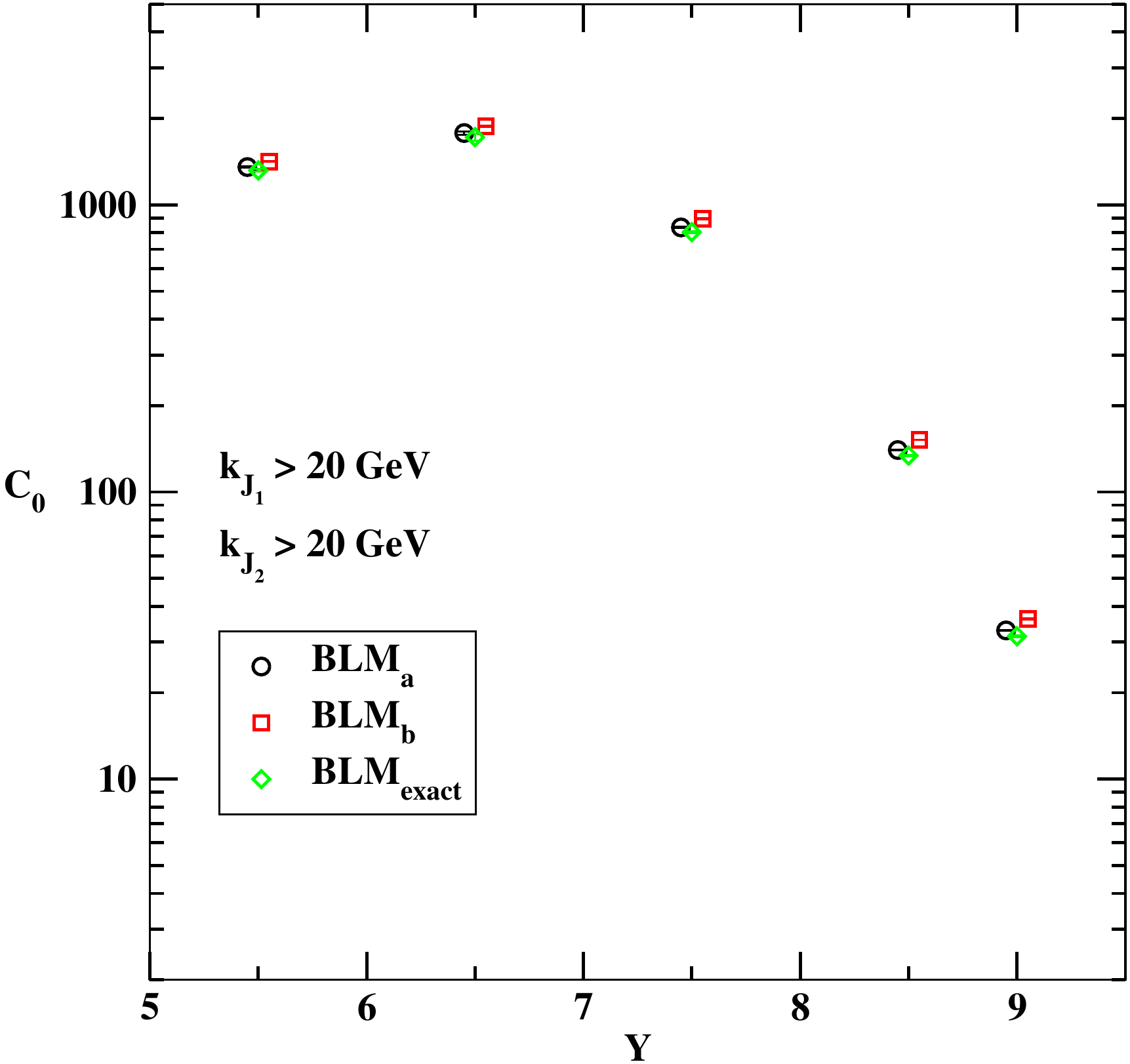}
    \includegraphics[scale=0.38]{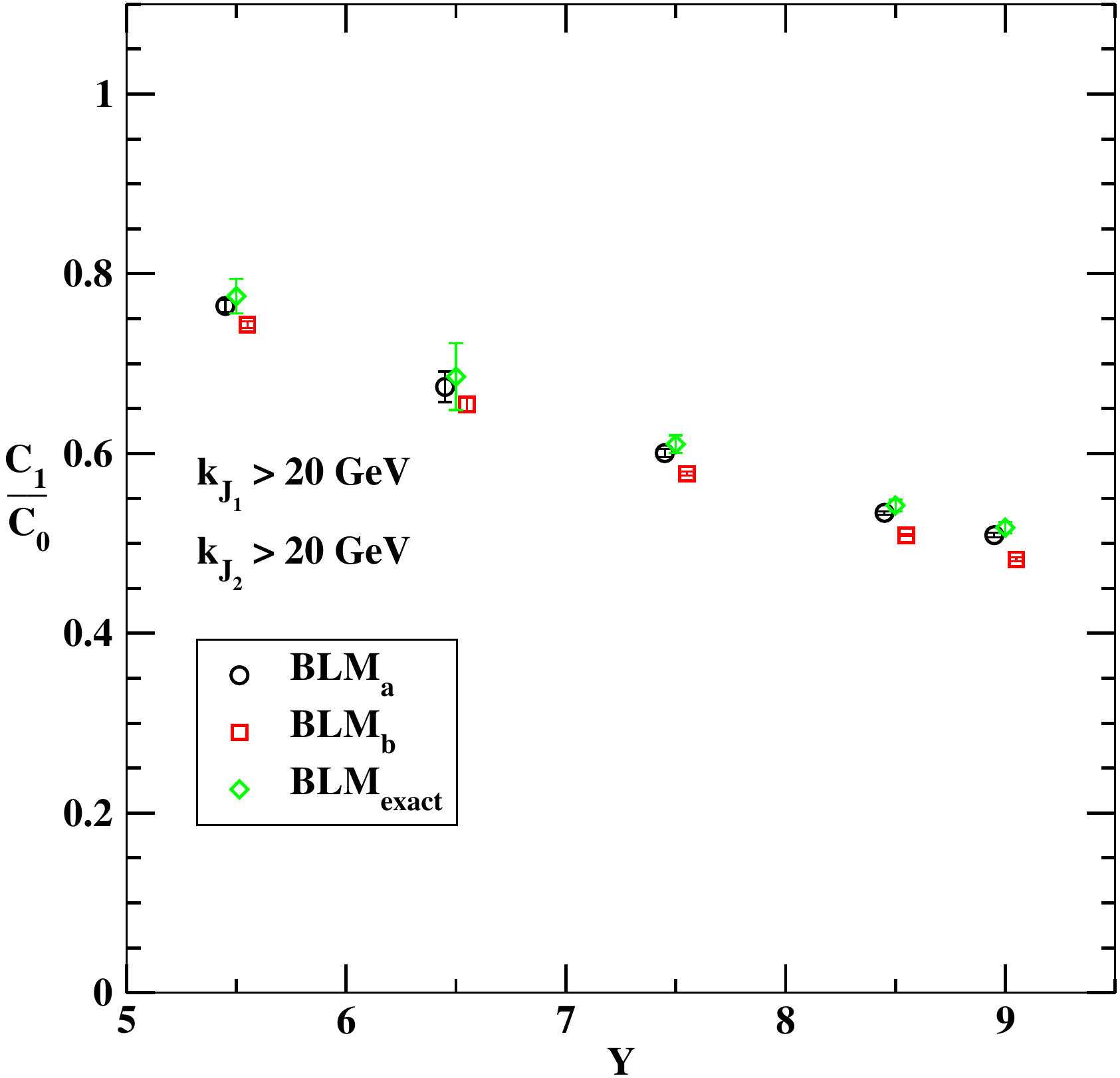}
 
    \includegraphics[scale=0.38]{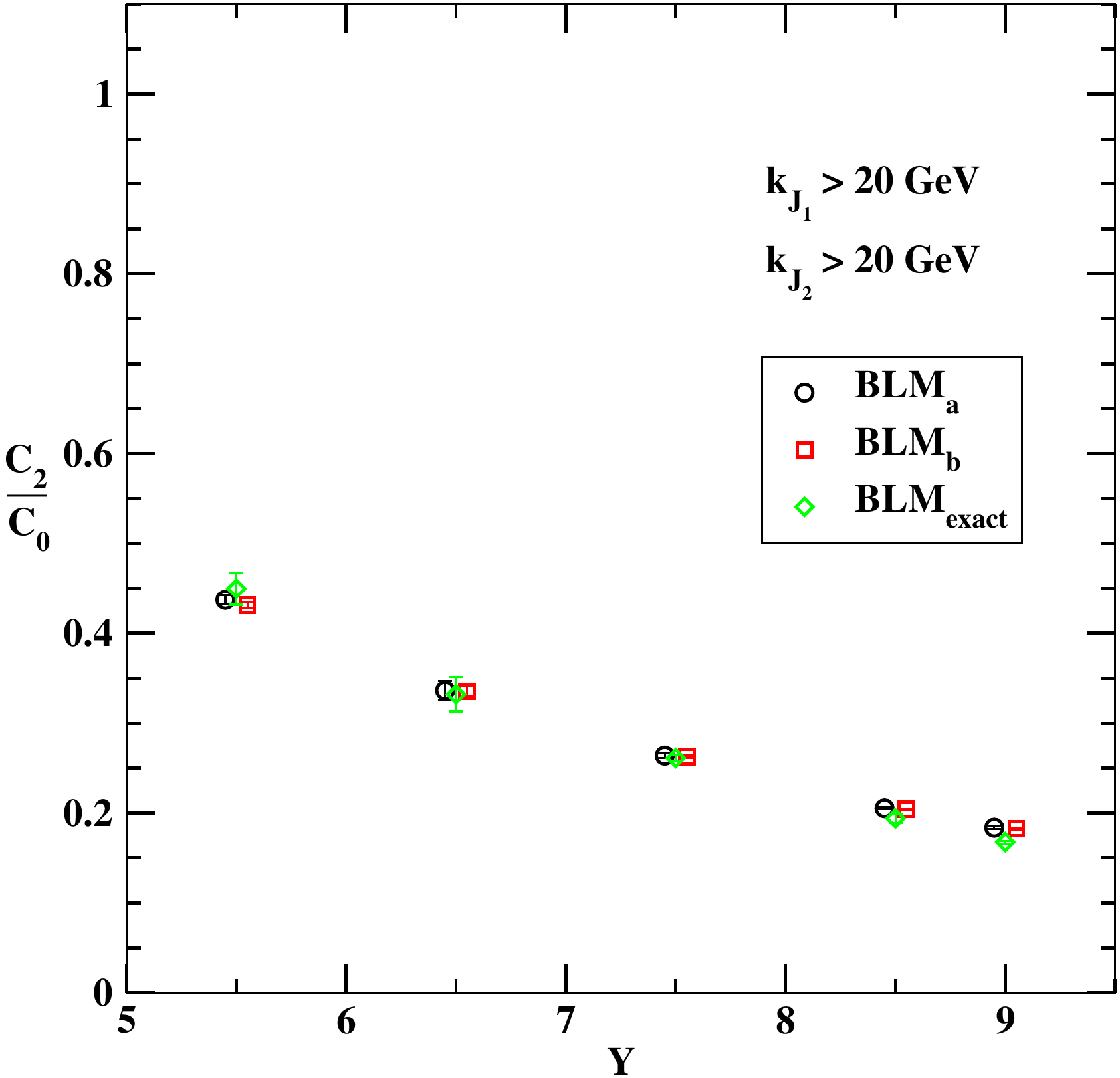}
    \includegraphics[scale=0.38]{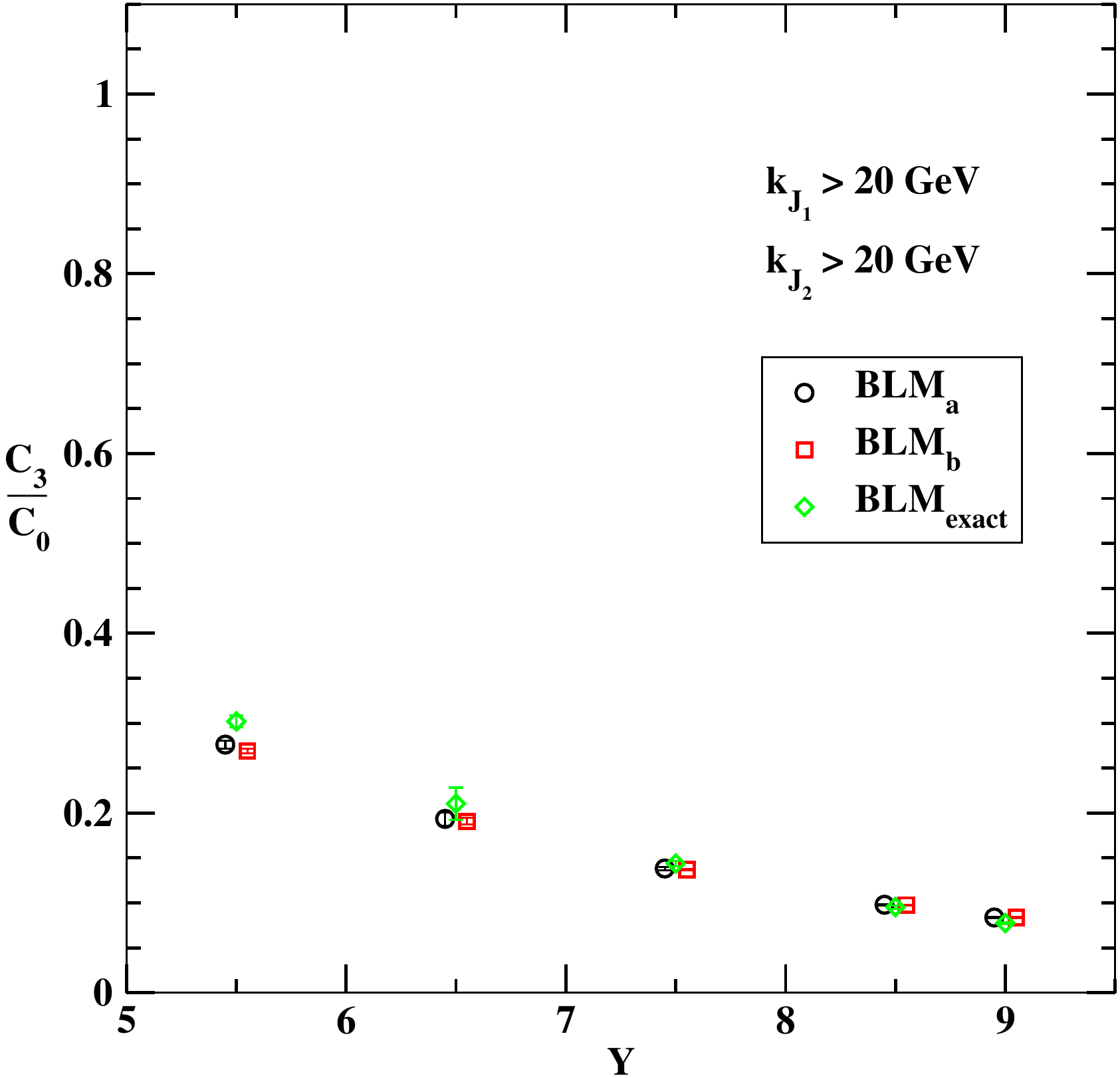}
 
    \includegraphics[scale=0.38]{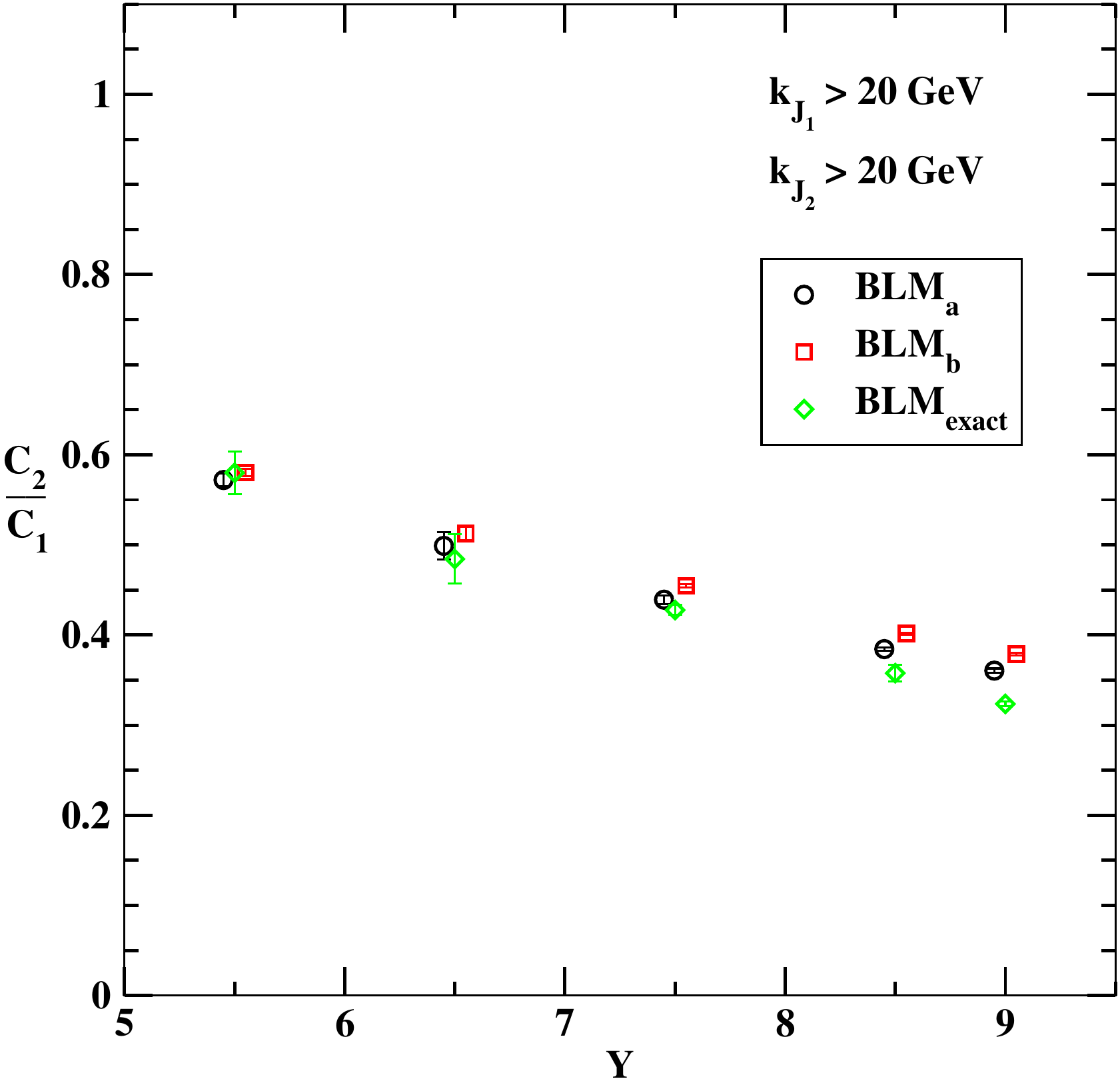}
    \includegraphics[scale=0.38]{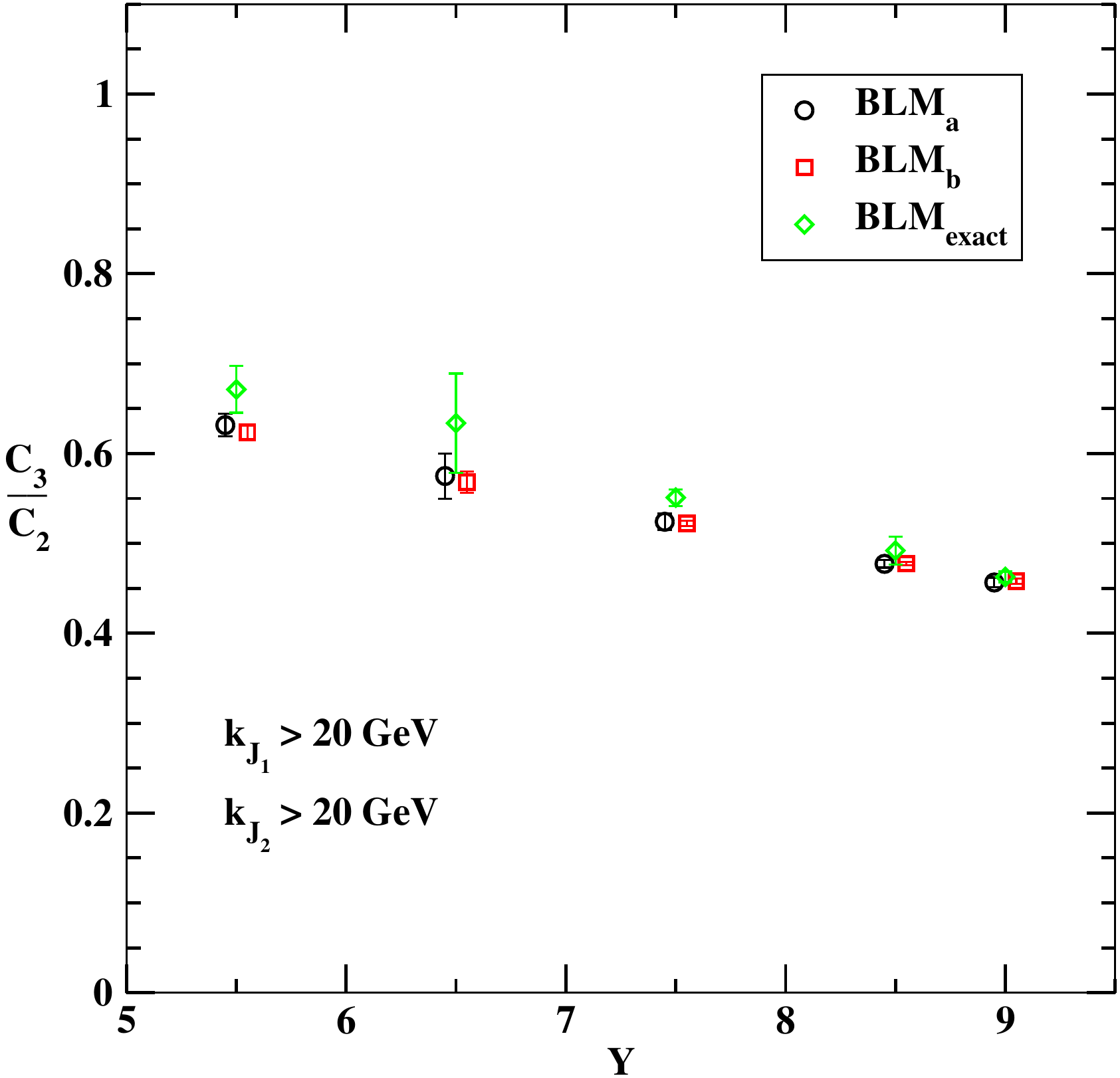}
  \caption[Rapidity veto effect for different BLM scale settings 
          in dijet production]
 {$Y$-dependence of $C_0$ and of several ratios $C_m/C_n$
  for $k_{J_1}^{\rm min}=k_{J_2}^{\rm min}=20$~GeV, $|y_{J_1}| > 2.5$ 
  and for $\sqrt s = 13$ TeV, from
  the three variants of the BLM method (data points have been slightly shifted
  along the horizontal axis for the sake of readability; see
  Table~\ref{tab:2020_2.5}).}
 \label{2020_2.5}
 \end{figure}

\begin{figure}[H]
 \centering
 
    \includegraphics[scale=0.38]{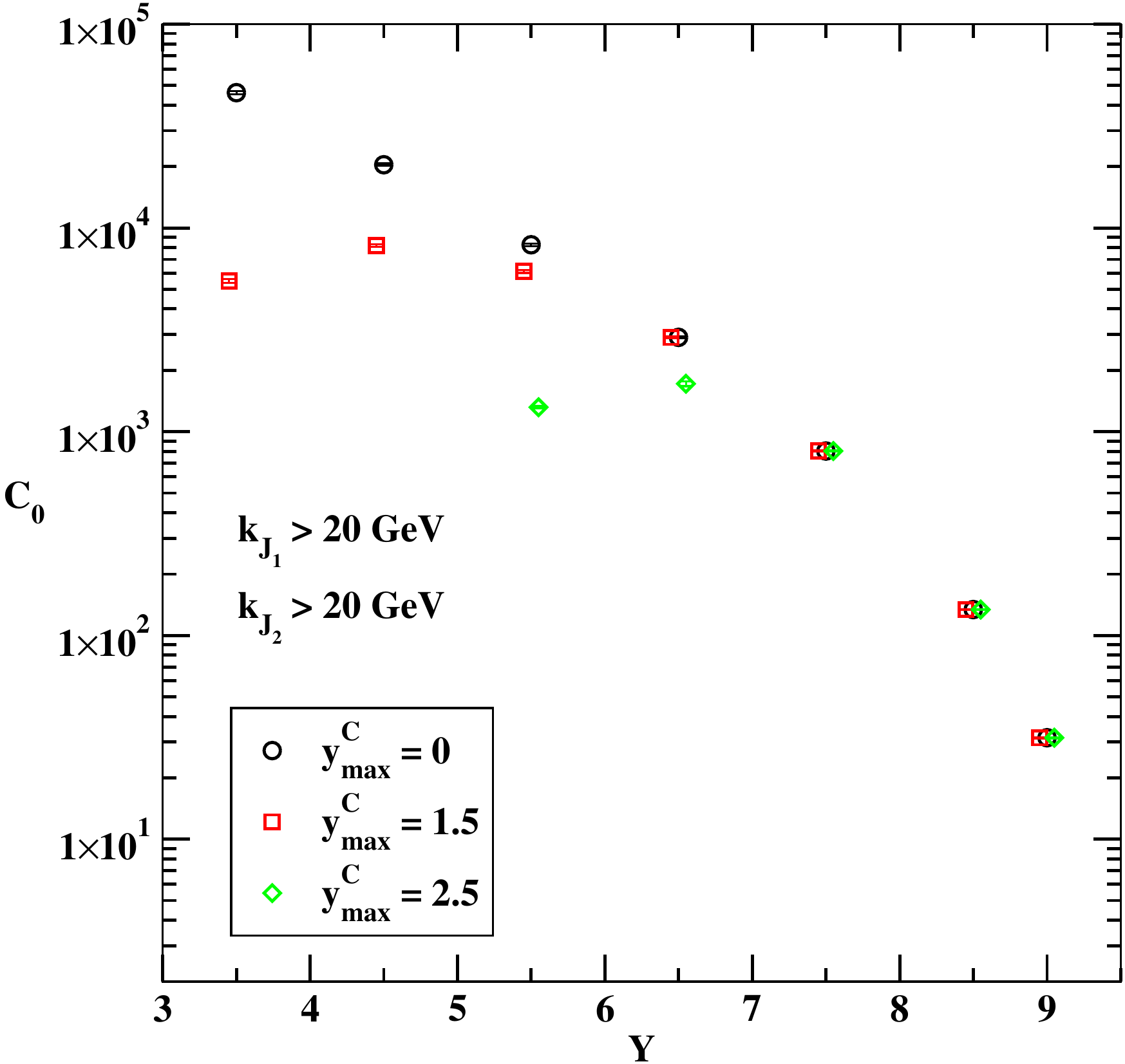}
    \includegraphics[scale=0.38]{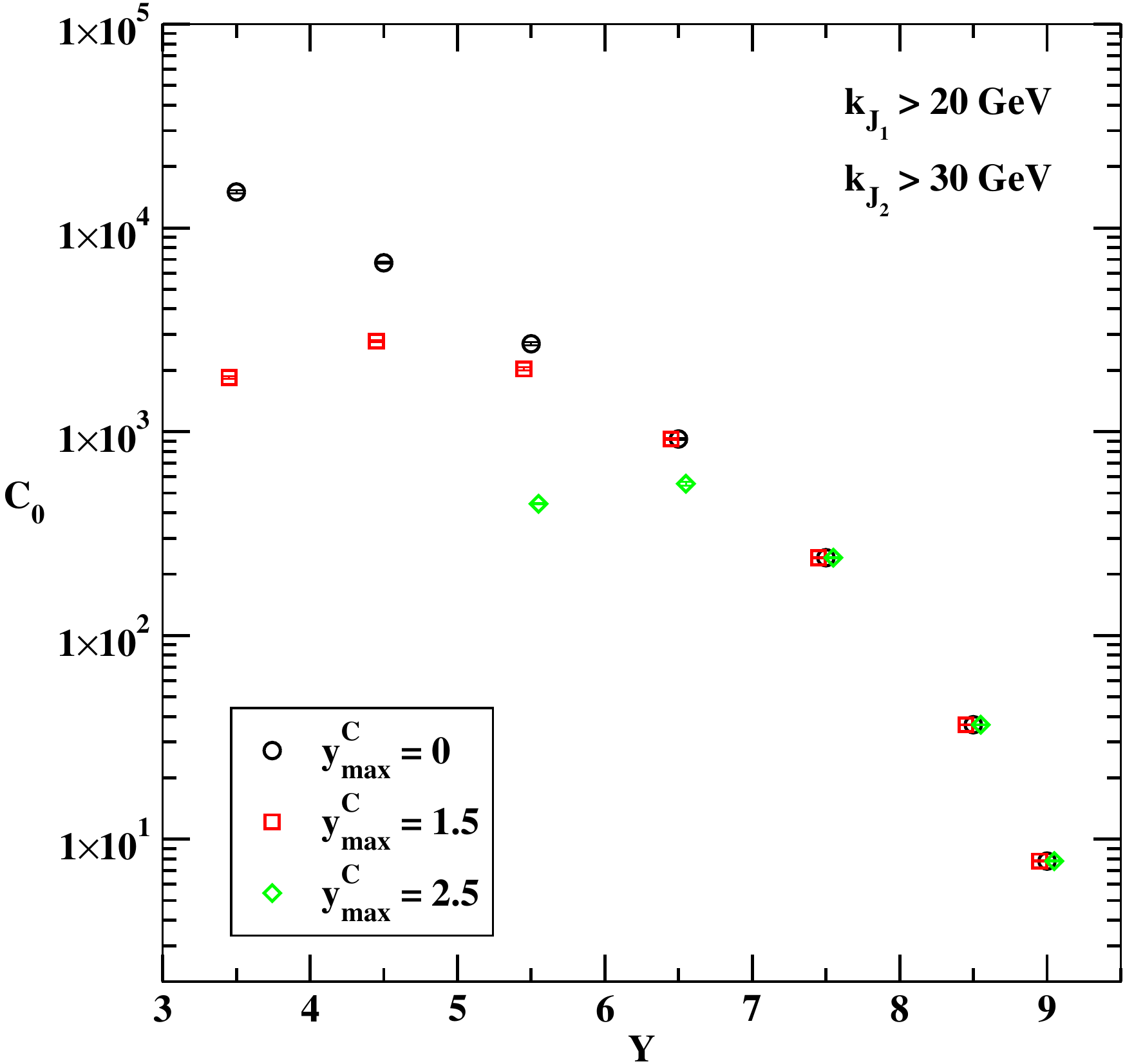}
 
    \includegraphics[scale=0.38]{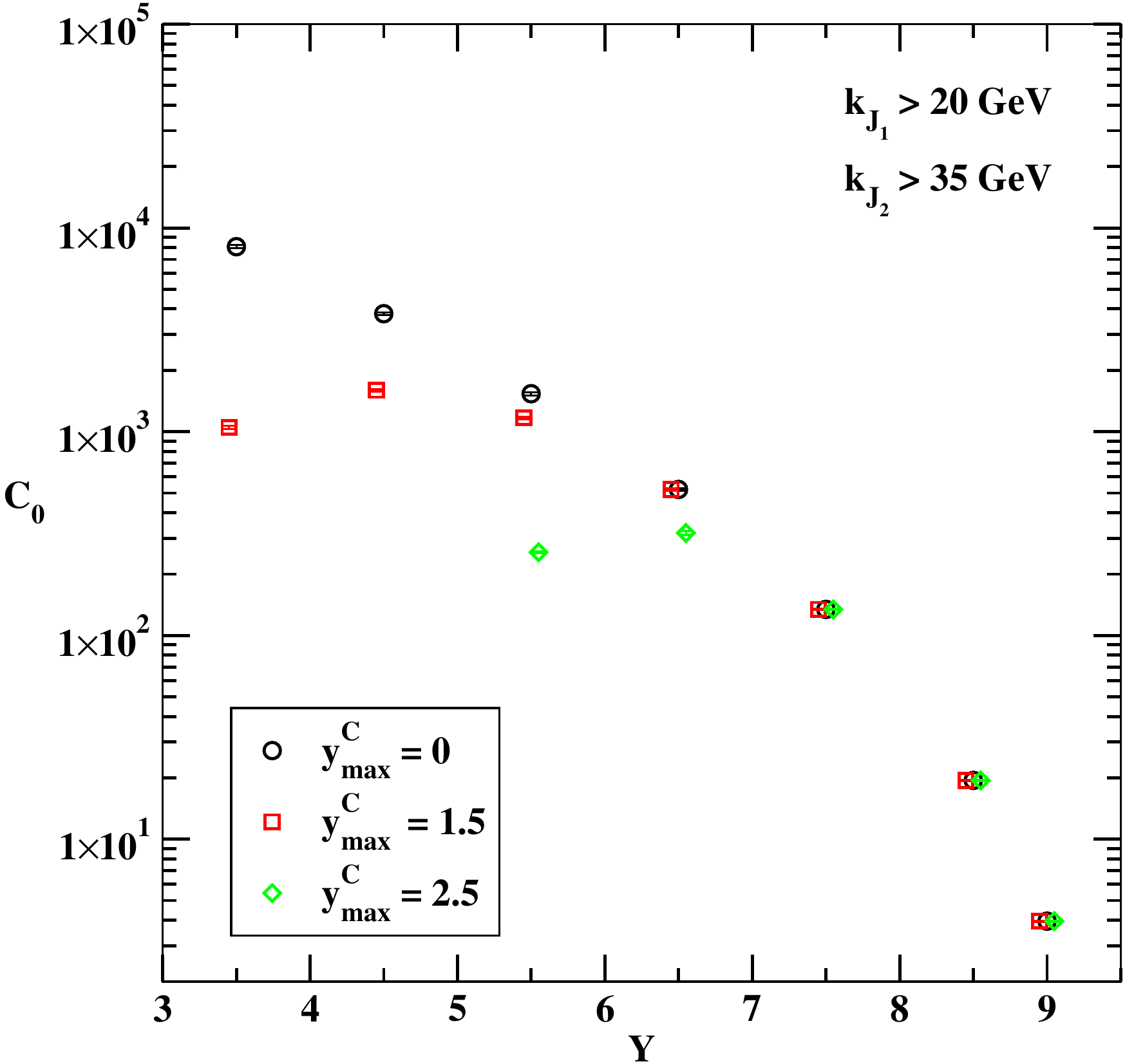}
    \includegraphics[scale=0.38]{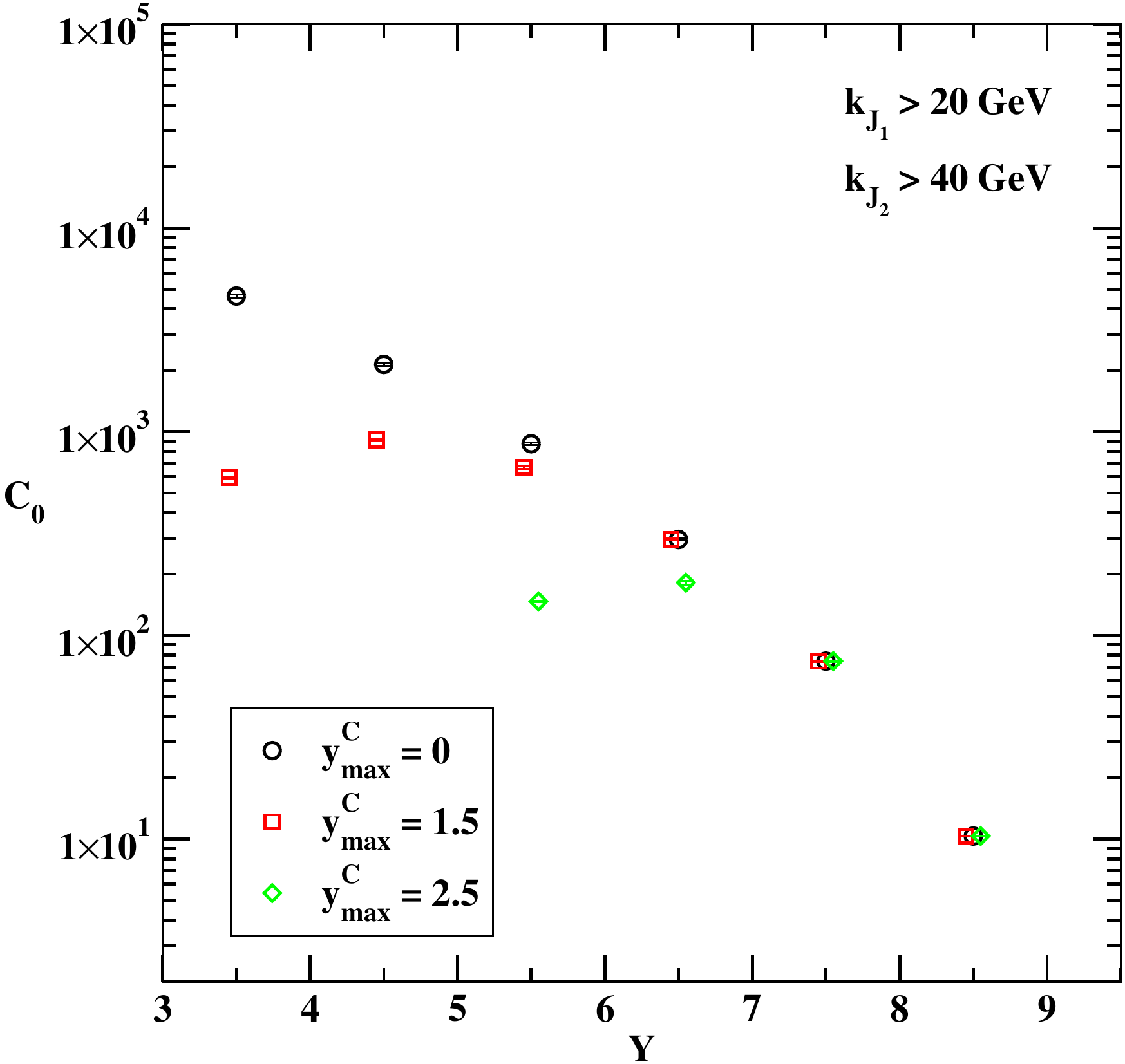}
 
    \includegraphics[scale=0.38]{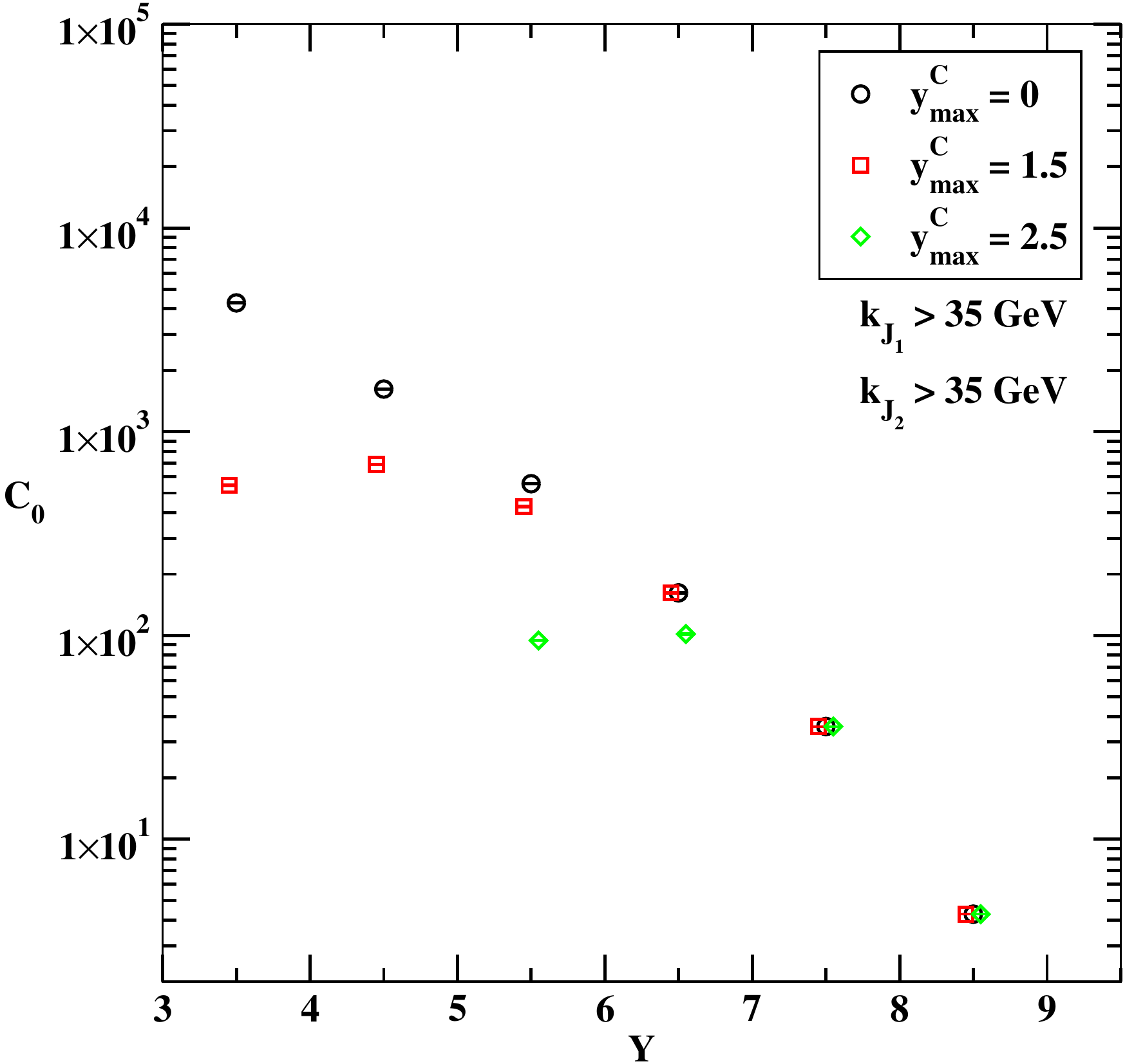}
  \caption[Rapidity veto effect on $C_0$ in dijet production]
 {$Y$-dependence of $C_0$ from the exact BLM method (Eq.~\ref{blm_exact-jets}), for all
  choices of the cuts on jet transverse momenta 
  and of the central rapidity region, and for $\sqrt s = 13$ TeV
  (data points have been slightly shifted along the horizontal axis
  for the sake of readability; see Table~\ref{tab:C0_e}).}
 \label{C0_e}
 \end{figure}

\begin{figure}[H]
 \centering
 
    \includegraphics[scale=0.38]{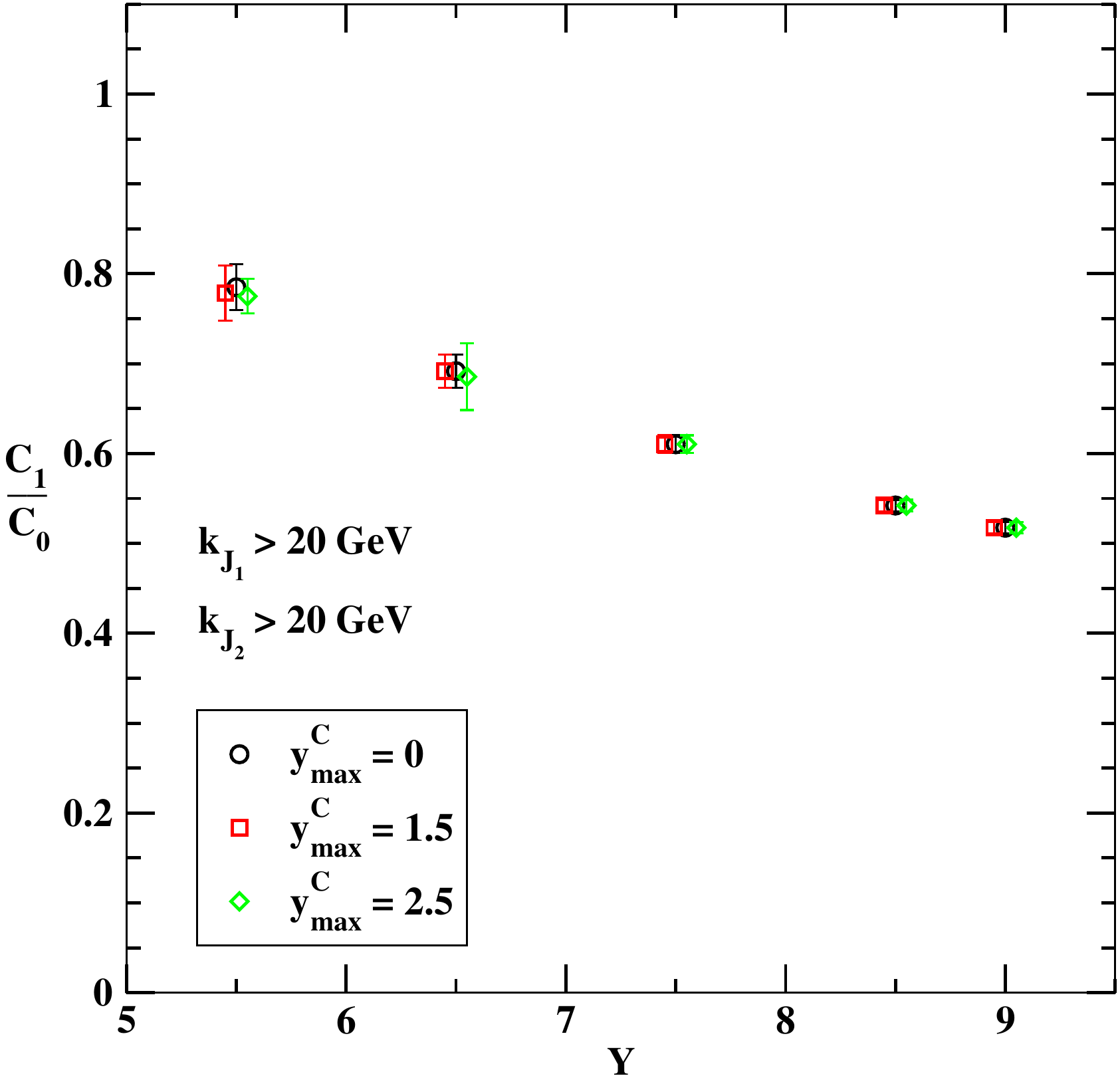}
    \includegraphics[scale=0.38]{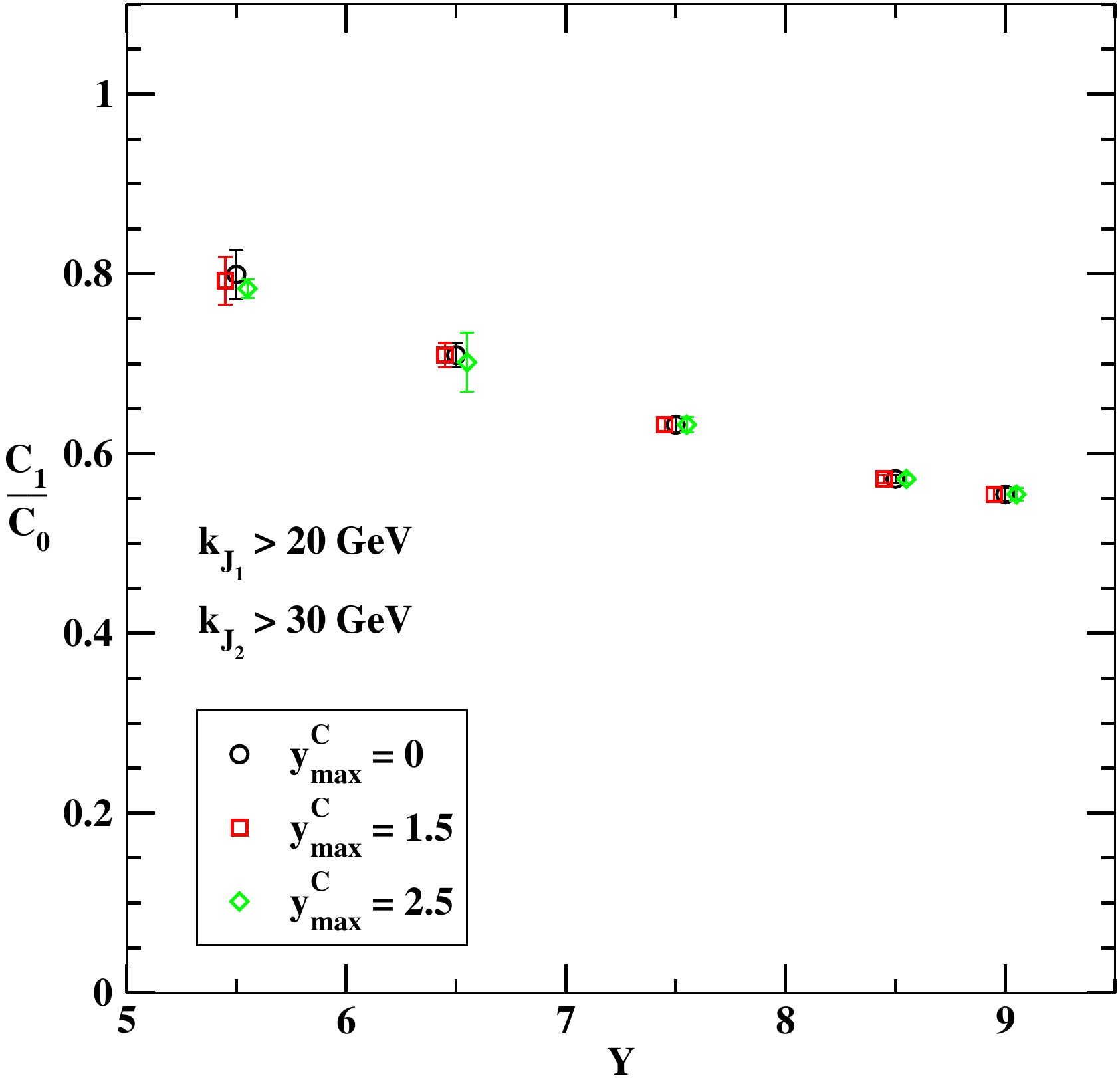}
 
    \includegraphics[scale=0.38]{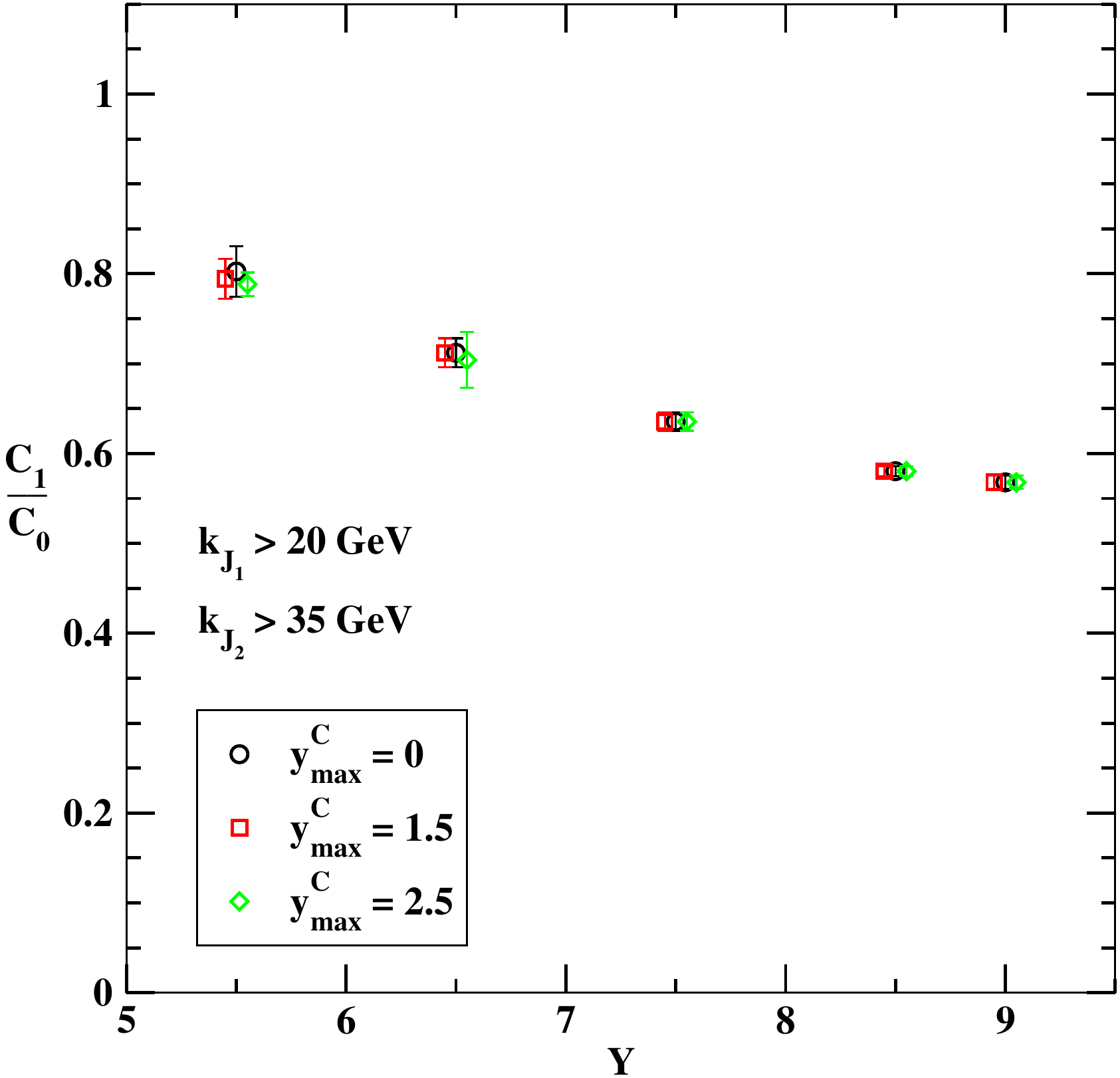}
    \includegraphics[scale=0.38]{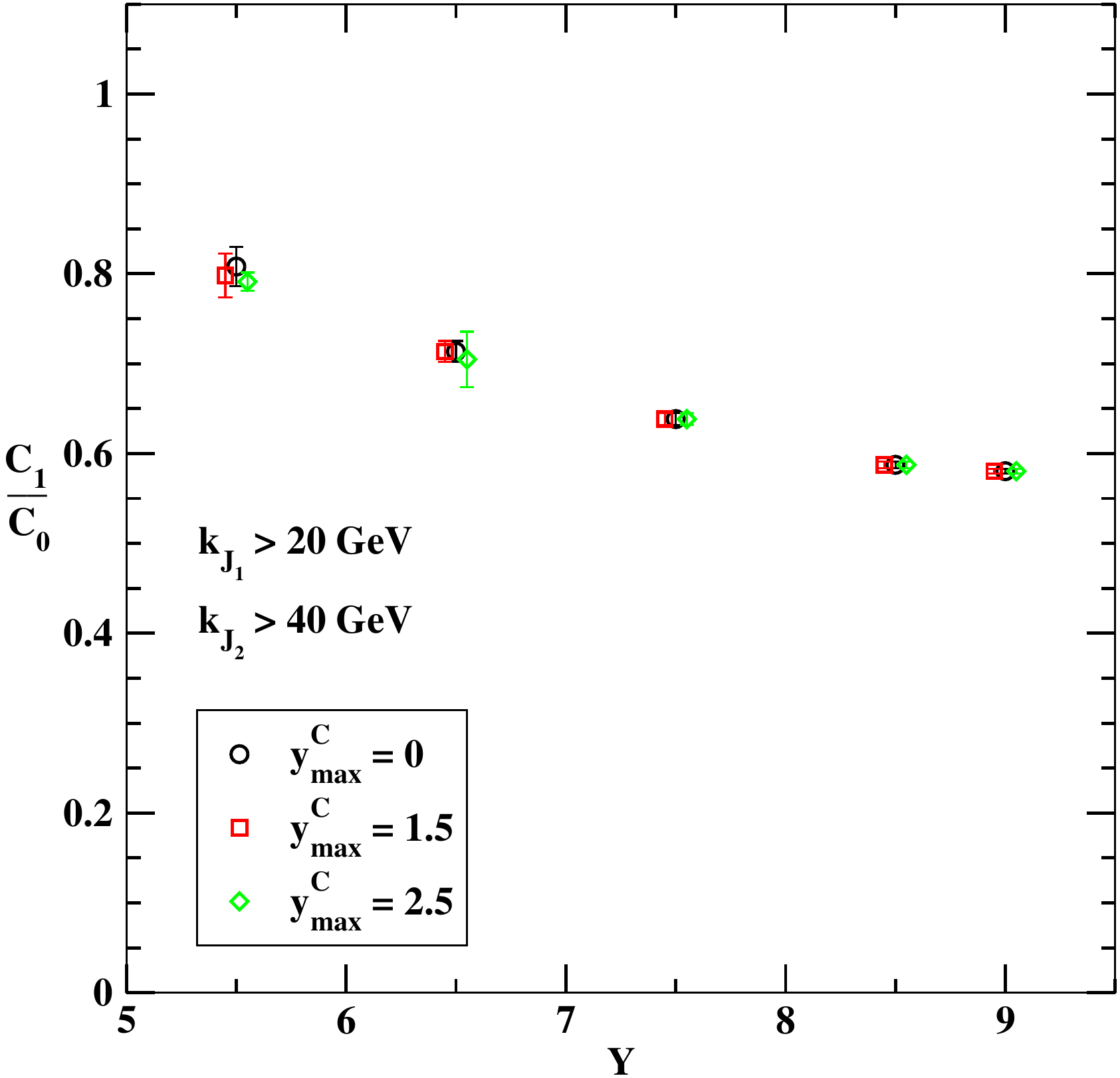}
 
    \includegraphics[scale=0.38]{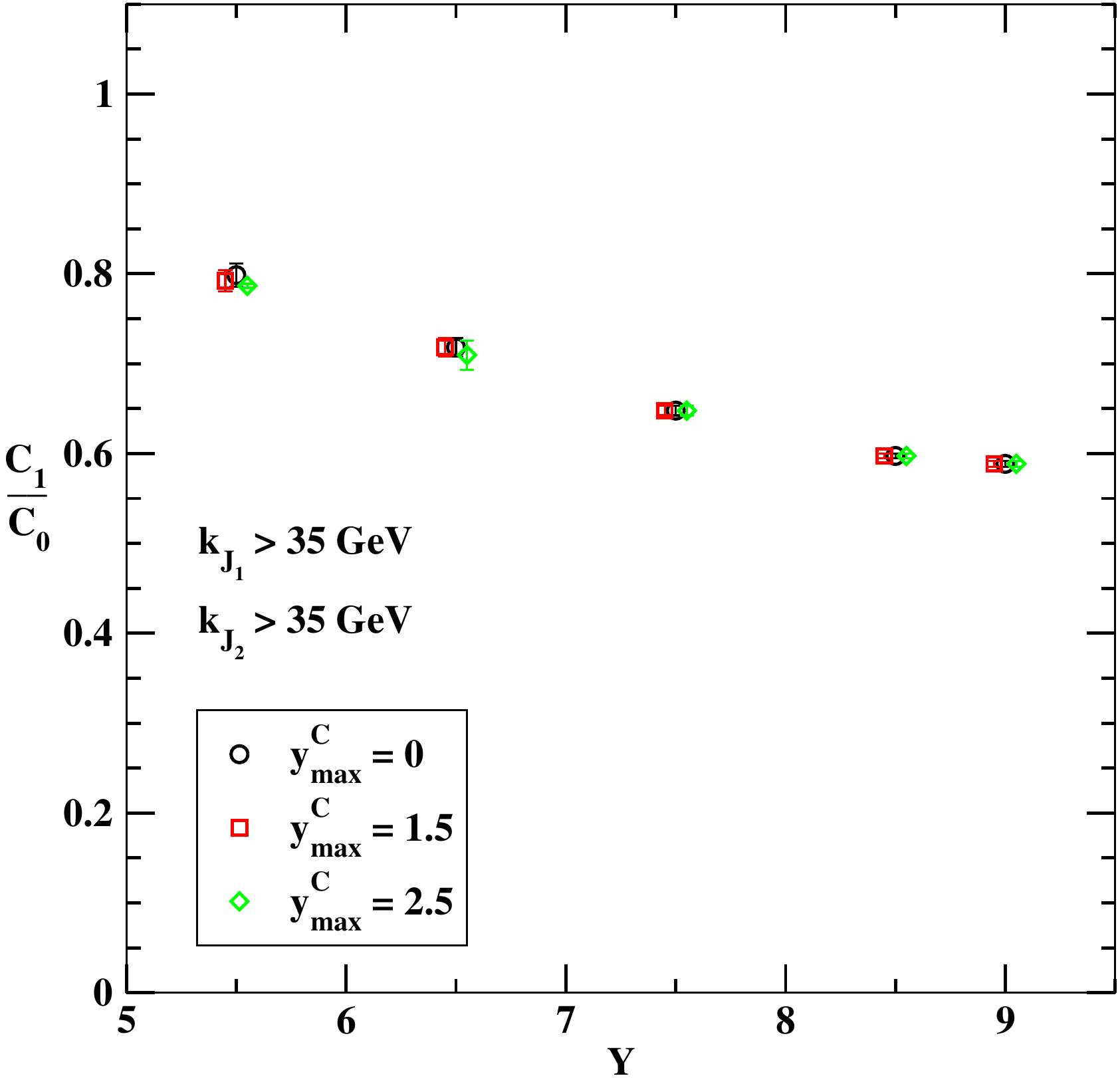}
 \caption[Rapidity veto effect on $R_{10}$ in dijet production]
 {$Y$-dependence of $C_1/C_0$ from the exact BLM method (Eq.~\ref{blm_exact-jets}),
  for all choices of the cuts on jet transverse momenta and of the central
  rapidity region, and for $\sqrt s = 13$ TeV 
  (data points have been slightly shifted along the horizontal
  axis for the sake of readability; see Table~\ref{tab:C1C0_e}).}
 \label{C1C0_e}
 \end{figure}

\begin{figure}[H]
 \centering
 
    \includegraphics[scale=0.38]{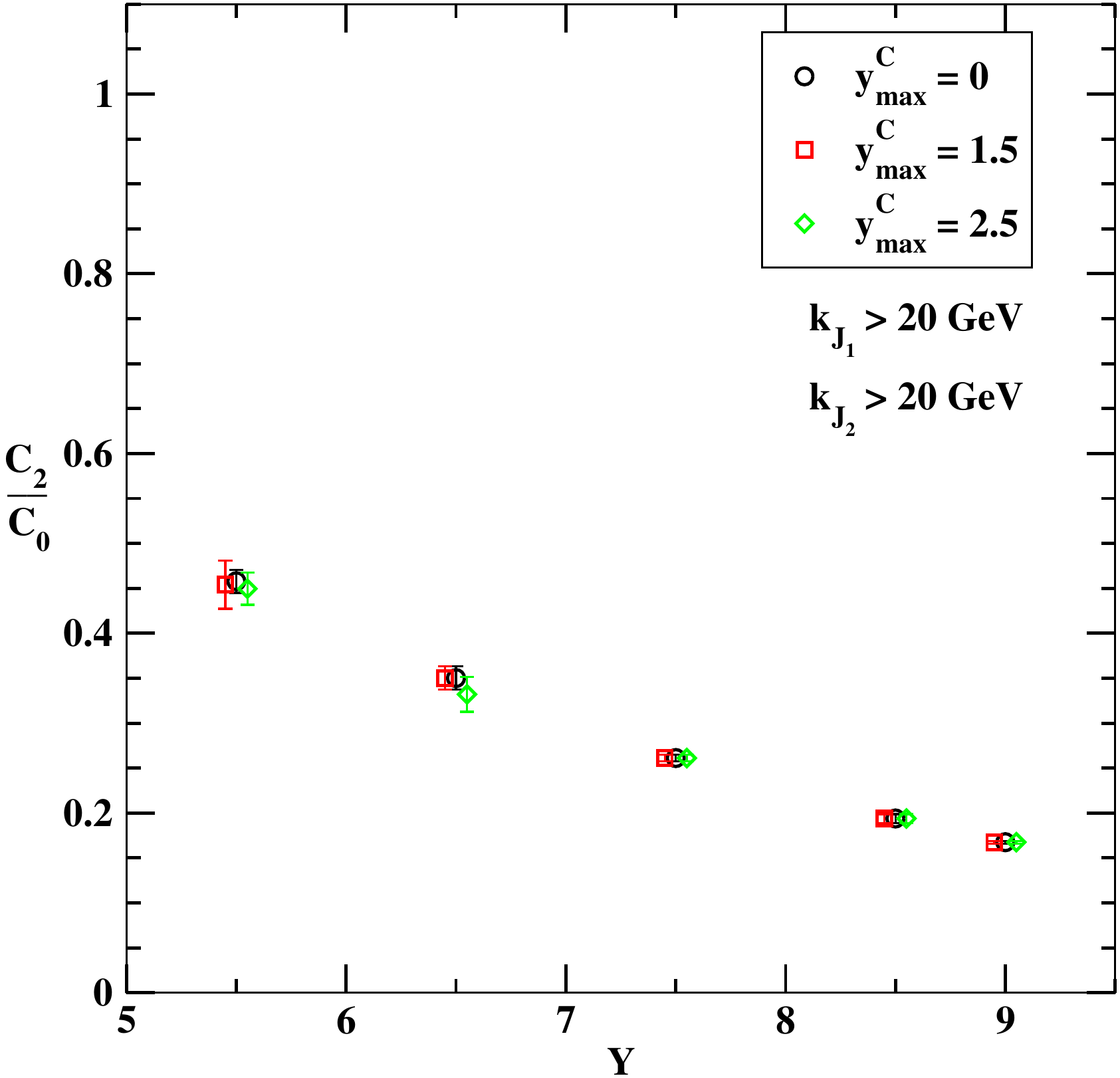}
    \includegraphics[scale=0.38]{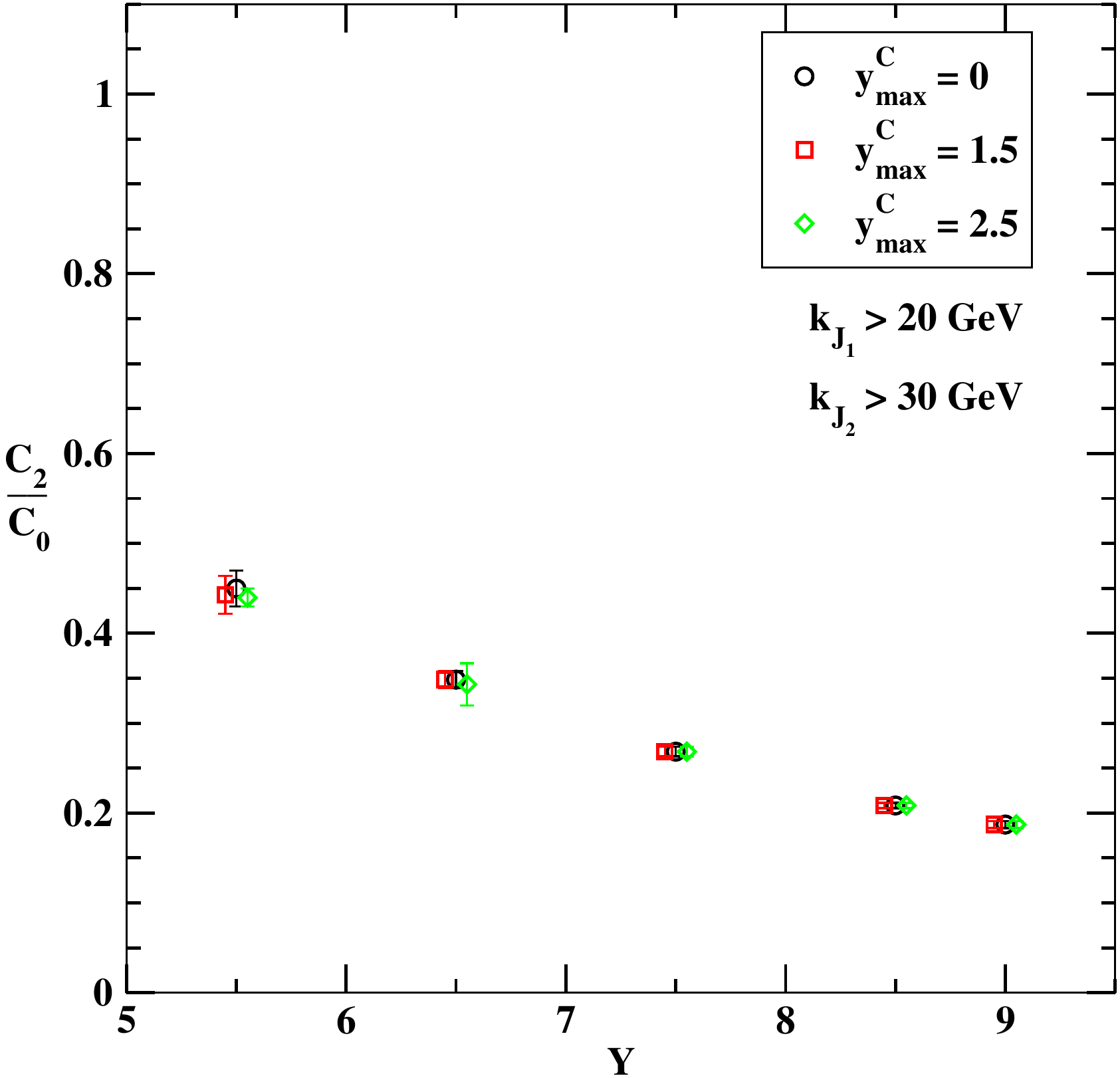}
 
    \includegraphics[scale=0.38]{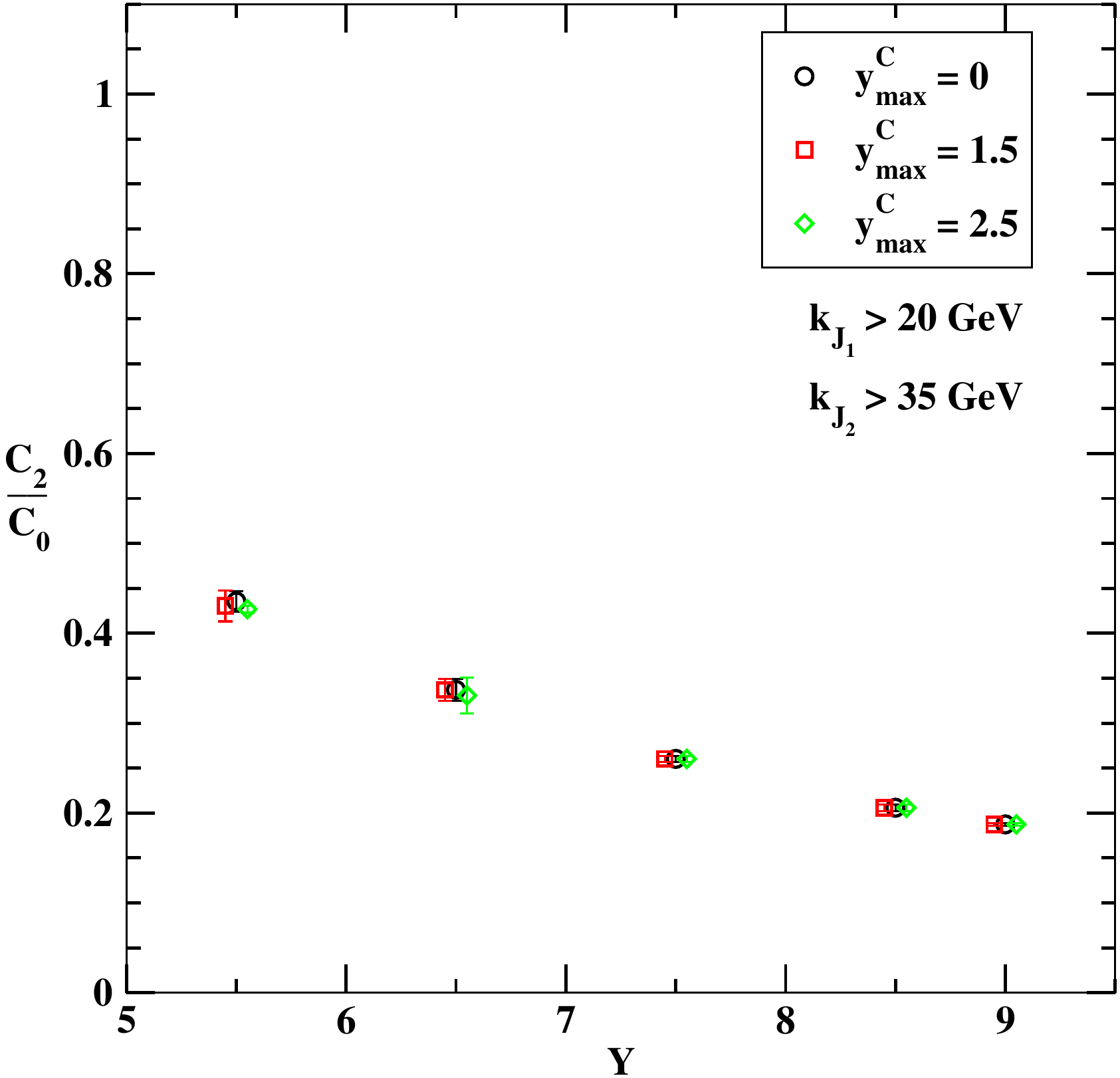}
    \includegraphics[scale=0.38]{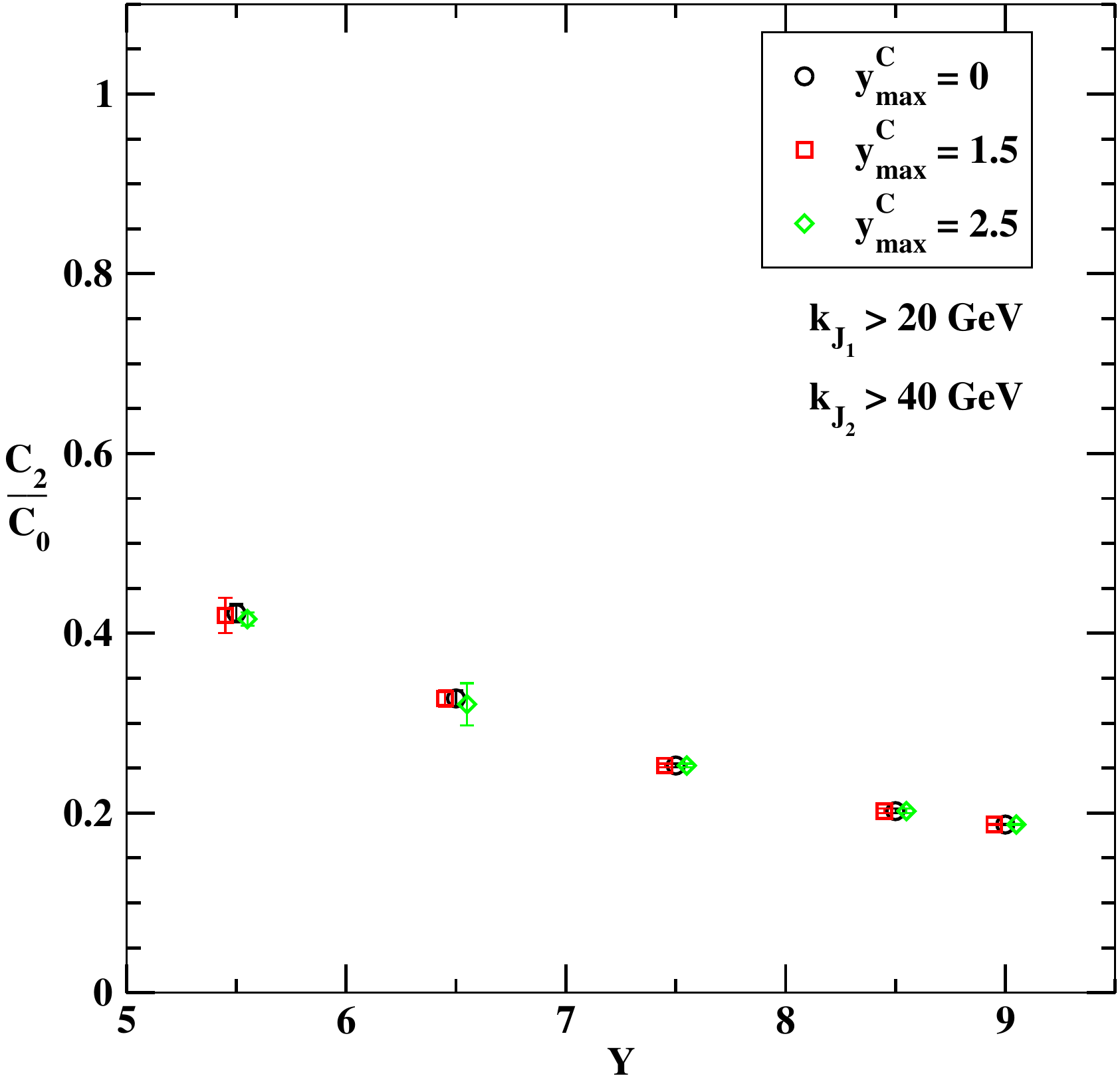}
 
    \includegraphics[scale=0.38]{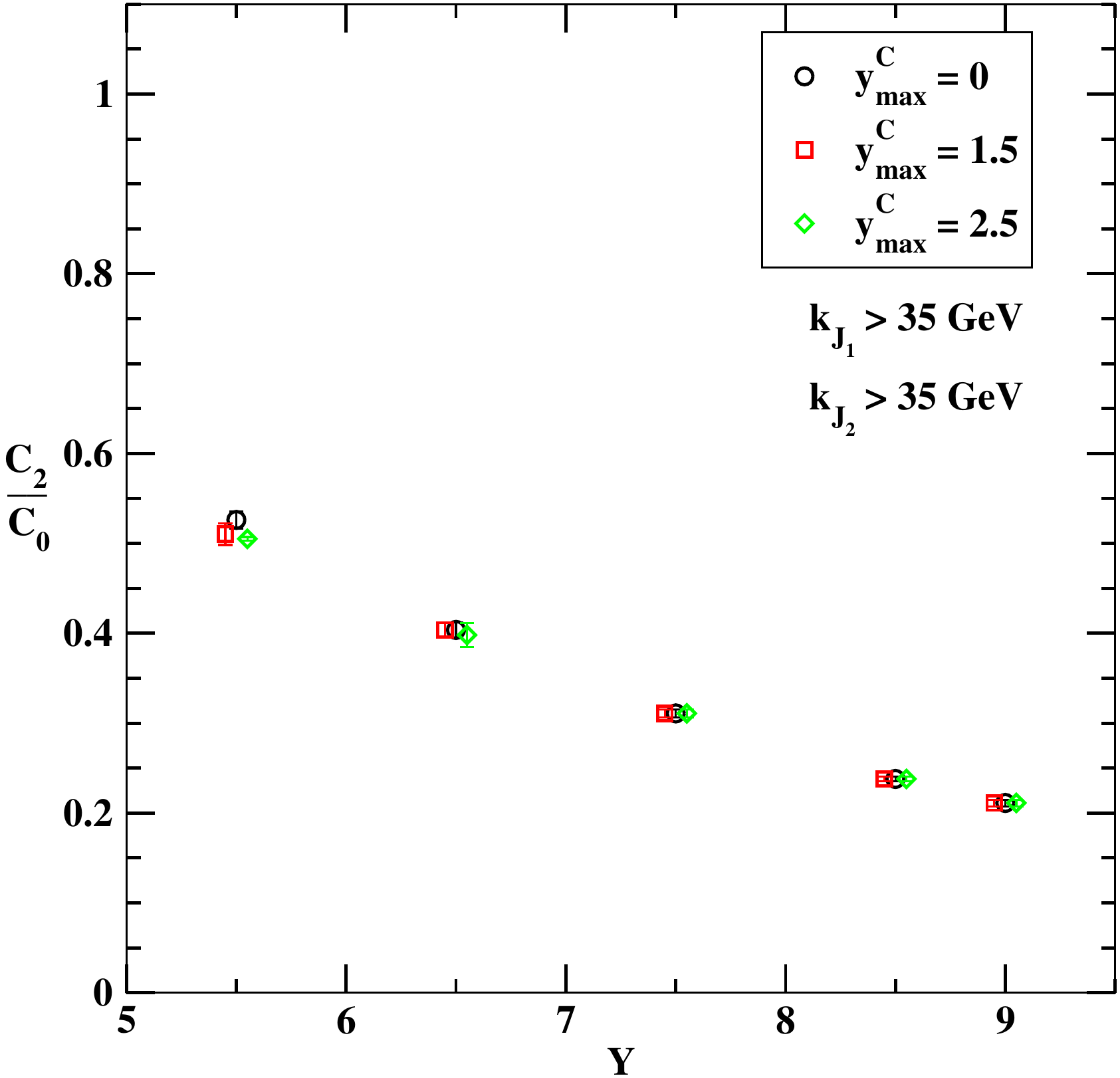}
 \caption[Rapidity veto effect on $R_{20}$ in dijet production]
 {$Y$-dependence of $C_2/C_0$ from the exact BLM method (Eq.~\ref{blm_exact-jets}),
  for all choices of the cuts on jet transverse momenta and of the central
  rapidity region, and for $\sqrt s = 13$ TeV 
  (data points have been slightly shifted along the horizontal
  axis for the sake of readability; see Table~\ref{tab:C2C0_e}).}
 \label{C2C0_e}
 \end{figure}

\begin{figure}[H]
 \centering
 
    \includegraphics[scale=0.38]{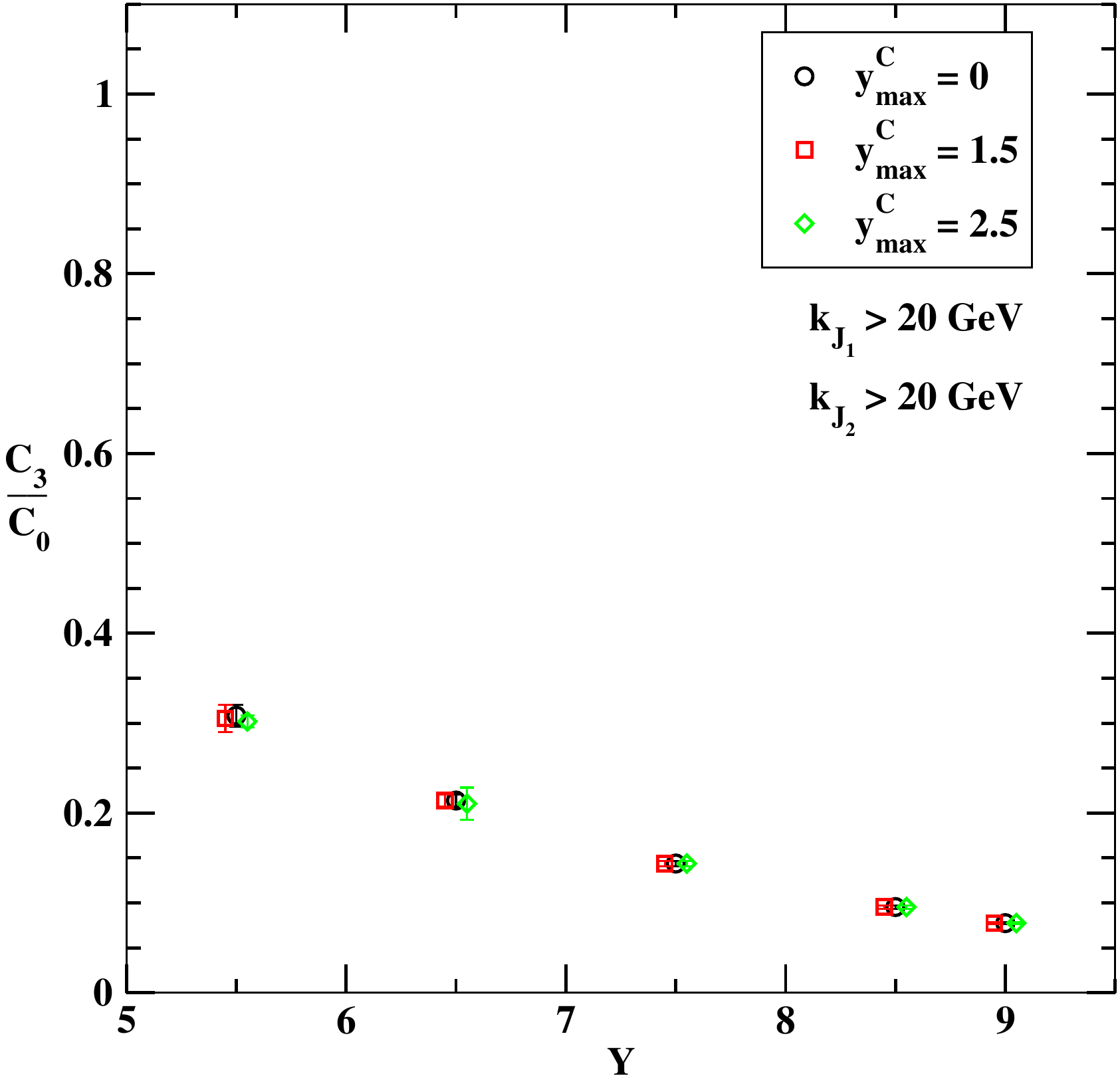}
    \includegraphics[scale=0.38]{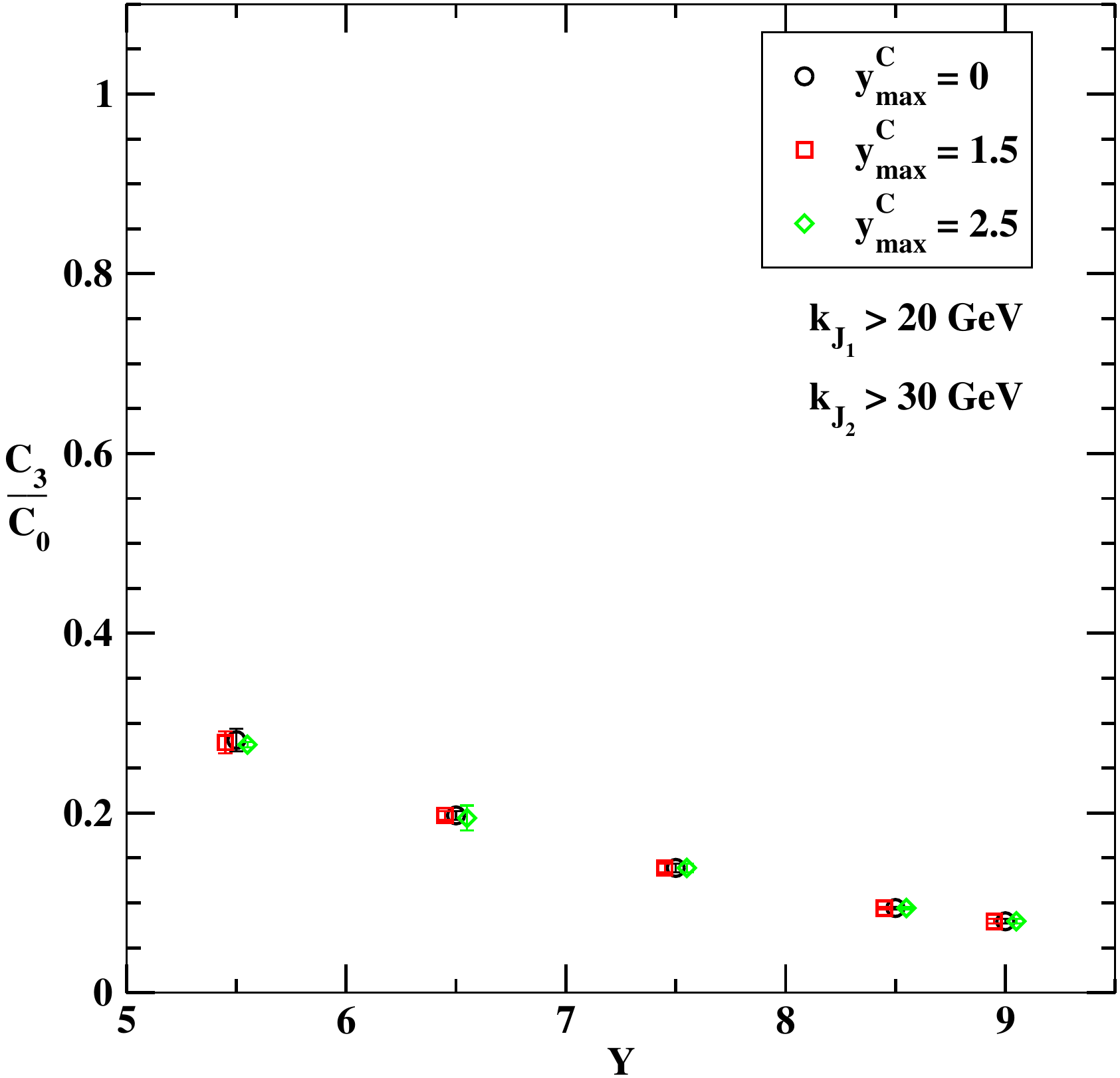}
 
    \includegraphics[scale=0.38]{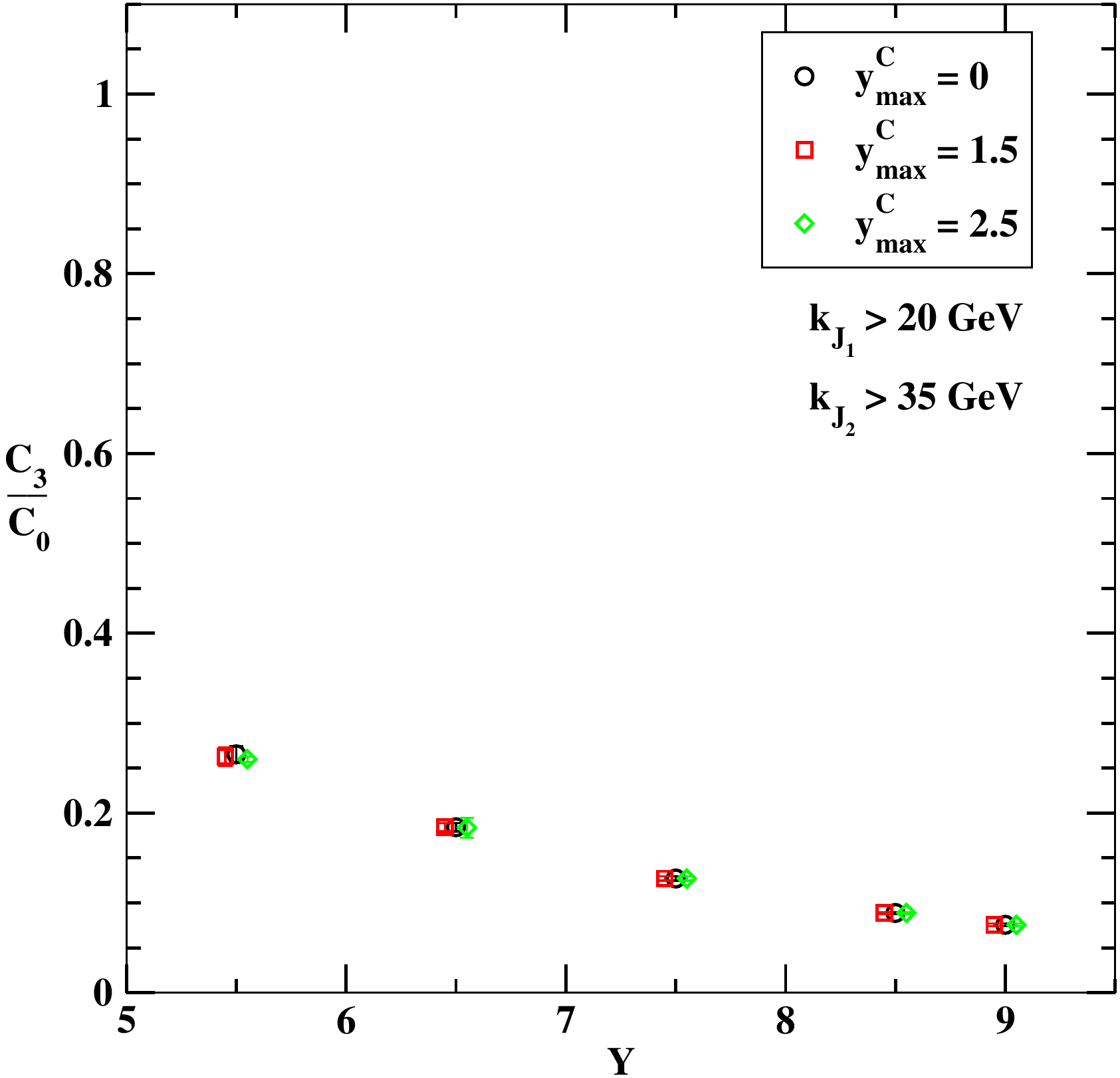}
    \includegraphics[scale=0.38]{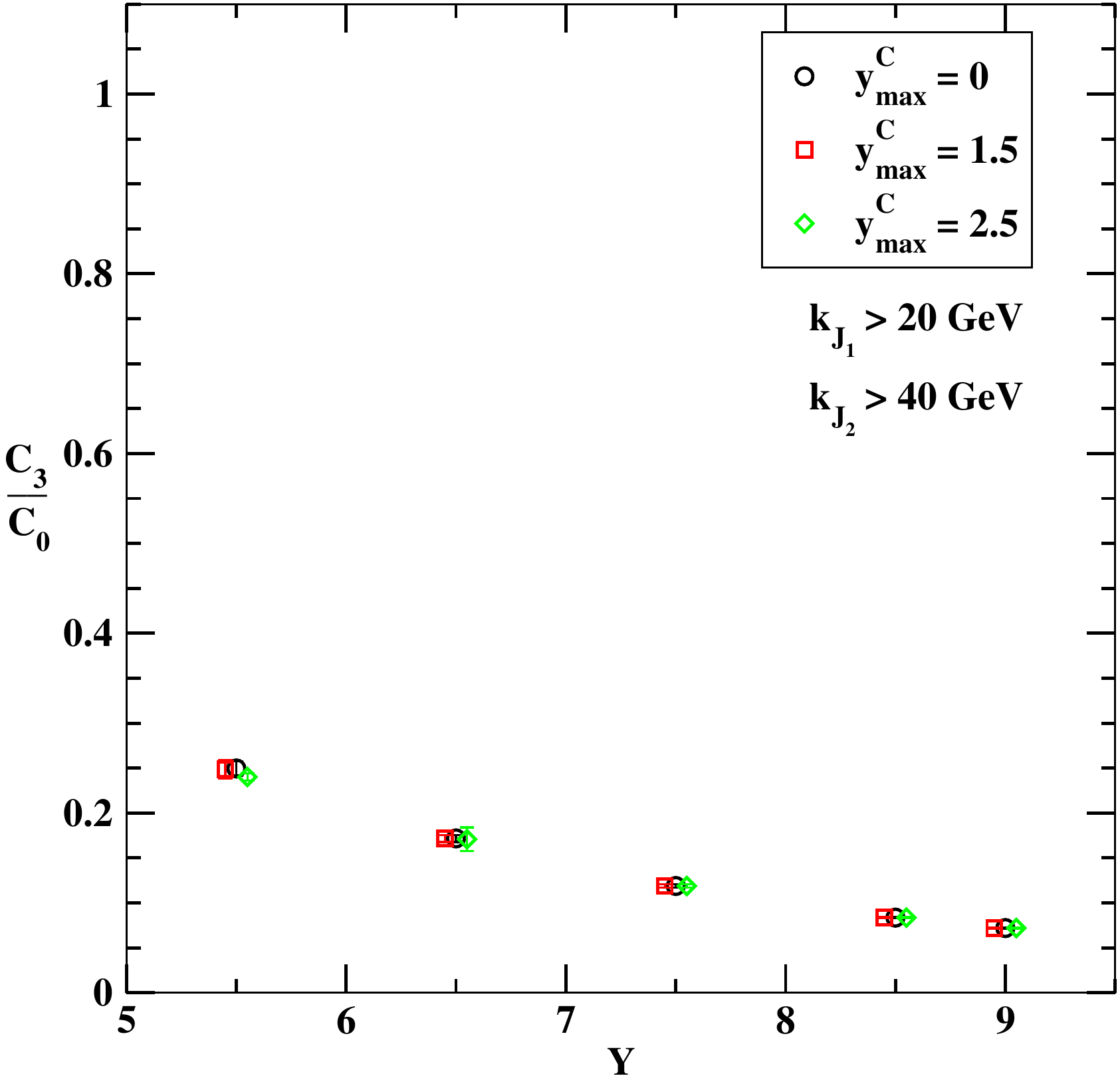}
 
    \includegraphics[scale=0.38]{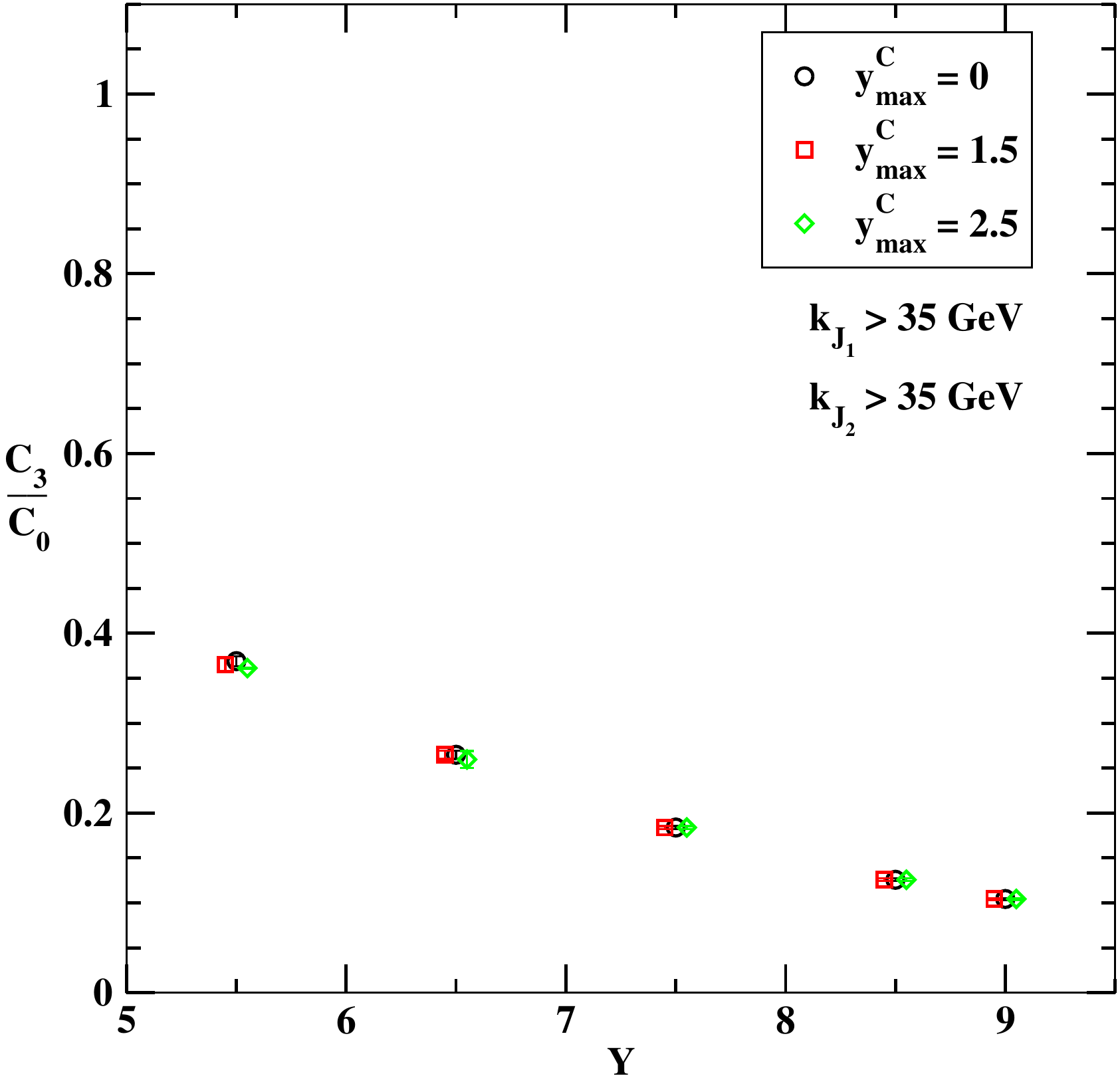}
  \caption[Rapidity veto effect on $R_{30}$ in dijet production]
  {$Y$-dependence of $C_3/C_0$ from the exact BLM method (Eq.~\ref{blm_exact-jets}),
  for all choices of the cuts on jet transverse momenta and of the central
  rapidity region, and for $\sqrt s = 13$ TeV 
  (data points have been slightly shifted along the horizontal
  axis for the sake of readability; see Table~\ref{tab:C3C0_e}).}
 \label{C3C0_e}
 \end{figure}

\begin{figure}[H]
 \centering
 
    \includegraphics[scale=0.38]{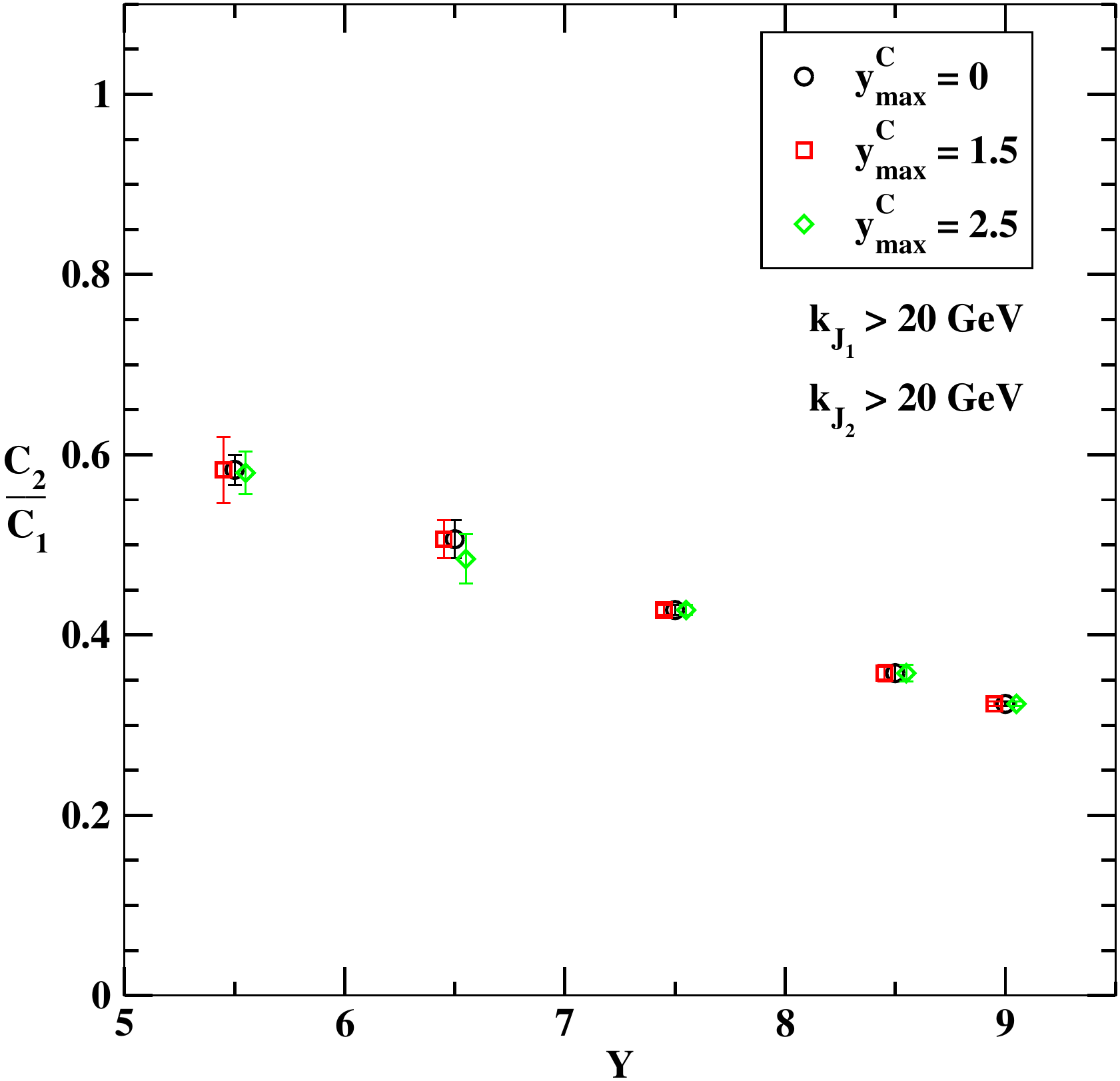}
    \includegraphics[scale=0.38]{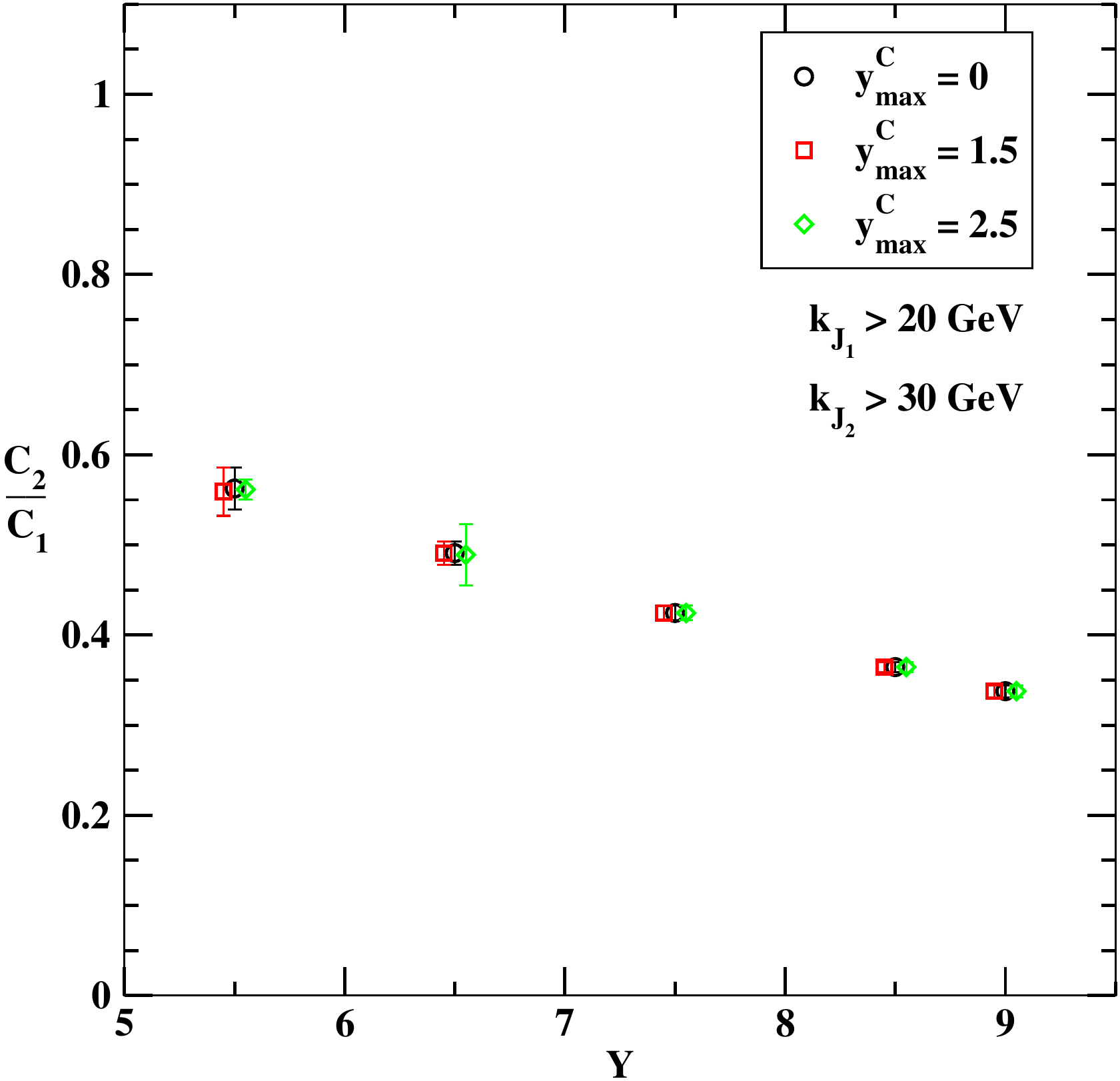}
 
    \includegraphics[scale=0.38]{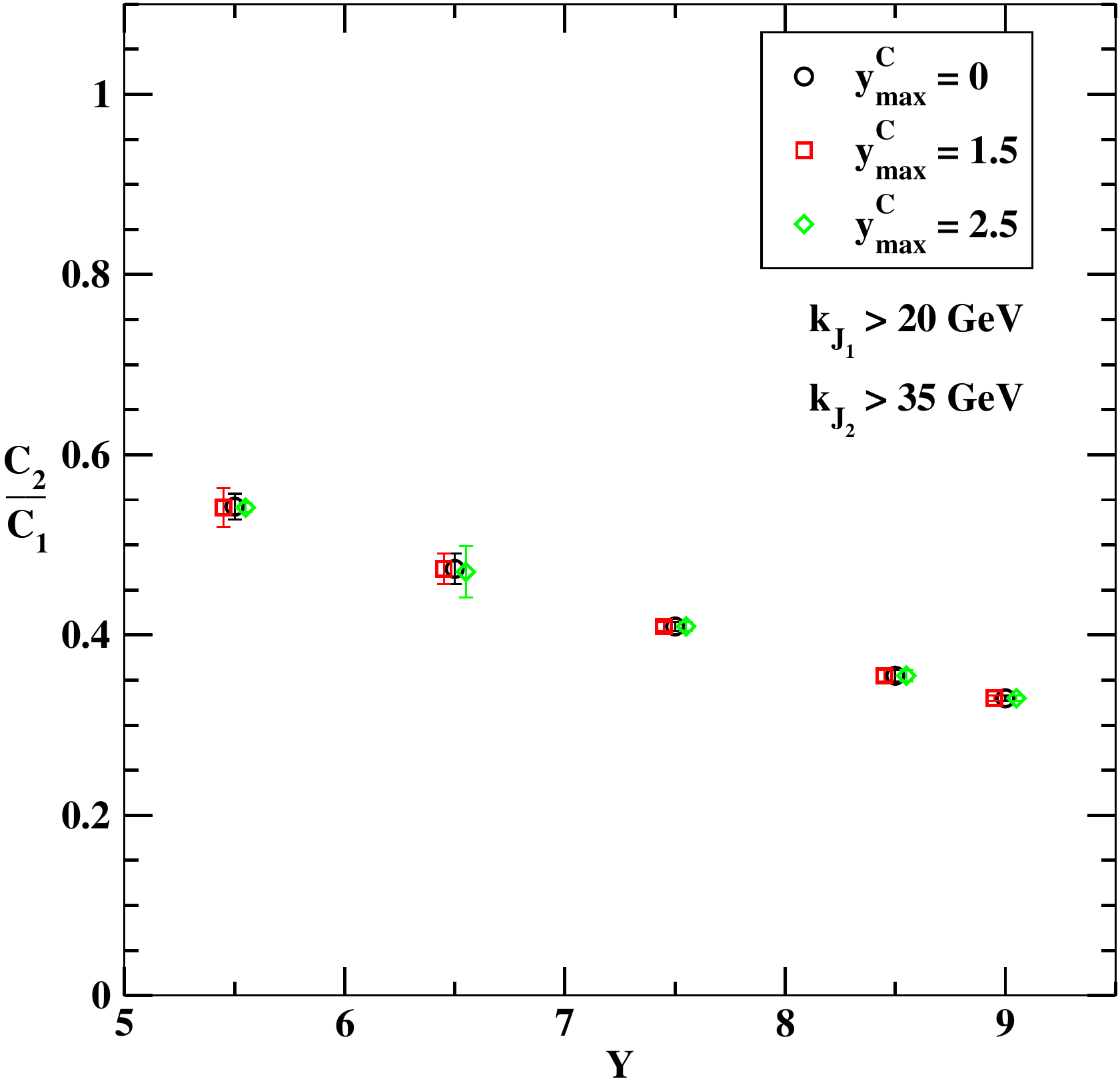}
    \includegraphics[scale=0.38]{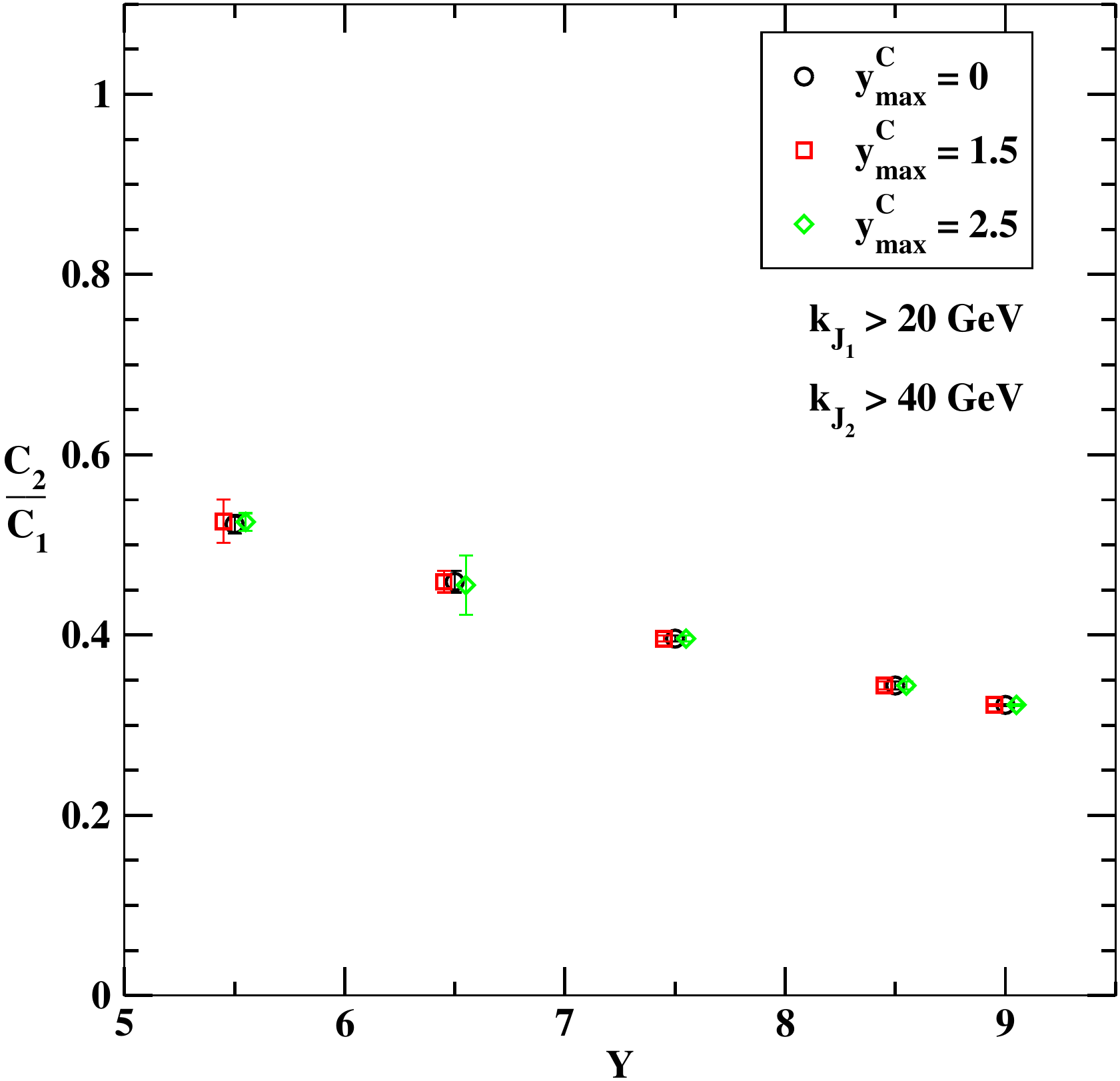}
 
    \includegraphics[scale=0.38]{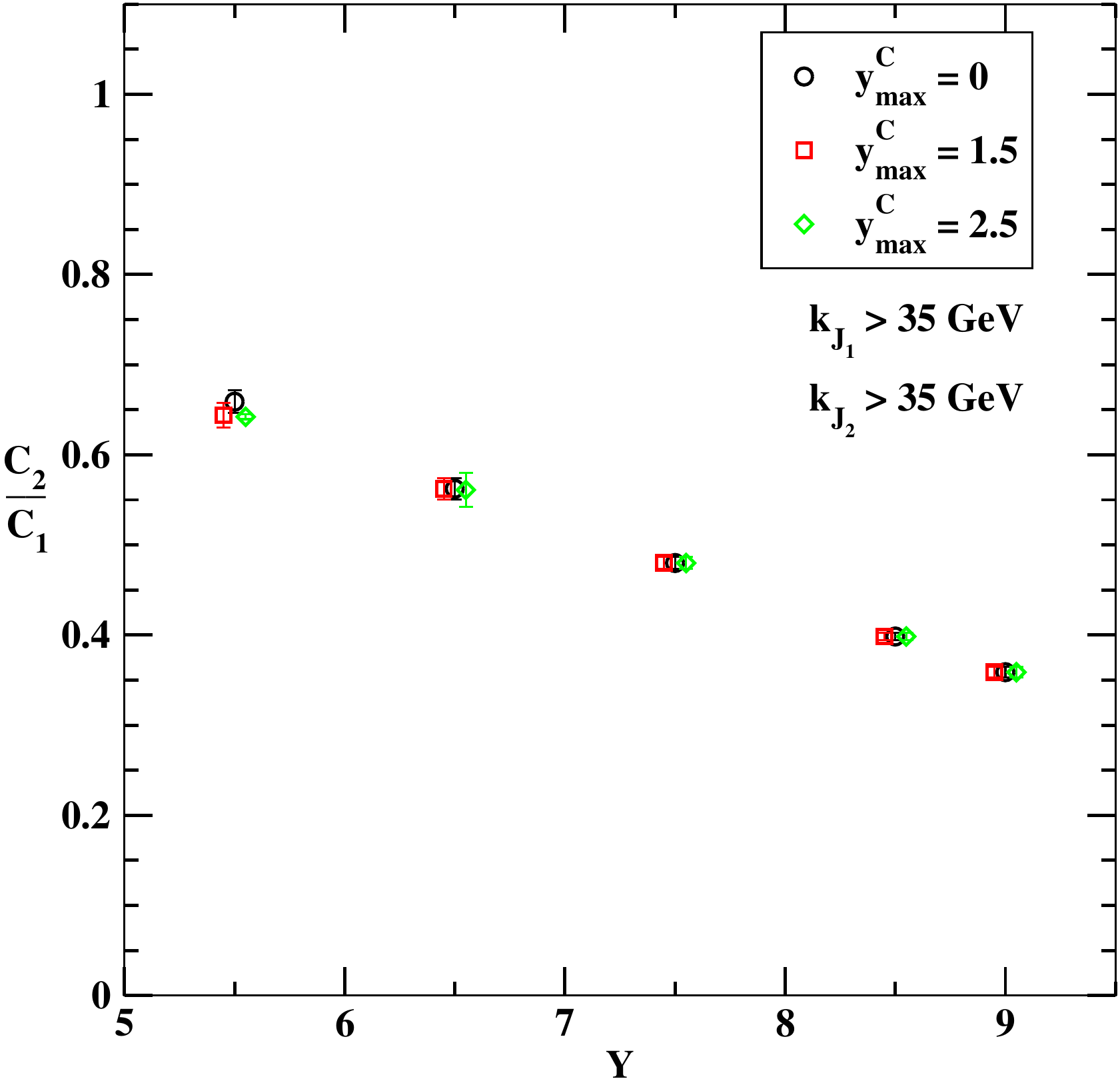}
 \caption[Rapidity veto effect on $R_{21}$ in dijet production]
 {$Y$-dependence of $C_2/C_1$ from the exact BLM method (Eq.~\ref{blm_exact-jets}),
  for all choices of the cuts on jet transverse momenta and of the central
  rapidity region, and for $\sqrt s = 13$ TeV 
  (data points have been slightly shifted along the horizontal
  axis for the sake of readability; see Table~\ref{tab:C2C1_e}).}
 \label{C2C1_e}
 \end{figure}

\begin{figure}[H]
 \centering
 
    \includegraphics[scale=0.38]{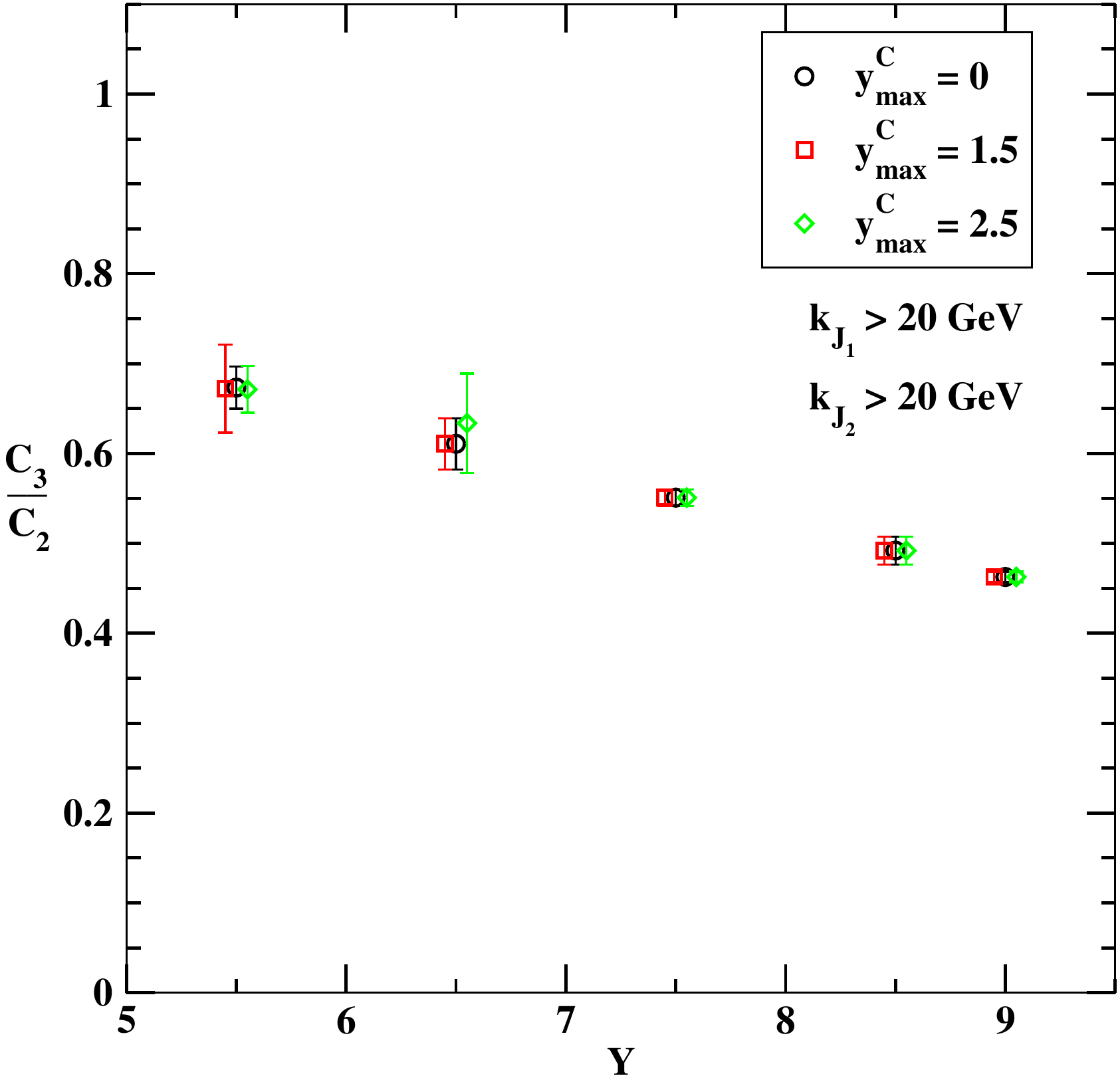}
    \includegraphics[scale=0.38]{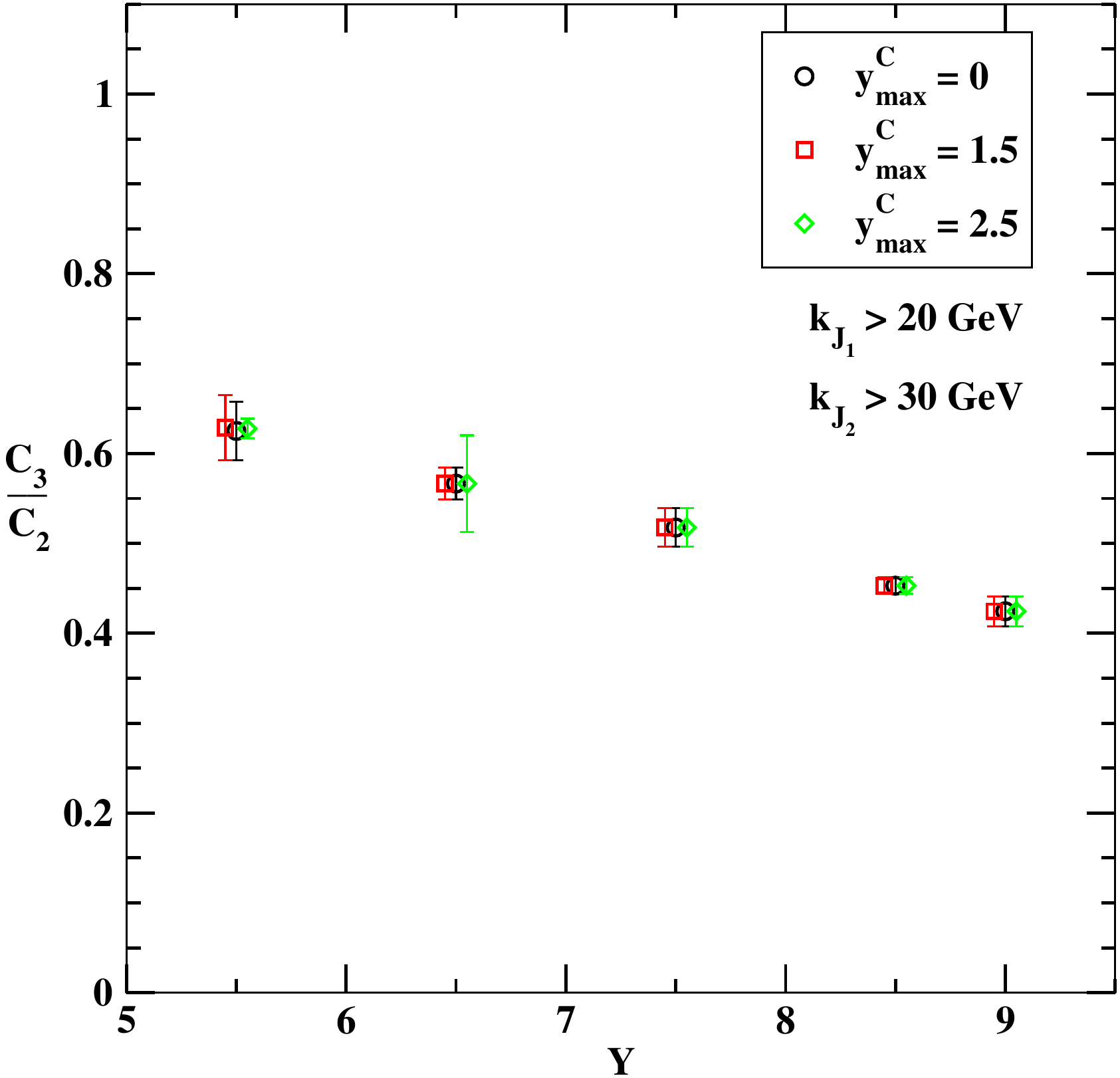}
 
    \includegraphics[scale=0.38]{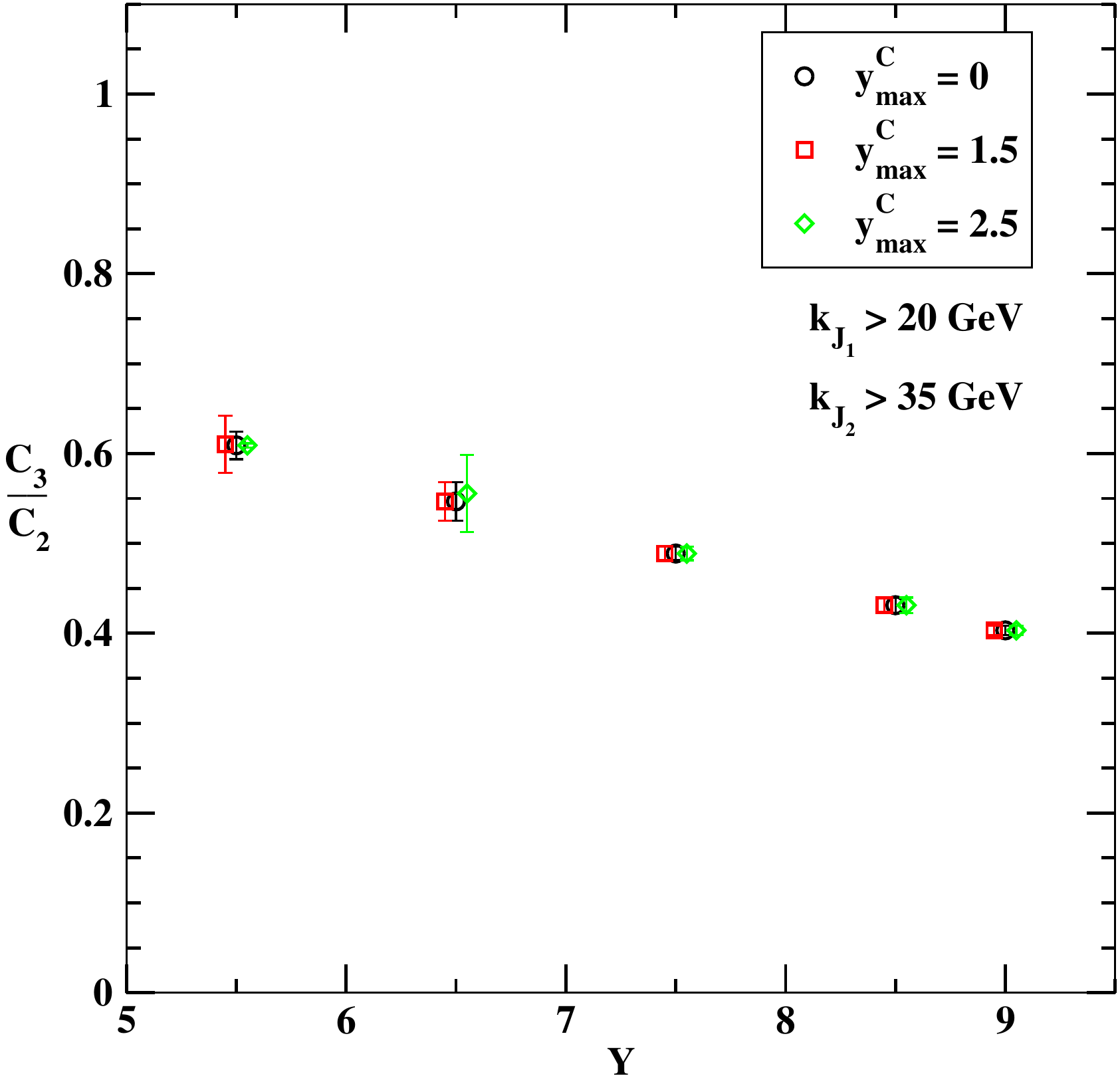}
    \includegraphics[scale=0.38]{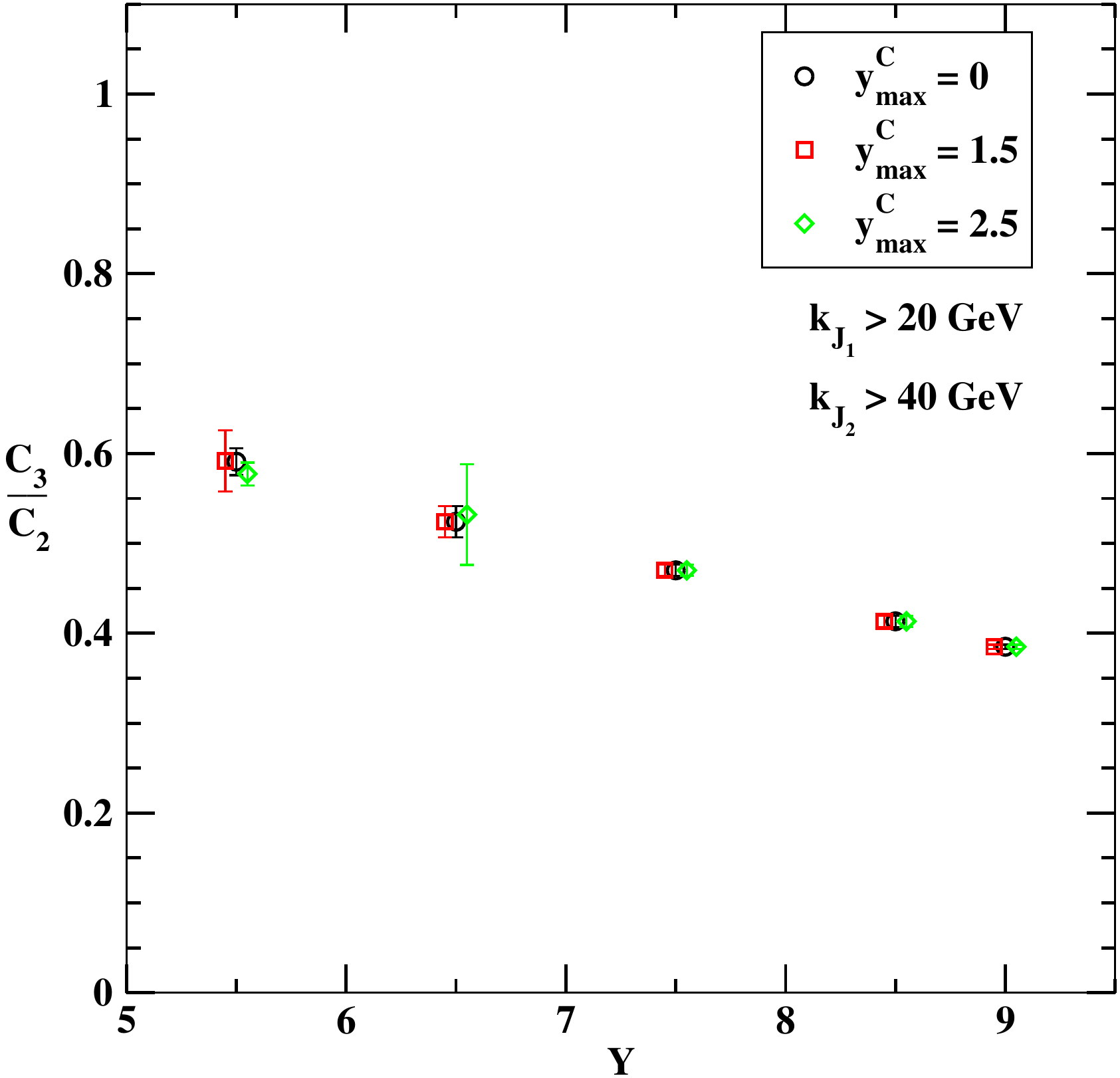}
 
    \includegraphics[scale=0.38]{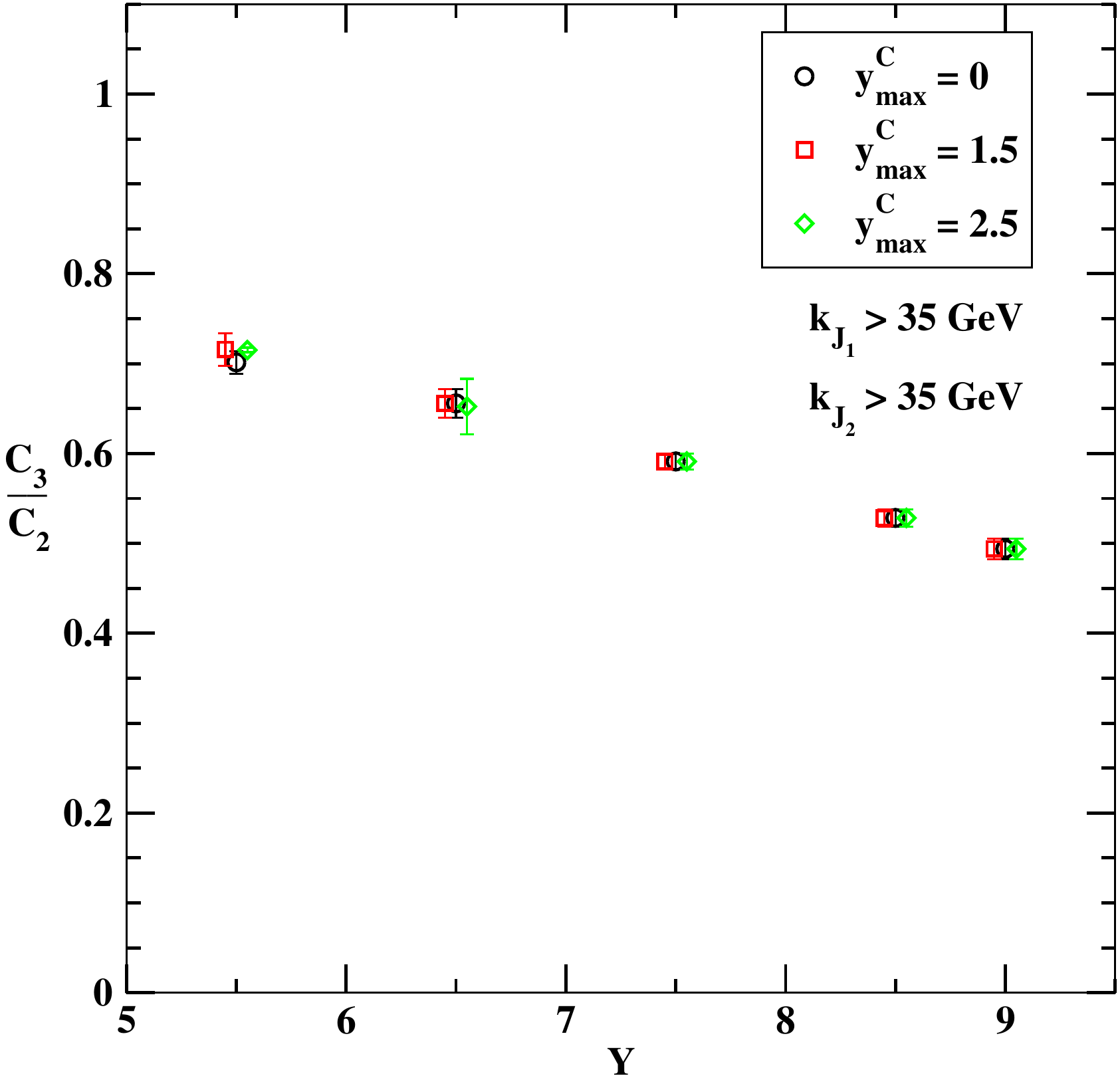}
 \caption[Rapidity veto effect on $R_{32}$ in dijet production]
 {$Y$-dependence of $C_3/C_2$ from the exact BLM method (Eq.~\ref{blm_exact-jets}),
  for all choices of the cuts on jet transverse momenta and of the central
  rapidity region, and for $\sqrt s = 13$ TeV 
  (data points have been slightly shifted along the horizontal
  axis for the sake of readability; see Table~\ref{tab:C3C2_e}).}
 \label{C3C2_e}
 \end{figure}

  We summarise our results in Tables~\ref{tab:2020_2.5}-\ref{tab:C3C2_e}
  and in Figs.~\ref{2020_2.5}-\ref{C3C2_e}. From Table~\ref{tab:2020_2.5}
  (and Fig.~\ref{2020_2.5}) we can see that the different variants of 
  implementation of the BLM method give predictions which deviate at the level 
  of $\sim 10\ $ for $C_0$ and at the level of $\sim 5\ $ for $R_{10}$, while 
  they basically agree within errors for all other ratios $R_{nm}$. For this 
  reason, all remaining Tables (and Figures) refer to the ``exact'' BLM case only.
  Table~\ref{tab:C0_e} (and Fig.~\ref{C0_e}) show, quite reasonably, that
  for all choices of the cuts on jet transverse momenta, the larger is
  $y^{\rm C}_{\rm max}$, the lower is the total cross section $C_0$, up the
  value of $Y$ is reached where the presence of cut of the central rapidity
  region becomes ineffective. All remaining Tables (and Figures) unanimously
  show that all ratios $R_{nm}$ remain unaffected by the cut on the central
  rapidity region, over the entire region of values of $Y$. This is obvious
  for the values of $Y$ large enough to be insensitive to the very presence of
  a non-zero $y^{\rm C}_{\rm max}$, but it is unexpectedly true also for the
  lower values of $Y$. 
  
  The latter point means that in our approach, {\it i.e.} NLA BFKL with BLM 
  optimisation, the cut on jet central rapidities leads to a proportional 
  reduction of both the total cross section, $C_0$, and the other coefficients 
  $C_1, C_2, C_3$, which parameterise the azimuthal angle distribution. In other 
  words in our approach, the central cut only reduces the value of the total 
  cross section, but does not affect the azimuthal angle distribution of dijets. 
  It would be very interesting to study whether such feature remains true also in 
  other approaches, both within the BFKL approach, but using different ide{\ae} 
  about the inclusion of the physics beyond NLA, and also in other, non-BFKL 
  schemes, like fixed-order DGLAP or approaches using $k_T$-factorisation~\cite{Catani:1990eg,Collins:1991ty,Gribov:1984tu,Levin:1991ry} for the central jet production.  
  
  \newpage
 

 \begin{table}[H]
 \centering
 \caption[Rapidity veto effect for different BLM scale settings 
          in dijet production]
 {$C_0$ [nb] and ratios $C_n/C_m$ 
  for $k_{J_1}^{\rm min}=k_{J_2}^{\rm min}=20$~GeV, 
  $y^{\rm C}_{\rm max}=2.5$, $\sqrt s = 13$ TeV, and 
  for the three variants of the BLM method (see Fig.~\ref{2020_2.5}).}
 \label{tab:2020_2.5}
 \begin{tabular}{c|c|lll}
 \toprule
           & $Y$ & BLM$_{a}$ & BLM$_{b}$ & BLM$_{\rm exact}$ \\
 \midrule
           & 5.5 & 1353.2(5.6) & 1413.2(3.2) & 1318(16)    \\
           & 6.5 & 1778(23)    & 1877(13)    & 1720(49)    \\
 $C_0$     & 7.5 & 834.6(2.8)  & 893.7(2.0)   & 803.4(6.6)  \\
           & 8.5 & 140.06(25)  & 152.03(18)  & 133.91(78)  \\
           & 9.0 & 32.97(10)   & 36.16(12)   & 31.46(20)   \\
 \midrule
           & 5.5 & 0.7641(68)  & 0.7434(37)  & 0.775(19)   \\
           & 6.5 & 0.674(17)   & 0.6546(87)  & 0.686(37)   \\
 $C_1/C_0$ & 7.5 & 0.6005(44)  & 0.5775(22)  & 0.6104(99)  \\
           & 8.5 & 0.5339(19)  & 0.5092(11)  & 0.5422(64)  \\
           & 9.0 & 0.5091(27)  & 0.4823(23)  & 0.5174(65)  \\
 \midrule
           & 5.5 & 0.4371(52)  & 0.4315(29)  & 0.450(18)   \\
           & 6.5 & 0.336(11)   & 0.3357(53)  & 0.3329(19)  \\
 $C_2/C_0$ & 7.5 & 0.2638(27)  & 0.2625(13)  & 0.2611(35)  \\
           & 8.5 & 0.2052(11)  & 0.20452(59) & 0.1939(49)  \\
           & 9.0 & 0.1835(14)  & 0.1827(11)  & 0.1674(14)  \\
 \midrule
           & 5.5 & 0.2761(45)  & 0.2691(26)  & 0.3019(68)  \\
           & 6.5 & 0.1934(74)  & 0.1907(37)  & 0.210(18)   \\
 $C_3/C_0$ & 7.5 & 0.1383(20)  & 0.13708(80) & 0.144(29)   \\
           & 8.5 & 0.09796(70) & 0.09765(31) & 0.095(17)   \\
           & 9.0 & 0.08378(90) & 0.08361(63) & 0.0775(13)  \\
 \midrule
           & 5.5 & 0.5721(71)  & 0.5804(42)  & 0.580(24)   \\
           & 6.5 & 0.499(15)   & 0.5128(76)  & 0.484(27)   \\
 $C_2/C_1$ & 7.5 & 0.4393(47)  & 0.4546(19)  & 0.4278(55)  \\
           & 8.5 & 0.3844(21)  & 0.4017(11)  & 0.3576(91)  \\
           & 9.0 & 0.3605(23)  & 0.3788(19)  & 0.3236(27)  \\
 \midrule
           & 5.5 & 0.632(13)   & 0.6236(74)  & 0.671(26)   \\
           & 6.5 & 0.575(25)   & 0.568(12)   & 0.634(55)   \\
 $C_3/C_2$ & 7.5 & 0.5241(93)  & 0.5221(32)  & 0.5509(92)  \\
           & 8.5 & 0.4773(41)  & 0.4775(18)  & 0.492(16)   \\
           & 9.0 & 0.4565(55)  & 0.4577(29)  & 0.4627(59)  \\
 \bottomrule
 \end{tabular}
 \end{table}

 \begin{table}[H]
 \vspace{-0.80cm}
 \centering
 \caption[Rapidity veto effect on $C_0$ in dijet production]
 {Values of $C_0$ [nb] from the exact BLM method (Eq.~\ref{blm_exact-jets}) at $\sqrt s = 13$ TeV, for all
  choices of the cuts on jet transverse momenta 
  and of the central rapidity region
  (see Fig.~\ref{C0_e}).}
 \label{tab:C0_e}
 \begin{tabular}{c|c|c|lll}
 \toprule
 $k_{J_1}^{\rm min}$ & $k_{J_2}^{\rm min}$ & $Y$  & $y^{\rm C}_{\rm max}=0$
                & $y^{\rm C}_{\rm max}=1.5$ & $y^{\rm C}_{\rm max}=2.5$ \\
 \midrule
        &        & 3.5 & 46100(950)  & 5498(110)   & -           \\
        &        & 4.5 & 20410(290)  & 8200(130)   & -           \\
        &        & 5.5 & 8270(130)   & 6120(110)   & 1318(16)    \\
 20 GeV & 20 GeV & 6.5 & 2902(31)    & 2902(31)    & 1720(49)    \\
        &        & 7.5 & 803.4(6.6)  & 803.4(6.6)  & 803.4(6.6)  \\
        &        & 8.5 & 133.91(78)  & 133.91(78)  & 133.91(78)  \\
        &        & 9.0 & 31.46(20)   & 31.46(20)   & 31.46(20)   \\
 \midrule
        &        & 3.5 & 15000(270)  & 1842(27)    & -           \\
        &        & 4.5 & 6734(73)    & 2779(33)    & -           \\
        &        & 5.5 & 2701(51)    & 2030(34)    & 442.3(3.4)  \\
 20 GeV & 30 GeV & 6.5 & 919.8(9.2)  & 919.8(9.2)  & 555(13)     \\
        &        & 7.5 & 240.8(1.6)  & 240.8(1.6)  & 240.8(1.6)  \\
        &        & 8.5 & 36.44(13)   & 36.44(13)   & 36.44(13)   \\
        &        & 9.0 & 7.801(53)   & 7.801(53)   & 7.801(53)   \\
 \midrule
        &        & 3.5 & 8090(160)   & 1050(20)    & -           \\
        &        & 4.5 & 3793(54)    & 1598(21)    & -           \\
        &        & 5.5 & 1534(26)    & 1169(16)    & 256.0(2.1)  \\
 20 GeV & 35 GeV & 6.5 & 520.6(6.2)  & 520.6(6.2)  & 318.5(6.9)  \\
        &        & 7.5 & 134.2(1.1)  & 134.2(1.1)  & 134.2(1.1)  \\
        &        & 8.5 & 19.422(98)  & 19.422(98)  & 19.422(98)  \\
        &        & 9.0 & 3.9601(23)  & 3.9601(23)  & 3.9601(23)  \\
 \midrule
        &        & 3.5 & 4627(86)    & 595.3(7.3)  & -           \\
        &        & 4.5 & 2137(31)    & 912(10)     & -           \\
        &        & 5.5 & 872(13)     & 668(10)     & 146.68(94)  \\
 20 GeV & 40 GeV & 6.5 & 295.4(2.7)  & 295.4(2.7)  & 181.6(4.1)  \\
        &        & 7.5 & 74.75(37)   & 74.75(37)   & 74.75(37)   \\
        &        & 8.5 & 10.362(30)  & 10.362(30)  & 10.362(30)  \\
        &        & 9.0 & 1.9980(45)  & 1.9980(45)  & 1.9980(45)  \\
 \midrule
        &        & 3.5 & 4286(36)    & 544.7(6.0)  & -           \\
        &        & 4.5 & 1618(13)    & 690.9(3.3)  & -           \\
        &        & 5.5 & 555.2(4.1)  & 429.0(3.6)  & 94.48(13)   \\
 35 GeV & 35 GeV & 6.5 & 161.8(1.2)  & 161.8(1.2)  & 101.5(1.1)  \\
        &        & 7.5 & 35.70(16)   & 35.70(16)   & 35.70(16)   \\
        &        & 8.5 & 4.2843(98)  & 4.2843(98)  & 4.2843(98)  \\
        &        & 9.0 & 0.7579(23)  & 0.7579(23)  & 0.7579(23)  \\
 \bottomrule
 \end{tabular}
 \end{table}
 
 \begin{table}[H]
 \vspace{-0.80cm}
 \centering
 \caption[Rapidity veto effect on $R_{10}$ in dijet production]
 {Values of $C_1/C_0$ from the exact BLM method (Eq.~\ref{blm_exact-jets}) 
  at $\sqrt s = 13$ TeV, for all
  choices of the cuts on jet transverse momenta 
  and of the central rapidity region
  (see Fig.~\ref{C1C0_e}).}
 \label{tab:C1C0_e}
 \begin{tabular}{c|c|c|lll}
 \toprule
 $k_{J_1}^{\rm min}$ & $k_{J_2}^{\rm min}$ & $Y$  & $y^{\rm C}_{\rm max}=0$
                & $y^{\rm C}_{\rm max}=1.5$ & $y^{\rm C}_{\rm max}=2.5$ \\
 \midrule
        &        & 3.5 & 0.988(37)   & 0.975(35)   & -           \\
        &        & 4.5 & 0.885(25)   & 0.874(27)   & -           \\
        &        & 5.5 & 0.785(25)   & 0.778(31)   & 0.775(19)   \\
 20 GeV & 20 GeV & 6.5 & 0.692(18)   & 0.692(18)   & 0.686(37)   \\
        &        & 7.5 & 0.6104(99)  & 0.6104(99)  & 0.6104(99)  \\
        &        & 8.5 & 0.5423(64)  & 0.5423(64)  & 0.5423(64)  \\
        &        & 9.0 & 0.5174(64)  & 0.5174(64)  & 0.5174(64)  \\
 \midrule
        &        & 3.5 & 1.004(31)   & 0.989(28)   & -           \\
        &        & 4.5 & 0.896(18)   & 0.886(20)   & -           \\
        &        & 5.5 & 0.799(27)   & 0.792(27)   & 0.783(10)   \\
 20 GeV & 30 GeV & 6.5 & 0.710(13)   & 0.710(13)   & 0.702(33    \\
        &        & 7.5 & 0.6321(83)  & 0.6321(83)  & 0.6321(83)  \\
        &        & 8.5 & 0.5717(45)  & 0.5717(45)  & 0.5717(45)  \\
        &        & 9.0 & 0.5543(70)  & 0.5543(70)  & 0.5543(70)  \\
 \midrule
        &        & 3.5 & 1.051(37)   & 1.005(33)   & -           \\
        &        & 4.5 & 0.907(24)   & 0.892(24)   & -           \\
        &        & 5.5 & 0.803(28)   & 0.795(22)   &  0.788(13)  \\
 20 GeV & 35 GeV & 6.5 & 0.712(16)   & 0.712(16)   &  0.704(31)  \\
        &        & 7.5 & 0.636(10)   & 0.636(10)   & 0.636(10)   \\
        &        & 8.5 & 0.5803(56)  & 0.5803(56)  & 0.5803(56)  \\
        &        & 9.0 & 0.5679(74)  & 0.5679(74)  & 0.5679(74)  \\
 \midrule
        &        & 3.5 & 1.043(35)   & 1.021(22)   & -           \\
        &        & 4.5 & 0.916(25)   & 0.899(20)   & -           \\
        &        & 5.5 & 0.808(22)   & 0.798(24)   & 0.791(10)   \\
 20 GeV & 40 GeV & 6.5 & 0.714(12)   & 0.714(12)   & 0.705(31)   \\
        &        & 7.5 & 0.6383(64)  & 0.6383(64)  & 0.6383(64)  \\
        &        & 8.5 & 0.5875(35)  & 0.5875(35)  & 0.5875(35)  \\
        &        & 9.0 & 0.5804(25)  & 0.5804(25)  & 0.5804(25)  \\
 \midrule
        &        & 3.5 & 0.963(16)   & 0.952(18)   & -           \\
        &        & 4.5 & 0.883(14)   & 0.8722(82)  & -           \\
        &        & 5.5 & 0.798(13)   & 0.792(12)   & 0.7866(22)  \\
 35 GeV & 35 GeV & 6.5 & 0.718(11)   & 0.718(11)   &  0.709(16)  \\
        &        & 7.5 & 0.6478(53)  & 0.6478(53)  & 0.6478(53)  \\
        &        & 8.5 & 0.5972(26)  & 0.5972(26)  & 0.5972(26)  \\
        &        & 9.0 & 0.5886(33)  & 0.5886(33)  & 0.5886(33)  \\
 \bottomrule
 \end{tabular}
 \end{table}
 
 \begin{table}[H]
 \vspace{-0.80cm}
 \centering
 \caption[Rapidity veto effect on $R_{20}$ in dijet production]
 {Values of $C_2/C_0$ from the exact BLM method (Eq.~\ref{blm_exact-jets})  
  at $\sqrt s = 13$ TeV, for all
  choices of the cuts on jet transverse momenta 
  and of the central rapidity region
  (see Fig.~\ref{C2C0_e}).}
 \label{tab:C2C0_e}
 \begin{tabular}{c|c|c|lll}
 \toprule
 $k_{J_1}^{\rm min}$ & $k_{J_2}^{\rm min}$ & $Y$ & $y^{\rm C}_{\rm max}=0$
                & $y^{\rm C}_{\rm max}=1.5$ & $y^{\rm C}_{\rm max}=2.5$ \\
 \midrule
        &        & 3.5 & 0.749(25)   & 0.730(30)   & -           \\
        &        & 4.5 & 0.594(23)   & 0.581(24)   & -           \\
        &        & 5.5 & 0.458(13)   & 0.454(27)   & 0.450(18)   \\
 20 GeV & 20 GeV & 6.5 & 0.350(13)   & 0.350(13)   & 0.332(19)   \\
        &        & 7.5 & 0.2611(35)  & 0.2611(35)  & 0.2611(35)  \\
        &        & 8.5 & 0.1939(49)  & 0.1939(49)  & 0.1939(49)  \\
        &        & 9.0 & 0.1674(14)  & 0.1674(14)  & 0.1674(14)  \\
 \midrule
        &        & 3.5 & 0.727(27)   & 0.719(26)   & -           \\
        &        & 4.5 & 0.575(15)   & 0.565(17)   & -           \\
        &        & 5.5 & 0.450(20)   & 0.443(21)   & 0.4398(98)  \\
 20 GeV & 30 GeV & 6.5 & 0.3483(94)  & 0.3483(94)  & 0.343(24)   \\
        &        & 7.5 & 0.2683(53)  & 0.2683(53)  & 0.2683(53)  \\
        &        & 8.5 & 0.2083(30)  & 0.2083(30)  & 0.2083(30)  \\
        &        & 9.0 & 0.1872(39)  & 0.1872(39)  & 0.1872(39)  \\
 \midrule
        &        & 3.5 & 0.750(22)   & 0.714(29)   & -           \\
        &        & 4.5 & 0.563(20)   & 0.555(20)   & -           \\
        &        & 5.5 & 0.435(11)   & 0.430(17)   & 0.4268(40)  \\
 20 GeV & 35 GeV & 6.5 & 0.337(12)   & 0.337(12)   & 0.331(20)   \\
        &        & 7.5 & 0.2602(32)  & 0.2602(32)  & 0.2602(32)  \\
        &        & 8.5 & 0.2059(37)  & 0.2059(37)  & 0.2059(37)  \\
        &        & 9.0 & 0.1874(15)  & 0.1874(15)  & 0.1874(15)  \\
 \midrule
        &        & 3.5 & 0.727(21)   & 0.710(19)   & -           \\
        &        & 4.5 & 0.560(17)   & 0.546(16)   & -           \\
        &        & 5.5 & 0.4225(99)  & 0.420(20)   & 0.4158(75)  \\
 20 GeV & 40 GeV & 6.5 & 0.3276(91)  & 0.3276(91)  & 0.321(23)   \\
        &        & 7.5 & 0.2528(22)  & 0.2528(22)  & 0.2528(22)  \\
        &        & 8.5 & 0.2021(26)  & 0.2021(26)  & 0.2021(26)  \\
        &        & 9.0 & 0.18712(7)  & 0.18712(7)  & 0.18712(7)  \\
 \midrule
        &        & 3.5 & 0.778(16)   & 0.766(16)   & -           \\
        &        & 4.5 & 0.642(12)   & 0.6321(85)  & -           \\
        &        & 5.5 & 0.5260(94)  & 0.510(12)   & 0.5051(20)  \\
 35 GeV & 35 GeV & 6.5 & 0.4038(86)  & 0.4038(86)  & 0.398(13)   \\
        &        & 7.5 & 0.3109(45)  & 0.3109(45)  & 0.3109(45)  \\
        &        & 8.5 & 0.2379(25)  & 0.2379(25)  & 0.2379(25)  \\
        &        & 9.0 & 0.2112(37)  & 0.2112(37)  & 0.2112(37)  \\
 \bottomrule
 \end{tabular}
 \end{table}
 
 \begin{table}[H]
 \vspace{-0.80cm}
 \centering
  \caption[Rapidity veto effect on $R_{30}$ in dijet production]
 {Values of $C_3/C_0$ from the exact BLM method (Eq.~\ref{blm_exact-jets}) 
  at $\sqrt s = 13$ TeV, for all
  choices of the cuts on jet transverse momenta 
  and of the central rapidity region
  (see Fig.~\ref{C3C0_e}).}
 \label{tab:C3C0_e}
 \begin{tabular}{c|c|c|lll}
 \toprule
 $k_{J_1}^{\rm min}$ & $k_{J_2}^{\rm min}$ & $Y$ & $y^{\rm C}_{\rm max}=0$
                & $y^{\rm C}_{\rm max}=1.5$ & $y^{\rm C}_{\rm max}=2.5$ \\
 \midrule
        &        & 3.5 & 0.593(22)   & 0.577(19)   & -           \\
        &        & 4.5 & 0.432(13)   & 0.425(14)   & -           \\
        &        & 5.5 & 0.308(12)   & 0.305(15)   & 0.3019(68)  \\
 20 GeV & 20 GeV & 6.5 & 0.2139(67)  & 0.2139(67)  & 0.210(18)   \\
        &        & 7.5 & 0.1439(29)  & 0.1439(29)  & 0.1439(29)  \\
        &        & 8.5 & 0.0954(17)  & 0.0954(17)  & 0.0954(17)  \\
        &        & 9.0 & 0.0775(13)  & 0.0775(13)  & 0.0775(13)  \\
 \midrule
        &        & 3.5 & 0.551(26)   & 0.544(14)   & -           \\
        &        & 4.5 & 0.3950(88)  & 0.3896(97)  & -           \\
        &        & 5.5 & 0.281(13)   & 0.278(12)   & 0.276(3)    \\
 20 GeV & 30 GeV & 6.5 & 0.1973(48)  & 0.1973(48)  & 0.194(14)   \\
        &        & 7.5 & 0.1389(49)  & 0.1389(49)  & 0.1389(49)  \\
        &        & 8.5 & 0.0944(13)  & 0.0944(13)  & 0.0944(13)  \\
        &        & 9.0 & 0.0795(25)  & 0.0795(25)  & 0.0795(25)  \\
 \midrule
        &        & 3.5 & 0.555(19)   & 0.528(15)   & -           \\
        &        & 4.5 & 0.377(11)   & 0.3724(94)  & -           \\
        &        & 5.5 & 0.2652(90)  & 0.263(10)   & 0.2599(30)  \\
 20 GeV & 35 GeV & 6.5 & 0.1842(48)  & 0.1842(48)  & 0.184(11)   \\
        &        & 7.5 & 0.1272(24)  & 0.1272(24)  & 0.1272(24)  \\
        &        & 8.5 & 0.0888(11)  & 0.0888(11)  & 0.0888(11)  \\
        &        & 9.0 & 0.0756(12)  & 0.0756(12)  & 0.0756(12)  \\
 \midrule
        &        & 3.5 & 0.529(18)   & 0.520(21)   & -           \\
        &        & 4.5 & 0.364(10)   & 0.3585(79)  & -           \\
        &        & 5.5 & 0.2496(80)  & 0.249(11)   & 0.2400(40)  \\
 20 GeV & 40 GeV & 6.5 & 0.1717(41)  & 0.1717(41)  & 0.171(13)   \\
        &        & 7.5 & 0.1188(18)  & 0.1188(18)  & 0.1188(18)  \\
        &        & 8.5 & 0.0836(66)  & 0.0836(66)  & 0.0836(66)  \\
        &        & 9.0 & 0.0720(52)  & 0.0720(52)  & 0.0720(52)  \\
 \midrule
        &        & 3.5 & 0.6478(76)  & 0.6360(95)  & -           \\
        &        & 4.5 & 0.4983(75)  & 0.4887(40)  & -           \\
        &        & 5.5 & 0.3690(55)  & 0.3652(69)  & 0.3613(84)  \\
 35 GeV & 35 GeV & 6.5 & 0.2648(47)  & 0.2648(47)  & 0.2596(93)  \\
        &        & 7.5 & 0.1838(17)  & 0.1838(17)  & 0.1838(17)  \\
        &        & 8.5 & 0.1257(16)  & 0.1257(16)  & 0.1257(16)  \\
        &        & 9.0 & 0.1043(12)  & 0.1043(12)  & 0.1043(12)  \\
 \bottomrule
 \end{tabular}
 \end{table}
 
 \begin{table}[H]
 \vspace{-0.80cm}
 \centering
  \caption[Rapidity veto effect on $R_{21}$ in dijet production]
 {Values of $C_2/C_1$ from the exact BLM method (Eq.~\ref{blm_exact-jets}) 
  at $\sqrt s = 13$ TeV, for all
  choices of the cuts on jet transverse momenta 
  and of the central rapidity region
  (see Fig.~\ref{C2C1_e}).}
 \label{tab:C2C1_e}
 \begin{tabular}{c|c|c|lll}
 \toprule
 $k_{J_1}^{\rm min}$ & $k_{J_2}^{\rm min}$ & $Y$ & $y^{\rm C}_{\rm max}=0$
                & $y^{\rm C}_{\rm max}=1.5$ & $y^{\rm C}_{\rm max}=2.5$ \\
 \midrule
        &        & 3.5 & 0.759(21)   & 0.749(28)   & -           \\
        &        & 4.5 & 0.671(26)   & 0.665(28)   & -           \\
        &        & 5.5 & 0.583(17)   & 0.583(37)   & 0.580(24)   \\
 20 GeV & 20 GeV & 6.5 & 0.506(21)   & 0.506(21)   & 0.484(27)   \\
        &        & 7.5 & 0.4278(55)  & 0.4278(55)  & 0.4278(55)  \\
        &        & 8.5 & 0.3576(91)  & 0.3576(91)  & 0.3576(91)  \\
        &        & 9.0 & 0.3236(27)  & 0.3236(27)  & 0.3236(27)  \\
 \midrule
        &        & 3.5 & 0.724(23)   & 0.727(25)   & -           \\
        &        & 4.5 & 0.642(15)   & 0.638(18)   & -           \\
        &        & 5.5 & 0.563(23)   & 0.559(27)   & 0.561(11)   \\
 20 GeV & 30 GeV & 6.5 & 0.491(13)   & 0.491(13)   & 0.489(34)   \\
        &        & 7.5 & 0.4245(83)  & 0.4245(83)  & 0.4245(83)  \\
        &        & 8.5 & 0.3644(54)  & 0.3644(54)  & 0.3644(54)  \\
        &        & 9.0 & 0.3377(67)  & 0.3377(67)  & 0.3377(67)  \\
 \midrule
        &        & 3.5 & 0.713(19)   & 0.710(24)   & -           \\
        &        & 4.5 & 0.622(21)   & 0.623(22)   & -           \\
        &        & 5.5 & 0.542(14)   & 0.542(22)   & 0.5414(50)  \\
 20 GeV & 35 GeV & 6.5 & 0.473(17)   & 0.473(17)   & 0.470(29)   \\
        &        & 7.5 & 0.4095(50)  & 0.4095(50)  & 0.4095(50)  \\
        &        & 8.5 & 0.3548(63)  & 0.3548(63)  & 0.3548(63)  \\
        &        & 9.0 & 0.3299(31)  & 0.3299(31)  & 0.3299(31)  \\
 \midrule
        &        & 3.5 & 0.697(18)   & 0.695(16)   & -           \\
        &        & 4.5 & 0.612(17)   & 0.607(17)   & -           \\
        &        & 5.5 & 0.523(10)   & 0.526(24)   & 0.5256(96)  \\
 20 GeV & 40 GeV & 6.5 & 0.459(12)   & 0.459(12)   & 0.455(33)   \\
        &        & 7.5 & 0.3960(33)  & 0.3960(33)  & 0.3960(33)  \\
        &        & 8.5 & 0.3441(45)  & 0.3441(45)  & 0.3441(45)  \\
        &        & 9.0 & 0.3224(12)  & 0.3224(12)  & 0.3224(12)  \\
 \midrule
        &        & 3.5 & 0.809(16)   & 0.805(14)   & -           \\
        &        & 4.5 & 0.728(14)   & 0.7247(98)  & -           \\
        &        & 5.5 & 0.659(13)   & 0.644(14)   & 0.6421(27)  \\
 35 GeV & 35 GeV & 6.5 & 0.562(12)   & 0.563(12)   & 0.561(19)   \\
        &        & 7.5 & 0.4799(67)  & 0.4799(67)  & 0.4799(67)  \\
        &        & 8.5 & 0.3984(40)  & 0.3984(40)  & 0.3984(40)  \\
        &        & 9.0 & 0.3588(61)  & 0.3588(61)  & 0.3588(61)  \\
 \bottomrule
 \end{tabular}
 \end{table}
 
 \begin{table}[H]
 \vspace{-0.80cm}
 \centering
  \caption[Rapidity veto effect on $R_{32}$ in dijet production]
 {Values of $C_3/C_2$ from the exact BLM method (Eq.~\ref{blm_exact-jets}) 
  at $\sqrt s = 13$ TeV, for all
  choices of the cuts on jet transverse momenta 
  and of the central rapidity region
  (see Fig.~\ref{C3C2_e}).}
 \label{tab:C3C2_e}
 \begin{tabular}{c|c|c|lll}
 \toprule
 $k_{J_1}^{\rm min}$ & $k_{J_2}^{\rm min}$ & $Y$ & $y^{\rm C}_{\rm max}=0$
                & $y^{\rm C}_{\rm max}=1.5$ & $y^{\rm C}_{\rm max}=2.5$ \\
 \midrule
        &        & 3.5 & 0.792(22)   & 0.790(26)   & -           \\
        &        & 4.5 & 0.727(30)   & 0.731(32)   & -           \\
        &        & 5.5 & 0.673(24)   & 0.672(49)   & 0.671(26)   \\
 20 GeV & 20 GeV & 6.5 & 0.611(28)   & 0.611(28)   & 0.634(55)   \\
        &        & 7.5 & 0.5509(92)  & 0.5509(92)  & 0.5509(92)  \\
        &        & 8.5 & 0.492(16)   & 0.492(16)   & 0.492(16)   \\
        &        & 9.0 & 0.4627(59)  & 0.4627(59)  & 0.4627(59)  \\
 \midrule
        &        & 3.5 & 0.758(37)   & 0.756(23)   & -           \\
        &        & 4.5 & 0.687(18)   & 0.689(22)   & -           \\
        &        & 5.5 & 0.625(32)   & 0.629(36)   & 0.628(11)   \\
 20 GeV & 30 GeV & 6.5 & 0.566(18)   & 0.567(18)   & 0.566(54)   \\
        &        & 7.5 & 0.518(22)   & 0.518(22)   & 0.518(22)   \\
        &        & 8.5 & 0.4530(93)  & 0.4530(93)  & 0.4530(93)  \\
        &        & 9.0 & 0.424(17)   & 0.424(17)   & 0.424(17)   \\
 \midrule
        &        & 3.5 & 0.741(19)   & 0.740(23)   & -           \\
        &        & 4.5 & 0.670(24)   & 0.671(23)   & -           \\
        &        & 5.5 & 0.609(15)   & 0.610(32)   & 0.6090(25)  \\
 20 GeV & 35 GeV & 6.5 & 0.547(21)   & 0.547(21)   & 0.555(43)   \\
        &        & 7.5 & 0.4887(75)  & 0.4887(75)  & 0.4887(75)  \\
        &        & 8.5 & 0.4312(86)  & 0.4312(86)  & 0.4312(86)  \\
        &        & 9.0 & 0.4033(50)  & 0.4033(50)  & 0.4033(50)  \\
 \midrule
        &        & 3.5 & 0.728(19)   & 0.732(32)   & -           \\
        &        & 4.5 & 0.650(18)   & 0.657(18)   & -           \\
        &        & 5.5 & 0.591(15)   & 0.592(34)   & 0.578(13)   \\
 20 GeV & 40 GeV & 6.5 & 0.524(17)   & 0.524(17)   & 0.532(56)   \\
        &        & 7.5 & 0.4700(63)  & 0.4700(63)  & 0.4700(63)  \\
        &        & 8.5 & 0.4134(62)  & 0.4134(62)  & 0.4134(62)  \\
        &        & 9.0 & 0.3850(25)  & 0.3850(25)  & 0.3850(25)  \\
 \midrule
        &        & 3.5 & 0.832(13)   & 0.830(11)   & -           \\
        &        & 4.5 & 0.776(14)   & 0.7731(94)  & -           \\
        &        & 5.5 & 0.701(13)   & 0.716(18)   & 0.7152(27)  \\
 35 GeV & 35 GeV & 6.5 & 0.656(16)   & 0.656(16)   & 0.652(31)   \\
        &        & 7.5 & 0.5912(88)  & 0.5912(88)  & 0.5912(88)  \\
        &        & 8.5 & 0.5284(96)  & 0.5284(96)  & 0.5284(96)  \\
        &        & 9.0 & 0.4939(11)  & 0.4939(11)  & 0.4939(11)  \\
 \bottomrule
 \end{tabular}
 \end{table}
 
 
 
 \section{Numerical specifics}
 \label{sec:mn-jets-numerics}
 
 In this Section the numerical specifics for all the calculations done in Sections~\ref{sec:mn-jets-BFKL-vs-DGLAP} and~\ref{sec:mn-jets-rapidity} are discussed.

  \subsection{Used tools}
  \label{sub:mn-jets-tools}
  
  All numerical calculations were implemented in \textsc{Fortran}, using
  the corresponding interfaces for the NLO MSTW 2008 PDFs~\cite{Martin:2009iq} and the two-loop
  running coupling with $\alpha_s\left(M_Z\right)=0.11707$ 
  with five quark flavours active.
  Now there exist updated PDF parameterisations, including 
  the NLO MMHT 2014 set~\cite{Harland-Lang:2014zoa}, which is the successor of the MSTW 2008 
  analysis. Here we continue to use MSTW 2008 PDFs
  because in our kinematical range the difference between MSTW 2008 and the 
  updated MMHT 2014 PDFs is very small. Also, we want to keep the 
  opportunity to compare our results at 13~TeV with our previous calculations at 
  7~TeV without introducing any other source of discrepancy related to the change of the PDF set.
  Numerical integrations and the computation of the polygamma functions 
  were performed using specific \textsc{CERN} program libraries~\cite{cernlib}.
  Furthermore, we used slightly modified versions of the \cod{Chyp}~\cite{chyp} 
  and \cod{Psi}~\cite{RpsiCody:1973} routines in order to perform the calculation
  of the Gauss hypergeometric function $_2F_1 $ and of the real part of the 
  $\psi$ function, respectively.

  \subsection{Uncertainty estimation}
  \label{sub:mn-jets-uncertainty}

  The are three main sources of uncertainty in our calculation:

  \begin{itemize} 

  \item The first source of uncertainty is the numerical 4-dimensional 
  integration over the variables $|\vec k_{J_1}|$, $|\vec k_{J_2}|$, $y_{J_1}$ and
  $\nu$ and was directly estimated by \cod{Dadmul} integration 
  routine~\cite{cernlib}.

  \item The second one is the one-dimensional integration over the longitudinal
  momentum fraction $\zeta$ entering the expression for the NLO 
  impact factors $c_{1,2}^{(1)}(n,\nu,|\vec k_{J_{1,2}}|, x_{J_{1,2}})$ (see Appendix~\hyperlink{app:jet-nlo-if-link}{B}).
  This integration was performed by using the \cod{WGauss} routine~\cite{cernlib}.
  At first, we fixed the best value of the input accuracy parameter \cod{EPS}
  by making comparisons between separate \textsc{Fortran} and 
  \textsc{Mathematica} calculations of the impact factor. Then, we verified
  that, under variations by factors of 10 or 1/10 of the \cod{EPS} parameter,
  the $C_n^{\rm BFKL}$ and $C_n^{\rm DGLAP}$ coefficients change by less than 
  1 permille.

  \item The third one is related to the upper cutoff in the integrations
  over $|\vec k_{J_1}|$, $|\vec k_{J_2}|$ and $\nu$. We fixed 
  $k_{J_1}^{\rm max}=k_{J_2}^{\rm max} = 60$ GeV as in Ref.~\cite{Ducloue:2013wmi}, where it was
  shown that the contribution to the integration from the omitted region
  is negligible. Concerning the $\nu$-integration, we fixed the upper 
  cutoff $\nu^{\rm max}=30$ for the calculation of the $C_n^{\rm BFKL}$ 
  coefficients, after verifying that a larger value does not change the 
  result in appreciable way. 

  The $C_n^{\rm DGLAP}$ coefficients show a more pronounced sensitivity
  to $\nu^{\rm max}$, due to the fact that the oscillations in the
  integrand in Eqs.~(\ref{casea-dglap-jets}) and~(\ref{caseb-dglap-jets}) are not dumped by the 
  exponential factor 
  as in the BFKL expressions~(\ref{casea}) and~(\ref{caseb}). For the same reason,
  the computational time of $C_n^{\rm DGLAP}$ is much larger than for 
  $C_n^{\rm BFKL}$. We found that the best compromise was to set 
  $\nu^{\rm max} = 50$. We checked in some sample cases, mostly at $Y=6$ and 9, 
  that, putting $\nu^{\rm max}$ at 60, ratios $C_m/C_n$ change always less than
  1\%, in spite of the fact that the single coefficients $C_n$ change
  in a more pronounced way.

  \end{itemize}

  Of the three main sources of uncertainty, the first one is, by far, the most
  significant, therefore the error bars of all data presented in this work
  are just those given by the \cod{Dadmul} integration. We checked, however,
  using some trial functions which mimic the behaviour of the true integrands
  involved in this work, that the error given by the \cod{Dadmul} integration
  is a large overestimate of the true one. We are therefore confident that
  our error estimation is quite conservative.

 \section{Summary} 
 \label{sec:mn-jets-summary}

 In this Chapter we have considered the Mueller--Navelet jet production process at the LHC at the two values for center-of-mass energy of $7$~TeV and $13$~TeV and for several and distinct kinematical configurations for the transverse momenta of the detected jets.

 First, we have shown how BFKL predictions~\cite{Caporale:2014gpa} taken at $\sqrt{s} = 7$~TeV, for {\it symmetric} cuts for the transverse
 momenta ($k_{J_{1,2} \rm \, min} = 35$~GeV) and with the renormalisation and the factorization scales optimised according to the BLM method, are in a very good agreement with experimental data~\cite{Khachatryan:2016udy}. 
 Then, we have addressed some of the open questions 
 raised in recent phenomenological works
 (see Section~\ref{sec:mn-jets-theory-vs-experiment}). 
  
 On one side~\cite{Celiberto:2015yba}, we compared predictions for several azimuthal correlations and ratios between them at $\sqrt{s} = 7$~TeV, both in full NLA BFKL approach and in fixed-order NLO DGLAP. Differently from current experimental analyses of the same process, 
 we have used {\it asymmetric} cuts for the transverse
 momenta of the detected jets. In particular, taking one of the cuts 
 at 35~GeV (as done by the CMS collaboration~\cite{Khachatryan:2016udy}) and the other at 
 45~GeV or 50~GeV, we can clearly see that predictions from BFKL and DGLAP 
 become separate for most azimuthal correlations and ratios between them,
 this effect being more and more visible as the rapidity gap between the 
 jets, $Y$, increases. In other words, in this kinematics the additional 
 undetected parton radiation between the jets which is present in the resummed 
 BFKL series, in comparison to just one undetected parton allowed by the NLO 
 DGLAP approach, makes its difference and leads to more azimuthal angle 
 decorrelation between the jets, in full agreement with the original proposal of 
 Mueller and Navelet.
 Another important benefit from the use of {\ti asymmetric} cuts, pointed out 
 in~\cite{Ducloue:2014koa}, is that the effect of violation of the 
 energy-momentum conservation in the NLA is strongly suppressed with respect 
 to what happens in the LLA.
 All these considerations persuade us to strongly suggest experimental collaborations to consider also {\ti asymmetric} cuts in jet transverse momenta in all future analyses of Mueller--Navelet jet production process.
  
 On the other side~\cite{Celiberto:2016ygs}, we performed the first analysis at $\sqrt{s} = 13$~TeV, studying the jet azimuthal correlations in five different configurations for the jet transverse momenta, which include \emph{asymmetric} cuts. Differently from all previous studies of the same kind, we considered in our analysis the effect of excluding the possibility that one of the two detected jets be produced in the central rapidity region. Central jets originate from small-$x$ partons, and the collinear approach for 
 the description of the Mueller--Navelet jet vertices may not hold at small $x$. The outcome of our analysis is that, for two reasonable ways to define the extension of the central region: $(i)$ the total cross section, $C_0$, is strongly reduced by the ``exclusion cuts'' in the range ($Y<5.5$) where they are effective; $(ii)$ on the other hand, in the same kinematics, the difference with respect to the case of no central rapidity exclusion is invisible in azimuthal correlations and in ratios between them. 
 We believe that it would be very interesting to confront these conclusions with the forthcoming LHC analysis at $13$~TeV. 
 
 As anticipated in Section~\ref{sub:bfkl-cross-section}, it would be interesting to figure out whether the growth with energy of the cross section, a characteristic feature of the BFKL resummation, can be observed in the Mueller--Navelet jet reaction. 
 First of all, the relevant observable for such investigation is  the $\phi$-averaged cross section ${\cal C}_0$, which, however, is affected by theoretical ambiguities coming from the choice of the representation (see Section~\ref{ssub:bfkl_cs_representation}) and the scale setting procedure (see Section~\ref{sec:bfkl_blm}). The possible experimental measurement of ${\cal C}_0$ would be extremely helpful in discriminating among the several NLA-equivalent options.  
 Even in case the theoretical ambiguities in the definition of ${\cal C}_0$ were cleared up, the search for the growth in the jet rapidity interval $Y$ of the cross section and the subsequent extraction of the Pomeron intercept would be hindered by the fact that,  
 due to collinear factorisation, the hadronic cross section embeds the parent proton PDFs which are responsible of its decrease for increasing $Y$. To switch off the role of PDFs and thus isolate the $Y$-dependence of the partonic cross section, one should select final-state configurations so as to keep $x$ constant in the PDFs. The price for that, however, is a too restrictive choice of the ranges for the jet transverse momenta. Indeed, since the dependence of $x$ on the rapiditity is exponential (Eq.~(\ref{Y-x_J})), one should consider ranges for the transverse momenta which are much larger than the ones available at the present and forthcoming LHC energies. 
 Furthermore, there is a principle difficulty in detecting at the LHC clear imprints of a genuine BFKL power asymptotic. As already stated at the end of Section~\ref{sub:mn-jets-DGLAP-results}, the limited number of undetected hard partons emitted with LHC kinematics makes BFKL hardly distinguishable from DGLAP when one goes from LLA to NLA, so that it is difficult to say something definite about the intercept. We can only compare our predictions for the cross section with data, to see whether our approach in its present status works or not. 
 Although these issues affect also the other hadroproduction processes investigated in this thesis, dihadron production (see Chapter~\ref{chap:dihadron}) has better chances to bring us closer to the Regge kinematics for partonic subprocesses, due to the fact that hadrons can be detected at the LHC at much smaller transverse momenta than jets.

\newpage
 
\setcounter{appcnt}{0}
\renewcommand{\theequation}{B.\arabic{appcnt}}
\setcounter{tmp}{2}
\clearpage
\hypertarget{app:jet-nlo-if-link}{}
\chapter*{Appendix~B}
\vspace{-0.5cm} 
\noindent
{\Huge \bf NLO impact factor for the small-cone forward jet}
\label{app:jet-nlo-if}
\addcontentsline{toc}{chapter}{\numberline{\Alph{tmp}}
 NLO impact factor for the small-cone forward jet}
\markboth{NLO impact factor for the small-cone forward jet}{}
\markright{APPENDIX B}{}
\vspace{1.3cm} 

In this Appendix the expression for the NLO 
correction to the forward jet impact factor in the small-cone limit is given (see Ref.~\cite{Caporale:2012ih} for further details). In the $(\nu,n)$ representation, we have:
\hypertarget{c11-jets}{}
\beq
\stepcounter{appcnt}
\label{c11-jets}
c_1^{(1)}(n,\nu,|\vec k|,x)=
\frac{1}{\pi}\sqrt{\frac{C_F}{C_A}}
\left(\vec k^{\,2} \right)^{i\nu-1/2}
\int\limits^1_{x}\frac{d\zeta}{\zeta}
\zeta^{-\bar\alpha_s(\mu_R)\chi(n,\nu)}
\eeq
\[
\left\{\sum_{a=q,\bar q} f_a \left(\frac{x}{ \zeta}\right)\left[\left(P_{qq}(\zeta)+\frac{C_A}{C_F}P_{gq}(\zeta)\right)
\ln\frac{\vec k^{\,2}}{\mu_F^2}\right.\right.
\]
\[
-2\zeta^{-2\gamma}\ln R\,
\left\{P_{qq}(\zeta)+P_{gq}(\zeta)\right\}-\frac{\beta_0}{2}
\ln\frac{\vec k^{\,2}}{\mu_R^2}\delta(1-\zeta)
\]
\[
+C_A\delta(1-\zeta)\left(\chi(n,\gamma)\ln\frac{s_0}{\vec k^{\,2}}
+\frac{85}{18}+\frac{\pi^2}{2}\right.
\]
\[
\left.+\frac{1}{2}\left(\psi^\prime
\left(1+\gamma+\frac{n}{2}\right)
-\psi^\prime\left(\frac{n}{2}-\gamma\right)-\chi^2(n,\gamma)\right)
\right)
\]
\[
+(1+\zeta^2)\left\{C_A\left(\frac{(1+\zeta^{-2\gamma})\,\chi(n,\gamma)}
{2(1-\zeta)_+}-\zeta^{-2\gamma}\left(\frac{\ln(1-\zeta)}{1-\zeta}\right)_+
\right)\right.
\]
\[
\left.+\left(C_F-\frac{C_A}{2}\right)\left[ \frac{\bar \zeta}{\zeta^2}I_2
-\frac{2\ln\zeta}{1-\zeta}
+2\left(\frac{\ln(1-\zeta)}{1-\zeta}\right)_+ \right]\right\}
\]
\[
+\delta(1-\zeta)\left(C_F\left(3\ln 2-\frac{\pi^2}{3}-\frac{9}{2}\right)
-\frac{5n_f}{9}\right)
+C_A\zeta+C_F\bar \zeta
\]
\[
\left. +\frac{1+\bar \zeta^2}{\zeta}
\left(C_A\frac{\bar \zeta}{\zeta}I_1+2C_A\ln\frac{\bar\zeta}{\zeta}
+C_F\zeta^{-2\gamma}(\chi(n,\gamma)-2\ln \bar \zeta)\right)\right]
\]
\[
+f_{g}\left(\frac{x}{ \zeta}\right)\frac{C_A}{C_F}
\left[
\left(P_{gg}(\zeta)+2 \,n_f \frac{C_F}{C_A}P_{qg}(\zeta)\right)
\ln\frac{\vec k^{\,2}}{\mu_F^2}
\right.
\]
\[
\left.
-2\zeta^{-2\gamma}\ln R \left(P_{gg}(\zeta)+2 \,n_f P_{qg}(\zeta)\right)
-\frac{\beta_0}{2}\ln\frac{\vec k^{\,2}}{4\mu_R^2}\delta(1-\zeta)
\right.
\]
\[
+\, C_A\delta(1-\zeta)
\left(
\chi(n,\gamma)\ln\frac{s_0}{\vec k^{\,2}}+\frac{1}{12}+\frac{\pi^2}{6}
\right.
\]
\[
\left.
+\frac{1}{2}\left(\psi^\prime\left(1+\gamma+\frac{n}{2}\right)
-\psi^\prime\left(\frac{n}{2}-\gamma\right)-\chi^2(n,\gamma)\right)
\right)
\]
\[
+\, 2 C_A (1-\zeta^{-2\gamma})\left(\left(\frac{1}{\zeta}-2
+\zeta\bar\zeta\right)\ln \bar \zeta + \frac{\ln (1-\zeta)}{1-\zeta}\right)
\]
\[
+ \, C_A\, \left[\frac{1}{\zeta}+\frac{1}{(1- \zeta)_+}-2+\zeta\bar\zeta\right]
\left((1+\zeta^{-2\gamma})\chi(n,\gamma)-2\ln\zeta+\frac{\bar \zeta^2}
{\zeta^2}I_2\right)
\]
\[
\left.\left.
+\, n_f\left[\, 2\zeta\bar \zeta \, \frac{C_F}{C_A} +(\zeta^2+\bar \zeta^2)
\left(\frac{C_F}{C_A}\chi(n,\gamma)+\frac{\bar \zeta}{\zeta}I_3\right)
-\frac{1}{12}\delta(1-\zeta)\right]\right]\right\} \; .
\]

Here $\bar \zeta=1-\zeta$, $\gamma=i\nu-1/2$, $P_{i j}(\zeta)$ are leading
order DGLAP kernels defined as 
\bea
\stepcounter{appcnt}
\label{DGLAP_kernels}
P_{gq}(z)&=&C_F\frac{1+(1-z)^2}{z}\;,\\ \nonumber
P_{qg}(z)&=&T_R\left[z^2+(1-z)^2\right]\;,\\ \nonumber
P_{qq}(z)&=&C_F\left( \frac{1+z^2}{1-z} \right)_+
          = C_F\left[ \frac{1+z^2}{(1-z)_+} +{3\over 2}\delta(1-z)\right]\;,\\ \nonumber
P_{gg}(z)&=&2C_A\left[\frac{1}{(1-z)_+} +\frac{1}{z} -2+z(1-z)\right]
          +\left({11\over 6}C_A-\frac{n_f}{3}\right)\delta(1-z) \nonumber \; ,
\eea
where $C_F$ is the Casimir operator associated with gluon emission from a quark, $C_F = (N_c^2-1)/(2N_c)$ 
and $T_R = 1/2$ is the colour factor associated with the splitting of a gluon into a quark-antiquark pair.
For the $I_{1,2,3}$ functions we have the results:
\[
I_2=
\frac{\zeta^2}{\bar \zeta^2}\left[
\zeta\left(\frac{{}_2F_1(1,1+\gamma-\frac{n}{2},2+\gamma-\frac{n}{2},\zeta)}
{\frac{n}{2}-\gamma-1}-
\frac{{}_2F_1(1,1+\gamma+\frac{n}{2},2+\gamma+\frac{n}{2},\zeta)}{\frac{n}{2}+
\gamma+1}\right)\right.
\]
\beq
\label{I2}
\stepcounter{appcnt}
\left.
+\zeta^{-2\gamma}
\left(\frac{{}_2F_1(1,-\gamma-\frac{n}{2},1-\gamma-\frac{n}{2},\zeta)}
{\frac{n}{2}+\gamma}-
\frac{{}_2F_1(1,-\gamma+\frac{n}{2},1-\gamma+\frac{n}{2},\zeta)}{\frac{n}{2}
-\gamma}\right)
\right.
\eeq
\[
\left.
+\left(1+\zeta^{-2\gamma}\right)\left(\chi(n,\gamma)-2\ln \bar \zeta \right)
+2\ln{\zeta}\right]\;,
\]

\beq
\stepcounter{appcnt}
\label{I1}
I_1=\frac{\bar \zeta}{2\zeta}I_2+\frac{\zeta}{\bar \zeta}\left[
\ln \zeta+\frac{1-\zeta^{-2\gamma}}{2}\left(\chi(n,\gamma)-2\ln \bar \zeta
\right)\right]\;,
\eeq

\beq
\stepcounter{appcnt}
\label{I3}
I_3=\frac{\bar \zeta}{2\zeta}I_2-\frac{\zeta}{\bar \zeta}\left[
\ln \zeta+\frac{1-\zeta^{-2\gamma}}{2}\left(\chi(n,\gamma)-2\ln \bar \zeta
\right)\right]\;.
\eeq

In Eq.~(\hyperlink{c11-jets}{B.1}) the \emph{plus-prescription} is introduced, which is defined as
\beq
\label{plus-prescription}
\stepcounter{appcnt}
\int\limits^1_a d \zeta \frac{F(\zeta)}{(1-\zeta)_+}
=\int\limits^1_a d \zeta \frac{F(\zeta)-F(1)}{(1-\zeta)}
-\int\limits^a_0 d \zeta \frac{F(1)}{(1-\zeta)}\; ,
\eeq
for any function $F(\zeta)$, regular at $\zeta=1$. Note that
\begin{equation}
\stepcounter{appcnt}
(1-\zeta)^{2\epsilon-1}=(1-\zeta)^{2\epsilon -1}_+ +\frac{1}{2\epsilon}
\delta(1-\zeta)
\end{equation}
\[
=
\frac{1}{2\epsilon}\delta(1-\zeta)+\frac{1}{(1-\zeta)_+}
+2\epsilon\left(\frac{\ln (1-\zeta)}{1-\zeta}\right)_+ +{\cal O}(\epsilon^2)
\; .
\]

The factor $\zeta^{-\bar\alpha_s(\mu_R)\chi(n,\nu)}$ appears in Eq.~(\hyperlink{c11-jets}{B.1})
due to extra contributions attributed to the jet vertices, as discussed after
Eq.~(29) of Ref.~\cite{Caporale:2012ih}.

\renewcommand{\theequation}
             {\arabic{chapter}.\arabic{equation}}
\chapter{Dihadron production}
\label{chap:dihadron}

\begin{figure}[t]
  \centering
    \includegraphics[scale=0.5]{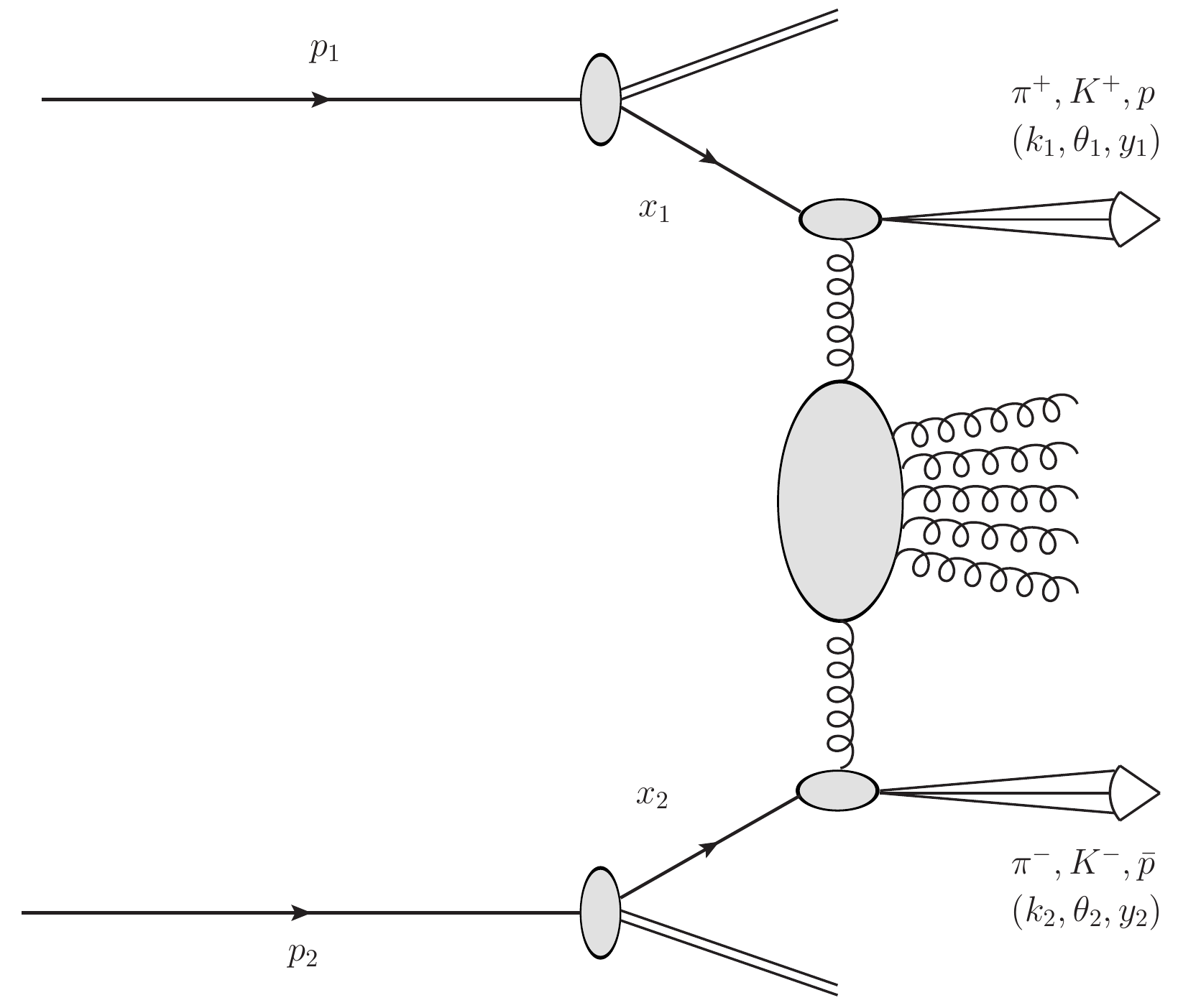}
  \caption[Inclusive dihadron production process in multi-Regge kinematics.]
  {Inclusive dihadron production process in multi-Regge kinematics.}
  \label{fig:dihadron}
 \end{figure}

In this Chapter the inclusive dihadron production 
\begin{eqnarray}
\label{process-dh}
{\rm p}(p_1) + {\rm p}(p_2) 
\to 
{\rm h}_1(k_1) + {\rm h}_2(k_2)+ {\rm X} 
\end{eqnarray}
is investigated, \emph{i.e.} when the two charged light hadrons: $\pi^{\pm}, K^{\pm}, p,\bar p$ 
with high transverse momenta and separated by a large interval of rapidity,
together with an undetected hadronic system X, are produced in the final state 
(see Fig.~\ref{fig:dihadron} for a schematic view).

This process is similar to the Mueller--Navelet jet production and shares
with it the underlying theoretical framework, the only obvious difference
lying in the vertices describing the dynamics in the proton fragmentation
region: instead of the proton-to-jet vertex, the vertex for the
proton to identified hadron transition is needed.
Such a vertex was considered in~\cite{hadrons} within NLA: it was shown
there that ultraviolet divergences are taken care of by the renormalisation
of the QCD coupling, soft and virtual infrared divergences cancel each
other, whereas the surviving infrared collinear ones are compensated by
the collinear counterterms related to the renormalisation of PDFs for
the initial proton and FFs describing the
detected hadron in the final state within collinear factorisation.
\footnote{The identified hadron production vertex  in the NLA  was found
  within the \emph{shockwave} approach (or \emph{colour glass condensate} effective theory)
  in~\cite{Chirilli:2012jd}. It was used there to study the
  single inclusive particle production at forward rapidities in proton-nucleus
  collisions; for recent developments of this line of research, see
  also~\cite{Iancu:2016vyg}. Unfortunately, the comparison  between  the
  results of~\cite{Chirilli:2012jd} and those of~\cite{hadrons} is not
  simple and straightforward, since the distribution of radiative corrections
  between the kernel and the impact factor is different in the shockwave and the
  BFKL frameworks. Non-trivial kernel and impact factor transformations are
  required for such a comparison. It certainly deserves a separate study,
  and the consideration of the process~(\ref{process-dh}) within both the
  shockwave and the BFKL resummation schemes seems the best possibility
   to this purpose.}
Hence, infrared-safe NLA predictions for observables related to this
process are amenable, thus making this process an additional 
clear channel to test the BFKL dynamics at the LHC. The reaction~(\ref{process-dh})
can be considered complementary to Mueller--Navelet jet production, since
hadrons can be detected at the LHC at much smaller values of the transverse
momentum than jets, thus giving access to a kinematical range outside the
reach of the Mueller--Navelet channel.

This Chapter is organised as follows. 
In Section~\ref{sec:dihadron-theory} we give the main formul{\ae} for cross section and azimuthal correlations (see Section~\ref{sub:dihadron-cs}), together with the BLM scale setting (see Section~\ref{sub:dihadron-blm}) and the final-state phase space integration (see Section~\ref{sub:dihadron-phase-space}). 
In Section~\ref{sec:dihadron-kernel} we present a comparison between full LLA and partial NLA BFKL predictions, \emph{i.e.} considering just the NLO kernel corrections and taking the hadron vertices at LLA. 
Full NLA predictions at 7 and 13 TeV and considering various realistic LHC kinematical constraint are given in Section~\ref{sec:dihadron-NLA}. Section~\ref{sec:dihadron-numerics} is devoted to the details on the numerical implementation, together with the study of the effects of using different PDF and FF parameterisations. The section Summary is drawn in Section~\ref{sec:dihadron-summary}.

The analysis given in this Chapter is based on 
the work done in Refs.~\cite{Celiberto:2016hae,Celiberto:2017ptm} and presented in 
Refs.~\cite{Celiberto:2016vhn,Celiberto:2016zgb,Celiberto:2017:lowx,Celiberto:2017:blois}.

 \section{Theoretical framework} 
 \label{sec:dihadron-theory}
 
 In this Section the BFKL cross section and the azimuthal corrections
 for the inclusive dihadron production process are presented.
 
  \subsection{Inclusive dihadron production}
  \label{sub:dihadron-if} 
 
  The process under investigation (see Eq.~(\ref{process-dh}) and Fig.~\ref{fig:dihadron}) is the inclusive production of a pair of identified
  hadrons featuring high transverse momenta, 
  $\vec k_1^2\sim \vec k_2^2 \gg \Lambda^2_{\rm QCD}$ 
  and separated by a large rapidity interval in high-energy proton-proton
  collisions. The protons' momenta $p_1$ and $p_2$ are taken as Sudakov
  vectors (see Eq.~(\ref{sudakov_general})) satisfying $p^2_1= p^2_2=0$ and $2 (p_1p_2) = s$, 
  so that the momentum of each hadron can be decomposed as
  \bea
  k_1&=& \alpha_1 p_1+ \frac{\vec k_1^2}{\alpha_1 s}p_2+k_{1\perp} \ , \quad
  k_{1\perp}^2=-\vec k_1^2 \ , \\ \nonumber 
  k_2&=& \alpha_2 p_2+ \frac{\vec k_2^2}{\alpha_2 s}p_1+k_{2\perp} \ , \quad
  k_{2\perp}^2=-\vec k_2^2 \ .
  \label{sudakov-dh}
  \eea
  
  In the center of mass system, the hadrons' longitudinal momentum fractions 
  $\alpha_{1,2}$ are connected to the respective 
  rapidities through the relations
  $y_1=\frac{1}{2}\ln\frac{\alpha_1^2 s}
  {\vec k_1^2}$, and $y_2=\frac{1}{2}\ln\frac{\vec k_2^2}{\alpha_2^2 s}$, 
  so that $dy_1=\frac{d\alpha_1}{\alpha_1}$, $dy_2=-\frac{d\alpha_2}{\alpha_2}$,
  and $Y=y_1-y_2=\ln\frac{\alpha_1\alpha_2 s}{|\vec k_1||\vec k_2|}$, here the
  space part of the four-vector $p_{1\parallel}$ being taken positive.
  
  By studying Mueller--Navelet jets, we probed the BFKL dynamics through a very inclusive process. Now we require that a couple of hadrons is always identified in the final state, considering so a less inclusive final-state reaction. 
  Following the course taken in the Mueller--Navelet case (see Section~\ref{sub:mn-jets-if}), we start from the NLO forward parton impact factors~\cite{Fadin:1999de,Fadin:1999df} (see Fig.~\ref{fig:if-parton}). 
  In order to allow the inclusive production of a given hadron, one of these integrations in the definition of parton impact factors is `opened' (see Fig.~\ref{fig:if-parton}). This means that the integration over the momentum of one of the intermediate-state partons is replaced by the convolution with a suitable FF.
  \begin{figure}[t]
   \centering
      \includegraphics[scale=0.85]{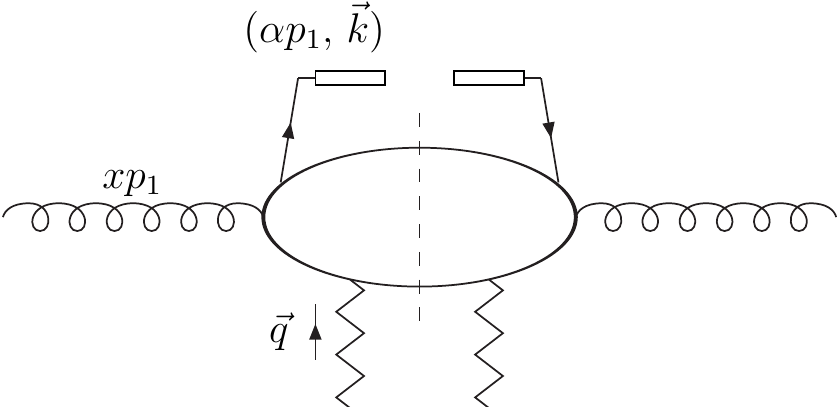}
      \includegraphics[scale=0.85]{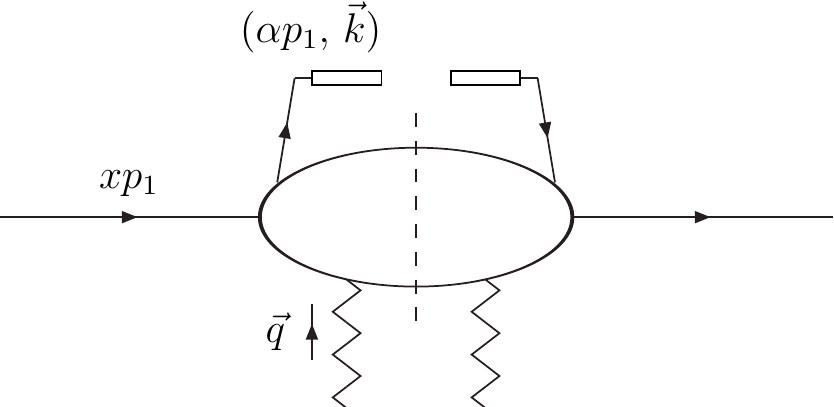}
    \caption[Identified-hadron impact factor]
    {Schematic view of the the vertex for the identified hadron production for the case of incoming gluon (left) or quark (right). 
    Here $p_1$ is the proton momentum, $x$ is the
    fraction of proton momentum carried by the gluon/quark, $\alpha_h p_1$ is the longitudinal momentum of the hadron $h$, $\vec k_h$ is the transverse hadron momentum and $\vec q$ is the transverse momentum of the incoming Reggeised gluon.}
   \label{fig:if-hadron}
  \end{figure}
  The expression for the identified-hadron impact factor 
  with LO accuracy is given below (see Fig.~\ref{fig:if-hadron}):
  \begin{equation}
  \label{if-hadron-lo}
   \frac{d\Phi_h^{(0)}}{\vec q^{2}}=
   2\pi\alpha_s\sqrt{\frac{2C_F}{C_A}}
   \frac{d\alpha_h d^2\vec k}
   {\vec k^{\, 2}}\int\limits^1_{\alpha_h} \frac{dx}{x}
   \, \delta^{(2)}\left(\vec k-\vec q\right)
  \end{equation}
  \[ \times \,
   \left(\frac{C_A}{C_F}f_g(x)
   D^h_g\left(\frac{\alpha_h}{x}\right)+\sum_{a=q,\bar q} f_a(x)
   D^h_a\left(\frac{\alpha_h}{x}\right)\right) \ ,
  \]
  where $D^h_a$ is the FF that describes non-perturbative, large-distance part of the transition from the parton $a$ produced with momentum $k$ and longitudinal fraction $x$ to a hadron with momentum fraction $\alpha_h$.
  As we did for the forward jet impact factor, the last step to do is to project Eq.~(\ref{if-hadron-lo}) onto the eigenfunctions (Eq.~(\ref{nuLLA})) of the LO BFKL kernel~(\ref{KLLA}), \emph{i.e.} transfer to the $(\nu,n)$-representation (see Section~\ref{sub:bfkl-cross-section}). The expression for the LO foward jet vertex will be given in Eq.~(\ref{c1-dh}) of Section~\ref{sub:dihadron-blm}. 
  In the NLO case, one-loop virtual corrections and two-body final-state contribution have to be considered. 
  The expression of the hadron vertex with NLO accuracy was calculated in Ref.~\cite{hadrons}. Its projection on the $(\nu,n)$-space (see Section~\ref{sub:bfkl-cross-section} for further details) is given in Eq.~(\ref{c11-dh}) and in Appendix~\hyperlink{app:hadron-nlo-if-link}{C}.

  \subsection{Dihadron cross section and azimuthal correlations}
  \label{sub:dihadron-cs}

  In QCD collinear factorisation the cross section 
  of the process~(\ref{process-dh}) reads
  \beq
  \frac{d\sigma}{d\alpha_1d\alpha_2d^2k_1d^2k_2}
  =
  \sum_{r,s=q,{\bar q},g}\int_0^1 dx_1 \int_0^1 dx_2\ f_r\left(x_1,\mu_F\right)
  \ f_s\left(x_2,\mu_F\right)
  \eeq
  \[ \times \,
  \frac{d{\hat\sigma}_{r,s}\left(\hat s,\mu_F\right)}
  {d\alpha_1d\alpha_2d^2k_1d^2k_2}\;,
  \]
  where the $r, s$ indices specify the parton types 
  (quarks $q = u, d, s, c, b$;
  antiquarks $\bar q = \bar u, \bar d, \bar s, \bar c, \bar b$; 
  or gluon $g$),
  $f_r\left(x, \mu_F \right)$ denotes the initial proton PDFs; 
  $x_{1,2}$ are the longitudinal fractions of the partons 
  involved in the hard subprocess, 
  while $\mu_F$ is the factorisation scale; 
  $d\hat\sigma_{r,s}\left(\hat s \right)$ is
  the partonic cross section and
  $x_1x_2s \equiv \hat s$ is the squared center-of-mass energy of the
  parton-parton collision subprocess.
  
  In the BFKL approach the cross section can be presented 
  (see Section~\ref{sub:bfkl-cross-section} for the details of the derivation)
  as the Fourier sum of the azimuthal coefficients ${\cal C}_n$, 
  having so:
  \beq
  \frac{d\sigma}
  {dy_1dy_2\, d|\vec k_1| \, d|\vec k_2|d\phi_1 d\phi_2}
  =\frac{1}{(2\pi)^2}\left[{\cal C}_0+\sum_{n=1}^\infty  2\cos (n\phi )\,
  {\cal C}_n\right]\, ,
  \eeq
  where $\phi=\phi_1-\phi_2-\pi$, with $\phi_{1,2}$ are the two hadrons' 
  azimuthal angles, while $y_{1,2}$ and $\vec k_{1,2}$ are their
  rapidities and transverse momenta, respectively. 
  The $\phi$-averaged cross section ${\cal C}_0$ 
  and the other coefficients ${\cal C}_{n\neq 0}$ are given by
  \beq\label{Cm}
   {\cal C}_n \equiv \int_0^{2\pi}d\phi_1\int_0^{2\pi}d\phi_2\,
   \cos[n(\phi_1-\phi_2-\pi)] \,
   \frac{d\sigma}{dy_1dy_2\, d|\vec k_1| \, d|\vec k_2|d\phi_1 d\phi_2}\;
  \eeq
  \[
   = \frac{e^Y}{s}
   \int_{-\infty}^{+\infty} d\nu \, \left(\frac{\alpha_1 \alpha_2 s}{s_0}
   \right)^{\bar \alpha_s(\mu_R)\left[\chi+\bar\alpha_s(\mu_R)
   \left( \bar\chi+\frac{\beta_0}{8 N_c}\chi\left(-\chi
   +\frac{10}{3}+\ln\frac{\mu_R^4}{\vec k_1^2\vec k_2^2}\right)\right)\right]}
  \]
  \[
   \times \alpha_s^2(\mu_R) c_1(n,\nu,|\vec k_1|, \alpha_1)
   c_2(n,\nu,|\vec k_2|,\alpha_2)\,
  \]
  \[
   \times \left[1
   +\alpha_s(\mu_R)\left(\frac{c_1^{(1)}(n,\nu,|\vec k_1|,
   \alpha_1)}{c_1(n,\nu,|\vec k_1|, \alpha_1)}
   +\frac{c_2^{(1)}(n,\nu,|\vec k_2|, \alpha_2)}{c_2(n,\nu,|\vec    k_2|,
   \alpha_2)}\right)\right.
  \]
  \[
   \left. + \bar\alpha_s^2(\mu_R) \ln\frac{\alpha_1 \alpha_2 s}{s_0}
   \frac{\beta_0}{8 N_c}\chi \left(2\ln \vec k_1^2 \vec k_2^2
   +i\frac{d\ln\frac{c_1(n,\nu)}{c_2(n,\nu)}}{d\nu}\right)\right]\;.
  \]
  \\
  Here $\bar \alpha_s(\mu_R) = N_c/\pi \, \alpha_s(\mu_R)$,  with $N_c$ the number of colours, $\chi = \chi\left(n,\nu\right)$
  is the LO BFKL characteristic function defined in~(\ref{KLLA}),
  $c_{1,2}(n,\nu)$ are the LO impact factors in the $\nu$-repre\-sen\-ta\-tion  (see Eq.~(\ref{if-hadron-lo}) for the corresponding expression in the momentum space), 
  that are given as an integral in the parton fraction $x$, containing
  the PDFs of the gluon and of the different quark/antiquark
  flavours in the proton, and the FFs of the detected hadron,
  \beq  
  \label{c1-dh}
  c_1(n,\nu,|\vec k_1|,\alpha_1) = 2 \sqrt{\frac{C_F}{C_A}}
  (\vec k_1^2)^{i\nu-1/2}\,\int_{\alpha_1}^1\frac{dx}{x}
  \left( \frac{x}{\alpha_1}\right)
  ^{2 i\nu-1} 
  \eeq
  \[ \times \,
  \left[\frac{C_A}{C_F}f_g(x)D_g^h\left(\frac{\alpha_1}{x}\right)
  +\sum_{a=q,\bar q}f_a(x)D_a^h\left(\frac{\alpha_1}{x}\right)\right]
  \]
  and
  \beq
  \label{c2-dh}
  c_2(n,\nu,|\vec k_2|,\alpha_2)=
  \biggl[c_1(n,\nu,|\vec k_2|,\alpha_2)\biggr]^* \;,
  \eeq
  while
  \beq
  \label{c11-dh}
  c_1^{(1)}(n,\nu,|\vec k_1|,\alpha_1)=
  2\sqrt{\frac{C_F}{C_A}}
  \left(\vec k_1^2\right)^{i\nu-\frac{1}{2}}\frac{1}{2\pi}
  \int_{\alpha_1}^1\frac{d x}{x}
  \int_{\frac{\alpha_1}{x}}^1\frac{d\zeta}{\zeta}
  \left(\frac{x\zeta}{\alpha_1}\right)^{2i\nu-1}
  \eeq
  \[ \times \,
  \left[
  \frac{C_A}{C_F}f_g(x)D_g^h\left(\frac{\alpha_1}{x\zeta}\right)C_{gg}
  \left(x,\zeta\right)+\sum_{a=q\bar q}f_a(x)D_a^h
  \left(\frac{\alpha_1}{x\zeta}
  \right)C_{qq}\left(x,\zeta\right)
  \right.
  \]
  \[ \times \,
  \left.D_g^h\left(\frac{\alpha_1}{x\zeta}\right)
  \sum_{a=q\bar q}f_a(x)C_{qg}
  \left(x,\zeta\right)+\frac{C_A}{C_F}f_g(x)\sum_{a=q\bar q}D_a^h
  \left(\frac{\alpha_1}{x\zeta}\right)C_{gq}\left(x,\zeta\right)
  \right]\, ,
  \]
  and 
  \beq\label{c21-dh}
  c_2^{(1)}(n,\nu,|\vec k_2|,\alpha_2)=\biggl[c_1^{(1)}(n,\nu,|\vec k_2|,
  \alpha_2)\biggr]^*
  \eeq
  are the NLO impact factor corrections in the $\nu$-representation. The
  expressions for the NLO coefficient functions $C_{ij}$ 
  in Eq.~(\ref{c11-dh}) are given in Appendix~\hyperlink{app:hadron-nlo-if-link}{C}.
  It is known~\cite{Caporale:2015uva} that contributions to the NLO impact
  factors  that are proportional to the QCD $\beta_0$-function are
  universally expressed in terms of the LO impact factors of the considered
  process, through the function $f\left(\nu\right)$, defined as follows:
  \begin{equation}
  \label{dh-f-nu}
  2\ln \mu_R^2+i\frac{d\ln\frac{c_1(n,\nu)}{c_2(n,\nu)}}{d\nu}
  =\ln\frac{\mu_R^4}{\vec k_1^2\vec k_2^2}
  \end{equation}
  \[
  -2 \frac{\int_{\alpha_1}^1\frac{d x}{x}\left(\frac{x}{\alpha_1}
  \right)^{2 i \nu-1}\log\left(\frac{x}{\alpha_1}\right)\left[\frac{C_A}{C_F}
  f_g(x)D_g^h\left(\frac{\alpha_1}{x}\right)+\sum_{a=q,\bar q}f_a(x)D_a^h
  \left(\frac{\alpha_1}{x}\right)\right]}
  {\int_{\alpha_1}^1\frac{d x}{x}\left( \frac{x}{\alpha_1}\right)^{2 i \nu-1}
  \left[\frac{C_A}{C_F}f_g(x)D_g^h\left(\frac{\alpha_1}{x}\right)+\sum_{a=q,\bar q}
  f_a(x)D_a^h\left(\frac{\alpha_1}{x}\right)\right]}
  \]
  \[
  -2  \frac{\int_{\alpha_2}^1\frac{d x}{x}\left( \frac{x}{\alpha_2}
  \right)^{-2 i \nu-1}\log\left(\frac{x}{\alpha_2}\right)\left[\frac{C_A}{C_F}
  f_g(x)D_g^h\left(\frac{\alpha_2}{x}\right)+\sum_{a=q,\bar q}f_a(x)D_a^h
  \left(\frac{\alpha_2}{x}\right)\right]}
  {\int_{\alpha_2}^1\frac{d x}{x}\left( \frac{x}{\alpha_2}\right)^{-2 i \nu-1}
  \left[\frac{C_A}{C_F}f_g(x)D_g^h\left(\frac{\alpha_2}{x}\right)+\sum_{a=q,\bar q}
  f_a(x)D_a^h\left(\frac{\alpha_2}{x}\right)\right]}
  \]
  \[
  \equiv \ln\frac{\mu_R^4}{\vec k_1^2\vec k_2^2} + 2f(\nu)\, .
  \]

  \subsection{Integration over the final-state phase space} 
  \label{sub:dihadron-phase-space}
  
  In order to match the actual LHC kinematical cuts, we integrate the
  coefficients over the phase space for two final-state hadrons,  
  \beq
  \label{Cm_int-dh}
  C_n= 
  \int_{y_{1}^{\rm min}}^{y_{1}^{\rm max}}dy_1
  \int_{y_{2}^{\rm min}}^{y_{2}^{\rm max}}dy_2\int_{k_{1}^{\rm min}}^{\infty}dk_1
  \int_{k_{2}^{\rm min}}^{\infty}dk_2
  \, {\cal C}_n \left(y_1,y_2,k_1,k_2 \right)\, .
  \eeq
  For the integrations over rapidities we consider two distinct ranges:
  \begin{enumerate}
  \item 
  $y_{1}^{\rm min}=-y_{2}^{\rm max}=-2.4$, 
  $y_{1}^{\rm max}=-y_{2}^{\rm min}=2.4$, 
  and $Y \leq 4.8$, \\
  typical for the identified hadron detection at the LHC; \,
  \item 
  $y_{1}^{\rm min}=-y_{2}^{\rm max}=-4.7$, 
  $y_{1}^{\rm max}=-y_{2}^{\rm min}=4.7$,
  and $Y \leq 9.4$, \\
  similar to those used in the CMS Mueller--Navelet jets analysis.
  \end{enumerate}
  As minimum transverse momenta we choose $k_{1}^{\rm min}=k_{2}^{\rm min}=5$~GeV,
  which are also realistic values for the LHC. We observe that the minimum
  transverse momentum in the CMS analysis~\cite{Khachatryan:2016udy} 
  of Mueller--Navelet jet production is much larger, 
  $k_{J}^{\rm min}=35$~GeV. In our calculations we use 
  the PDF set NLO MSTW 2008~\cite{Martin:2009iq} 
  with two different NLO parameterisations for hadron FFs:  
  AKK~\cite{Albino:2008fy} and HKNS~\cite{Hirai:2007cx} 
  (see Section~\ref{sec:dihadron-numerics} for a related discussion). 
  In the results presented below we sum over the production of charged light 
  hadrons: $\pi^{\pm}, K^{\pm}, p,\bar p$.

  \subsection{BLM scale setting} 
  \label{sub:dihadron-blm}

  In order to find the values of the BLM scales, we introduce the ratios of the 
  BLM to the ``natural'' scale suggested by the kinematical of the process, 
  $\mu_N=\sqrt{|\vec k_{1}||\vec k_{2}|}$, so that $m_R=\mu_R^{\rm BLM}/\mu_N$, 
  and look for the values of $m_R$ such that Eq.~(\ref{c_nnnbeta}) is satisfied. 

  Then we plug these scales into our expression for the \emph{integrated coefficients}
  in the BLM scheme (for the derivation see Section~\ref{sec:bfkl_blm} and Ref.~\cite{Caporale:2015uva}):
  \beq
  \label{dh-eq}
  C_n =
  \int_{y_{1}^{\rm min}}^{y_{1}^{\rm max}}dy_1
  \int_{y_{2}^{\rm min}}^{y_{2}^{\rm max}}dy_2\int_{k_{1}^{\rm min}}^{\infty}dk_1
  \int_{k_{2}^{\rm min}}^{\infty}dk_2
  \,
  \int\limits^{\infty}_{-\infty} d\nu 
  \frac{e^Y}{s}\,
  \eeq
  \[ \times \,
  e^{Y \bar \alpha^{\rm MOM}_s(\mu^{\rm BLM}_R)\left[\chi(n,\nu)
  +\bar \alpha^{\rm MOM}_s(\mu^{\rm BLM}_R)\left(\bar \chi(n,\nu) +\frac{T^{\rm conf}}
  {3}\chi(n,\nu)\right)\right]}
  \]
  \[ \times \,
   \left(\alpha^{\rm MOM}_s (\mu^{\rm BLM}_R)\right)^2 c_1(n,\nu)c_2(n,\nu)
  \]
  \[ \times \,
  \left[1+\bar \alpha^{\rm MOM}_s(\mu^{\rm BLM}_R)\left\{\frac{\bar c^{(1)}_1(n,\nu)}
  {c_1(n,\nu)}+\frac{\bar c^{(1)}_2(n,\nu)}{c_2(n,\nu)}+\frac{2T^{\rm conf}}{3}
  \right\} \right] \, ,
  \]
  with $T^{\rm conf}$ defined in~(\ref{T_Tbeta_Tconf}).
  The coefficient $C_0$ gives the total cross sections and the ratios
  $C_n/C_0 = \langle\cos(n\phi)\rangle$ determine the values of the mean cosines,
  or azimuthal correlations, of the produced hadrons. In Eq.~(\ref{dh-eq}), 
  $\bar \chi(n,\nu)$ is the eigenvalue of NLO BFKL kernel given in Eq.~(\ref{ch11}), whereas $\bar c^{(1)}_{1,2}$ are the NLA parts of the hadron vertices~\cite{hadrons}. 

  We give predictions for $C_n$ by fixing the factorisation scale
  $\mu_F$ in three different ways:
  \begin{enumerate}
  \item 
  $\mu_F = \mu_R = \mu_R^{\rm BLM}$;
  \item
  $(\mu_F)_{1,2} = |\vec k_{1,2}|$;
  \item
  $\mu_F = \mu_R = \mu_N=\sqrt{|\vec k_{1}||\vec k_{2}|}$.
  \end{enumerate}
  Note that the option 3., which correspond to ``natural '' scale 
  selection for both $\mu_F$ and $\mu_R$, is used only in the full LLA calculations given in Section~\ref{sec:dihadron-kernel}.
  
\begin{figure}[H]
   \centering
   \includegraphics[scale=0.38]{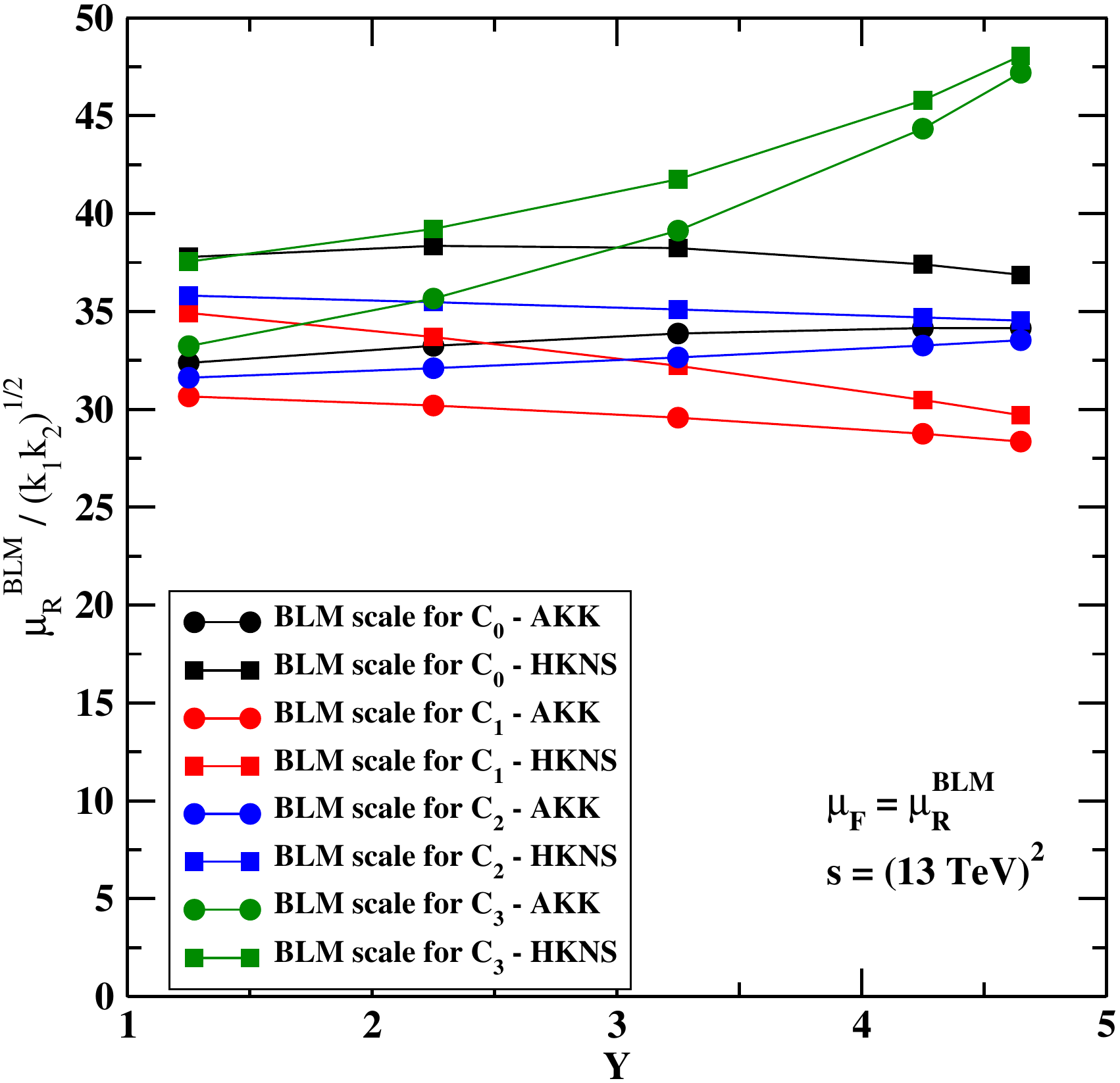} 
   \includegraphics[scale=0.38]{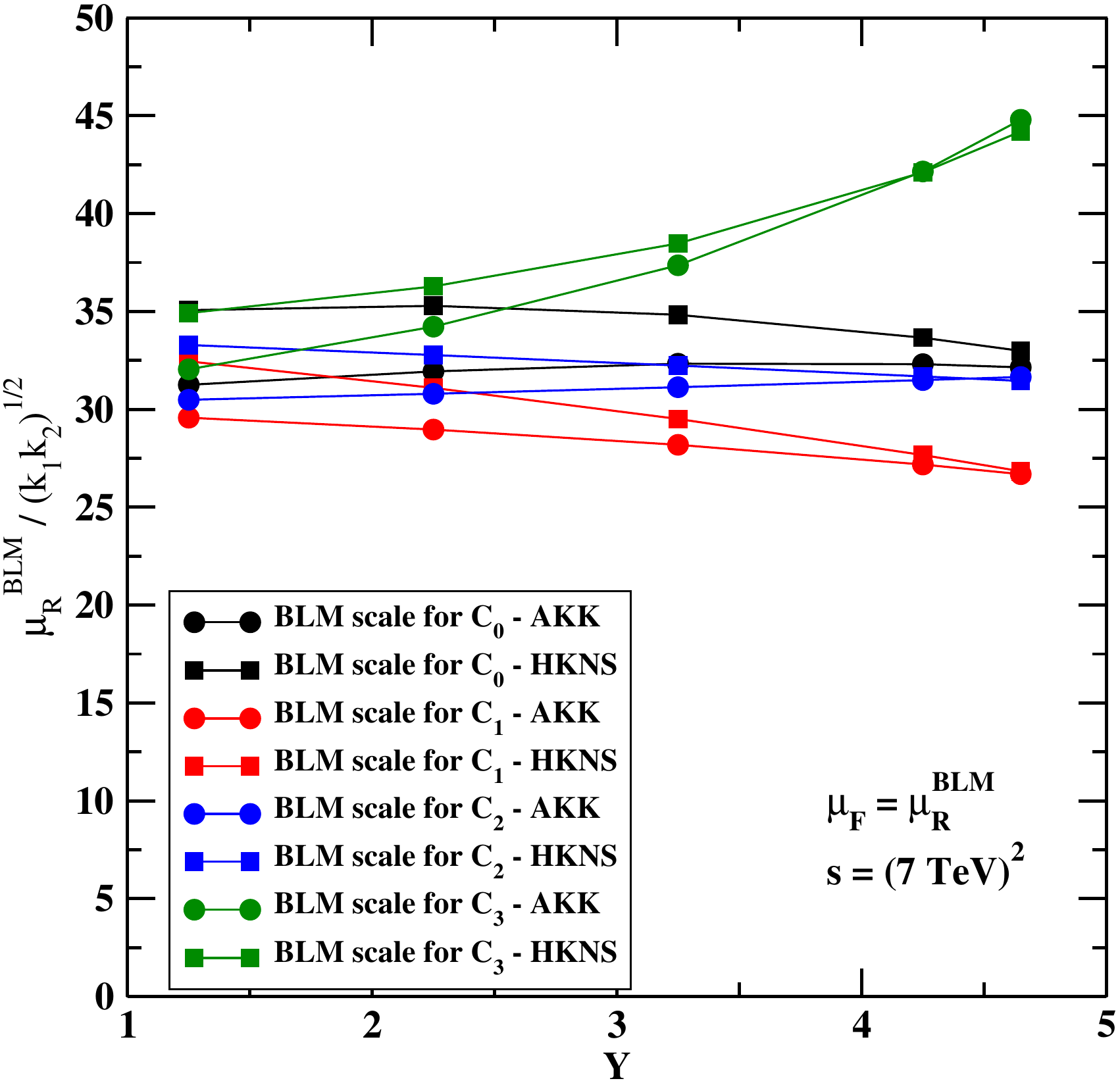}
   
   \includegraphics[scale=0.38]{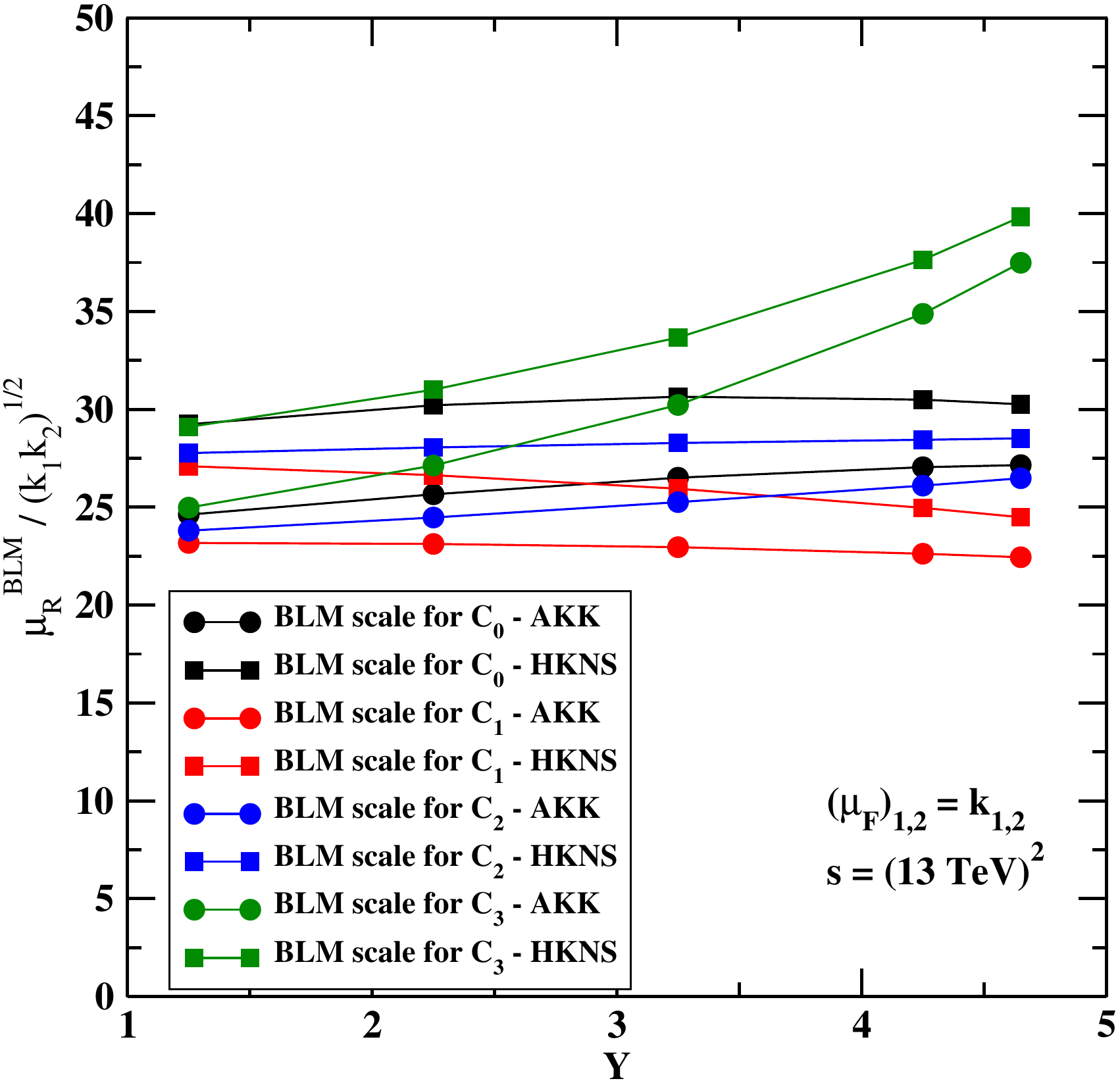} 
   \includegraphics[scale=0.38]{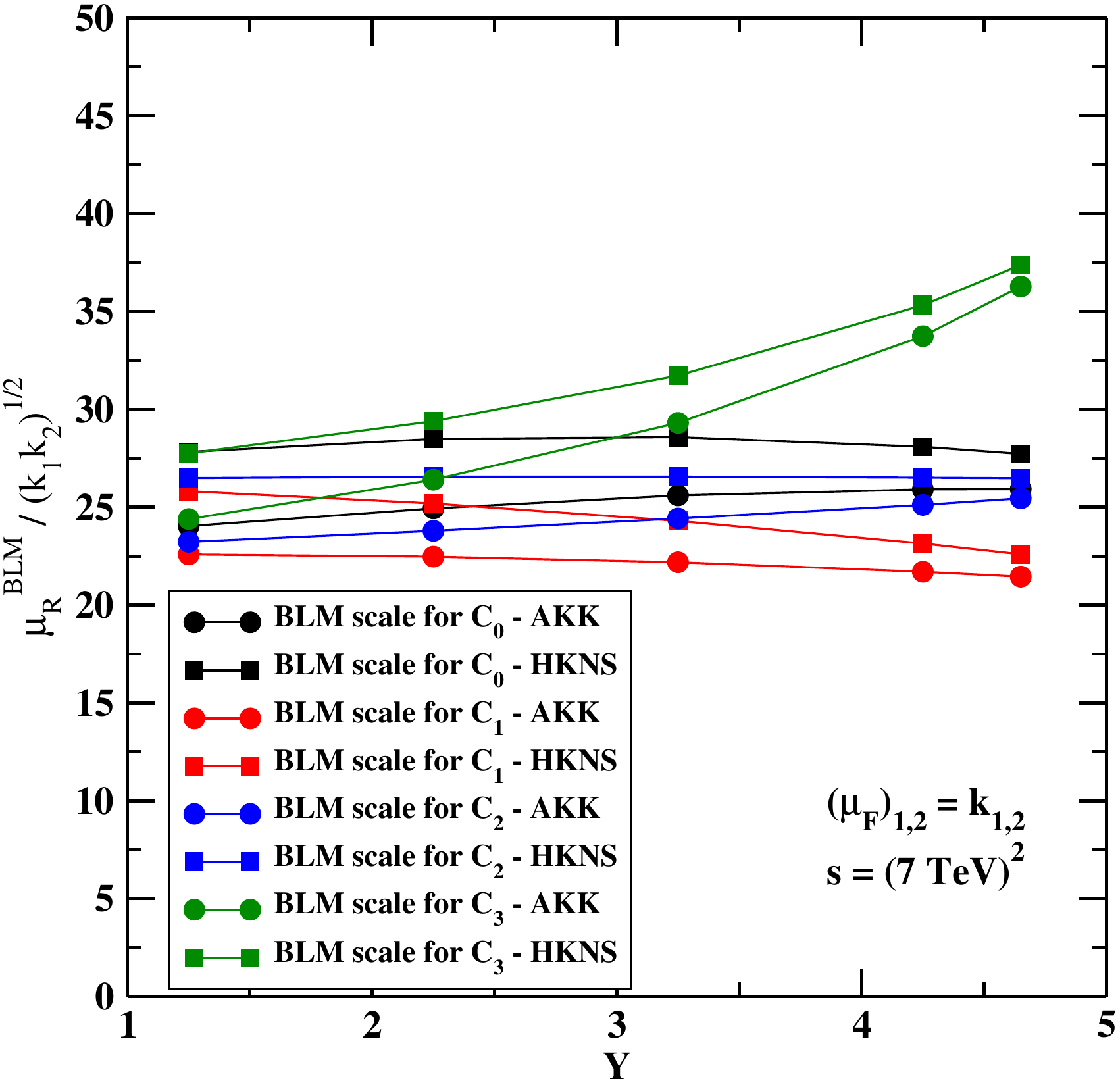}
   \caption[BLM scales for dihadron production]
    {BLM scales for the dihadron production 
    {\it versus} the rapidity interval $Y$ 
    for the $\phi$-averaged cross section $C_0$ and
    for the azimuthal coefficients $C_{1,2,3}$.
    The two choices for the factorisation scale 
    $\mu_F = \mu^{\rm BLM}_R$ and $(\mu_F)_{1,2} = |\vec k_{1,2}|$
    are considered, while the center-of-mass energy 
    takes the values $\sqrt{s} = 7,13$ TeV.}
   \label{blm_scales-dh}
  \end{figure}

  In Fig.~\ref{blm_scales-dh} we present the $Y$-dependence 
  of $m_R$ at $\sqrt{s} = 7, 13$ TeV, for the first few values of $n$, and for $Y \leq 4.8$.
  We obtain rather large numbers, $m_R\sim 35$. These  values are larger than those obtained previously for 
  similar scale ratios in the case of the Mueller--Navelet jet production process. 
  The difference may be attributed to the fact that, in the case of dihadron 
  production, we have an additional branching of the parton momenta (described 
  by the detected hadron FFs), and typical transverse momenta of the partons 
  participating in the hard scattering turn to be considerably larger than 
  $|\vec k_{1,2}|$, the momenta of the hadrons detected in the final state.
  We found that typical value of the fragmentation fraction is 
  $z=\alpha_h/x\sim 0.4$, which explains the main source of the difference 
  between the values of the BLM scales in the case of dijet and dihadron
  production. Another source is related to the difference in the function 
  $f(\nu)$, defined in Eq.~(\ref{dh-f-nu}), which appears in the expression for the jet-
  and hadron-vertex in these two reactions, and enters also the definition of 
  the BLM scale: $f(\nu)$ is zero for the jets and non-zero in the dihadron case.
  The typical $m_R$ values obtained for $4.8 < Y \leq 9.4$ are not very different from those shown here (Fig.~\ref{blm_scales-dh}), except that for $n=3$ we got values four to five times larger than in the region $Y\leq 4.8$. 
  All calculations are done in the MOM scheme. For comparison, we present results 
  for the $\phi$-averaged cross section $C_0$ in the $\overline{\rm MS}$ scheme 
  (as implemented in Eq.~(\ref{Cm}))
  for $\sqrt{s} = 7, 13$ TeV and for $Y \leq 4.8, 9.4$.
  In this case, we choose ``natural'' values for $\mu_R$, {\it i.e.} 
  $\mu_R = \mu_N = \sqrt{|\vec k_{1}||\vec k_{2}|}$, and the option 2.,
  {\it i.e.} $(\mu_F)_{1,2} = |\vec k_{1,2}|$ for the factorisation scale. 

 \section{A first phenomenological analysis} 
 \label{sec:dihadron-kernel}
 
 In this Section the first stage of our analysis, with  
 the implementation of a partial NLA BFKL in which we take only 
 the higher-order corrections coming from the kernel and 
 neglecting the NLA parts of hadron 
 vertices, {\it i.e.} putting $\bar c^{(1)}_{1,2}=0$ 
 in Eq.~(\ref{dh-eq}). 
 
\begin{figure}[H]
  \vspace{-0.55cm}
  \centering
  \includegraphics[scale=0.38]{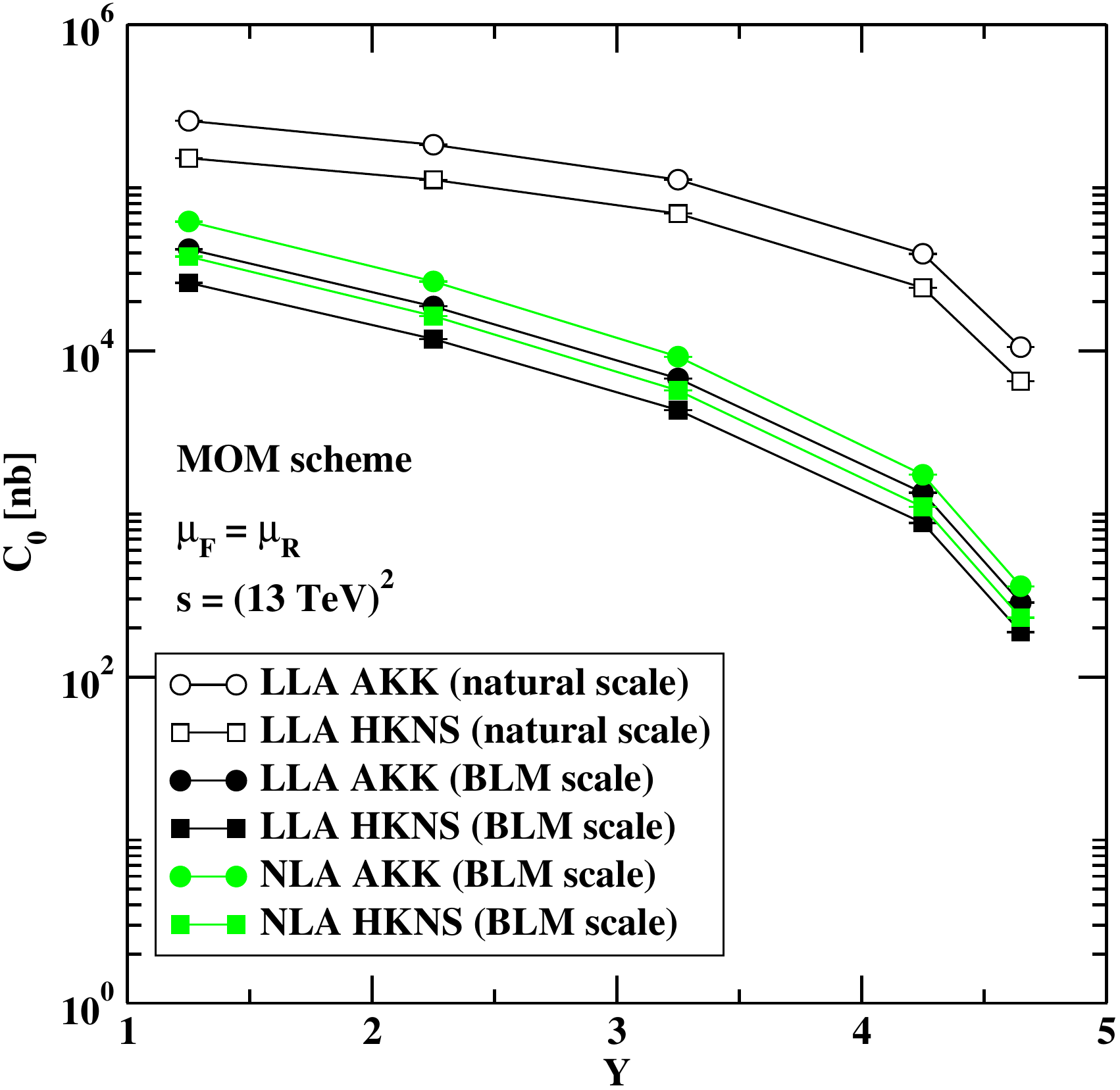}
  \includegraphics[scale=0.38]{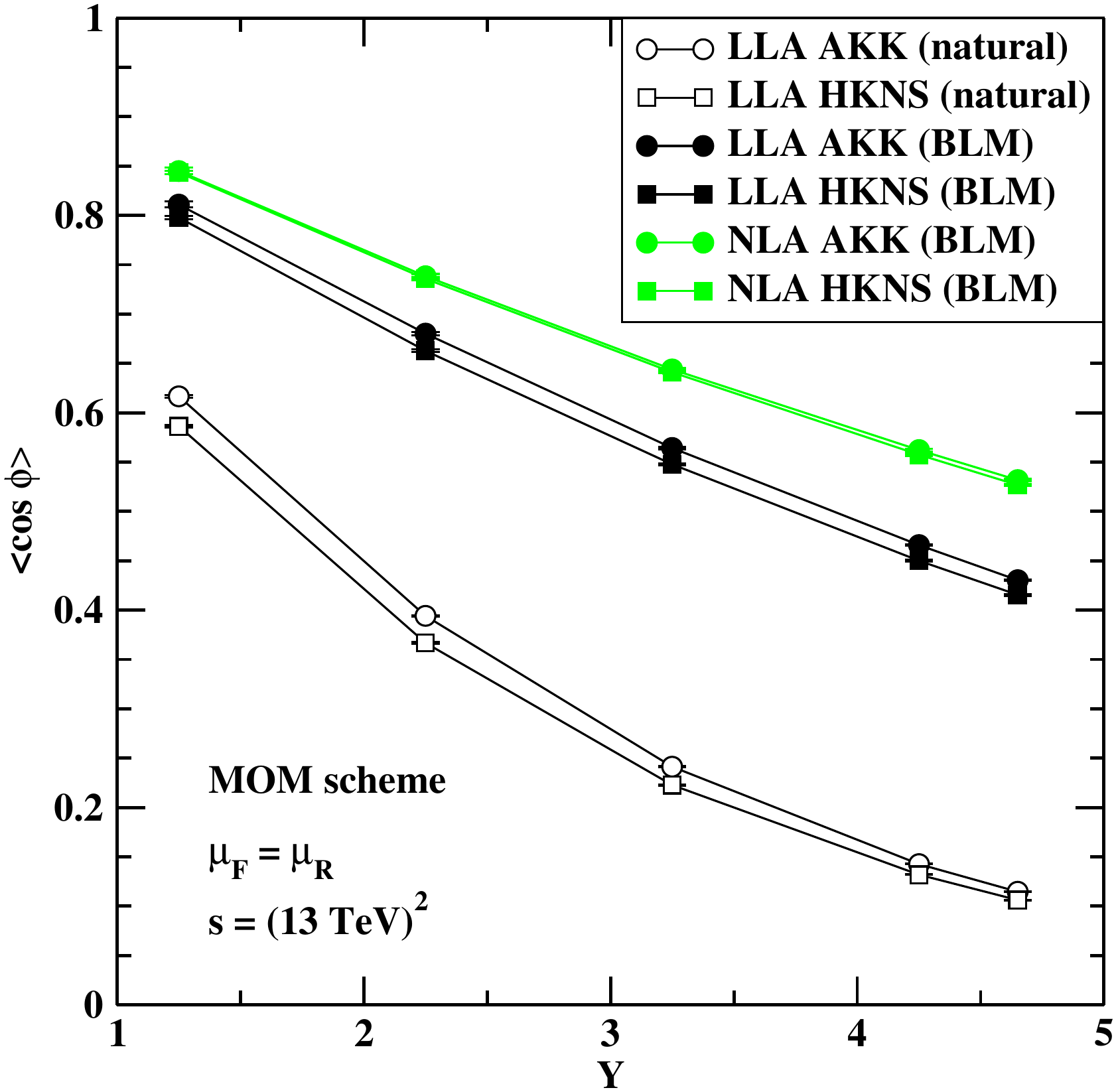}

  \includegraphics[scale=0.38]{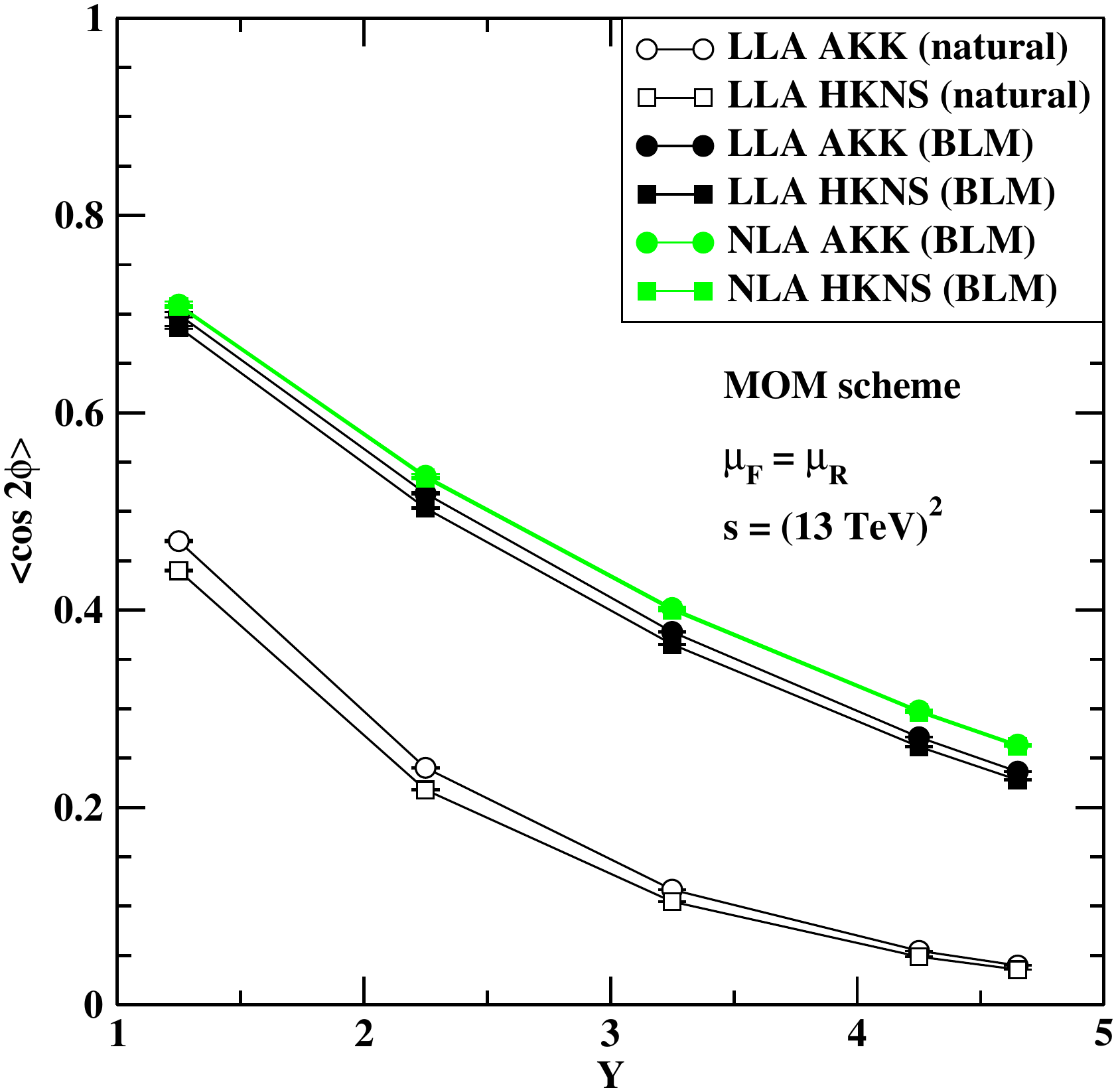}
  \includegraphics[scale=0.38]{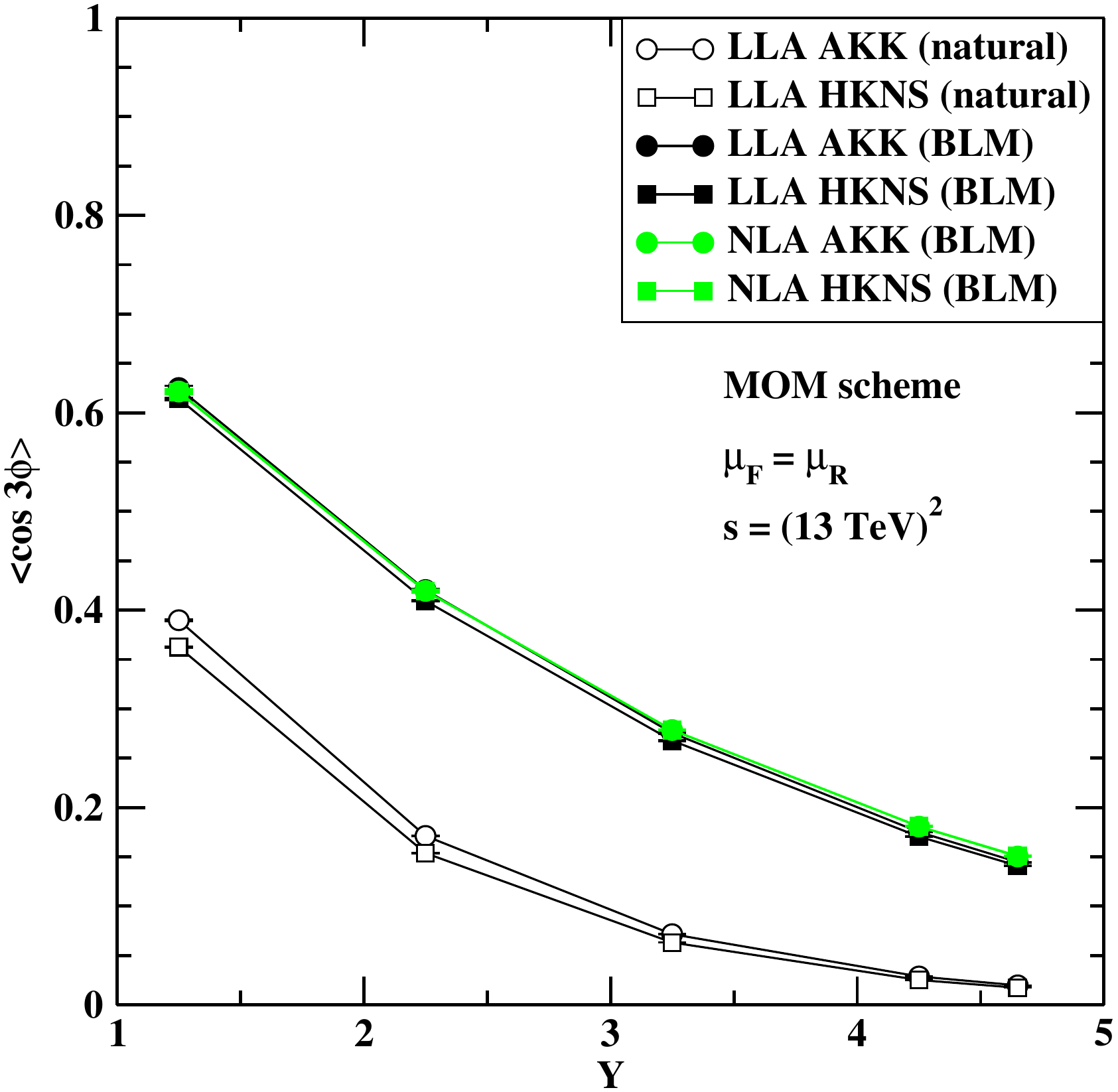}

  \includegraphics[scale=0.38]{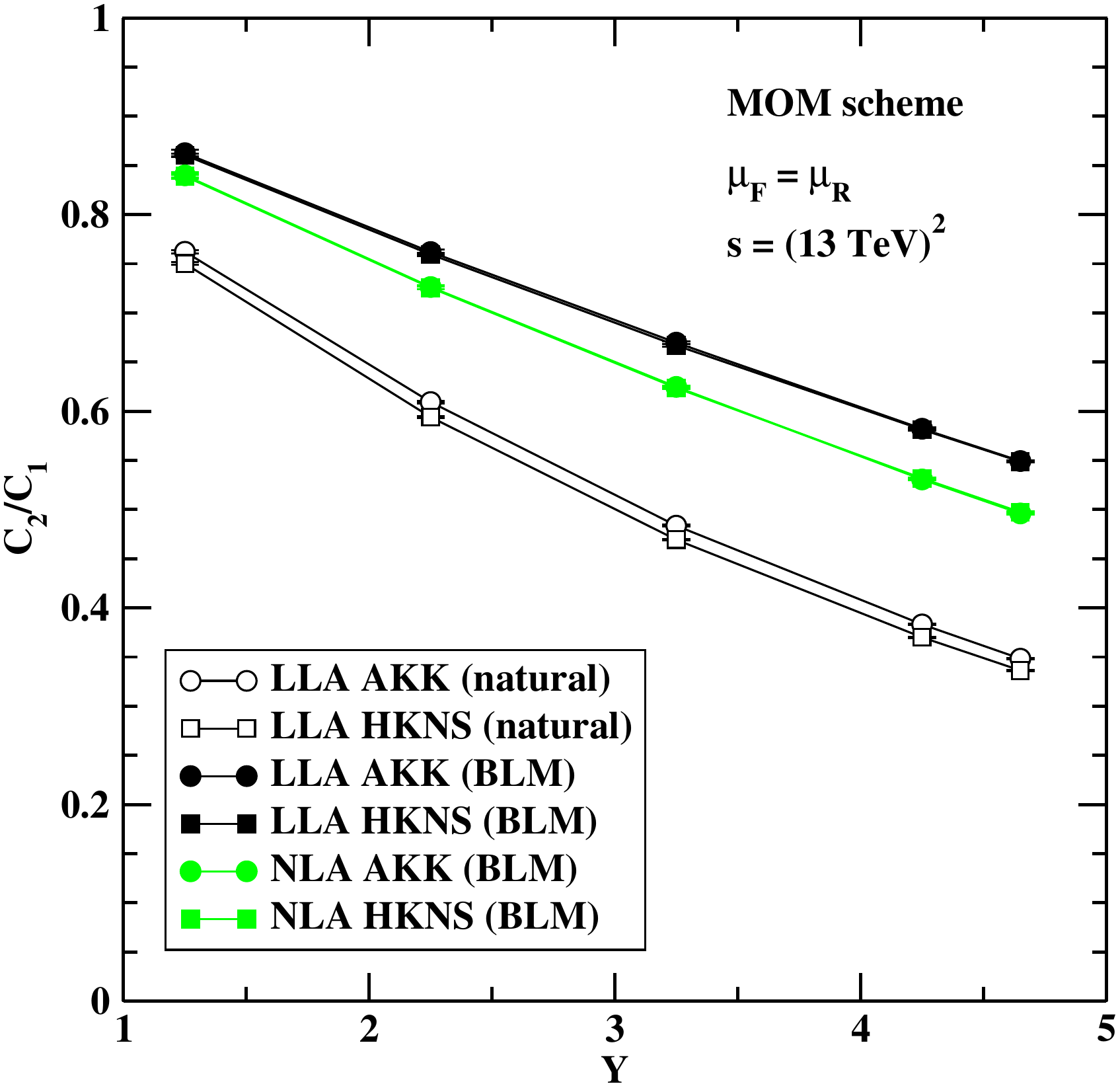}
  \includegraphics[scale=0.38]{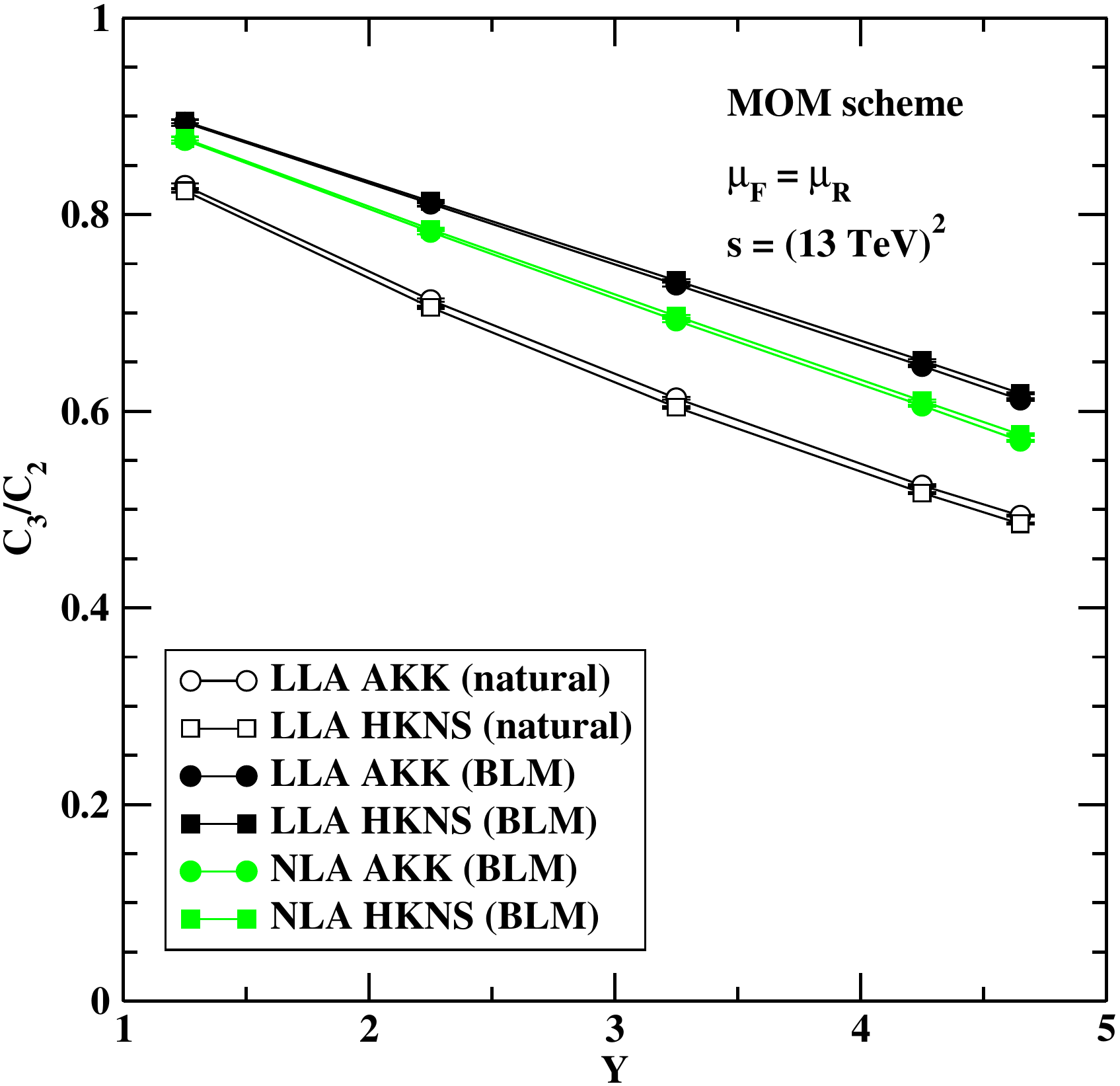}
  \caption[NLA kernel predictions for dihadron production at 13 TeV]
  {Cross section and azimuthal ratios for dihadron 
   production for $\mu_F = \mu^{\rm BLM}_R$, 
   $\sqrt{s} = 13$ TeV, and $Y \leq 4.8$.
   Here ``LLA'' means pure leading logarithmic approximation, 
   while ``NLA kernel'' means inclusion of the NLA corrections 
   from the kernel only.
   See the text for the definition of ``natural'' and ``BLM'' scales.}
  \label{fig:blm13nlk}
 \end{figure}

\begin{figure}[H]
  \centering
  \includegraphics[scale=0.38]{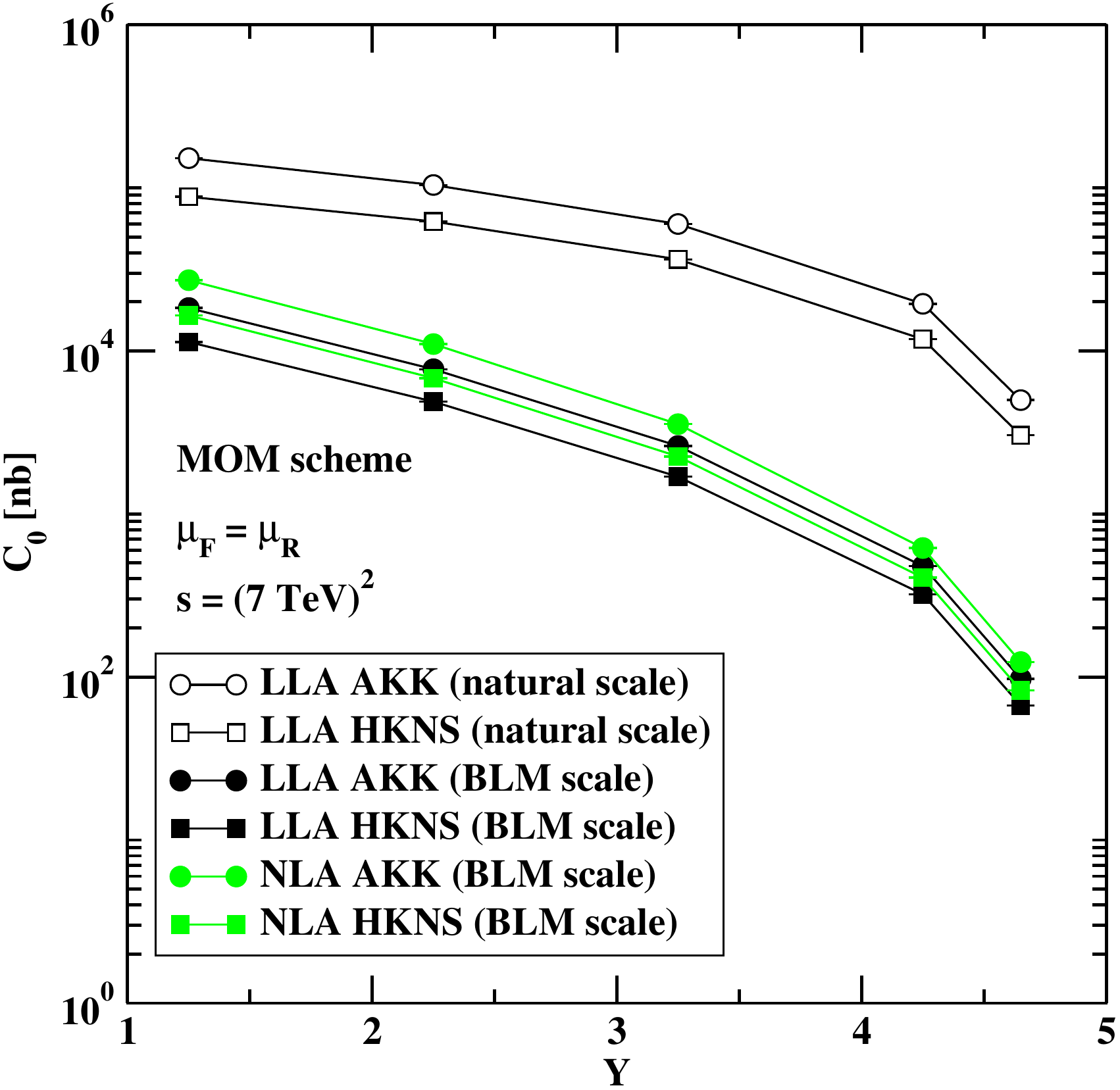}
  \includegraphics[scale=0.38]{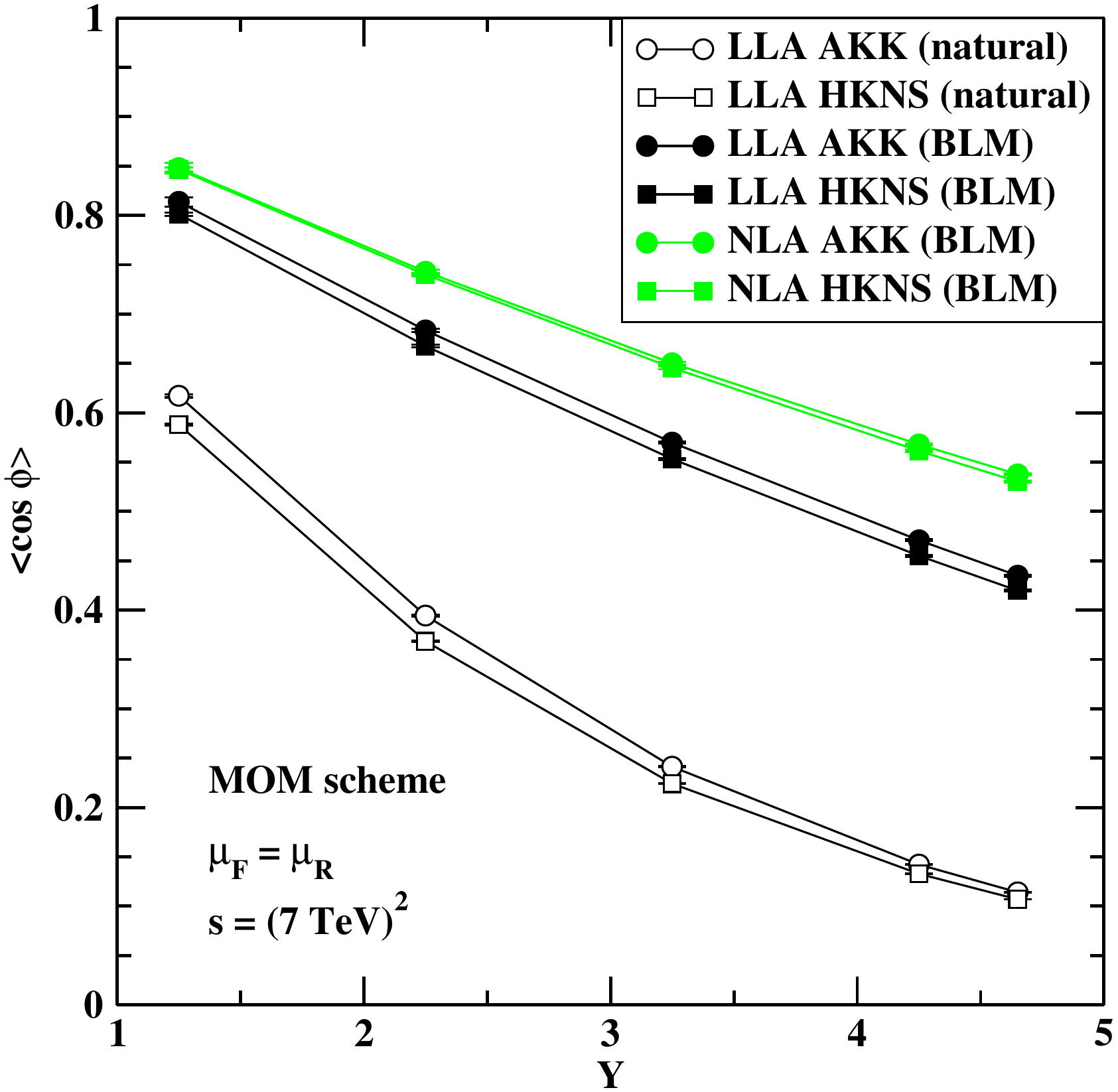}

  \includegraphics[scale=0.38]{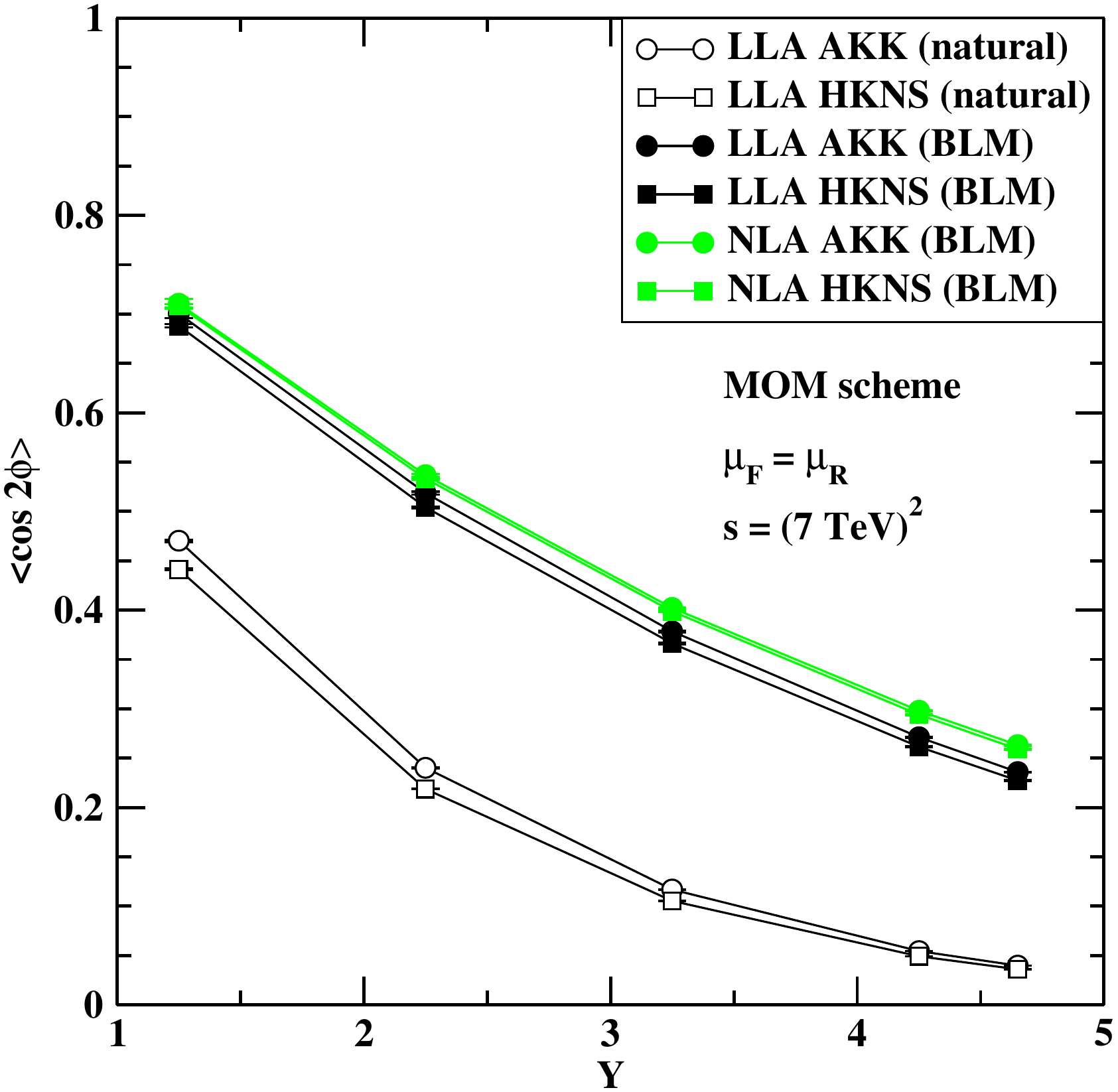}
  \includegraphics[scale=0.38]{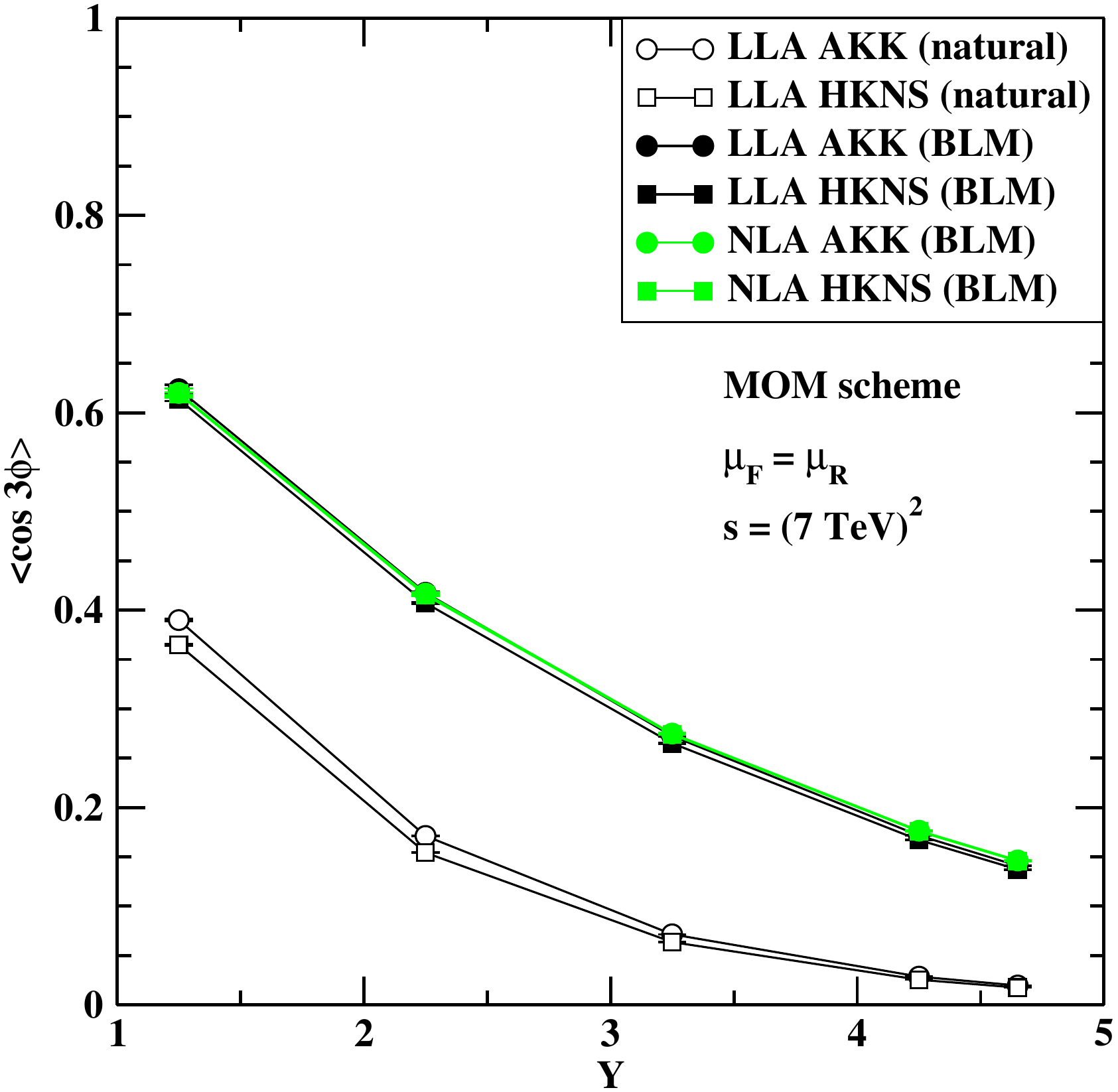}

  \includegraphics[scale=0.38]{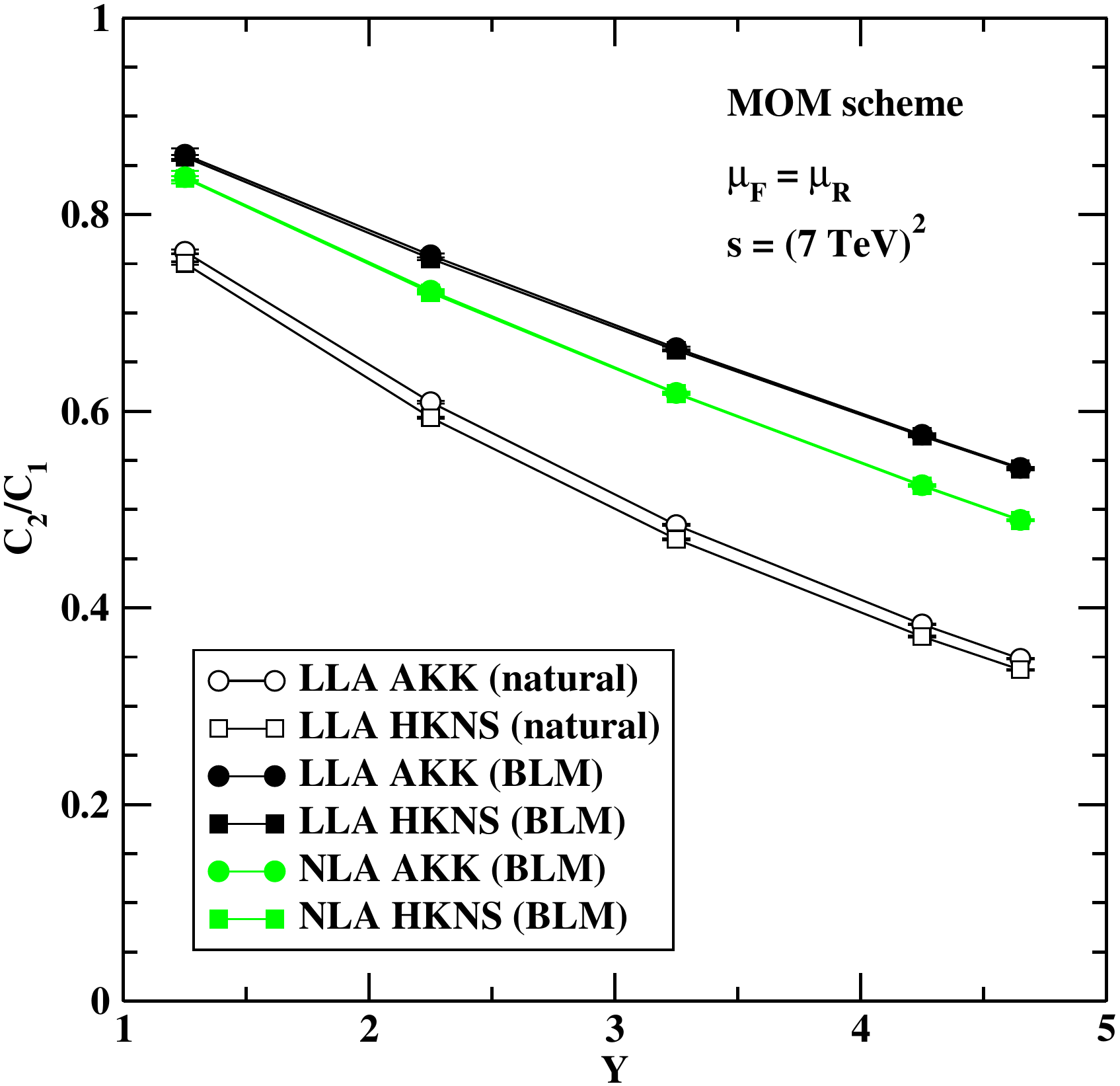}
  \includegraphics[scale=0.38]{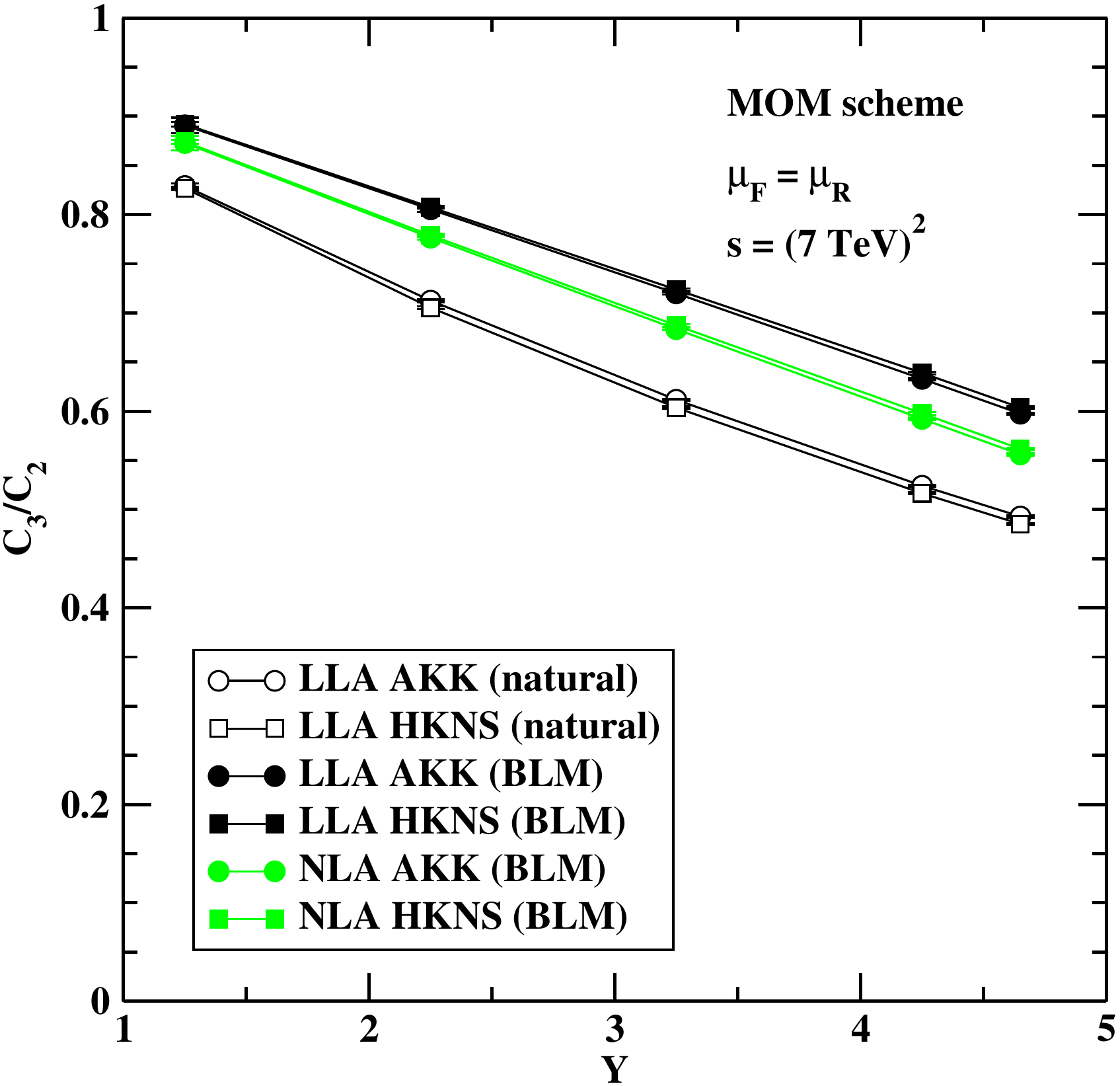}
  \caption[NLA kernel predictions for dihadron production at 7 TeV]
  {Cross section and azimuthal ratios for dihadron production 
   for $\mu_F = \mu^{\rm BLM}_R$, $\sqrt{s} = 7$ TeV, and $Y \leq 4.8$.
   Here ``LLA'' means pure leading logarithmic approximation, 
   while ``NLA kernel'' means inclusion of the NLA corrections 
   from the kernel only.
   See the text for the definition of ``natural'' and ``BLM'' scales.}
  \label{fig:blm7nlk}
 \end{figure}

 We found that the difference between our predictions for the $C_m/C_n$ ratios at $\sqrt{s}=13$~TeV (Fig.~\ref{fig:blm13nlk}) and the ones at $\sqrt{s}=13$~TeV (Fig.~\ref{fig:blm7nlk}), 
 is not larger than $3\%$.
 We see on the first caption of Figs.~\ref{fig:blm13nlk} and~\ref{fig:blm7nlk}.
 the sizeable difference between predictions of the $\phi$-averaged cross section $C_0$ in two cases of selected FFs, AKK and HKNS, which  
 means that the FFs are not well constrained in the required kinematical region. 
 In a similar range the difference between $\pi^{\pm}$ and $K^{\pm}$ AKK and 
 HKNS FFs was recently discussed in Ref.~\cite{Yang:2015avi}. Our calculation 
 with the AKK FFs gives bigger cross sections, whereas the difference between 
 AKK and HKNS for the azimuthal ratios $C_m/C_n$ is small, since the FFs uncertainties 
 are largely cancelled in the coefficient ratios describing the azimuthal-angle
 correlations. 
 Our predictions for dihadron production calculated in LLA with the use of 
 the ``natural'' scale $\mu_N$ and our NLA results obtained with the
 BLM scale setting are different: with NLA BLM we got much lower values of the 
 cross sections  and considerably larger 
 predictions for the $C_m/C_n$. 
 Plots of Figs.~\ref{fig:blm13nlk} 
 and~\ref{fig:blm7nlk} show that the LLA results 
 with BLM scales lie closer to the NLA BLM ones than LLA results with 
 ``natural'' scales. The difference between NLA BLM and LLA with BLM scale 
 predictions is due to the account of NLA corrections to the BFKL kernel in the former. 
 This clearly represents a reliability test for the BLM method.
 
 \section{Full NLA BFKL calculation} 
 \label{sec:dihadron-NLA}
 
 Int this section the first analysis for dihadron production in the full NLA BFKL accuracy is presented.
 
 We checked that in our numerical analysis the essential values of $x$ are 
 not too small, $x\sim [10^{-3}\div 10^{-2}]$, and even bigger in the case of the
 larger $Y \leq 9.4$. This justifies our use of 
 PDFs with the standard DGLAP evolution. Note that our process is not a 
 low-$x$ one, and similarly to the Mueller--Navelet jet production, we are 
 dealing with a dilute partonic system. Therefore possible saturation effects 
 are not important here, and the BFKL dynamics appears only through resummation 
 effects in the hard scattering subprocesses, without influence on the 
 PDF evolution.
 
\begin{figure}[H]
 \centering

   \includegraphics[scale=0.38]{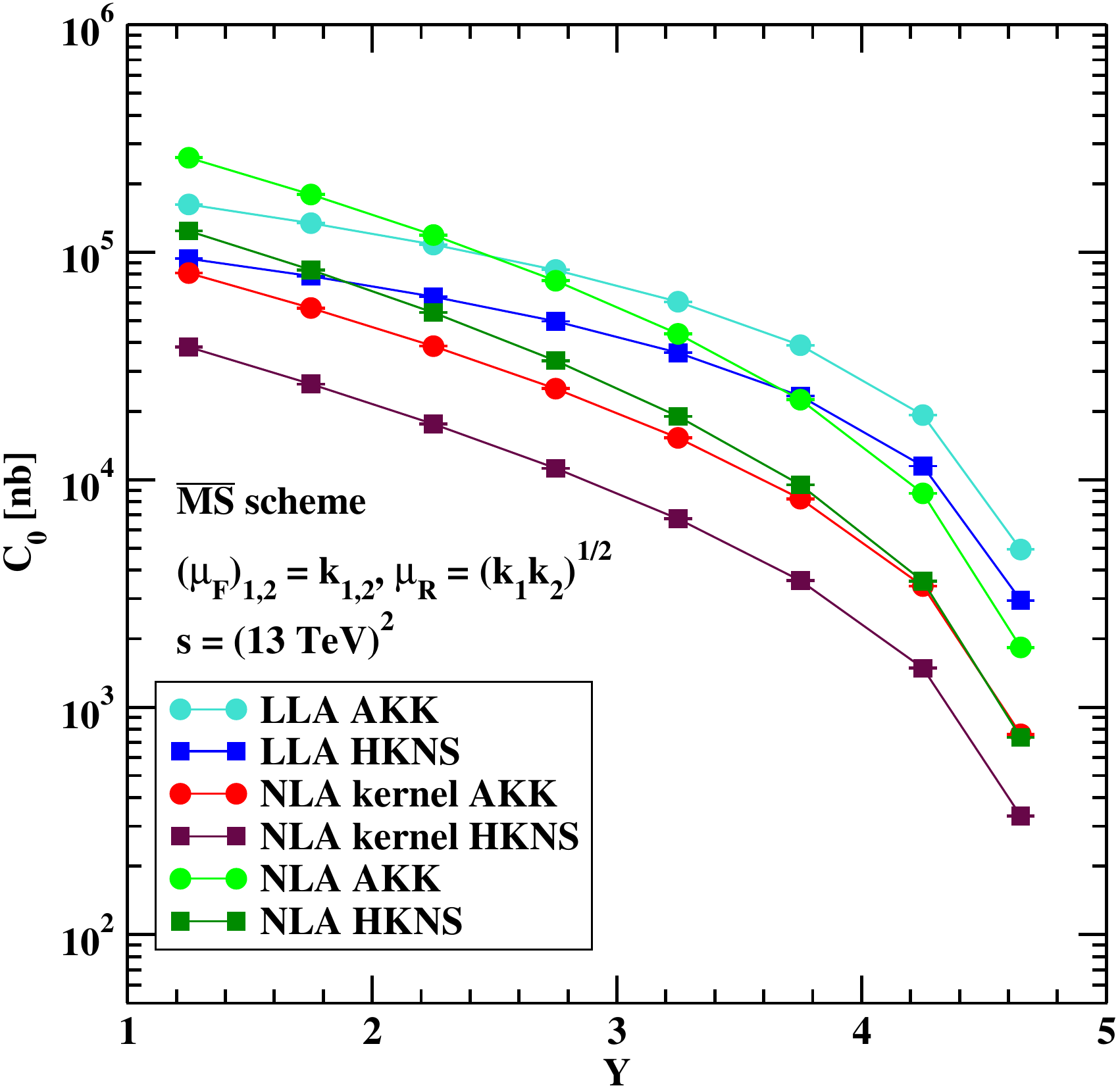}
   \includegraphics[scale=0.38]{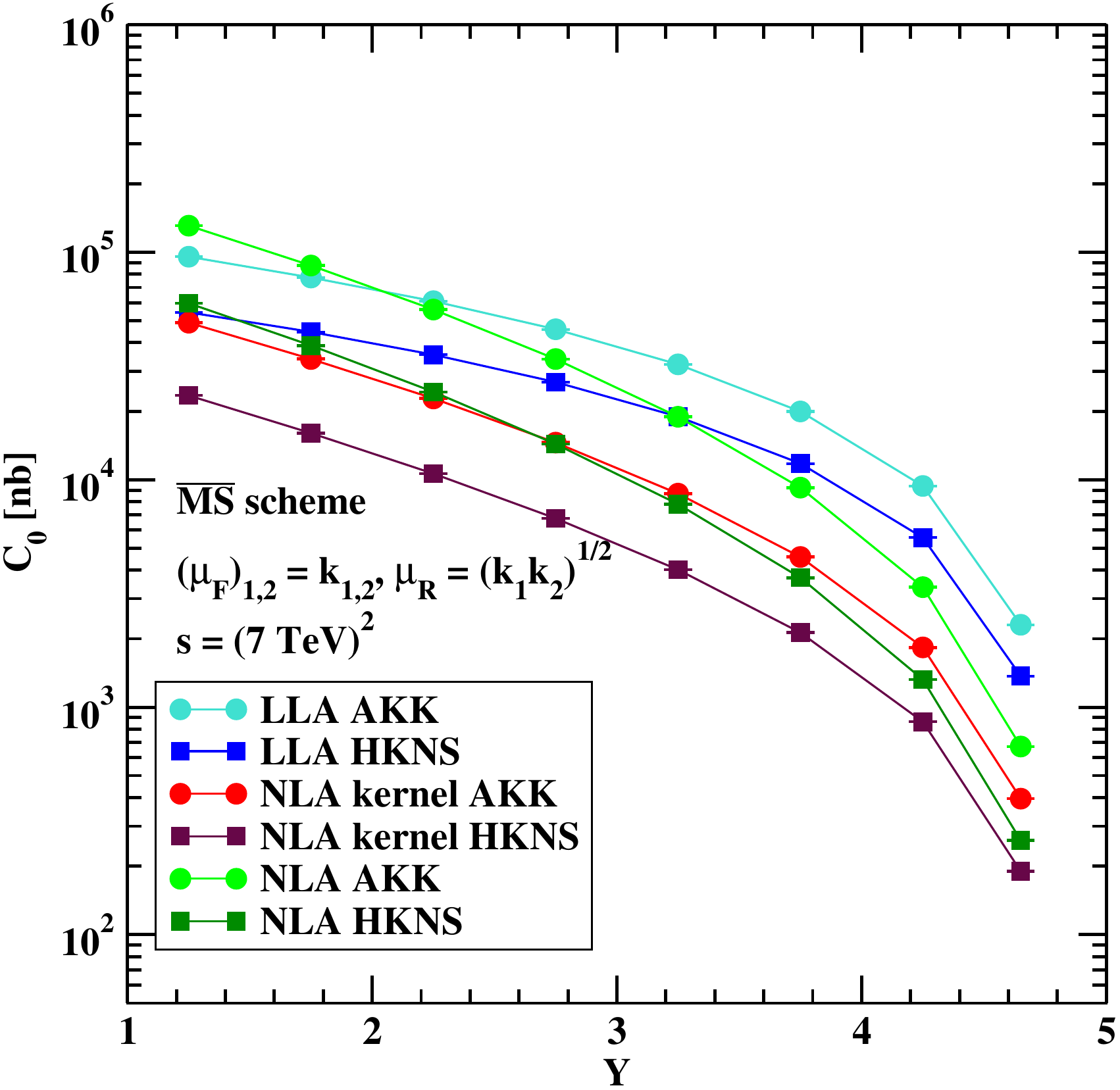}

   \includegraphics[scale=0.38]{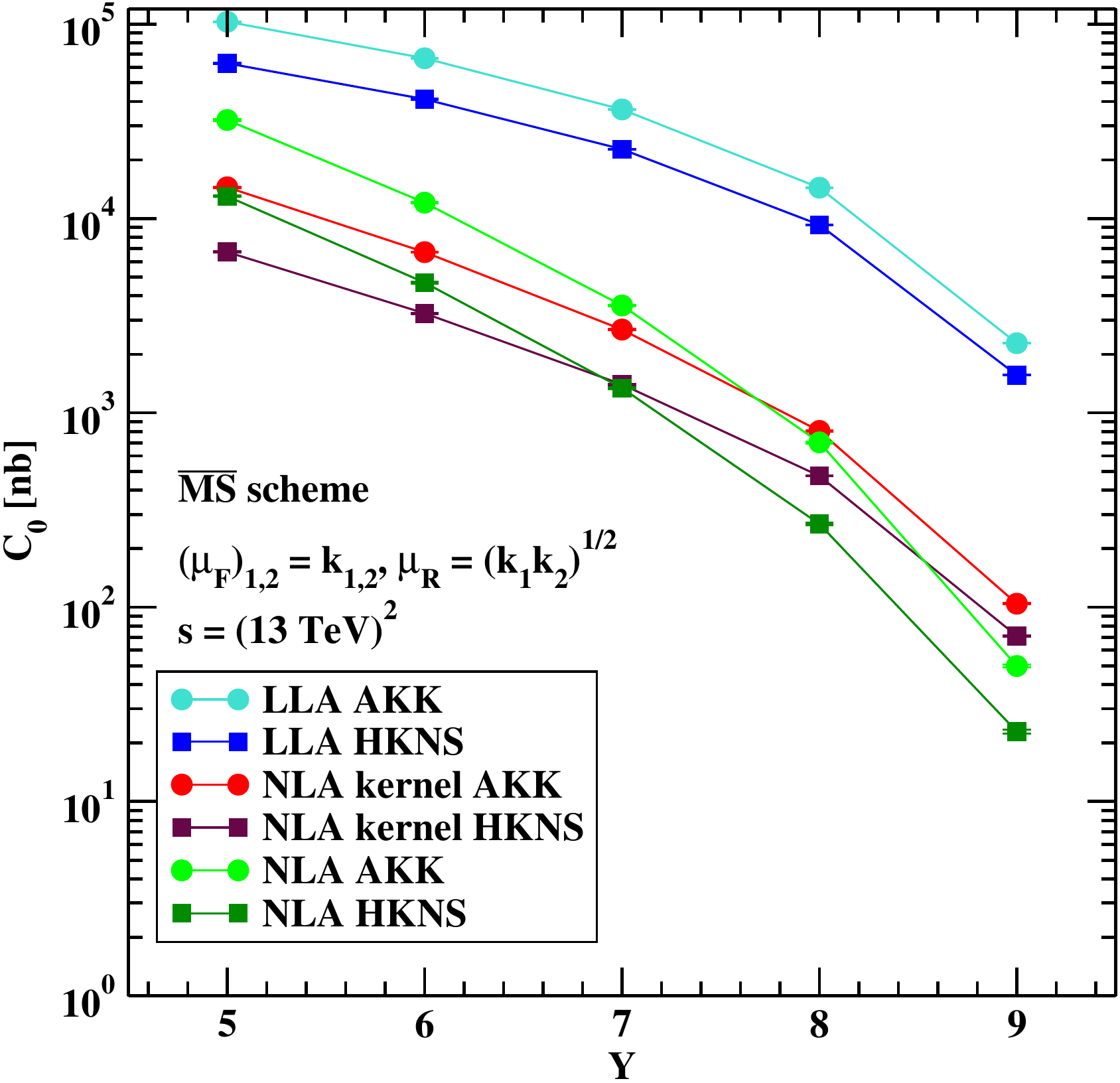}
   \includegraphics[scale=0.38]{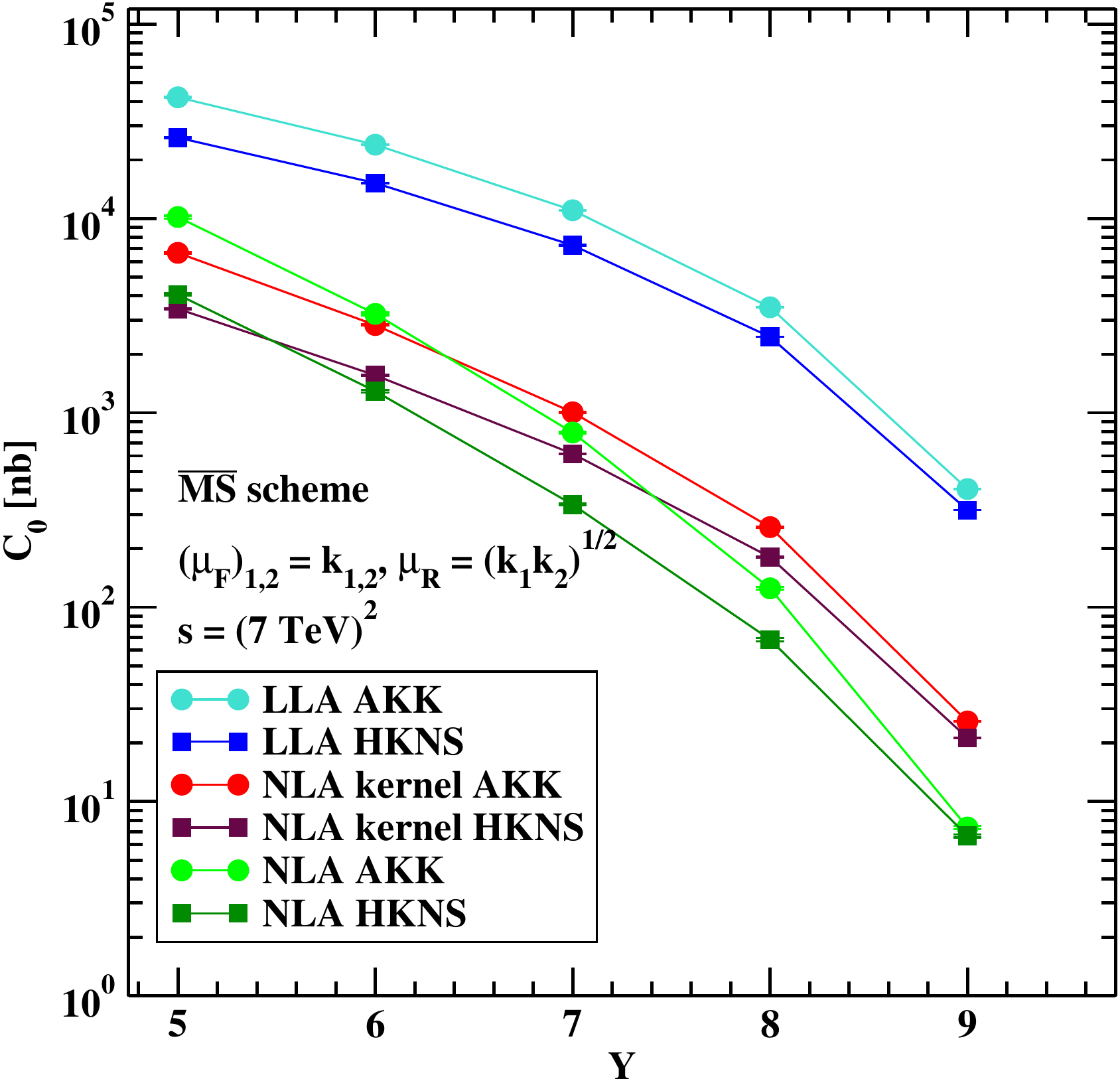}

 \caption[Full NLA predictions of $C_0$ for dihadron production 
          in the $\overline{\rm MS}$ scheme]
 {$Y$-dependence of $C_0$ in the $\overline{\rm MS}$ scheme 
  (as implemented in Eq.~(\ref{Cm})) at ``natural'' scales for
  $\mu_R$ and $\mu_F$, $\sqrt{s} = 7, 13$ TeV, and in the two cases 
  of $Y \leq 4.8$ and $Y \leq 9.4$.  Here and in the following figure
  captions ``LLA'' means pure leading logarithmic approximation, 
  ``NLA kernel'' means inclusion of the NLA corrections 
  from the kernel only, ``NLA'' stands for full inclusion 
  of NLA corrections, {\it i.e.} both from
  the kernel and the hadron vertices.}
 \label{fig:C0MSbNS}
 \end{figure}

 In Fig.~\ref{fig:C0MSbNS} we show our results for $C_0$ in the
 $\overline{\rm MS}$ scheme (as implemented in Eq.~(\ref{Cm})) for 
  we already specified above the scale settings  $\sqrt{s} = 7, 13$ TeV,
 and in the two cases of $Y \leq 4.8$ and $Y \leq 9.4$.
 We clearly see that NLA corrections become negative 
 with respect to the LLA prediction when $Y$ grows.
 Besides, it is interesting to note that the full NLA approach predicts larger
 values for the cross sections in comparison to the case where only NLA
 corrections to the BFKL kernel are taken into account. It means that the
 inclusion into the analysis of the NLA corrections to the hadron vertices
 makes the predictions for the cross sections somewhat bigger and partially
 compensates the large negative effect from the NLA corrections to the BFKL
 kernel.
 
\begin{figure}[H]
 \centering

   \includegraphics[scale=0.38]{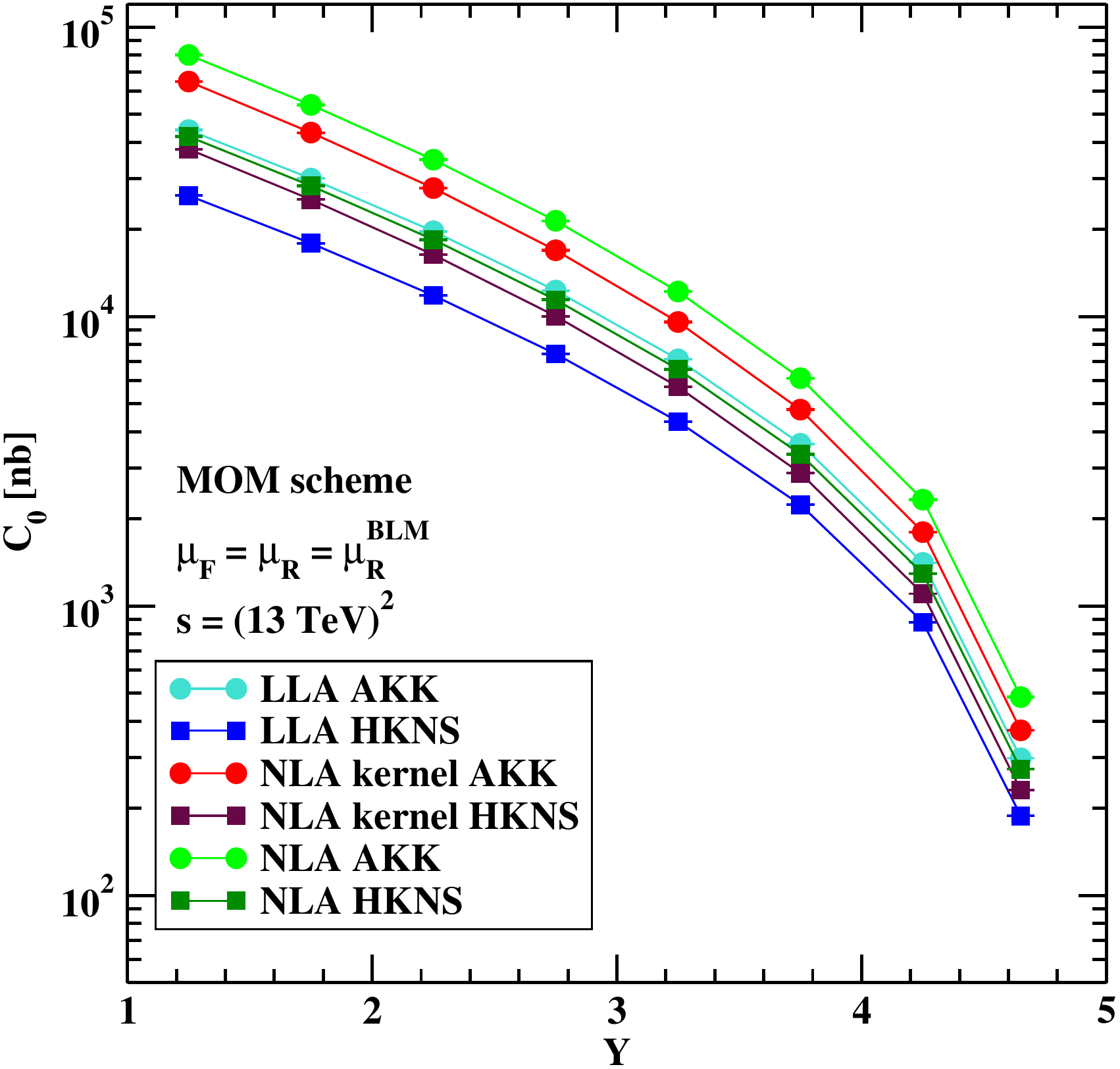}
   \includegraphics[scale=0.38]{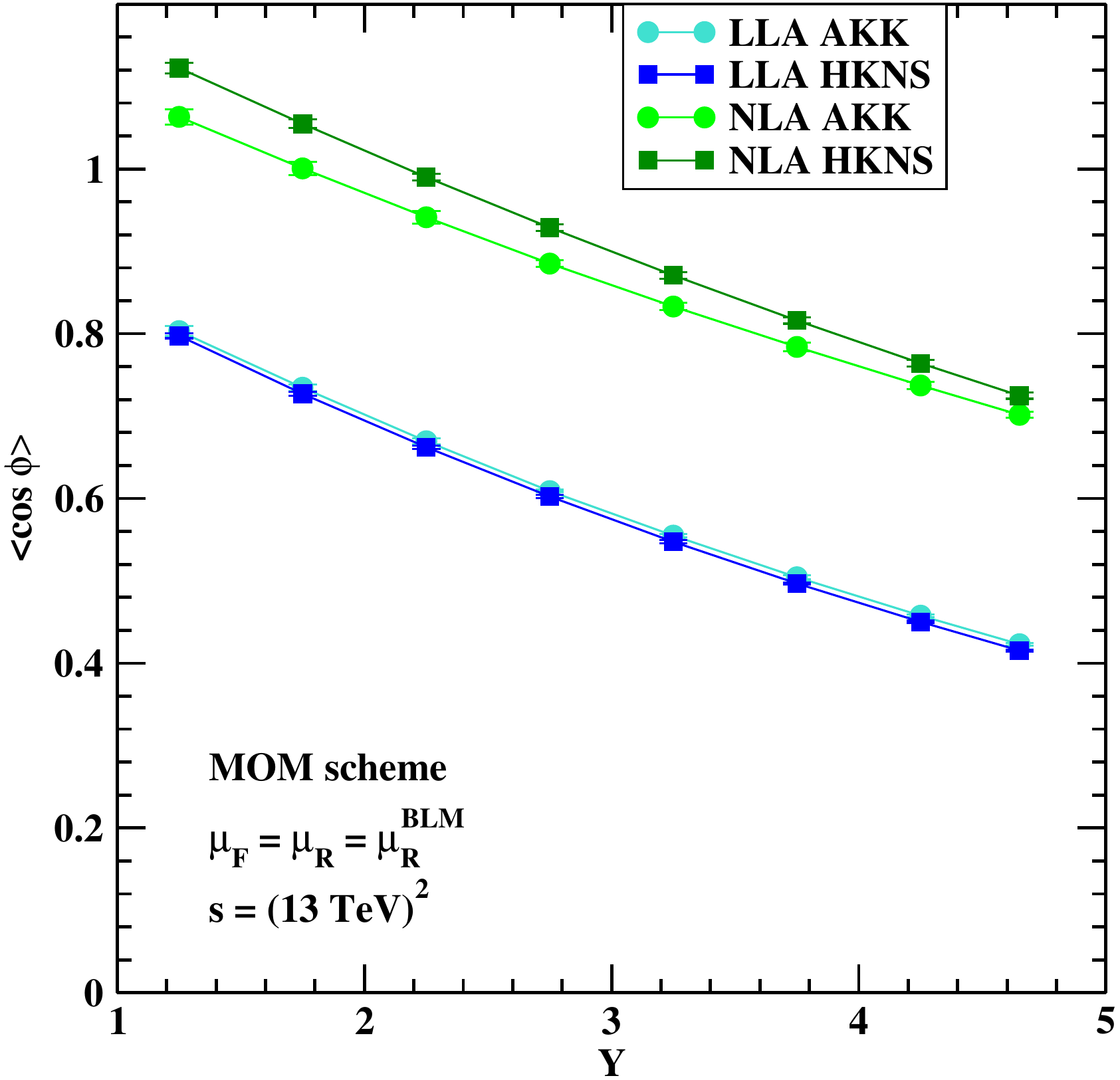}

   \includegraphics[scale=0.38]{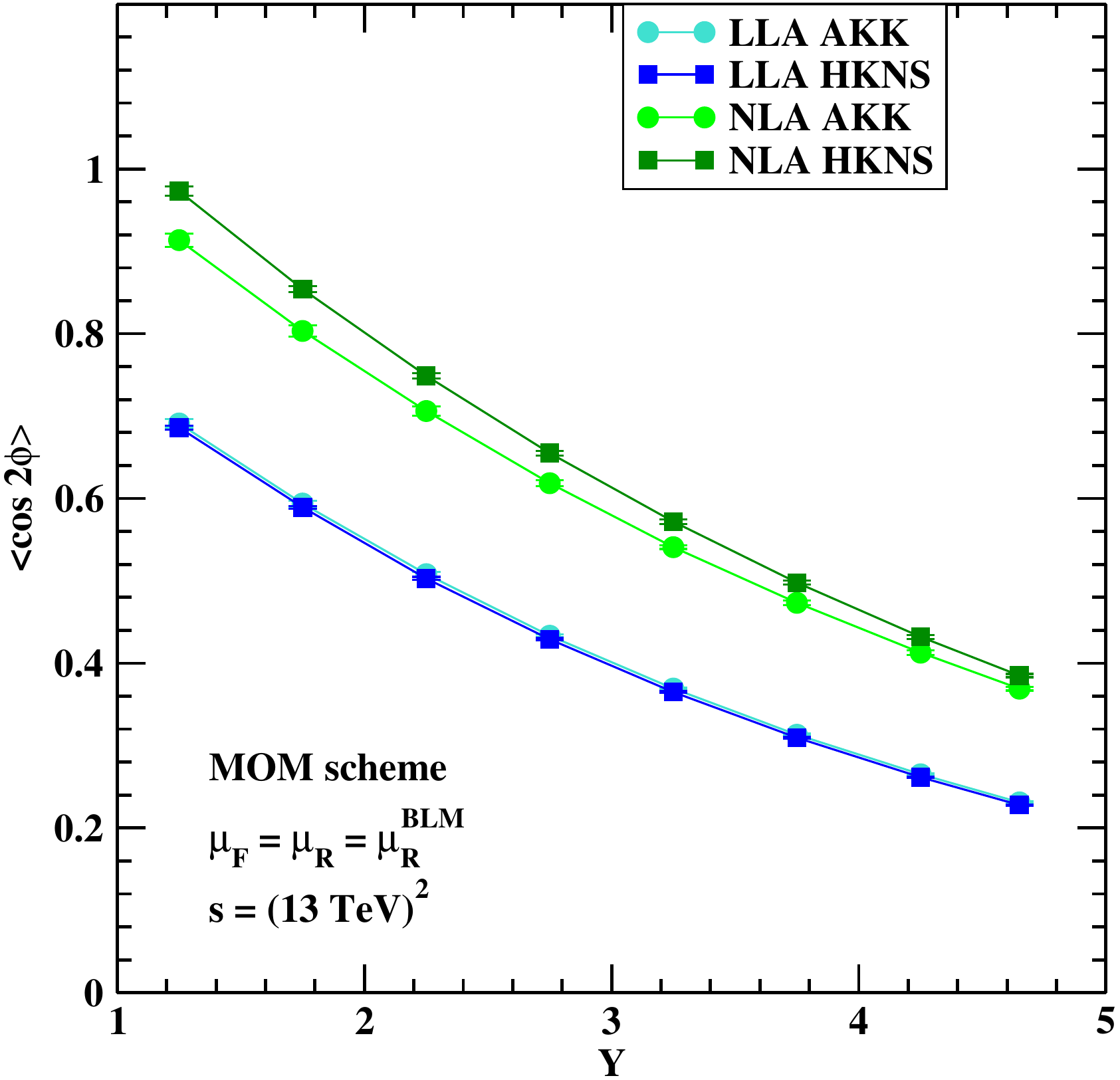}
   \includegraphics[scale=0.38]{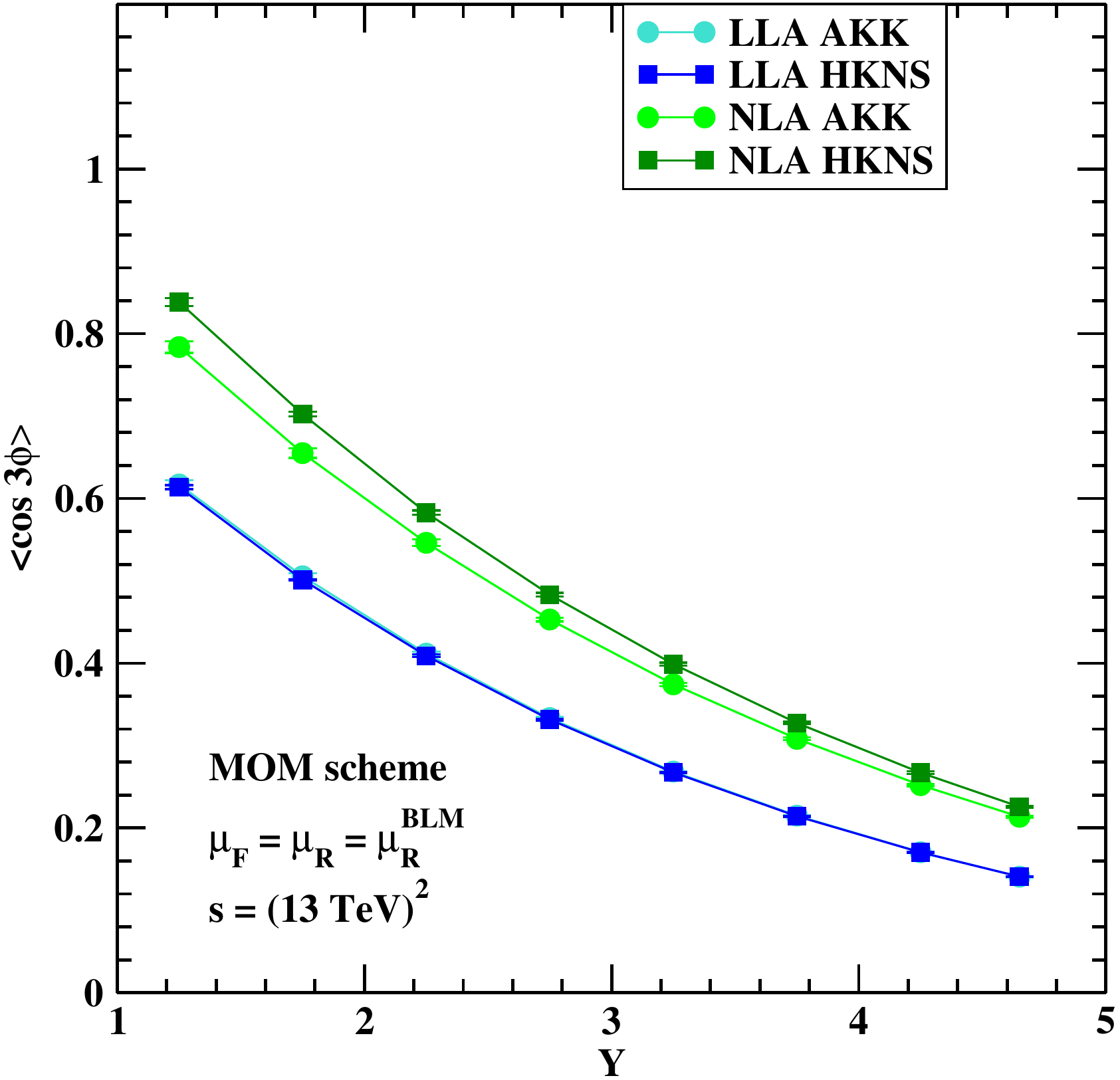}

   \includegraphics[scale=0.38]{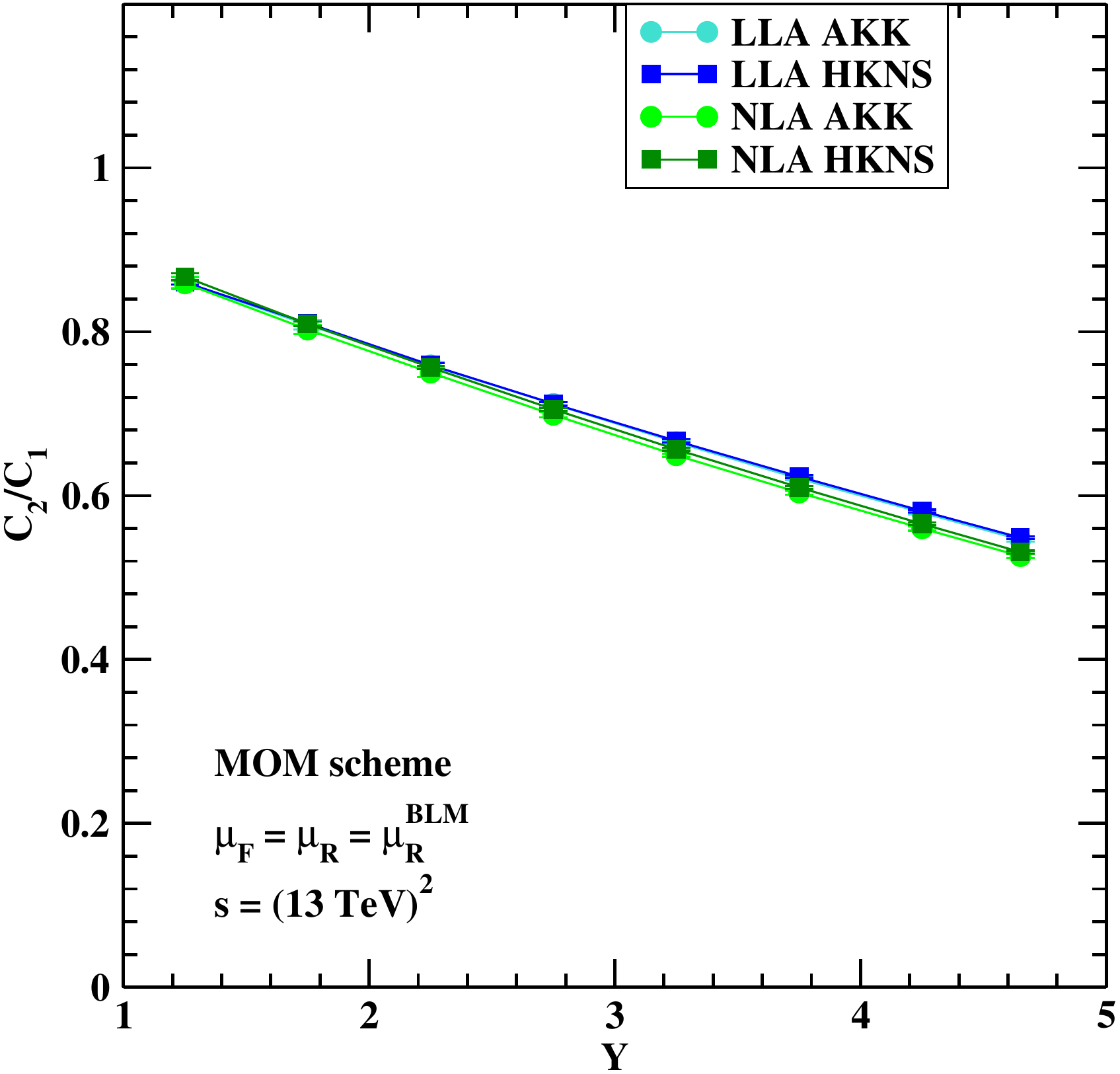}
   \includegraphics[scale=0.38]{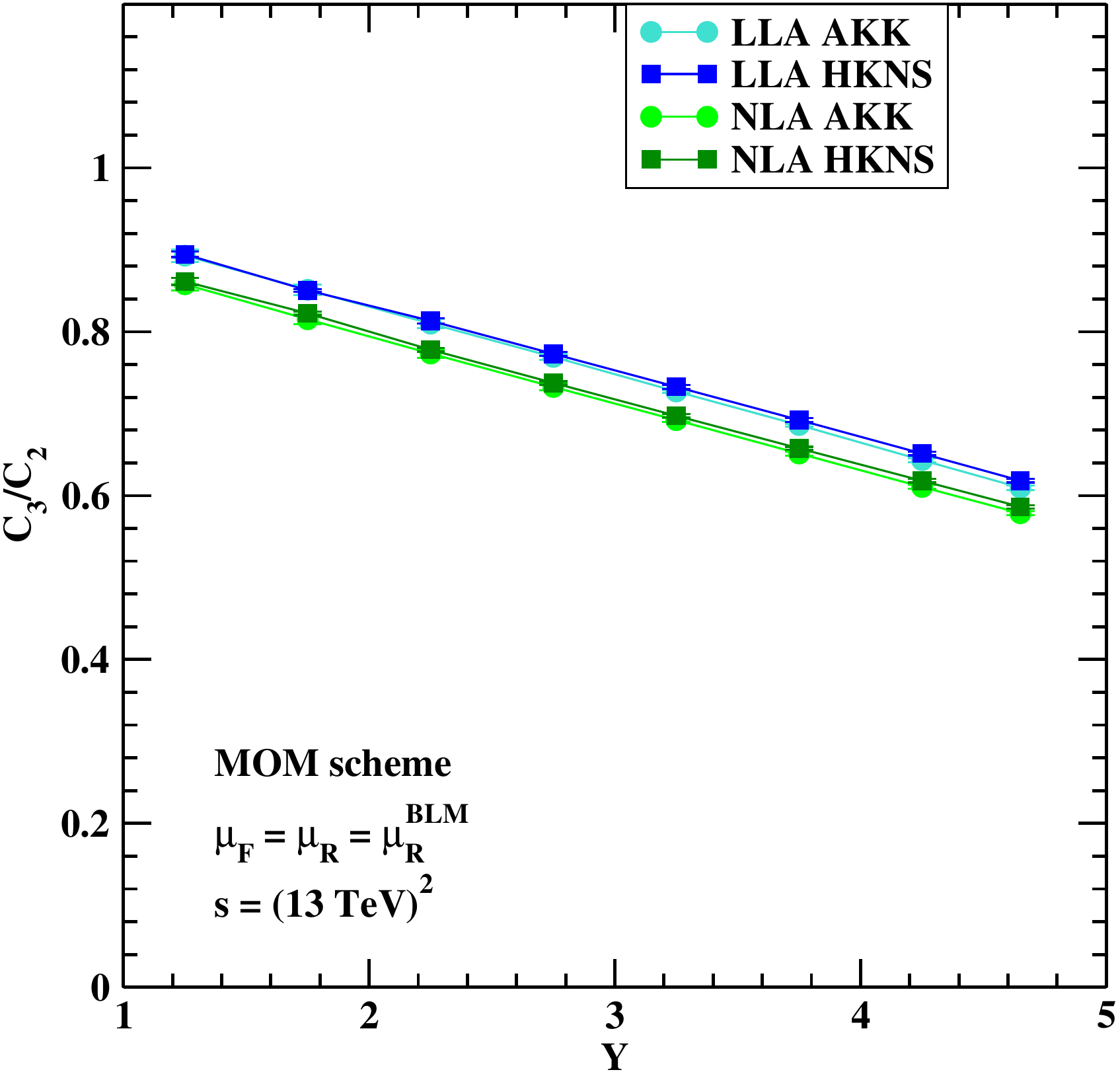}
 \caption[Full NLA predictions for dihadron production 
          for $\mu_F = \mu^{\rm BLM}_R$, $\sqrt{s} = 13$ TeV, 
          and $Y \leq 4.8$]
 {$Y$-dependence of $C_0$ and of several ratios $C_m/C_n$ for 
  $\mu_F = \mu^{\rm BLM}_R$, $\sqrt{s} = 13$ TeV, and $Y \leq 4.8$.}
 \label{fig:blm13}
 \end{figure}

\begin{figure}[H]
 \centering

   \includegraphics[scale=0.38]{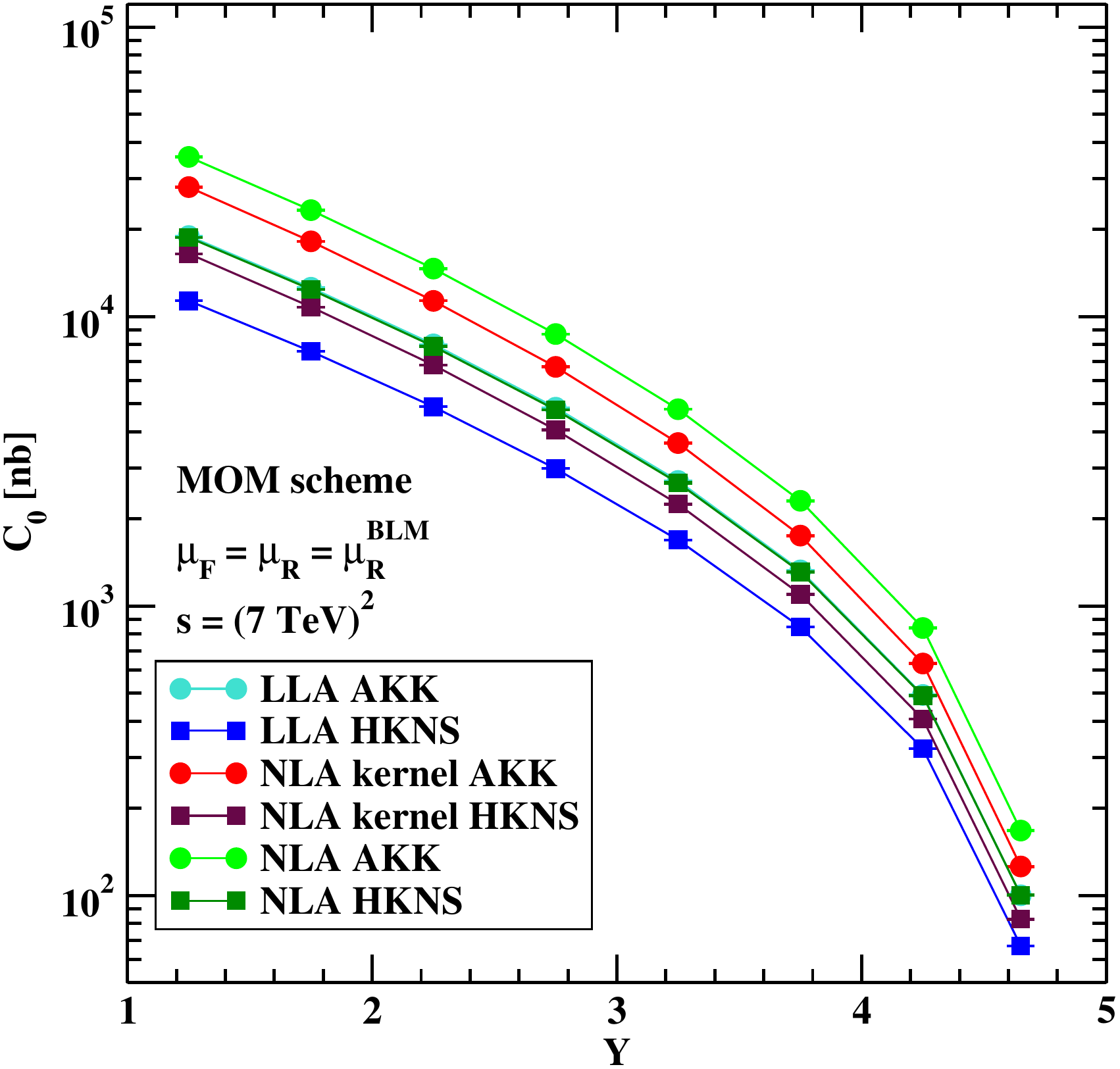}
   \includegraphics[scale=0.38]{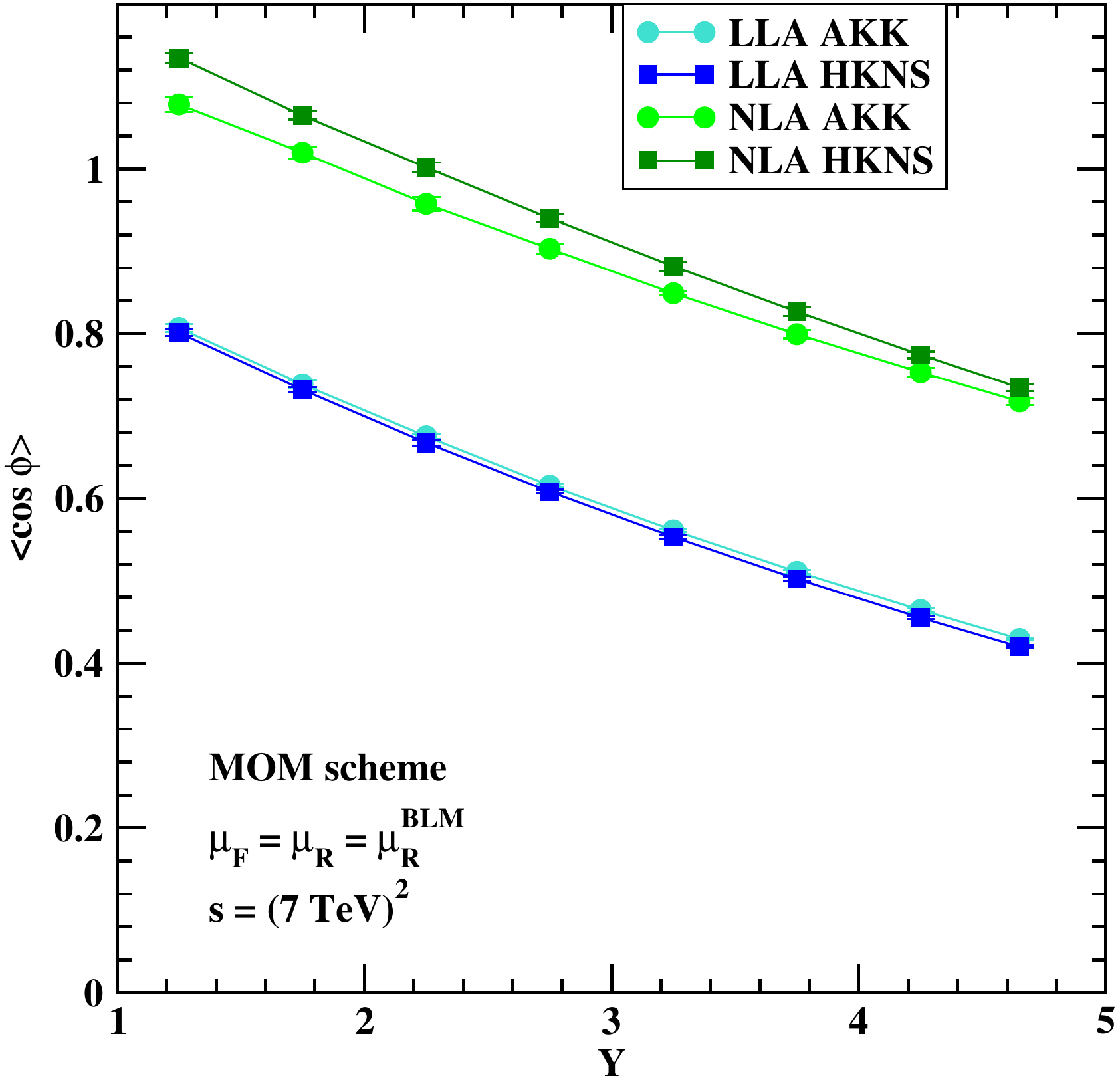}

   \includegraphics[scale=0.38]{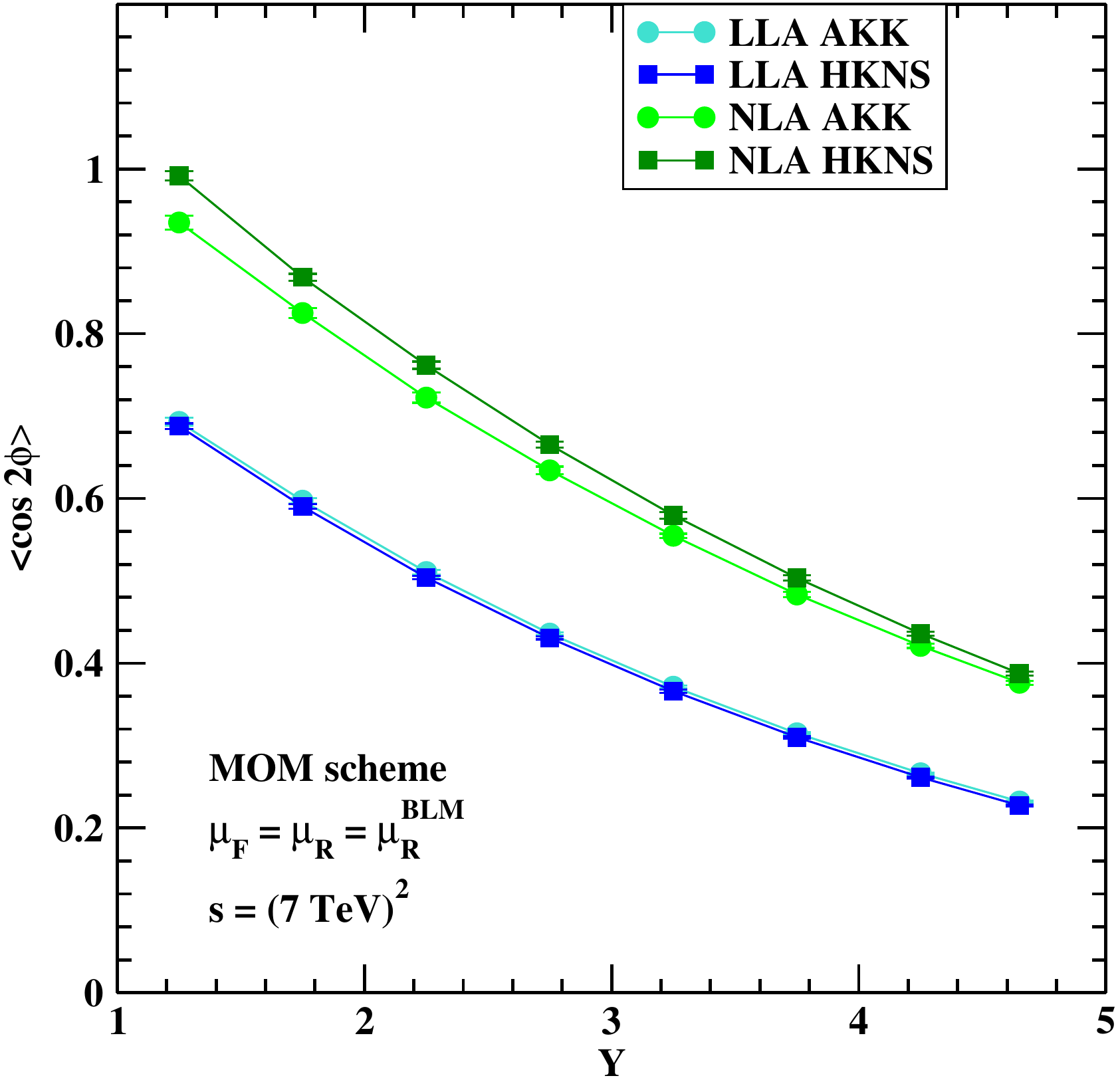}
   \includegraphics[scale=0.38]{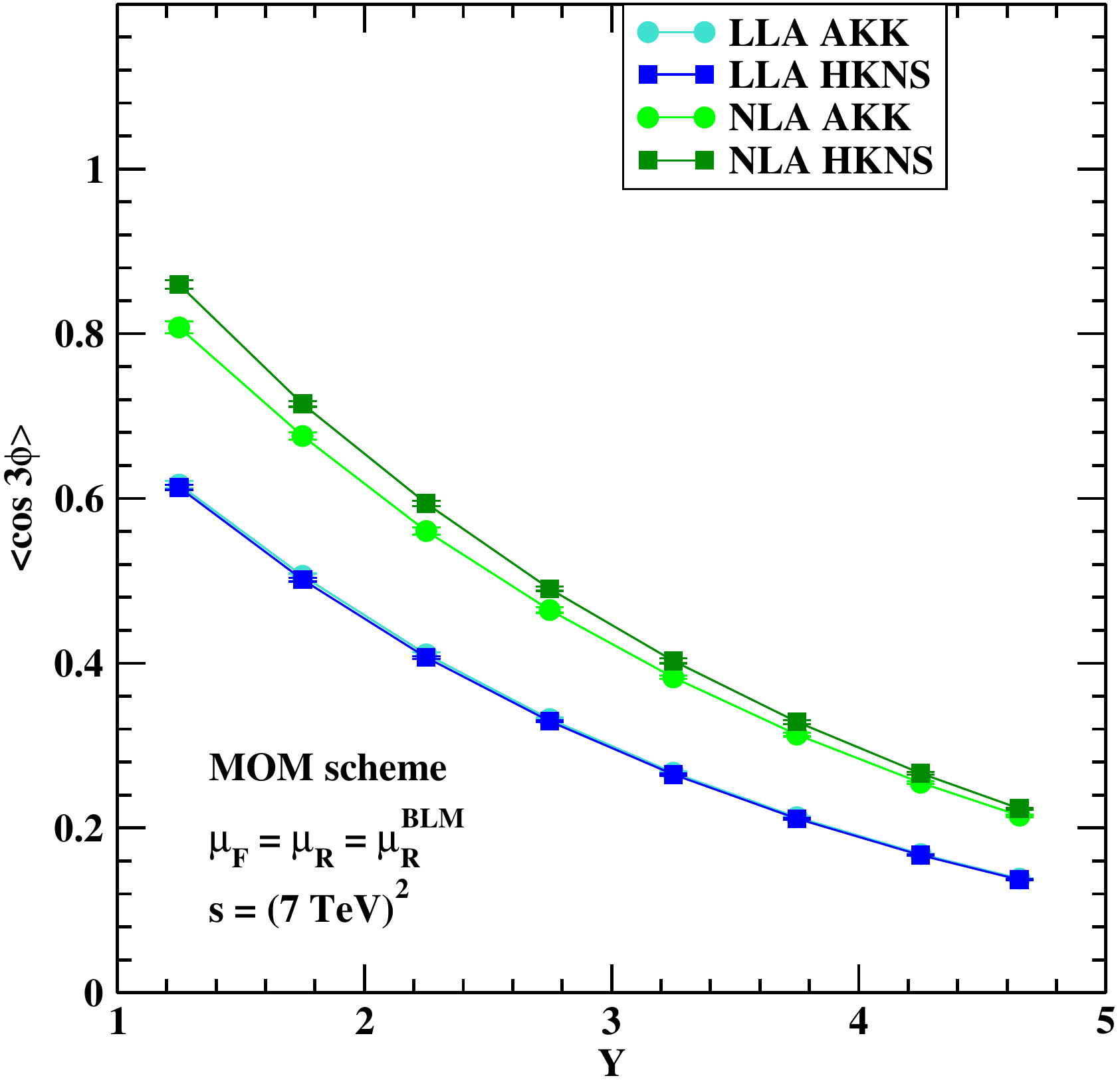}

   \includegraphics[scale=0.38]{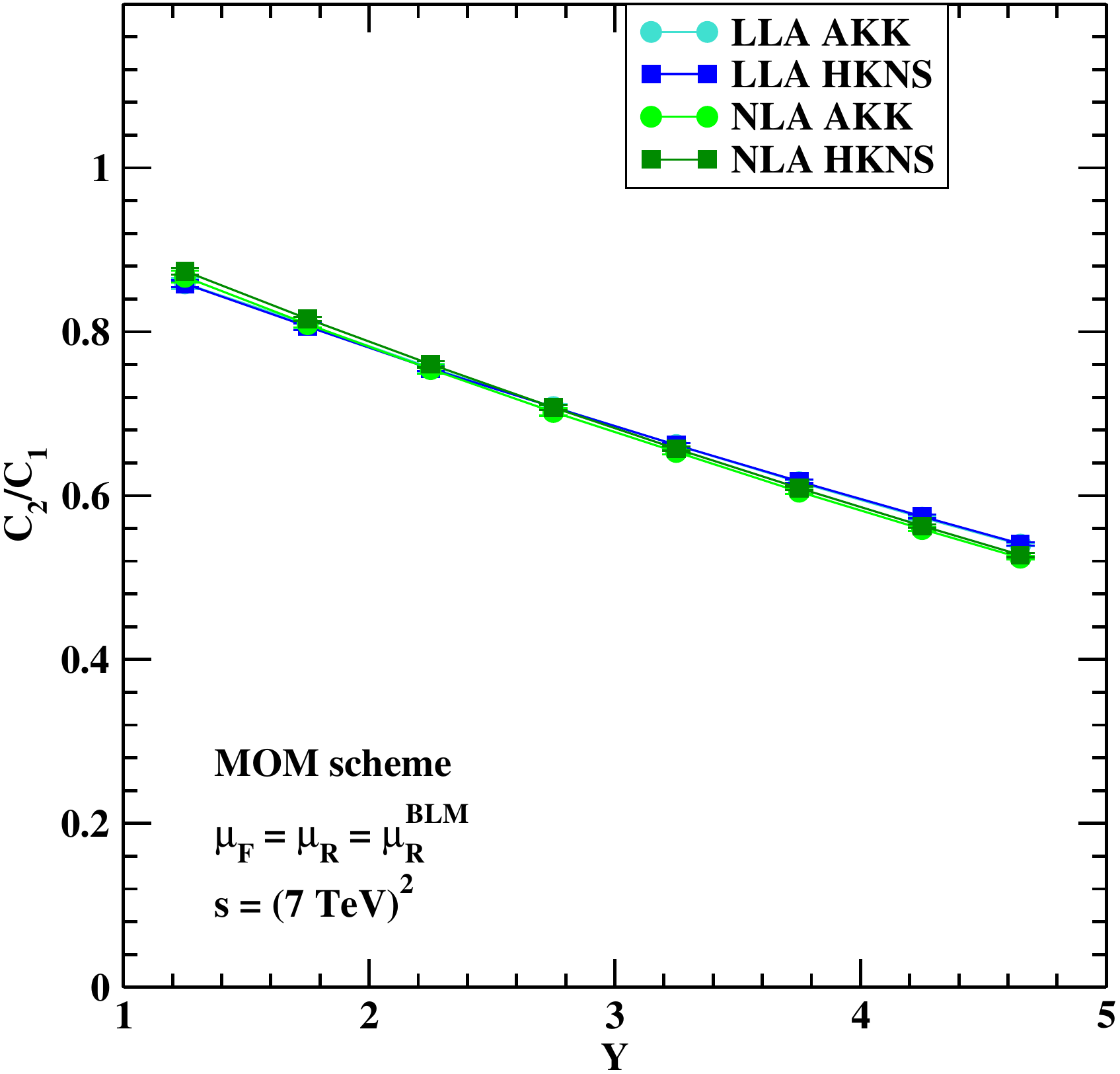}
   \includegraphics[scale=0.38]{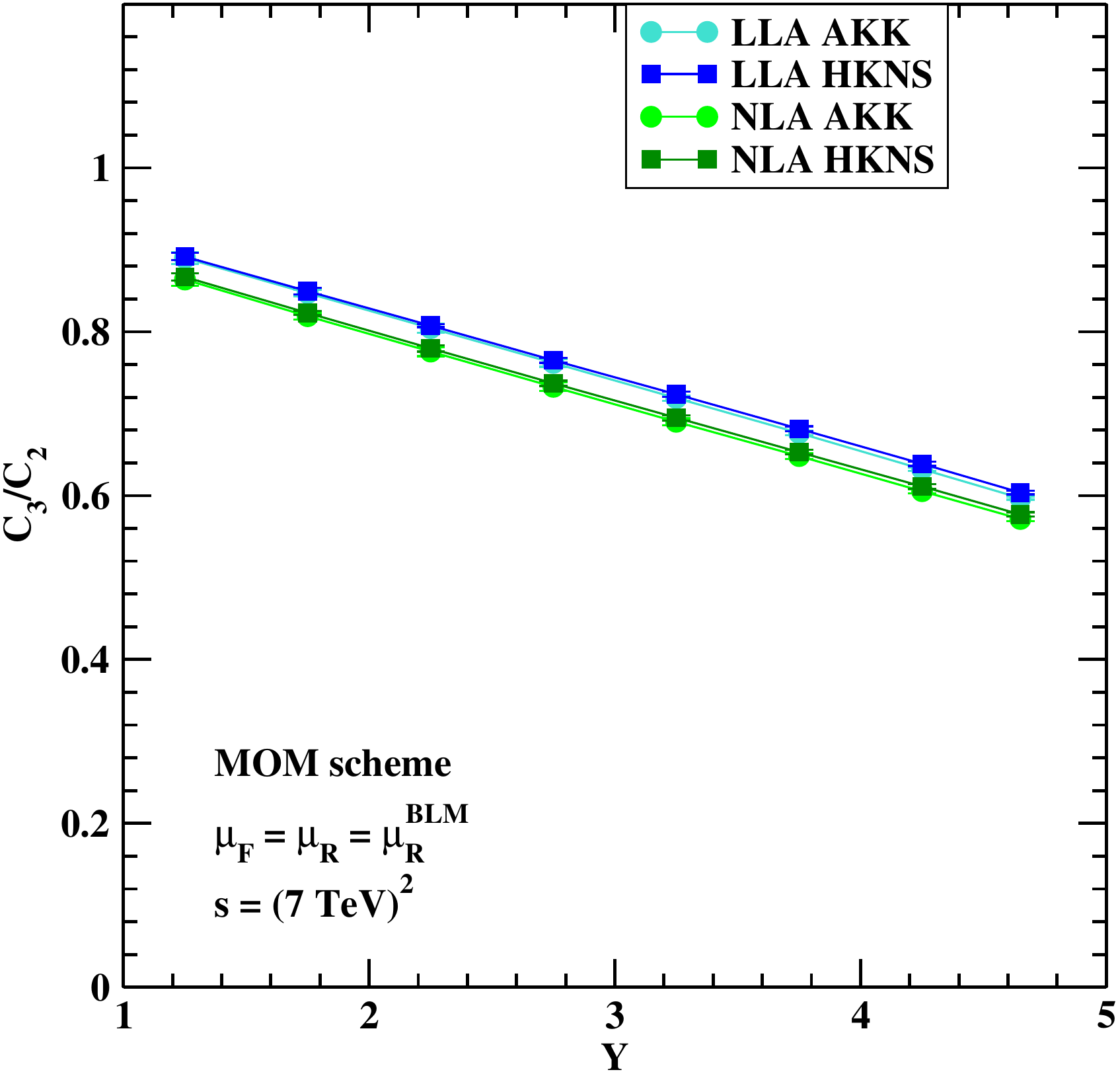}
 \caption[Full NLA predictions for dihadron production 
          for $\mu_F = \mu^{\rm BLM}_R$, $\sqrt{s} = 7$ TeV, 
          and $Y \leq 4.8$]
 {$Y$-dependence of $C_0$ and of several ratios $C_m/C_n$ for 
  $\mu_F = \mu^{\rm BLM}_R$, $\sqrt{s} = 7$ TeV, and $Y \leq 4.8$.}
 \label{fig:blm7}
 \end{figure}

 The other results we presented below are obtained using BLM in the MOM scheme,
 as it is given in Eq.~(\ref{dh-eq}). 
 In Figs.~\ref{fig:blm13}~and~\ref{fig:blm7} we present our results 
 for $C_0$ and for several ratios $C_m/C_n$ at $\sqrt{s}=13$ and $7$ TeV,
 respectively; $\mu_F$ is set equal to $\mu^{\rm BLM}_R$,
 while $Y \leq 4.8$. It is worth to note that in this case the NLA corrections 
 to $C_0$ are positive, so they increase the value 
 of the $\phi$-averaged cross section at all values of $Y$. 
 This is the result of the combination of two distinct effects:
 on one side, we already saw in Ref.~\cite{Celiberto:2016hae} 
 that changing the renormalisation scheme produces 
 a non-exponentiated extra factor in Eq.~(\ref{dh-eq}) proportional
 to $T^{\rm conf}$, and that is positive. On the other side, we found that 
 the $C_{gg}$ coefficient in Eq.~(\ref{c11-dh})
 gives a large and positive contribution to the NLO impact factor. 
 We see also that NLA corrections increase the azimuthal correlations: 
 $C_1/C_0$, $C_2/C_0$, and $C_3/C_0$, while their effect is small with respect
 to LLA predictions in their ratios, $C_2/C_1$ and $C_3/C_2$. 
 The value of $C_1/C_0$ for $Y \leq 2.75$ in some cases exceeds $1$.
 We consider this as an effect due to the fact that, at very small $Y$,
 which corresponds to the small values of partonic subenergies $\hat s$, 
 we are crossing the applicability limit of the BFKL approach, which
 systematically neglects any contributions that are suppressed by the powers of 
 $\hat s$.  
 
\begin{figure}[H]
 \centering

   \includegraphics[scale=0.38]{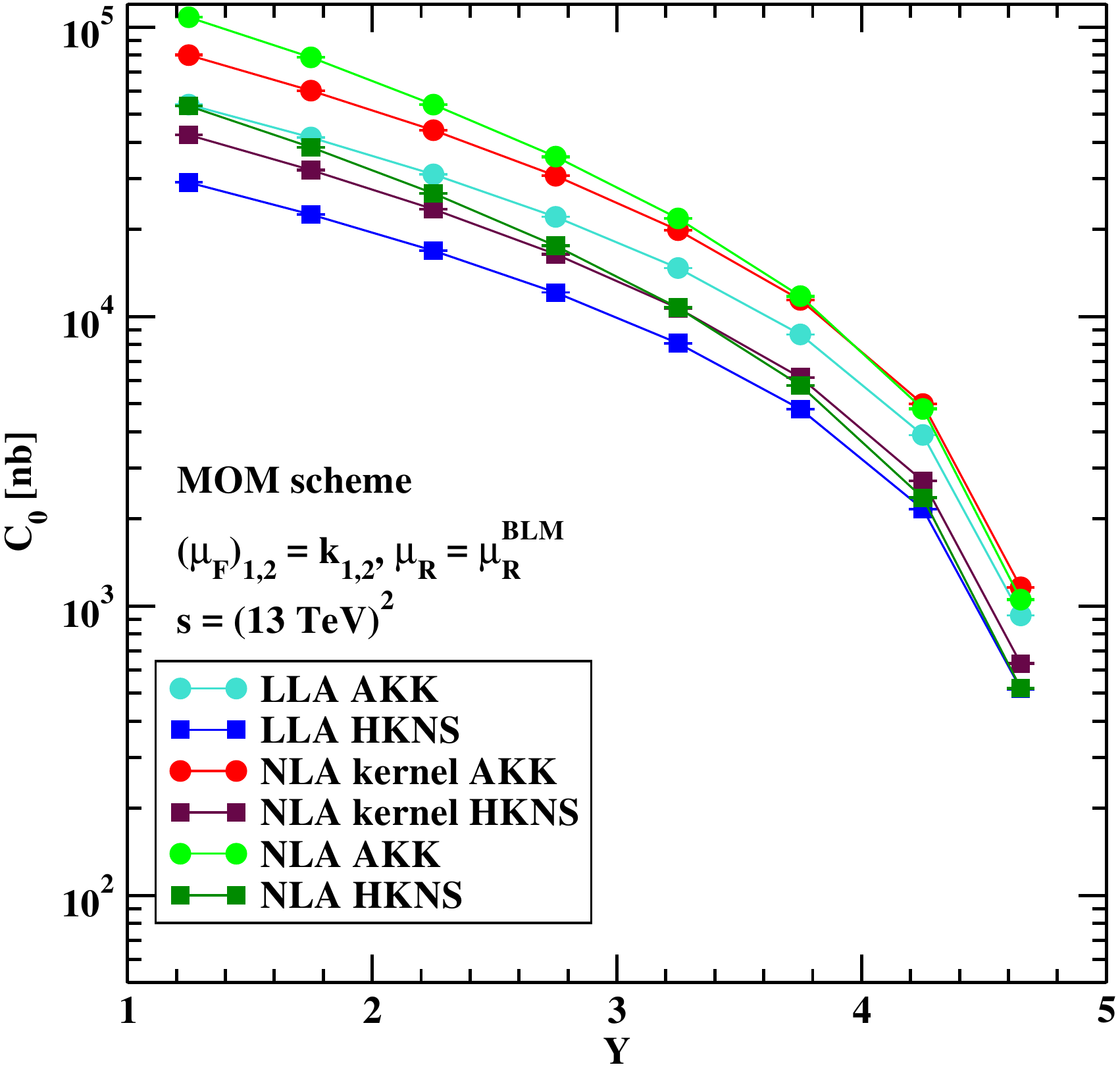}
   \includegraphics[scale=0.38]{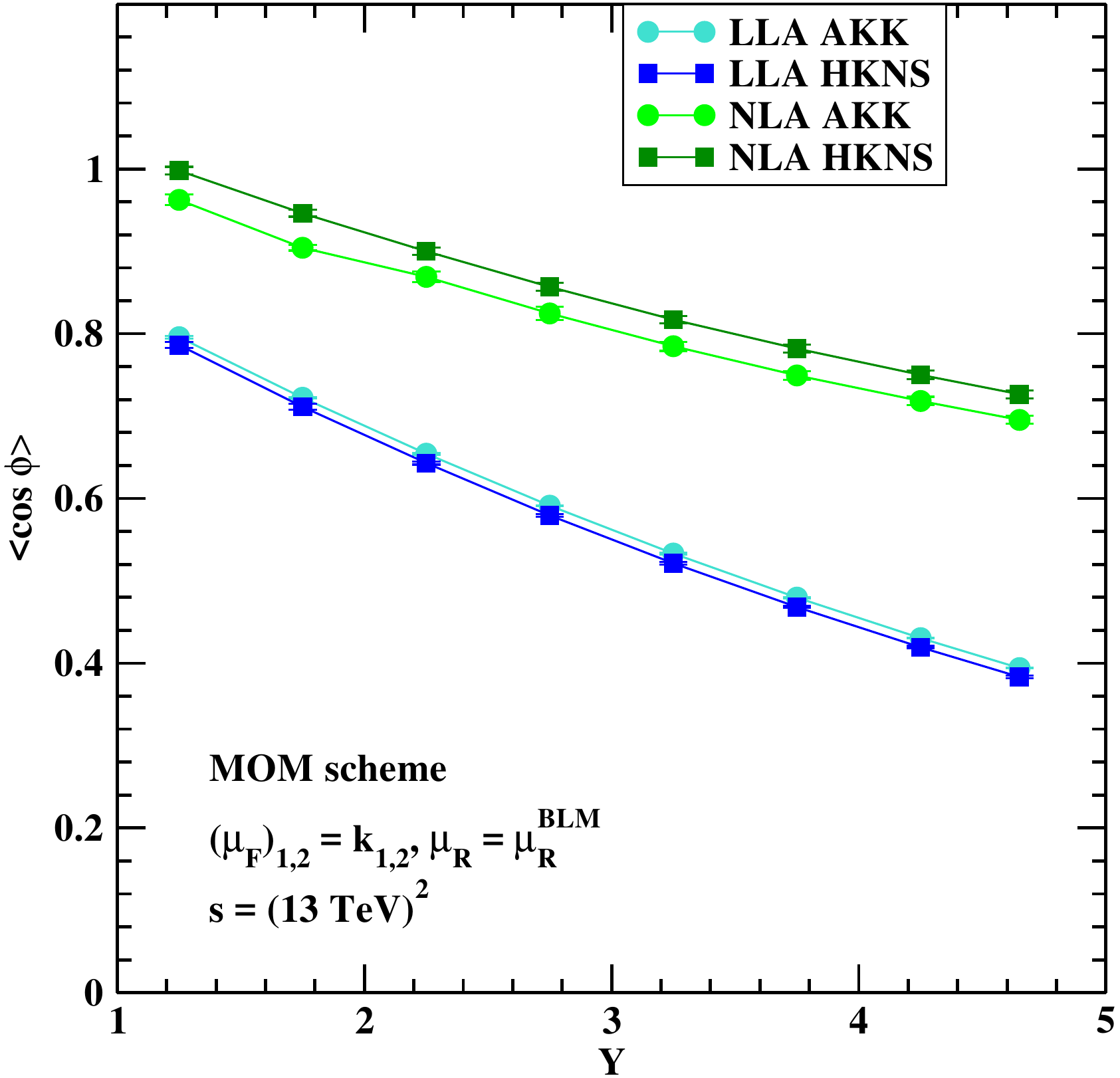}

   \includegraphics[scale=0.38]{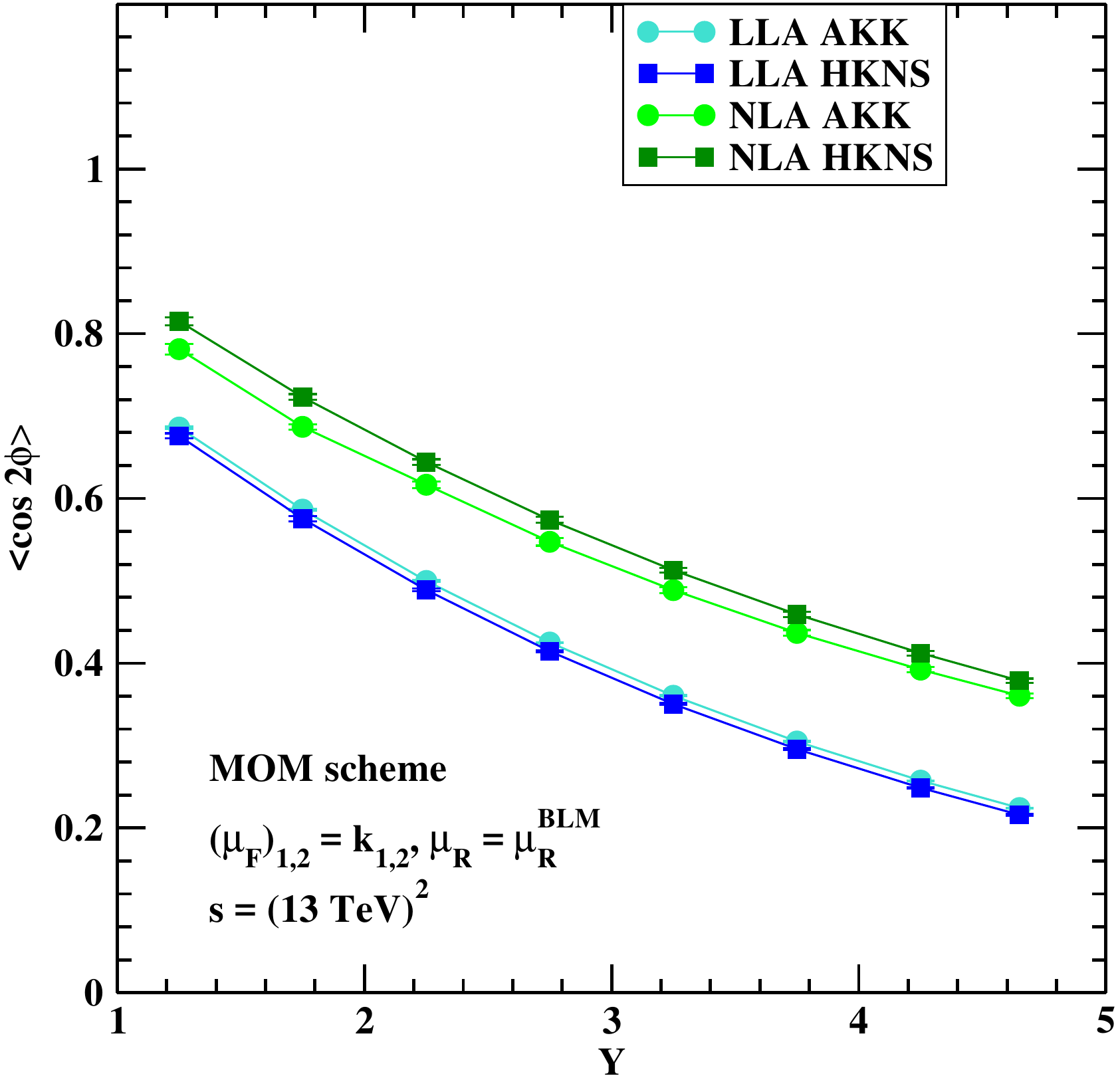}
   \includegraphics[scale=0.38]{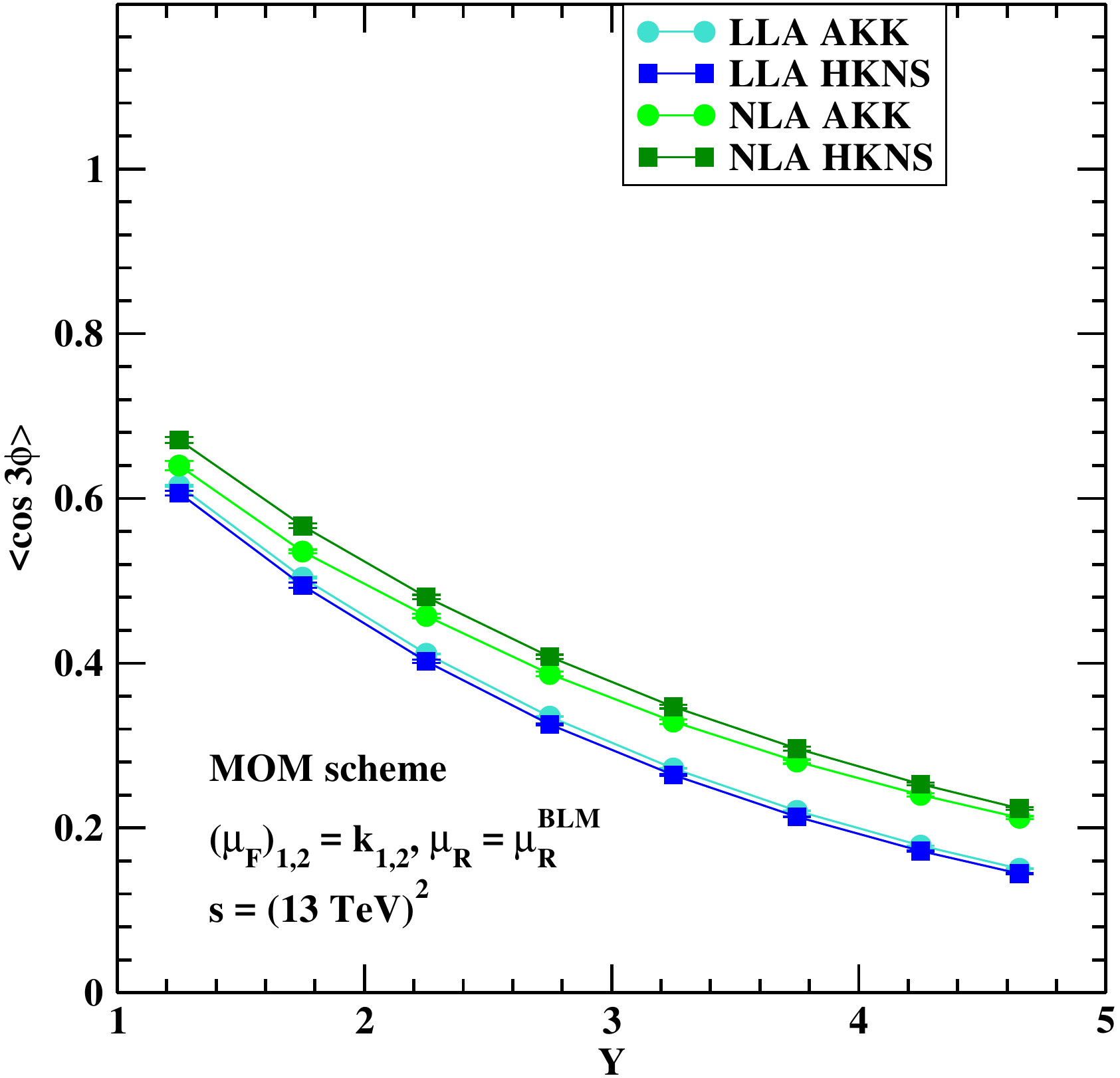}

   \includegraphics[scale=0.38]{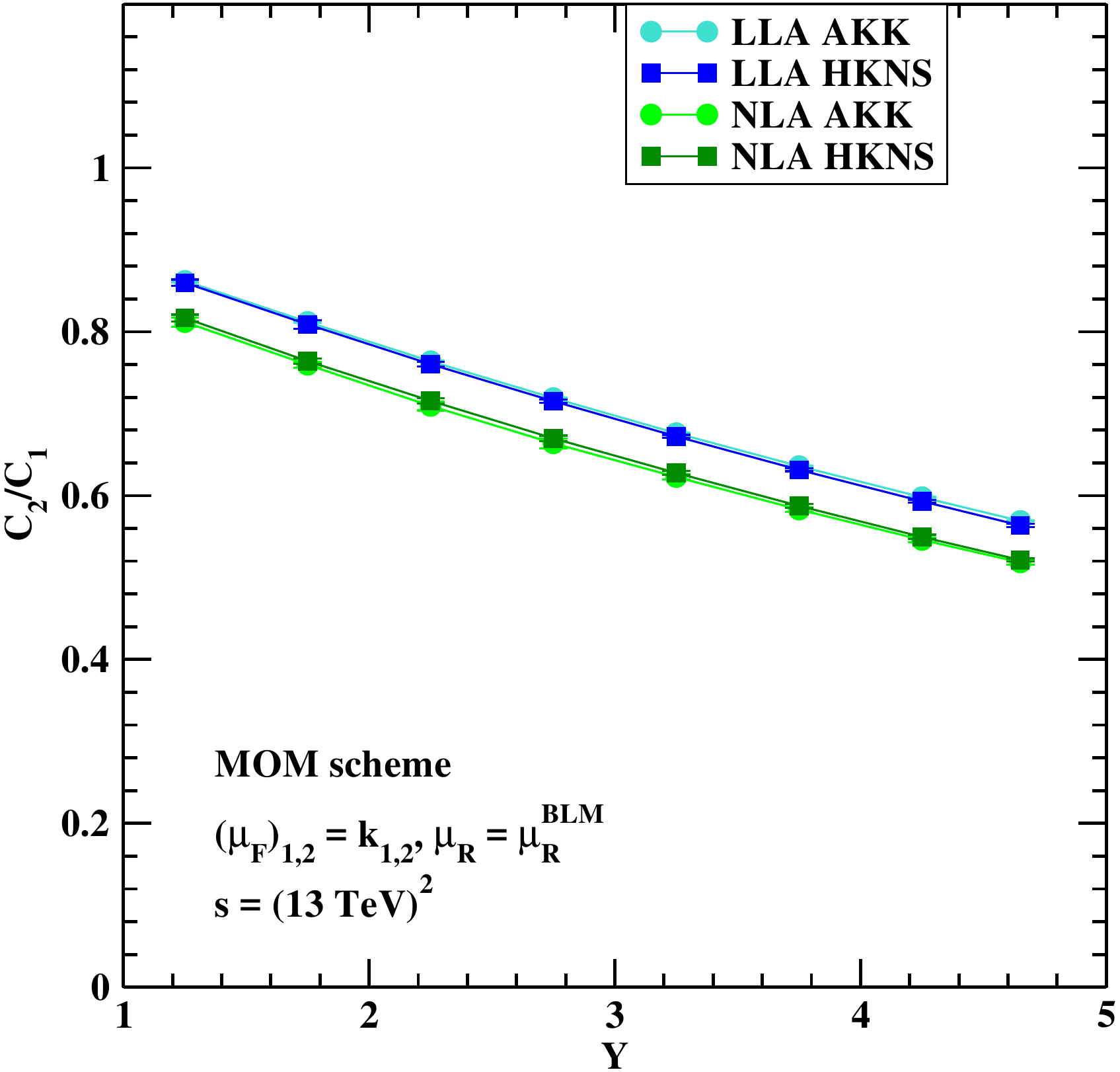}
   \includegraphics[scale=0.38]{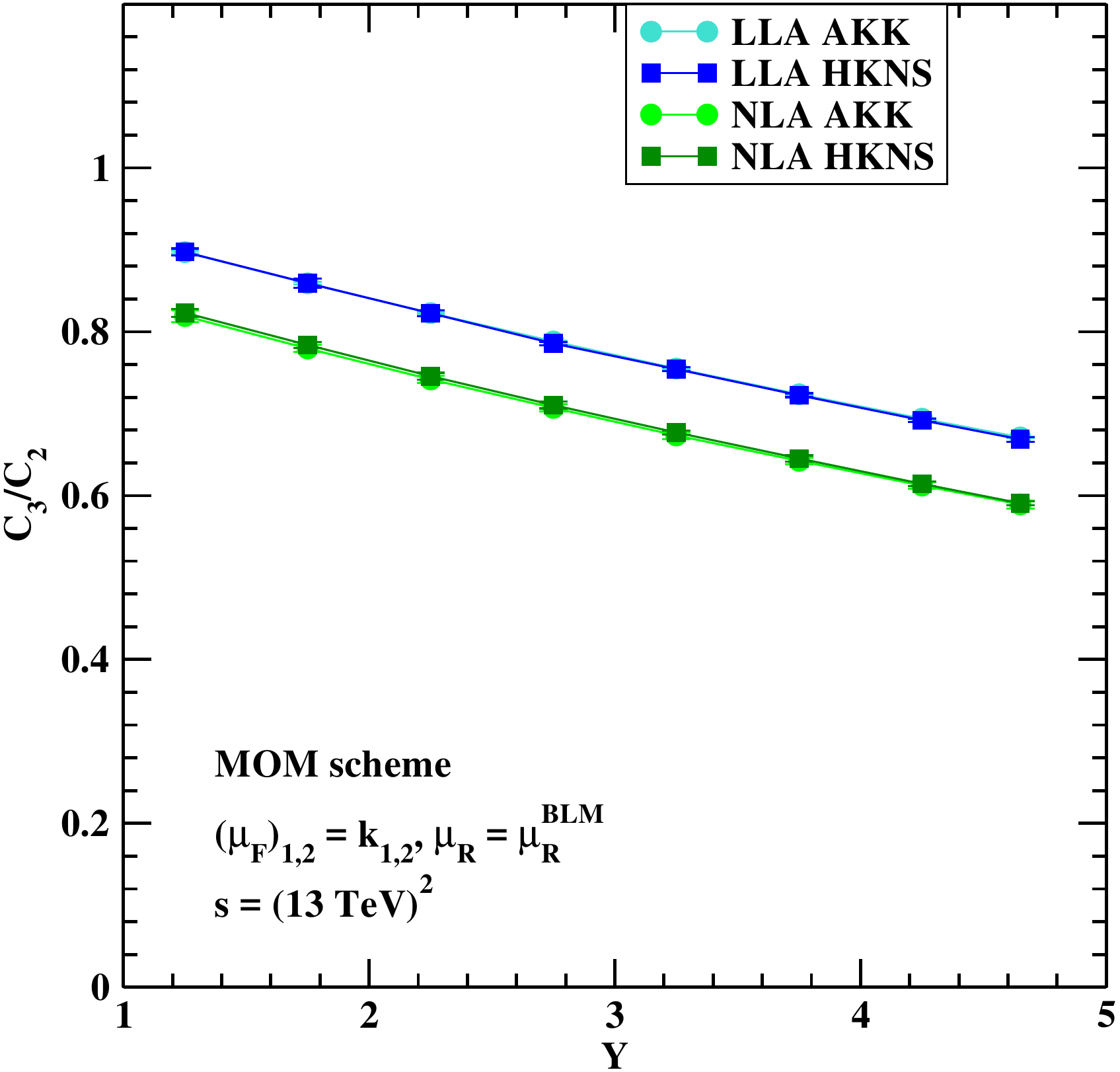}
 \caption[Full NLA predictions for dihadron production 
          for $(\mu_F)_{1,2} = |\vec k_{1,2}|$, $\sqrt{s} = 13$ TeV, 
          and $Y \leq 4.8$]
 {$Y$-dependence of $C_0$ and of several ratios $C_m/C_n$ for 
  $(\mu_F)_{1,2} = |\vec k_{1,2}|$, $\sqrt{s} = 13$ TeV, and $Y \leq 4.8$.}
 \label{fig:ns13}
 \end{figure}

\begin{figure}[H]
 \centering

   \includegraphics[scale=0.38]{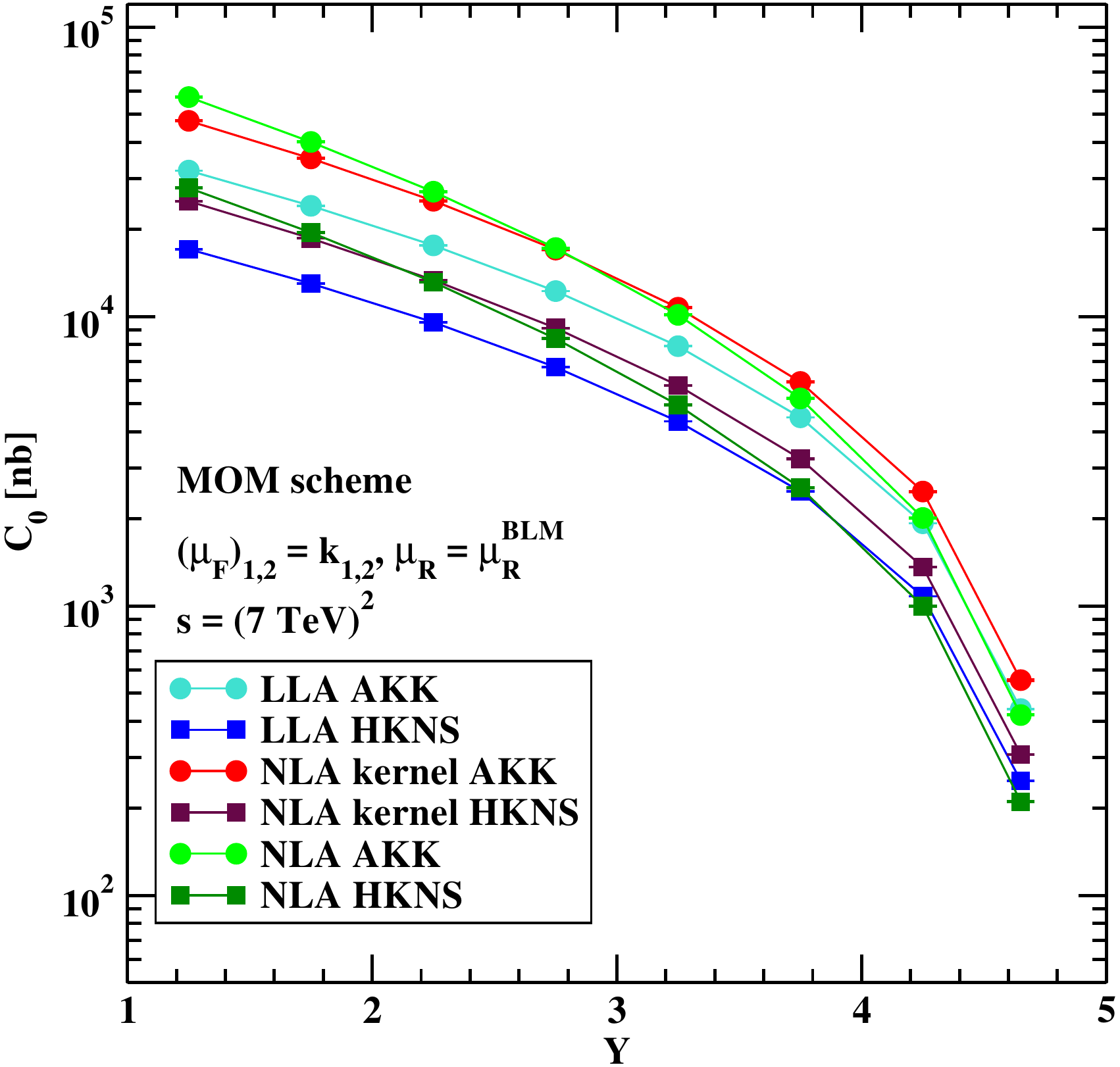}
   \includegraphics[scale=0.38]{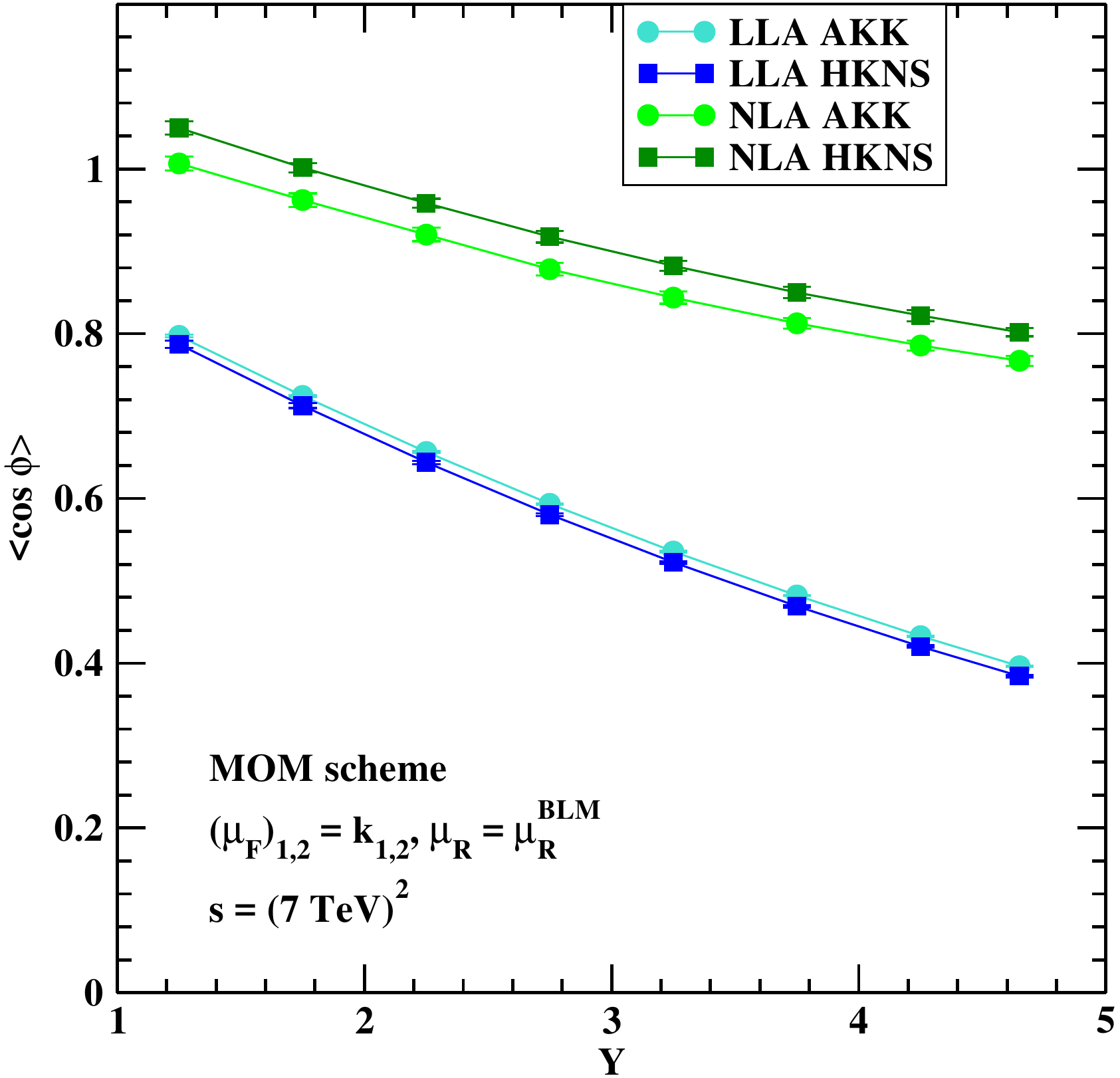}

   \includegraphics[scale=0.38]{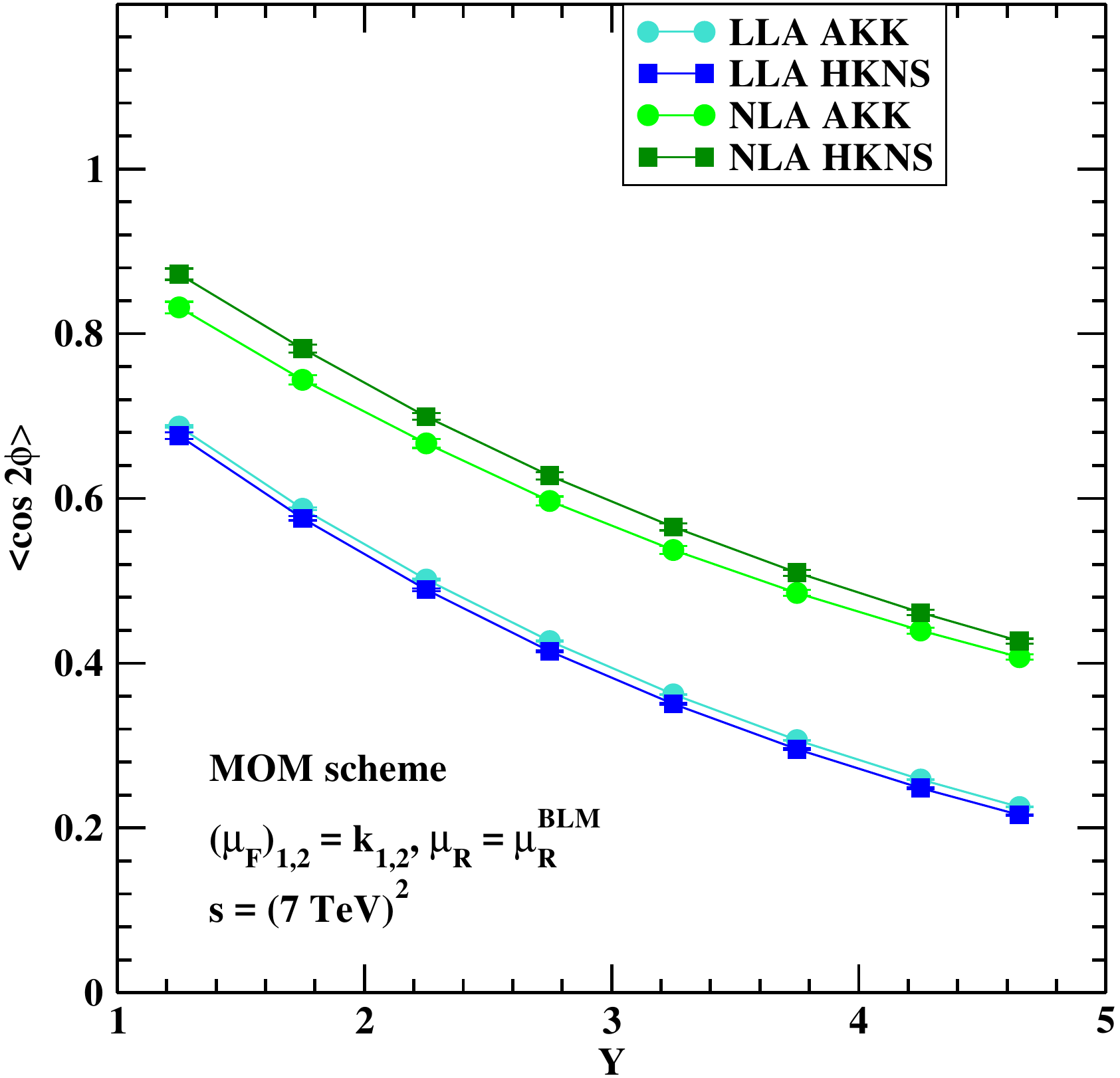}
   \includegraphics[scale=0.38]{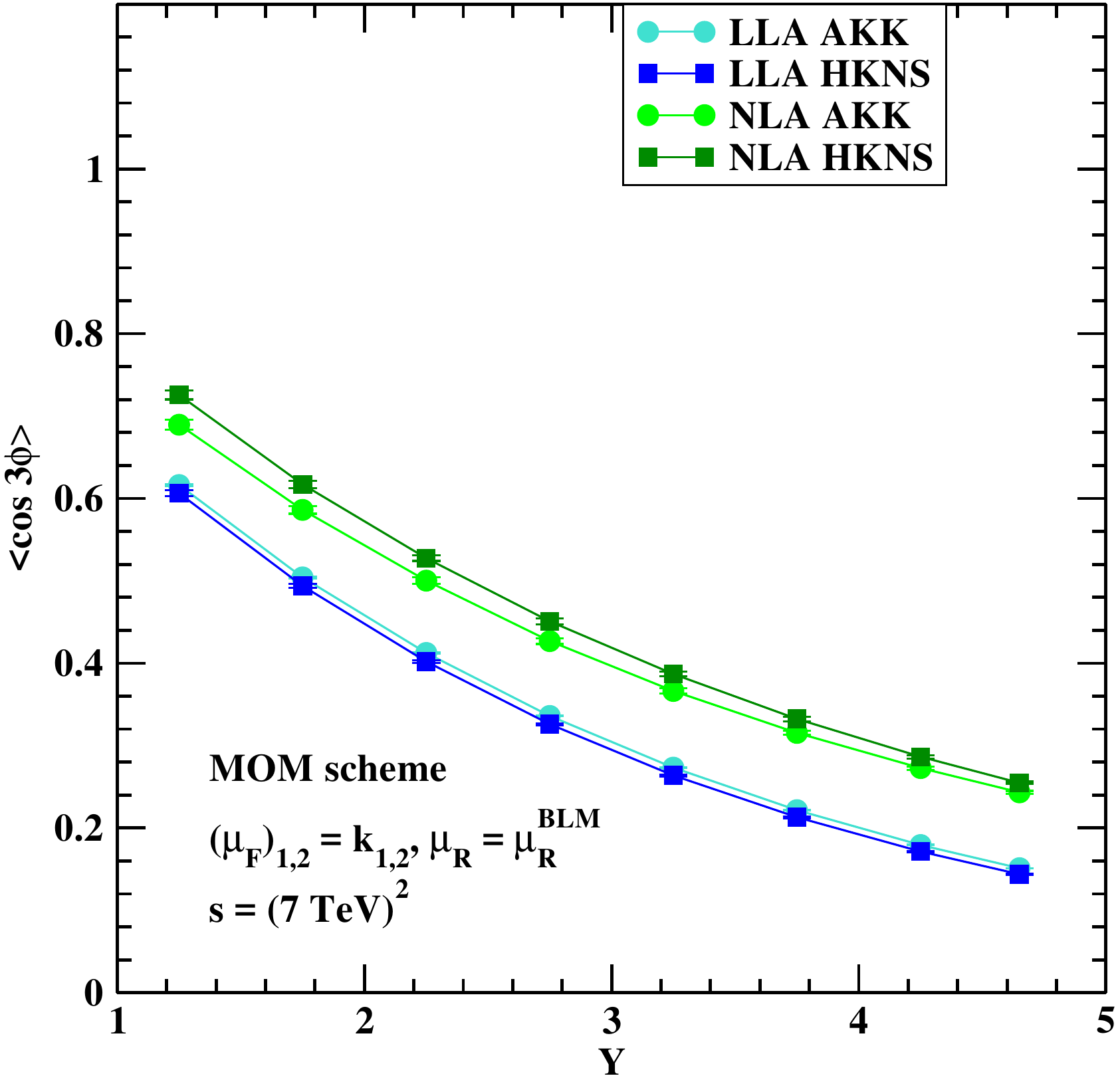}

   \includegraphics[scale=0.38]{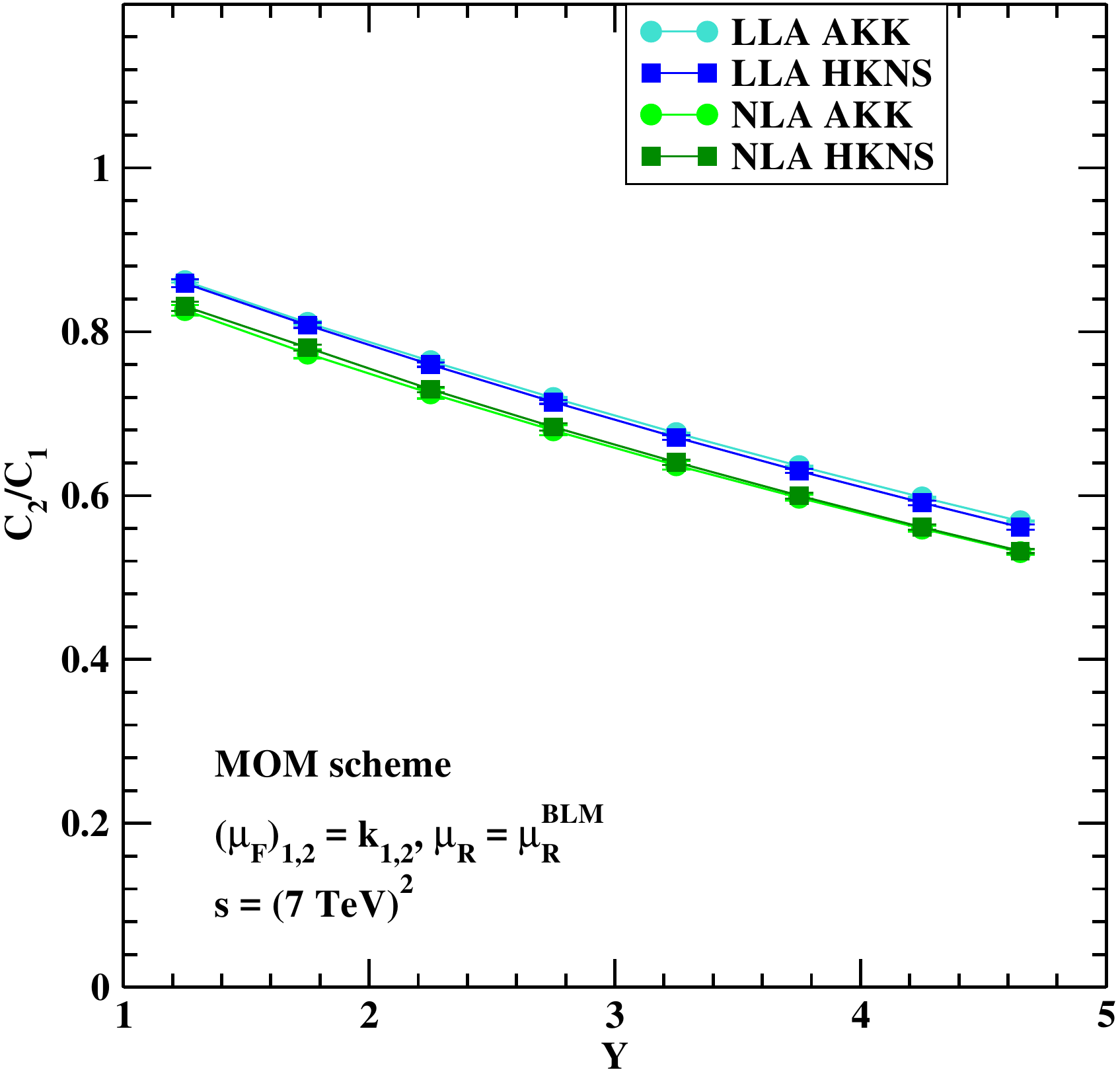}
   \includegraphics[scale=0.38]{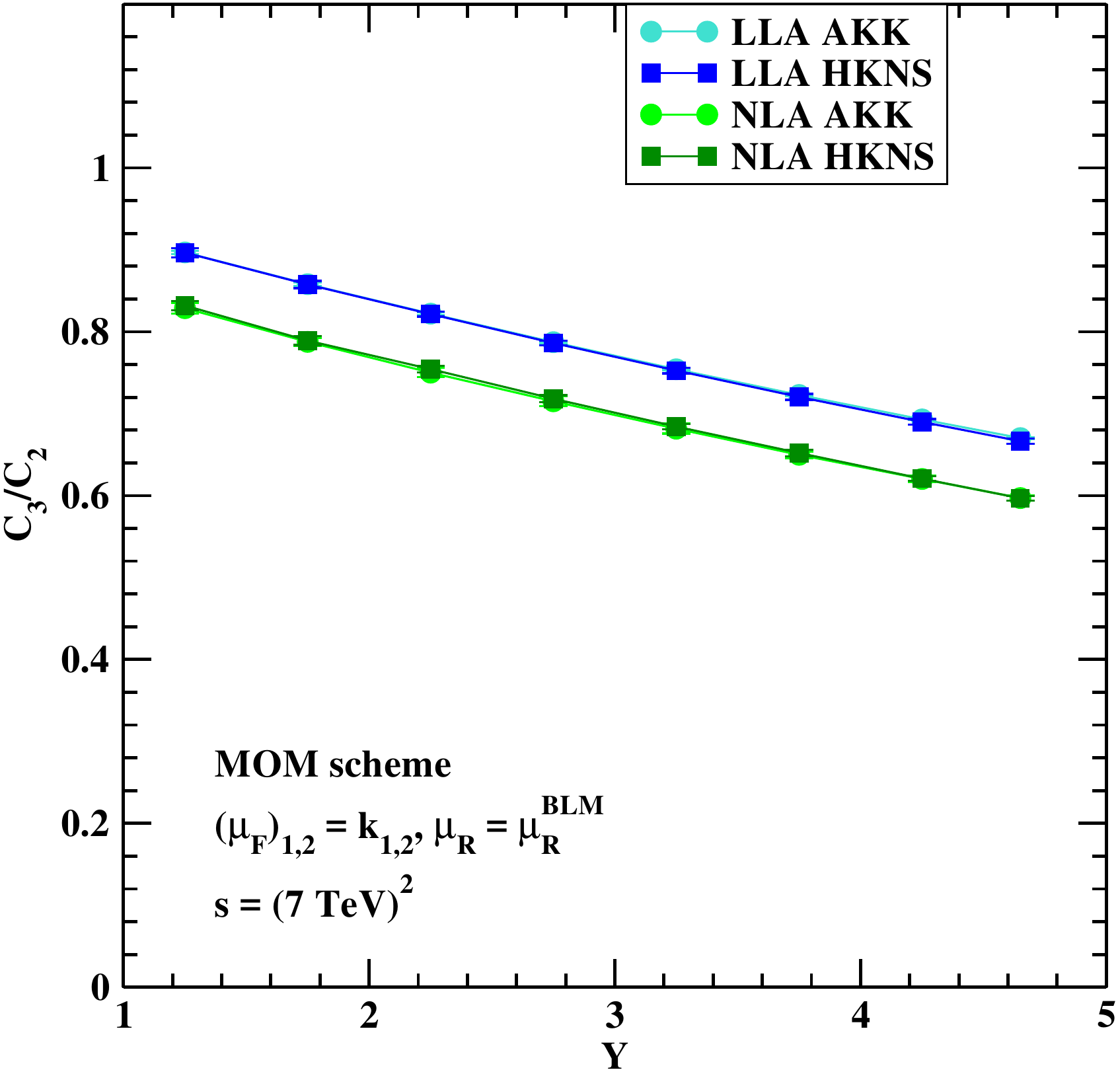}
 \caption[Full NLA predictions for dihadron production 
          for $(\mu_F)_{1,2} = |\vec k_{1,2}|$, $\sqrt{s} = 7$ TeV, 
          and $Y \leq 4.8$]
 {$Y$-dependence of $C_0$ and of several ratios $C_m/C_n$ for 
  $(\mu_F)_{1,2} = |\vec k_{1,2}|$, $\sqrt{s} = 7$ TeV, and $Y \leq 4.8$.}
 \label{fig:ns7}
 \end{figure}

 For comparison, we show in Figs.~\ref{fig:ns13}~and~\ref{fig:ns7} 
 the results for the same observables 
 with the choice of $(\mu_F)_{1,2} = |\vec k_{1,2}|$.
 The patterns we have found are very similar to the previous ones, 
 but we see that the effect of having $C_1/C_0$ larger than $1$
 at small $Y$ is reduced. 
 Furthermore, NLA corrections are negative for larger $Y$ values.
 On the basis of this, we may conclude that, in the $Y \leq 4.8$ kinematical
 regime, the choice of ``natural'' scales for $\mu_F$ stabilises the results. 
 
\begin{figure}[H]
 \centering

   \includegraphics[scale=0.38]{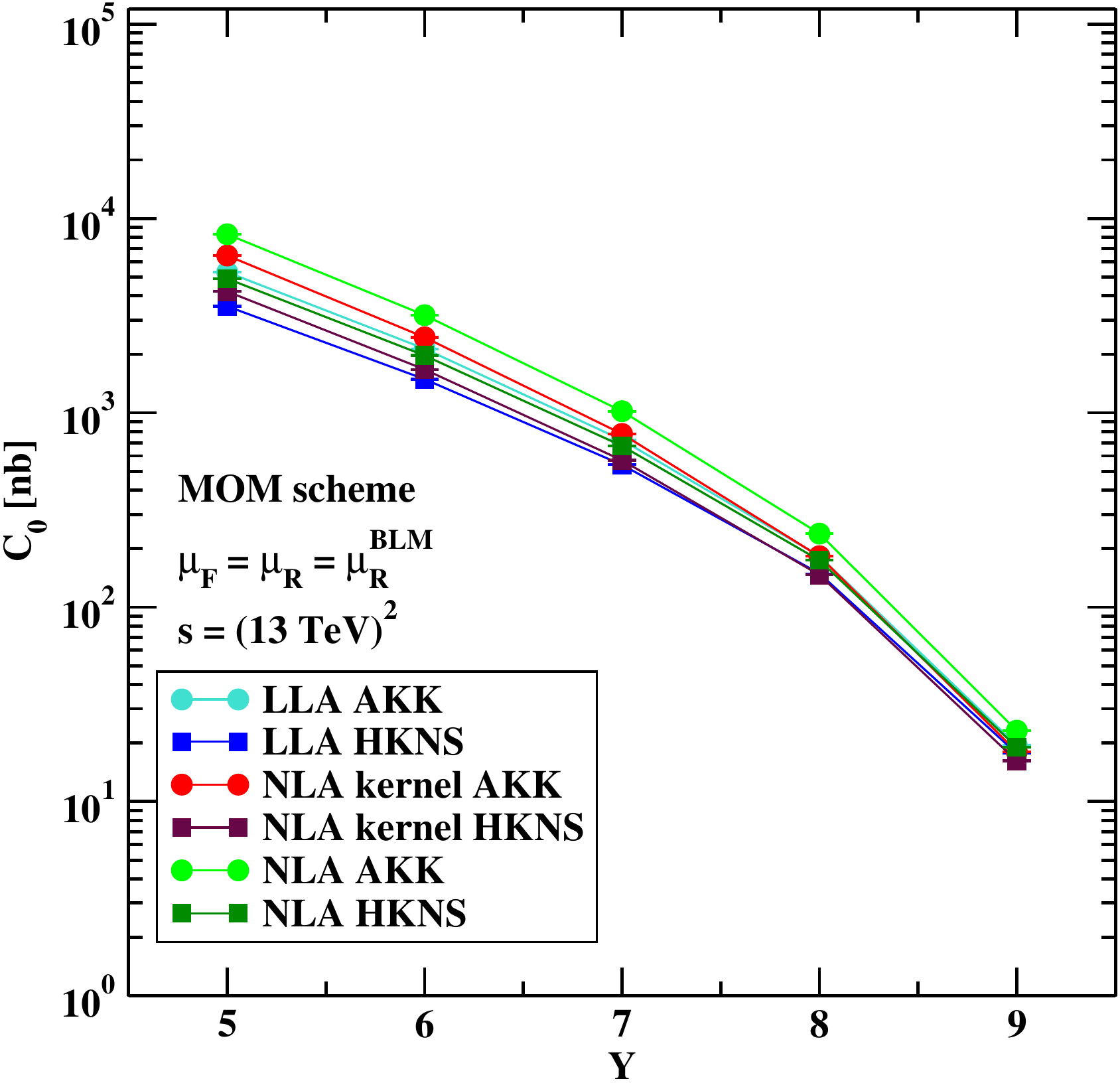}
   \includegraphics[scale=0.38]{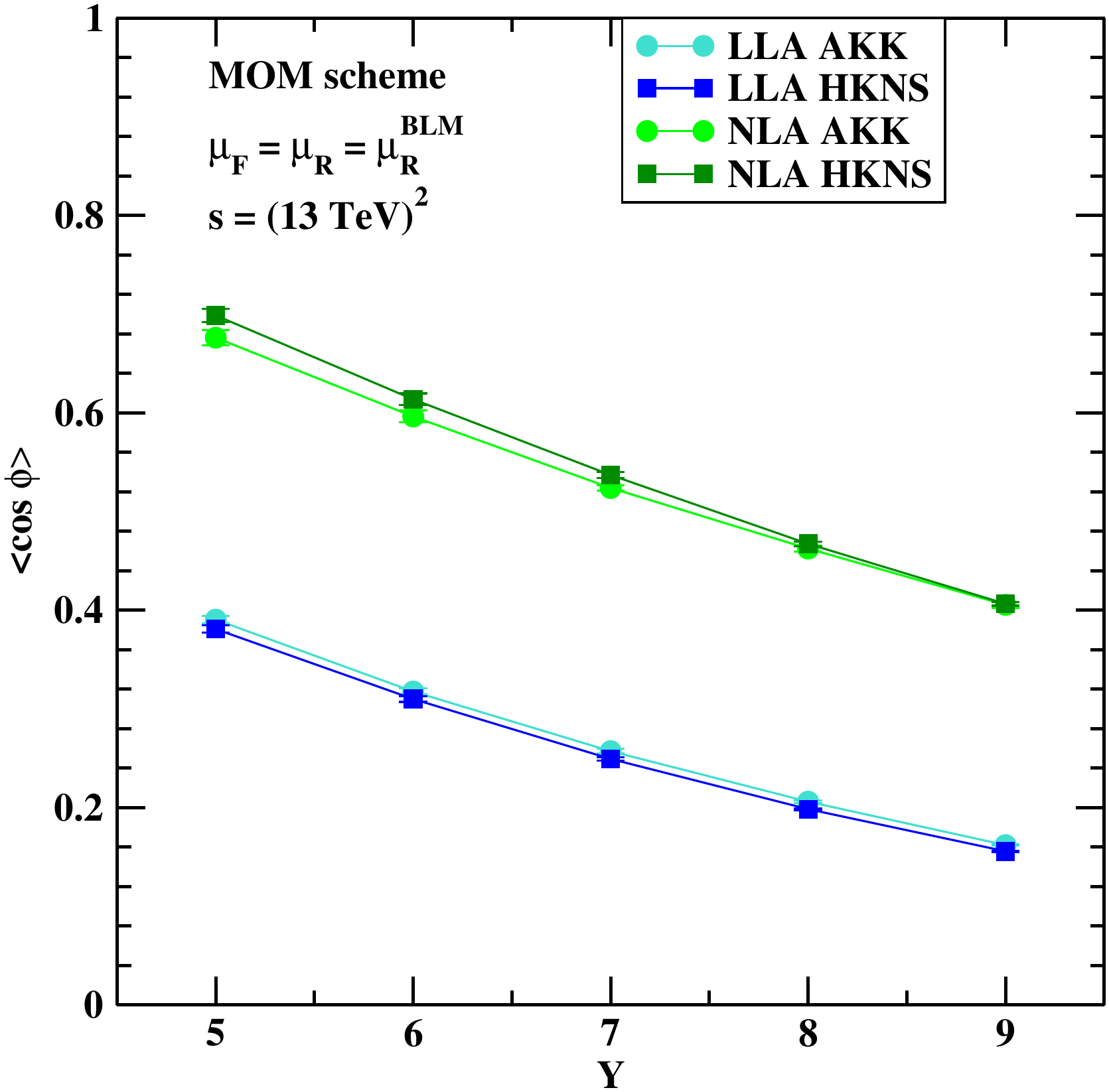}

   \includegraphics[scale=0.38]{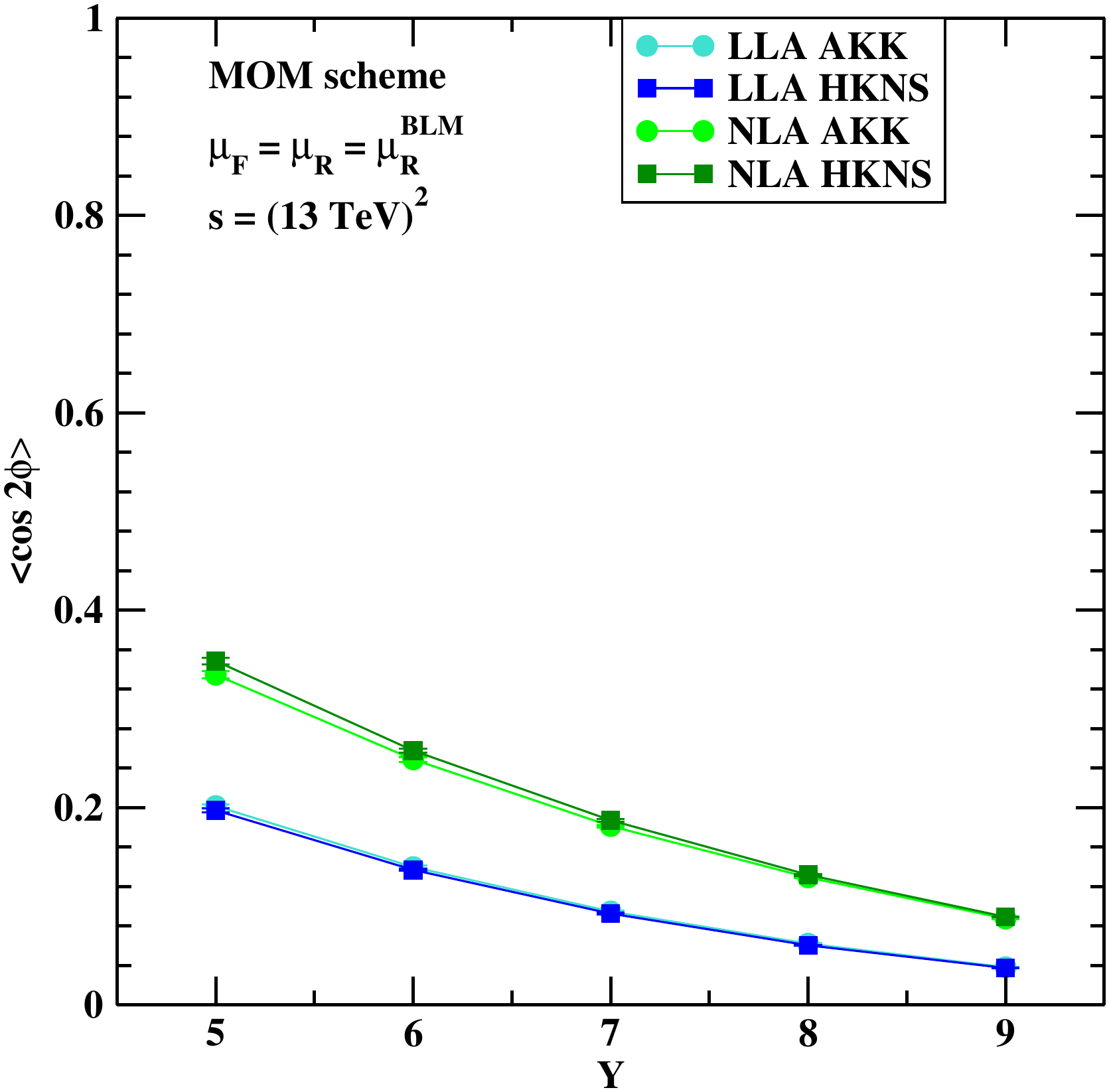}
   \includegraphics[scale=0.38]{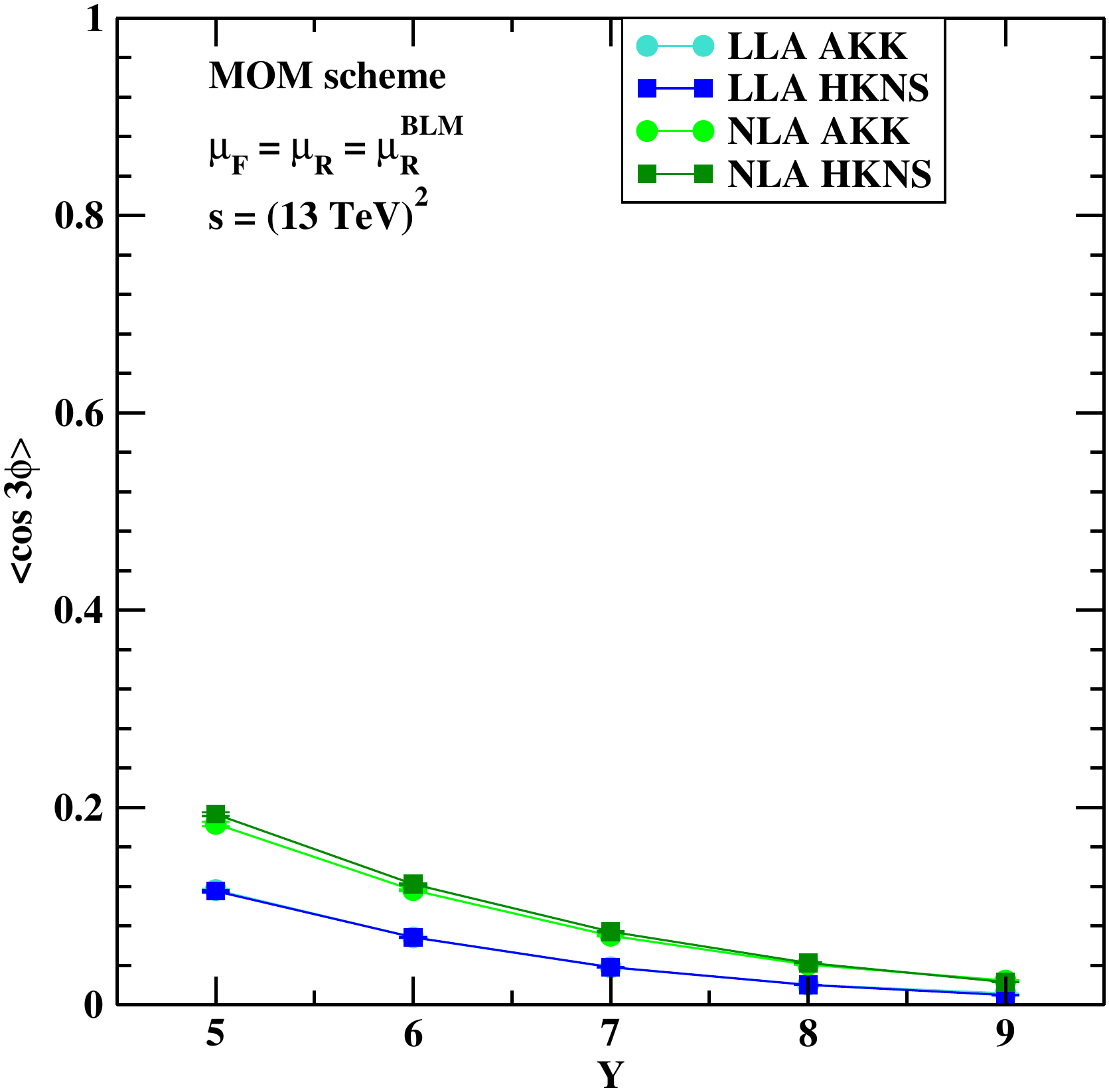}

   \includegraphics[scale=0.38]{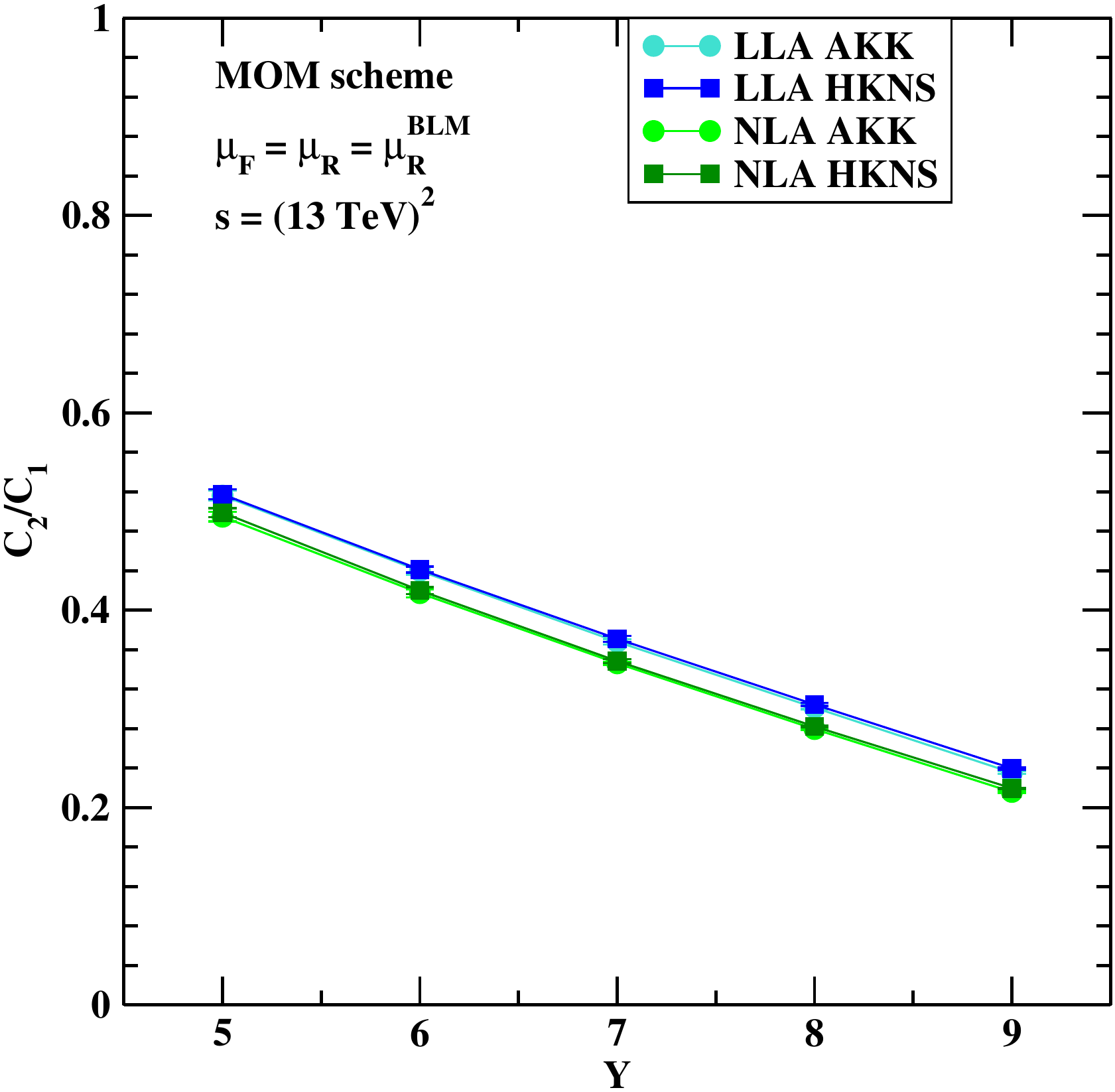}
   \includegraphics[scale=0.38]{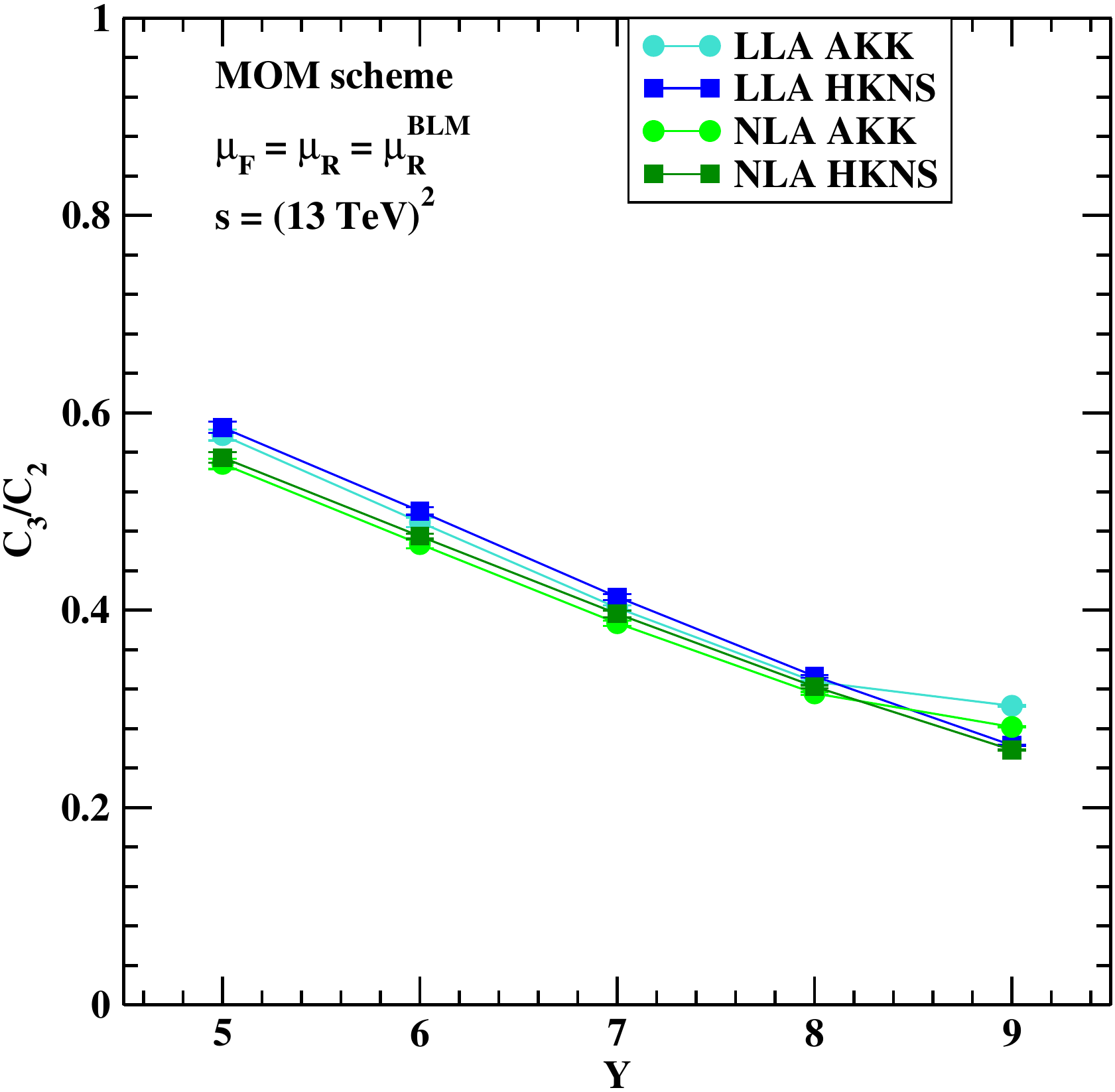}
 \caption[Full NLA predictions for dihadron production 
          for $\mu_F = \mu^{\rm BLM}_R$, $\sqrt{s} = 13$ TeV, 
          and $Y \leq 9.4$]
 {$Y$-dependence of $C_0$ and of several ratios $C_m/C_n$ for 
  $\mu_F = \mu^{\rm BLM}_R$, $\sqrt{s} = 13$ TeV, and $Y \leq 9.4$.}
 \label{fig:blmLY13}
 \end{figure}

\begin{figure}[H]
 \centering

   \includegraphics[scale=0.38]{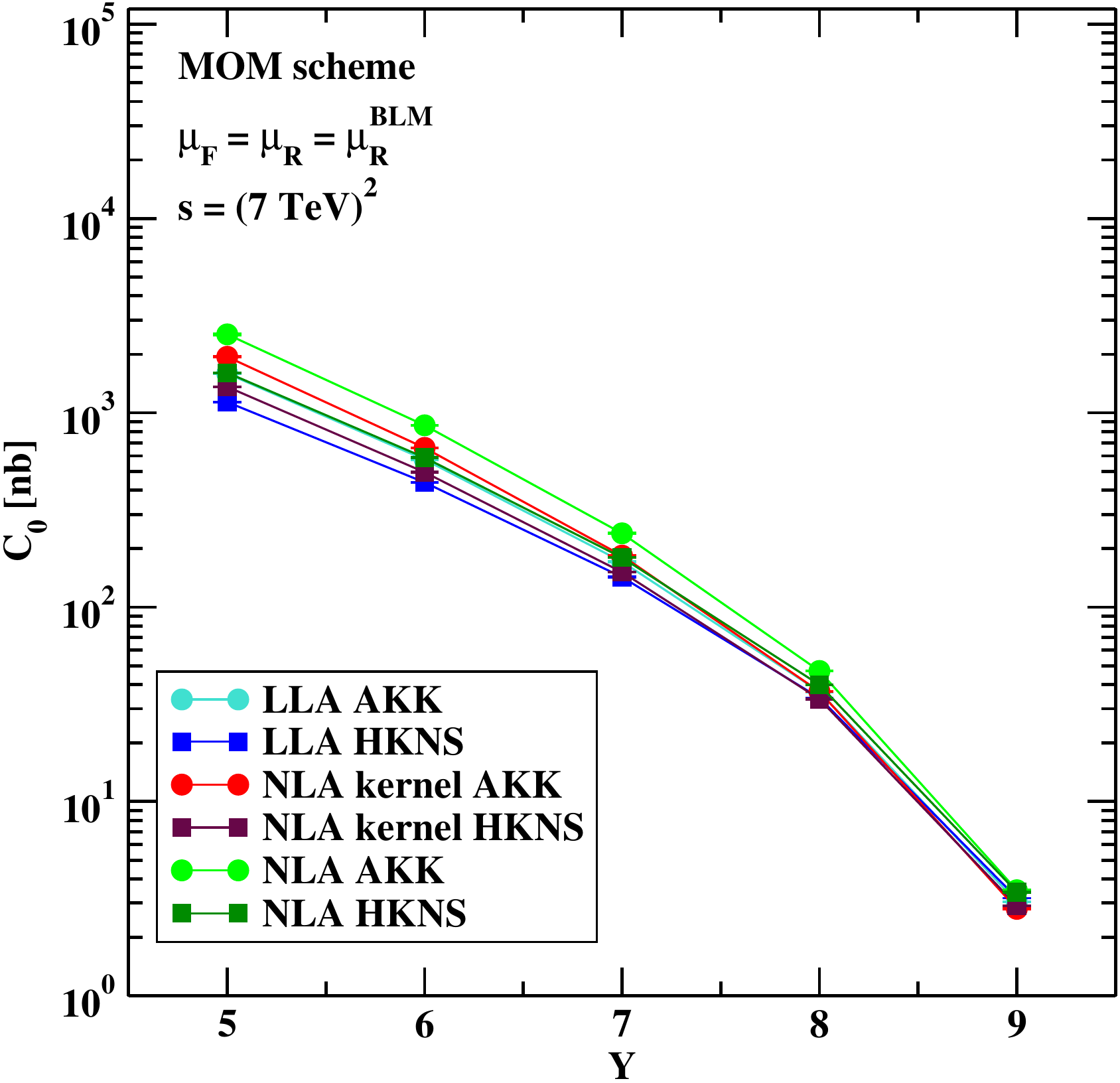}
   \includegraphics[scale=0.38]{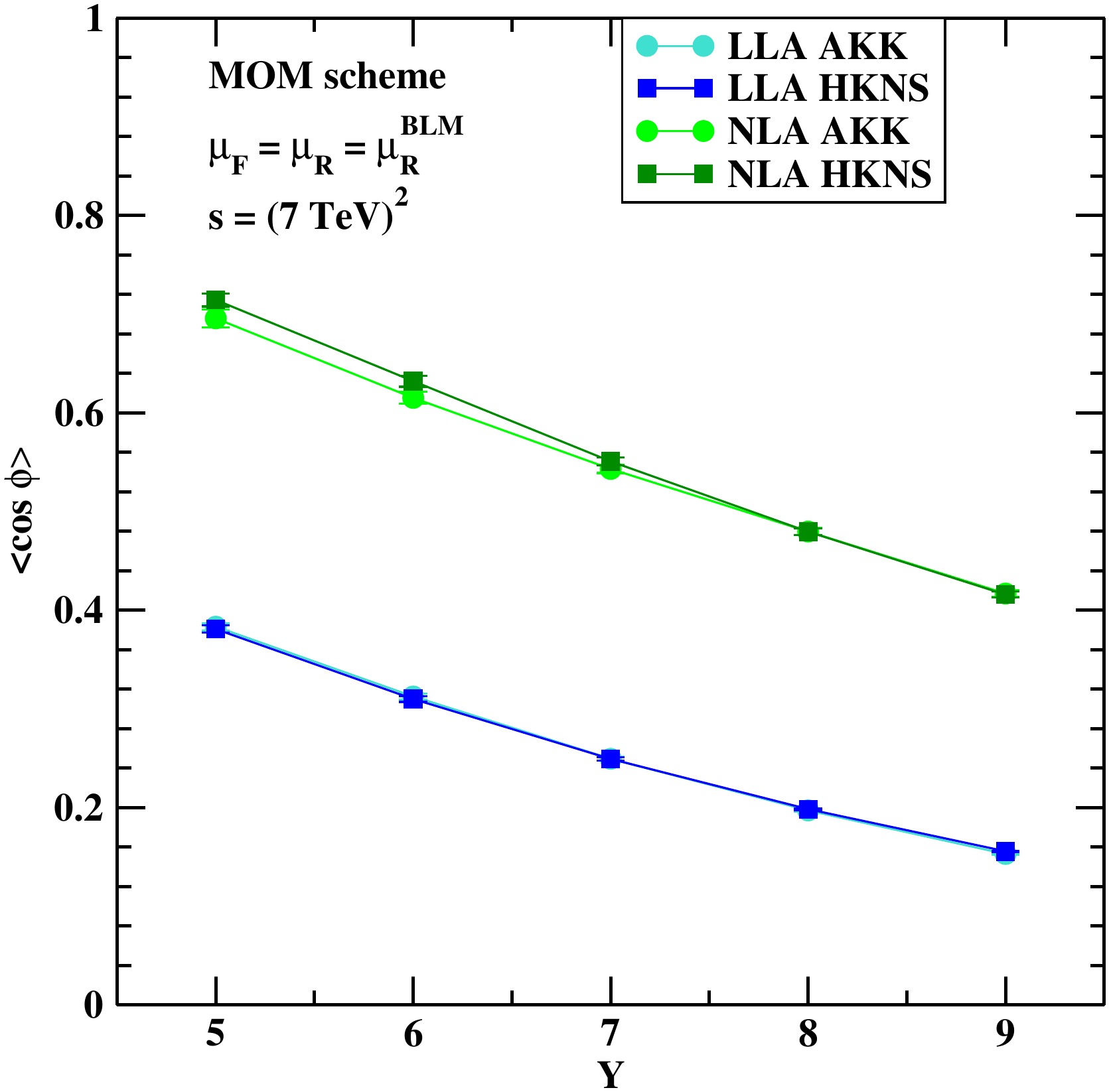}

   \includegraphics[scale=0.38]{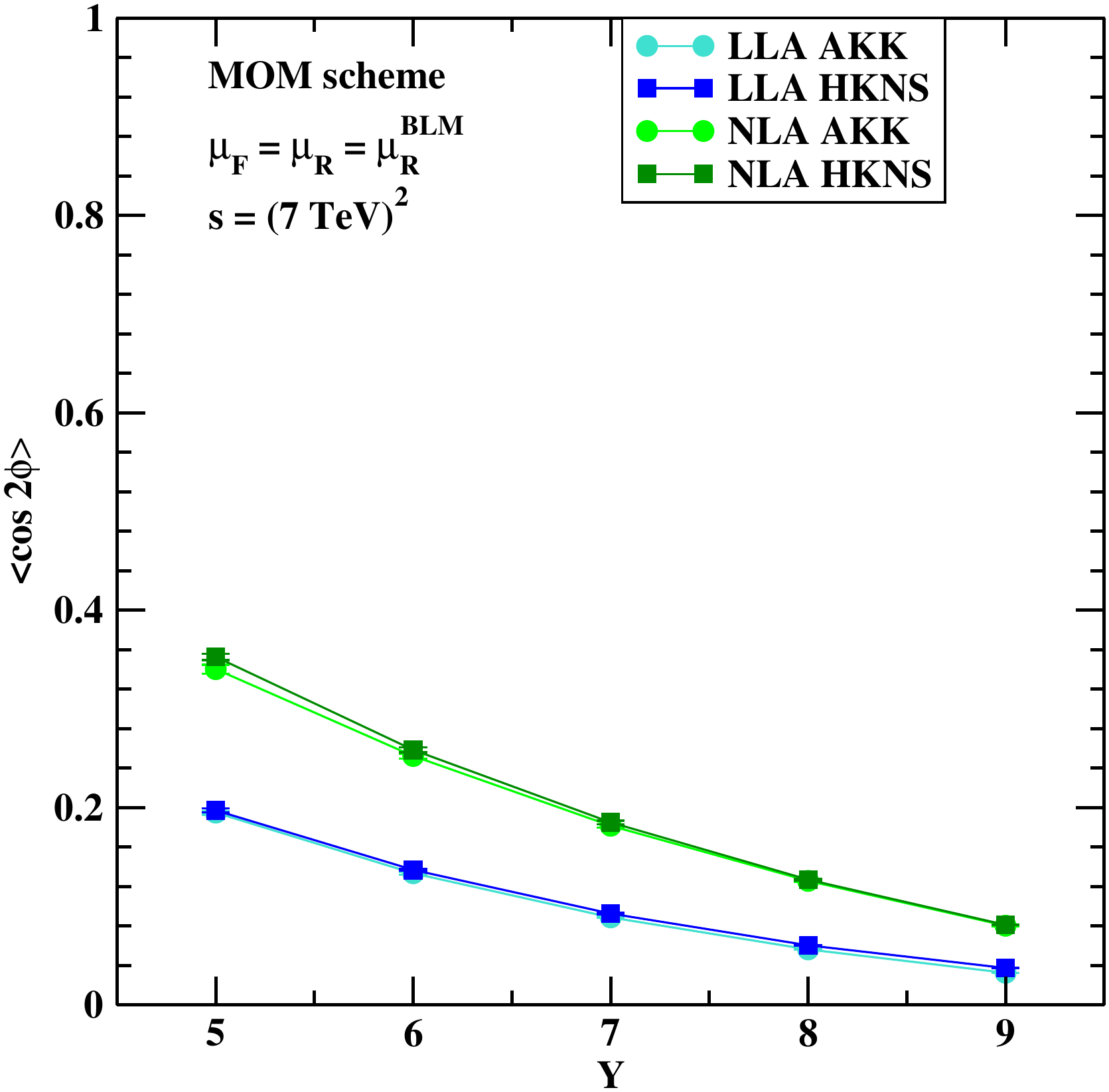}
   \includegraphics[scale=0.38]{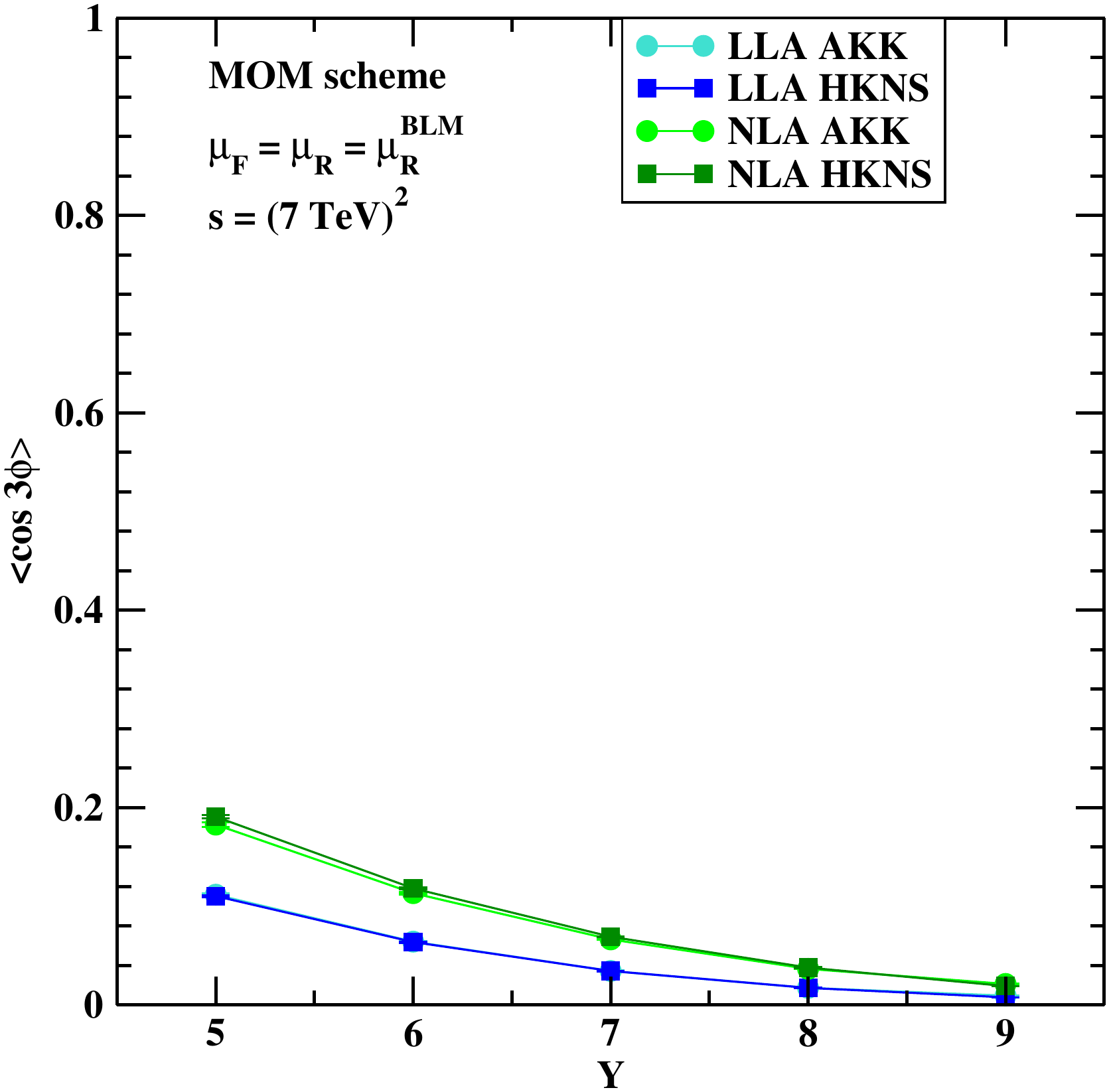}

   \includegraphics[scale=0.38]{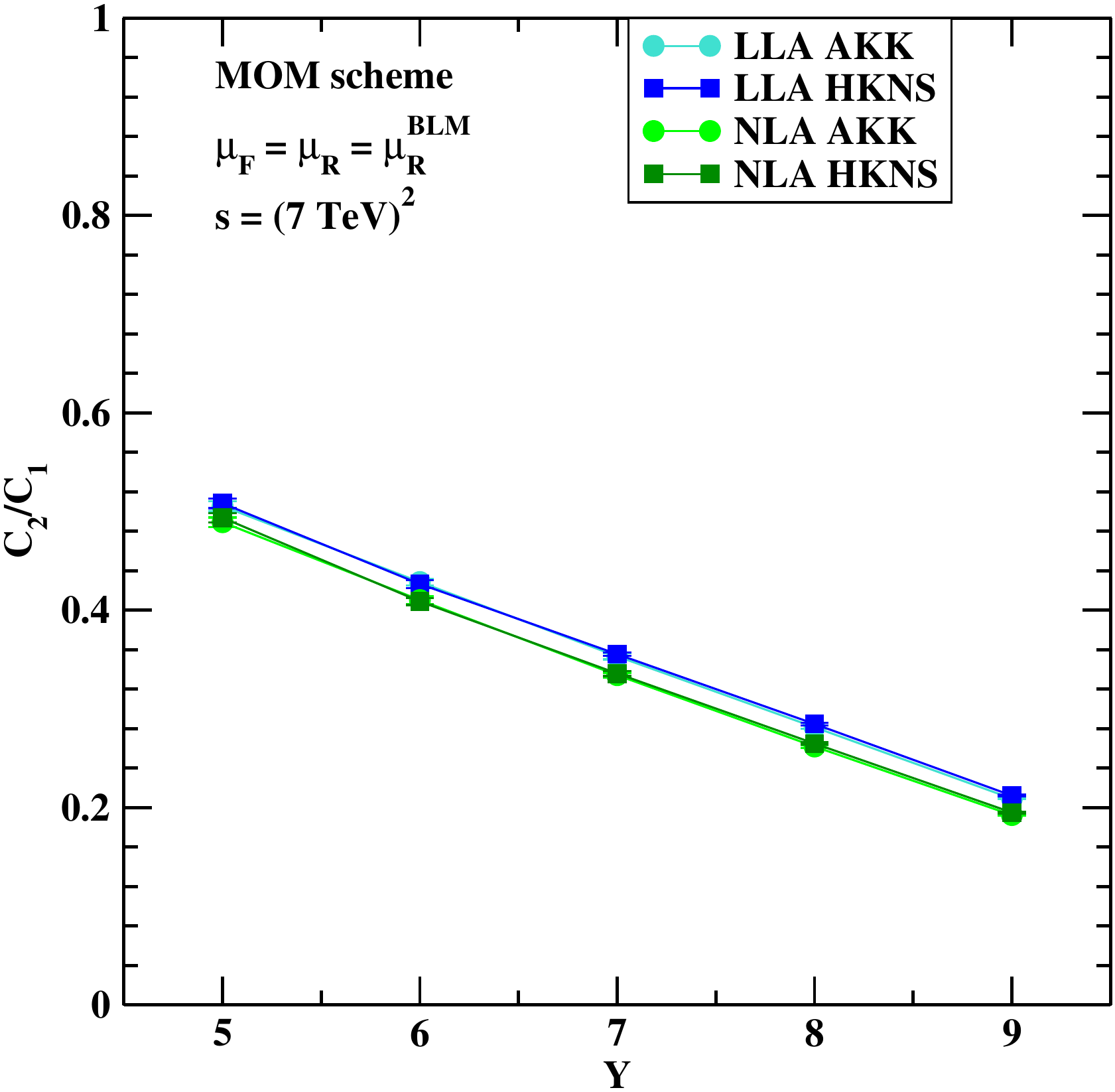}
   \includegraphics[scale=0.38]{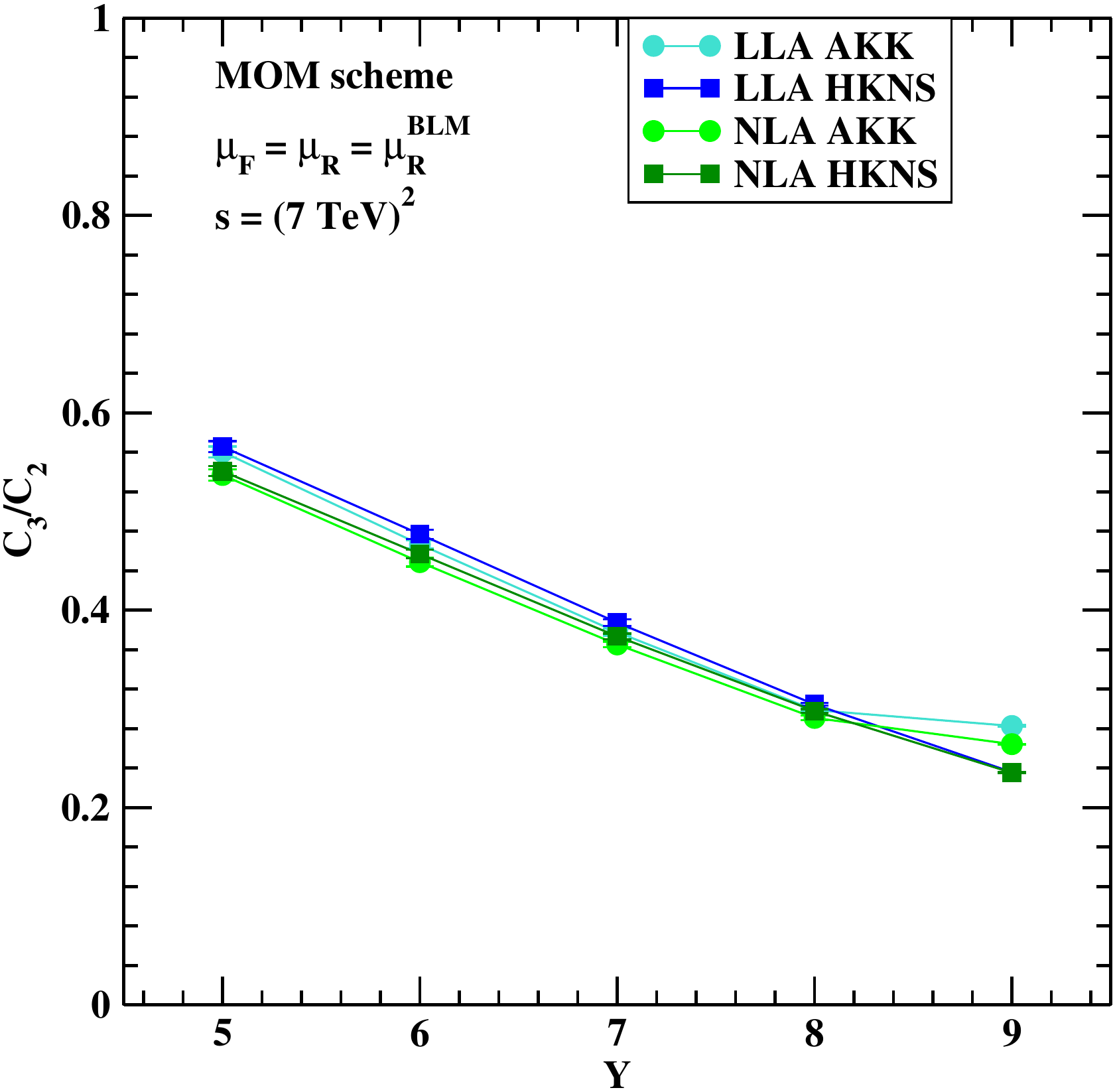}
 \caption[Full NLA predictions for dihadron production 
          for $\mu_F = \mu^{\rm BLM}_R$, $\sqrt{s} = 7$ TeV, 
          and $Y \leq 9.4$]
 {$Y$-dependence of $C_0$ and of several ratios $C_m/C_n$ for 
  $\mu_F = \mu^{\rm BLM}_R$, $\sqrt{s} = 7$ TeV, and $Y \leq 9.4$.}
 \label{fig:blmLY7}
 \end{figure}

 In Figs.~\ref{fig:blmLY13}~and~\ref{fig:blmLY7} we present our results 
 for $C_0$ and for several ratios $C_m/C_n$
 at $\sqrt{s}=13$ and $7$ TeV respectively;
 $\mu_F$ is set equal to $\mu^{\rm BLM}_R$,
 while $Y$ lies on a larger range, {\it i.e.} $Y \leq 9.4$.
 
\begin{figure}[H]
 \centering

   \includegraphics[scale=0.38]{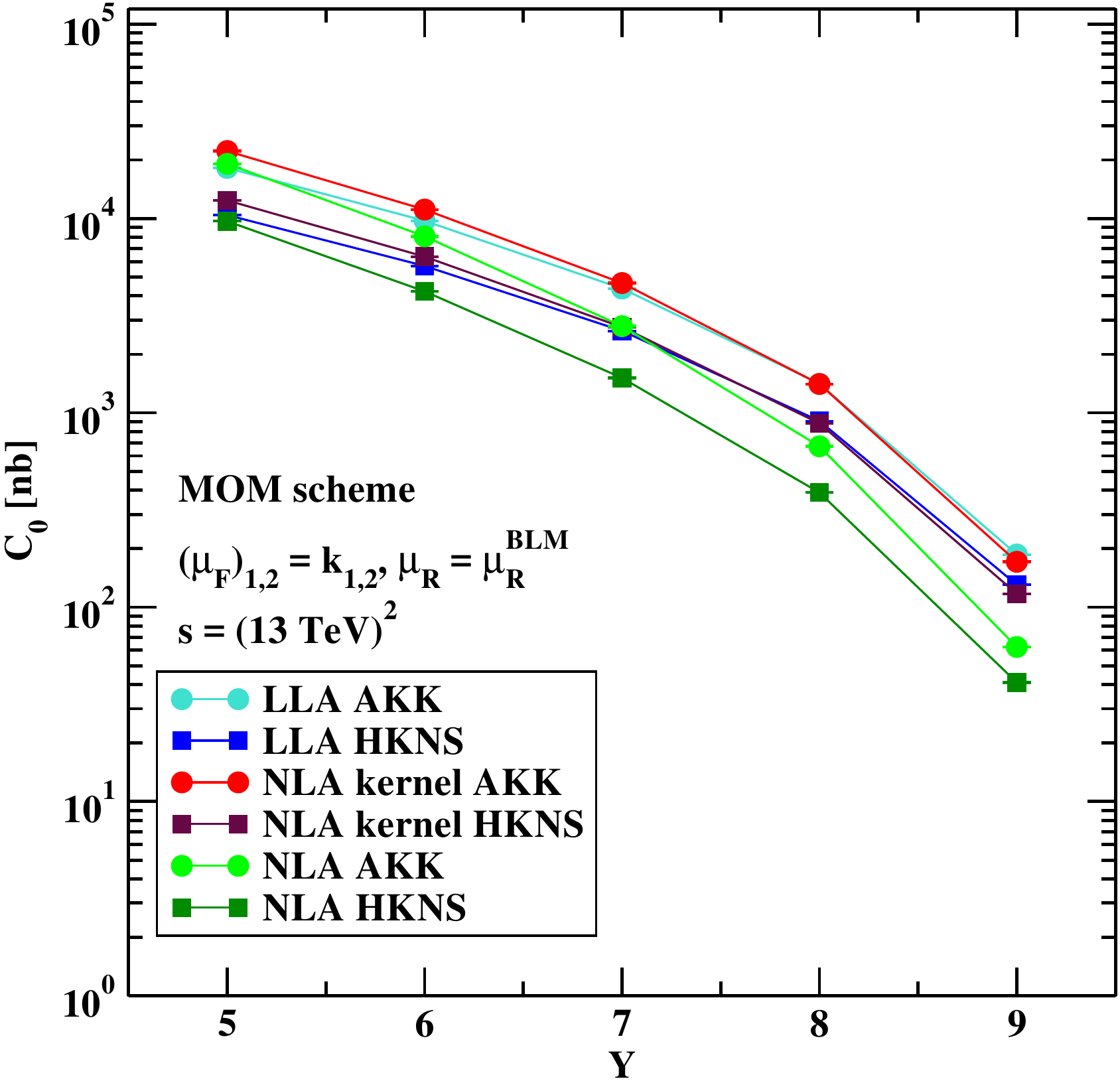}
   \includegraphics[scale=0.38]{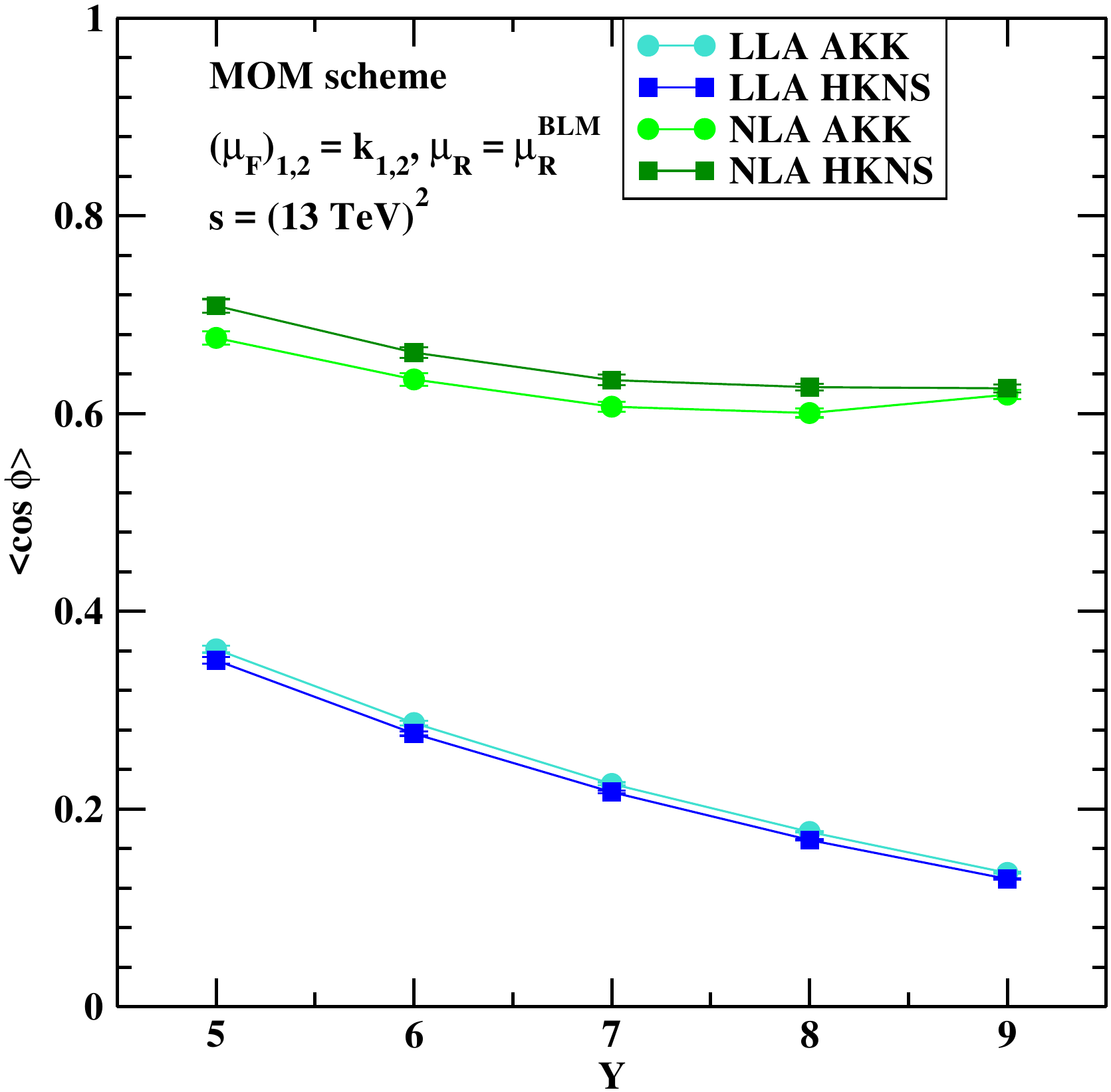}

   \includegraphics[scale=0.38]{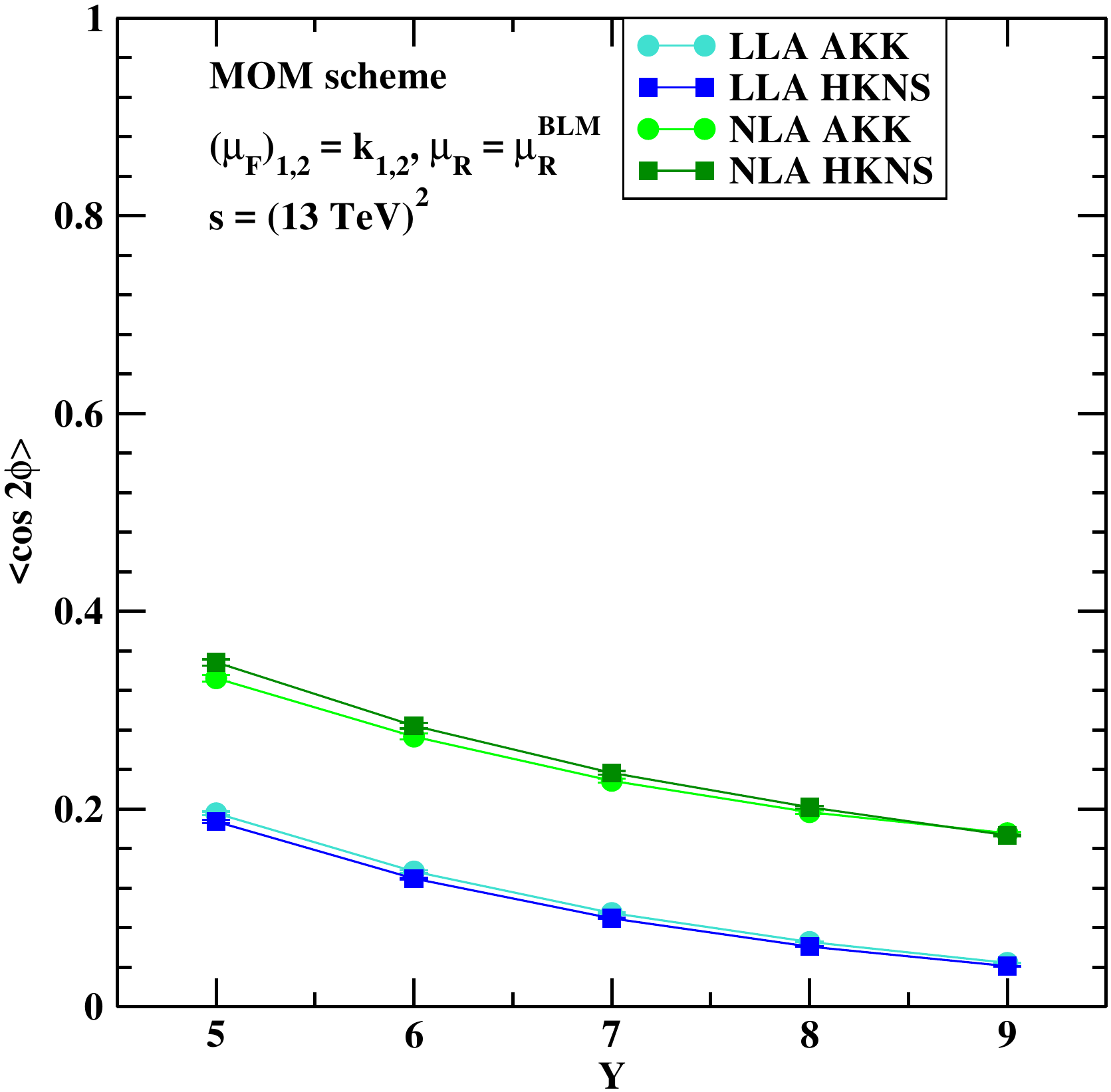}
   \includegraphics[scale=0.38]{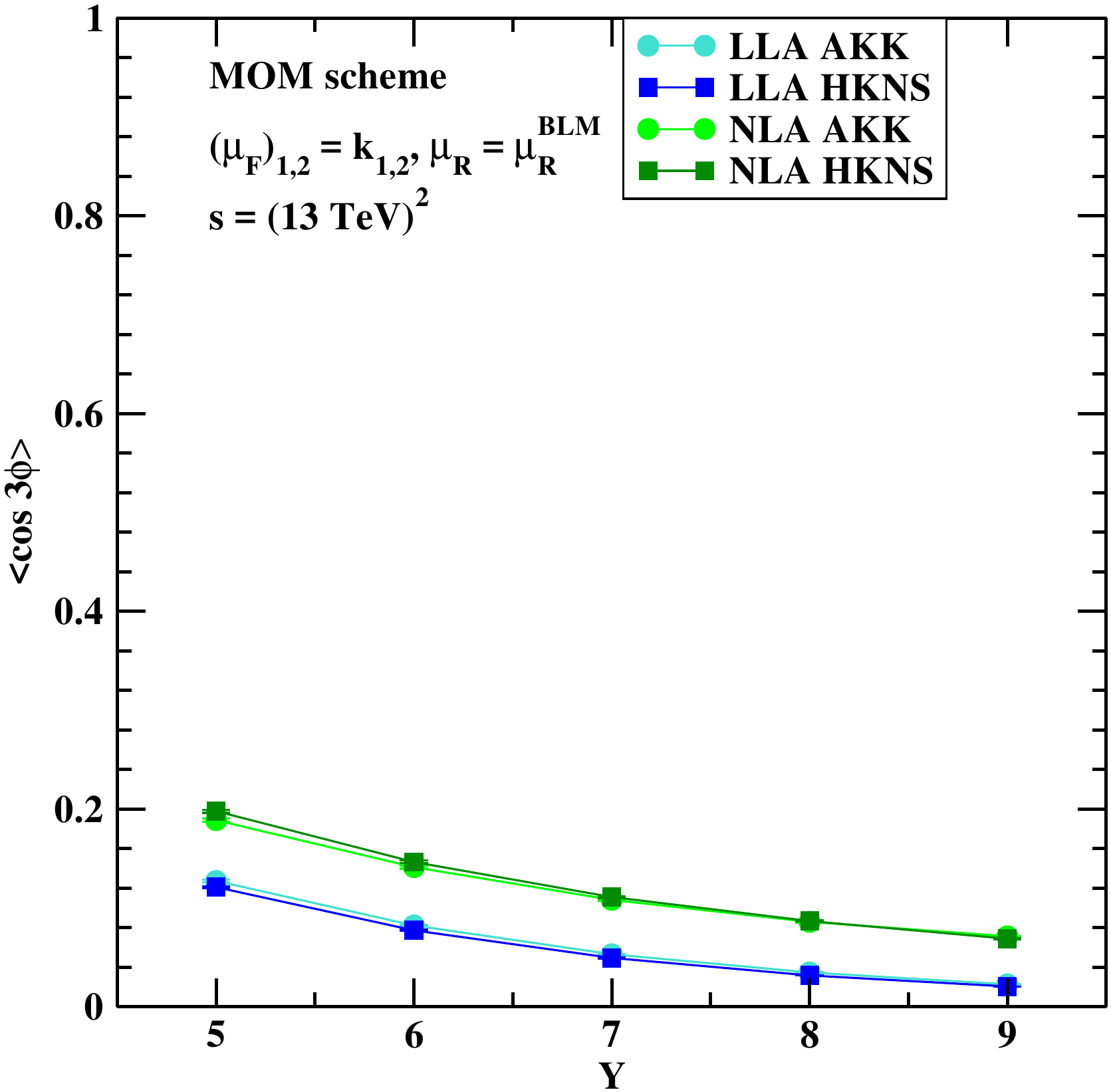}

   \includegraphics[scale=0.38]{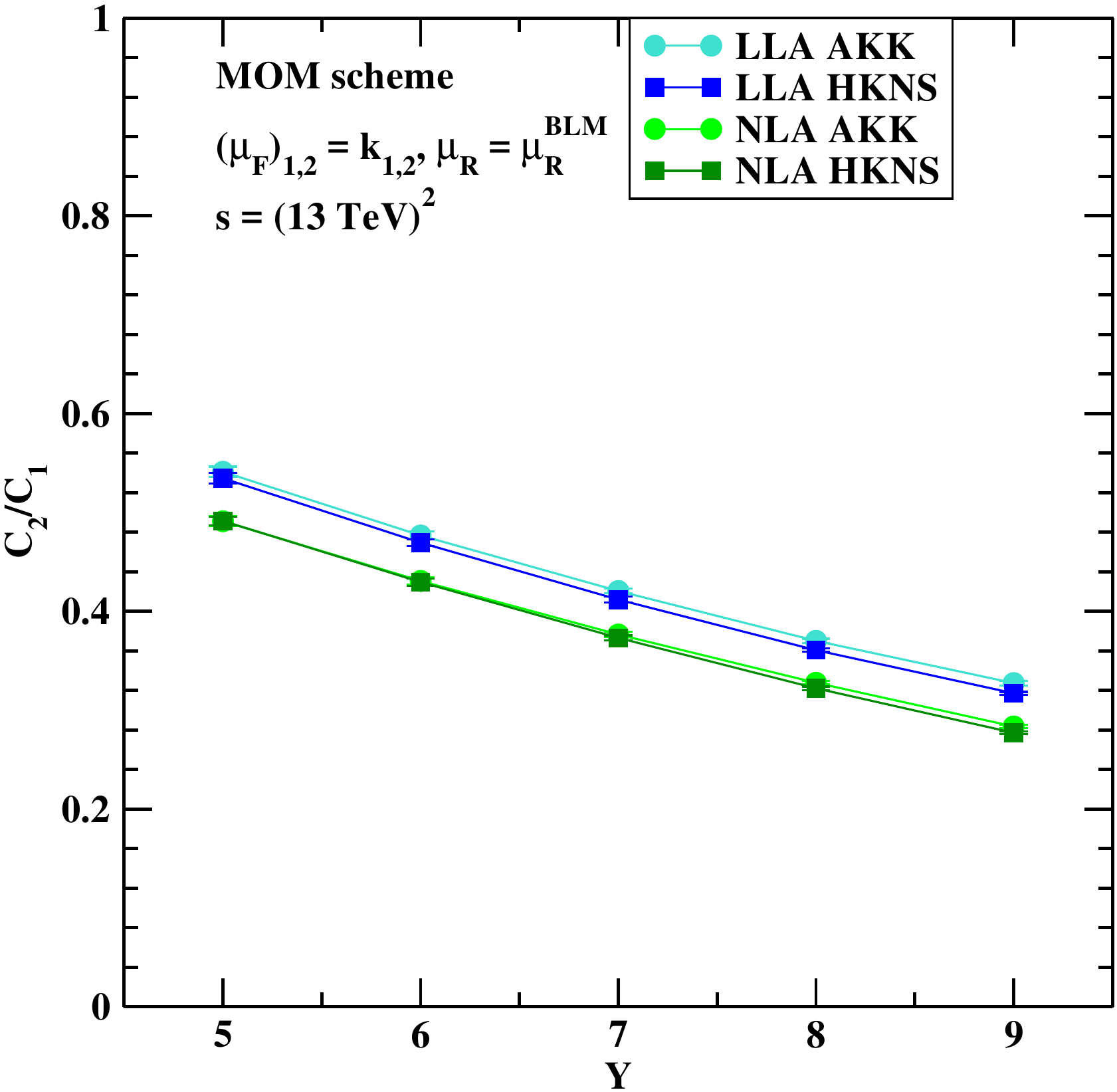}
   \includegraphics[scale=0.38]{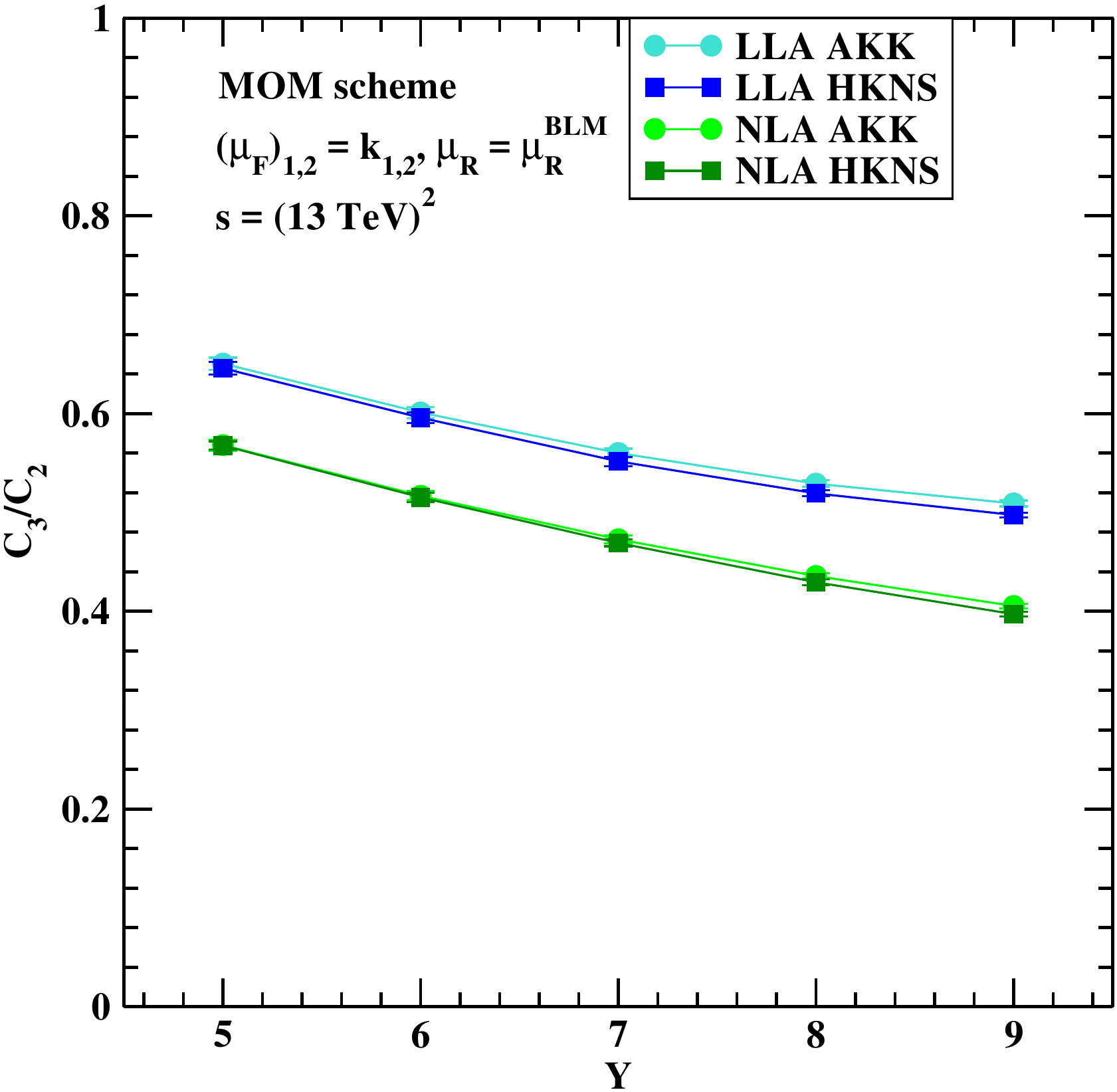}
 \caption[Full NLA predictions for dihadron production 
          for $(\mu_F)_{1,2} = |\vec k_{1,2}|$, $\sqrt{s} = 13$ TeV, 
          and $Y \leq 9.4$]
 {$Y$-dependence of $C_0$ and of several ratios $C_m/C_n$ for 
  $(\mu_F)_{1,2} = |\vec k_{1,2}|$, $\sqrt{s} = 13$ TeV, and $Y \leq 9.4$.}
 \label{fig:nsLY13}
 \end{figure}

\begin{figure}[H]
 \centering

   \includegraphics[scale=0.38]{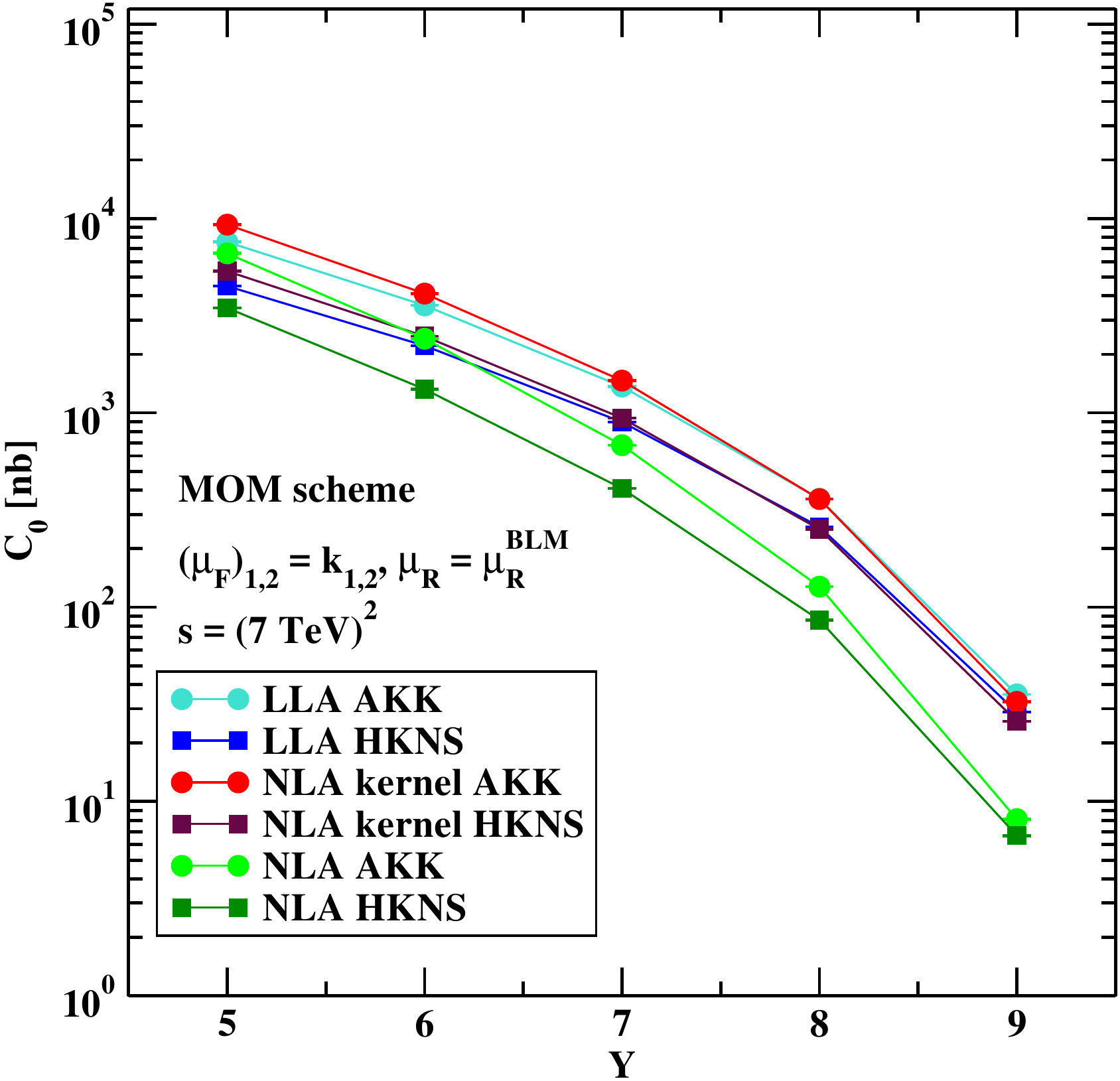}
   \includegraphics[scale=0.38]{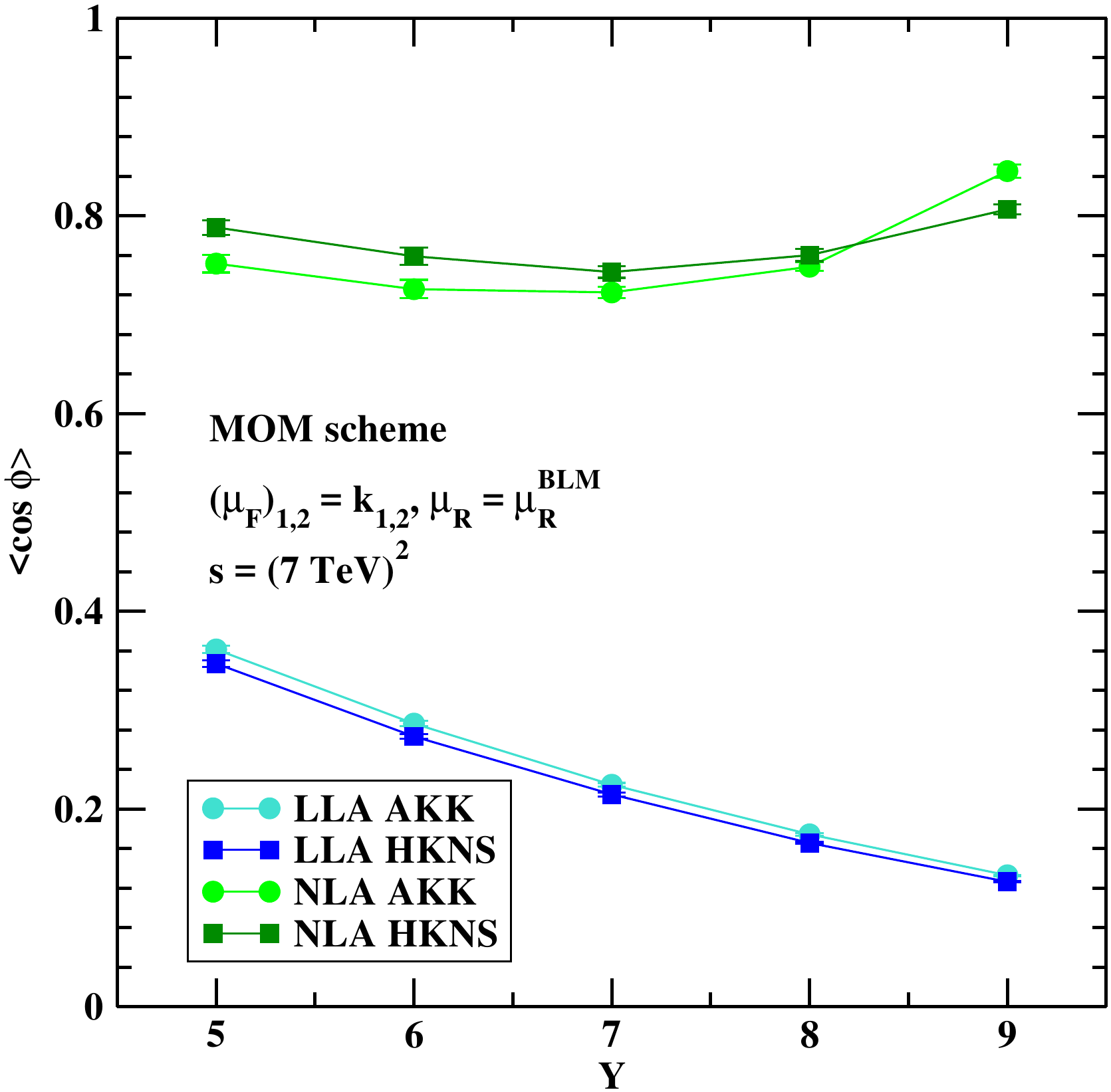}

   \includegraphics[scale=0.38]{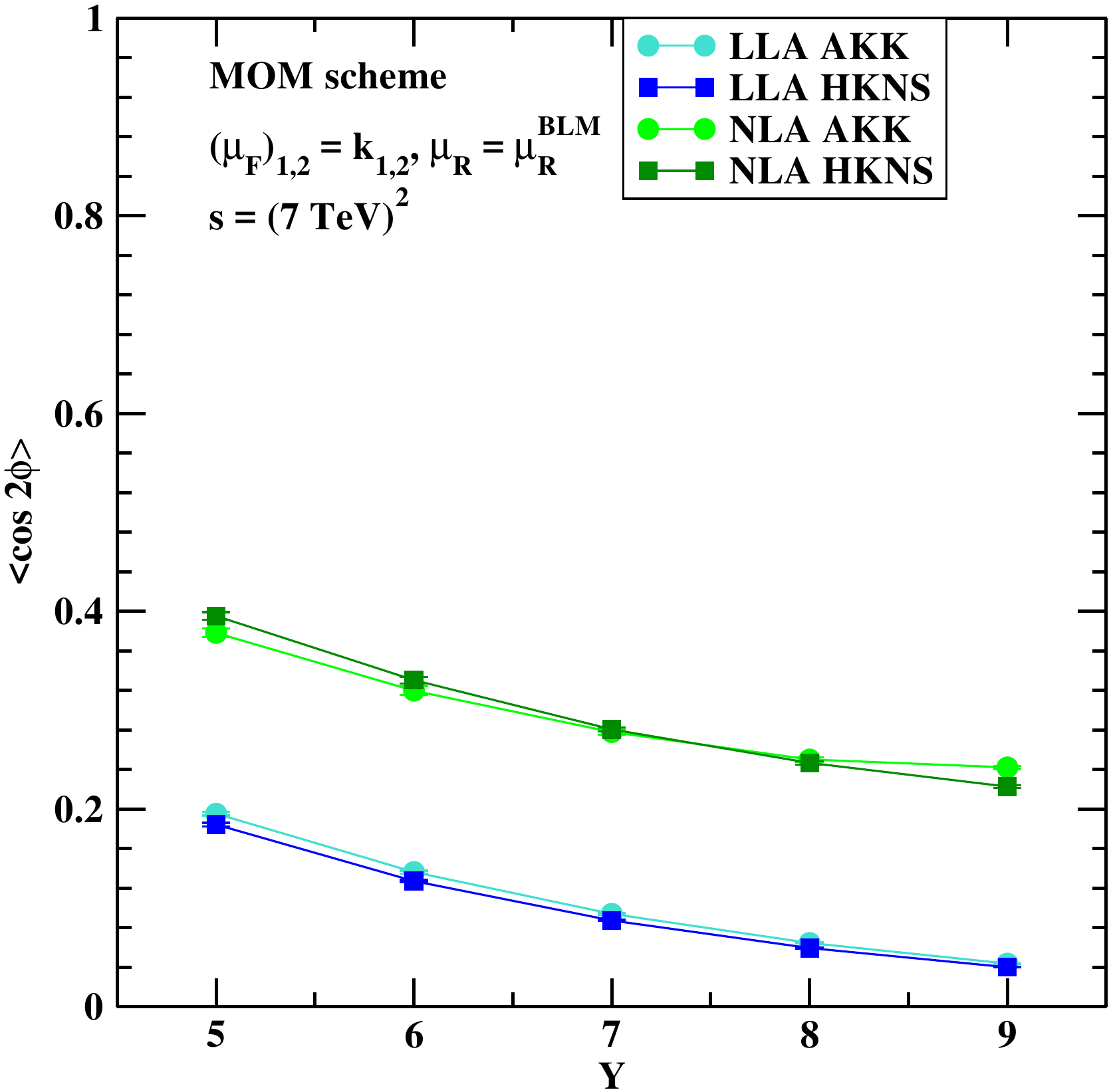}
   \includegraphics[scale=0.38]{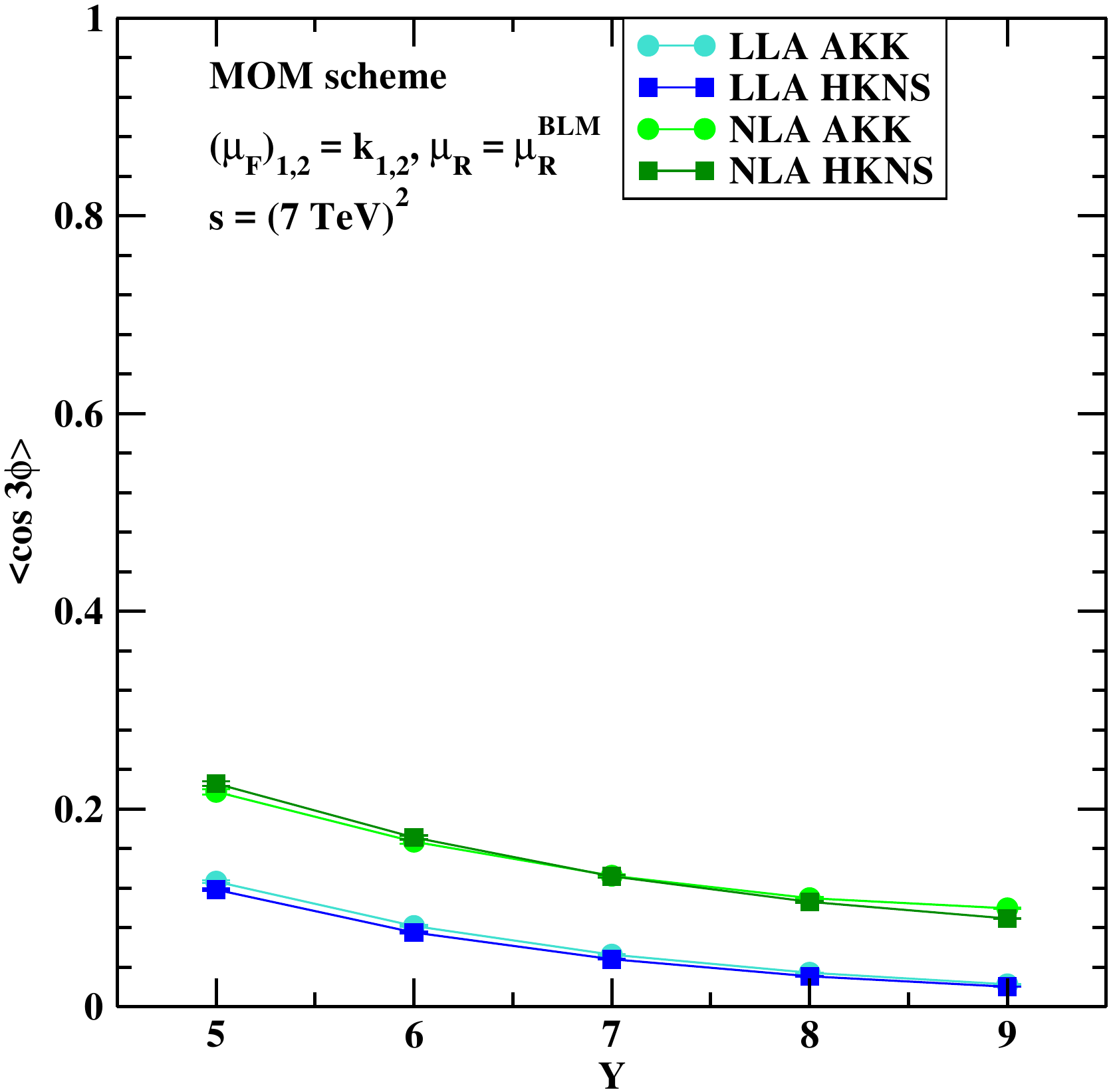}

   \includegraphics[scale=0.38]{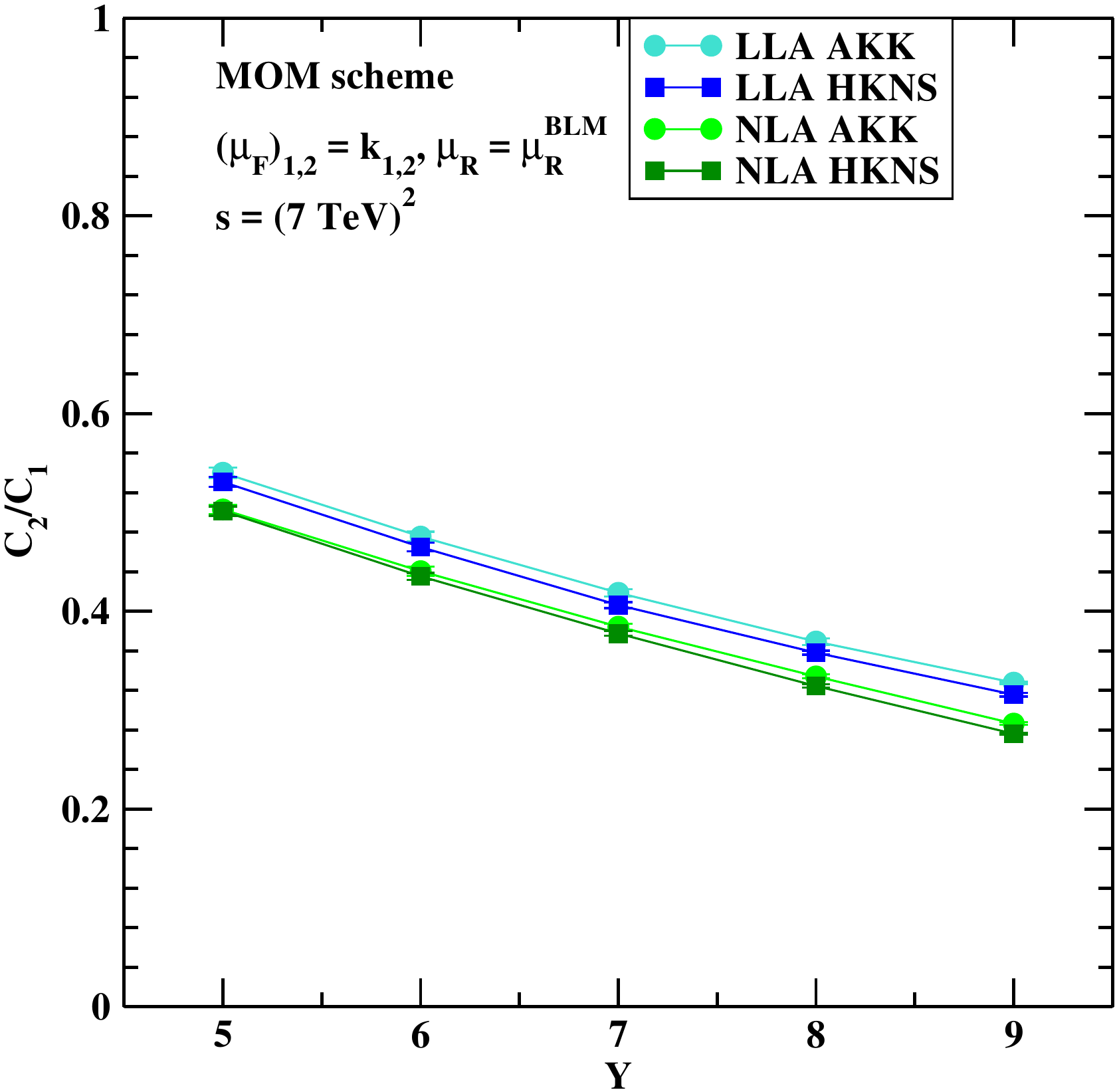}
   \includegraphics[scale=0.38]{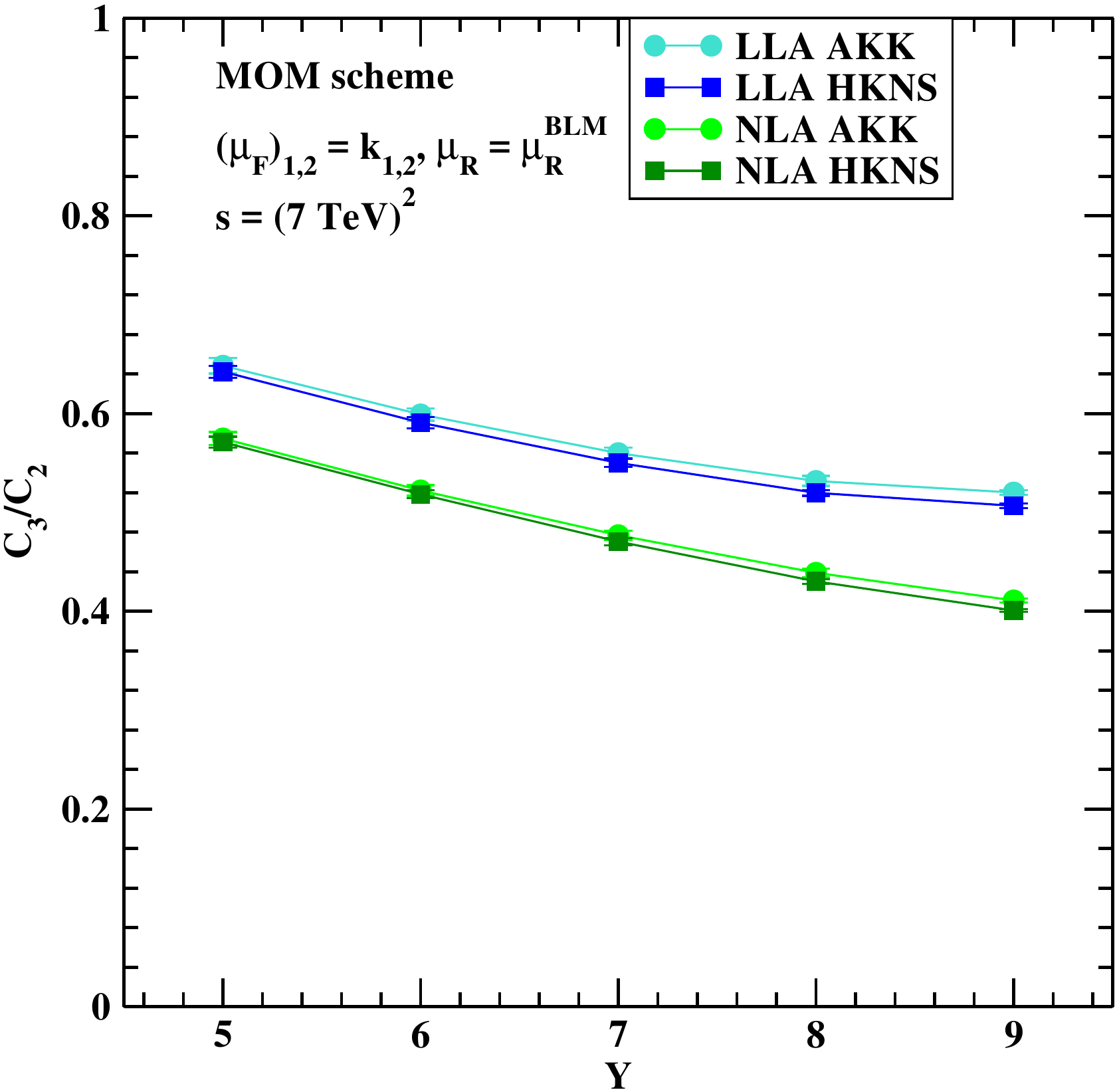}
 \caption[Full NLA predictions for dihadron production 
          for $(\mu_F)_{1,2} = |\vec k_{1,2}|$, $\sqrt{s} = 7$ TeV, 
          and $Y \leq 9.4$]
 {$Y$-dependence of $C_0$ and of several ratios $C_m/C_n$ for 
  $(\mu_F)_{1,2} = |\vec k_{1,2}|$, $\sqrt{s} = 7$ TeV, and $Y \leq 9.4$.}
 \label{fig:nsLY7}
 \end{figure}

 For comparison, we show in Figs.~\ref{fig:nsLY13}~and~\ref{fig:nsLY7} 
 the results for the same observables 
 with the choice of $(\mu_F)_{1,2} = |\vec k_{1,2}|$.
 We clearly see that, in the case of larger rapidity intervals $Y$ and with the
 ``natural'' choice for the factorisation scale, the situation is different in
 comparison to the $\mu_F=\mu_R^{BLM}$ choice: the NLA corrections to the cross
 section $C_0$ are negative, while the pattern of $C_1/C_0$ shows a somewhat
 unexpected ``turn-up'' at large $Y$, and these effects are more pronounced for
 the lower LHC energy,  $\sqrt{s}=7 \ \rm{TeV}$.
 Such a sensitivity to the factorisation scale setting may be an indication of
 the fact that with the increase of $Y$ values we are moving towards the
 threshold region, where the energy of detected dihadron system becomes
 comparable with  $\sqrt{s}$. In this situation the FFs and PDFs are probed
 in regions that are close to the end-points of their definitions, where they
 exhibit large dependence on the factorisation scale. From the physical site,  
 in this kinematics the undetected hard-gluon radiation is getting restricted
 and only radiation of soft gluons is allowed.
 Soft-gluon radiation can not change the kinematics of the hard subprocess,
 therefore one expects restoration of the correlation of the detected hadrons
 in the relative azimuthal angle when we  approach  the threshold region.    
 It is well known that in this situation large \emph{threshold double logarithms}~\cite{Jager:2004jh,Kidonakis:1998bk,deFlorian:2007fv,Catani:1996yz} 
 appear in the perturbative series, and such contributions have to be resummed to all orders. 
 Threshold logarithms appear when the parent parton has just enough energy to produce the identified particle in the final state and the unobserved recoiling partonic final state.
 Resummation in the kinematics where both threshold and BFKL
 logarithms are important is an interesting task, but it goes well beyond
 the scope of the present study. Here we just note that pure BFKL
 predictions in the region of largest $Y$ become rather sensitive to the
 choice of the factorisation scale. 
 
\begin{figure}[H]
 \centering

   \includegraphics[scale=0.38]{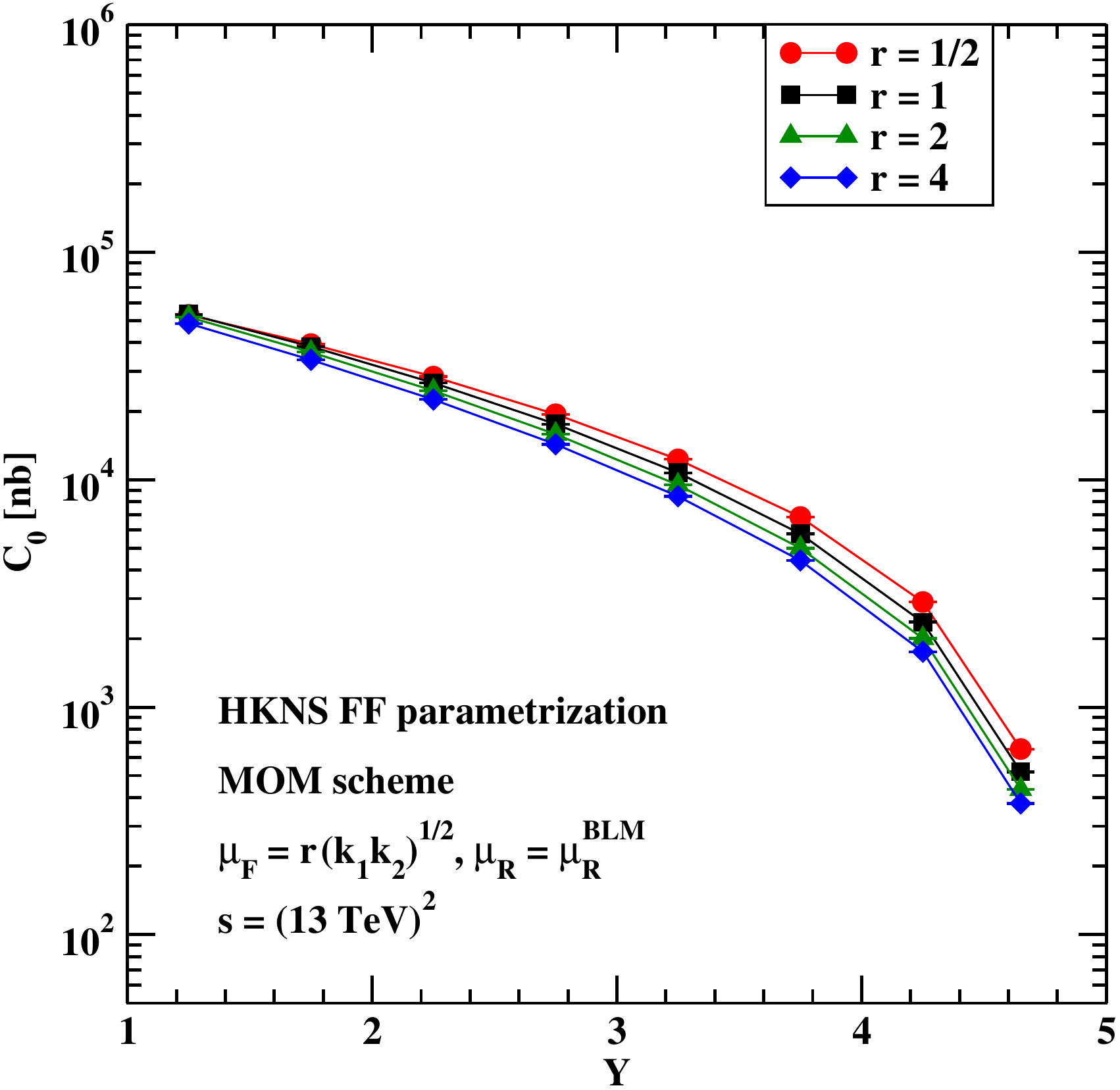}
   \includegraphics[scale=0.38]{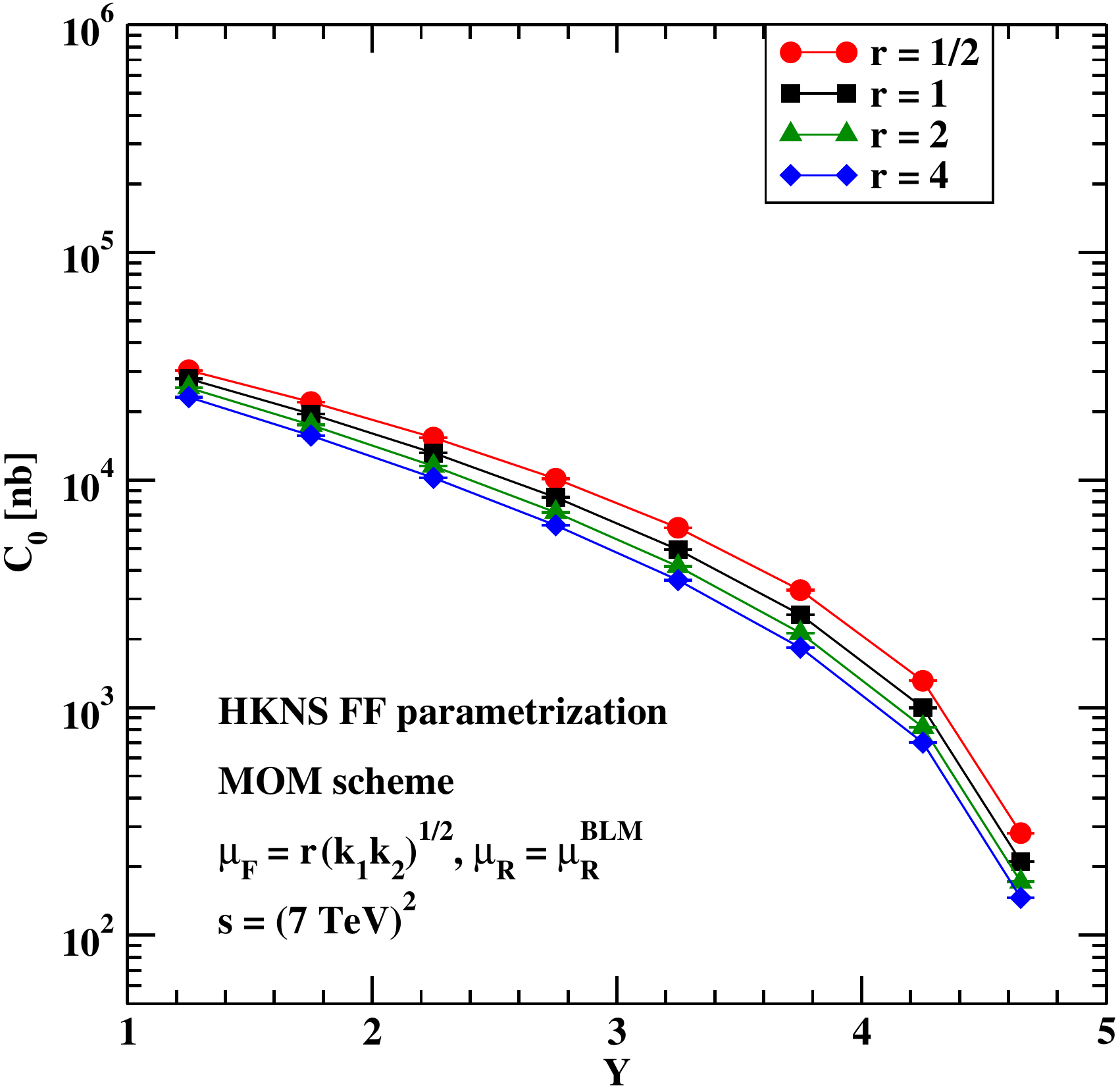}

   \includegraphics[scale=0.38]{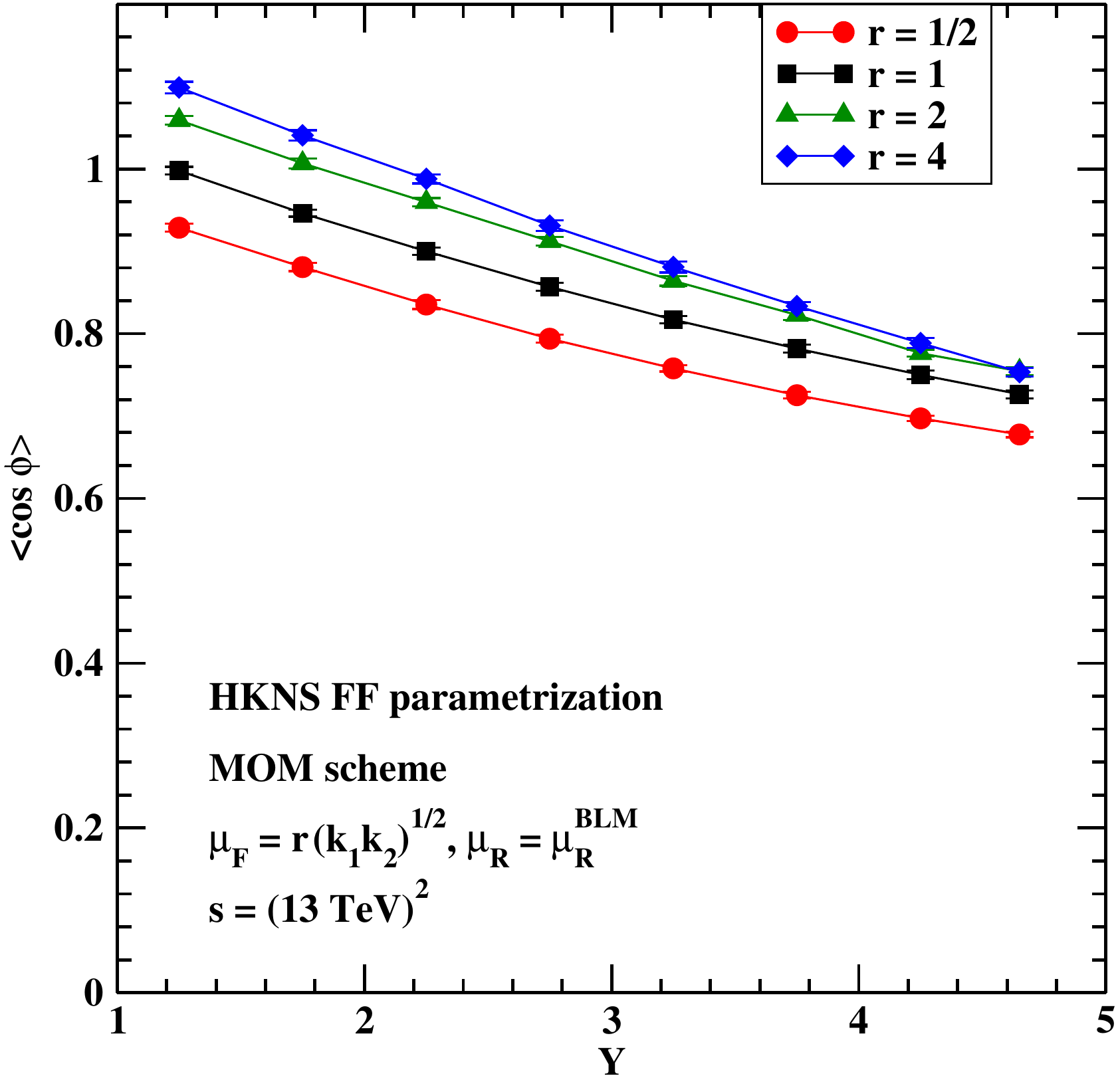}
   \includegraphics[scale=0.38]{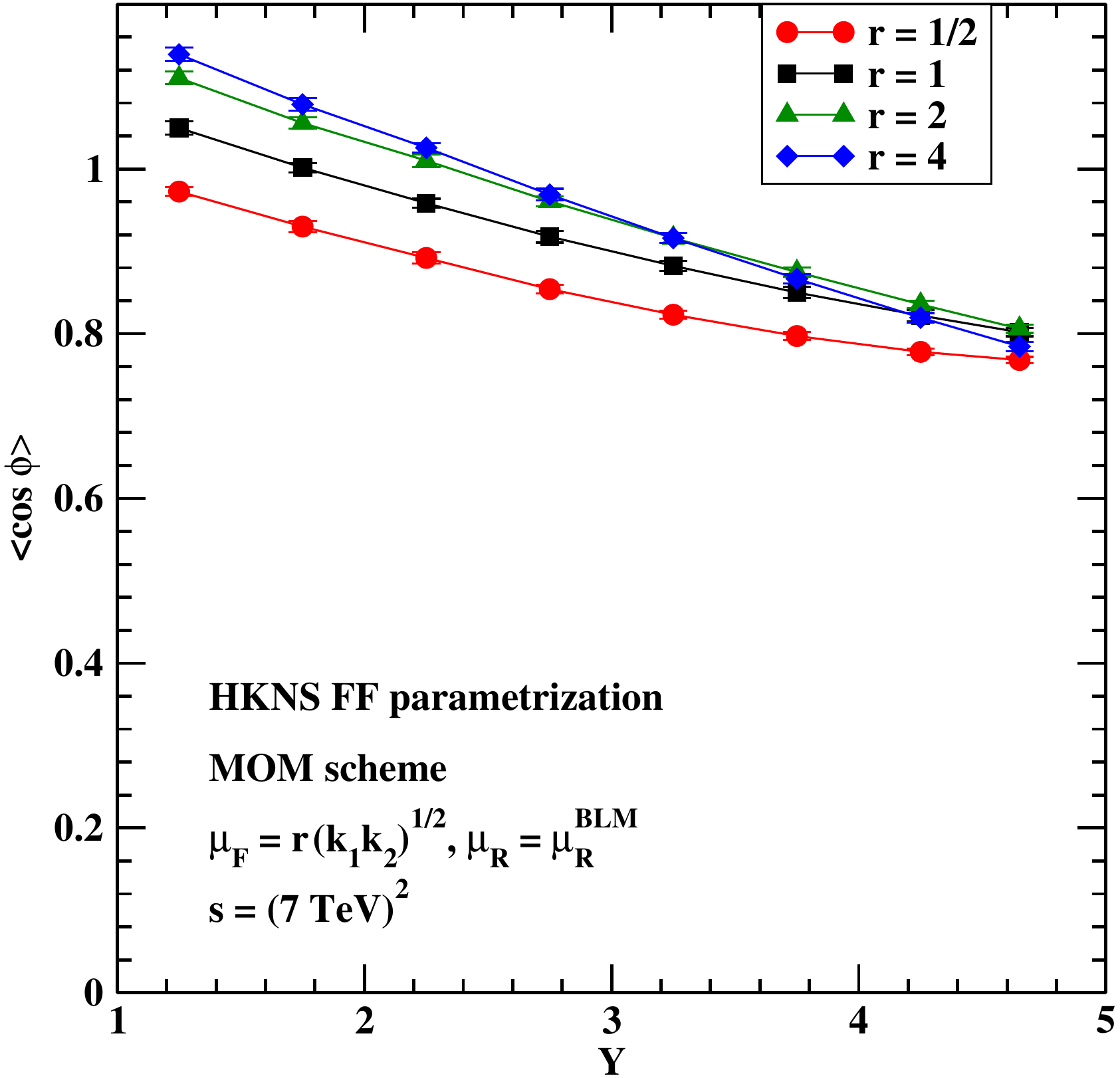}
 \caption[Full NLA predictions for dihadron production 
          for $\mu_ = r \sqrt{|\vec k_1| |\vec k_2|}$, with $r = 1/2, 1, 2, 4$, and $Y \leq 4.8$]
 {$Y$-dependence of $C_0$ and of $C_1/C_0$ for 
  $\mu_F = r \sqrt{|\vec k_1| |\vec k_2|} $, with $r = 1/2, 1, 2, 4$, and
  $Y \leq 4.8$.}
 \label{fig:dh-muf}
 \end{figure}

 To better assess the factorization scale dependence, we have considered also the case when $\mu_F$ is varied around its ``natural'' value $\sqrt{|\vec k_1| |\vec k_2|}$ by a factor $r$ taking values in the range 1/2 to four. In Fig.~\ref{fig:dh-muf}, as a selection of our results, we present the plots for $C_0$ and $C_1/C_0$ at a squared center-of-mass energy of 7 and 13~TeV
 for the rapidity region $Y\leq 4.8$ and the HKNS parametrization of the fragmentation functions.
 
 It is worth to note that the general features
 of our predictions for dihadron production are rather similar to those
 obtained earlier for the Mueller--Navelet jet process. Although the BFKL 
 resummation leads to the growth with energy of the partonic subprocess
 cross sections, the convolution of the latter with the proton PDFs makes
 the net effect of a decrease with $Y$ of our predictions. This is due to 
 the fact that, at larger values of $Y$, PDFs are probed effectively at
 larger values of $x$, where they fall very fast.
 For the dihadron azimuthal correlations we predict a decreasing behaviour
 with $Y$. That originates from the increasing amount of hard undetected
 parton radiation in the final state allowed by the growth of the partonic 
 subprocess energy. 
 
 \section{Numerical specifics}
 \label{sec:dihadron-numerics}

  \subsection{Used tools}
  \label{sub:dihadron-tools}
   
  All the numerical calculations presented
  in Section~\ref{sec:dihadron-kernel} and 
  in Section~\ref{sec:dihadron-NLA} 
  were performed in \textsc{Fortran}, 
  choosing a two-loop running coupling setup 
  with $\alpha_s\left(M_Z\right)=0.11707$ and five quark flavours.
  It is known that potential sources of uncertainty 
  could be due to the particular PDF and FF parameterisations used. 
  For this reason, preliminary tests were done by using 
  three different NLO PDF sets, expressly:  
  MSTW~2008~\cite{Martin:2009iq}, 
  MMHT~2014~\cite{Harland-Lang:2014zoa}, 
  and CT~2014~\cite{Dulat:2015mca}, 
  and convolving them with the three following NLO FF routines: 
  AKK~\cite{Albino:2008fy}, 
  DSS~\cite{deFlorian:2007aj,deFlorian:2007ekg}, 
  and HNKS~\cite{Hirai:2007cx}. 
  Our tests have shown no significant discrepancy 
  when different PDF sets are used in our kinematical range. 
  In view of this result, in the final calculations 
  the MSTW 2008 PDF set (which was successfully used 
  in various analyses of inclusive semi-hard processes at the LHC, 
  including our previous studies of Mueller--Navelet jets) 
  was selected, together with the FF interfaces mentioned above.
  The results with the DSS routine are not shown, 
  since they would be hardly distinguishable from those with 
  the HKNS parameterisation.

  Specific \textsc{CERN} program libraries~\cite{cernlib} 
  were used to evaluate the azimuthal coefficients 
  given in Eq.~(\ref{dh-eq}), 
  which requires a complicated 8-dimensional numerical integration 
  (the expressions for $\bar c^{(1)}_{1,2}$ contain 
  an additional longitudinal fraction integral in 
  comparison to the formul{\ae} for the LLA vertices, 
  given in Eqs.~(\ref{c1-dh}) 
  and~(\ref{c2-dh})).
  Furthermore, slightly modified versions of the \cod{Chyp}~\cite{chyp} 
  and \cod{Psi}~\cite{RpsiCody:1973} routines were used to calculate 
  the Gauss hypergeometric function $_2F_1$ 
  and the real part of the $\psi$ function, respectively.

  \subsection{Uncertainty estimation}
  \label{sub:dihadron-uncertainty}
   
  The most significant uncertainty comes from the numerical 
  4-dimensional integration over the two transverse momenta 
  $|\vec k_{1,2}|$, the rapidity $y_1$, and over $\nu$.  
  Its effect was directly estimated by \cod{Dadmul} 
  integration routine~\cite{cernlib}.
  The other three sources of uncertainty, which are respectively:
  the one-dimensional integration over the parton fraction $x$
  needed to perform the convolution between PDFs and FFs 
  in the LO/NLO impact factors (see Eq.~(\ref{c1-dh})~and~(\ref{c11-dh})),
  the one-dimensional integration over 
  the longitudinal momentum fraction
  $\zeta$ in the NLO impact factor correction (see Eqs.~(\ref{c11-dh})),
  and the upper cutoff in the numerical integrations over
  $|\vec k_{1,2}|$ and $\nu$, are negligible with respect to
  the first one. For this reason the error bars 
  of all predictions presented in this work 
  are just those given by the \cod{Dadmul} routine.

 \section{Summary} 
 \label{sec:dihadron-summary}
 
 We studied the inclusive dihadron production
 process at the LHC within the BFKL approach,   
 giving the first complete phenomenological predictions 
 for cross sections and azimuthal correlation momenta 
 in the full NLA approximation.
 We implemented the exact version of the BLM optimisation procedure, which 
 requires the choice of renormalisation scale $\mu_R=\mu_R^{BLM}$ such that   
 it makes completely vanish the NLA terms proportional to the QCD
 $\beta$-function. 
 This procedure leads to rather large values of the scale $\mu_R^{BLM}$ and
 it allows to minimise the size of the NLA corrections in our observables.
 We considered two center-of-mass energies, $\sqrt s = 7, 13$ TeV, 
 and two different ranges for the rapidity interval between the two hadrons
 in the final state, $Y \leq 4.8$ and $Y \leq 9.4$, 
 which are typical for the last CMS analyses.
 The first rapidity range we investigated, $Y\leq 4.8$, may look to 
 be not large enough for the dominance of BFKL dynamics. But we see, however, 
 that in this range there are large NLA BFKL corrections, thus indicating that
 the BFKL resummation is playing here a non-trivial role. To clarify the issue  
 it would be very interesting to confront our predictions with the results of 
 fixed-order NLO DGLAP calculations. But this would require new numerical 
 analysis in our semi-hard kinematical range, because the existing NLO DGLAP 
 results cover the hard kinematical range for the energies of fixed target 
 experiments, see for instance Refs.~\cite{Owens:2001rr,Almeida:2009jt}.
 
 As for the hadron's transverse momenta, we imposed the \emph{symmetric} lower
 cutoff: $|\vec k_{1,2}|\geq 5$~GeV. 
 Considering a region of lower hadron transverse momenta, 
 say $|\vec k_{1,2}| \geq 2$~GeV, would lead to even larger values of the cross
 sections. 
 But it should be noted that in our calculation we use the BFKL method together 
 with leading-twist collinear factorisation, which means that we 
 are systematically neglecting power-suppressed corrections. 
 Therefore, going to smaller transverse momenta we would enter a region 
 where higher-twist effects must be important. 
 
 The general features of our predictions for dihadron production are rather similar to those obtained earlier for the Mueller-Navelet jet process. In particular, we observe that the account of NLA BFKL terms leads to much less azimuthal angle decorrelation with increasing $Y$ in comparison to LLA BFKL calculations. As for the difference between the Mueller-Navelet jet and dihadron production processes, we would mention the fact that, contrary to the jets' case, the full account of NLA terms leads in dihadron production to an increase of our predictions for the cross sections in comparison to the LLA BFKL calculation.
 
 We considered the effect of using different parameterisation sets 
 for the PDFs and the FFs, that could potentially give rise 
 to uncertainties which, in principle, are not negligible. 
 We did some preliminary tests devoted to gauge 
 the effect of using different PDF routines, 
 showing that it leads to no significant difference in the results.
 Then, we investigated the $Y$-behaviour of our observables 
 by using two different FF parameterisations. 
 Our calculation with the AKK FFs gives bigger cross sections, while the
 difference between AKK and HKNS is small, since the FFs uncertainties are mostly
 wiped out in the azimuthal ratios.
 
 We studied the effect of using two different choices 
 for the factorisation scale, $\mu_F = \mu_R^{\rm BLM}$ and
 $(\mu_F)_{1,2} = |\vec k_{1,2}|$, whereas $\mu_R = \mu_R^{\rm BLM}$
 runs at BLM scales.
 We see  some  difference in predictions within these two approaches,
 especially for larger values of $Y$ and at the smaller value of the energy
  $\sqrt{s}=7 \ \rm{TeV}$. In this region, the kinematical restriction for
 the undetected hard gluon radiation may start to be important, requiring
 resummation of threshold double logarithms together with BFKL logarithms of energy.  
 In this case, the phase space available for gluon
 bremsstrahlung vanishes, so that only soft and collinear emission is allowed, resulting in large
 logarithmic corrections to the partonic cross section.
 This issue maybe a physical reason for the observed strong dependence on the
 factorisation scale choice in our pure BFKL approach, and it definitely
 deserves a further study.

\newpage
 
\setcounter{appcnt}{0}
\renewcommand{\theequation}{C.\arabic{appcnt}}
\setcounter{tmp}{3}
\clearpage
\hypertarget{app:hadron-nlo-if-link}{}
\chapter*{Appendix~C}
\vspace{-0.5cm} 
\noindent
{\Huge \bf NLO impact factor for the identified hadron}
\label{app:hadron-nlo-if}
\addcontentsline{toc}{chapter}{\numberline{\Alph{tmp}}
 NLO impact factor for the identified hadron}
\markboth{NLO impact factor for the identified hadron}{}
\markright{APPENDIX C}{}
\vspace{1.3cm} 

In this Appendix the expressions for the NLO coefficient functions $C_{ij}$ 
in Eq.~(\ref{c11-dh}) are given (see Ref.~\cite{hadrons} for further details). 
In particular, we have:

\bea
\stepcounter{appcnt}
&&
C_{gg}\left(x,\zeta\right) =  P_{gg}(\zeta)\left(1+\zeta^{-2\gamma}\right)
\ln \left( \frac {\vec k_h^2 x^2 \zeta^2 }{\mu_F^2 \alpha_h^2}\right)
-\frac{\beta_0}{2}\ln \left( \frac {\vec k_h^2 x^2 \zeta^2 }
{\mu^2_R \alpha_h^2}\right)
\\
&&
+ \, \delta(1-\zeta)\left[C_A \ln\left(\frac{s_0 \, \alpha_h^2}{\vec k^2_h \,
x^2 }\right) \chi(n,\gamma)
- C_A\left(\frac{67}{18}-\frac{\pi^2}{2}\right)+\frac{5}{9}n_f
\right.
\nonumber \\
&&
\left.
+\frac{C_A}{2}\left(\psi^\prime\left(1+\gamma+\frac{n}{2}\right)
-\psi^\prime\left(\frac{n}{2}-\gamma\right)
-\chi^2(n,\gamma)\right) \right]
\nonumber \\
&&
+ \, C_A \left(\frac{1}{\zeta}+\frac{1}{(1-\zeta)_+}-2+\zeta\bar\zeta\right)
\nonumber \\
&&
\times \,\left(\chi(n,\gamma)(1+\zeta^{-2\gamma})-2(1+2\zeta^{-2\gamma})\ln\zeta
+\frac{\bar \zeta^2}{\zeta^2}I_2\right)
\nonumber \\
&&
+ \, 2 \, C_A (1+\zeta^{-2\gamma})
\left(\left(\frac{1}{\zeta}-2+\zeta\bar\zeta\right) \ln\bar\zeta
+\left(\frac{\ln(1-\zeta)}{1-\zeta}\right)_+\right) \ ,
\nonumber
\eea

\bea
\stepcounter{appcnt}
&&
C_{gq}\left(x,\zeta\right) =  P_{qg}(\zeta)\left(\frac{C_F}{C_A}
+\zeta^{-2\gamma}\right)
\ln \left( \frac {\vec k_h^2 x^2 \zeta^2 }{\mu_F^2 \alpha_h^2}\right)
\\
&&
+ \, 2 \, \zeta \bar\zeta \, T_R \, \left(\frac{C_F}{C_A}+\zeta^{-2\gamma}
\right)
+\, P_{qg}(\zeta)\, \left(\frac{C_F}{C_A}\, \chi(n,\gamma)+2 \zeta^{-2\gamma}\,
\ln\frac{\bar\zeta}{\zeta} + \frac{\bar \zeta}{\zeta}I_3\right) \ ,
\nonumber
\eea

\bea
\stepcounter{appcnt}
&&
C_{qg}\left(x,\zeta\right) =  P_{gq}(\zeta)\left(\frac{C_A}{C_F}
+\zeta^{-2\gamma}\right)
\ln \left( \frac {\vec k_h^2 x^2 \zeta^2 }{\mu_F^2 \alpha_h^2}\right)
\\
&&
+ \zeta\left(C_F\zeta^{-2\gamma}+C_A\right)
+ \, \frac{1+\bar \zeta^2}{\zeta}\left[C_F\zeta^{-2\gamma}(\chi(n,\gamma)
-2\ln\zeta)
+2C_A\ln\frac{\bar \zeta}{\zeta} + \frac{\bar \zeta}{\zeta}I_1\right]
 \ ,
\nonumber
\eea

\bea
\stepcounter{appcnt}
&&
C_{qq}\left(x,\zeta\right) =  P_{qq}(\zeta)\left(1+\zeta^{-2\gamma}\right)
\ln \left( \frac {\vec k_h^2 x^2 \zeta^2 }{\mu_F^2 \alpha_h^2}\right)
-\frac{\beta_0}{2}\ln \left( \frac {\vec k_h^2 x^2 \zeta^2 }{\mu^2_R
\alpha_h^2}\right)
\label{final-e}
\\
&&
+ \, \delta(1-\zeta)\left[C_A \ln\left(\frac{s_0 \, \alpha_h^2}{\vec k^2_h \,
x^2 }\right) \chi(n,\gamma)
+ C_A\left(\frac{85}{18}+\frac{\pi^2}{2}\right)-\frac{5}{9}n_f - 8\, C_F
\right.
\nonumber \\
&&
\left.
+\frac{C_A}{2}\left(\psi^\prime\left(1+\gamma+\frac{n}{2}\right)
-\psi^\prime\left(\frac{n}{2}-\gamma\right)
-\chi^2(n,\gamma)\right) \right] + \, C_F \,\bar \zeta\,
(1+\zeta^{-2\gamma})
\nonumber \\
&&
+  \left(1+\zeta^2\right)\left[C_A (1+\zeta^{-2\gamma})\frac{\chi(n,\gamma)}
{2(1-\zeta )_+}
+\left(C_A-2\, C_F(1+\zeta^{-2\gamma})\right)\frac{\ln \zeta}{1-\zeta}
\right]
\nonumber\\
&&
+\, \left(C_F-\frac{C_A}{2}\right)\left(1+\zeta^2\right)
\left[2(1+\zeta^{-2\gamma})\left(\frac{\ln (1-\zeta)}{1-\zeta}\right)_+
+ \frac{\bar \zeta}{\zeta^2}I_2\right] \; ,
\nonumber
\eea
with the plus-prescription defined in Eq.~(\ref{plus-prescription}).

Here $\gamma=i\nu-1/2$, while 
$P_{i j}(\zeta)$ are leading
order DGLAP kernels defined 
in Appendix~\hyperlink{app:jet-nlo-if-link}{B}.
The expressions for the $I_{1,2,3}$ functions 
are given in Appendix~\hyperlink{app:jet-nlo-if-link}{B}.

\renewcommand{\theequation}
             {\arabic{chapter}.\arabic{equation}}
\chapter{Three-jet production}
\label{chap:3j}

In the last two Chapters we investigated semi-hard processes with two objects (jets or charged light hadrons) always tagged in the final state. We started from the expression
of the forward parton impact factors~\cite{Fadin:1999de,Fadin:1999df}, `opening' one of the integrations over the intermediate-state phase space to allow one parton to generate the detected object in the final state. Thus, we obtained the expressions for the process-dependent vertex, which have to be convoluted (Eq.~(\ref{sigma-ff})) with the universal Green's function in order to get the cross section for the considered processes. 

Inclusive multi-jet production represents a further step towards the study of BFKL dynamics in a much more exclusive way. 
While in the two-body case we modified the expression of the parton impact factors in order to allow the detection of two objects in the fragmentation region of the respective parent proton, in the $n$-body case we need to suitably generalise our formalism to account for the emission of extra particles in more central regions covered by the LHC detectors. 
The first advance in this direction, presented in this Chapter, is to propose new observables associated to the inclusive production of three jets: two of them are the original Mueller--Navelet jets, while the third one is tagged in central regions of rapidity.

When a jet central in rapidity is emitted in the final state, 
it is possible to single out an extra gluon emission by extracting its emission probability from the BFKL kernel. For further details and a more specific discussion, 
we refer to Refs.~\cite{Bartels:2006hg,Schwennsen:2007hs} and to the preliminary discussion given in Section~\ref{sub:3jets-nlk-3}, respectively.
The three-jet cross section can be constructed in this way:
\begin{align}
\label{sigma_3j_start}
&
\sigma^{\rm 3-jet}(s)=
\frac{1}{(2\pi)^{2}}
\int\frac{d^{2}\vec q_1}{\vec q_1^{\,\, 2}}
\int\frac{d^{2}\vec q_2}{\vec q_2^{\,\, 2}} 
\Phi_1(\vec q_1,s_0)
\Phi_2(-\vec q_2,s_0)
\\ & \nonumber \times \,
\int d^{2}\vec{q_A}
\int d^{2}\vec{q_B}
\int\limits^{\delta +i\infty}_{\delta
-i\infty}\frac{d\omega}{2\pi i}\left(\frac{s}{s_0}\right)^\omega
G_\omega (\vec q_1, \vec q_A) 
\\ & \nonumber \times \,
\Phi_C(-\vec q_A,\vec q_B,s_0)
\int\limits^{\delta +i\infty}_{\delta
-i\infty}\frac{d\omega^\prime}{2\pi i}\left(\frac{s}{s_0}\right)^{\omega^\prime}
G_\omega^\prime (\vec q_B, \vec q_2) \; ,
\end{align}
where $\Phi_{1,2}$ are the two impact factors which describe the two forward/backward jets (as in the Mueller--Navelet case), while $\Phi_C$ is the central-jet emission vertex~\cite{Bartels:2006hg}. By selecting one emission to be exclusive we have factorised the Green's function into two components. Each of them connects one of the external jets to the central one.

\begin{figure}[t]
  \centering
  \vspace{-.5cm}
  \includegraphics[scale=0.45]{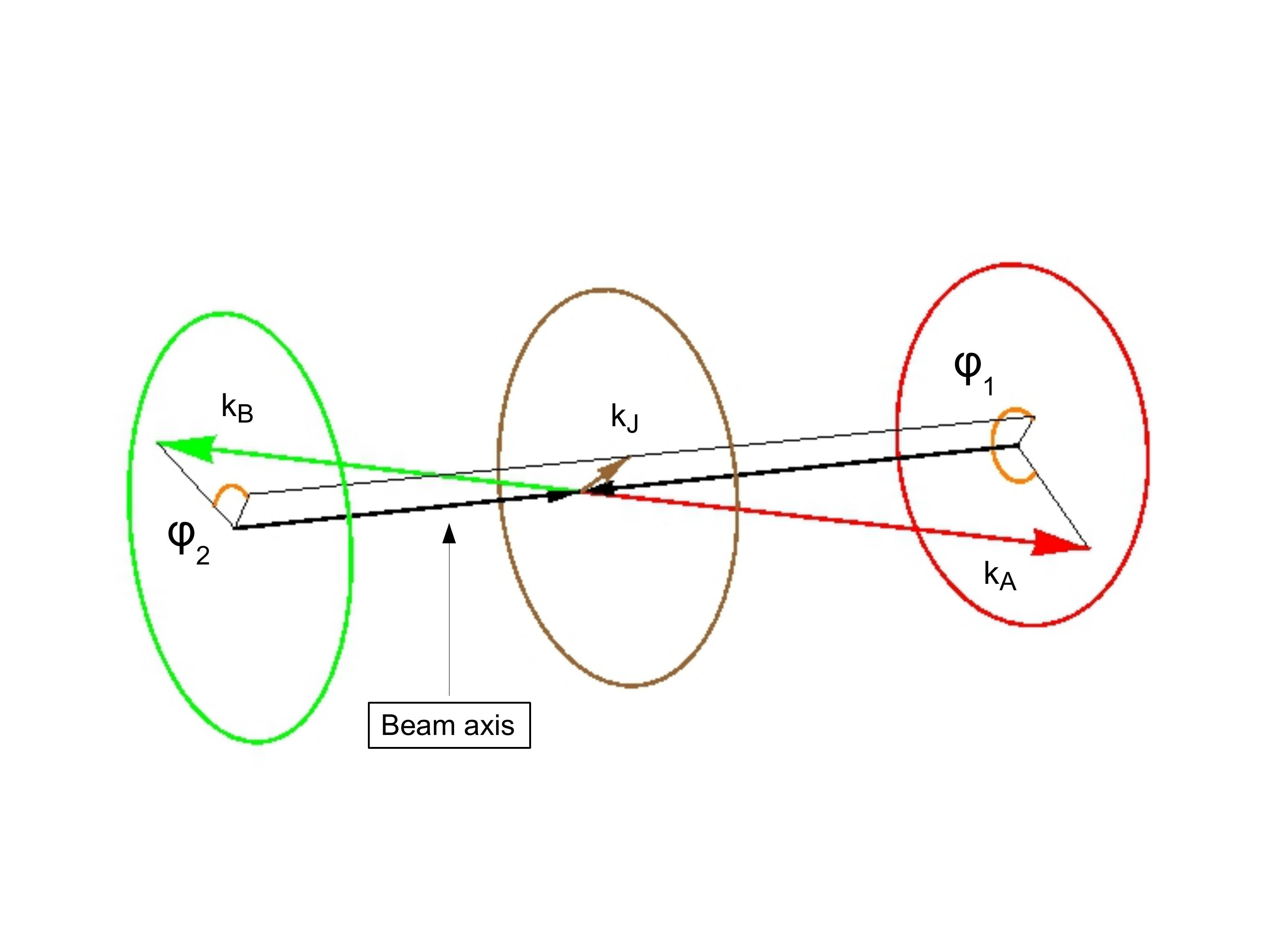}
  \caption[A three-jet event in a generic detector]
  {Representation of a three-jet event 
   in a generic detector. 
   All three circles are perpendicular to the beam axis.}
  \label{fig:3jdetector}
 \end{figure}

We will give predictions for the new azimuthal correlation momenta defined as
\begin{eqnarray}
{\cal R}^{M N}_{P Q} =\frac{ \langle \cos{(M \, \phiaj)} \cos{(N \, \phijb)} \rangle}{\langle \cos{(P \, \phiaj)} \cos{(Q \, \phijb)} \rangle} \, , 
\label{Rmnpq_intro}
\end{eqnarray}
where $\phiaj$ and $\phijb$ are, respectively, the azimuthal-angle difference between the first and the second (central) jet and between this one and the third jet (see Fig.~\ref{fig:3jdetector}). 
The distribution ratios defined in Eq.~(\ref{Rmnpq_intro}) generalise the $R_{MN}$ ratios typical of two-body final-state processes, as the previously discussed Mueller--Navelet jet (see Chapter~\ref{chap:mn-jets}) and dihadron (see Chapter~\ref{chap:dihadron}) production processes, by showing an extra dependence on transverse momentum and rapidity of the central jet. 
Cross sections are calculated using collinear factorisation to produce the two most forward/backward jets, taking the convolution of the partonic cross section, which follows the BFKL dynamics, with collinear PDFs included in the forward jet vertex. These two Mueller--Navelet jet vertices are linked to the centrally produced jet via two BFKL Green's functions. To simplify our predictions, we integrate over the momenta of all produced jets, using current LHC experimental cuts.

This Chapter is organised as follows: 
In Section~\ref{sec:3j-theory} the main formul{\ae} are given, including a first analysis at partonic level; in Section~\ref{sec:3j-hadronic} hadronic level predictions are presented, with the inclusion of NLA BFKL corrections together with BLM scales and for three different kinematical configurations 
(see Sections~\ref{sub:3jets-nlk-1}, \ref{sub:3jets-nlk-2} and \ref{sub:3jets-nlk-3}). The section Summary is given in~\ref{sec:4j-summary}.

The analysis given in this Chapter is based on 
the work done in Refs.~\cite{Caporale:2015vya,Caporale:2016soq,Caporale:2016zkc} and presented in 
Refs.~\cite{Celiberto:2016vhn,Caporale:2016oxl,Chachamis:2016lyi,Caporale:2016pqe,Caporale:2016djm,Caporale:2016lnh,Chachamis:2017:dis}.

 \section{A new way to probe BFKL} 
 \label{sec:3j-theory}
 

 \begin{figure}[t]
  \centering
  \includegraphics[scale=0.5]{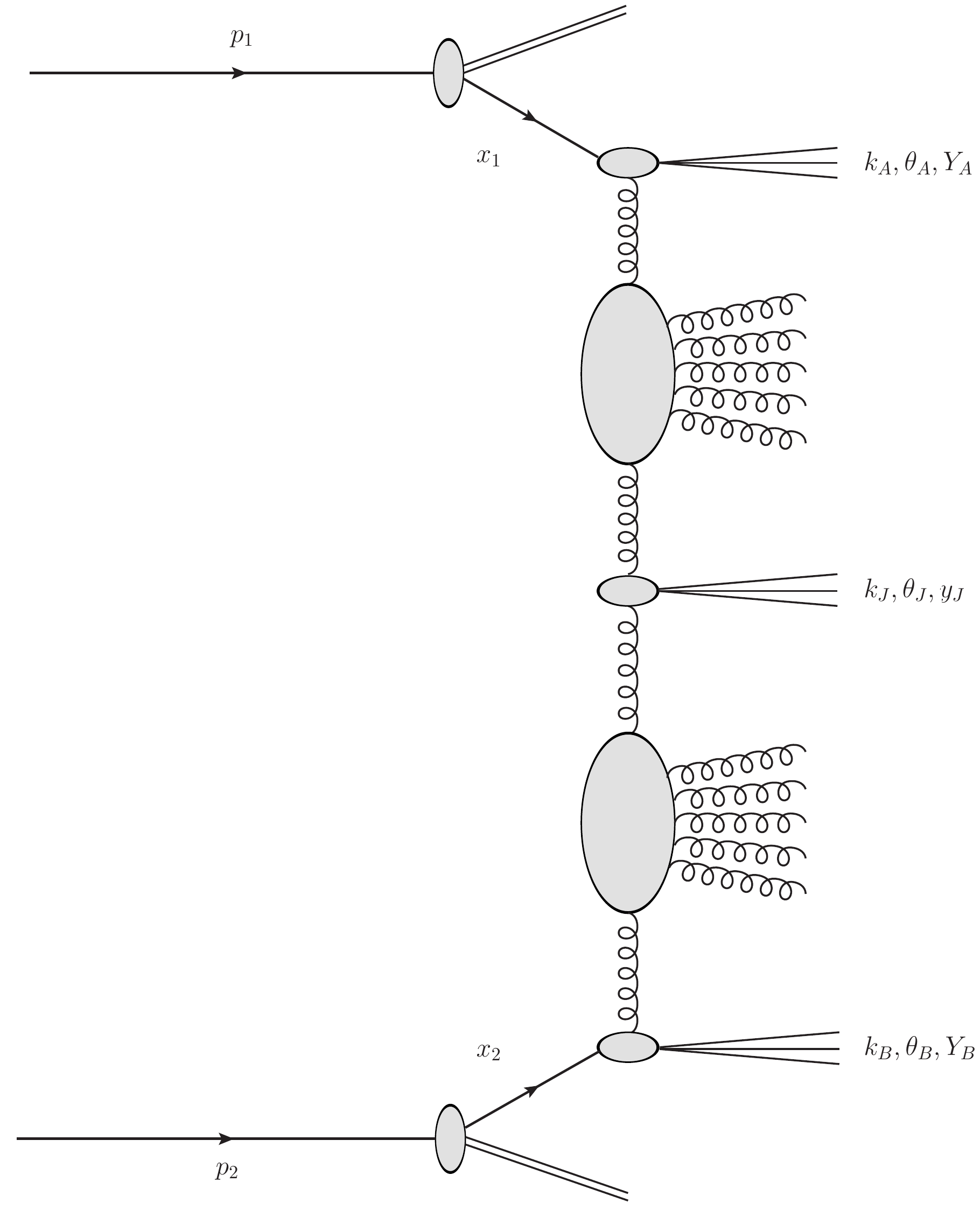}
  \caption[Inclusive three-jet production process 
           in multi-Regge kinematics]
   {Inclusive three-jet production process 
    in multi-Regge kinematics.}
  \label{fig:3j}
 \end{figure}

  \subsection{The three-jet cross section}
  \label{sub:3j-cs}
  
  The process under investigation 
  (see Figs.~\ref{fig:3jdetector} and~\ref{fig:3j})
  is the production of two forward/backward jets,
  both characterised by high transverse momenta $\vec{k}_{A,B}$ 
  and well separated in rapidity, 
  together with a third jet produced in the
  central rapidity region and with possible associated 
  minijet production. 
  This corresponds to 
  \begin{equation}
  \label{process-3j}
  {\rm p}(p_1)   \, + \, {\rm p} (p_2)  \, \to \,
  {\rm j}_A(k_A) \, + \, {\rm j}_C(k_J) \, + \, 
  {\rm j}_B(k_B) \, + \, {\rm minijets} \; ,
  \end{equation}
  where ${\rm j}_A$ is the forward jet 
  with transverse momentum $\vec{k}_A$
  and rapidity $Y_A$, ${\rm j}_B$ 
  is the backward jet with transverse momentum $\vec{k}_B$
  and rapidity $Y_B$ and ${\rm j}_C$ 
  is the central jet with transverse momentum $\vec{k}_J$
  and rapidity $y_J$.
  
  In collinear factorisation the cross section 
  for the process~(\ref{process-3j}) reads
  \begin{align}
  \label{dsigma_pdf_convolution}
   & \frac{d\sigma^{3-{\rm jet}}}
        {dk_A \, dY_A \, d\theta_A \, 
         dk_B \, dY_B \, d\theta_B \, 
         dk_J \, dy_J d\theta_J} 
   \\ \nonumber & 
   = 
   \sum_{r,s=q,{\bar q},g}\int_0^1 dx_1 \int_0^1 dx_2
   \ f_r\left(x_1,\mu_F\right)
   \ f_s\left(x_2,\mu_F\right) \;
   d\stjp\left(\hat{s},\mu_F\right) \;,
  \end{align}
  where the $r, s$ indices specify the parton types 
  (quarks $q = u, d, s, c, b$;
  antiquarks $\bar q = \bar u, \bar d, \bar s, \bar c, \bar b$; 
  or gluon $g$),
  $f_{r,s}\left(x, \mu_F \right)$ are the initial proton PDFs; 
  $x_{1,2}$ represent the longitudinal fractions 
  of the partons involved 
  in the hard subprocess; 
  $d\stjp\left(\hat{s}, \mu_F \right)$ 
  is the partonic cross section for the production of jets and
  $x_1x_2s \equiv \hat{s}$ is the squared center-of-mass energy
  of the hard subprocess (see Fig.~\ref{fig:3j}). 
  The BFKL dynamics enters in the cross section 
  for the partonic hard subprocess $d\stjp$ 
  (in the form of two forward Green's functions $\varphi$ 
  to be described in a while), which can be presented as 
  (from here we start to use the notation 
  $k_{A,B,Y} \equiv |\vec{k}_{A,B,Y}|$, which holds in the following):
  \begin{equation}
   \frac{d \stjp}{d k_J d\theta_J d y_J} =  
   \frac{\bar{\alpha}_s }{\pi k_J} 
   \int d^2 \vec{p}_A \int d^2 \vec{p}_B \, 
   \delta^{(2)} \left(\vec{p}_A 
   + \vec{k}_J- \vec{p}_B\right)
   \label{Onejetemission}
  \end{equation}
  \[ \times \,
   \varphi \left(\vec{k}_A,\vec{p}_A,Y_A - y_J\right) 
   \varphi \left(\vec{p}_B,\vec{k}_B,y_J - Y_B\right) 
   \; , 
  \]
  where $\bar \alpha_s = N_c/\pi \, \alpha_s$, with $N_c$ the number of colours in QCD.
  In order to lie within MRK, we have considered the ordering in the rapidity of the produced particles $Y_A > y_J > Y_B$, while $k_J^2$ is always 
  above the experimental resolution scale.
  $\varphi\left(\vec{p},\vec{q},0\right)$ is a suitable redefinition of the BFKL Green's function, which now encode also the momentum two-dimensional 
  delta coming from the LO jet function (Eq.~(\ref{jetF0})). 
  In this way, some pieces of the LO jet vertex (Eq.~(\ref{if-jet-lo})) are encoded in a very useful expression for the Green function, which holds at LLA. 
  
  It is possible algebraically manipulate the expression given 
  in Eq.~(\ref{Onejetemission}) in order to find distinct BFKL features.
  First of all, one can integrate the two-dimensional 
  delta function in Eq.~(\ref{Onejetemission}), to obtain
  \begin{align}
   \frac{d\stjp}{d^2 \vec{k}_J dy_J}
   =\frac{\bar{\alpha}_s}{\pi k_J^2}\int dp^2 d\theta \;\varphi(\vec{k}_A,\vec{p},Y_A-y_J)\varphi(\vec{p}+\vec{k}_J,\vec{k}_B,y_J-Y_B)
   \label{Onejetemission_nodelta} \; .
  \end{align}  
  Then, the Green's function can be expanded in Fourier components of the azimuthal angle to write:
  \begin{align}
   \frac{d\stjp}{d^2 \vec{k}_J dy_J}
   =\frac{\bar{\alpha}_s}{\pi k_J^2}
    \sum_{m,n=-\infty}^{+\infty} 
    e^{i(m\theta_A-n\theta_B)} \:\:        
    \Omega_{m,n}\left(\vec{k_A},\vec{k_B},Y_A,Y_B,\vec{k_J},\theta_J,y_J\right) \; ,
   \label{Onejetemission_omega_mn} 
  \end{align}
  where
  \begin{align}\label{omega_mn_1}
   &
   \Omega_{m,n}\left(\vec{k_A},\vec{k_B},Y_A,Y_B,\vec{k_J},\theta_J,y_J\right)
   \\ & \nonumber =  
   \int_0^{+\infty} dp \, p 
   \int_0^{2\pi} d\theta 
   e^{i \left\lbrack n\arctan\left
                     (\frac{p\sin\theta+k_J\sin\theta_J} 
                           {p\cos\theta+k_J\cos\theta_J}
                           \right)-m\theta \right\rbrack}
   \varphi_{m}\left(|\vec{k_A}|,|\vec{p_A}|,Y_A-y_J\right)
   \\ & \nonumber \times \,
   \varphi_{n} 
   \left(
    \sqrt{p^2+k_J^2+2 |\vec{p}| |\vec{k}_J|
           \cos\left(\theta-\theta_J\right)},
    |\vec{p_B}|,y_J-Y_B
   \right) \; .
  \end{align}
    Here $\varphi_{i}$ is the $i$-th azimuthal component 
  of the Green's function obtained after the projection on the 
  $(\nu,n)$-space, whose LLA and NLA expressions are given respectively 
  in Eqs.~(\ref{phinLO}) and~(\ref{phinNLO}).
  Using the relation 
  $\arctan \alpha = 
  \frac{i}{2} \ \ln\left(\frac{1-i\alpha}{1+i\alpha}\right)$, 
  which holds for any real $\alpha$, one has  
  \begin{align}\label{omega_mn_2}
   &
   \Omega_{m,n}\left(\vec{k_A},\vec{k_B},Y_A,Y_B,\vec{k_J},\theta_J,y_J\right)
   \\ & \nonumber = 
   \int_0^{+\infty} dp \, p 
   \int_0^{2\pi} d\theta
   \: e^{-im\theta} \,
   \left(
    \frac{p e^{i\theta}  + k_J e^{i\theta_J}} 
         {p e^{-i\theta} + k_J e^{-i\theta_J}}
   \right)^\frac{n}{2}
   \\ & \nonumber \times \,
   \varphi_{m}\left(|\vec{k_A}|,|\vec{p_A}|,Y_A-y_J\right)
  \\ & \nonumber \times \,
   \varphi_{n}
   \left(
    \sqrt{p^2+k_J^2+2 |\vec{p}| |\vec{k}_J|
           \cos\left(\theta-\theta_J\right)},
    |\vec{p_B}|,y_J-Y_B
   \right) \; .
  \end{align}
  The dependence on $\theta_J$ can be factorised out 
  by making the change of variable 
  $\theta - \theta_J \to \theta$. 
  The final expression for $\Omega_{m,n}$ reads:
  \begin{align}\label{omega_mn_final}
   &
   \Omega_{m,n}\left(\vec{k_A},\vec{k_B},Y_A,Y_B,\vec{k_J},\theta_J,y_J\right)
   = \\ & \nonumber
   \: e^{i(n-m)\theta_J} \,
   \int_0^{+\infty} dp \, p 
   \int_0^{2\pi} d\theta \:
   \frac{e^{-im\theta} \,
         \left(
          p e^{i\theta} + k_J  
         \right)^n} 
         {\sqrt{\left(p^2 + k_J^2 + 2 |\vec{p}| |\vec{k}_J| \cos\theta\right)^n}}
   \\ & \nonumber
   \varphi_{m}\left(|\vec{k_A}|,|\vec{p_A}|,Y_A-y_J\right)
   \varphi_{n}
   \left(
    \sqrt{p^2+k_J^2+2 |\vec{p}| |\vec{k}_J| \cos\theta},
    |\vec{p_B}|,y_J-Y_B
   \right) \; .
  \end{align}
  
  In Sections~\ref{sub:3j-1cos} and~\ref{sub:3j-2cos}
  we investigate the properties of two 
  new, generalised and suitable BFKL observables 
  that the developed formalism allows us to define.

  \subsection{One-cosine projection: \emph{\`A la} Mueller--Navelet}
  \label{sub:3j-1cos}

   \subsubsection{One-cosine azimuthal correlations}
   \label{ssub:3j-1cos-theory}
   
   The first step is to 
   integrate over the azimuthal angle 
   of the central jet 
   and over the difference in azimuthal angle 
   between the two forward jets, 
   $\Delta \phi = \theta_A - \theta_B - \pi$, 
   to define a quantity similar to the usual Mueller--Navelet case, 
   {\it i.e.},  
   \begin{align}
   &
   \int_0^{2 \pi} d \Delta \phi \, \cos{\left(M \Delta \phi \right)} 
   \int_0^{2 \pi} d \theta_J\,
   \frac{d\stjp}{d k_J d\theta_J d y_J} \\ & =  
   \frac{\bar{\alpha}_s}{2\pi} \sum_{L=0}^M 
    \int_{0}^\infty dp^2 \int_0^{2 \pi} d \theta 
    \frac{(-1)^M \left( \begin{array}{c}\hspace{-.2cm}M \\
   \hspace{-.2cm}L\end{array} \hspace{-.18cm}\right)
    \left(k_J^2\right)^{\frac{L-1}{2}}  
    \, \left(p^2\right)^{\frac{M-L}{2}}
    \cos{\left(L \, \theta\right)}}
    { \sqrt{\left(p^2 + k_J^2+ 2 |\vec{p}| |\vec{k}_J| 
    \cos{\theta}\right)^{M}}}\nonumber\\ &  
    \times \,\varphi^{\rm (LLA)}_{M} \left(|\vec{k_A}|,|\vec{p}|,Y_A-y_J\right)
   \nonumber\\ &
    \times \,\varphi^{\rm (LLA)}_{M} \left(\sqrt{p^2+ k_J^2 + 2 |\vec{p}| |\vec{k}_J|} 
   \cos{\theta},|\vec{k_B}|,y_J-Y_B\right) \; ,
   \nonumber
   \end{align}
  where $\varphi^{\rm (LLA)}_{n}$ is the Green's function at LLA
  \begin{align}\label{phinLO}
   \varphi^{\rm (LLA)}_{n} \left(|\vec{k}|,|\vec{q}|,y\right) \; &= \; 
   2 \, \int_0^\infty d \nu   
   \cos{\left(\nu \ln{\frac{k^2}{q^2}}\right)}  
   \frac{e^{\bar{\alpha}_s  \chi(n,\nu) y}}
        {\pi \sqrt{k^2 q^2} } \; .
   \end{align}
   It is worth to note that, since the LO jet vertex (Eq.~(\ref{c1})) does not depend on $\nu$, 
   we can let the Green's function encode the $\nu$-integration.
   This second, suitable redefinition of the Green's function 
   is no more valid when NLO jet vertices (Eq.~(\hyperlink{c11-jets}{B.1})) are considered.
   One of the experimental observables 
   we want to highlight here corresponds to the mean 
   value of the cosine of $\Delta \phi$ in the recorded events:
   \begin{eqnarray}
   \langle \cos{\left(M \left(\Delta \phi\right) \right)} 
   \rangle_{d\stjp} = 
   \frac{\int_0^{2 \pi} d \Delta \phi \, 
   \cos{\left(M \Delta \phi \right)} 
   \int_0^{2 \pi} d \theta_J\, 
   \frac{d \stjp}{d^2 \vec{k}_J d y_J}}
   {\int_0^{2 \pi} d \Delta \phi 
   \int_0^{2 \pi} d \theta_J\,
   \frac{d \stjp}{d^2 \vec{k}_J d y_J}} \; .
   \end{eqnarray}
   The perturbative stability 
   (including renormalisation scale dependence) 
   of the mean value defined above 
   can be significantly improved (see Ref.~\cite{Caporale:2013uva} 
   for a related discussion) if the contribution coming from the zero conformal spin, which corresponds 
   to the index $n=0$ in Eq.~(\ref{phinLO}), is removed. 
   This can be achieved by defining the ratios
   \begin{equation}
   {\cal R}_{N}^M =
   \frac{\langle \cos{\left(M \left(\Delta \phi\right) \right)} 
         \rangle_{d\stjp}}
   {\langle \cos{\left(N \left(\Delta \phi\right) \right)} 
         \rangle_{d\stjp}} \; ,
   \label{RMN}
   \end{equation}
   where we consider $M,N$ as positive integers.

   \subsubsection{Numerical analysis}
   \label{ssub:3j-1cos-results}
   
   The observables defined in Eq.~(\ref{RMN}) 
   allow us to perform different kinds of analysis 
   at the partonic level.
   A first study, 
   where the transverse momenta 
   of the forward jets are fixed 
   to $k_A=35$ GeV and $k_B = 38$ GeV, 
   done in Ref.~\cite{Caporale:2015vya}, is presented in this Section. 
   
   \begin{figure}[t]
    \centering
     \includegraphics[scale=0.35]{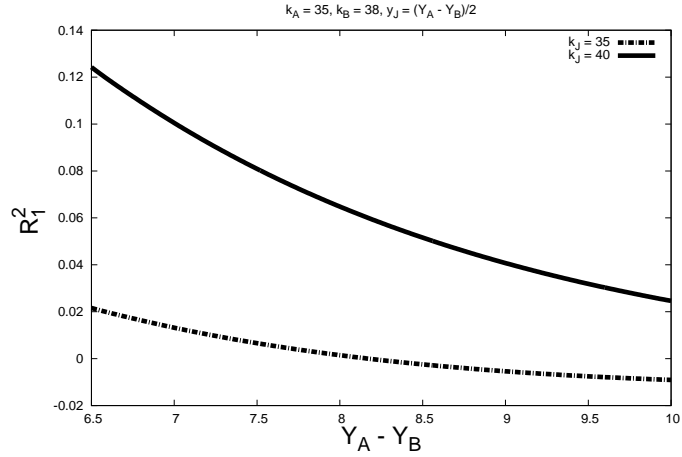}
    \caption[\emph{\`A la} Mueller--Navelet observables
             in three-jet production]
     {A study of the ratio ${\cal R}_{1}^2$ 
      as defined in Eq.~(\ref{RMN}) 
      for fixed values of the $p_t$ 
      of the two forward jets and two values of the $p_t$ 
      of the tagged central jet, 
      as a function of the rapidity difference 
      between the two forward jets for the rapidity 
      of the central jet chosen as $y_J = (Y_A - Y_B)/2$.}
    \label{R21}
   \end{figure}
   
   The rapidity of the central jet is also fixed 
   to be one half of the rapidity difference 
   between the two forward jets: $y_J = (Y_A - Y_B)/2$ 
   simply because this allows us to connect 
   with the well-known Mueller--Navelet jets. 
   In this way one can study, {\it e.g.}, 
   the behaviour of 
   the ratio ${\cal R}_{1}^2$ in Fig.~\ref{R21} 
   for two values of the transverse momentum 
   of the central jet $k_J = 35, 40$ GeV.
   We see that this ratio decreases 
   as a function of $Y_A-Y_B$. 
   This is a consequence of the increase of the available phase space 
   for inclusive minijet radiation 
   and that the $n=1$ component decreases 
   which energy slower that the $n=2$ one. 

   In the BFKL formalism one has that the larger the $n$ 
   the slower the evolution with rapidity differences. 
   This is very important since it is distinct 
   from other approaches where QCD coherence 
   is introduced as it was shown in Ref.~\cite{Chachamis:2011rw}.

  \subsection[Two-cosine projection: generalised correlations]
             {Two-cosine projection: generalised azimuthal correlations}
  \label{sub:3j-2cos}

   \subsubsection{Two-cosine azimuthal correlations}
   \label{ssub:3j-2cos-theory}

   In this Section new observables,
   whose associated distributions have a very different behaviour 
   to the ones characteristic of the Mueller--Navelet case, are proposed. 
   These new distributions are defined using the projections 
   on the two relative azimuthal angles formed 
   by each of the forward jets with the central jet, 
   \begin{align}\label{phi_ajb}
    \phiaj &= \theta_A - \theta_J - \pi \\\nonumber 
    \phijb &= \theta_J - \theta_B - \pi \; ,
   \end{align}
   in the form
   \begin{equation}
   \int_0^{2 \pi} d \theta_A \int_0^{2 \pi} d \theta_B 
   \int_0^{2 \pi} d \theta_J 
   \cos{\left(M \phiaj \right)} \cos{\left(N \phijb \right)}
   \frac{d^3 \sigma^{3-{\rm jet}}}{d k_J d\theta_J d y_J} \end{equation}
   \[ =
   \bar{\alpha}_s \sum_{L=0}^{N} 
   \left( \begin{array}{c}
   \hspace{-.2cm}N \\
   \hspace{-.2cm}L\end{array} \hspace{-.18cm}\right)
    \left(k_J^2\right)^{\frac{L-1}{2}} 
   \int_{0}^\infty d p^2 \, \left(p^2\right)^{\frac{N-L}{2}} \]
   \[ \times \,
   \int_0^{2 \pi}  d \theta    \frac
   {   (-1)^{M+N} \cos{ \left(M \theta\right)} 
   \cos{\left((N-L) \theta\right)}
   }{
    \sqrt{\left(p^2 + k_J^2+ 2 |\vec{p}| |\vec{k}_J| 
    \cos{\theta}\right)^{N}}
   }
   \]
   \[ \times \,
   \varphi^{\rm (LLA)}_{M} \left(|\vec{k_A}|,|\vec{p}|,Y_A-y_J\right)
   \]
   \[ \times \,
   \varphi^{\rm (LLA)}_{N} \left(\sqrt{p^2+ k_J^2 + 2 |\vec{p}| |\vec{k}_J|
   \cos{\theta}},|\vec{k_B}|,y_J-Y_B\right) \; .
   \nonumber
   \]
   
   The experimentally relevant observable is the mean value 
   in the selected events of the two cosines, {\it i.e.}
   \begin{equation}
   \langle \cos{\left(M \phiaj \right)}  
   \cos{\left(N \phijb \right)}
   \rangle_{d\stjp}
   \end{equation}
   \[ =
   \frac{\int_0^{2 \pi} d \theta_A 
   d \theta_B d \theta_J \cos{\left(M \phiaj \right)}  
   \cos{\left(N \phijb \right)}
   \frac{d^3 \sigma^{3-{\rm jet}}}{d^2 \vec{k}_J d y_J} }
   {\int_0^{2 \pi} d \theta_A d \theta_B d \theta_J 
   \frac{d^3 \sigma^{3-{\rm jet}}}{d^2 \vec{k}_J d y_J} } \; .
   \]
   As done in Section~\ref{ssub:3j-1cos-theory}, in order to have optimal perturbative convergence 
   and eliminate collinear contamination, 
   one can remove the contributions from zero conformal spin 
   by defining the ratios:
   \begin{eqnarray}
   {\cal R}_{PQ}^{MN} &=& \frac{\langle \cos{\left(M \phiaj \right)}  
   \cos{\left(N \phijb \right)}
   \rangle_{d\stjp}}{\langle \cos{\left(P \phiaj \right)}  
   \cos{\left(Q \phijb \right)}
   \rangle_{d\stjp}}
   \label{RMNPQ}
   \end{eqnarray}
   and consider $M,N,P,Q >0$ as integer numbers.

   \subsubsection{Numerical analysis}
   \label{ssub:3j-2cos-results}
   
\begin{figure}[H]
   \centering
    \vspace{-0.5cm}
     \includegraphics[scale=0.35]{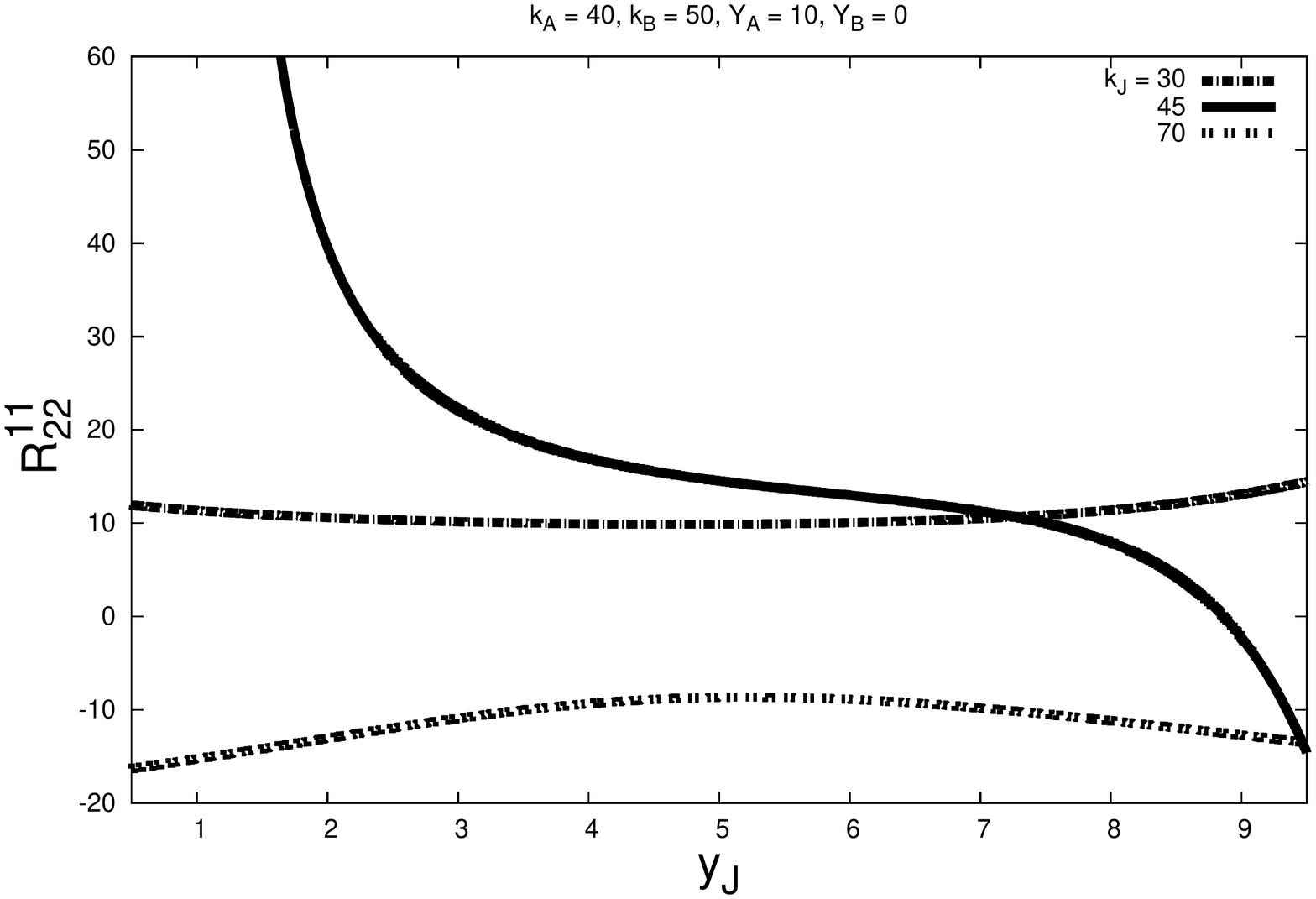}\\
    \vspace{-1.2cm}
     \includegraphics[scale=0.35]{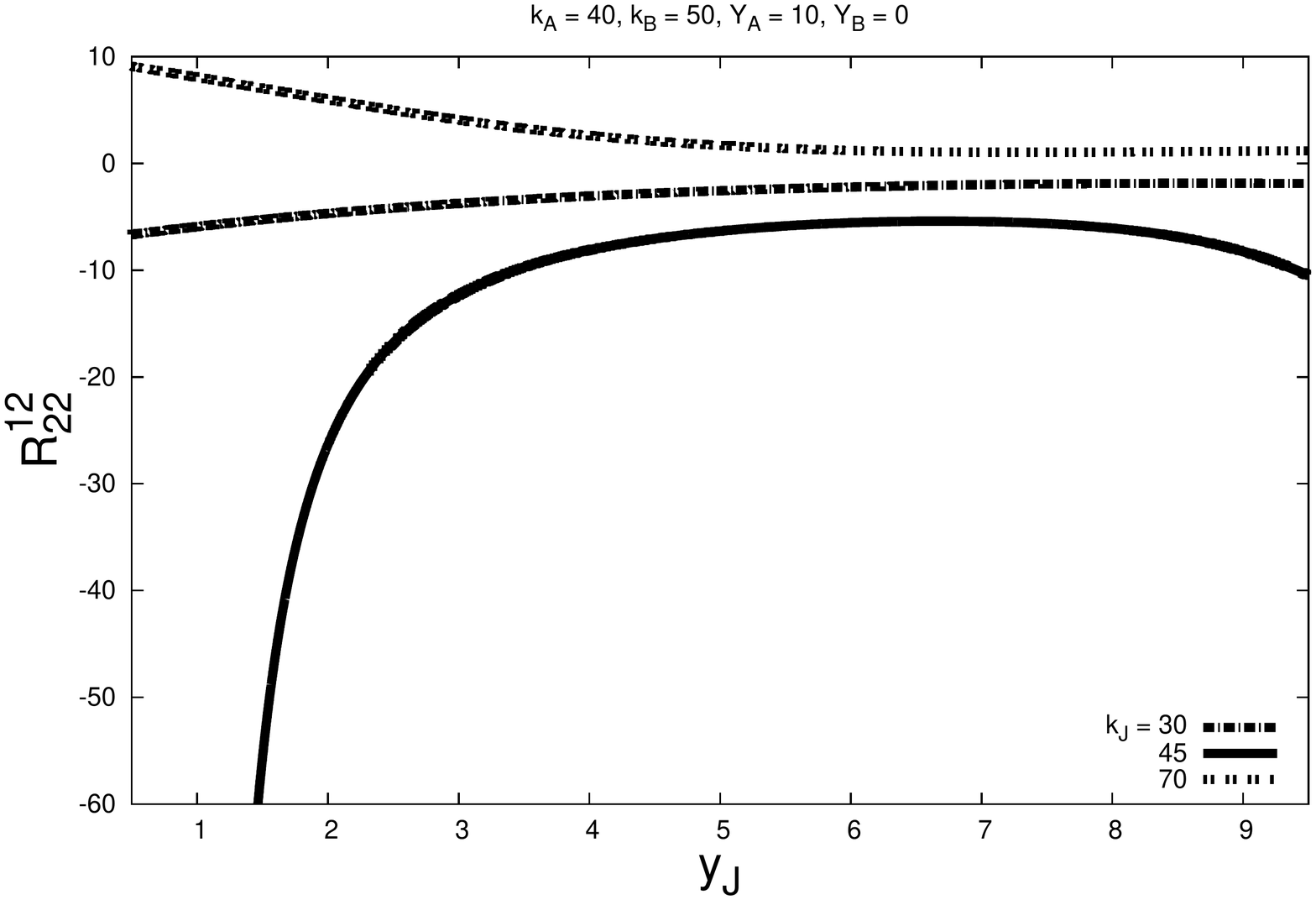}\\
    \vspace{-1.2cm}
     \includegraphics[scale=0.35]{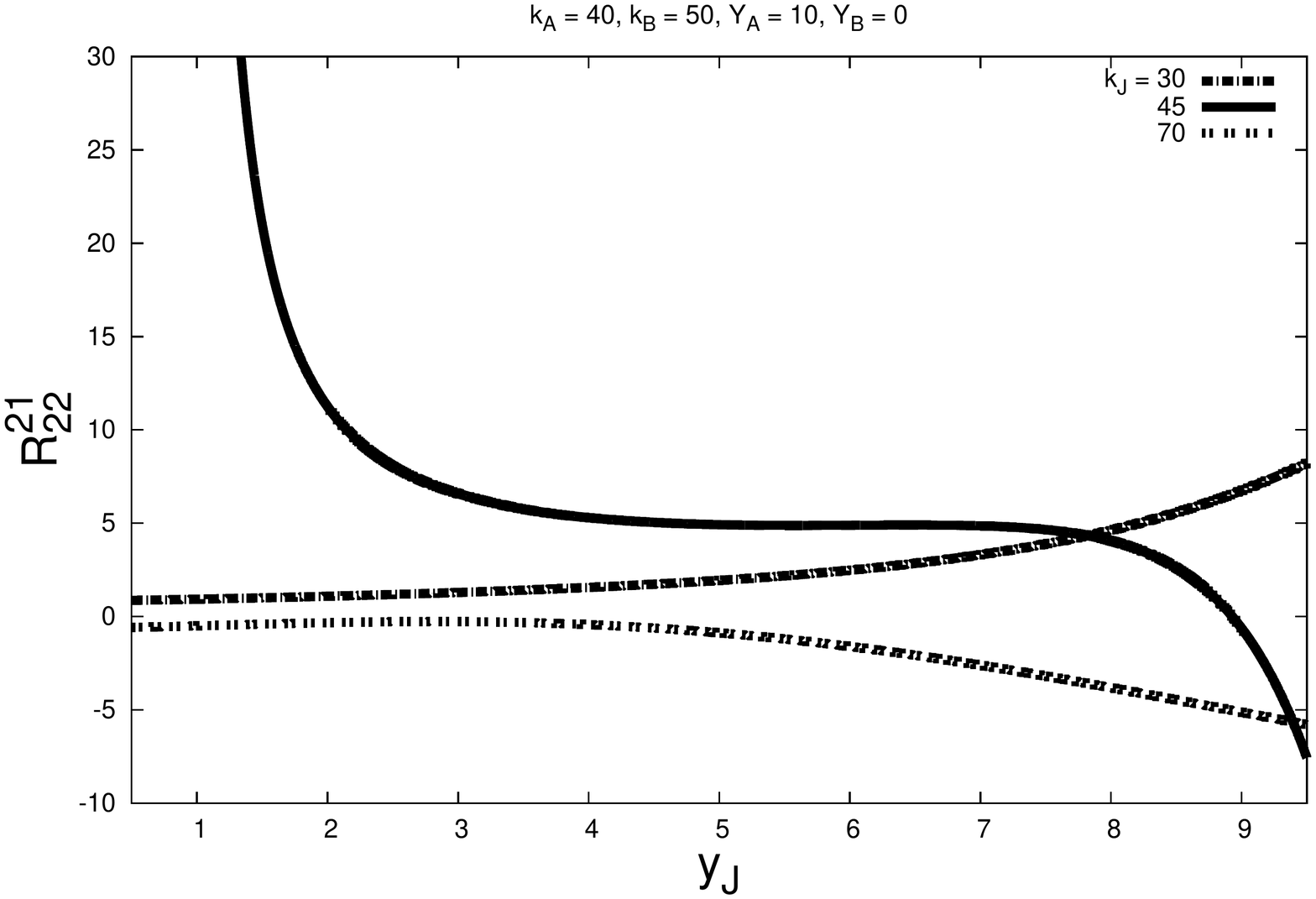}
    \vspace{-0.7cm}
    \caption[Partonic three-jets: BFKL azimuthal ratios]
    {A study of the ratios ${\cal R}_{22}^{11}$, 
     ${\cal R}_{22}^{12}$ and ${\cal R}_{22}^{21}$ 
     as defined in Eq.~(\ref{RMNPQ}) 
     for fixed values of the $p_t$ of the two forward jets 
     and three values of the $p_t$ of the tagged central jet, 
     as a function of the rapidity of the central jet $y_J$.}
    \label{RMNPQfigure}
   \end{figure}

   Working with fixed kinematics, it is possible to investigate many momenta configurations 
   (see Ref.~\cite{Caporale:2015vya}). 
   As an example, the ratios 
   ${\cal R}_{22}^{11}$, 
   ${\cal R}_{22}^{12}$ and ${\cal R}_{22}^{21}$ 
   are presented in Fig.~\ref{RMNPQfigure},
   the momenta of the forward jets 
   being fixed 
   to $k_A=40$ GeV and $k_B=50$ GeV and 
   their rapidities to $Y_A=10$ and $Y_B=0$. 
   For the transverse momentum of the central jet 
   the three values $k_J= 30, 45, 70$ GeV 
   are chosen, 
   the rapidity of the central jet $y_J$ 
   is allowed to take values in the range 
   in between the two rapidities of the forward jets. 
   These distributions are proving 
   the fine structure of the QCD radiation 
   in the high-energy limit. They gauge the relative weights 
   of each conformal spin contribution to the total cross section.
   
   \newpage
   
 \section{Hadronic level predictions} 
 \label{sec:3j-hadronic}
 
 In Section~\ref{sec:3j-theory} the theoretical setup 
 for the three-jet production in the BFKL approach was built up, and a first study at the partonic level was given. Here we continue and extend our analysis by giving predictions for the generalised azimuthal correlations at the hadronic level. The inclusion of the NLA contributions coming from the higher correction to the BFKL kernel is considered.

  \subsection[A more phenomenological analysis: inclusion of NLA]
             {A more phenomenological analysis: 
              inclusion of NLA kernel corrections} 
  \label{sub:3j-nlk}

  Using the definition of the LO jet vertex (Eq.~(\ref{c1})), 
  the cross section 
  given in Eq.~(\ref{dsigma_pdf_convolution}) 
  can be rewritten, according to Eq.~(\ref{sigma_3j_start}), as 
  \begin{equation}
   \frac{d\sigma^{3-{\rm jet}}}
        {dk_A \, dY_A \, d\theta_A \, 
         dk_B \, dY_B \, d\theta_B \, 
         dk_J \, dy_J d\theta_J} 
  \end{equation}
  \[ =
   \frac{8 \pi^3 \, C_F \, \asb^3}{N_c^3} \, 
   \frac{x_{J_A} \, x_{J_B}}{k_A \, k_B \, k_J} \,
   \int d^2 \vec{p}_A \int d^2 \vec{p}_B \,
   \delta^{(2)} \left(\vec{p}_A + \vec{k}_J- \vec{p}_B\right) \,
  \]
  \[ \times \, 
   \left(\frac{N_c}{C_F}f_g(x_{J_A},\mu_F)
   +\sum_{r=q,\bar q}f_r(x_{J_A},\mu_F)\right) \,
  \]
  \[ \times \, 
   \left(\frac{N_c}{C_F}f_g(x_{J_B},\mu_F)
   +\sum_{s=q,\bar q}f_s(x_{J_B},\mu_F)\right)
  \]
  \[ \times \,
   \varphi \left(\vec{k}_A,\vec{p}_A,Y_A - y_J\right) 
   \varphi \left(\vec{p}_B,\vec{k}_B,y_J - Y_B\right) \, .
  \]
  In MRK characteristic ordering in rapidity 
  is achieved by imposing that $Y_A > y_J > Y_B$, while $k_J^2$ is always 
  above the experimental resolution scale.
  $x_{J_{A,B}}$ are the longitudinal momentum fractions
  of the two external jets, linked to the respective rapidities 
  $Y_{{A,B}}$ by the relation 
  $x_{{A,B}} = k_{A,B} \, e^{\, \pm \, Y_{{A,B}}} / \sqrt{s}$. 

\begin{figure}[t]
    \includegraphics[scale=0.35]{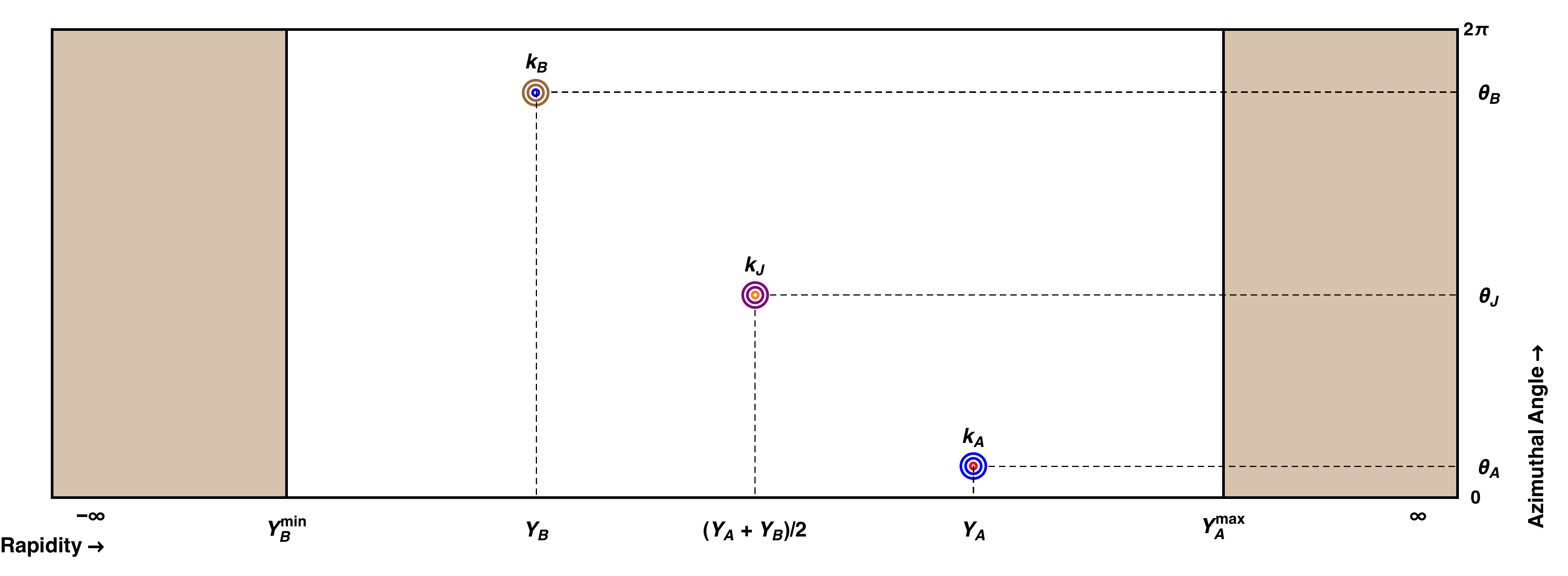}
   \caption[Lego plot for a three-jet event with fixed kinematics
            for the central jet rapidity]
    {A primitive lego plot 
     depicting a three-jet event. $k_A$ 
     is a forward jet with large positive
     rapidity $Y_A$ and azimuthal angle $\theta_A$, $k_B$ 
     is a forward jet with large negative
     rapidity $Y_B$ and azimuthal angle $\theta_B$ and $k_J$ 
     is a central jet with rapidity $y_J$ 
     and azimuthal angle $\theta_J$. 
     The fade-brown areas to the left and right
     highlight the regions in rapidity 
     which are not covered by the standard detectors.
    }
  \label{fig:lego1}
  \end{figure}

  Our goal is to study observables for which the BFKL approach will be distinct from other formalisms and also 
  rather insensitive to possible higher-order corrections. 
  Following the course taken in Section~\ref{sub:3j-2cos}, 
  we focus on new quantities 
  whose associated distributions are different from the ones which
  characterise the Mueller--Navelet case, though still related 
  to the azimuthal-angle correlations by projecting 
  differential cross section
  on the two relative azimuthal angles between each external jet
  and the central one $\Delta\phi_{\widehat{AJ},\widehat{JB}}$ 
  defined in Eq.~(\ref{phi_ajb})
  (see also Fig.~\ref{fig:lego1}). 
  Taking into account the factors coming from the jet vertices, 
  it is possible to rewrite the
  projection of the differential cross section 
  on the azimuthal-angle differences
  in the form
  \begin{align}
   \label{lo-nlo}
   & \int_0^{2 \pi} d \theta_A \int_0^{2 \pi} d \theta_B \int_0^{2 \pi} 
   d \theta_J \cos{\left(M \phiaj \right)} \,
              \cos{\left(N \phijb \right)}\\
   &  
   \frac{d\sigma^{3-{\rm jet}}}
        {dk_A \, dY_A \, d\theta_A \, 
         dk_B \, dY_B \, d\theta_B \, 
         dk_J \, dy_J d\theta_J}  
   \nonumber \\
   & =  
   \frac{8 \pi^4 \, C_F \, \asb^3}{N_C^3} \, 
   \frac{x_{J_A} \, x_{J_B}}{k_A \, k_B} 
   \left(\frac{N_C}{C_F}f_g(x_{J_A},\mu_F) \,
   +\sum_{r=q,\bar q}f_r(x_{J_A},\mu_F)\right) \,
   \nonumber \\ 
   & \times \,
   \left(\frac{N_C}{C_F}f_g(x_{J_B},\mu_F)
   +\sum_{s=q,\bar q}f_s(x_{J_B},\mu_F)\right) \, 
   \sum_{L=0}^{N} 
   \left( \begin{array}{c}
   \hspace{-.2cm}N \\
   \hspace{-.2cm}L\end{array} \hspace{-.18cm}\right)
   \left(k_J^2\right)^{\frac{L-1}{2}}
   \nonumber \\ 
   & \times \,
   \int_{0}^\infty d p^2 \, \left(p^2\right)^{\frac{N-L}{2}} \,
   \int_0^{2 \pi}  d \theta    \frac
   {(-1)^{M+N} \cos{ \left(M \theta\right)} \cos{\left((N-L) \theta\right)}}
   {\sqrt{\left(p^2 + k_J^2+ 2 |\vec{p}| |\vec{k}_J| \cos{\theta}\right)^{N}}}
   \nonumber \\ 
   & \times \,
   \varphi^{(LLA,NLA)}_{M} \left(|\vec{k_A}|,|\vec{p}|,Y_A-y_J\right)
   \nonumber \\ 
   & \times \, 
   \varphi^{(LLA,NLA)}_{N} \left(\sqrt{p^2+ k_J^2 + 2 |\vec{p}| |\vec{k}_J| \cos{\theta}},vec{k_B},y_J-Y_B\right). \nonumber
  \end{align}
  In this expression the Green's function 
  is either at LLA ($\varphi^{\rm (LLA)}$), 
  whose expression is given in Eq.~(\ref{phinLO}),  
  or at NLA ($\varphi^{\rm (NLA)}$) accuracy. 
  In particular, at NLA it reads
  \begin{equation}
   \varphi^{(NLA)}_{n} \left(|\vec{k}|,|\vec{q}|,y\right)  
   =2 \int_0^\infty d \nu   
   \cos{\left(\nu \ln{\frac{k^2}{q^2}}\right)}  
   \frac{e^{\bar{\alpha}_s \left(  \chi(n,\nu) 
   + \bar{\alpha}_s \chi^{(1)}(n,\nu)  \right) Y}}
        {\pi \sqrt{k^2 q^2} } \, , \label{phinNLO}
  \end{equation}
  with $\chi(n,\nu)$ and $\chi^{(1)}(n,\nu)$ given in Eqs.~(\ref{KLLA}) and~(\ref{ch11}), respectively. 
  
  The experimental observables we initially proposed 
  are based on the partonic-level
  average values (with $M,N$ being positive integers)
  \begin{equation}
  \label{Cmn}
   {\cal C}_{MN} \, = \,
   \langle \cos{\left(M \phiaj \right)}  
   \cos{\left(N \phijb \right)}
   \rangle \,
  \end{equation}
  \[ =
    \frac{\int_0^{2 \pi} 
   d \theta_A d \theta_B d \theta_J 
   \cos{\left(M \phiaj \right)}  
   \cos{\left(N \phijb \right)}
   d\sigma^{3-{\rm jet}} }{\int_0^{2 \pi} 
   d \theta_A d \theta_B d \theta_J 
   d\sigma^{3-{\rm jet}} } \; ,
  \]
  whereas,
  in order to provide testable predictions for
  the current and future experimental data, we introduce 
  the hadronic-level values $C_{MN}$ 
  after integrating ${\cal C}_{M,N}$ 
  over the momenta of the tagged jets,
  as we will see in the following Sections.
  
  From a more theoretical perspective, 
  it is important to have as good as possible perturbative 
  stability in our 
  predictions (see also Section~\ref{ssub:3j-2cos-theory}). 
  This can be achieved by removing the contribution stemming  
  from the zero conformal spin, 
  which corresponds to the index $n=0$ 
  in Eqs.~(\ref{phinLO}) and~(\ref{phinNLO}).
  We, therefore, introduce the ratios
  \begin{eqnarray}
  \label{RPQMN}
  R_{PQ}^{MN} \, = \, \frac{C_{MN}}{C_{PQ}}
  \label{RmnqpNew}
  \end{eqnarray}
  which are free from any $n=0$ dependence, 
  as long as $M, N , P, Q >0$. 
  The postulate that Eq.~(\ref{RmnqpNew}) generally describes 
  observables with good perturbative 
  stability is under scrutiny 
  in Sections~\ref{sub:3jets-nlk-1}, 
  \ref{sub:3jets-nlk-2}, and \ref{sub:3jets-nlk-3}, 
  where we compare LLA and NLA results.

  \subsection{BLM scale setting}
  \label{sub:3j-blm}
  
  In order to make an appropriate choice 
  of the renormalisation scale $\mu_R$, 
  the BLM prescription \cite{BLM,BLM_2,BLM_3,BLM_4,BLM_5} is used. 
  As explained in Section~\ref{sec:bfkl_blm}, it consists of using the MOM scheme and choosing the scale $\mu_R$ 
  such that the $\beta_0$-dependence of a given observable vanishes.
  Applying the BLM prescription leads to the modification 
  of the exponent in Eq.~(\ref{phinNLO}) in the following way:
  \begin{equation}
  \bar{\alpha}_s \left(  \chi(n,\nu) 
  + \bar{\alpha}_s \chi^{(1)}(n,\nu)  \right) Y \,\, \to \,\, 
  \bar{\alpha}_s \left(  \chi(n,\nu) 
  \left(1 + \frac{\alpha_s}{\pi} T  \right) 
  + \bar{\alpha}_s \chi^{(1)}(n,\nu)  \right) Y \, ,
  \end{equation}
  with $T$ given in~(\ref{T_Tbeta_Tconf}). 
  Note that this way to implement BLM is a generalisation 
  of the case (b), given in Eq.~(\ref{caseb}).  
  In the three-jet case, we remove the $\beta_0$ dependent factors from 
  the NLA objects present in Eq.~(\ref{lo-nlo}), 
  \emph{i.e.} the NLA Green's functions.
  
  Following this procedure, the renormalisation scale $\mu_R$ 
  is fixed at the value
  \begin{equation}
   ( \mu_R^{\rm BLM})^2=k_{A}k_{B}\ 
   \exp\left[\frac{1+4I}{3}+\frac{1}{2} 
   \chi(n,\nu)\right] \, .
  \end{equation}
  In our numerical analysis we consider two cases. 
  In one, we set $\mu_R=\mu_R^{\rm BLM}$ 
  only in the exponential factor of the Green's function 
  $\varphi_n$, while we let the argument of the 
  $\bar{\alpha}_s^3$ in Eq.~(\ref{lo-nlo}) 
  to be at the ``natural'' scale $\sqrt{k_{A}k_{B}}$, that is,
  $\bar{\alpha}_s^3 \sqrt{k_{A}k_{B}}$.
  In the second case, we fix 
  $\mu_R=\mu_R^{\rm BLM}$ everywhere in Eq.~(\ref{lo-nlo}). 
  These two cases lead 
  in general to two different but similar values for our 
  NLA predictions and wherever we present plots
  we fill the space in between so that we end up having 
  a band instead of a single curve for the
  NLA observables.
  The band represents the uncertainty that comes 
  into play after using the BLM prescription since
  there is no unambiguous way to apply it.

  \subsection[Fixed rapidity kinematics for the central jet]
             {Fixed rapidity kinematics for the central jet}
  \label{sub:3jets-nlk-1}           
             
  In this Section, results 
  for three generalised ratios, $R^{12}_{22}$, 
  $R^{12}_{33}$ and $R^{22}_{33}$ are presented, 
  assuming that the central jet 
  is fixed in rapidity at $y_J = (Y_A+Y_B)/2$
  (see Fig.~\ref{fig:lego1}). 
  In particular,
  \begin{equation}
  \label{Cmn_int}
   C_{MN} =
   \int_{Y_A^{\rm min}}^{Y_A^{\rm max}} \hspace{-0.25cm} dY_A
   \int_{Y_B^{\rm min}}^{Y_B^{\rm max}} \hspace{-0.25cm} dY_B
   \int_{k_A^{\rm min}}^{k_A^{\rm max}} \hspace{-0.25cm} dk_A
   \int_{k_B^{\rm min}}^{k_B^{\rm max}} \hspace{-0.25cm} dk_B
   \int_{k_J^{\rm min}}^{k_J^{\rm max}} \hspace{-0.25cm} dk_J
   \delta\left(Y_A - Y_B - Y\right) {\cal C}_{MN} \; ,
  \end{equation}
  where the forward jet rapidity is taken in the
  range delimited by $0 < Y_A < 4.7$,
  the backward jet rapidity in the range  $-4.7 < Y_B < 0$,
  while their difference 
  $Y \equiv Y_A - Y_B$ is kept fixed at definite 
  values in the range $5.5 < Y < 9$.
  
  It is possible to study the ratios  $R_{PQ}^{MN}$ in Eq.~(\ref{RmnqpNew}) 
  as functions of the 
  rapidity difference Y between the most forward 
  and the most backward jets 
  for a set of characteristic values of $M, N, P, Q$ 
  and for two different 
  center-of-mass energies: $\sqrt s = 7$ and $\sqrt s = 13$ TeV. 
  Since we are integrating over $k_A$ and $k_B$,  
  we have the opportunity to impose either 
  \emph{symmetric} or \emph{asymmetric} kinematical
  cuts, as it has been previously done in
  Mueller--Navelet studies.
  Here, and in the next two Sections, 
  we choose to study the \emph{asymmetric} cut which presents certain
  advantages over the \emph{symmetric} one 
  (see Section~\ref{sec:mn-jets-BFKL-vs-DGLAP} and  Refs.~\cite{Ducloue:2013wmi,Celiberto:2015yba}).
  To be more precise, we set
  $k_A^{\rm min} = 35$ GeV, $k_B^{\rm min} = 50$ GeV, 
  $k_A^{\rm max} = k_B^{\rm max}  = 60$ GeV
  throughout the whole analysis.
  
  In order to be as close as possible to the 
  characteristic rapidity ordering of the MRK, 
  we set the value of the central jet rapidity 
  such that it is equidistant to $Y_A$ and $Y_B$ by imposing
  the condition $y_J = (Y_A + Y_B)/2$. 
  Moreover, since the tagging of
  a central jet permits us to extract 
  more exclusive information from our
  observables, we allow three possibilities 
  for the transverse momentum
  $k_J$, that is, $20\, \mathrm{GeV} < k_J < 35\, \mathrm{GeV}$ (bin-1),
  $35 \,\mathrm{GeV} < k_J < 60\, \mathrm{GeV}$ (bin-2) and
  $60\, \mathrm{GeV} < k_J < 120\, \mathrm{GeV}$ (bin-3). 
  Keeping in mind that 
  the forward/backward jets have transverse momenta in the range
  $\left[35 \,\mathrm{GeV}, 60 \,\mathrm{GeV}\right]$, 
  restricting the value
  of $k_J$ within these three bins allows us to see how the ratio
  $R_{PQ}^{MN}$ changes behaviour 
  depending on the relative size of the
  central jet momentum when compared to the forward/backward ones. 
  Throughout the whole Section~\ref{sec:4j-hadronic},
  we will keep the same setup regarding
  bin-1, bin-2 and bin-3 which roughly correspond to the cases
  of $k_J$ being `smaller' than, `similar' to and `larger' than
  $k_A$, $k_B$, respectively. 
  
  Finally, apart from the functional dependence of the ratios on $Y$
  we will also show the relative corrections when we go from LLA to NLA.
  To be more precise, we define 
  \begin{eqnarray}
  \delta x(\%) = \left(
   \text{res}^{\rm(LLA)} - 
   \frac{\text{res}^{\rm (BLM-1)}+\text{res}^{\rm (BLM-2)}}{2}
   \right) \frac{1}{ \text{res}^{\rm(LLA)}}\,.
   \label{corrections}
  \end{eqnarray}
  $\text{res}^{\rm(BLM-1)}$ is the BLM NLA result 
  for $\mu_R=\mu_R^{\rm BLM}$ 
  only in the Green's function 
  while the cubed term of the strong coupling in Eq.~(\ref{lo-nlo})
  actually reads $\bar{\alpha}_s^3 = \bar{\alpha}_s^3(\sqrt{k_{A}k_{B}})$).
  $\text{res}^{\rm (BLM-2)}$ is
  the BLM NLA result  for  $\mu_R=\mu_R^{\rm BLM}$ everywhere in 
  Eq.~(\ref{lo-nlo}), therefore, 
  $\bar{\alpha}_s^3 = \bar{\alpha}_s^3(\mu_R^{\rm BLM})$,
  as was previously discussed in Section~\ref{sub:3j-nlk}.
  
\begin{figure}[p]
  \newgeometry{left=-10cm,right=1cm}
  \centering

     \hspace{-16.25cm}
     \includegraphics[scale=0.28]{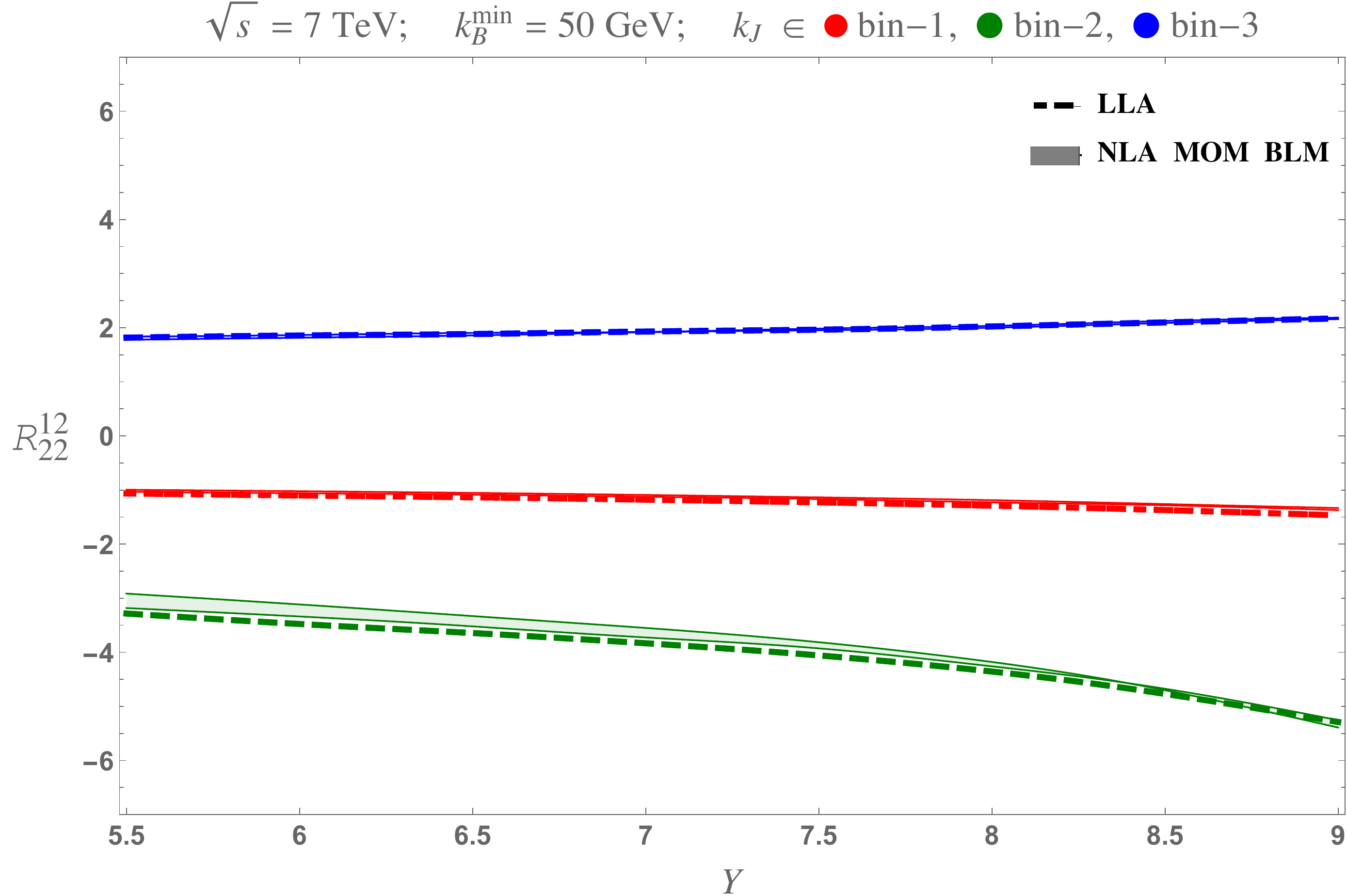}
     \includegraphics[scale=0.28]{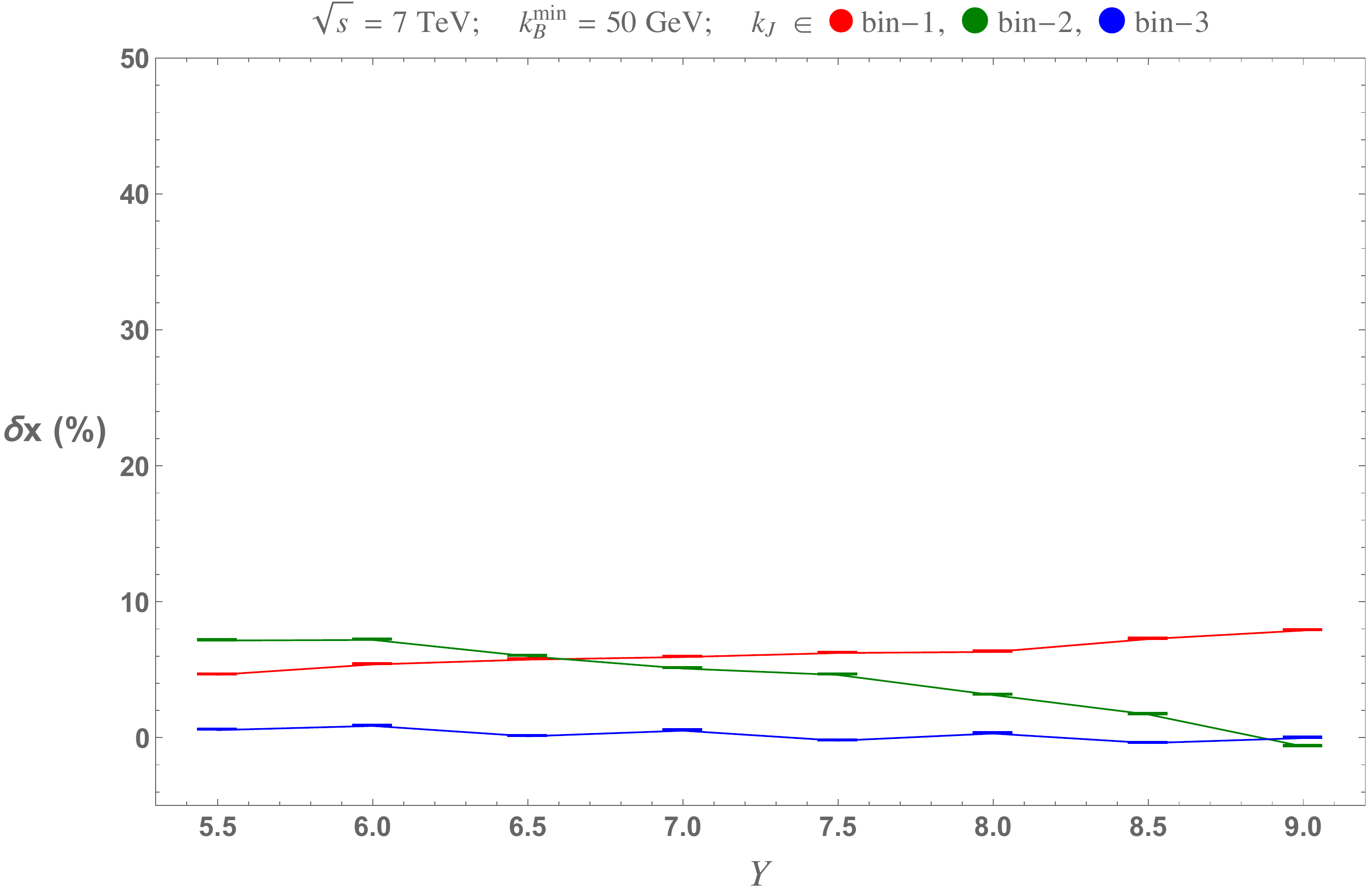}
     \vspace{1cm}

     \hspace{-16.25cm}
     \includegraphics[scale=0.28]{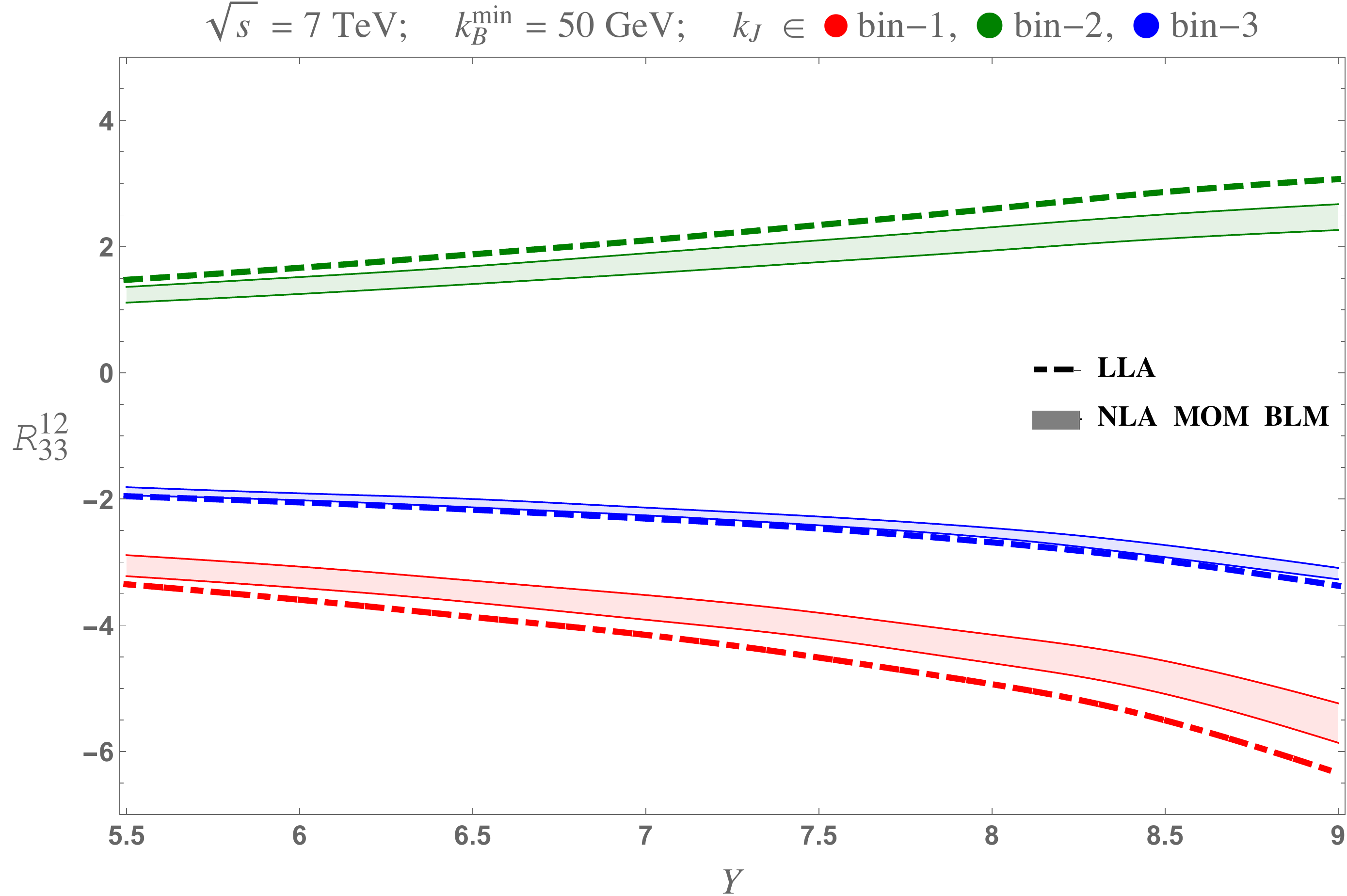}
     \includegraphics[scale=0.28]{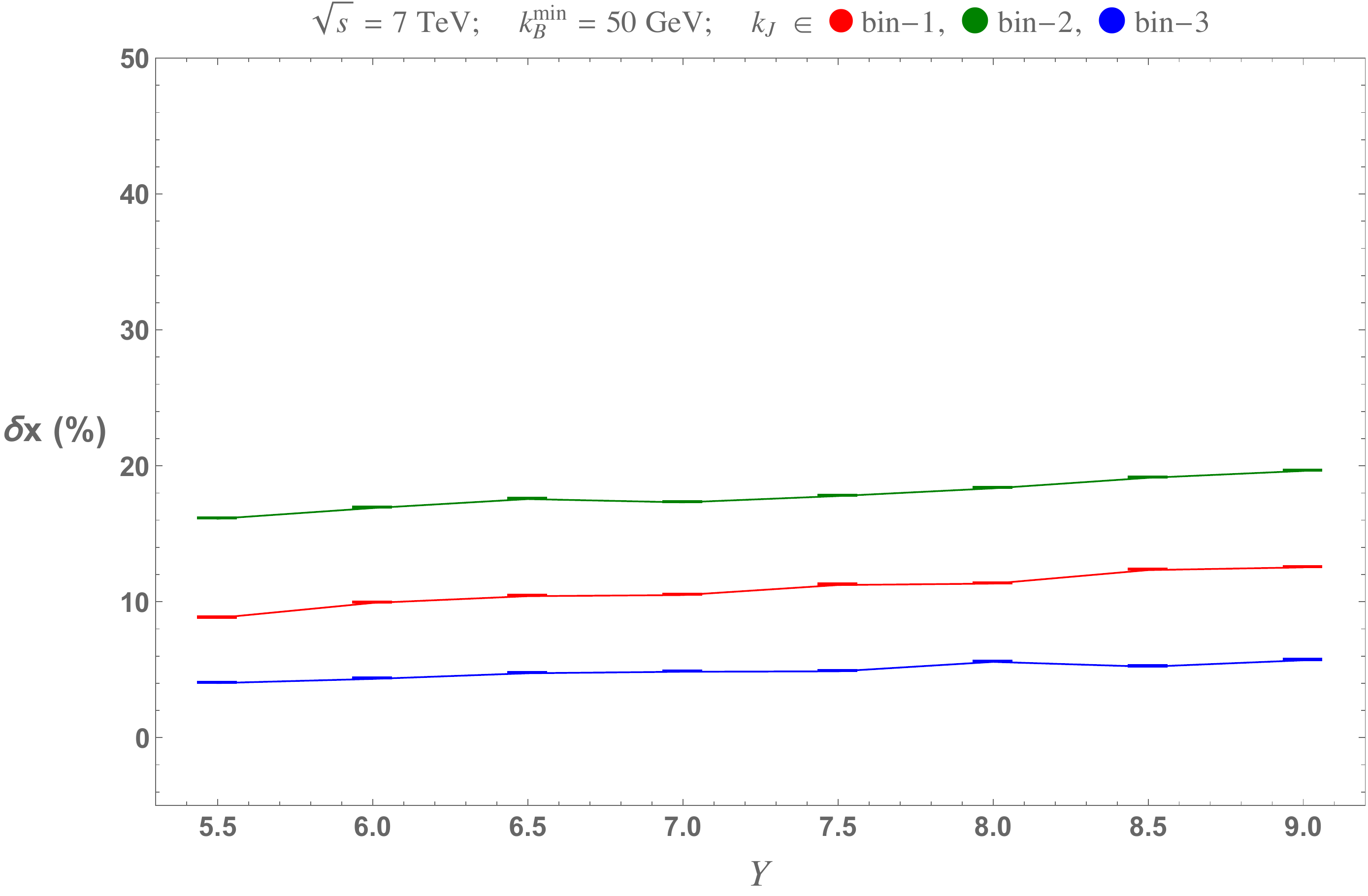}
     \vspace{1cm}

     \hspace{-16.25cm}   
     \includegraphics[scale=0.28]{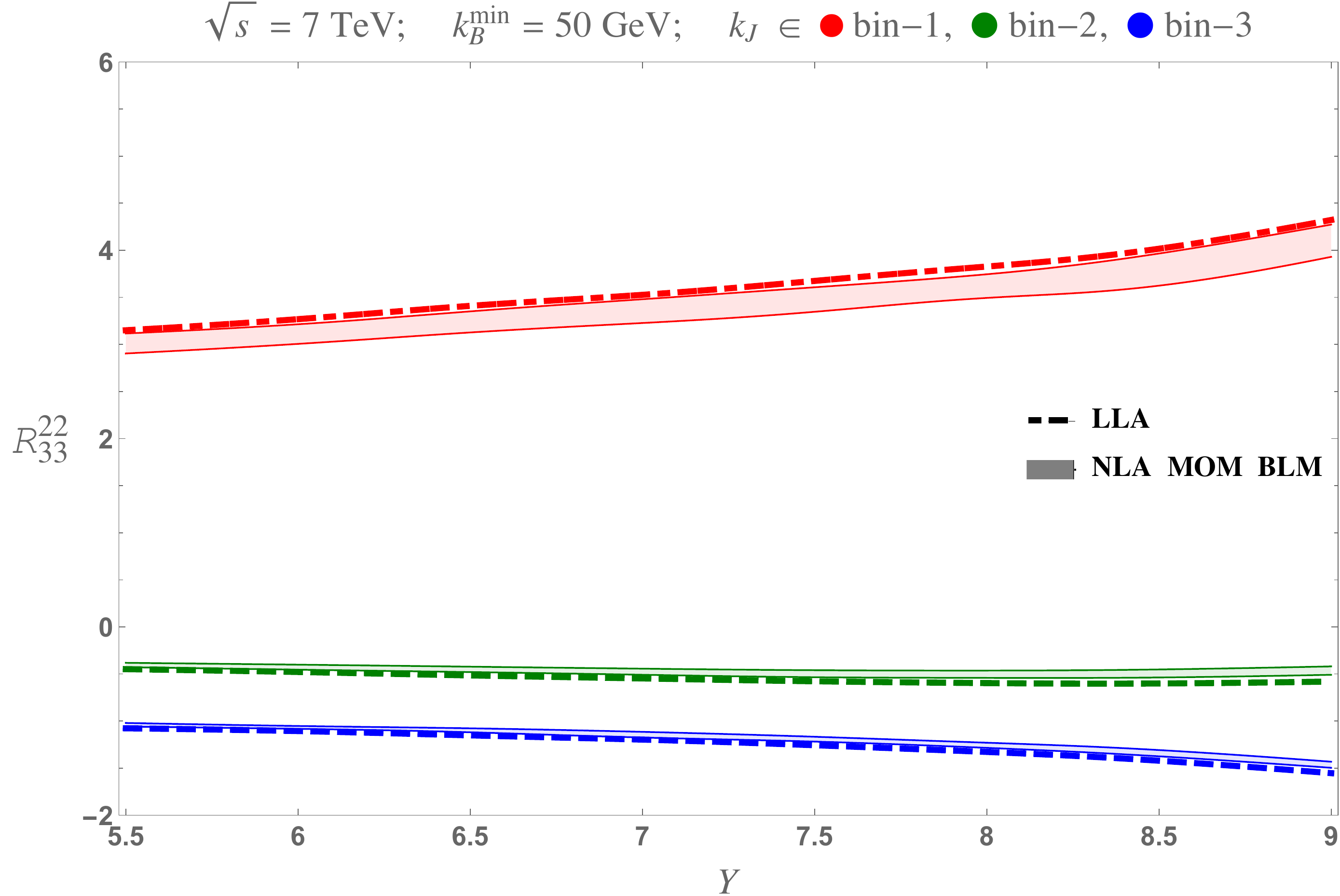}
     \includegraphics[scale=0.28]{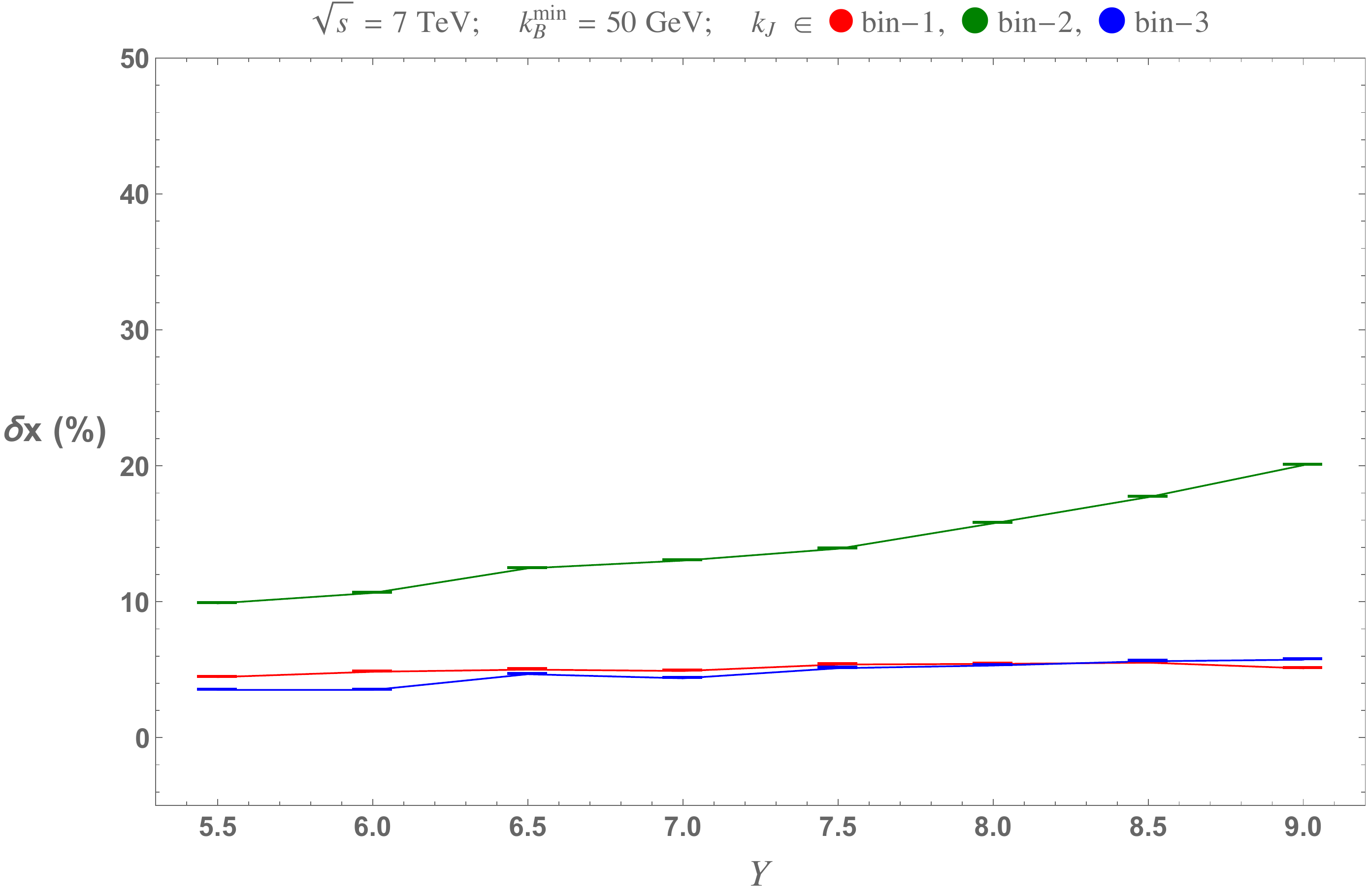}

  \restoregeometry
  \caption[LLA and NLA
           $R^{12}_{22}$, 
           $R^{12}_{33}$, and 
           $R^{22}_{33}$ 
           at $\sqrt s = 7$ TeV]
  {LLA and NLA
   $R^{12}_{22}$, $R^{12}_{33}$, and $R^{22}_{33}$ 
   at $\sqrt s = 7$ TeV with $y_J$ fixed
   (left) and the relative NLA to LLA corrections  (right).} 
  \label{fig:7-first}
  \end{figure}

\begin{figure}[p]
  \newgeometry{left=-10cm,right=1cm}
  \vspace{-2cm}
  \centering

     \hspace{-16.25cm}
     \includegraphics[scale=0.28]{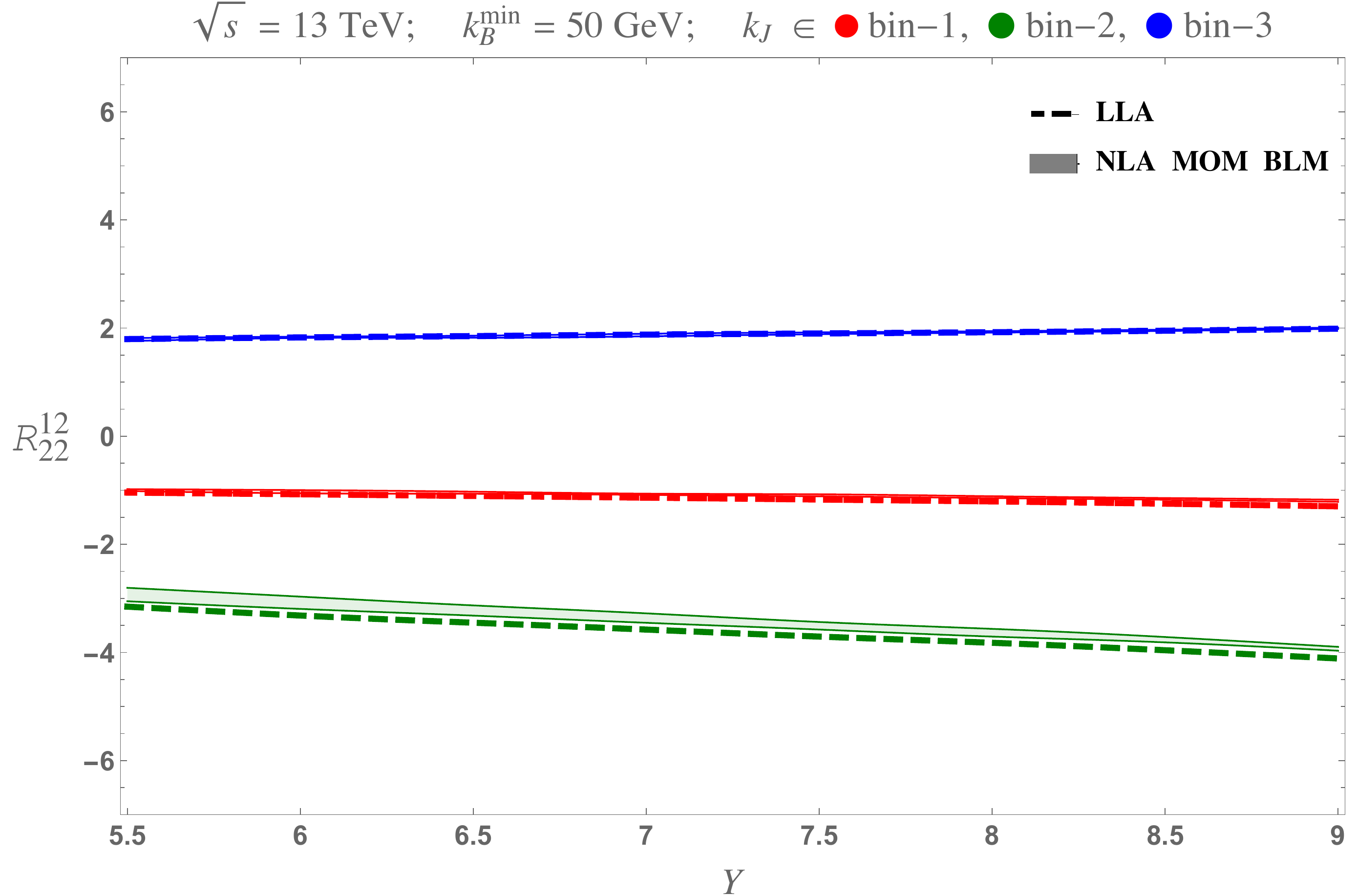}
     \includegraphics[scale=0.28]{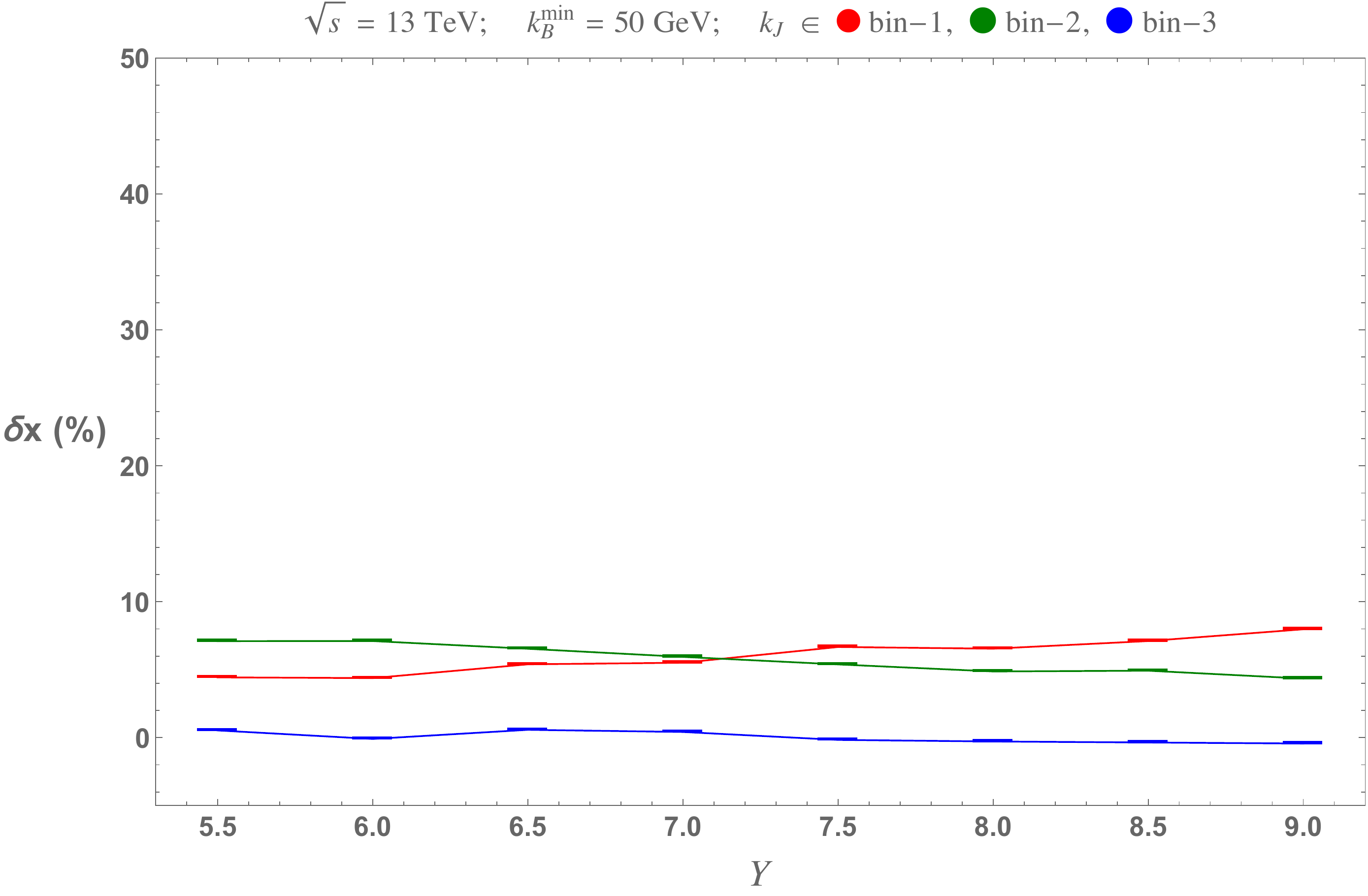}
     \vspace{1cm}

     \hspace{-16.25cm}
     \includegraphics[scale=0.28]{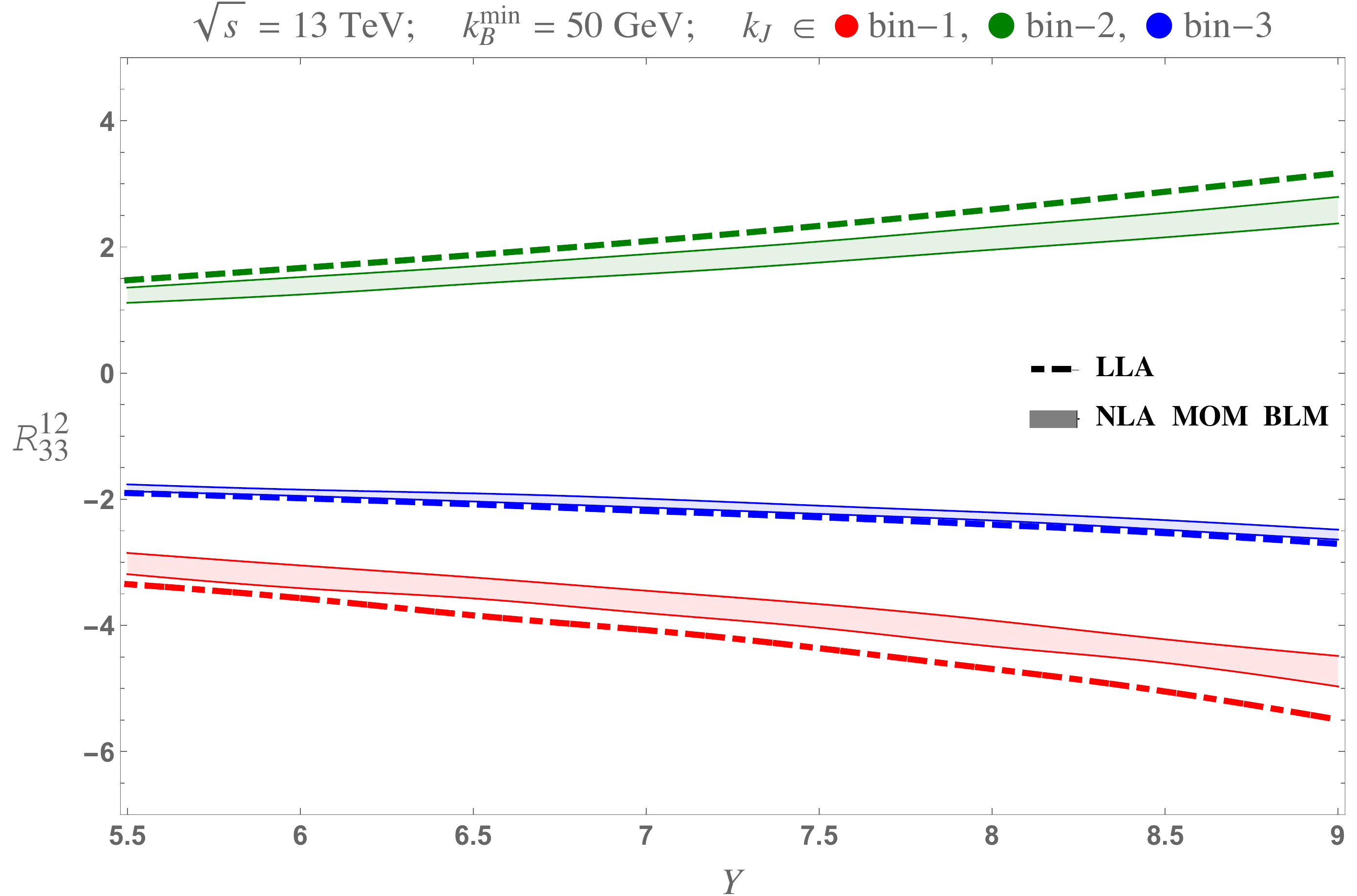}
     \includegraphics[scale=0.28]{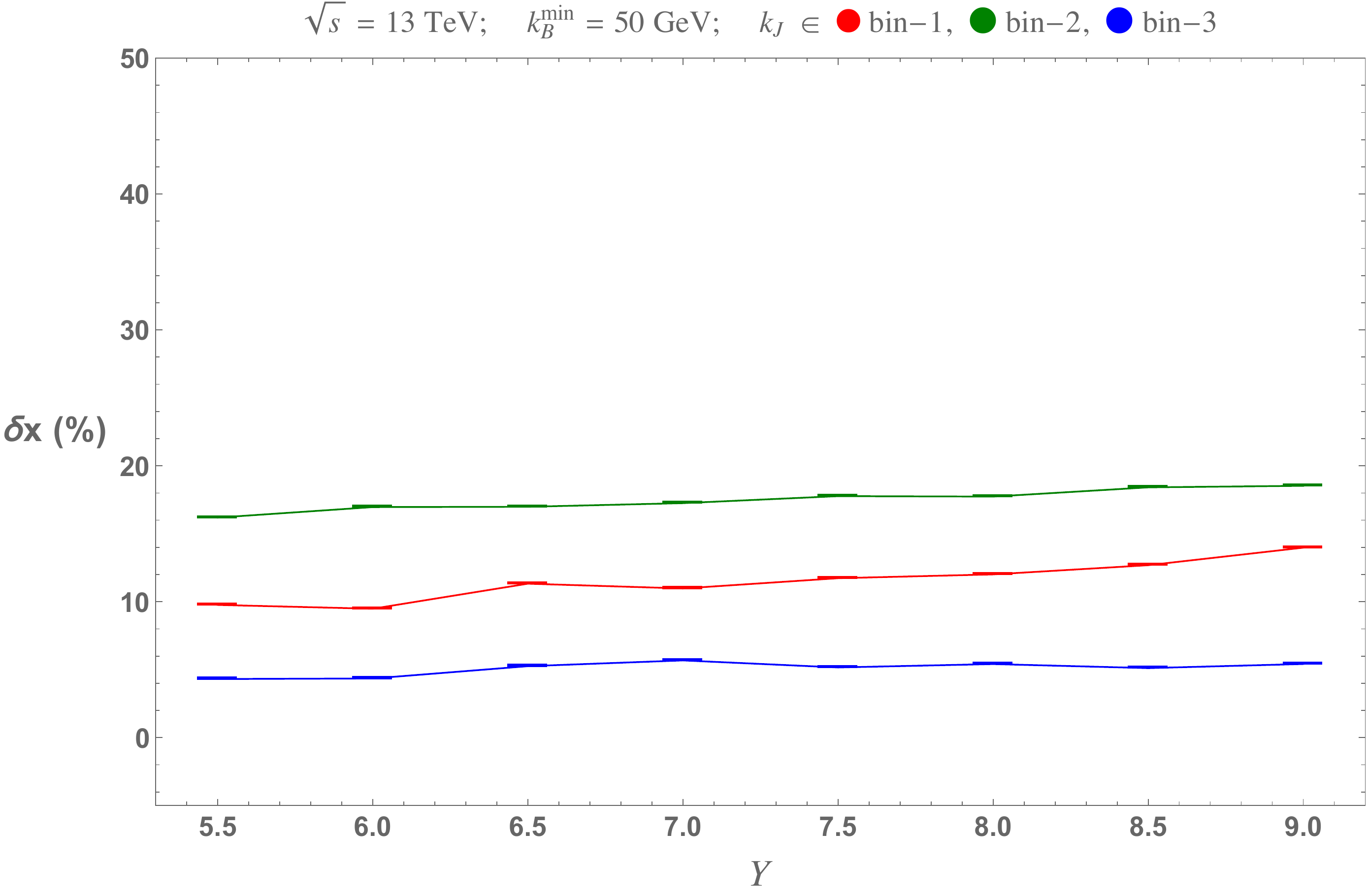}
     \vspace{1cm}

     \hspace{-16.25cm}   
     \includegraphics[scale=0.28]{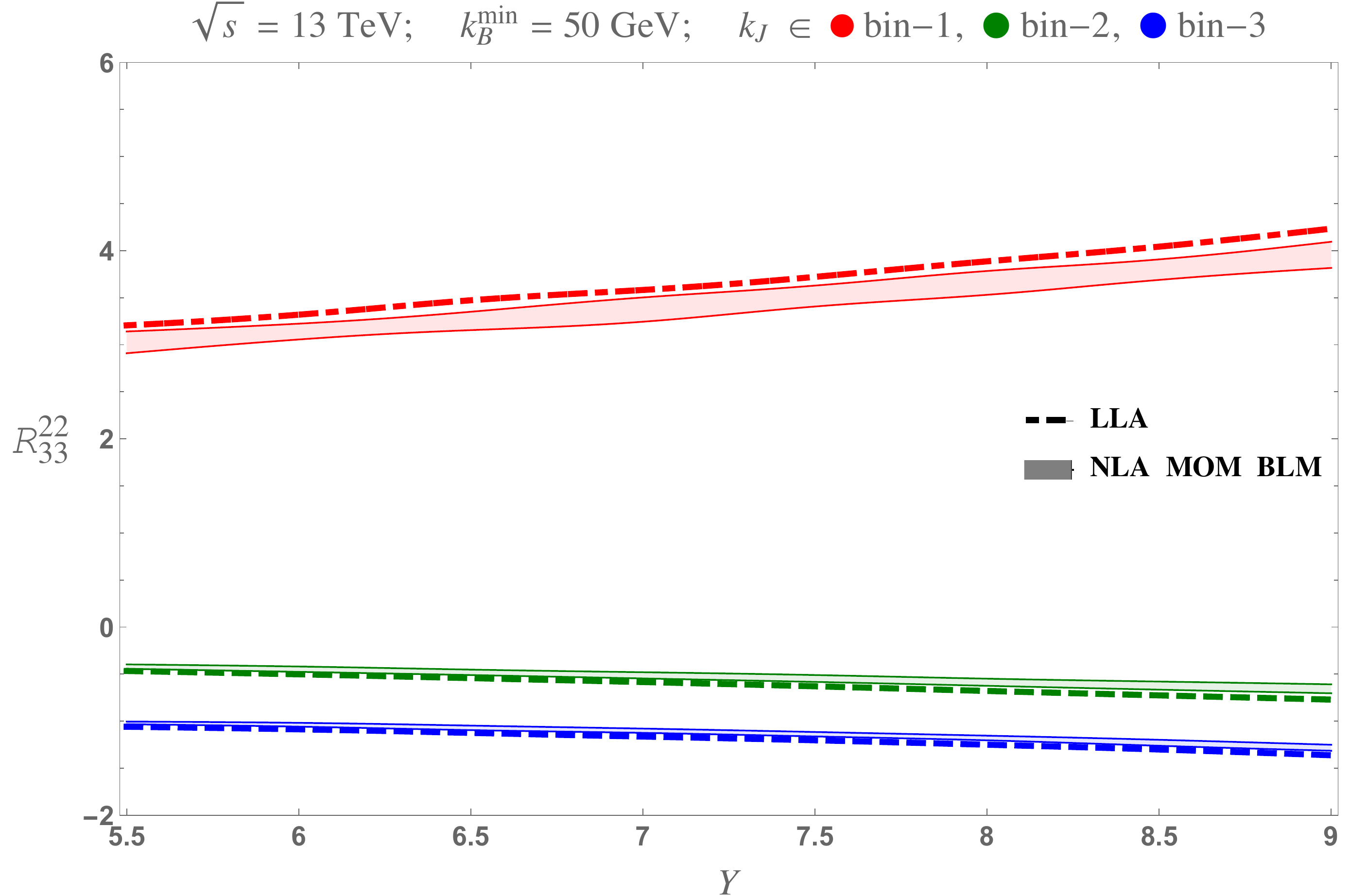}
     \includegraphics[scale=0.28]{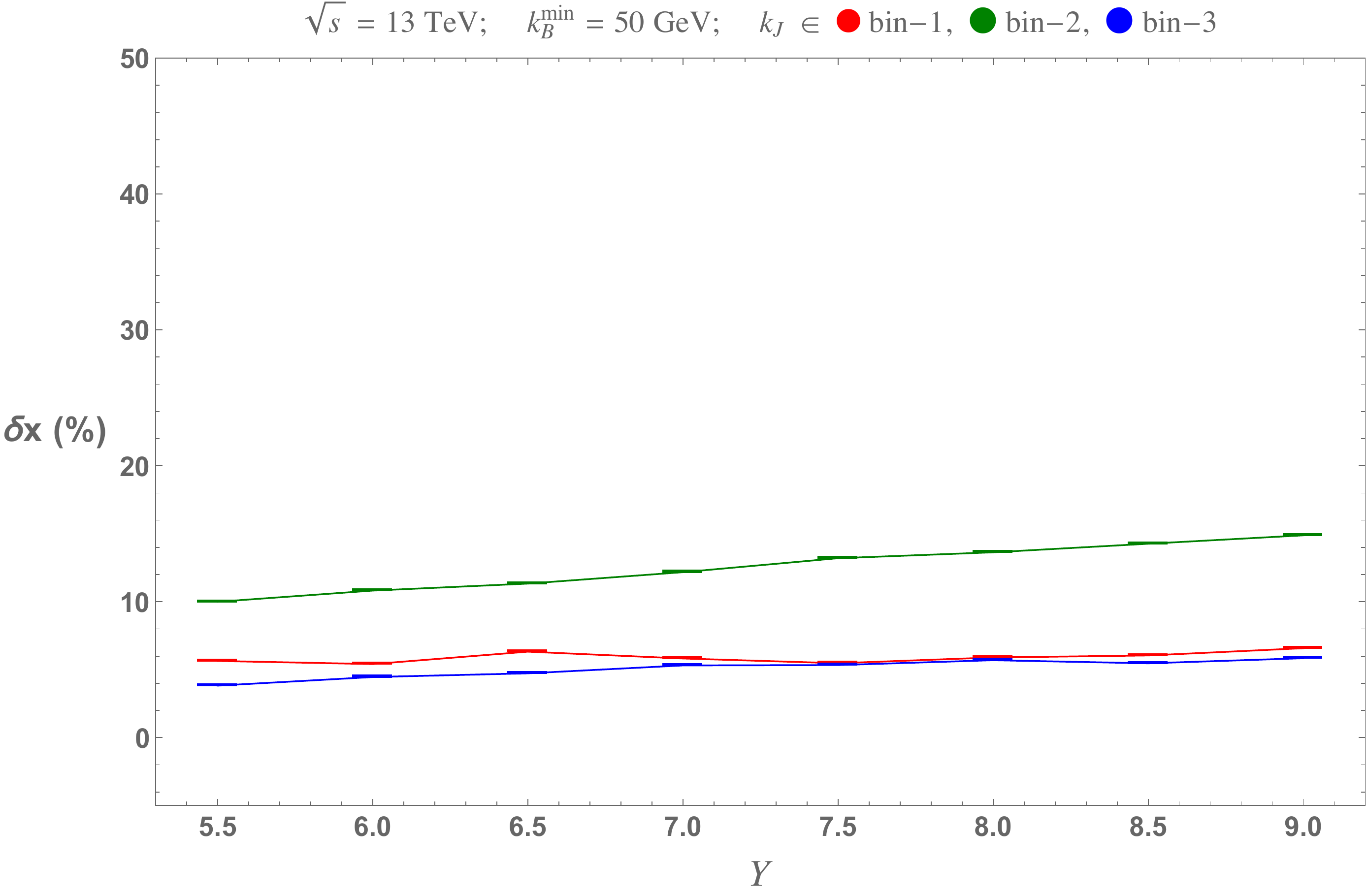}

  \restoregeometry
  \caption[LLA and NLA
           $R^{12}_{22}$, 
           $R^{12}_{33}$, and 
           $R^{22}_{33}$ 
           at $\sqrt s = 13$ TeV]
   {$Y$-dependence of the LLA and NLA
    $R^{12}_{22}$, $R^{12}_{33}$, and $R^{22}_{33}$ 
    at $\sqrt s = 13$ TeV with $y_J$ fixed
    (left) and the relative NLA to LLA corrections (right).} 
  \label{fig:13-first}
  \end{figure}

  In the following, we present our results 
  for $R^{12}_{22}$, $R^{12}_{33}$ and $R^{22}_{33}$,
  with $y_J = (Y_A+Y_B)/2$,
  collectively in Fig.~\ref{fig:7-first} 
  ($\sqrt{s} = 7$ TeV) and Fig.~\ref{fig:13-first} ($\sqrt{s} = 13$ TeV),
  In the left column  we  are showing plots 
  for $R^{MN}_{PQ}(Y)$ whereas to the right we are showing 
  the corresponding $\delta x(\%)$ between LLA and NLA corrections.
  The LLA results are represented 
  with dashed lines whereas the NLA ones with a continuous band.
  The boundaries of the band are the two different curves 
  we obtain by the two different approaches in
  applying the BLM prescription. 
  Since there is no definite way to choose one in favour of the other,
  we allow for any possible value in between 
  and hence we end up with a band.
  In many cases, as we will see in the following, 
  the two boundaries are so close that the band
  almost degenerates into a single curve.
  The red curve (band) corresponds to $k_J$ bounded in bin-1,
  the green curve (band) to $k_J$  bounded in bin-2 
  and finally the blue curve (band)
  to $k_J$ bounded in bin-3. For the $\delta x(\%)$ plots 
  we only have three 
  curves, one for each of the three different bins of $k_J$.

  A first observation from inspecting Figs.~\ref{fig:7-first} 
  and~\ref{fig:13-first}
  is that the dependence of the different observables on the rapidity
  difference between $k_A$ and $k_B$ is rather smooth.
  $R^{12}_{22}$ (top row in Figs.~\ref{fig:7-first} and~\ref{fig:13-first})
  at $\sqrt{s} = 7$ TeV and for $k_J$ 
  in bin-1 and bin-3 exhibits an almost linear behaviour 
  with $Y$ both at LLA and NLA, whereas at $\sqrt{s} = 13$ TeV 
  the linear behaviour is extended also for
  $k_J$ in bin-2. The difference  between the NLA BLM-1 and BLM-2 values
  is small, to the point that the blue and the red bands 
  collapse into a single line which in addition
  lies very close to the LLA results.
  When  $k_J$ is restricted in bin-2 (green curve/band), 
  the uncertainty from applying the BLM
  prescription in two different ways seems to be larger.
  The relative NLA corrections at both colliding energies 
  are very modest ranging from
  close to $1\%$ for $k_J$ in bin-3 to less than $10\%$ for $k_J$ 
  in the other two bins.
  $R^{12}_{33}$ (middle row in Figs.~\ref{fig:7-first} 
  and~\ref{fig:13-first})
  compared to $R^{12}_{22}$,
  shows a larger difference between BLM-1 and BLM-2 values for $k_J$
  in bin-1 and bin-2. The `green' corrections 
  lower the LLA estimate
  whereas the `red' ones  make 
  the corresponding LLA estimate less negative.
  The corrections are generally below $20\%$, 
  in particular, `blue' $\sim 5\%$,
  `red' $\sim 10\%$ and `green' $\sim 20\%$.
  Finally, $R^{22}_{33}$ (bottom row 
  in Figs.~\ref{fig:7-first} and~\ref{fig:13-first}) also
  shows a larger difference between BLM-1 and BLM-2 values for $k_J$
  in bin-1 and less so for $k_J$ in bin-2. 
  Here, the `red' corrections lower the LLA estimate
  whereas the `green' ones make 
  the corresponding LLA estimate less negative.
  The corrections are smaller than the ones for $R^{12}_{33}$ 
  and somehow larger 
  than the corrections for $R^{12}_{22}$, 
  specifically, `blue' $\sim 5\%$,
  `red' $\sim 5\%$ and `green' $\sim 15\%$. 
  Noticeably, while for $R^{12}_{22}$ and
  $R^{12}_{33}$ the corrections are very similar 
  at $\sqrt{s} = 7$ and $\sqrt{s} = 13$ TeV,
  the `green' $R^{22}_{33}$ 
  receives larger corrections at $\sqrt{s} = 7$ TeV.
  One important conclusion we would like to draw 
  after comparing Figs.~\ref{fig:7-first} and~\ref{fig:13-first}
  is that, in general, for most of the observables there are no
  striking changes when we increase the colliding energy from
  7 to 13 TeV. This indicates that a sort 
  of asymptotic regime has been approached 
  for the kinematical configurations included in our analysis. 
  It also tells us that our observables
  are really as insensitive as possible to effects
  which have their origin outside the BFKL dynamics and which 
  normally cannot be isolated  ({\it e.g.} 
  influence from the PDFs) with a possible exclusion
  at the higher end of the plots, when $Y \sim 8.5-9$. 
  There, some of the observables 
  and by that we mean the `red', `green'
  or `blue' cases of  $R^{12}_{22}$, $R^{12}_{33}$ and $R^{22}_{33}$,
  exhibit a more curved rather than linear behaviour 
  with $Y$ at $\sqrt{s} = 7$ TeV.

  \subsection[Dependence on the central-jet rapidity bin]
             {Dependence of the generalised azimuthal correlations
              on the central-jet rapidity bin}
  \label{sub:3jets-nlk-2} 
  
  \begin{figure}[H]
   \includegraphics[scale=0.35]{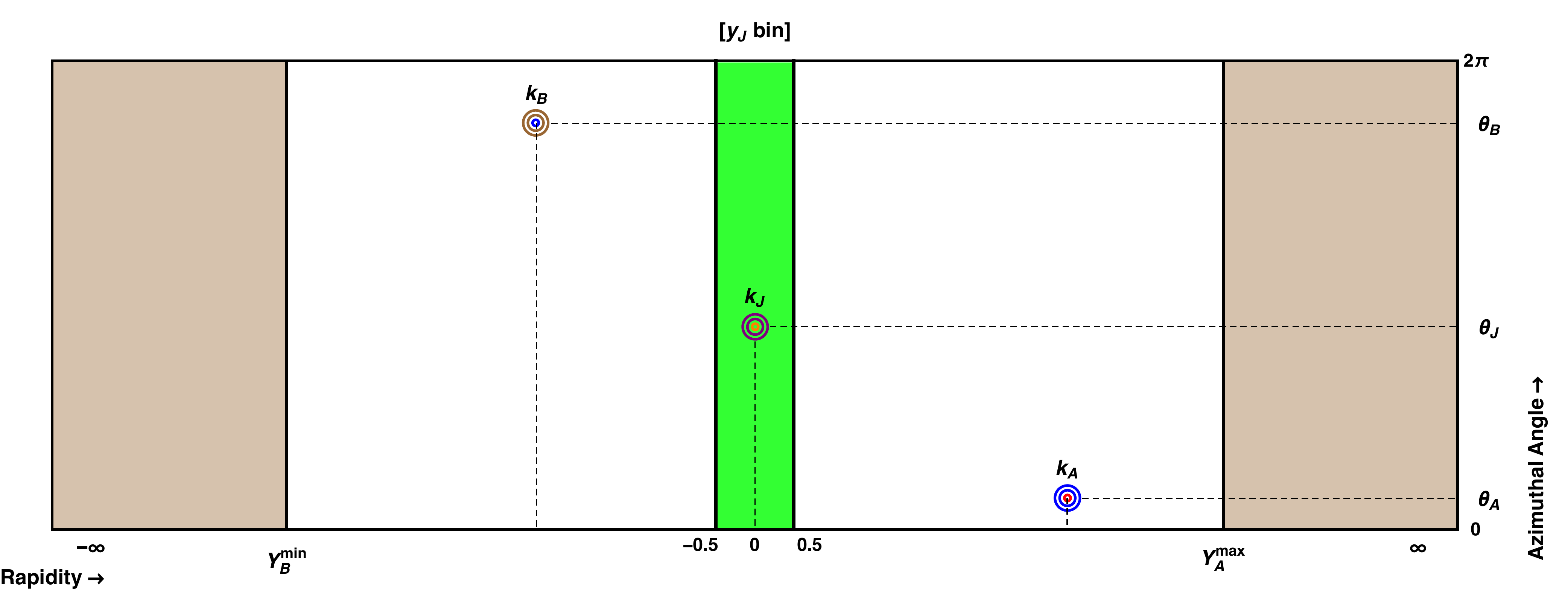}
   \caption[Lego plot for a three-jet event with integration 
            over a central-jet rapidity bin]
   {A primitive lego plot 
   depicting a three-jet event similar to Fig.~\ref{fig:lego1}. 
   Here, however, the central jet can take any value 
   in the rapidity range $-0.5 < y_J < 0.5$.
   }
   \label{fig:lego2}
  \end{figure}
  
\begin{figure}[p]
  \newgeometry{left=-10cm,right=1cm}
  \centering
  
     \hspace{-16.25cm}
     \includegraphics[scale=0.28]{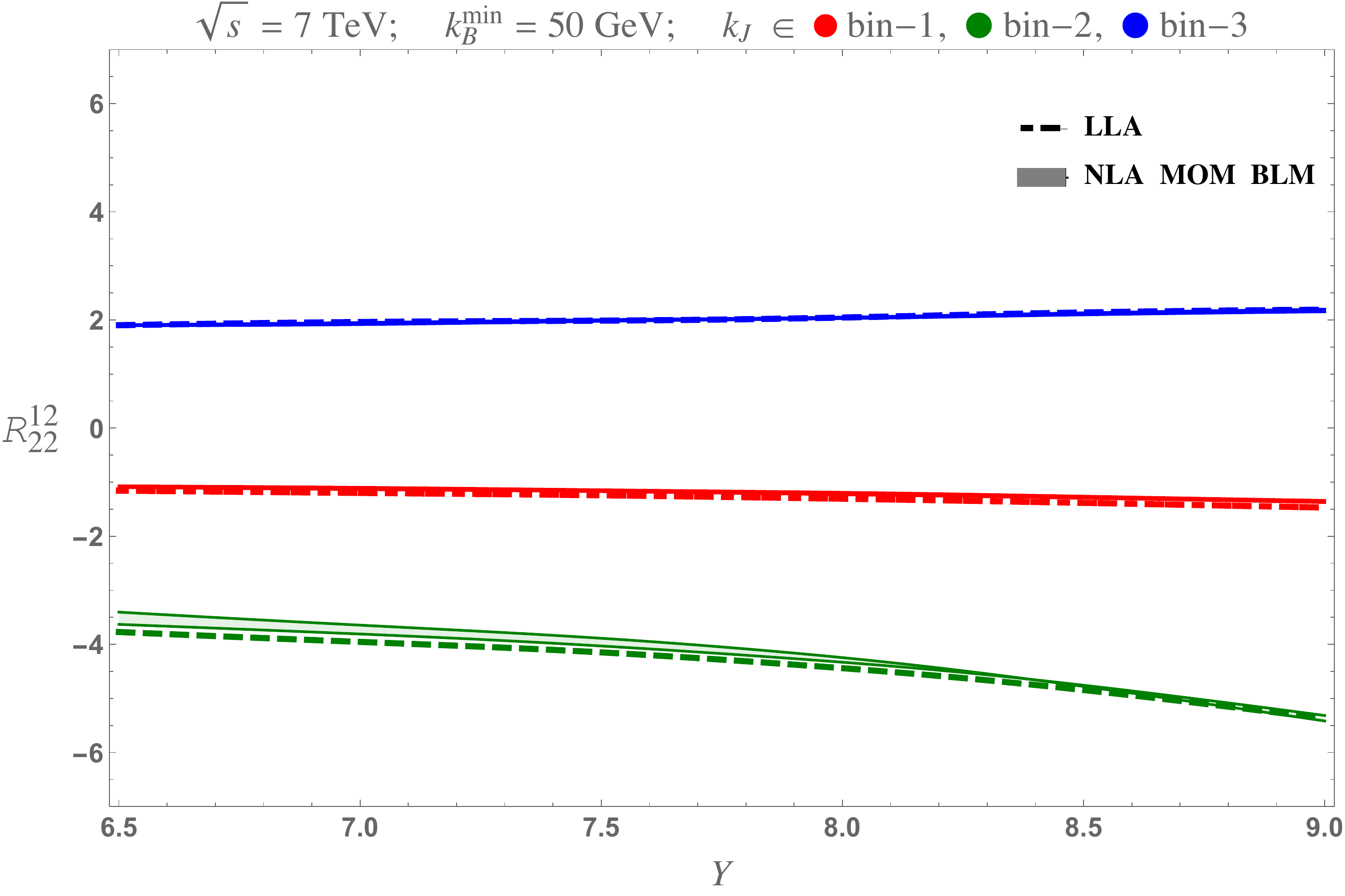}
     \includegraphics[scale=0.28]{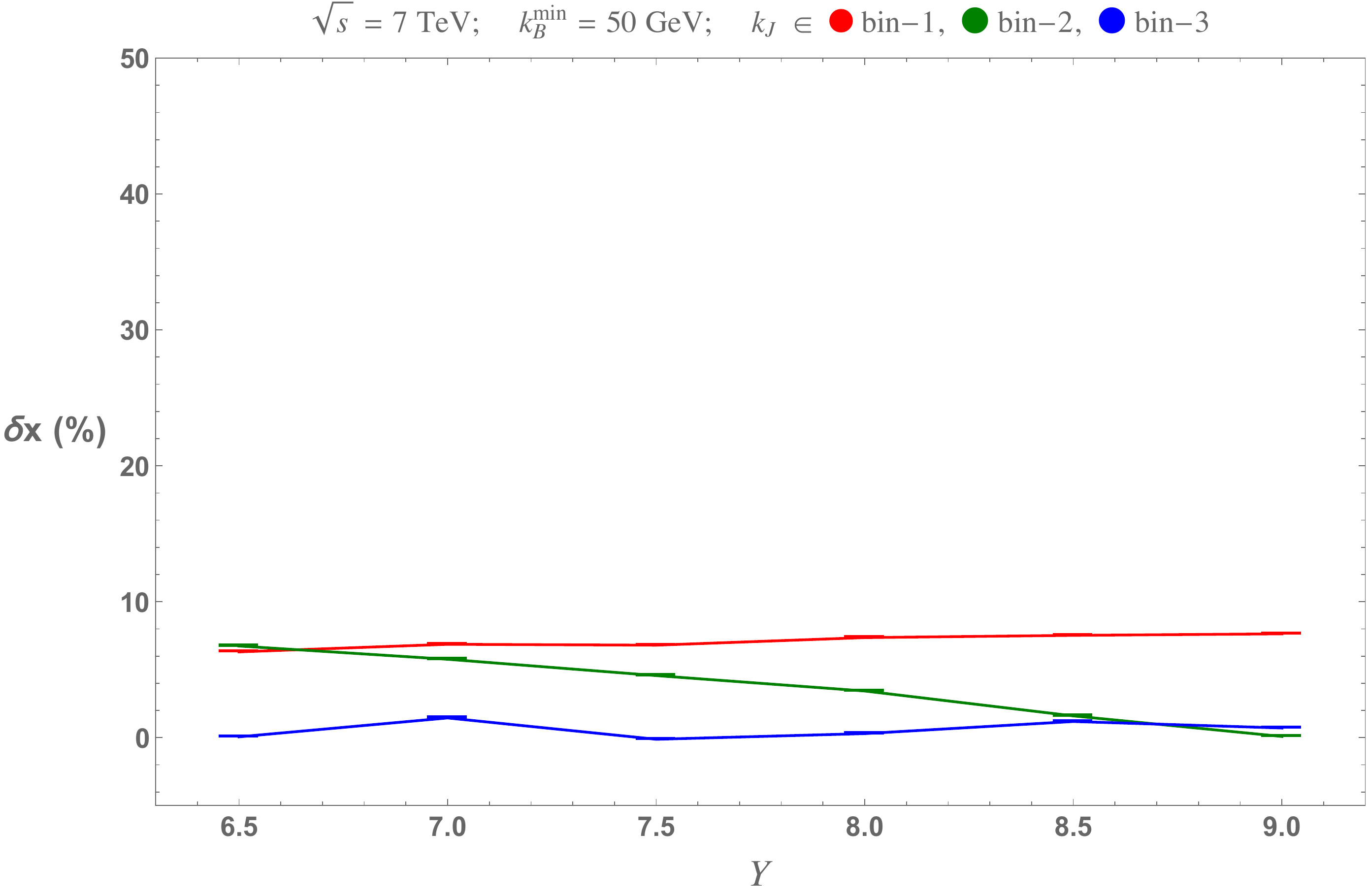}
     \vspace{1cm}
  
     \hspace{-16.25cm}
     \includegraphics[scale=0.28]{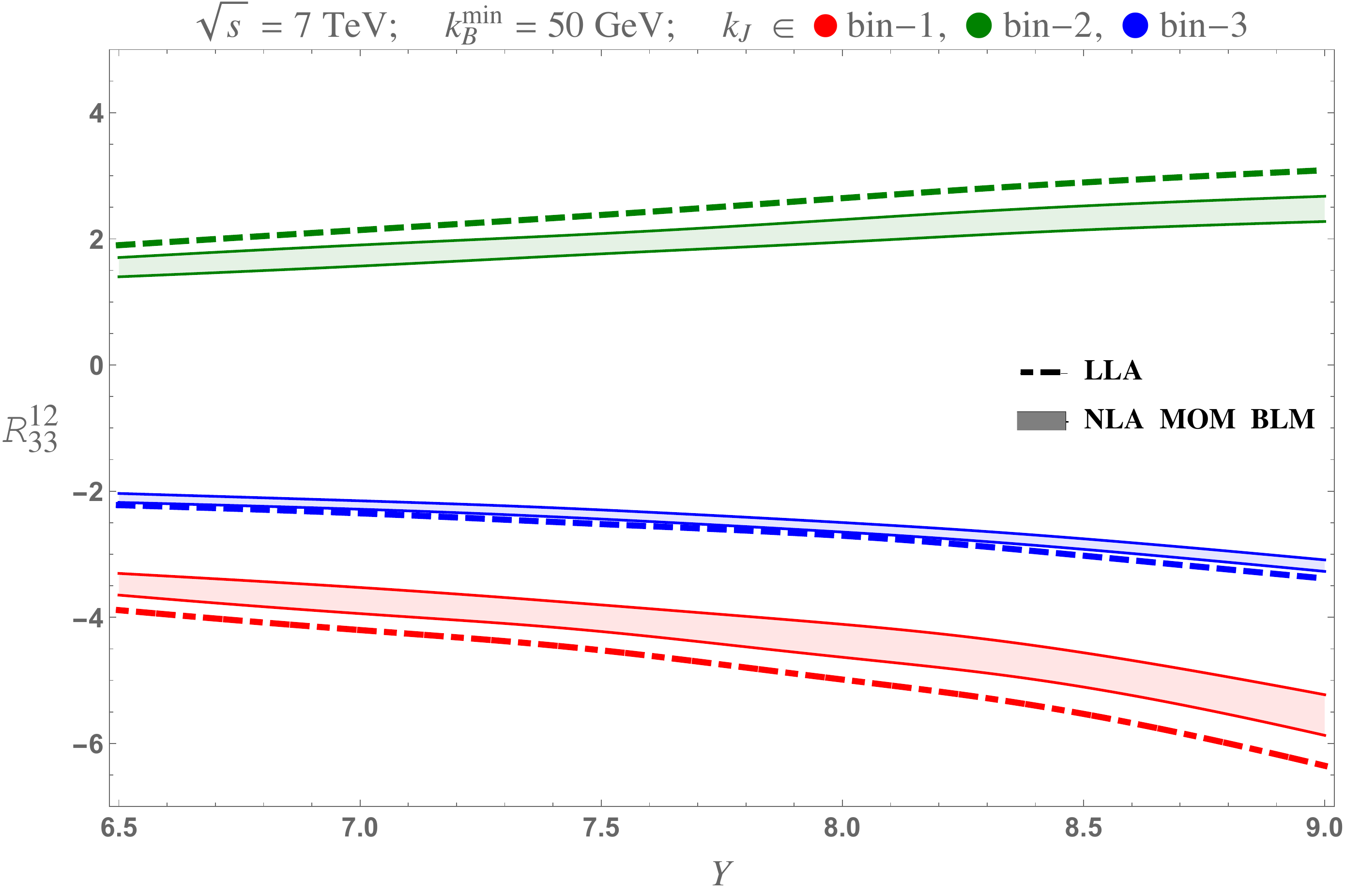}
     \includegraphics[scale=0.28]{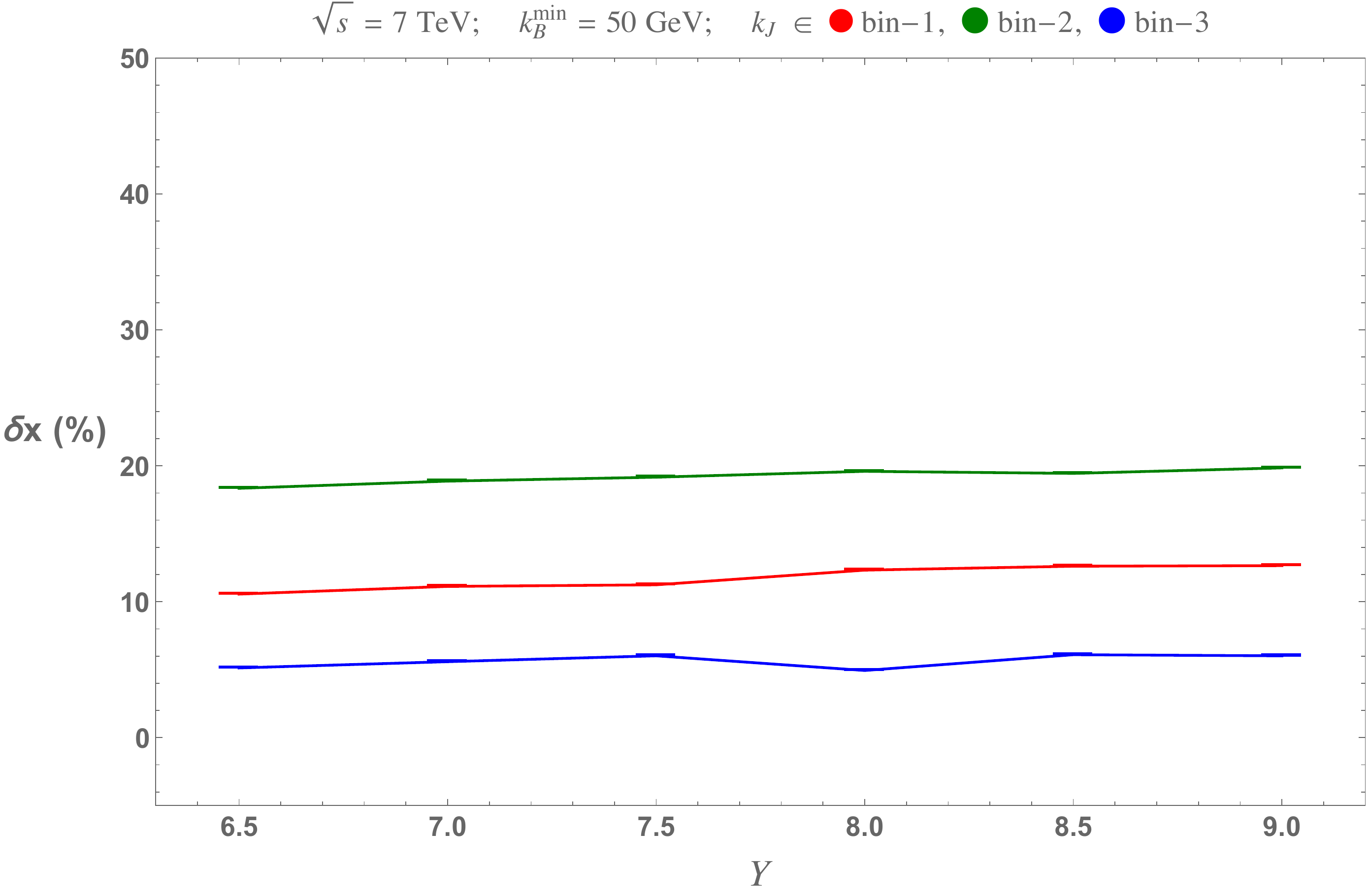}
     \vspace{1cm}
  
     \hspace{-16.25cm}   
     \includegraphics[scale=0.28]{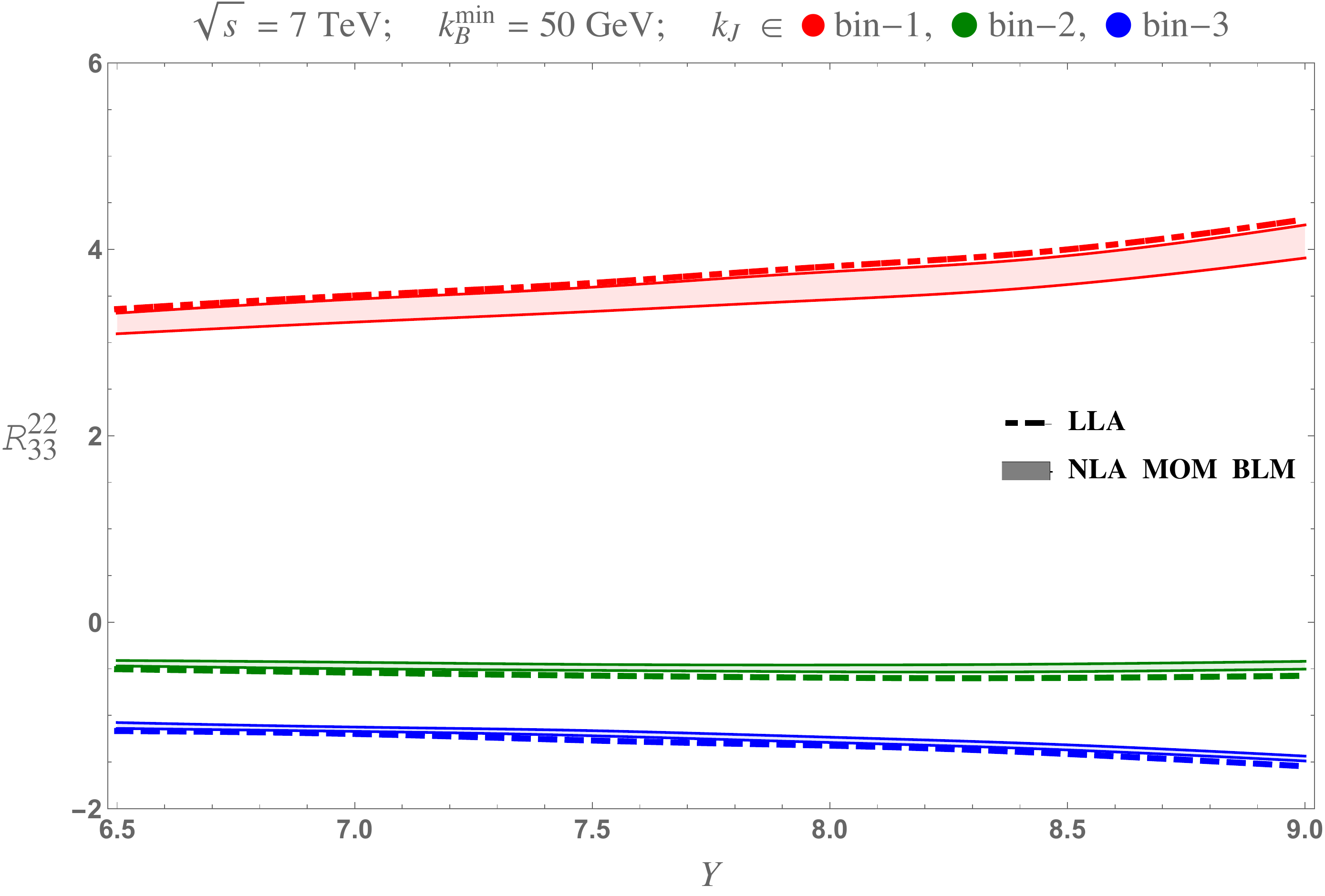}
     \includegraphics[scale=0.28]{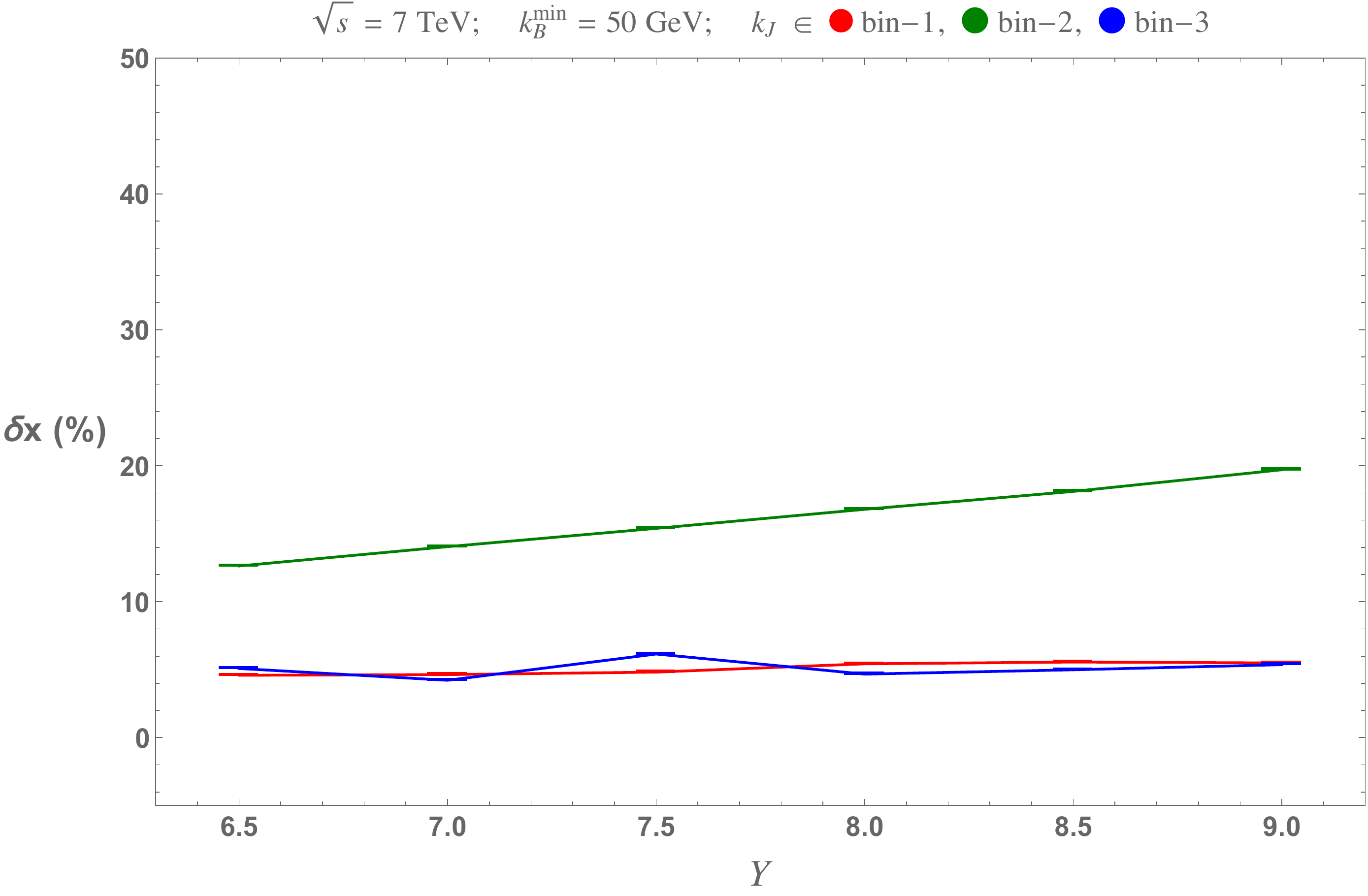}
  
  \restoregeometry
  \caption[LLA and NLA
           $^{\text{(i)}}R^{12}_{22}$, 
           $^{\text{(i)}}R^{12}_{33}$, and $^{\text{(i)}}R^{22}_{33}$ 
           at $\sqrt s = 7$ TeV]
   {LLA and NLA
    $^{\text{(i)}}R^{12}_{22}$, 
    $^{\text{(i)}}R^{12}_{33}$, and $^{\text{(i)}}R^{22}_{33}$ 
    at $\sqrt s = 7$ TeV with $y_J$ integrated over a central
    rapidity bin (left) and the relative NLA to LLA corrections (right).} 
  \label{fig:7-second}
  \end{figure}

\begin{figure}[p]
  \newgeometry{left=-10cm,right=1cm}
  \centering
  
     \hspace{-16.25cm}
     \includegraphics[scale=0.28]{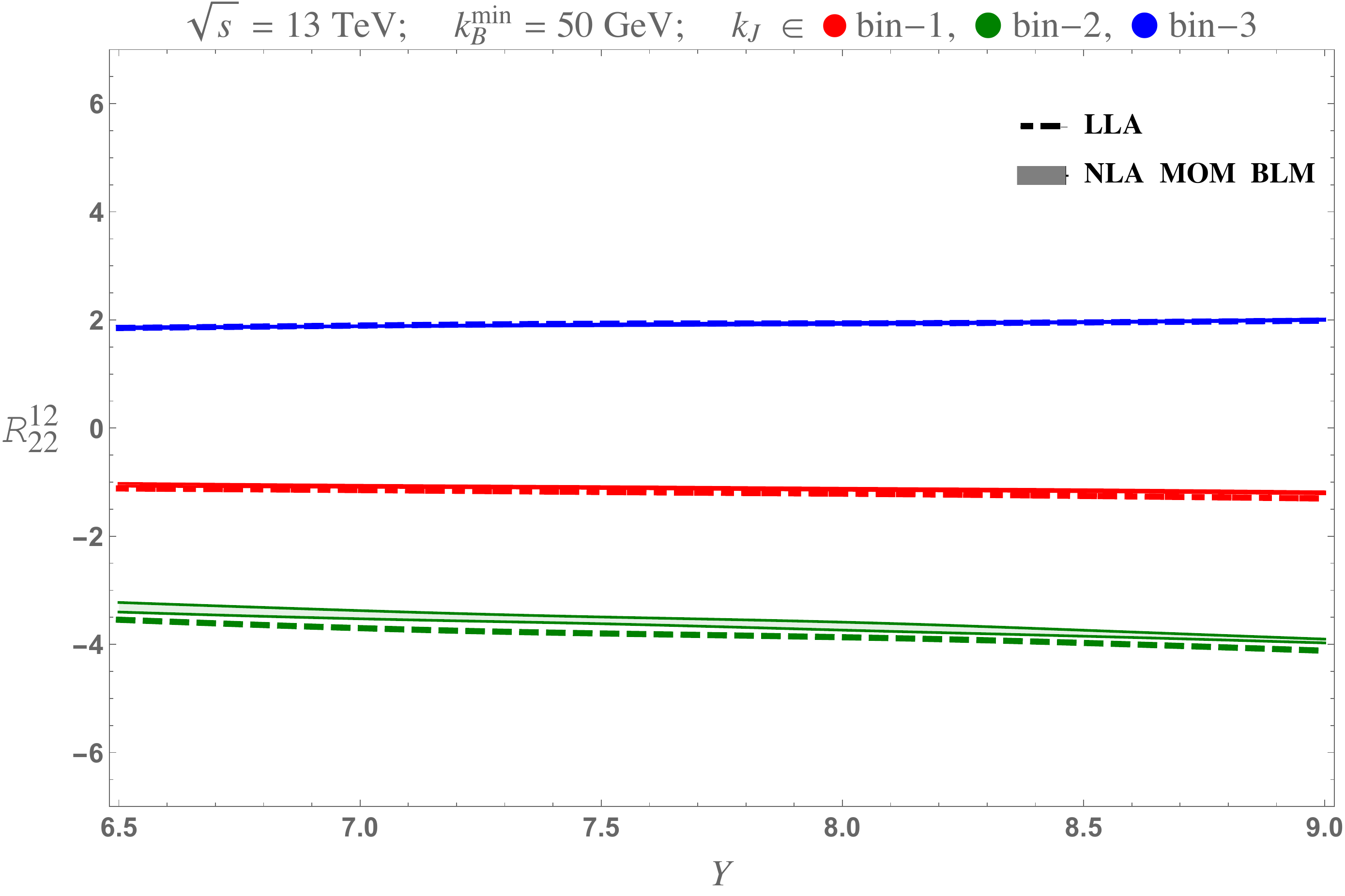}
     \includegraphics[scale=0.28]{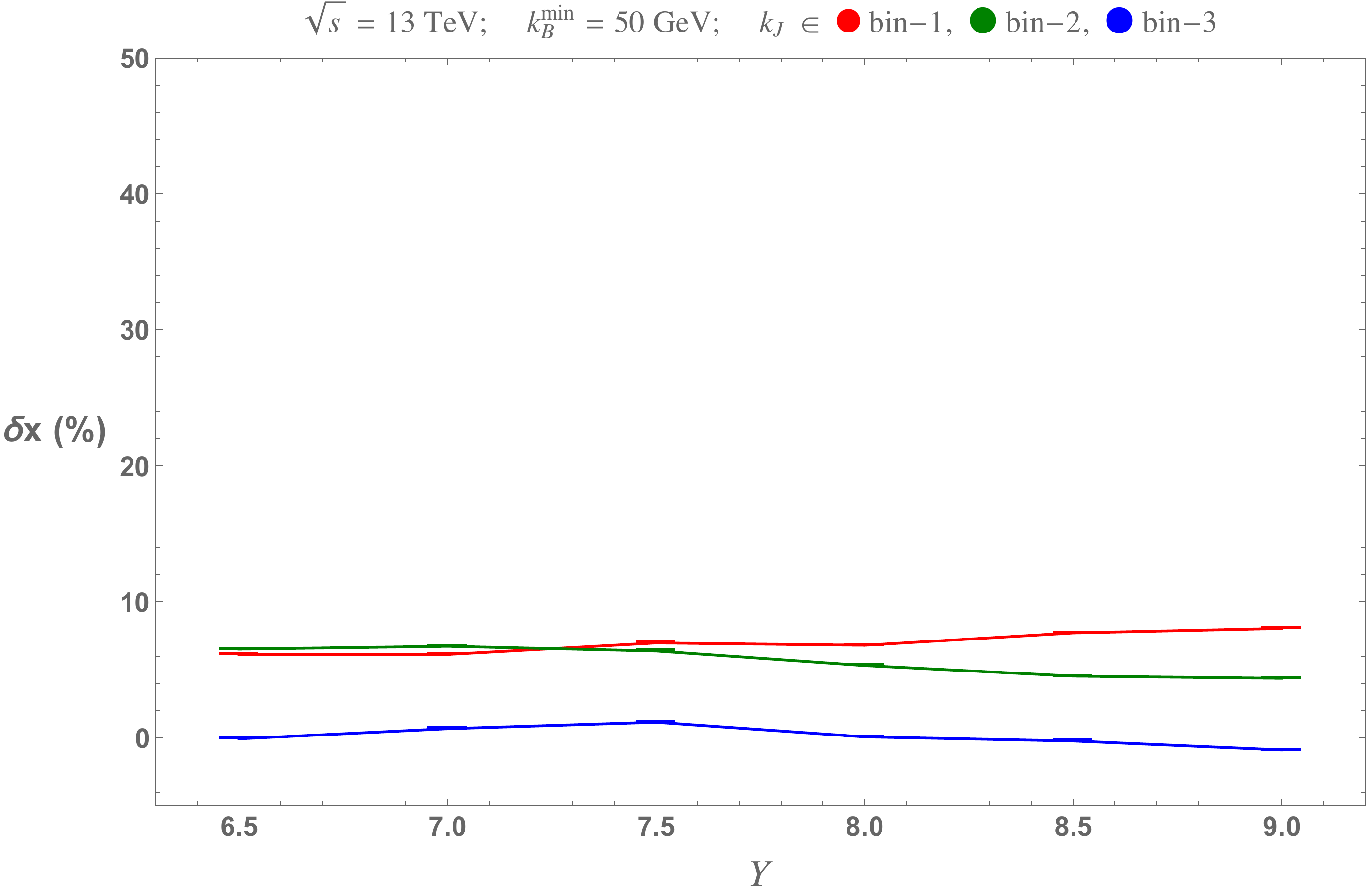}
     \vspace{1cm}
  
     \hspace{-16.25cm}
     \includegraphics[scale=0.28]{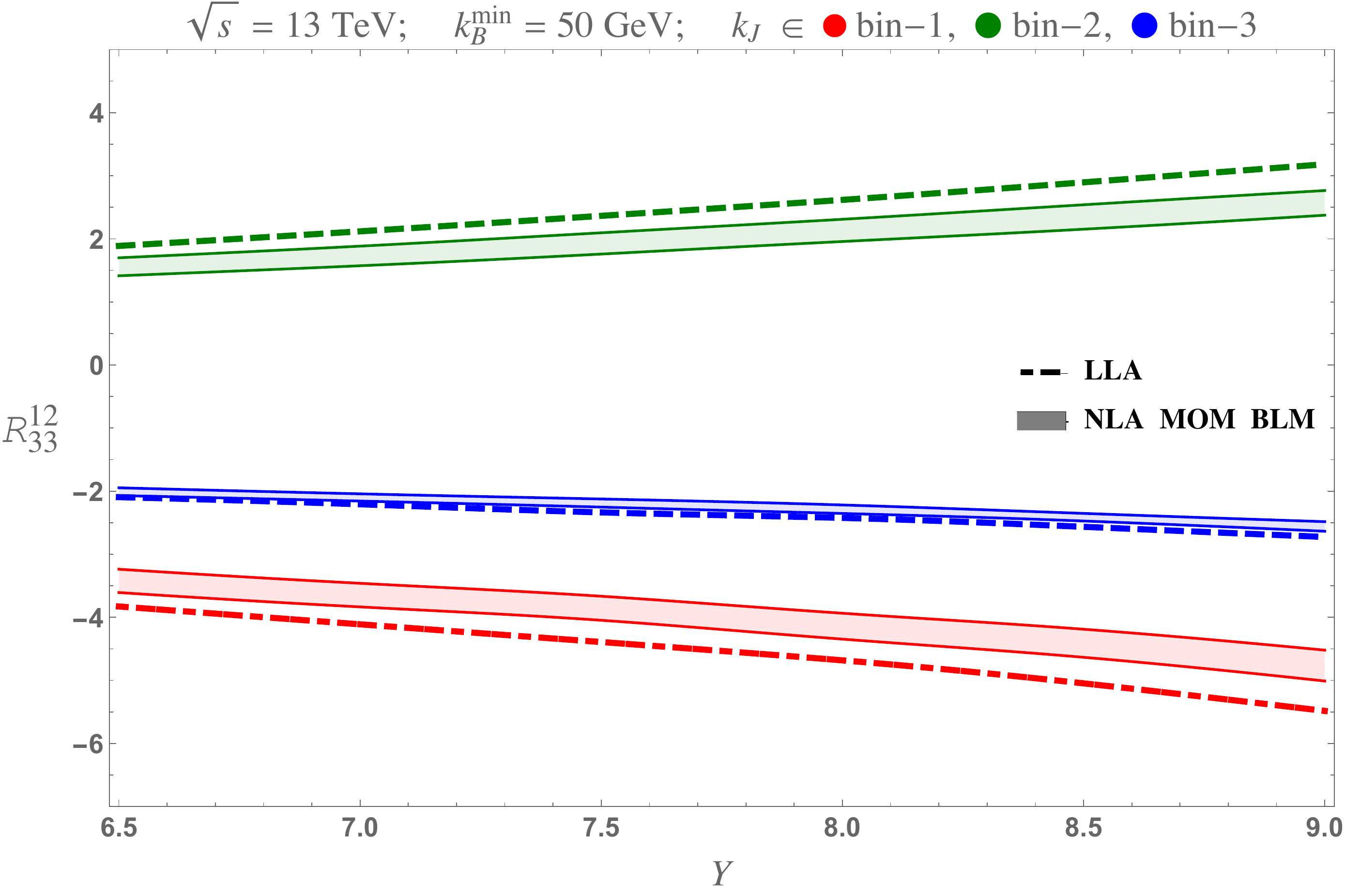}
     \includegraphics[scale=0.28]{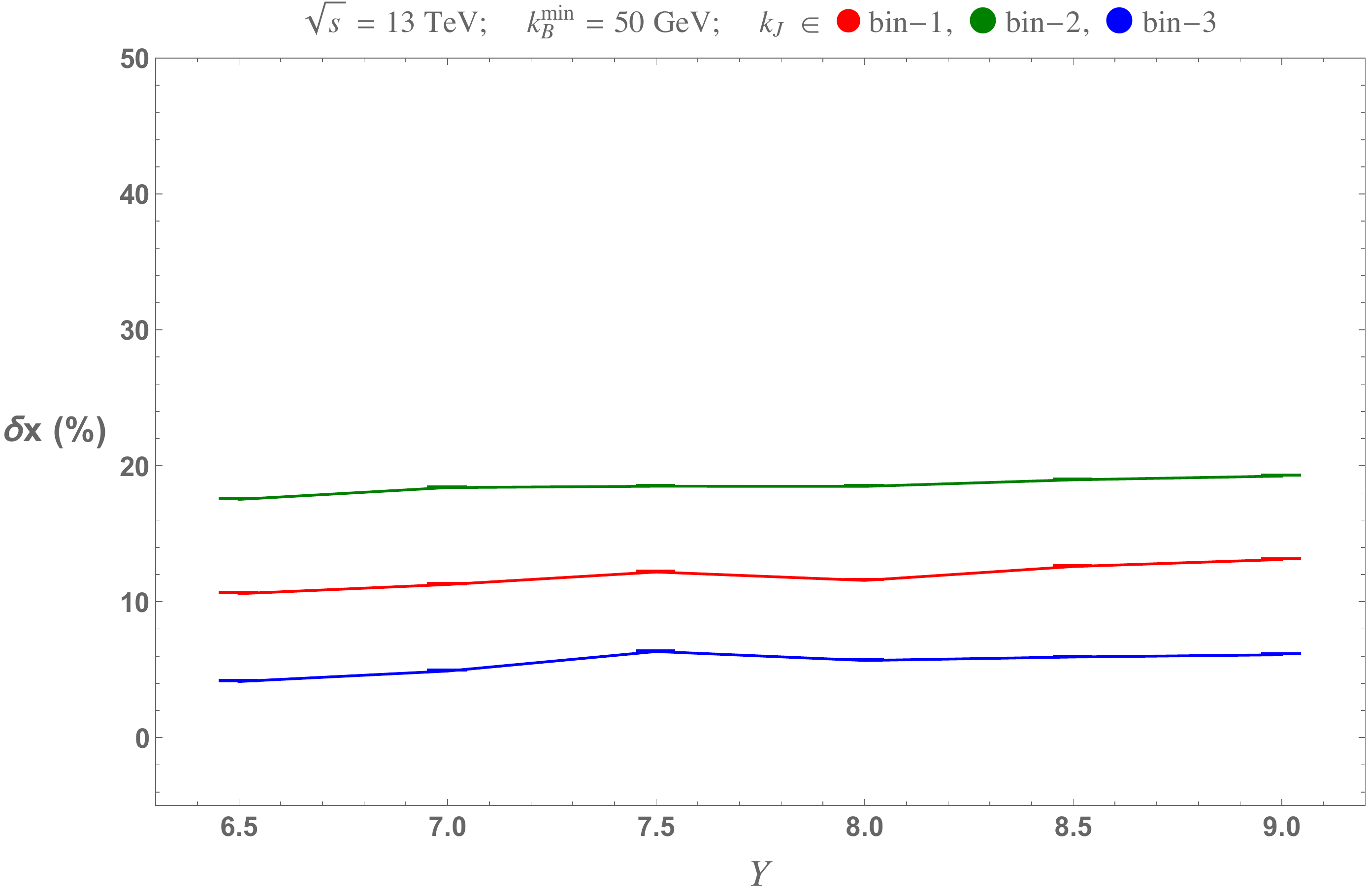}
     \vspace{1cm}
  
     \hspace{-16.25cm}   
     \includegraphics[scale=0.28]{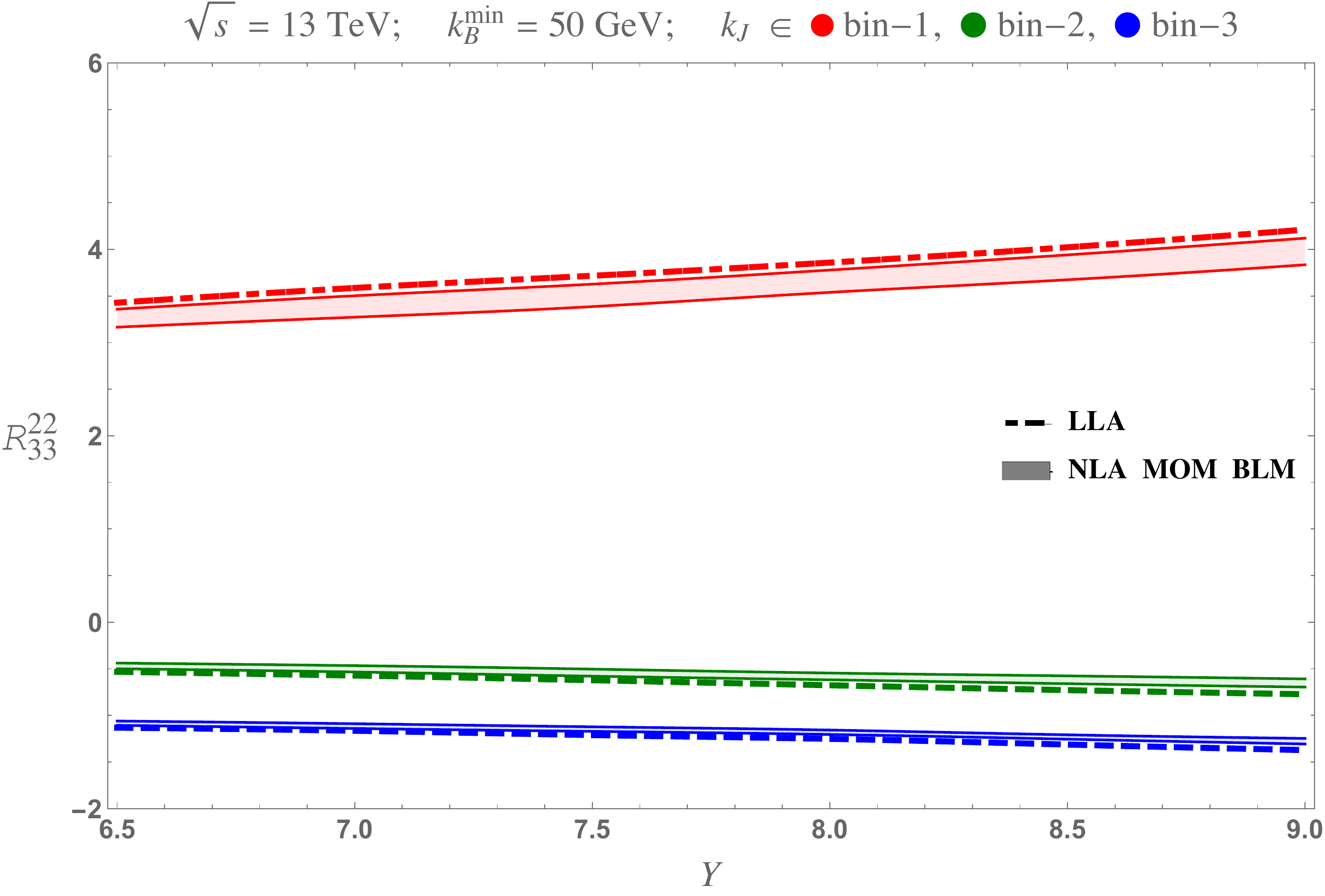}
     \includegraphics[scale=0.28]{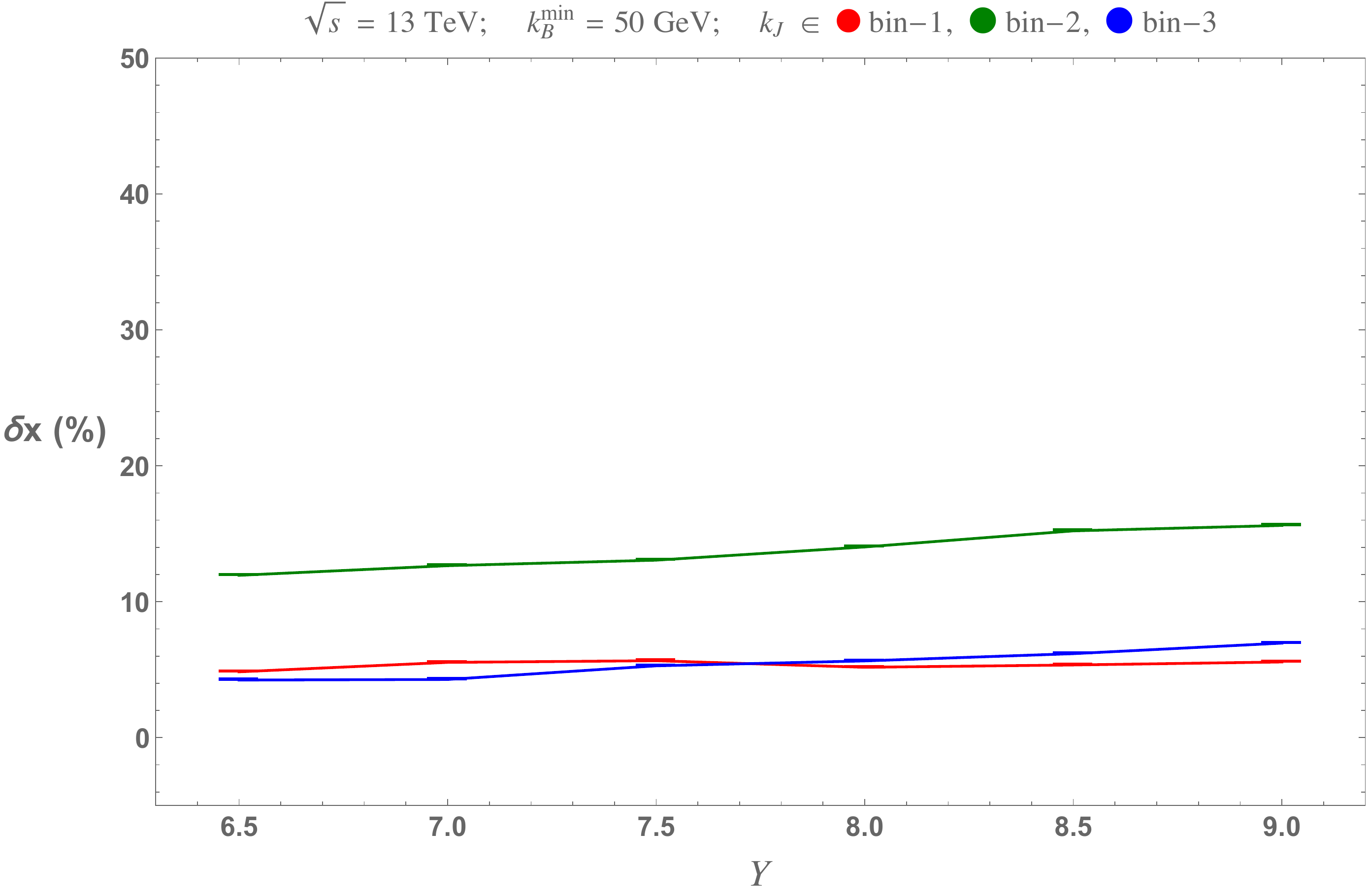}
  
  \restoregeometry
  \caption[LLA and NLA
           $^{\text{(i)}}R^{12}_{22}$, 
           $^{\text{(i)}}R^{12}_{33}$, and $^{\text{(i)}}R^{22}_{33}$ 
           at $\sqrt s = 13$ TeV]
   {LLA and NLA
    $^{\text{(i)}}R^{12}_{22}$, 
    $^{\text{(i)}}R^{12}_{33}$, and $^{\text{(i)}}R^{22}_{33}$ 
    at $\sqrt s = 13$ TeV with $y_J$ integrated over a central
    rapidity bin (left) and the relative NLA to LLA corrections (right).} 
  \label{fig:13-second}
  \end{figure}

  In this Section, everything is kept the same 
  as in Section~\ref{sub:3jets-nlk-2} 
  with the exemption
  of the allowed values for $y_J$ (see Fig.~\ref{fig:lego2}). 
  While in the previous Section $y_J = (Y_A+Y_B)/2$,
  here $y_J$ is not anymore dependent on the rapidity difference between
  the outermost jets, $Y$, and is allowed to take values in a 
  rapidity bin around $y_J = 0$.
  In particular, $-0.5 < y_J < 0.5$, which in turn 
  means that an additional integration
  over $y_J$ needs to be considered in Eq.~(\ref{Cmn_int}) 
  with $y_J^{\rm min} = -0.5$
  and  $y_J^{\rm max} = 0.5$:
  \begin{equation}
  \label{Cmn_int2}
   C_{MN}^{\text{(i)}} =
   \int_{y_J^{\rm min}}^{y_J^{\rm max}} \hspace{-0.25cm} dy_J
   \int_{Y_A^{\rm min}}^{Y_A^{\rm max}} \hspace{-0.25cm} dY_A
   \int_{Y_B^{\rm min}}^{Y_B^{\rm max}} \hspace{-0.25cm} dY_B
   \int_{k_A^{\rm min}}^{k_A^{\rm max}} \hspace{-0.25cm} dk_A
   \int_{k_B^{\rm min}}^{k_B^{\rm max}} \hspace{-0.25cm} dk_B
   \int_{k_J^{\rm min}}^{k_J^{\rm max}} \hspace{-0.25cm} dk_J
   \delta\left(Y_A - Y_B - Y\right) {\cal C}_{MN},
  \end{equation}
  We define our observables $^{\text{(i)}}R_{PQ}^{MN}$:
  \begin{equation}
  ^{\text{(i)}}R_{PQ}^{MN} 
  \, = \, 
  \frac{C_{MN}^{\text{(i)}}}{C_{PQ}^{\text{(i)}}}\,.
  \label{RmnqpNew2}
  \end{equation}
  The results for the $^{\text{(i)}}R^{12}_{22}$, 
  $^{\text{(i)}}R^{12}_{33}$ and $^{\text{(i)}}R^{22}_{33}$ 
  are shown in Figs.~\ref{fig:7-second} and~\ref{fig:13-second}.
  We notice immediately that Fig.~\ref{fig:7-first} 
  is very similar to the 
  integrated over $y_J$ observables 
  Fig.~\ref{fig:7-second} and
  the same holds for Figs.~\ref{fig:13-first} 
  and~\ref{fig:13-second}. 
  Therefore, we will not discuss here the individual 
  behaviours of $R^{12}_{22}$, $R^{12}_{33}$ and $R^{22}_{33}$
  with $Y$, neither the $\delta x(\%)$ corrections, 
  since this would only mean to repeat
  the discussion of Section~\ref{sub:3jets-nlk-1}.
  We would like only to note that the striking similarity 
  between Fig.~\ref{fig:7-first}
  and Fig.~\ref{fig:7-second} and between Fig.~\ref{fig:13-first}
  and Fig.~\ref{fig:13-second} was to be expected 
  if we remember that the 
  partonic-level quantities $\mathcal{R}_{PQ}^{MN}$ 
  do not change noticeably
  if we vary the position in rapidity of the central jet, 
  as long as the position remains
  `sufficiently' central (see Ref.~\cite{Caporale:2015vya}). 
  This property is very important and we will discuss it 
  more in the next Section.
  Here, we should stress that the observables 
  as presented in this Section
  can be readily compared to experimental data.

  \subsection[Dependence on bins for all three jets]
             {Dependence of the generalised azimuthal correlations
              on a forward-, backward- and central-rapidity bin}
  \label{sub:3jets-nlk-3}
  
    \begin{figure}[H]
   \includegraphics[scale=0.35]{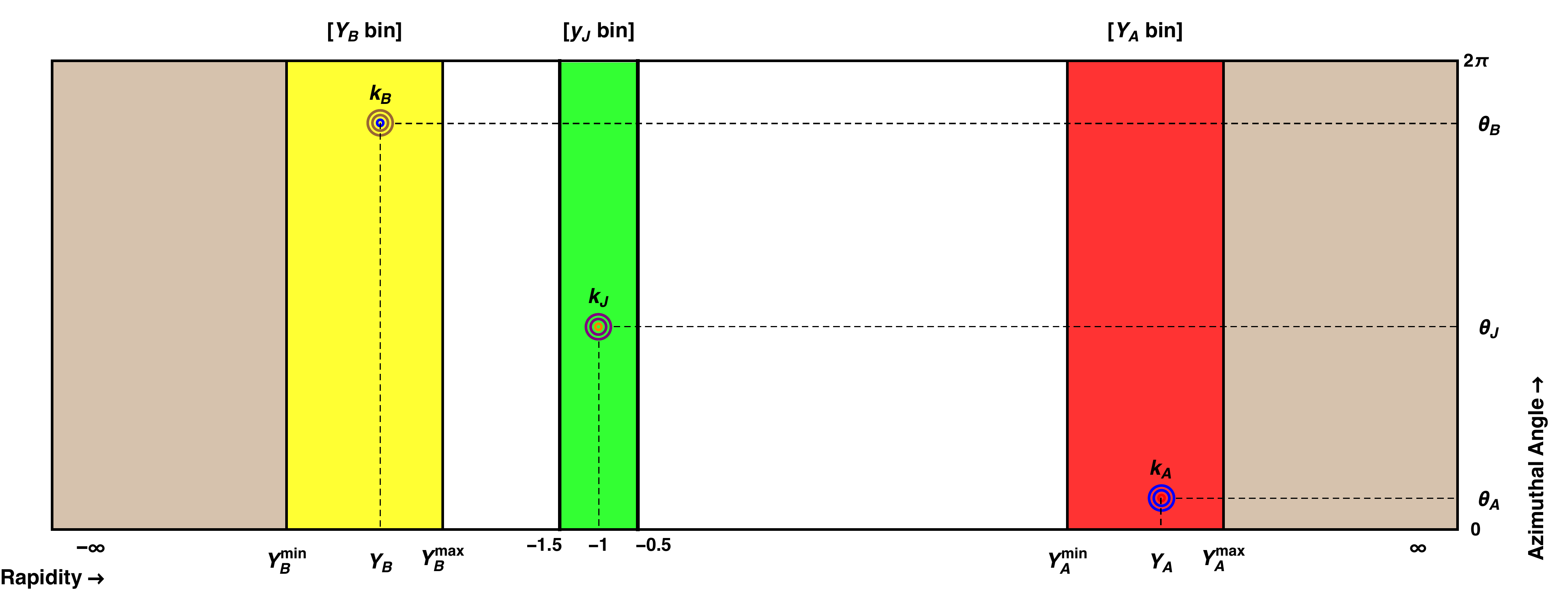}
   \caption[Lego plot for a three-jet event with integration 
            over bin on all jet rapidities]
   {A primitive lego plot 
   depicting a three-jet event similar to Figs.~\ref{fig:lego1} and~\ref{fig:lego2}. 
   Here, however, the rapidity of the central jet 
   can take any value in the distinct ranges 
   $y_i-0.5 < y_J < y_i+0.5$, where
   $y_i$ is the central value of the rapidity bin 
   with $y_i = \{-1, -0.5, 0, 0.5, 1\}$.
   In this figure, $y_i = -1$.
   Moreover, $Y = Y_A - Y_B$ is not anymore fixed. 
   Instead, the forward jet has a rapidity restricted in
   the red bin whereas the backward jet in the yellow bin. 
   }
   \label{fig:lego3}
   \end{figure}
  
  In this Section, an alternative kinematical configuration 
  (see Fig.~\ref{fig:lego3}) for the 
  generalised ratios $R_{PQ}^{MN}$ is presented, 
  whose choice relies on two reasons.
  The first one is to offer a different setup for which the comparison 
  between theoretical
  predictions and experimental data might be easier, 
  compared to the previous
  Section. The second one, to demonstrate that 
  the generalised ratios do capture
  the Bethe--Salpeter characteristics~\cite{Salpeter:1951sz} of the BFKL radiation. 
  The latter needs
  a detailed explanation. 
  
  Let us assume that we have a gluonic ladder exchanged 
  in the $t$-channel between a
  forward jet (at rapidity $Y_A$) 
  and a backward jet (at rapidity $Y_B$) 
  accounting for minijet activity
  between the two jets. By gluonic ladder here 
  we mean the Green's function 
  $\varphi \left(\vec{p}_A,\vec{p}_B,Y_A - Y_B\right) $, 
  where $\vec{p}_A$ and $\vec{p}_B$ are the Reggeised momenta 
  connected to the forward
  and backward jet vertex respectively. 
  It is known that the following relation holds 
  for the Green's function:
  \begin{eqnarray}
  \hspace{-.3cm}
  \varphi \left(\vec{p}_A,\vec{p}_B,Y_A - Y_B\right) &=& 
  \int d^2 \vec{k} \, 
  \varphi \left(\vec{p}_A,\vec{k},Y_A - y\right) 
   \varphi \left(\vec{k},\vec{p}_B,y - Y_B\right).
   \label{BasicRelation0}
  \end{eqnarray}
  In other words, one may `cut' the gluonic ladder 
  at any rapidity $y$ between 
  $Y_A$ and $Y_B$ and then integrate
  over the Reggeised momentum $\vec{k}$ that flows
  in the $t$-channel, to recover the initial ladder.
  Which value of $y$ one chooses to `cut' the ladder at is irrelevant.
  Therefore, observables directly connected 
  to a realisation of the r.h.s of Eq.~(\ref{BasicRelation0})
  should display this $y$-independence.
  
  In our study actually, we have a very similar picture
  as the one described in the r.h.s of Eq.~(\ref{BasicRelation0}). 
  The additional element is that we do not only `cut' 
  the gluonic ladder but
  we also `insert' a jet vertex for the central jet. 
  This means that the $y$-independence we discussed
  above should be present in one form or another. 
  To be precise, we do see the $y$-independence behaviour
  but now we have to consider the additional constraint 
  that $y$ cannot take any extreme values,
  that is, it cannot be close to $Y_A$ or $Y_B$. 
  For a more detailed discussion
  of Eq.~(\ref{BasicRelation0}), we refer the
  reader to Appendix~\hyperlink{app:y-link}{D}), 
  here we will proceed to present our numerical results.
  
  The kinematical setup now is different 
  than in the previous Sections. 
  We allow $Y_A$ and $Y_B$
  to take values such that 
  $(Y_A^{\text{min}} = 3) < Y_A < (Y_A^{\text{max}} = 4.7)$ 
  and
  $(Y_B^{\text{min}} = -4.7) < Y_B < (Y_B^{\text{max}} = -3)$.
  Moreover, we allow for
  the rapidity of the central jet to take values 
  in five distinct rapidity bins of unit width, that is,
  $y_i-0.5 < y_J<y_i+0.5$, with $y_i = \{-1, -0.5, 0, 0.5, 1\}$ 
  and we define the
  coefficients $C_{MN}^{\rm (i)}(y_i)$ as function of $y_i$:
  \begin{equation}
  \label{Cmn_int3}
   C_{MN}^{\text{(i)}}(y_i) =
   \int_{y_i-0.5}^{y_i+0.5} \hspace{-0.25cm} dy_J
   \int_{Y_A^{\rm min}}^{Y_A^{\rm max}} \hspace{-0.25cm} dY_A
   \int_{Y_B^{\rm min}}^{Y_B^{\rm max}} \hspace{-0.25cm} dY_B
   \int_{k_A^{\rm min}}^{k_A^{\rm max}} \hspace{-0.25cm} dk_A
   \int_{k_B^{\rm min}}^{k_B^{\rm max}} \hspace{-0.25cm} dk_B
   \int_{k_J^{\rm min}}^{k_J^{\rm max}} \hspace{-0.25cm} dk_J\,\,
  {\mathcal C}_{MN}.
  \end{equation}
  
\begin{figure}[p]
  \newgeometry{left=-10cm,right=1cm}
  \centering
  
     \hspace{-16.25cm}
     \includegraphics[scale=0.28]{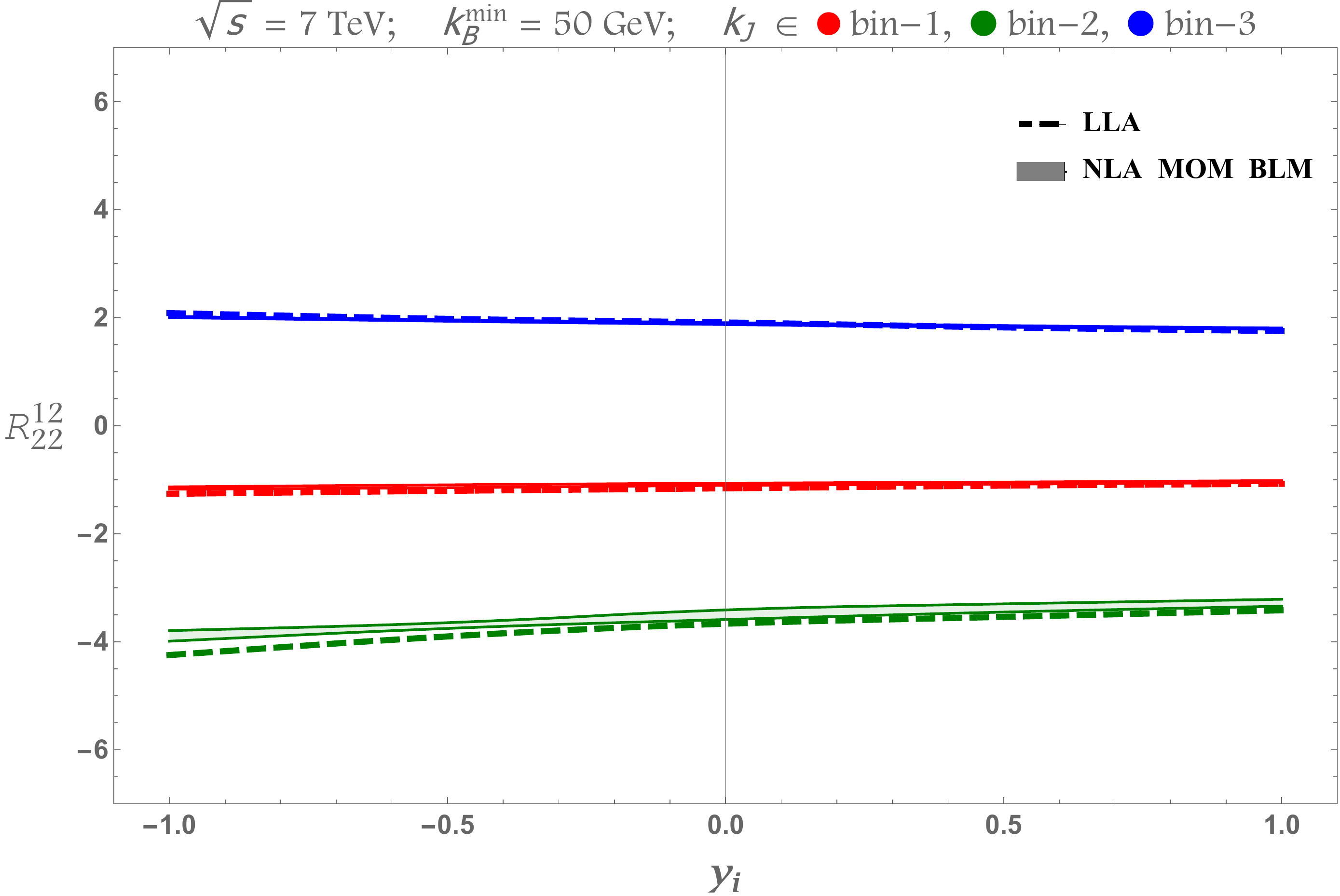}
     \includegraphics[scale=0.28]{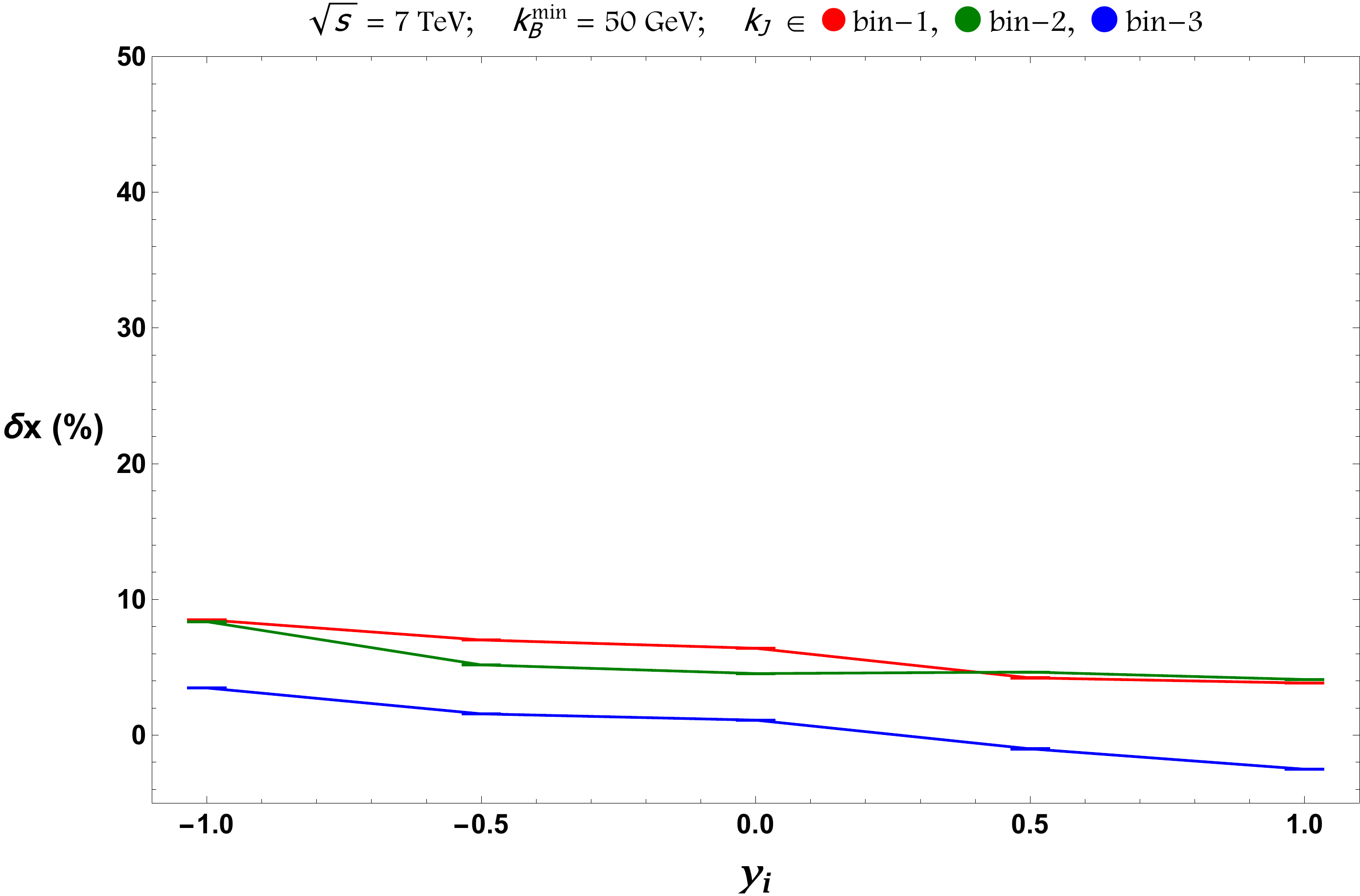}
     \vspace{1cm}
  
     \hspace{-16.25cm}
     \includegraphics[scale=0.28]{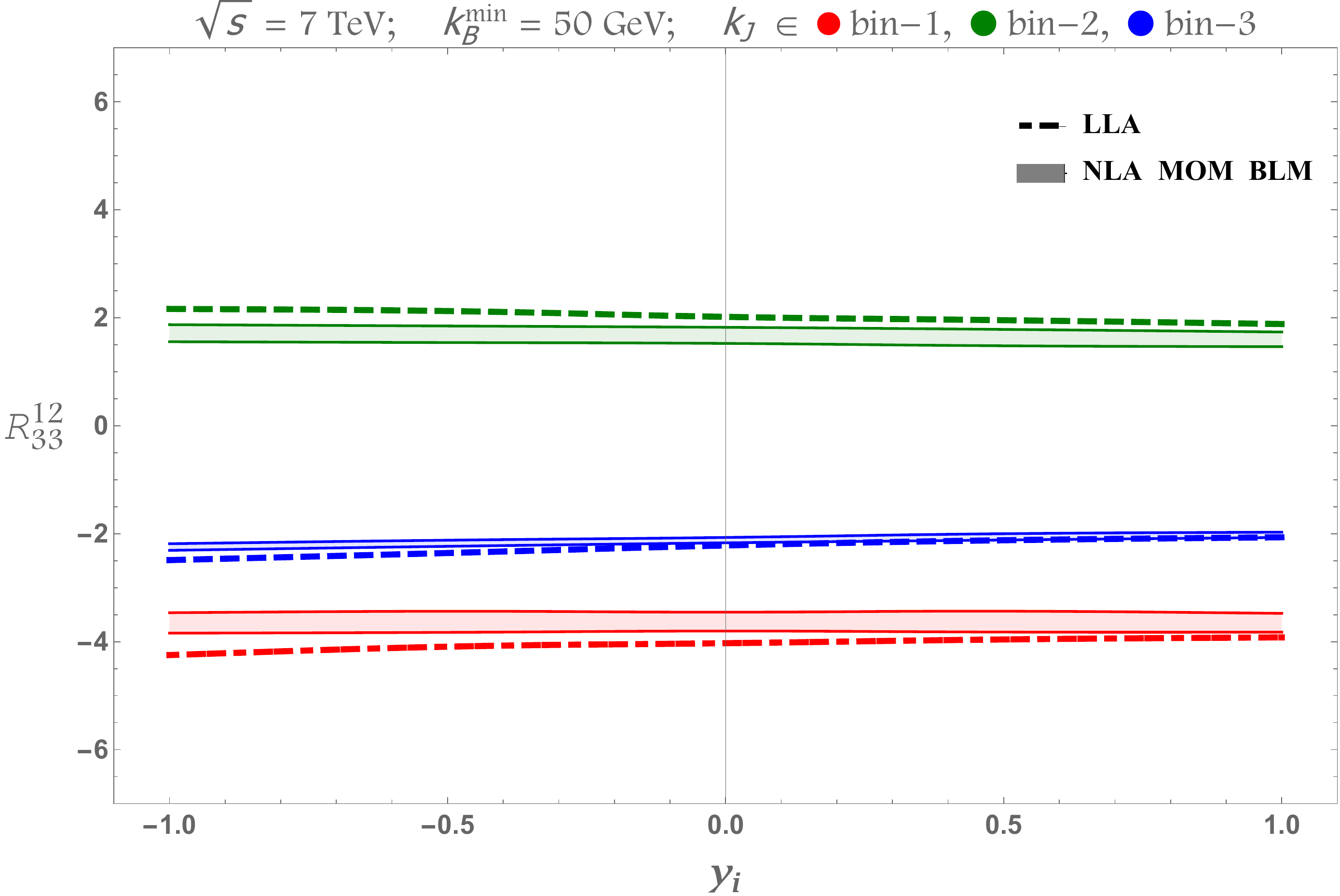}
     \includegraphics[scale=0.28]{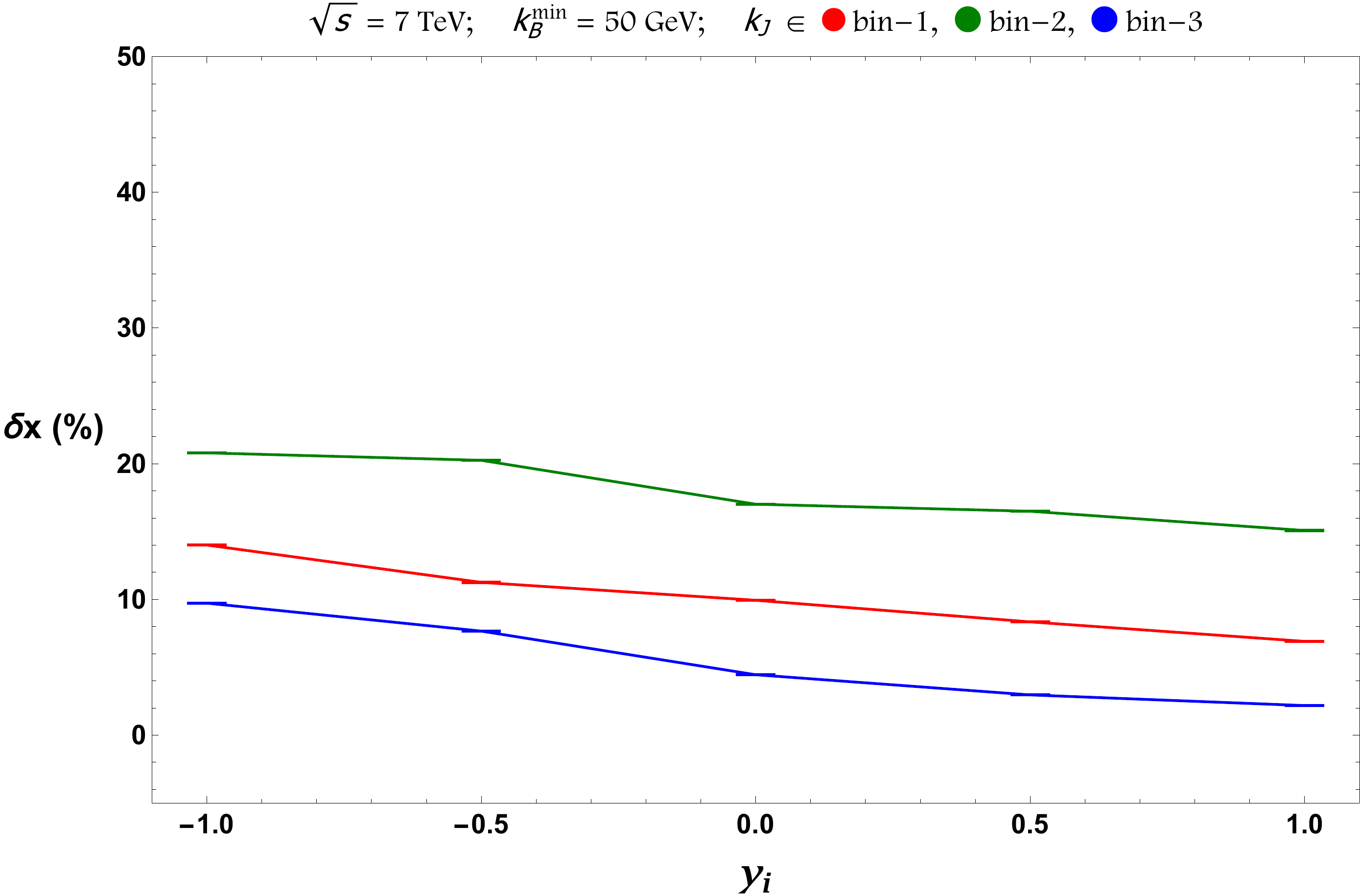}
     \vspace{1cm}
  
     \hspace{-16.25cm}   
     \includegraphics[scale=0.28]{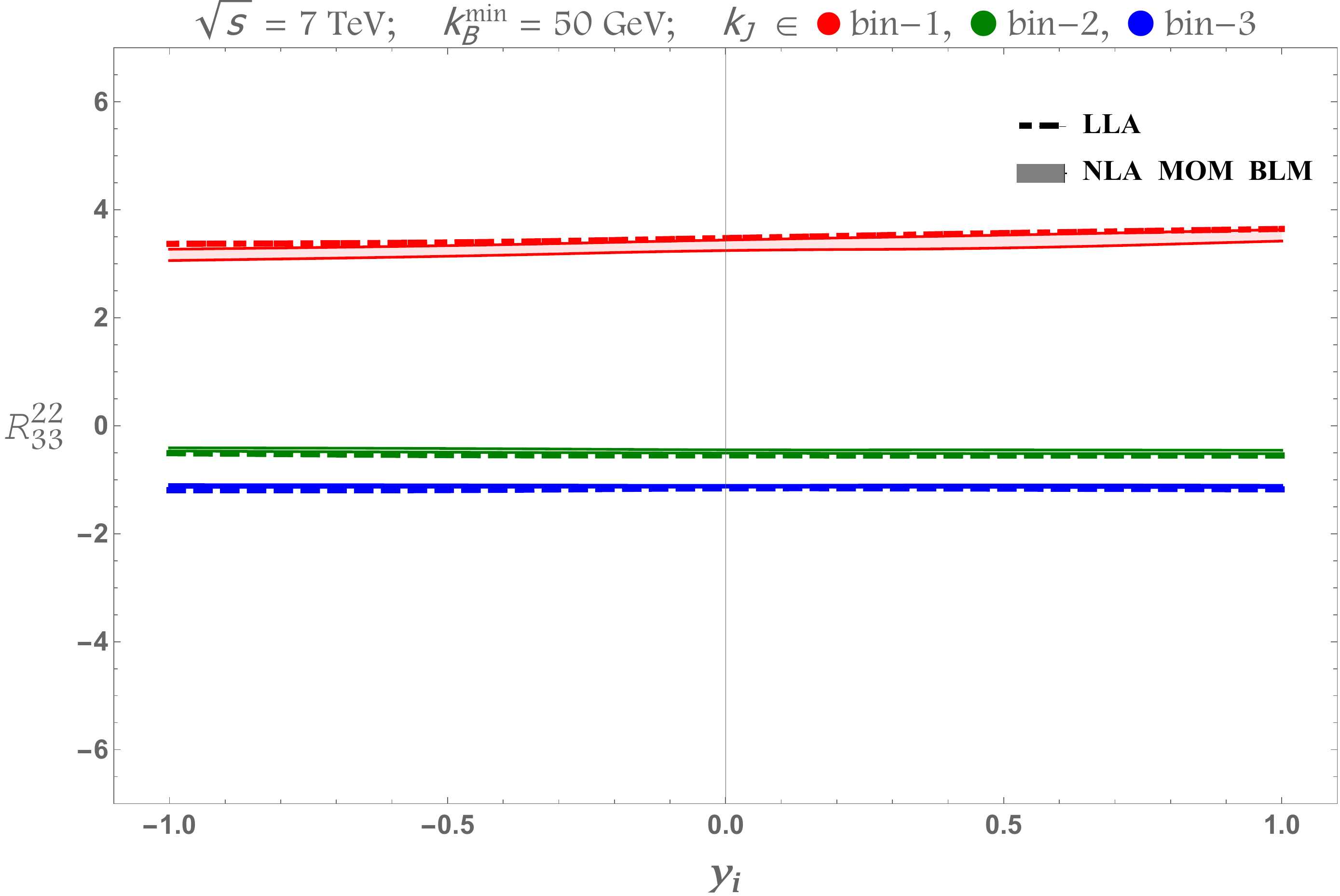}
     \includegraphics[scale=0.28]{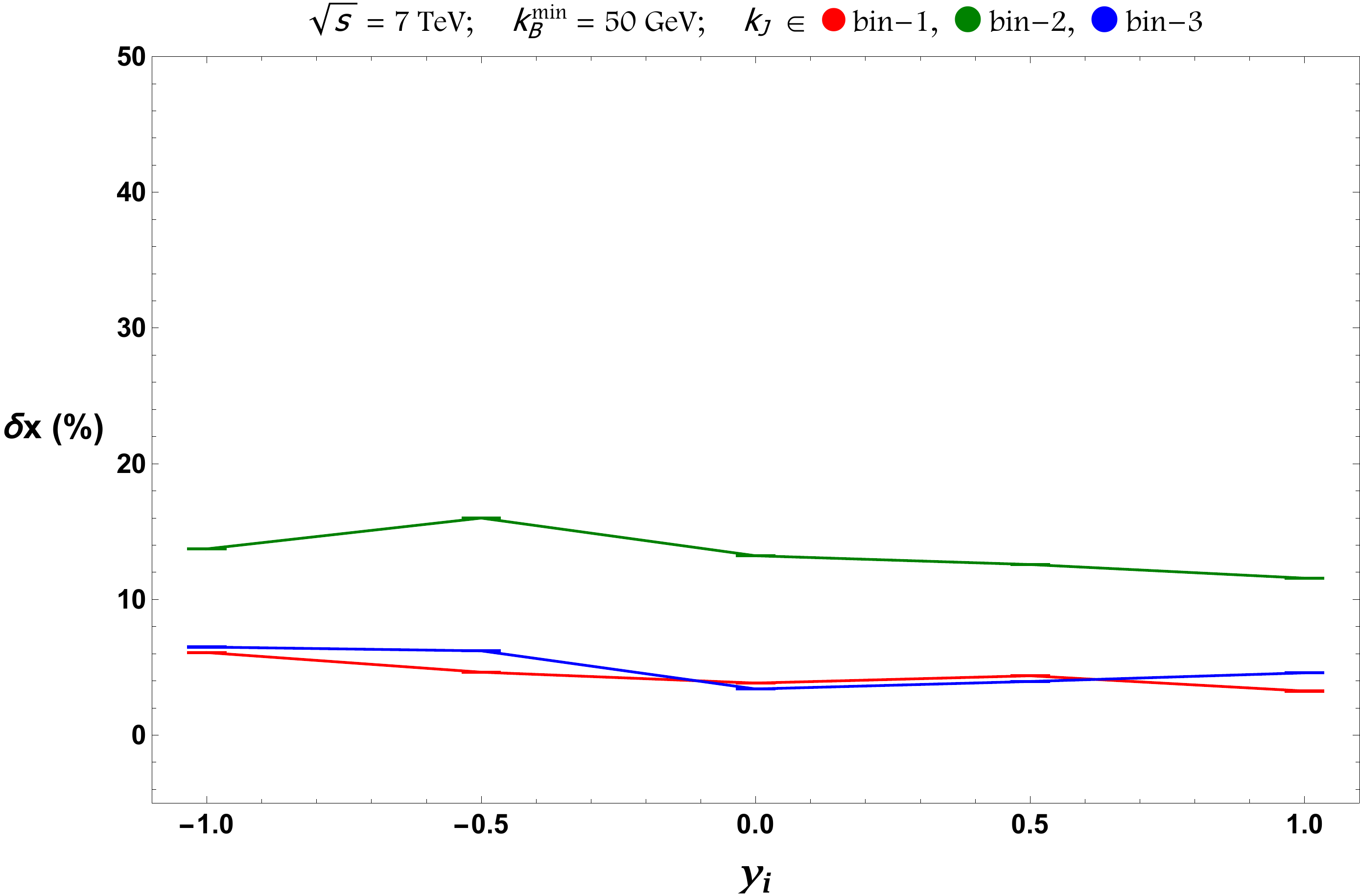}
  
  \restoregeometry
  \caption[LLA and NLA
           $^{\text{(i)}}R^{12}_{22}(y_i)$, 
           $^{\text{(i)}}R^{12}_{33}(y_i)$, and  
           $^{\text{(i)}}R^{22}_{33}(y_i)$ 
           at $\sqrt s = 7$ TeV]
   {$y_i$-dependence of the LLA and NLA
    $^{\text{(i)}}R^{12}_{22}(y_i)$, 
    $^{\text{(i)}}R^{12}_{33}(y_i)$ and 
    $^{\text{(i)}}R^{22}_{33}(y_i)$ 
    at $\sqrt s = 7$ TeV (left) and the relative NLA 
    to LLA corrections  (right).} 
  \label{fig:7-third}
  \end{figure}

\begin{figure}[p]
  \newgeometry{left=-10cm,right=1cm}
  \centering
  
     \hspace{-16.25cm}
     \includegraphics[scale=0.28]{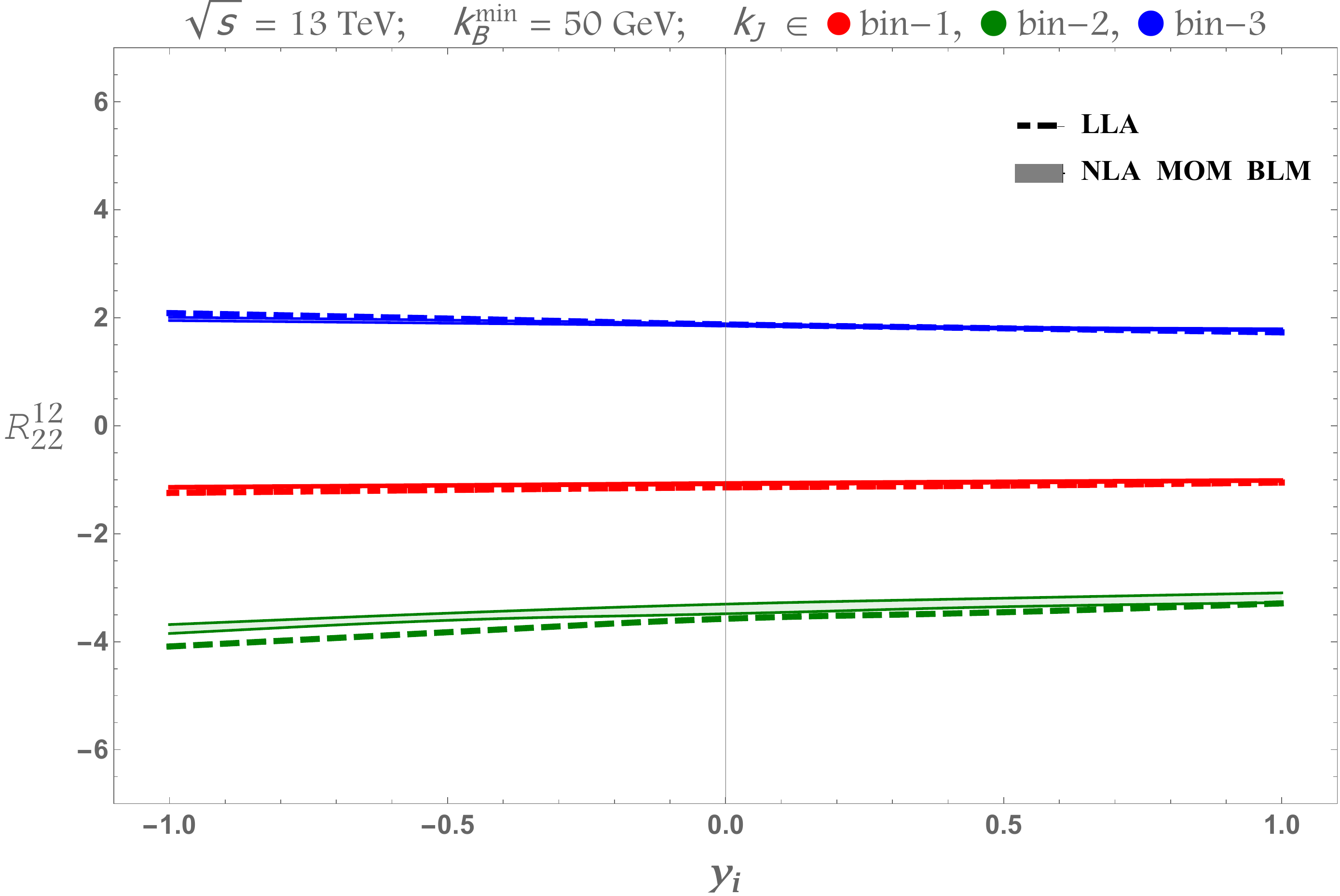}
     \includegraphics[scale=0.28]{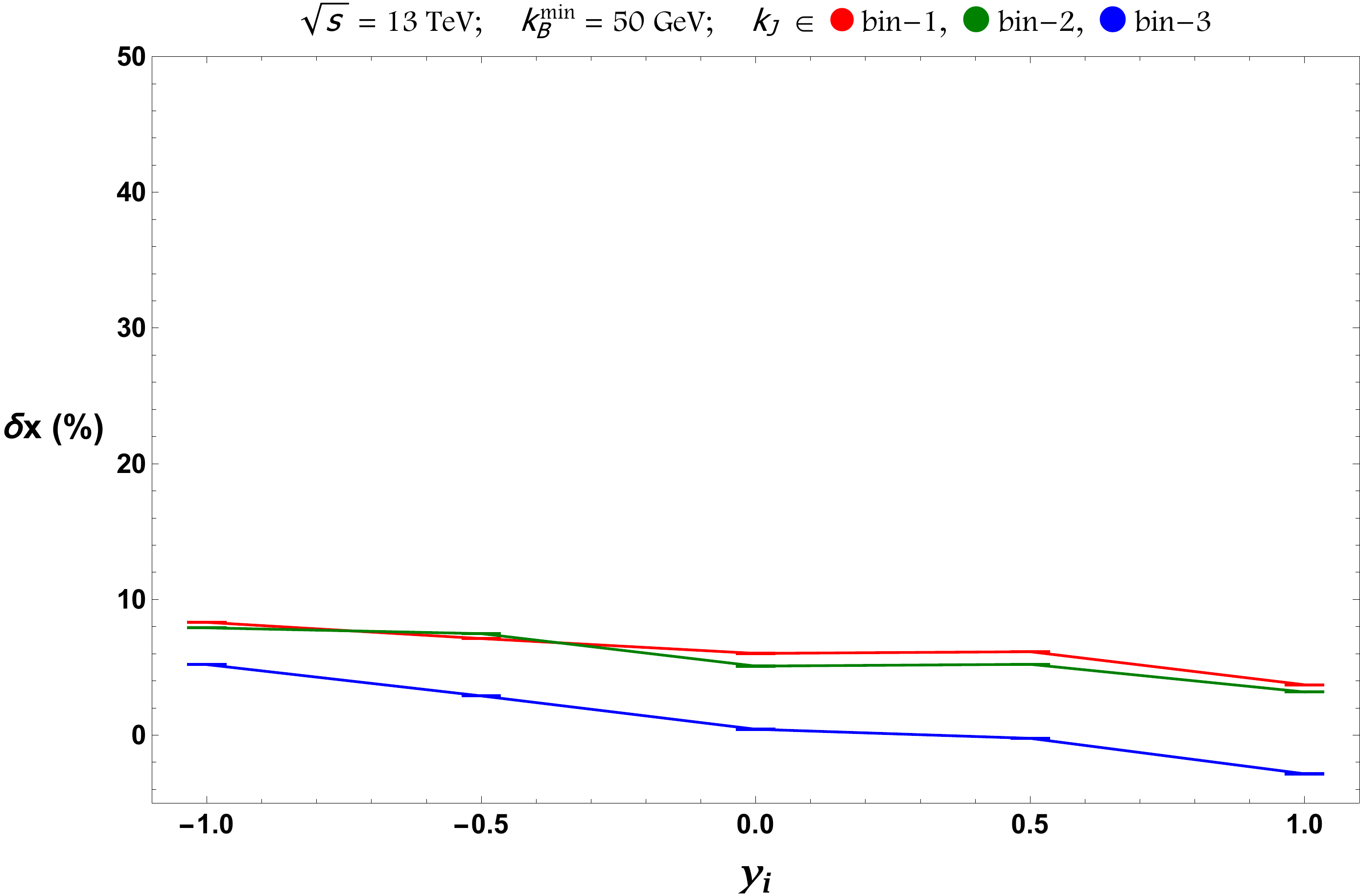}
     \vspace{1cm}
  
     \hspace{-16.25cm}
     \includegraphics[scale=0.28]{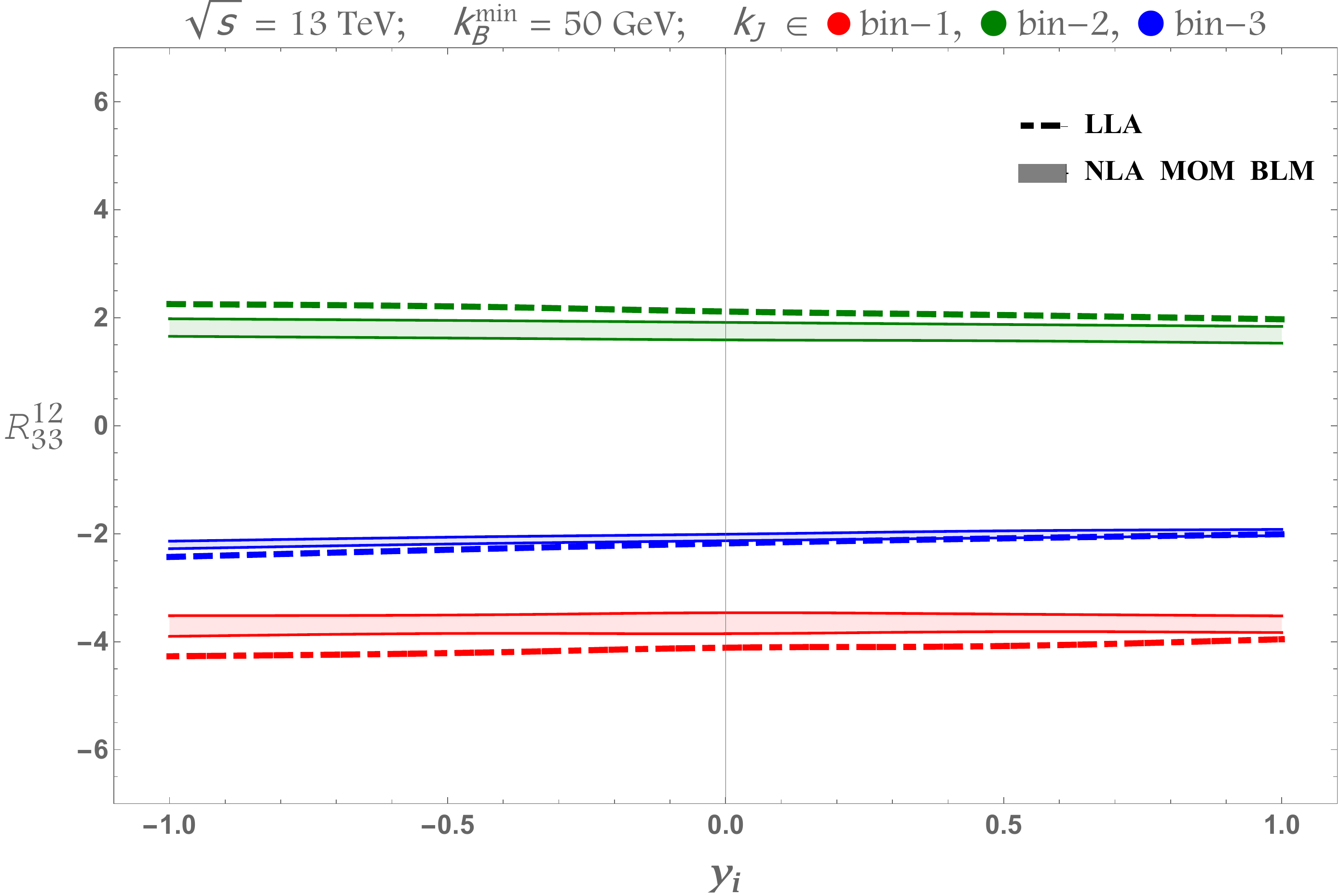}
     \includegraphics[scale=0.28]{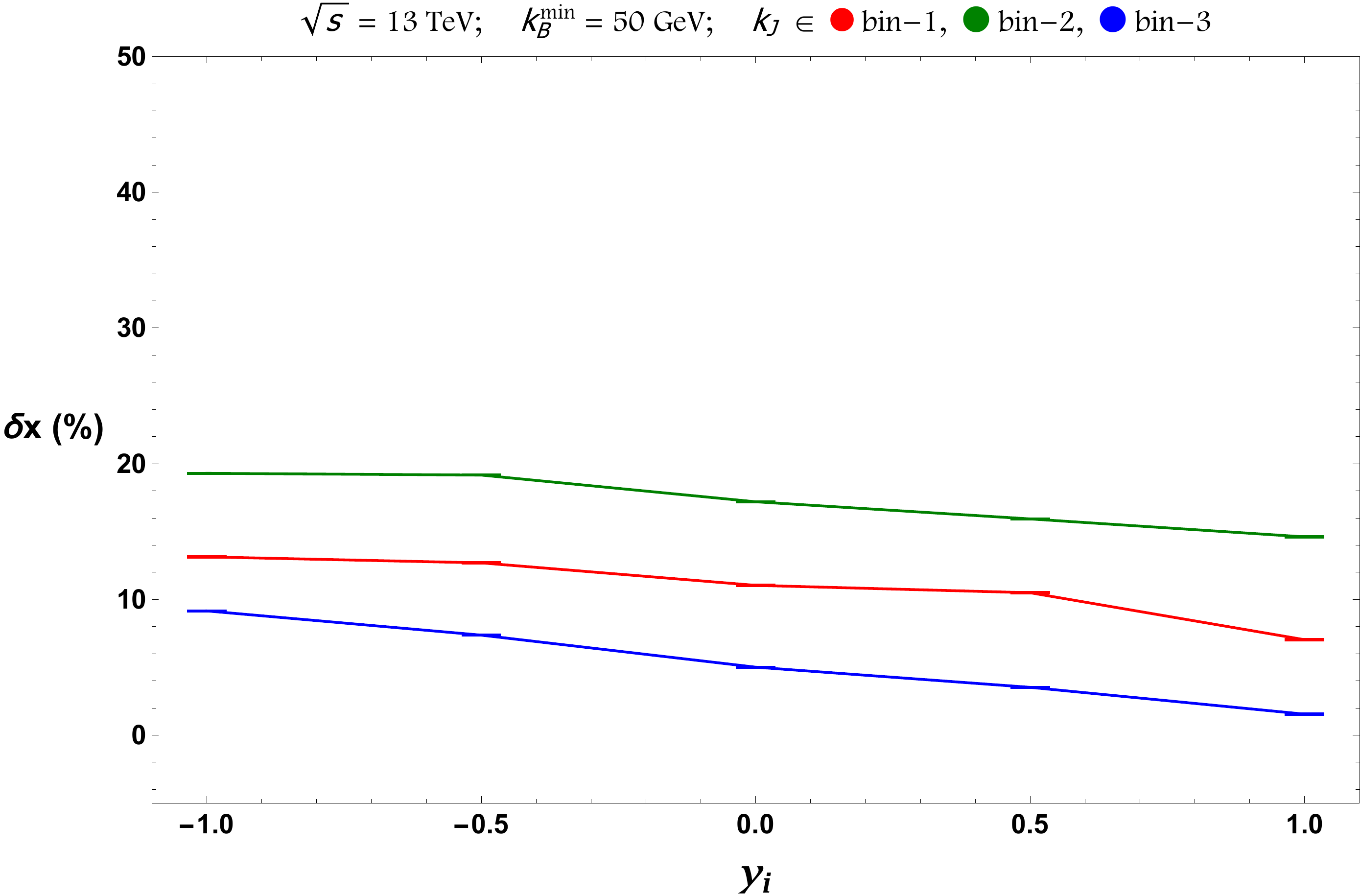}
     \vspace{1cm}
  
     \hspace{-16.25cm}   
     \includegraphics[scale=0.28]{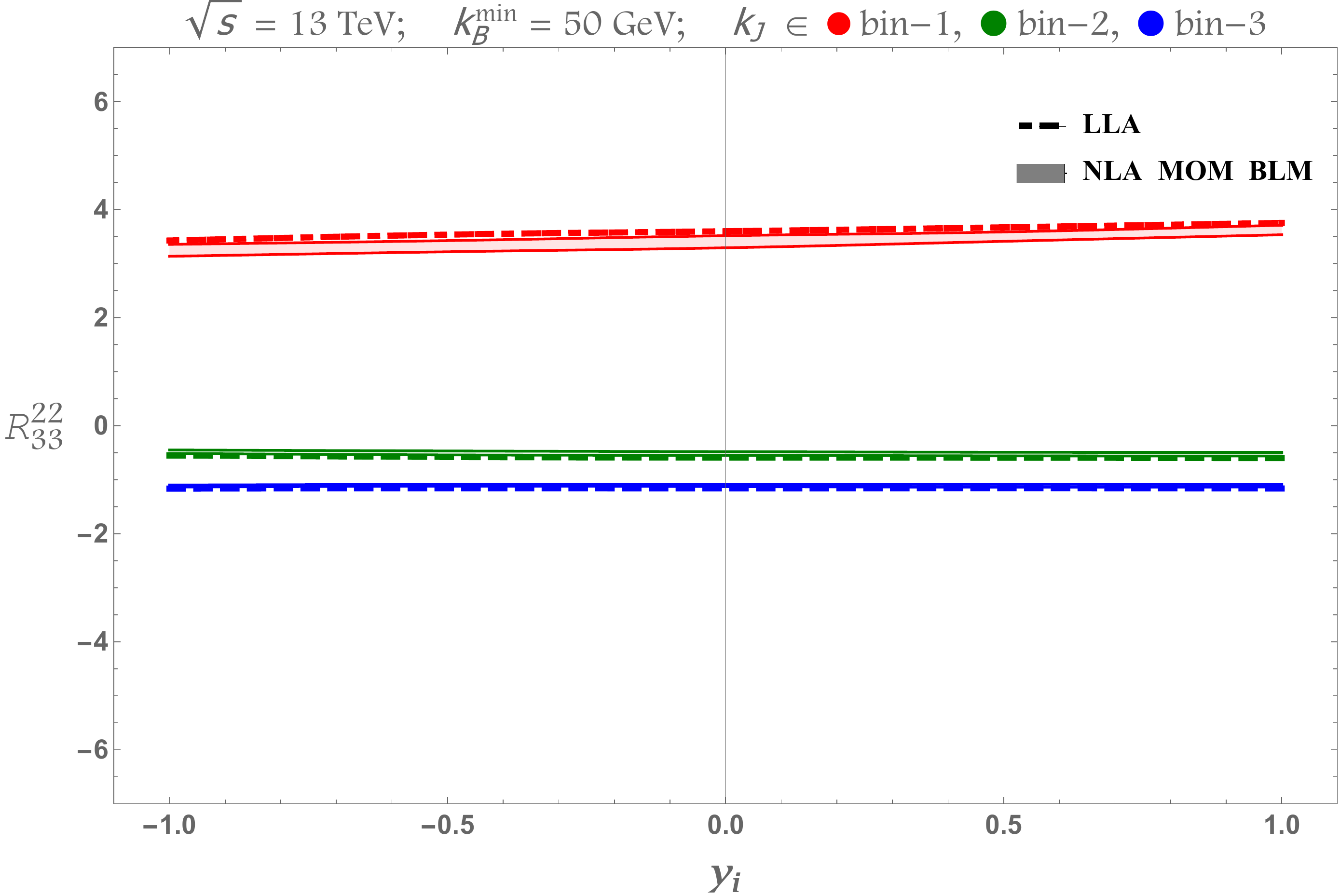}
     \includegraphics[scale=0.28]{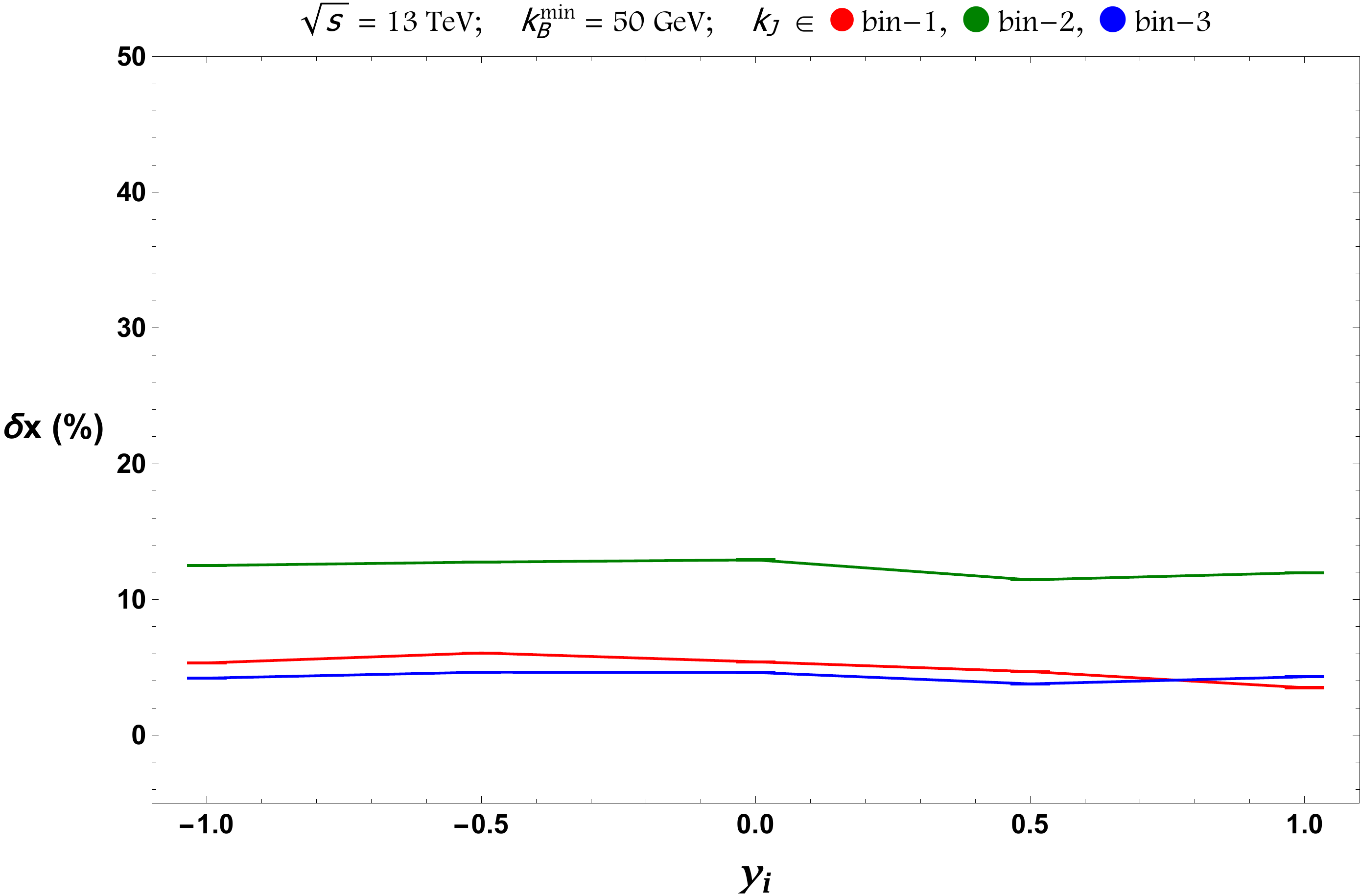}
  
  \restoregeometry
  \caption[LLA and NLA
    $^{\text{(i)}}R^{12}_{22}$, 
    $^{\text{(i)}}R^{12}_{33}$, and 
    $^{\text{(i)}}R^{22}_{33}$ 
    at $\sqrt s = 13$ TeV]
   {$y_i$-dependence of the LLA and NLA
    $^{\text{(i)}}R^{12}_{22}$, 
    $^{\text{(i)}}R^{12}_{33}$, and $^{\text{(i)}}R^{22}_{33}$ 
    at $\sqrt s = 13$ TeV (left) and the relative NLA 
    to LLA corrections (right).} 
  \label{fig:13-third}
  \end{figure}

  We denote our observables by $^{\text{(i)}}R_{PQ}^{MN}$, 
  which are now functions of $y_i$ instead of $Y$:
  \begin{eqnarray}
  \label{iRPQMNyi}
  R_{PQ}^{MN}(y_i) 
  \, = \, 
  \frac{C_{MN}^{\text{(i)}}(y_i)}{C_{PQ}^{\text{(i)}}(y_i)}\,.
  \label{RmnqpNew3}
  \end{eqnarray}
  We present our results in Figs.~\ref{fig:7-third} 
  and~\ref{fig:13-third}.
  We see that indeed, the $y_i$-dependence 
  of the three ratios is very weak.
  Moreover, the similarity between 
  the $\sqrt s = 7$ TeV and $\sqrt s = 13$ TeV
  plots is more striking that in the previous Sections.
  The relative NLA to LLA corrections seem 
  to be slightly larger here than in the previous Sections.
  We would like to stress once more that the results 
  in this Section are readily
  comparable to the experimental data 
  once the same cuts are applied in the 
  experimental analysis.
  
  \subsection{Numerical tools}
  \label{sub:3j-numerics}
 
  The numerical computation of the $\mathcal{R}^{MN}_{PQ}$ ratios
  presented in Sections~\ref{sub:3jets-nlk-1}, 
  \ref{sub:3jets-nlk-2} and \ref{sub:3jets-nlk-3} 
  were done both in \textsc{Fortran} 
  and in \textsc{Mathematica} (mainly for cross-checks).
  The NLO MSTW 2008 PDF sets~\cite{Martin:2009iq} were used 
  \footnote{
   Other potential sources of uncertainty could be due to the particular PDF sets one uses.
   One can still argue though that
   the uncertainty due to different PDF sets does not need to be ascertained before
   one has gauged how large are the full beyond the LLA corrections to the partonic-level ratios, since
   it will be overshadowed by the latter. Indeed, from first tries we see no significant
   difference in the results when we work with different PDF sets and therefore
   we do not offer any dedicated analysis on that here.
   }
  and a two-loop running coupling setup 
  with $\alpha_s\left(M_Z\right)=0.11707$ 
  was chosen with five quark flavours active. 
  An extensive use of the integration routine 
  \cod{Vegas}~\cite{VegasLepage:1978} was made, as implemented 
  in the \cod{Cuba} library~\cite{Cuba:2005,ConcCuba:2015}.
  Furthermore, \cod{Quadpack} library~\cite{Quadpack:book:1983}
  and a slightly modified version 
  of the \cod{Psi}~\cite{RpsiCody:1973} routine were used.

 \section{Summary} 
 \label{sec:3j-summary}
 
 The first complete phenomenological analysis 
 for the inclusive three-jet production 
 was presented. New azimuthal-angle-dependent 
 observables, the $R^{MN}_{PQ}$ ratios, were defined first at the partonic level, 
 then extended and studied at the hadronic level, 
 taking in the account of NLA BFKL corrections.
 Two colliding energies, $\sqrt s = 7, 13$ TeV, 
 together with an \emph{asymmetric} kinematical cut with respect to the transverse momentum
 of the forward ($k_A$) and backward ($k_B$) jets 
 were considered. 
 In addition, an
 extra condition regarding the value of the transverse momentum $k_J$ of the central jet was taken up,
 dividing the allowed region for $k_J$ into three sub-regions: $k_J$ smaller than
 $k_{A,B}$, $k_J$ similar to $k_{A,B}$, and $k_J$ larger than $k_{A,B}$.
 
 For a proper study at full NLA, one needs to consider the NLO jet vertices
 and the NLA Green's functions, 
 with the latter being expected to be of higher relevance. 
 BLM prescription, which has been successful in previous phenomenological analyses~\cite{Khachatryan:2016udy}, was used to choose the values of the renormalisation scale $\mur$.
 It was shown how the $R^{MN}_{PQ}$ ratios
 change when we vary the rapidity
 difference Y between $k_A$ and $k_B$ from 5.5 to 9 units for a fixed $y_J$ and
 from 6.5 to 9 units for $-0.5 < y_J < 0.5$. 
 Both the LLA and NLA results were presented, 
 along with plots that show the relative size of the 
 NLA corrections compared to the LLA ones.
 An alternative kinematical setup were also investigated, where
 $Y_A$ and $Y_B$ are allowed to take
 values such that $3 < Y_A <  4.7$ and
 $-4.7 < Y_B < -3$, while
 the rapidity of the central jet takes values in five distinct rapidity bins of unit width, that is,
 $y_i-0.5 < y_J<y_i+0.5$, with $y_i = \{-1, -0.5, 0, 0.5, 1\}$. 
 In this alternative setup,
 we presented the behaviour of our observables
 as functions of $y_i$.
 
 The general conclusion is that the NLA corrections are moderate and our proposed
 observables exhibit a good perturbative stability. Furthermore, we see that for a wide
 range of rapidities, the changes we notice when going from $7$ TeV to $13$ TeV are
 small which makes us confident that
 these generalised ratios pinpoint the crucial characteristics
 of the BFKL dynamics regarding
 the azimuthal behaviour of the hard jets in inclusive three-jet production.
 
 It would be very interesting to have an experimental analysis for these observables using previous and current LHC data. We have the strong belief that such an analysis will help us gauge the applicability of the BFKL dynamics in phenomenological studies at present colliding energies.

\newpage 
 
\setcounter{appcnt}{0}
\renewcommand{\theequation}{D.\arabic{appcnt}}
\setcounter{tmp}{4}
\clearpage
\hypertarget{app:y-link}{}
\chapter*{Appendix~D}
\vspace{-0.5cm} 
\noindent
{\Huge \bf $y_J$-independent integrated distributions}
\label{app:y}
\addcontentsline{toc}{chapter}{\numberline{\Alph{tmp}}
 $y_J$-independent integrated distributions}
\markboth{$y_J$-independent integrated distributions}{}
\markright{APPENDIX D}{}
\vspace{1.3cm} 

In this Appendix it is shown that Eq.~(\ref{BasicRelation0})
is fulfilled in our normalisations.
We introduce the notation $t = \ln{k^2}$ 
to write the Green's function in the form
\begin{align}\label{eq:app-y-1}
\stepcounter{appcnt}
& \varphi \left(t_A,t_B,\theta_A,\theta_B,Y\right)  
\\ \nonumber =  
& \frac{e^{-\frac{t_A+t_B}{2}}}{\pi^2} 
\sum_{n=-\infty}^\infty e^{i n \left(\theta_A - \theta_B\right)} 
\int_0^\infty d \nu   
\cos{\left(\nu \left(t_A-t_B\right)\right)} \, 
e^{\bar{\alpha}_s  \chi(n,\nu) Y}\;.
\end{align}
Making use of $d k = \frac{1}{2} e^{\frac{t}{2}} dt$ and $k \, dk \, d \theta = \frac{e^t}{2} d \theta$ we then want to 
show that
\begin{align}\label{eq:app-y-2}
\stepcounter{appcnt}
& \varphi \left(t_A,t_B,\theta_A,\theta_B,Y\right) 
  \\ \nonumber & =
  \int_0^{2 \pi} d \theta 
\int_{-\infty}^\infty dt \, \frac{ e^t}{2} 
\varphi \left(t_A,t,\theta_A,\theta,y\right)\varphi \left(t,t_B,\theta,\theta_B,Y-y\right) \\\nonumber
& = \int_0^{2 \pi} d \theta 
\int_{-\infty}^\infty dt \, \frac{ e^t}{2} \frac{e^{-\frac{t_A+t}{2}} }{\pi^2} 
\sum_{m=-\infty}^\infty e^{i m \left(\theta_A - \theta\right)}
\\\nonumber
& \times \,
\int_0^\infty d \nu   
\cos{\left(\nu \left(t_A-t\right)\right)} \, e^{\bar{\alpha}_s  \chi(m,\nu) y} \\\nonumber
& \times \, 
\frac{e^{-\frac{t+t_B}{2}} }{\pi^2} 
\sum_{n=-\infty}^\infty e^{i n \left(\theta - \theta_B\right)} 
\int_0^\infty d \mu   
\cos{\left(\mu \left(t-t_B\right)\right)} \, e^{\bar{\alpha}_s  \chi(n,\nu) (Y-y)}\;.
\end{align}
The integration over $\theta$ generates a $\delta_m^n$ contribution:
\begin{align}\label{eq:app-y-3}
\stepcounter{appcnt}
& \varphi \left(t_A,t_B,\theta_A,\theta_B,Y\right) = 
\frac{e^{-\frac{t_A+t_B}{2}} }{\pi^3}  \sum_{n=-\infty}^\infty 
e^{i n  \left(\theta_A - \theta_B\right)}
\int_0^\infty d \nu   
\, e^{\bar{\alpha}_s  \chi(n,\nu) y} \\\nonumber
& \times \,
\int_0^\infty d \mu   
\, e^{\bar{\alpha}_s  \chi(n,\nu) (Y-y)}
\int_{-\infty}^\infty dt  \cos{\left(\nu \left(t_A-t\right)\right)} \cos{\left(\mu \left(t-t_B\right)\right)}\;.
\end{align}

It can be shown that
\begin{align}\label{eq:app-y-4}
\stepcounter{appcnt}
&\int_{-\infty}^\infty dt  \cos{\left(\nu \left(t_A-t\right)\right)} \cos{\left(\mu \left(t-t_B\right)\right)}\\\nonumber 
& = \pi \left( \cos{ (\nu t_A - \mu t_B)}  \delta (\nu-\mu)
+ \cos{ ( \nu t_A+ \mu t_B)} \delta (\nu+\mu) \right)\;, 
\end{align}
which can be used to write Eq.~(\ref{eq:app-y-3}) as
\begin{align}\label{eq:app-y-5}
\stepcounter{appcnt}
& \varphi \left(t_A,t_B,\theta_A,\theta_B,Y\right)
\\\nonumber 
& = \frac{e^{-\frac{t_A+t_B}{2}} }{2 \pi^2}  \sum_{n=-\infty}^\infty 
e^{i n  \left(\theta_A - \theta_B\right)}
\int_{-\infty}^\infty d \nu   
\int_0^\infty d \mu   
\, e^{\bar{\alpha}_s  \chi(n,\nu) (Y-y)} e^{\bar{\alpha}_s  \chi(n,\nu) y} 
\\\nonumber
& \times \, \left( \cos{ (\nu t_A - \mu t_B)}  \delta (\nu-\mu)
+ \cos{ ( \nu t_A+ \mu t_B)} \delta (\nu+\mu)
 \right)\;,
\end{align}
and, finally,
\begin{align}\label{eq:app-y-6}
\stepcounter{appcnt}
& \varphi \left(t_A,t_B,\theta_A,\theta_B,Y\right)
\\\nonumber
& = \frac{e^{-\frac{t_A+t_B}{2}} }{\pi^2}  \sum_{n=-\infty}^\infty 
 e^{i n  \left(\theta_A - \theta_B\right)}   
\int_0^\infty d \mu   
 \, e^{\bar{\alpha}_s  \chi(n,\nu) Y}\cos{ (\mu (t_A -t_B))} \;,
\end{align}
which is the same as our initial representation for $\varphi$ in Eq.~(\ref{eq:app-y-1}).
The relation in Eq.~(\ref{eq:app-y-2}) is remarkable because it holds for 
any rapidity $y$.

\renewcommand{\theequation}
             {\arabic{chapter}.\arabic{equation}}
\chapter{Four-jet production}
\label{chap:4j}

In this Chapter we extend the discussion of Chapter~\ref{chap:3j} to the case of four jets, which represents our ultimate way to probe the BFKL dynamics through more exclusive processes. 
We have shown at the beginning of the last Chapter how our formalism can be extended to allow for the tagging of an extra central jet, by picking up its emission probability from the BFKL kernel. We got, as result (Eq.~(\ref{sigma_3j_start})), an expression for the three-jet cross section in the form of a double convolution of two Green's functions with three jet vertices, two of them describing the emission of the respective forward/backward jets (\emph{\`a la} Mueller--Navelet), while the other one being characteristic of the central-jet emission~\cite{Bartels:2006hg}.
This procedure can be generalised to the study of $n$-jet production processes, in which we cut $n-1$ times the original Green's function to permit the tagging of $n-2$ extra jets in the central rapidity regions of the detectors. This allows us to extend our discussion by investigating the four-jet production in MRK.

In our analysis we consider the emission of four jets in the final state: one in the forward direction with rapidity $Y_A$, one in the backward direction with rapidity $Y_B$ and both well-separated in rapidity from the each other, $Y = Y_A -Y_B$ large, along with two more jets tagged in more central regions of the detectors such that the relative rapidity separation between any two neighbouring jets is actually $Y/3$ respecting thus the MRK
ordering. 
We define and study new generalised azimuthal correlations, 
\begin{eqnarray}
{\cal R}^{M N L}_{P Q R} =\frac{ \langle \cos{(M \, \phi_1)} \cos{(N \, \phi_2)} \cos{(L \, \phi_3)} \rangle}{\langle \cos{(P \, \phi_1)} \cos{(Q \, \phi_2)} \cos{(R \, \phi_3)} \rangle} \; , 
\label{Rmnlpqr:intro}
\end{eqnarray}
where $\phi_1$, $\phi_2$ and $\phi_3$ are
the azimuthal-angle differences between neighbouring in rapidity jets. 
In this way we can investigate even more differential 
distributions in the transverse momenta, azimuthal angles and 
rapidities of the two central jets, for fixed values of the 
four momenta of the two forward (originally Mueller--Navelet) jets. The main observable 
${\cal R}^{M N L}_{P Q R}$ proposed at parton level is the extension 
of the three-jet one, defined in Eq.~(\ref{Rmnpq_intro}), using three cosines instead of two in numerator and denominator.

We make use of the collinear factorisation scheme
to produce the two most forward/backward jets and we convolute the partonic differential cross section, 
which is described by the BFKL dynamics, with collinear parton distribution functions.
As done in Section~\ref{chap:3j}, we include in our computation the forward jet vertex.
Three BFKL Green's functions link these two Mueller--Navelet jet-vertices with the 
more centrally produced jets.

This Chapter is organised as follows: 
In Section~\ref{sec:4j-theory} the main formul{\ae} are given; in Section~\ref{sec:4j-partonic} a first, parton level study is presented; in Section~\ref{sec:4j-hadronic} the first phenomenological analysis at LLA is shown, using realistic LHC kinematical cuts for the final-state phase space integration. The section Summary is given in~\ref{sec:4j-summary}.

The analysis given in this Chapter is based on 
the work done in Refs.~\cite{Caporale:2015int,Caporale:2016xku} and presented in 
Refs.~\cite{Celiberto:2016vhn,Caporale:2016pqe,Caporale:2016djm,Caporale:2016vxt}.

 \section{Theoretical framework} 
 \label{sec:4j-theory}
 
 In this Section the BFKL cross section 
 for the four-jet production process is presented, 
 the main focus lying on the definition
 of new, generalised and suitable BFKL observables.
 
 \begin{figure}[t]
 \centering
 \includegraphics[scale=0.5]{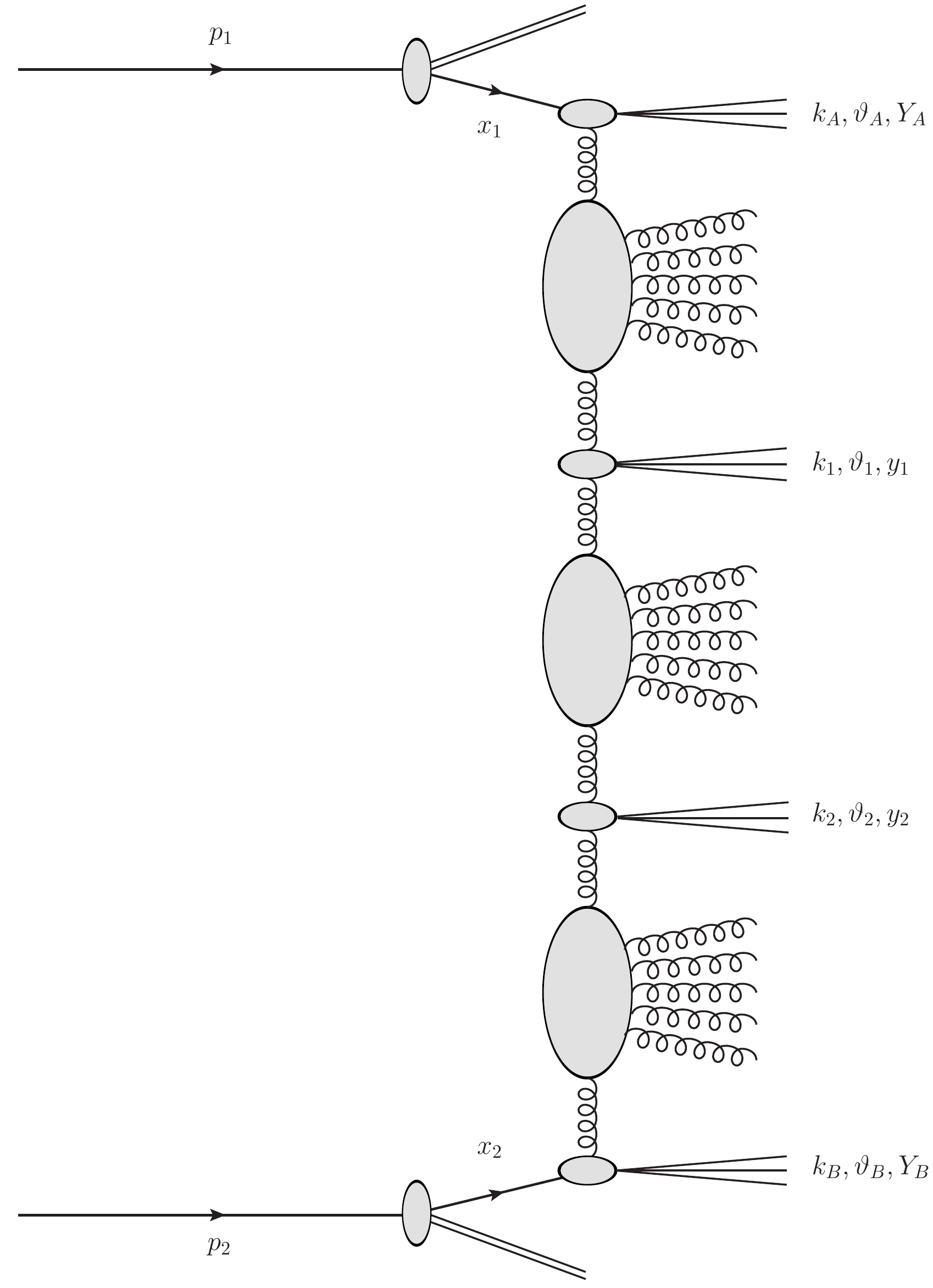}
 \caption[Inclusive four-jet production process 
          in multi-Regge kinematics]
 {Inclusive four-jet production process 
  in multi-Regge kinematics.}
 \label{fig:4j}
\end{figure}

  \subsection{The four-jet cross section}
  \label{sub:4j-cs}
  
\begin{figure}[t]
   \centering
   \includegraphics[scale=0.35]{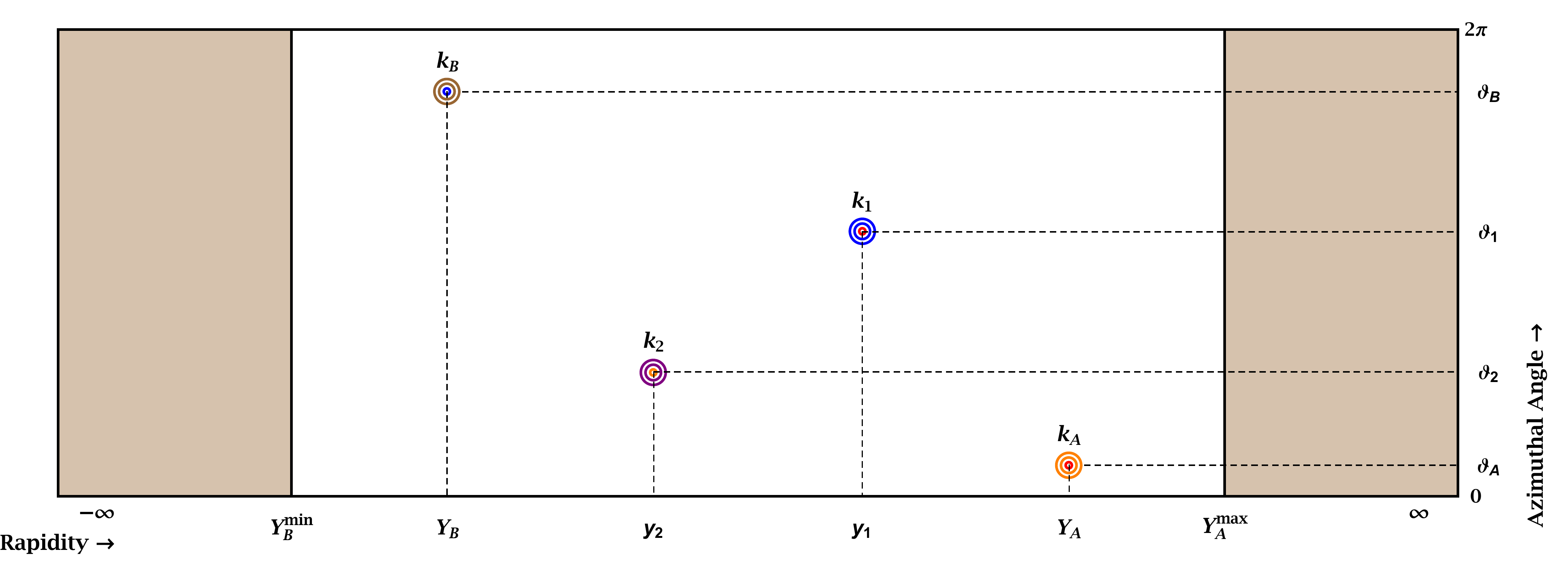}
   \caption[A primitive lego plot depicting a four-jet event]
    {A primitive lego plot 
     depicting a four-jet event. $k_A$ is a forward jet with 
     large positive
     rapidity $Y_A$ and azimuthal angle $\vartheta_A$, $k_B$ 
     is a backward jet with 
     large negative
     rapidity $Y_B$ and azimuthal angle $\vartheta_B$ and
     $k_1$ and $k_2$ are two jets with azimuthal angles $\vartheta_1$ 
     and $\vartheta_2$ respectively
     and rapidities $y_1$ and $y_2$
     such that $Y_A - y_1 \sim y_1- y_2 \sim y_2 - Y_B$. 
    }
   \label{fig:lego4j}
  \end{figure}

  The process under exam (see Figs.~\ref{fig:4j} and~\ref{fig:lego4j}) 
  is the production of two forward/backward jets, 
  both characterised by high transverse momenta $\vec{k}_{A,B}$ 
  and well separated in rapidity, together with two more jets
  produced in the central rapidity region 
  and with possible associated mini-jet production: 
  \begin{eqnarray}
  \label{process-4j}
  {\rm p}(p_1)   \, + \, {\rm p} (p_2)  \, \to \,  
  {\rm j}_A(k_A) \, + \, {\rm j}_1(k_1) \, + \, 
  {\rm j}_2(k_2) \, + \, {\rm j}_B(k_B) \, + \, 
  {\rm minijets} \; ,
  \end{eqnarray}
  where ${\rm j}_A$ is the forward jet 
  with transverse momentum $\vec{k}_A$
  and rapidity $Y_A$, ${\rm j}_B$ 
  is the backward jet with transverse momentum $\vec{k}_B$
  and rapidity $Y_B$, and with ${\rm j}_{1,2}$   
  being the two central jets with transverse momenta $\vec{k}_{1,2}$ and rapidities $y_{1,2}$, such that
  $Y_A > y_1 > y_2 > Y_B$ 
  according to the ordering characteristic of MRK. 
  
  The cross section for the inclusive 
  four-jet production process~(\ref{process-4j}) 
  reads in collinear factorisation
  \begin{equation}
  \label{dsigma_pdf_convolution-4j}
   \frac{d\sigma^{4-{\rm jet}}}
        {dk_A \, dY_A \, d\vartheta_A \, 
         dk_B \, dY_B \, d\vartheta_B \, 
         dk_1 \, dy_1 d\vartheta_1  \, 
         dk_2 \, dy_2 d\vartheta_2}
  \end{equation}
  \[ = 
   \sum_{r,s=q,{\bar q},g}
   \int_0^1 dx_1 \int_0^1 dx_2
   \ f_{r}\left(x_1,\mu_F\right)
   \ f_{s}\left(x_2,\mu_F\right) \;
   d\sfjp\left(\hat{s},\mu_F\right) \; ,
  \]
  where $r,s$ characterise the partons 
  (gluon $g$; quarks $q = u, d, s, c, b$;
  antiquarks $\bar q = \bar u, \bar d, \bar s, \bar c, \bar b$),
  $f_{r,s}\left(x, \mu_F \right)$ 
  are the parton distribution functions of the protons; 
  $x_{1,2}$ represent the longitudinal fractions 
  of the partons involved 
  in the hard subprocess; 
  $d\sfjp\left(\hat{s}, \mu_F \right)$ 
  is the partonic cross section for the production of jets and
  $x_1x_2s \equiv \hat{s}$ 
  is the partonic squared center-of-mass energy 
  (see Fig.~\ref{fig:4j}). The cross section 
  for the partonic hard subprocess $d\sfjp$ 
  can be presented as (from here we start to use the notation 
  $k_{A,B,1,2} \equiv |\vec{k}_{A,B,1,2}|$, which holds in the following):
  \begin{align}\label{d6sigma}
   & 
   \frac{d\sfjp}
        {d^2\vec{k_1} dy_1 d^2\vec{k_2} dy_2} =
   \frac{\asb^2}{\pi^2 k_1^2 k_2^2}
   \int d^2\vec{p_A} \int d^2\vec{p_B} 
   \int d^2\vec{p_1} \int d^2\vec{p_2}
   \\ & \nonumber 
   \times \, 
   \delta^{(2)}\left(\vec{p_A}+\vec{k_1}-\vec{p_1}\right)
   \delta^{(2)}\left(\vec{p_B}-\vec{k_2}-\vec{p_2}\right)
   \\ & \nonumber 
   \times \, 
   \varphi\left(\vec{k_A},\vec{p_A},Y_A-y_1\right)
   \varphi\left(\vec{p_1},\vec{p_2},y_1-y_2\right)
   \varphi\left(\vec{p_B},\vec{k_B},y_2-Y_B\right) \; ,
  \end{align}
  where $\bar \alpha_s = N_c/\pi \, \alpha_s$, with $N_c$ the number of colours.
  $\varphi$ are the BFKL Green's functions suitably redefined 
  as explained in Section~\ref{sec:3j-theory}.
  Integrating one of the two two-dimensional
  delta functions in Eq.~(\ref{d6sigma}), we have  
  \begin{align}\label{d6sigma_nodelta}
   &
   \frac{d\sfjp}
        {d^2\vec{k_1} dy_1 d^2\vec{k_2} dy_2}    
   =
   \frac{\asb^2}{\pi^2 k_1^2 k_2^2}
   \int d^2\vec{p_A} \int d^2\vec{p_B}
   \varphi\left(\vec{k_A},\vec{p_A},Y_A-y_1\right)
   \\ & \nonumber \times \, 
   \varphi\left(\vec{p_A}+\vec{k_1},\vec{p_B}-\vec{k_2},y_1-y_2\right)
   \varphi\left(\vec{p_B},\vec{k_B},y_2-Y_B\right) \; .
  \end{align}
  We can expand the Green's function (whose expression is taken at LLA) in Fourier components 
  on the respective azimuthal angles and write
  \begin{align}\label{d6sigma_omega_mnl}
   &
   \frac{d\sfjp}
        {d^2\vec{k_1} dy_1 d^2\vec{k_2} dy_2}
   = 
   \frac{\asb^2}{\pi^2 k_1^2 k_2^2}
   \sum_{m,n,l=-\infty}^{+\infty} e^{i(m\vartheta_A-l\vartheta_B)}         
   \\ & \nonumber \times \, 
   \Omega_{m,n,l}\left(\vec{k_A},\vec{k_B},Y_A,Y_B,\vec{k_1},\vec{k_2},
                       \vartheta_1,\vartheta_2,y_1,y_2\right) \; ,
  \end{align}
  where
  \begin{align}\label{omega_mnl_1}
   &
   \Omega_{m,n,l}\left(\vec{k_A},\vec{k_B},Y_A,Y_B,\vec{k_1},\vec{k_2},
                       \vartheta_1,\vartheta_2,y_1,y_2\right)
   \\ & \nonumber =  
   \int_0^{+\infty} dp_A \, p_A \int_0^{+\infty} dp_B \, p_B
   \int_0^{2\pi} d\phi_A \int_0^{2\pi} d\phi_B
   \: e^{in(\phi_1-\phi_2)} \, e^{i(l\phi_B-m\phi_A)}
   \\ & \nonumber \times \, 
   \varphi^{\rm (LLA)}_{m}\left(|\vec{k_A}|,|\vec{p_A}|,Y_A-y_1\right)
   \varphi^{\rm (LLA)}_{l}\left(|\vec{p_B}|,|\vec{k_B}|,y_2-Y_B\right)
   \\ & \nonumber \times \, 
   \varphi^{\rm (LLA)}_{n}\left(|\vec{p_A}+\vec{k_1}|,
                                |\vec{p_B}-\vec{k_2}|,
                                y_1-y_2\right)
  \end{align}
  and 
  \begin{align}\label{args}
   &
   \phi_1 = \arctan \left(
   \frac{p_A \sin\phi_A + k_1 \sin\vartheta_1}
        {p_A \cos\phi_A + k_1 \cos\vartheta_1} \right)
   \\ & \nonumber
   \phi_2 = \arctan \left(
   \frac{p_B \sin\phi_B - k_2 \sin\vartheta_2}
        {p_B \cos\phi_A - k_2 \cos\vartheta_2} \right) \; ;  
  \end{align} 
  Here $\varphi^{\rm (LLA)}_{i}$ is the $i$-th azimuthal component 
  of the LLA Green's function given in Eq.~(\ref{phinLO}).
  It is possible to use the relation 
  $\arctan \alpha = 
  \frac{i}{2} \ \ln\left(\frac{1-i\alpha}{1+i\alpha}\right)$,  
  with $\alpha$ being a real number, to write
  \begin{align}\label{omega_mnl_2}
   &
   \Omega_{m,n,l}\left(\vec{k_A},\vec{k_B},Y_A,Y_B,\vec{k_1},\vec{k_2},
                       \vartheta_1,\vartheta_2,y_1,y_2\right)
   \\ & \nonumber = 
   \int_0^{+\infty} dp_A \, p_A \int_0^{+\infty} dp_B \, p_B
   \int_0^{2\pi} d\phi_A \int_0^{2\pi} d\phi_B
   \: e^{-im\phi_A} \, e^{il\phi_B}
   \\ & \nonumber
   \left(\frac{p_A e^{i\phi_A}+k_1 e^{i\vartheta_1}}
                 {p_A e^{-i\phi_A}+k_1 
                  e^{-i\vartheta_1}}\right)^\frac{n}{2}
   \left(\frac{p_B e^{-i\phi_B}-k_2 e^{-i\vartheta_2}}
                 {p_A e^{i\phi_B}-k_2 
                 e^{i\vartheta_2}}\right)^\frac{n}{2}
   \\ & \nonumber \times \, 
   \varphi^{\rm (LLA)}_{m}
     \left(|\vec{k_A}|,|\vec{p_A}|,Y_A-y_1\right)
   \varphi^{\rm (LLA)}_{l}
     \left(|\vec{p_B}|,|\vec{k_B}|,y_2-Y_B\right)
   \\ & \nonumber \times \, 
   \varphi^{\rm (LLA)}_{n}\left
     (\tilde{p}_{\widehat{A1}}
     ,\tilde{p}_{\widehat{2B}}
     ,y_1-y_2\right) \; ,
  \end{align}
  where
  \begin{align}
   & 
   \tilde{p}_{\widehat{A1}} = \sqrt{p_A^2+k_1^2+2 |\vec{p}_A| |\vec{k}_1| \cos\left(\phi_A-\vartheta_1\right)} \; ,
   \\ \nonumber &
   \tilde{p}_{\widehat{2B}} = \sqrt{p_B^2+k_2^2-2 |\vec{p}_B| |\vec{k}_2| \cos\left(\phi_B-\vartheta_2\right)} \; .
  \end{align}
  Making the double change of variables 
  $\phi_A - \vartheta_1 \to \phi_A$ and 
  $\phi_B - \vartheta_2 \to \phi_B$, 
  we obtain the final expression for $\Omega_{m,n,l}$:
  \begin{align}\label{omega_mnl_final}
   &
   \Omega_{m,n,l}\left(\vec{k_A},\vec{k_B},Y_A,Y_B,\vec{k_1},\vec{k_2},
                       \vartheta_1,\vartheta_2,y_1,y_2\right)
   \\ & \nonumber = 
   e^{i(n-m)\vartheta_1} \ e^{-i(n-l)\vartheta_2} 
   \int_0^{+\infty} dp_A \, p_A \int_0^{+\infty} dp_B \, p_B
   \int_0^{2\pi} d\phi_A \int_0^{2\pi} d\phi_B
   \\ & \nonumber \times \, 
   \frac{e^{-im\phi_A} \, e^{il\phi_B} \,
         \left(p_A e^{i\phi_A}+k_1\right)^n \,
         \left(p_B e^{-i\phi_B}-k_2\right)^n}
        {
         \sqrt{\left(p_A^2+k_1^2+2 |\vec{p}_A| |\vec{k}_1| 
         \cos\left(\phi_A\right)\right)^n}
         \
         \sqrt{\left(p_B^2+k_2^2-2 |\vec{p}_B| |\vec{k}_2| 
         \cos\left(\phi_B\right)\right)^n}
        }
   \\ & \nonumber \times \, 
   \varphi^{\rm (LLA)}_{m}
     \left(|\vec{k_A}|,|\vec{p_A}|,Y_A-y_1\right)
   \varphi^{\rm (LLA)}_{l}
     \left(|\vec{p_B}|,|\vec{k_B}|,y_2-Y_B\right)
   \\ & \nonumber \times \, 
   \varphi^{\rm (LLA)}_{n}\left
     (\tilde{p}_{1}
     ,\tilde{p}_{2}
     ,y_1-y_2\right) \,
  \end{align}
  with
  \begin{align}
   & 
   \tilde{p}_{1} = \sqrt{p_A^2+k_1^2+2 |\vec{p}_A| |\vec{k}_1| \cos\left(\phi_A\right)} \; ,
   \\ \nonumber &
   \tilde{p}_{2} = \sqrt{p_B^2+k_2^2-2 |\vec{p}_B| |\vec{k}_2| \cos\left(\phi_B\right)} \; ,
  \end{align}
  in which the dependence on 
  $\vartheta_1$ and $\vartheta_2$ 
  has been successfully factorised out.

  \subsection{The four-jet azimuthal correlations: partonic level}
  \label{sub:4j-2cos-partonic}
  
  Following the course taken for the inclusive three-jet production (see Sections~\ref{sub:3j-1cos} and \ref{sub:3j-2cos}), 
  our goal is to define and study 
  the behaviour of observables for which the BFKL approach 
  will show distinct features with respect to other formalisms and, 
  if possible, are also quite insensitive to higher-order corrections. 
  We start with the study of a quantity similar 
  to the usual Mueller--Navelet case such that 
  we integrate over the azimuthal angles 
  of the two central jets and over the difference in azimuthal angle 
  between the two forward jets, 
  $\Delta\phi = \vartheta_A - \vartheta_B - \pi$, 
  to define
  \begin{align} \label{Dth_th1th2_d6sigma}
   &  
   \int_0^{2\pi} d\Delta\phi \cos\left(M\Delta\phi\right)
   \int_0^{2\pi} d\vartheta_1 \int_0^{2\pi} d\vartheta_2
   \frac{d\sigma^{\rm 4-jet}\left(\vec{k_A},\vec{k_B},Y_A-Y_B\right)}
        {dk_1 dy_1 d\vartheta_1 dk_2 d\vartheta_2 dy_2}     
   \\ & \nonumber 
   =
   \frac{4 \asb^2}{k_1 k_2}
    \left(e^{iM\pi} \, 
     \tilde{\Omega}_{M}(\vec{k_A},\vec{k_B},Y_A,Y_B,
     \vec{k_1},\vec{k_2},
                             y_1,y_2)
     +    c.c.
    \right)
  \end{align}
  where 
  \begin{align}\label{omega_n_tilde}
   &    
   \tilde{\Omega}_{n}(\vec{k_A},\vec{k_B},Y_A,Y_B,\vec{k_1},\vec{k_2},
                      y_1,y_2)
   \\ & \nonumber
   =
   \int_0^{+\infty} dp_A \, p_A \int_0^{+\infty} dp_B \, p_B
   \int_0^{2\pi} d\phi_A \int_0^{2\pi} d\phi_B
   \\ & \nonumber \times \, 
   \frac{\left(p_A+k_1 e^{-i\phi_A}\right)^n \,
         \left(p_B-k_2 e^{i\phi_B}\right)^n}
        {
         \sqrt{\left(p_A^2+k_1^2+2 |\vec{p}_A| |\vec{k}_1| \cos\phi_A\right)^n}
         \
         \sqrt{\left(p_B^2+k_2^2-2 |\vec{p}_B| |\vec{k}_2| \cos\phi_B\right)^n}
        }
   \\ & \nonumber \times \, 
   \varphi^{\rm (LLA)}_{n}\left(|\vec{k_A}|,|\vec{p_A}|,Y_A-y_1\right)
   \varphi^{\rm (LLA)}_{n}\left(|\vec{p_B}|,|\vec{k_B}|,y_2-Y_B\right)
   \\ & \nonumber \times \, 
   \varphi^{\rm (LLA)}_{n}\left
     (\sqrt{p_A^2+k_1^2+2 |\vec{p}_A| |\vec{k}_1| \cos\phi_A}
     ,\sqrt{p_B^2+k_2^2-2 |\vec{p}_B| |\vec{k}_2| \cos\phi_B}
     ,y_1-y_2\right) \; .
  \end{align}

  The associated experimental observable corresponds to the mean value
  of the cosine of $\Delta\phi$ in the recorded events:
  \begin{align}\label{C0}
   & 
   \left\langle
   \cos(M\Delta\phi)
   \right\rangle_{d\sfjp}
   \\ & 
   \nonumber 
   =
   \frac{\int_0^{2\pi} d\Delta\phi \cos(M\Delta\phi)
         \int_0^{2\pi} d\vartheta_1 \int_0^{2\pi} d\vartheta_2
         \frac{d\sigma^{\rm 4-jet}}
              {dk_1 dy_1 d\vartheta_1 dk_2 d\vartheta_2 dy_2}}
        {\int_0^{2\pi} d\Delta\phi 
         \int_0^{2\pi} d\vartheta_1 \int_0^{2\pi} d\vartheta_2
         \frac{d\sigma^{\rm 4-jet}}
              {dk_1 dy_1 d\vartheta_1 dk_2 d\vartheta_2 dy_2}}.
   \end{align}
  In order to improve the perturbative stability of our predictions  
  (see Ref.~\cite{Caporale:2013uva} for a related discussion) 
  it is convenient to remove 
  the contribution from the zero conformal spin 
  by defining the ratios
  \begin{equation}\label{Rmn}
   \mathcal{R}^M_N =
   \frac{\left\langle\cos(M\Delta\phi)
   \right\rangle_{d\sfjp}}
        {\left\langle\cos(N\Delta\phi)
        \right\rangle_{d\sfjp}}
  \end{equation}
  where $M,N$ are positive integers. 
  
  The next step now is to propose new observables, 
  different from those characteristic 
  of the Mueller--Navelet case though still related 
  to azimuthal-angle projections. 
  Let us first define  the
  following azimuthal-angle differences:
  \begin{align}\label{phi123}
   & \phi_1 = \vartheta_A-\vartheta_1-\pi \; , \\ \nonumber
   & \phi_2 = \vartheta_1-\vartheta_2-\pi \; , \\ \nonumber
   & \phi_3 = \vartheta_2-\vartheta_B-\pi \; .
  \end{align}
  Then we define
  \begin{align}\label{projections_1} 
   & \mathcal{C}_{MNL} =  
   \int_0^{2\pi} d\vartheta_A \int_0^{2\pi} d\vartheta_B
   \int_0^{2\pi} d\vartheta_1 \int_0^{2\pi} d\vartheta_2
   \cos\left(M\left(\phi_1\right)\right)
   \cos\left(N\left(\phi_2\right)\right)
   \\ & \nonumber \times \, 
   \cos\left(L\left(\phi_3\right)\right)
   \frac{d\sigma^{\rm 4-jet}\left(\vec{k_A},\vec{k_B},Y_A-Y_B\right)}
   {dk_1 dy_1 d\vartheta_1 dk_2 d\vartheta_2 dy_2} \, ,
  \end{align}
  where we consider $M$, $N$, $L > 0$ and integer.
  After a bit of algebra we have
  \begin{align}\label{projections_3}
   & \mathcal{C}_{MNL} = 
   \frac{2\pi^2 \asb^2}{k_1 k_2} \, (-1)^{M + N + L} \:
        (  \tilde{\Omega}_{M,N,L} + \tilde{\Omega}_{M,N,-L} +
            \tilde{\Omega}_{M,-N,L}  
   \\     & \nonumber + \tilde{\Omega}_{M,-N,-L} +
            \tilde{\Omega}_{-M,N,L} + \tilde{\Omega}_{-M,N,-L} +
            \tilde{\Omega}_{-M,-N,L} + \tilde{\Omega}_{-M,-N,-L})
  \end{align}
  with
  \begin{align}\label{omega_mnl_tilde}
   &
   \tilde{\Omega}_{m,n,l}
   = 
   \int_0^{+\infty} dp_A \, p_A \int_0^{+\infty} dp_B \, p_B
   \int_0^{2\pi} d\phi_A \int_0^{2\pi} d\phi_B
   \\ & \nonumber \times \, 
   \frac{e^{-im\phi_A} \, e^{il\phi_B} \,
         \left(p_A e^{i\phi_A}+k_1\right)^n \,
         \left(p_B e^{-i\phi_B}-k_2\right)^n}
        {
         \sqrt{\left(p_A^2+k_1^2+2 |\vec{p}_A| |\vec{k}_1| \cos\phi_A\right)^n}
         \
         \sqrt{\left(p_B^2+k_2^2-2 |\vec{p}_B| |\vec{k}_2| \cos\phi_B\right)^n}
        }
   \\ & \nonumber \times \, 
   \varphi^{\rm (LLA)}_{m}\left(|\vec{k_A}|,|\vec{p_A}|,Y_A-y_1\right)
   \varphi^{\rm (LLA)}_{l}\left(|\vec{p_B}|,|\vec{k_B}|,y_2-Y_B\right) 
   \\ & \nonumber \times \, 
   \varphi^{\rm (LLA)}_{n}\left
     (\sqrt{p_A^2+k_1^2+2 |\vec{p}_A| |\vec{k}_1| \cos\phi_A}
     ,\sqrt{p_B^2+k_2^2-2 |\vec{p}_B| |\vec{k}_2| \cos\phi_B}
     ,y_1-y_2\right).
   \end{align}
  In order to drastically reduce the dependence 
  on collinear configurations we can remove 
  the zero conformal spin contribution by defining 
  the following ratios:
  \begin{align}\label{Rmnlpqr}
   &
   \mathcal{R}^{MNL}_{PQR}
   =
   \frac{\left\langle\cos(M(\phi_1))
                     \cos(N(\phi_2))
                     \cos(L(\phi_3))
         \right\rangle_{d\sfjp}}
        {\left\langle\cos(P(\phi_1))
                     \cos(Q(\phi_2))
                     \cos(R(\phi_3))
         \right\rangle_{d\sfjp}}
  \end{align}
  with integer $M,N,L,P,Q,R > 0$.

 \section{Partonic level analysis} 
 \label{sec:4j-partonic}
 
\begin{figure}[H]
  \centering
     \includegraphics[scale=0.55]{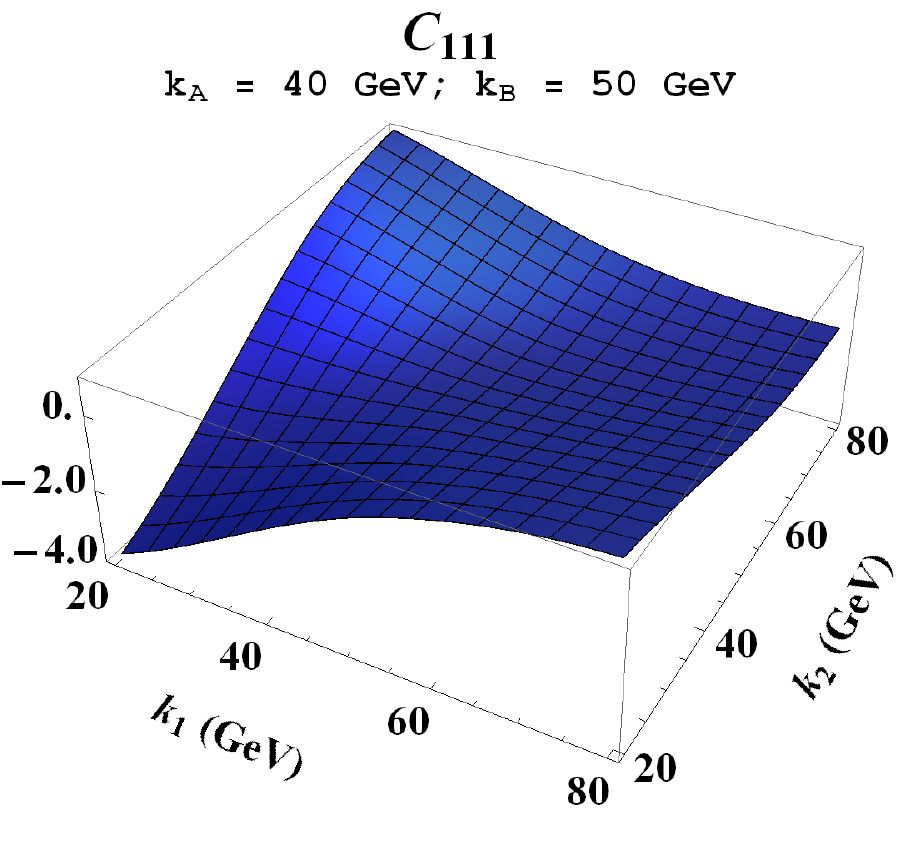}
     \includegraphics[scale=0.55]{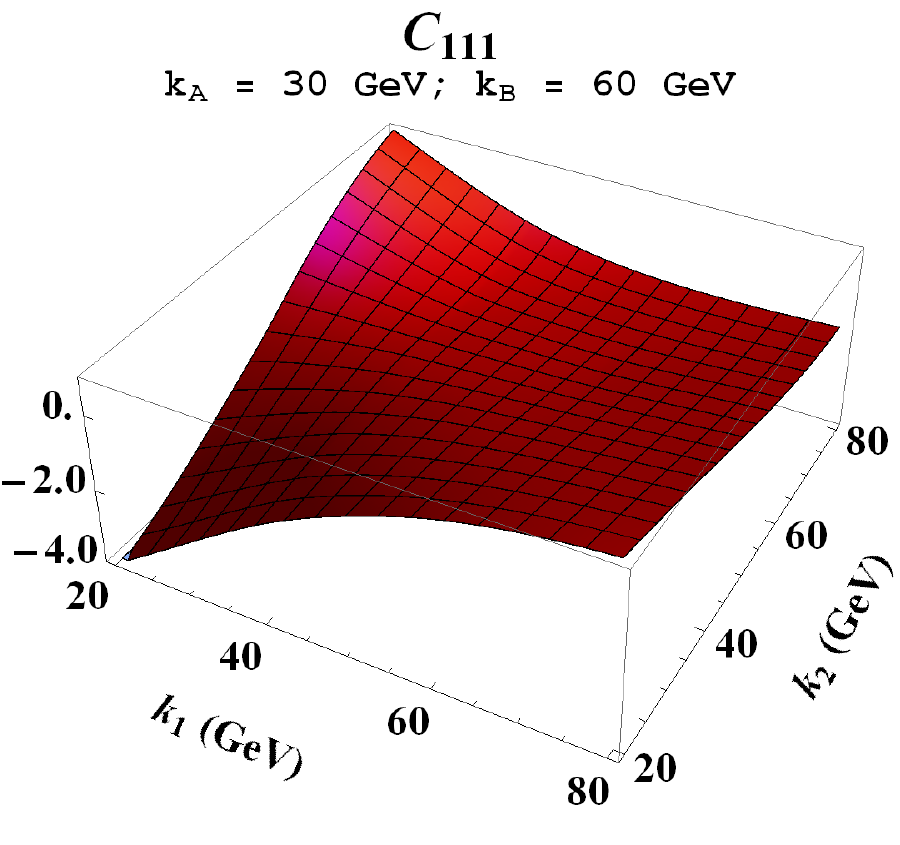}
  
     \includegraphics[scale=0.55]{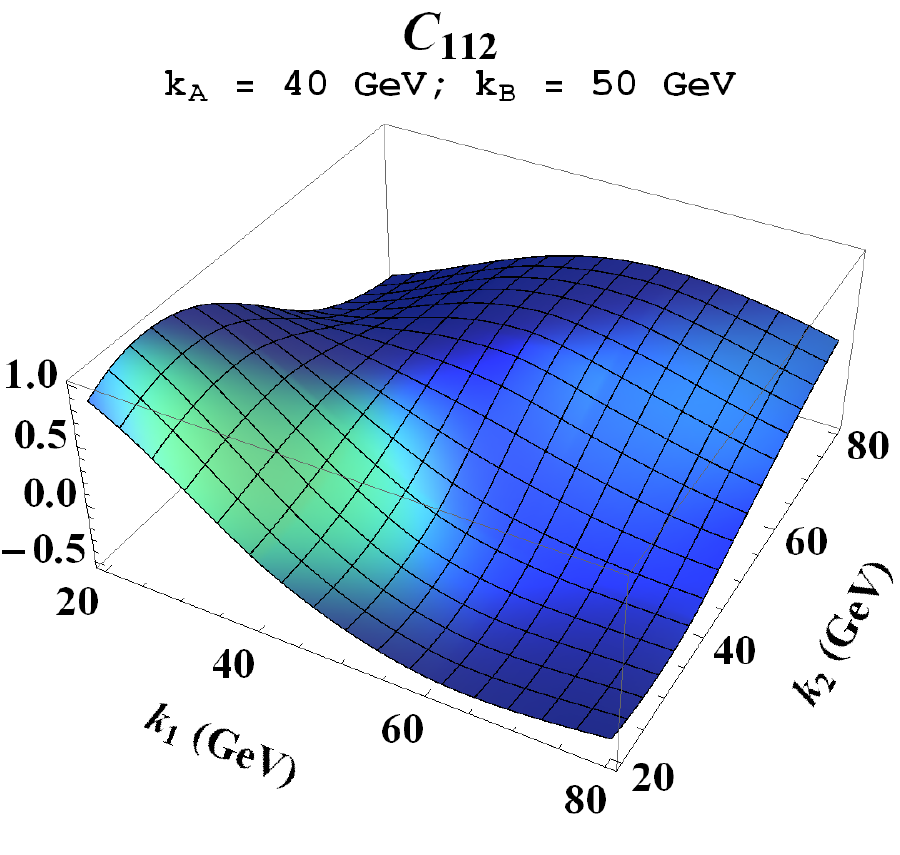}
     \includegraphics[scale=0.55]{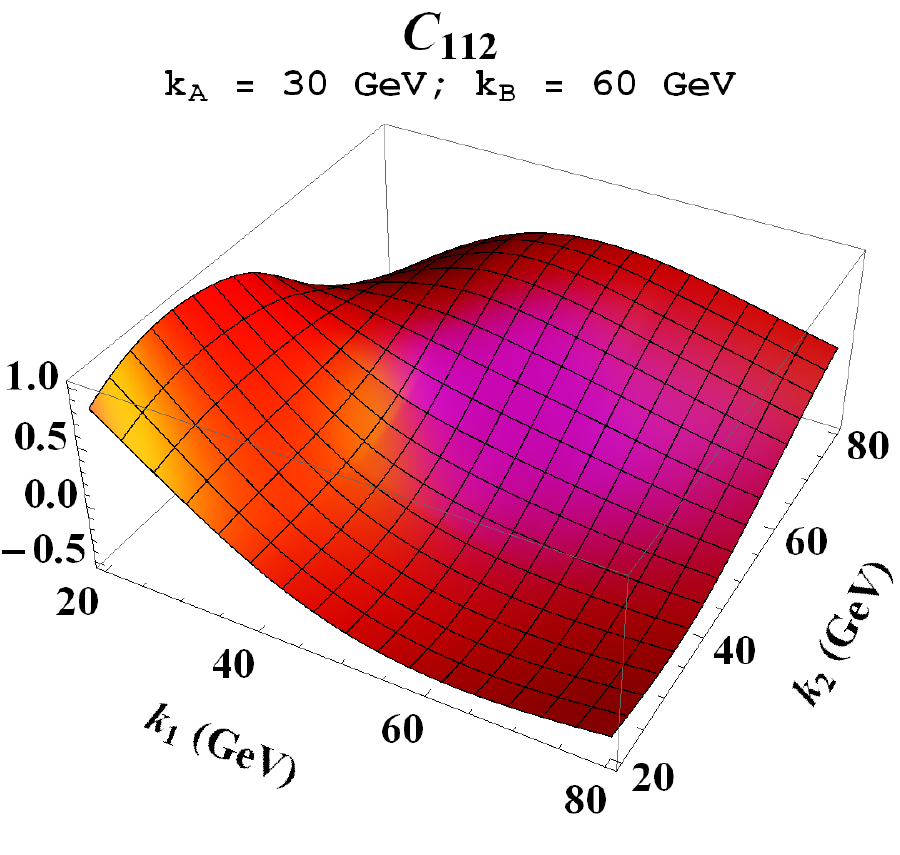}
     
     \includegraphics[scale=0.55]{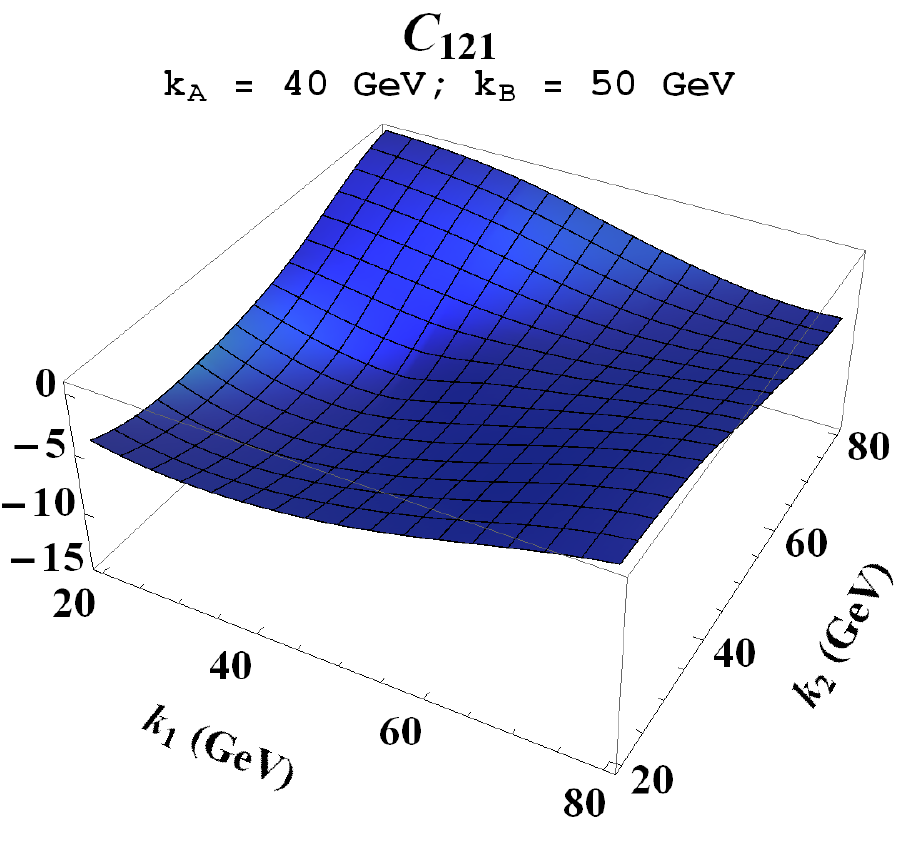}
     \includegraphics[scale=0.55]{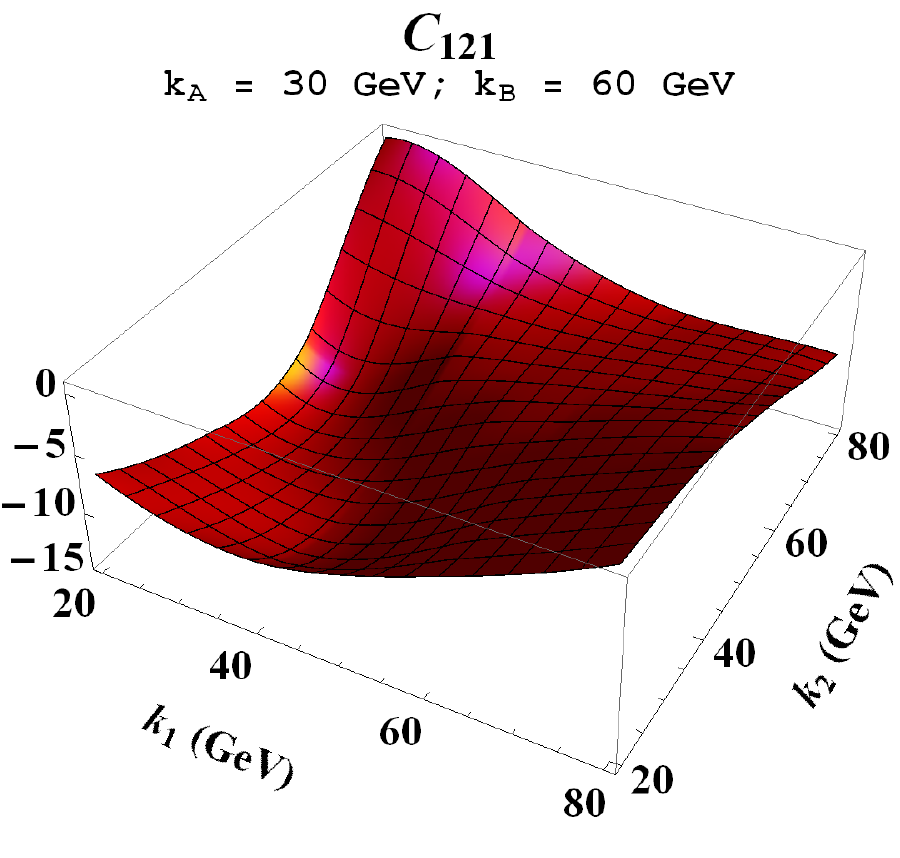}
     
     \includegraphics[scale=0.60]{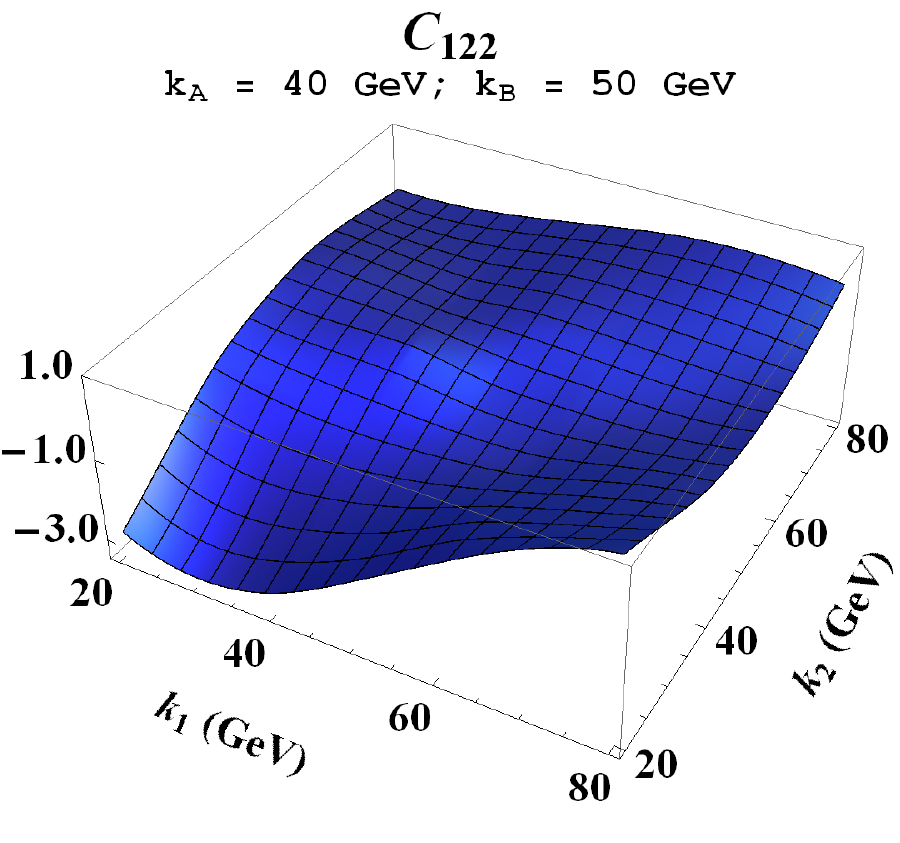}
     \includegraphics[scale=0.60]{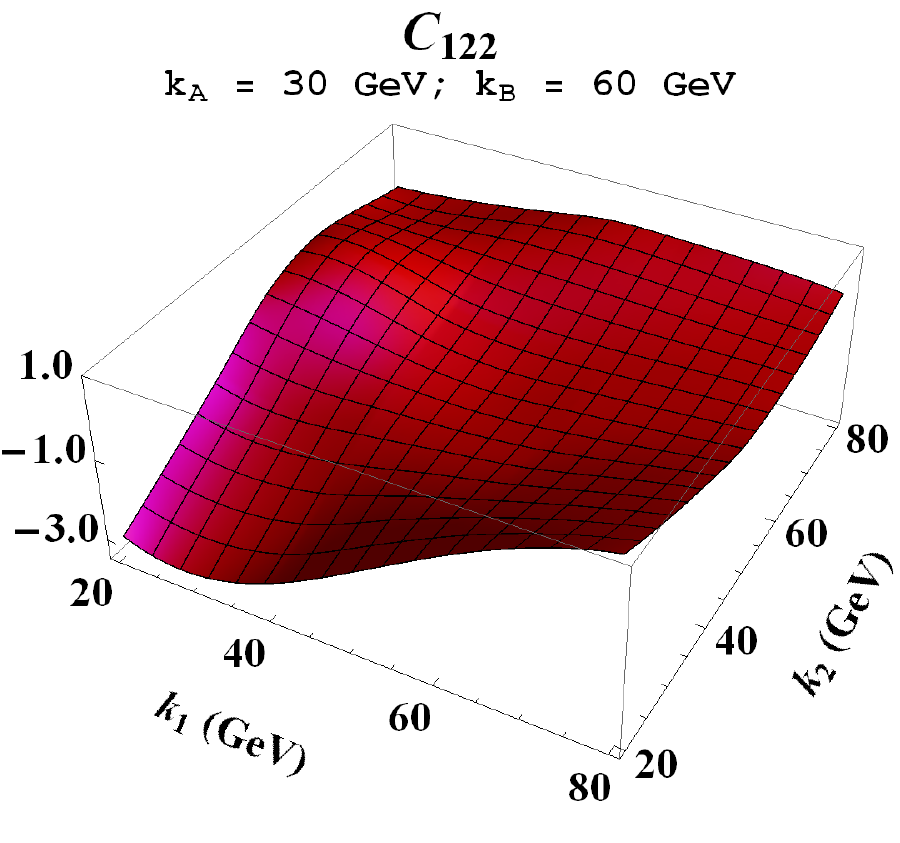}
  
  \caption[Partonic four-jets: BFKL azimuthal coefficients (1)]
  {$k_{1,2}$-dependence of the normalised 
  ${\cal C}_{111}$, $C_{112}$, ${\cal C}_{121}$
  and ${\cal C}_{122}$ for the two selected cases of forward jet 
  transverse momenta $k_A$ and $k_B$.}
  \label{C1nl}
 \end{figure}

\begin{figure}[H]
  \centering
     \includegraphics[scale=0.55]{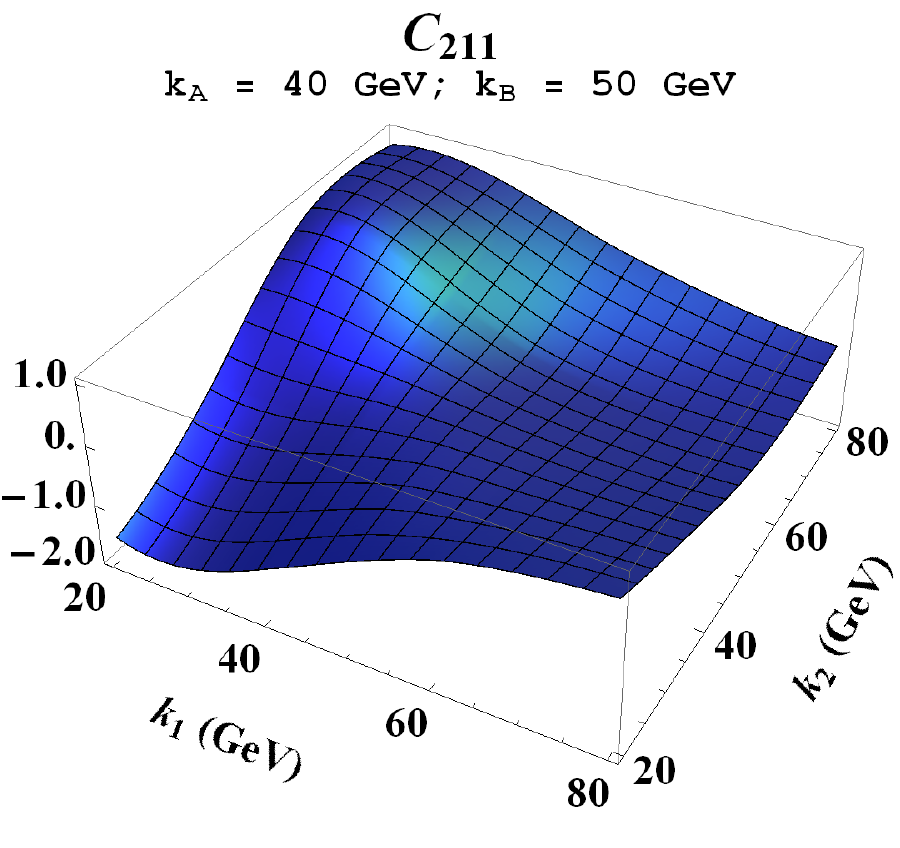}
     \includegraphics[scale=0.55]{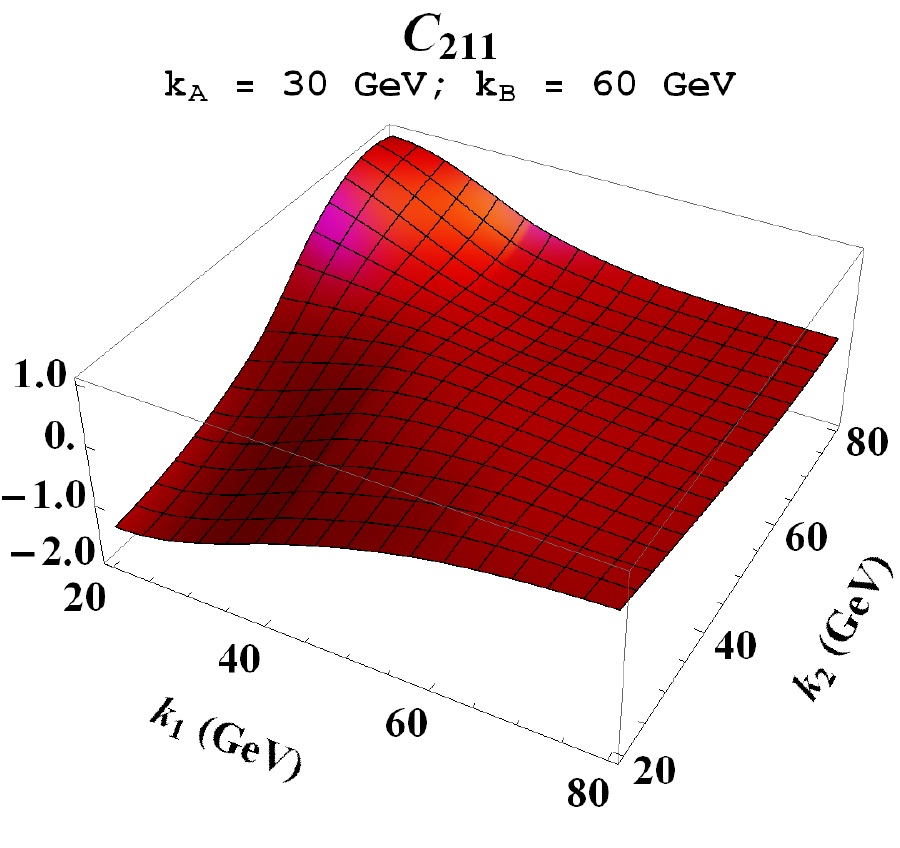}
  
     \includegraphics[scale=0.55]{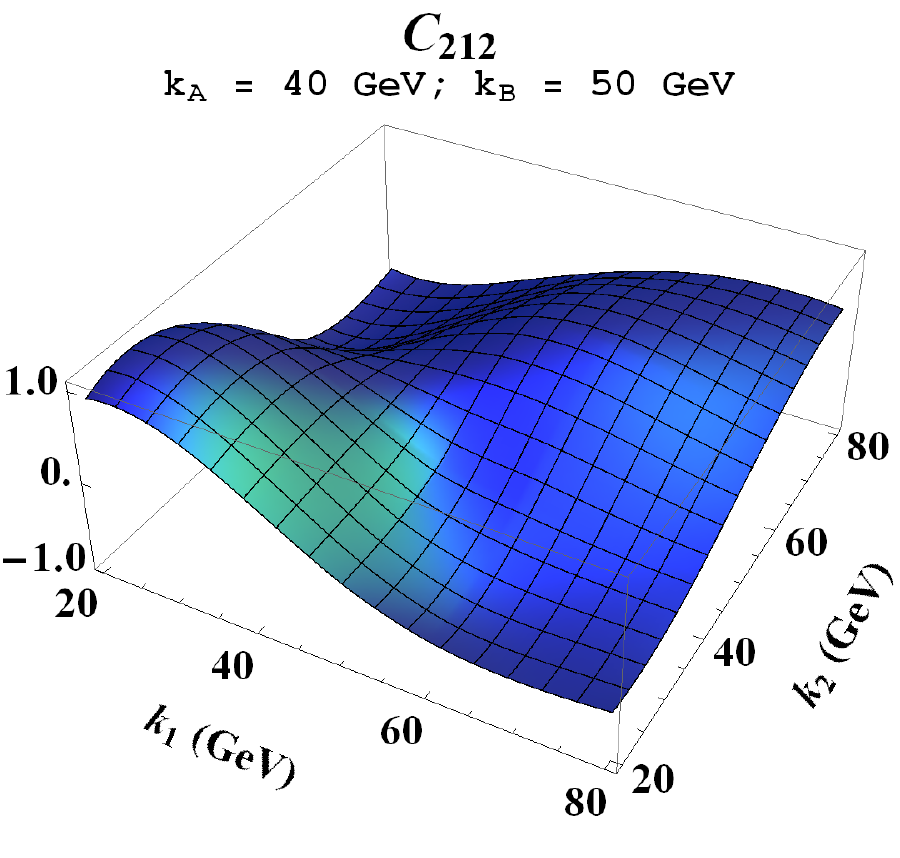}
     \includegraphics[scale=0.55]{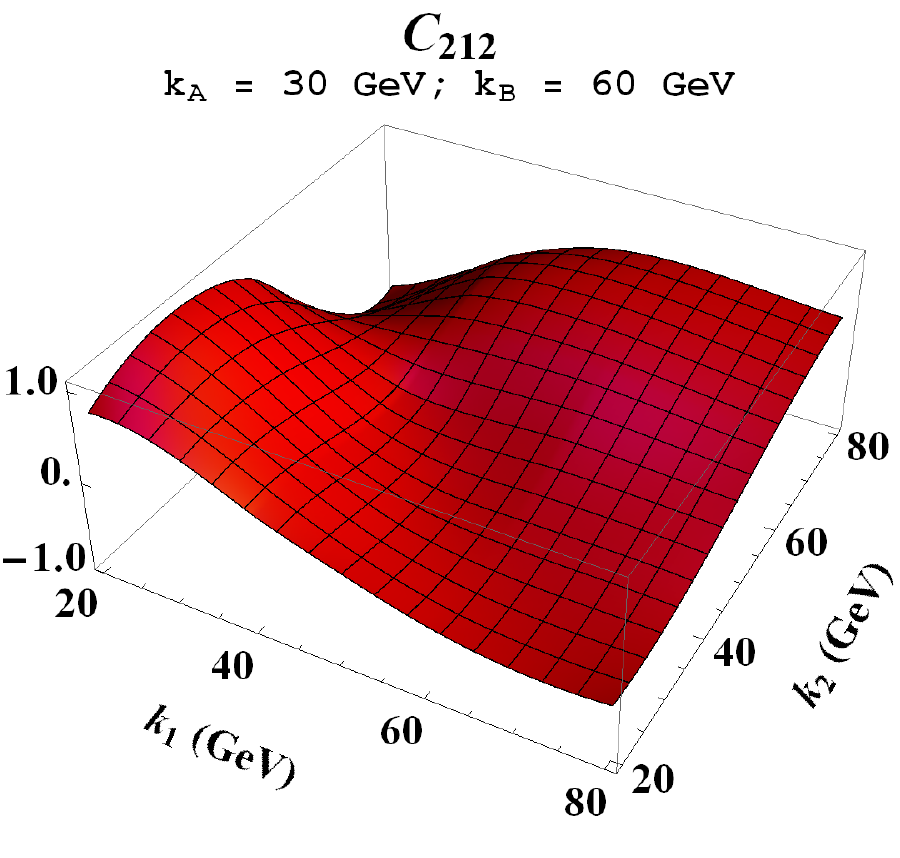}
     
     \includegraphics[scale=0.55]{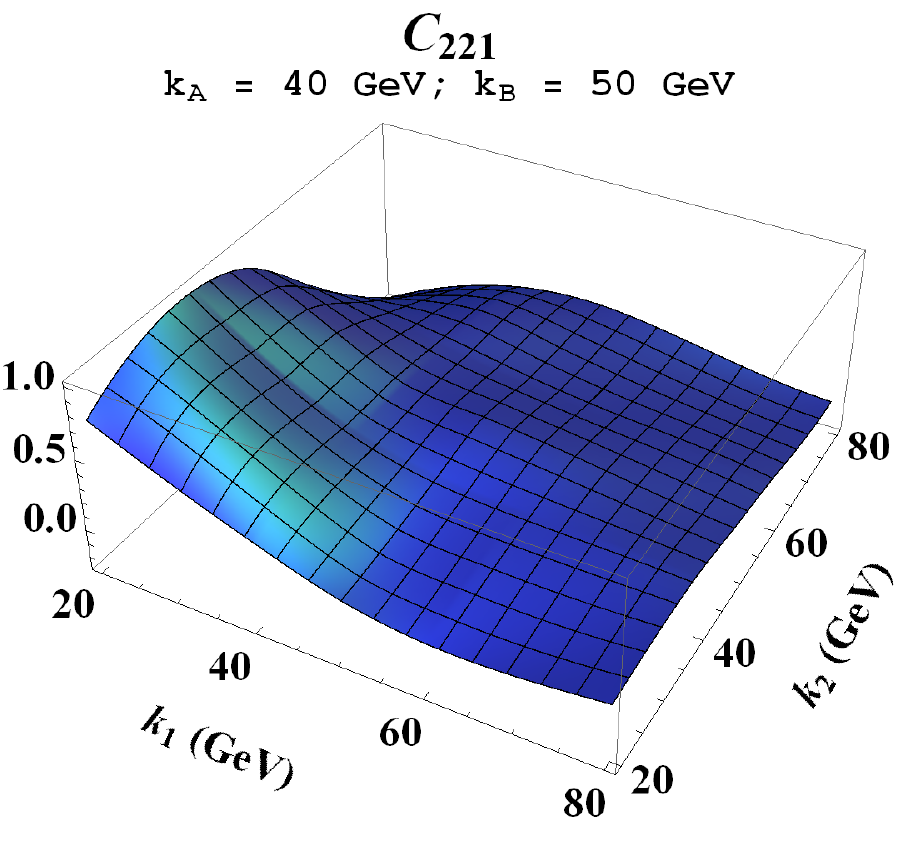}
     \includegraphics[scale=0.55]{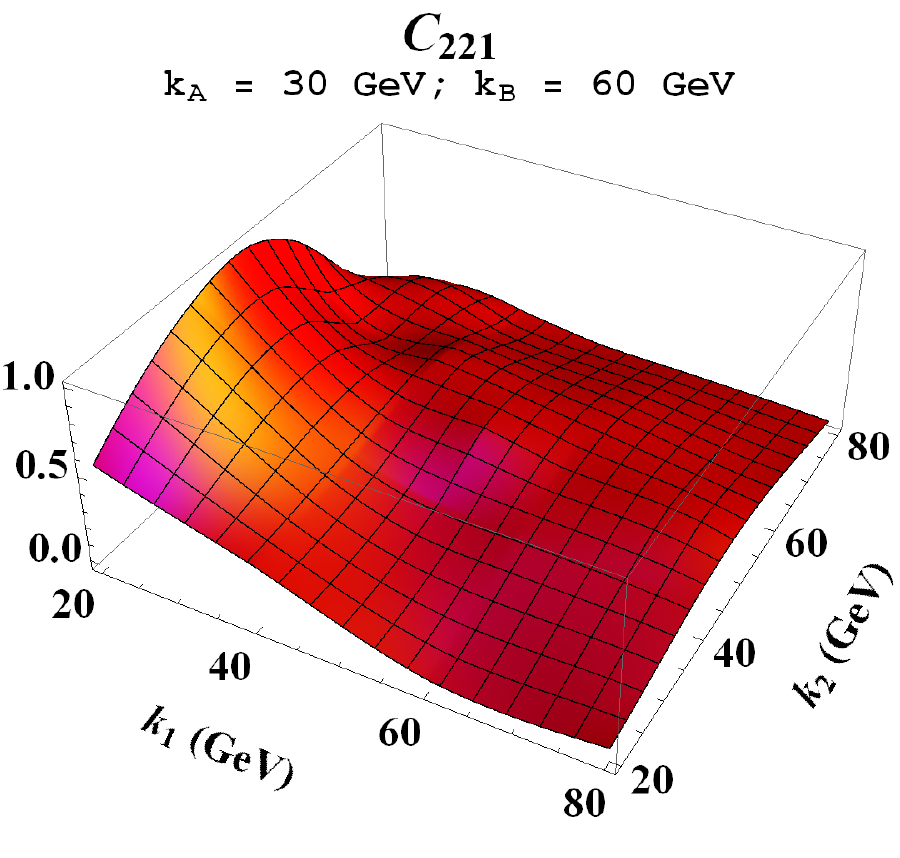}
     
     \includegraphics[scale=0.60]{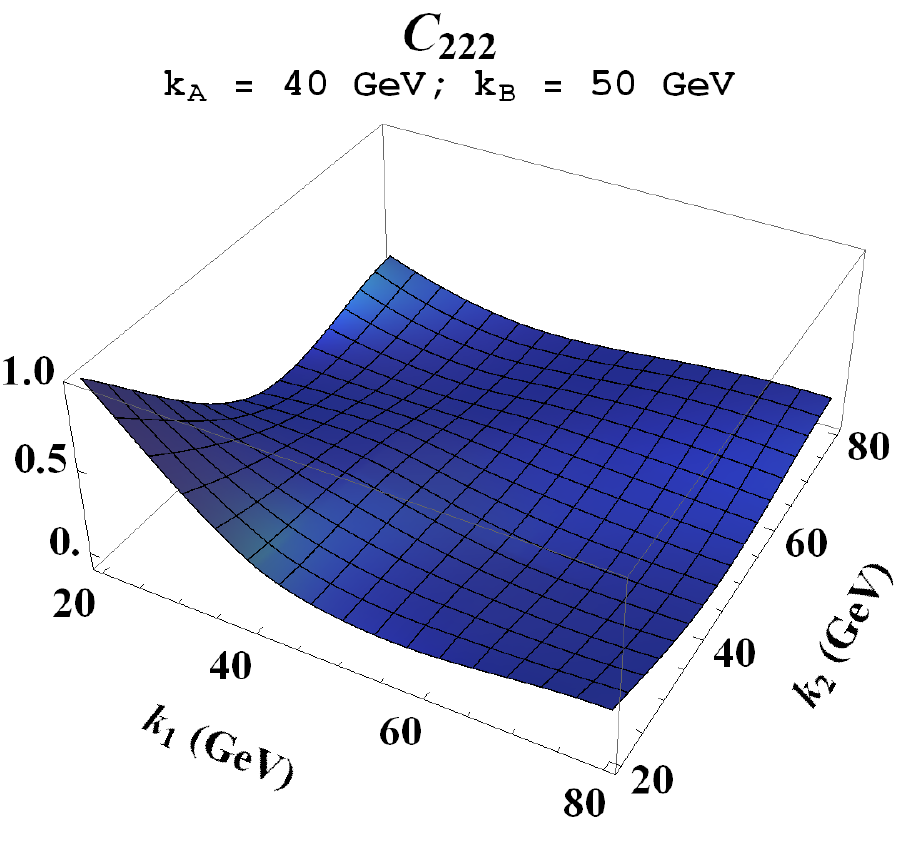}
     \includegraphics[scale=0.60]{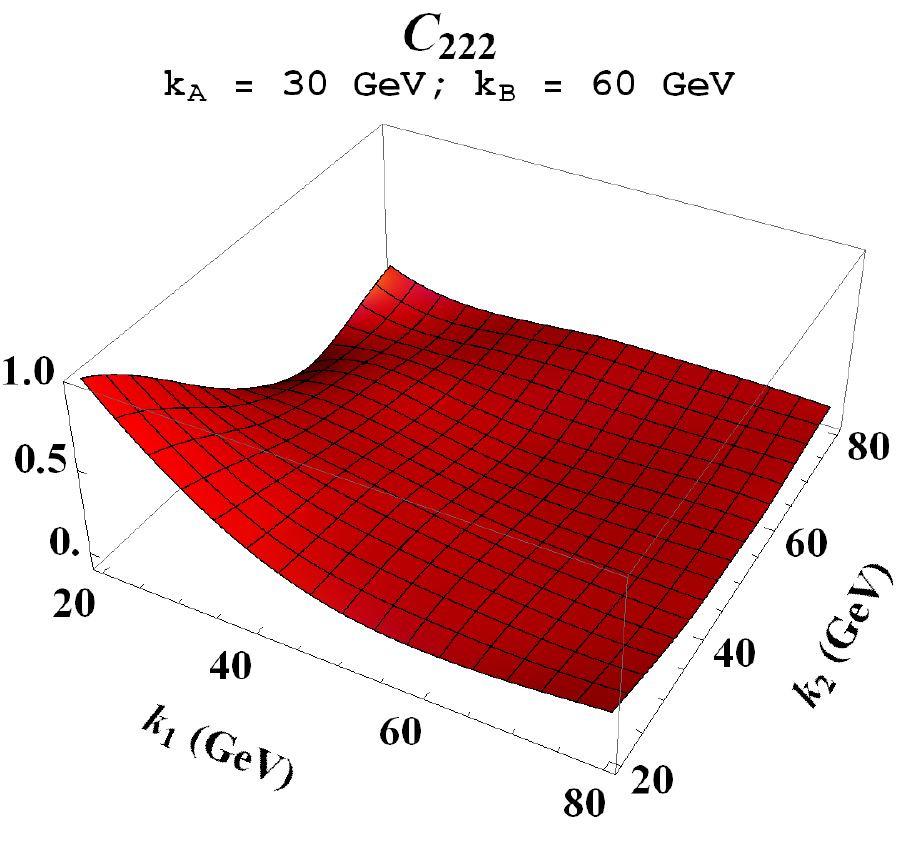}
  
  \caption[Partonic four-jets: BFKL azimuthal coefficients (2)]
  {$k_{1,2}$-dependence of the normalised 
  ${\cal C}_{211}$, ${\cal C}_{212}$, ${\cal C}_{221}$
  and ${\cal C}_{222}$ for the two selected cases of forward jet 
  transverse momenta $k_A$ and $k_B$.}
  \label{C2nl}
 \end{figure}

\begin{figure}[H]
  \centering
     \includegraphics[scale=0.60]{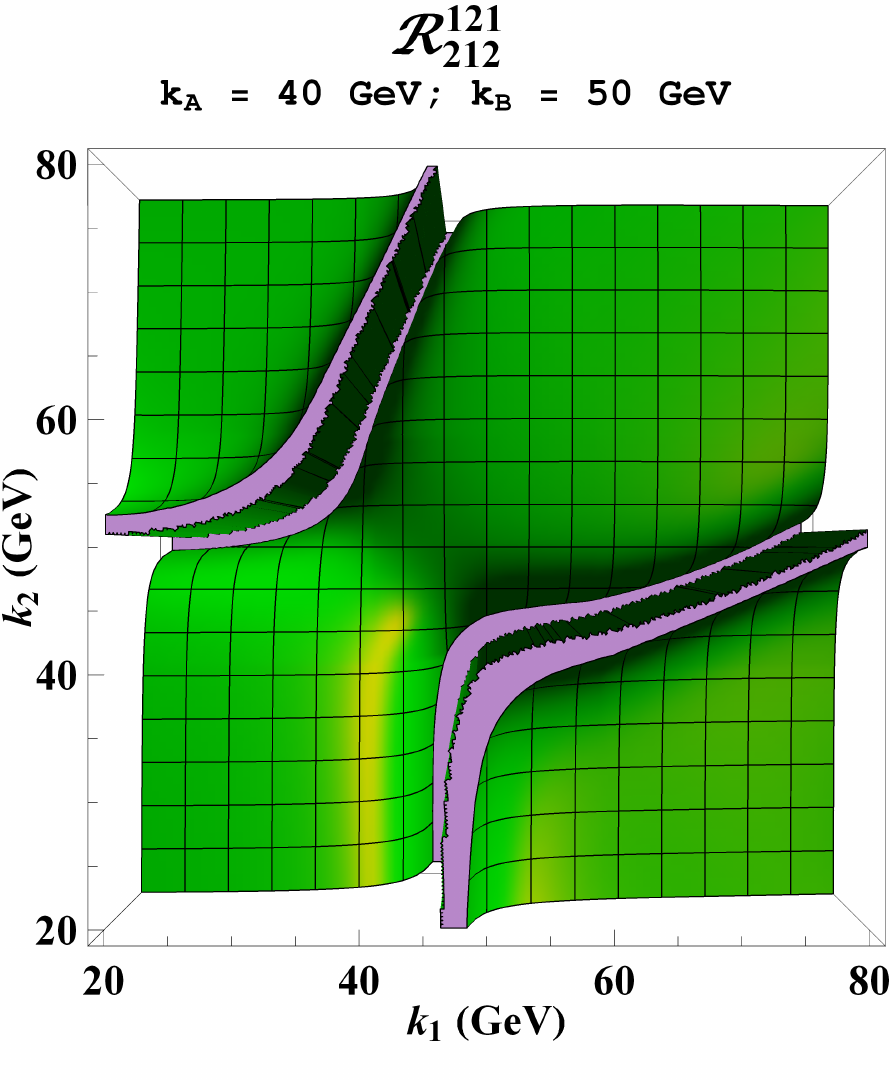}
     \includegraphics[scale=0.60]{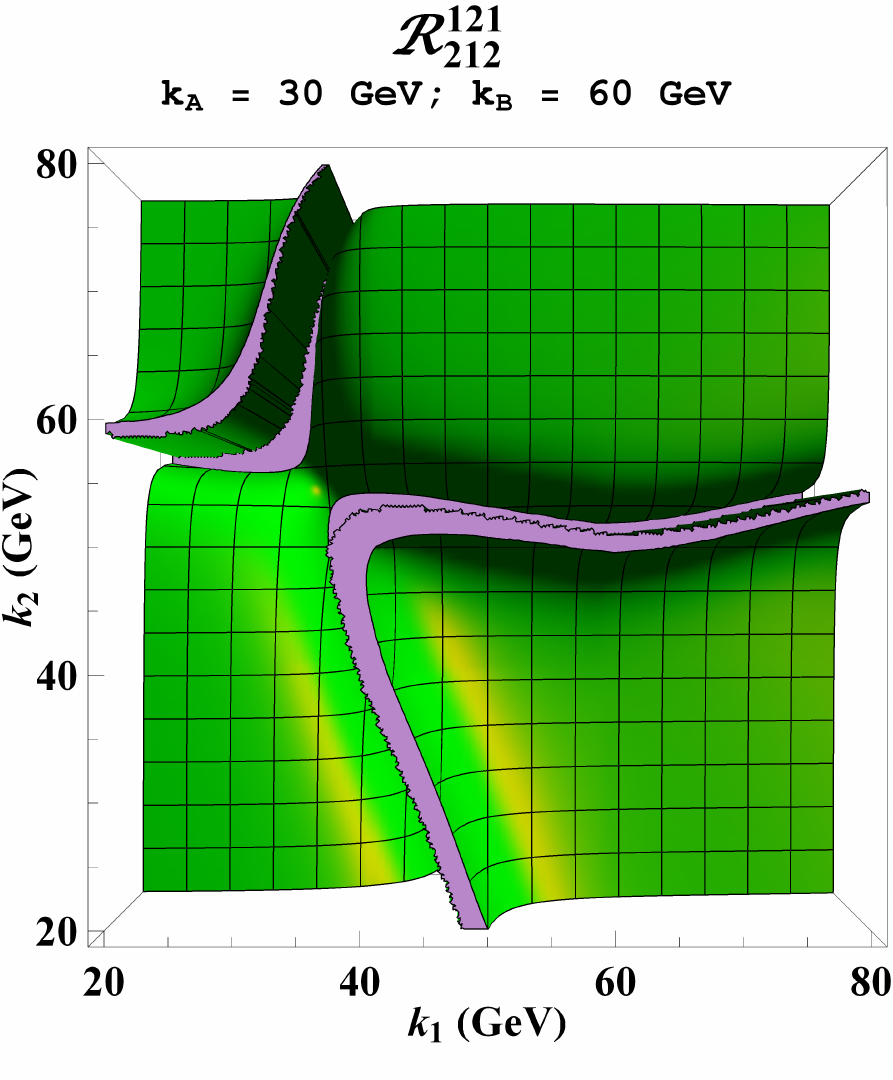}
  
     \includegraphics[scale=0.60]{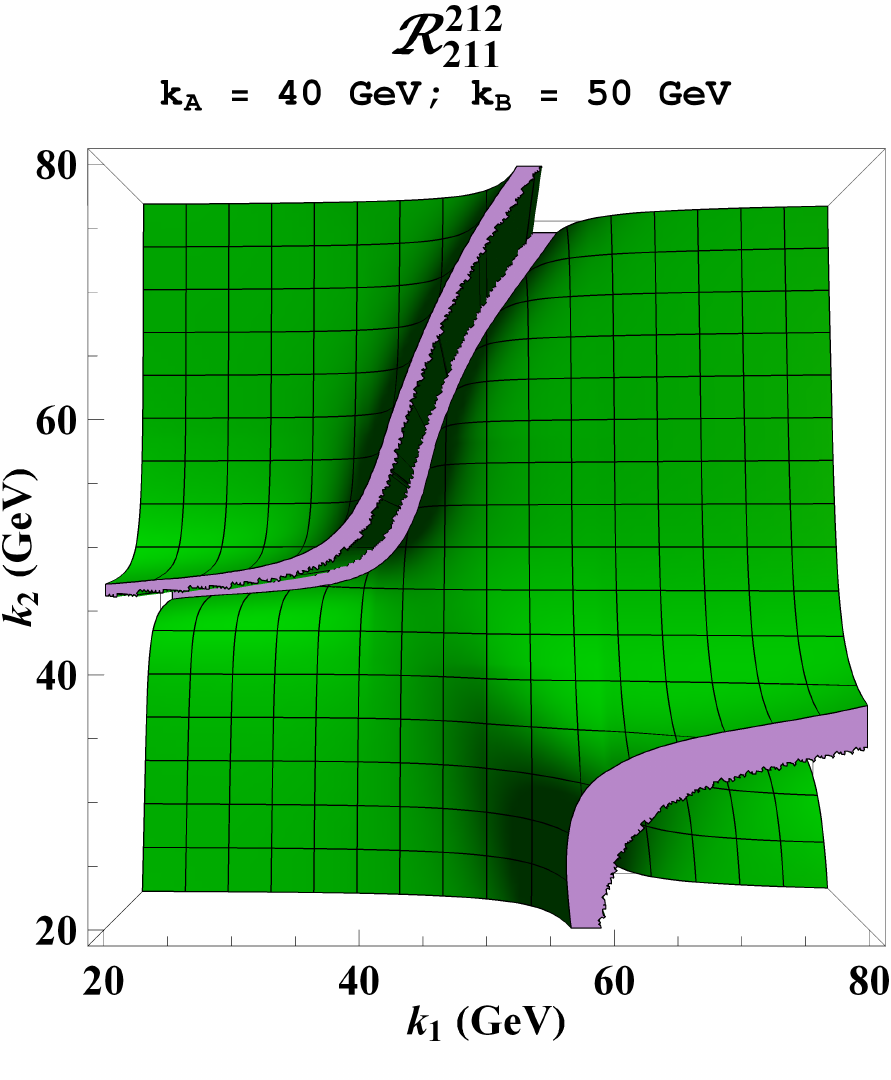}
     \includegraphics[scale=0.60]{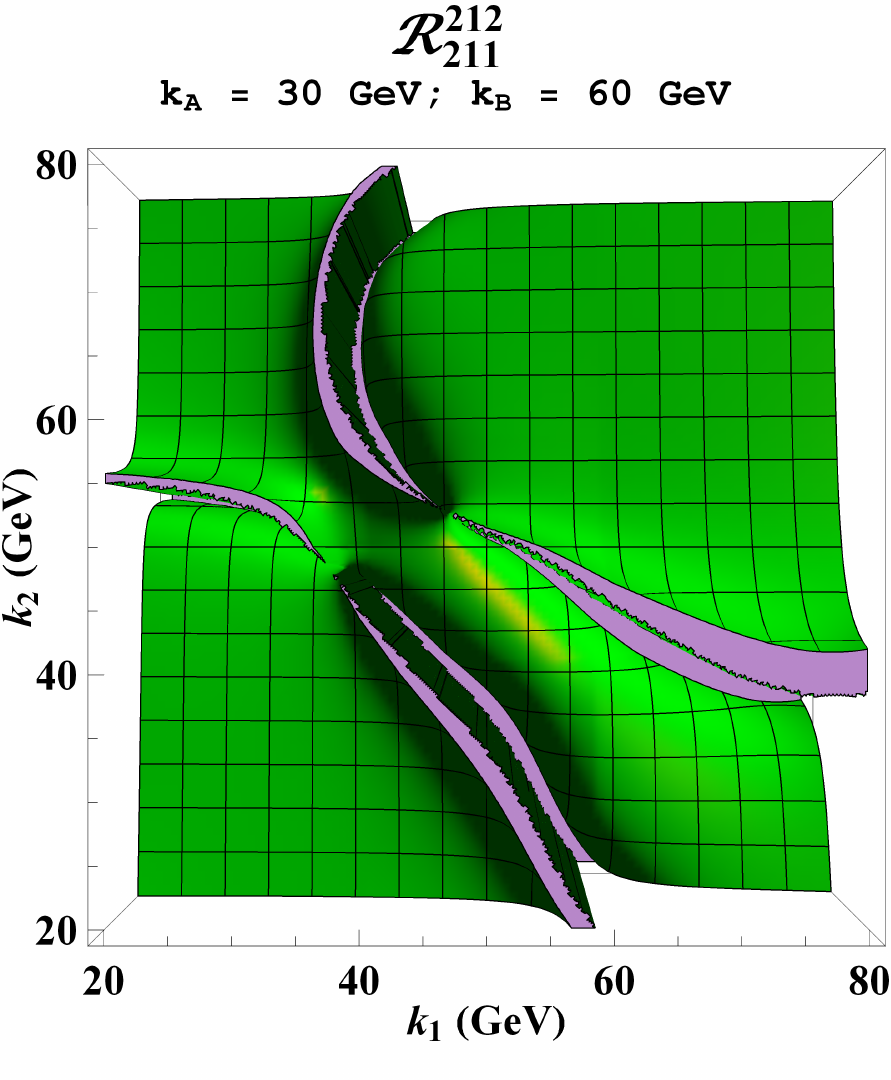}
     
     \includegraphics[scale=0.60]{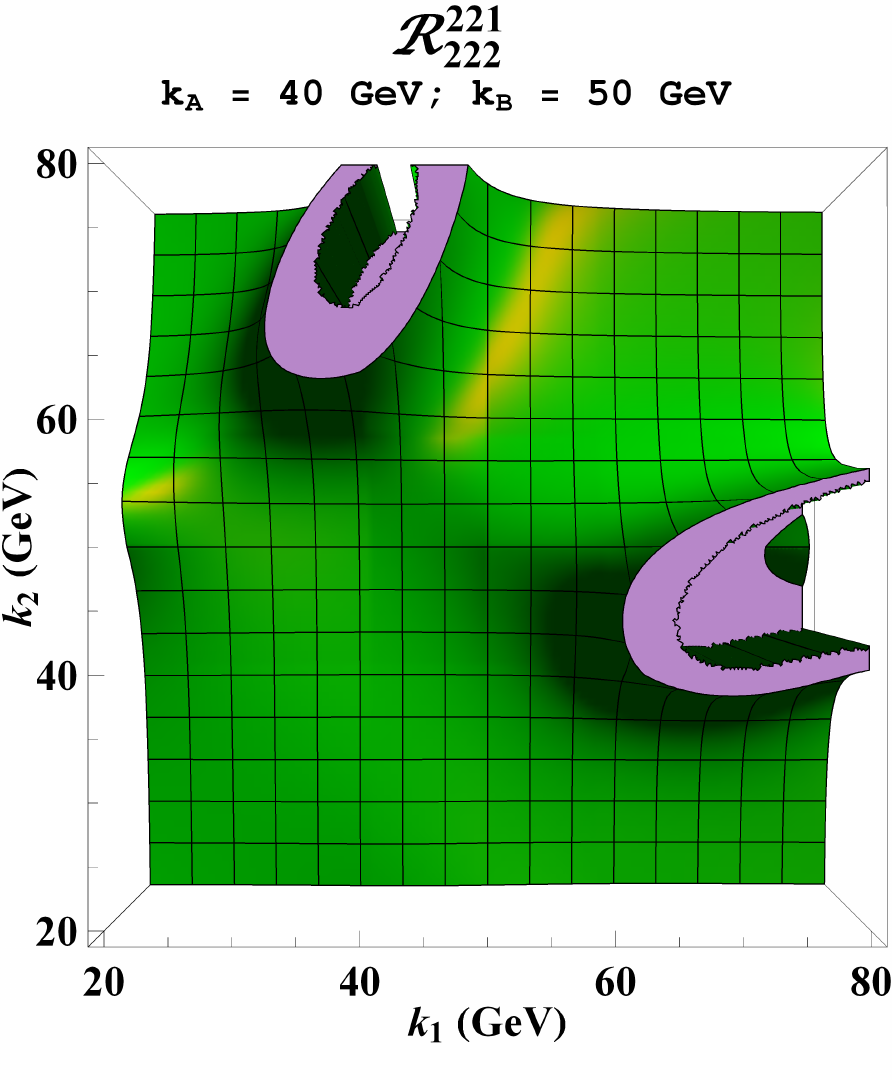}
     \includegraphics[scale=0.60]{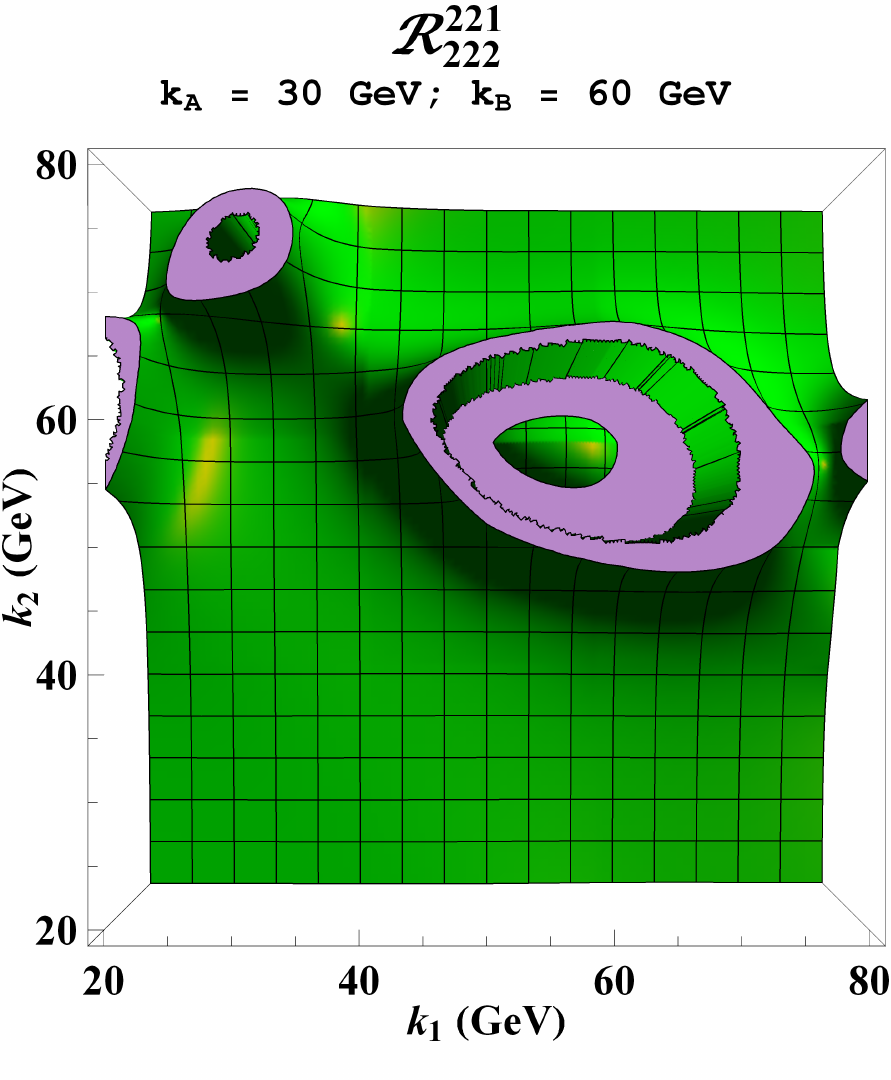}
  
  \caption[Partonic four-jets: BFKL azimuthal ratios (1)]
  {$k_{1,2}$-dependence of $\mathcal{R}^{121}_{212}$, 
  $\mathcal{R}^{212}_{211}$ and $\mathcal{R}^{221}_{222}$ 
  for the two selected cases of forward/backward jets  
  transverse momenta $k_A$ and $k_B$.}
  \label{fig:ratios_1}
 \end{figure}

\begin{figure}[H]
  \vspace{-0.35cm}
  \centering
     \includegraphics[scale=0.60]{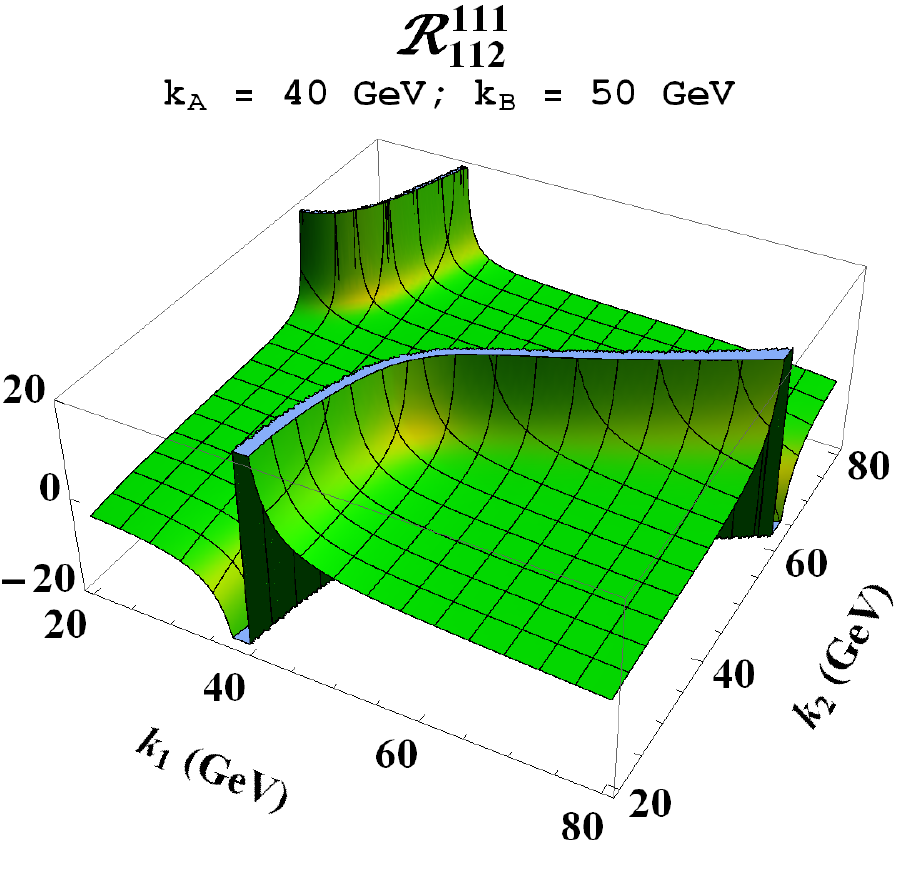}
     \includegraphics[scale=0.60]{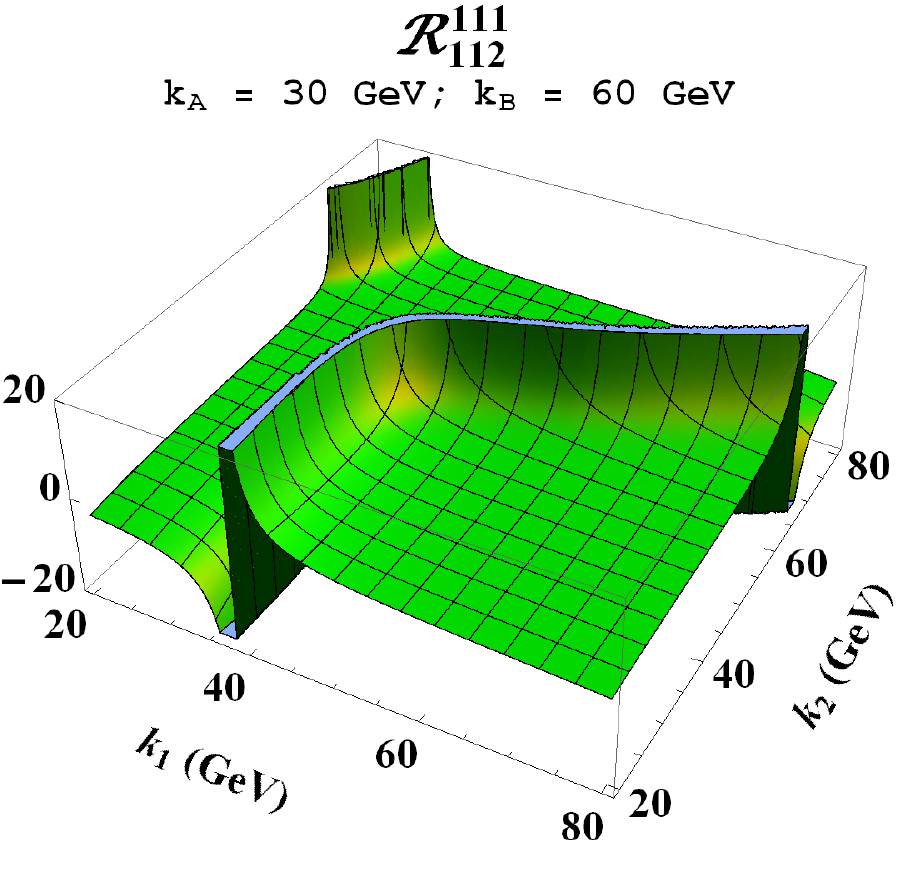}
  
     \includegraphics[scale=0.55]{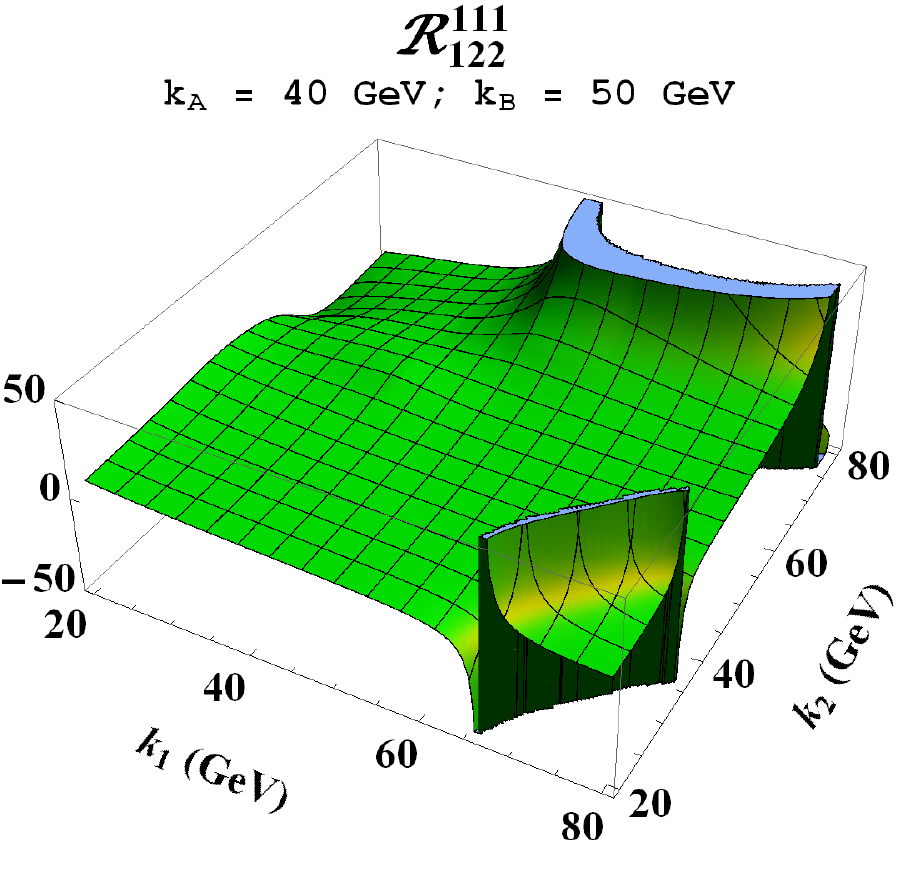}
     \includegraphics[scale=0.55]{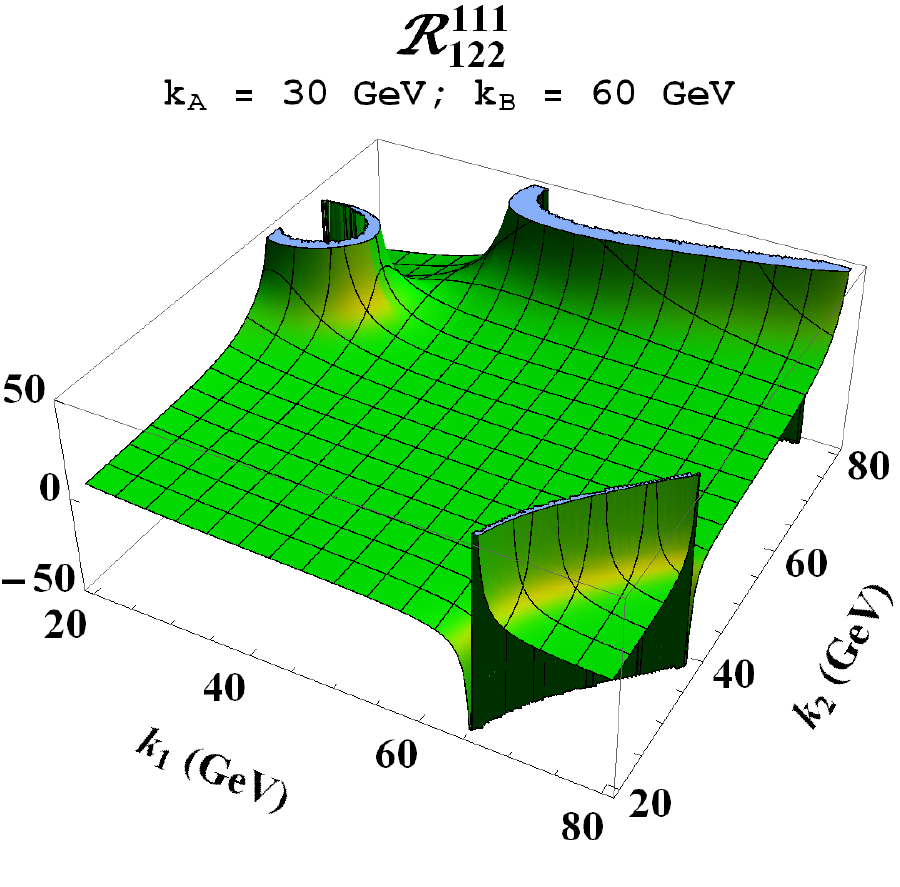}
     
     \includegraphics[scale=0.55]{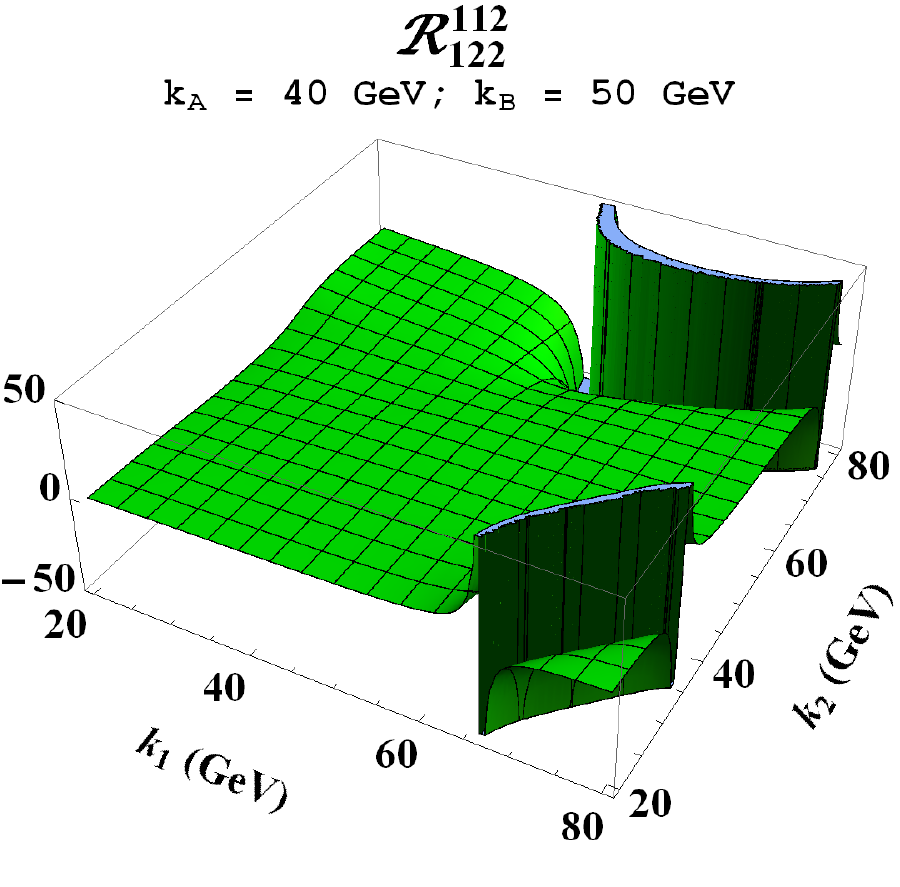}
     \includegraphics[scale=0.55]{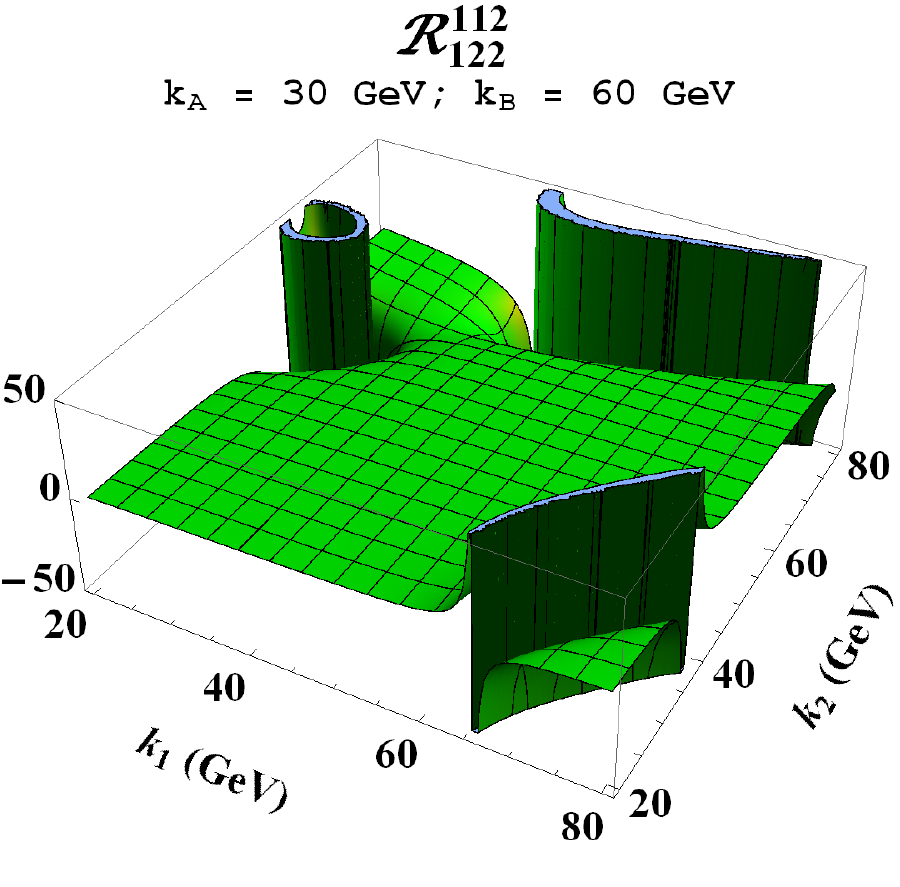}
     
     \includegraphics[scale=0.55]{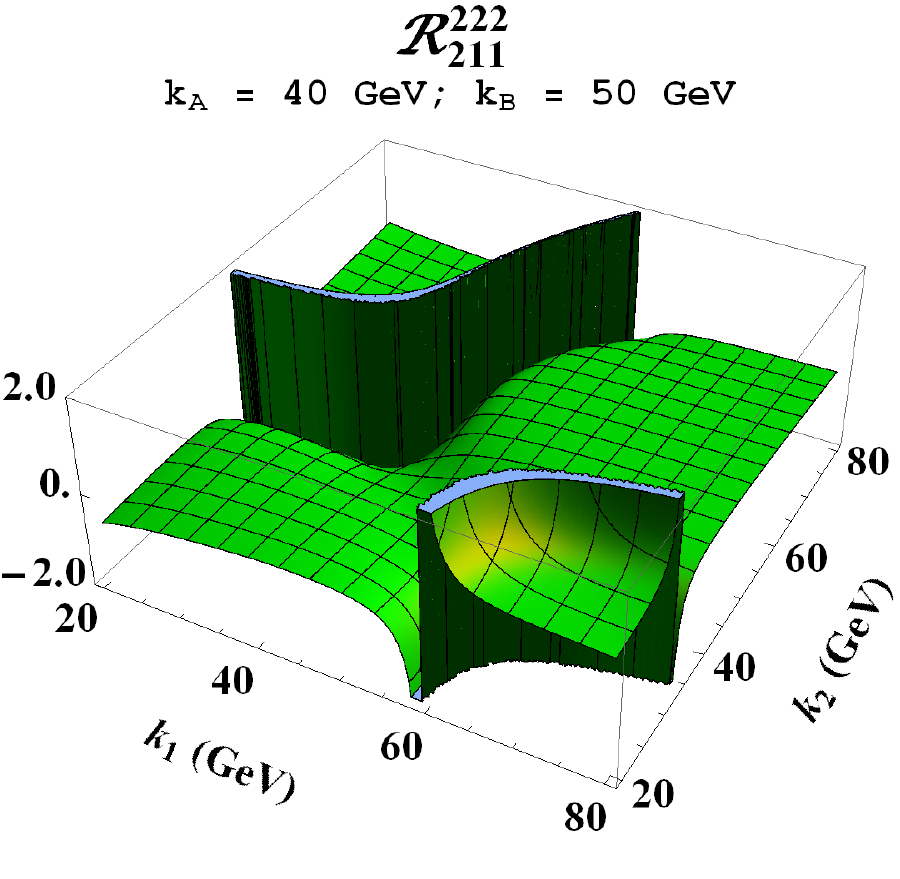}
     \includegraphics[scale=0.55]{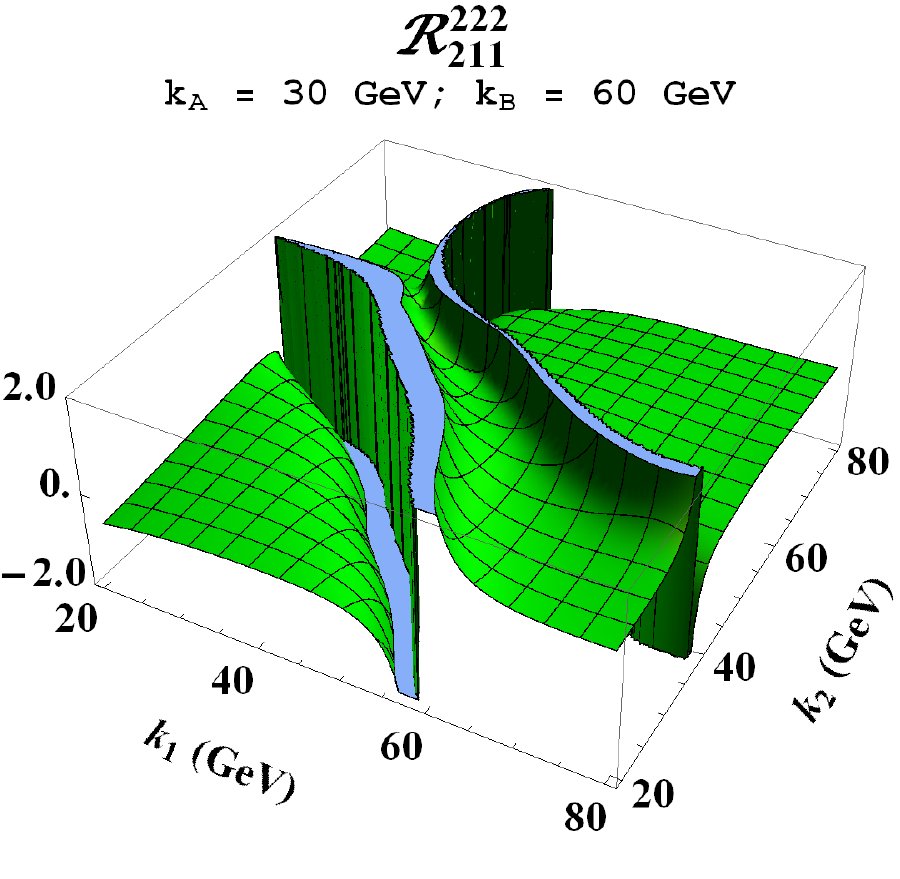}
  
  \caption[Partonic four-jets: BFKL azimuthal ratios (2)]
  {$k_{1,2}$-dependence of 
  $\mathcal{R}^{111}_{112}$, $\mathcal{R}^{111}_{122}$, 
  $\mathcal{R}^{112}_{122}$ and $\mathcal{R}^{222}_{211}$ 
  for the two selected cases of forward/backward jets  
  transverse momenta $k_A$ and $k_B$.}
  \label{fig:ratios_2}
 \end{figure}

 In this section the behaviour of our observables 
 is investigated in many different momenta configurations. 
 In order to cover two characteristic 
 cases, namely $ k_A \sim k_B $ and $ k_A < k_B $ 
 (or equivalently $k_A > k_B$)
 the following two fixed configurations for
 the transverse momenta of the forward jets
 have been chosen: $\left( k_A, k_B \right)$ =  $(40, 50)$  
 and $\left( k_A, k_B \right)$ = $(30, 60)$ GeV. 
 The rapidities of the four tagged jets are fixed to the values 
 $Y_A = 9$, $y_1 = 6$, $Y_2 = 3$, and $Y_B = 0$ 
 whereas the two inner jets can have transverse momenta 
 in the range $20 < k_{1,2} < 80$ GeV.
 In Fig.~\ref{C1nl} the results for the normalised
 coefficients ${\cal C}_{111}$, ${\cal C}_{112}$, 
 ${\cal C}_{121}$ and ${\cal C}_{122}$ are shown, 
 after they are divided by their respective maximum.
 The distributions are quite similar 
 for the two configurations here chosen 
 ($\left( k_A, k_B \right)$ = $(40, 50)$, $(30, 60)$ GeV) 
 apart from the coefficient ${\cal C}_{121}$ 
 which is quite more negative for the latter configuration 
 when the transverse momentum of the first central jet, $k_1$, is low. 
 Further coefficients, normalised as above, 
 are calculated in Fig.~\ref{C2nl} for the cases 
 ${\cal C}_{211}$, ${\cal C}_{212}$, ${\cal C}_{221}$
 and ${\cal C}_{222}$. Again they are rather similar 
 with the exception of  ${\cal C}_{221}$ at low $p_t$ 
 of one of the centrals jets with largest rapidity.
 Since these coefficients change sign 
 on the parameter space here studied, 
 it is clear that for the associated ratios 
 $\mathcal{R}^{MNL}_{PQR}$ 
 there will be some lines of singularities. 
 We have investigated $\mathcal{R}^{121}_{212}$,  
 $\mathcal{R}^{212}_{211}$ and $\mathcal{R}^{221}_{222}$ 
 in Fig.~\ref{fig:ratios_1}.
 In this case the configurations   
 $\left( k_A, k_B \right)$ = $(40, 50)$, $(30, 60)$ GeV 
 behave quite differently. This is due to the variation 
 of the position of the zeroes of those coefficients ${\cal C}_{MNP}$ 
 chosen as denominators in these quantities. 
 It would be very interesting to test 
 if these singularity lines are present 
 in any form in the LHC experimental data. 
 A further set of ratios, $\mathcal{R}^{111}_{112}$, 
 $\mathcal{R}^{111}_{122}$, $\mathcal{R}^{112}_{122}$ 
 and $\mathcal{R}^{222}_{211}$, with their characteristic 
 singular lines, is presented in Fig.~\ref{fig:ratios_2}.
 In general, a very weak dependence on variations 
 of the rapidity of the more central jets $y_{1,2}$
 is found for all the observables here presented. 

 \section{Hadronic level predictions} 
 \label{sec:4j-hadronic}
 
 In order to perform a more phenomenological analysis 
 it is needed to give predictions at the hadronic level
 by considering observables built up 
 from the hadronic cross section 
 (see Eq.~(\ref{dsigma_pdf_convolution-4j})).
 
  \subsection{The four-jet azimuthal correlations: hadronic level}
  \label{sub:4j-2cos-hadronic}
  
  Making use of the expression for the jet 
  vertex in the LO approximation (Eq.~\ref{c1}), 
  the hadronic cross section for the process~(\ref{process-4j}) reads
  \begin{align}
   & \frac{d\sigma^{4-{\rm jet}}}
        {dk_A \, dY_A \, d\vartheta_A \, 
         dk_B \, dY_B \, d\vartheta_B \, 
         dk_1 \, dy_1 d\vartheta_1 \, 
         dk_2 \, dy_2 d\vartheta_2} 
   \\ & = 
   \frac{16 \pi^4 \, C_F \, \asb^4}{N_C^3} \, 
   \frac{x_{J_A} \, x_{J_B}}{k_A \, k_B \, k_1\, k_2} \,
   \int d^2 \vec{p}_A \int d^2 \vec{p}_B  \int d^2 \vec{p}_1 
   \int d^2 \vec{p}_2 \,
   \nonumber \\ & \times \, 
   \delta^{(2)} \left(\vec{p}_A + \vec{k}_1- \vec{p}_1\right) 
   \delta^{(2)} \left(\vec{p}_B - \vec{k}_2- \vec{p}_2\right) \,
   \nonumber \\ & \times \,  
   \left(\frac{N_C}{C_F}f_g(x_{J_A},\mu_F)
   +\sum_{r=q,\bar q}f_r(x_{J_A},\mu_F)\right) \,
   \nonumber \\ & \times \,  
   \left(\frac{N_C}{C_F}f_g(x_{J_B},\mu_F)
   +\sum_{s=q,\bar q}f_s(x_{J_B},\mu_F)\right)
   \nonumber \\ & \times \, \nonumber  
   \varphi \left(\vec{k}_A,\vec{p}_A,Y_A - y_1\right) 
   \varphi \left(\vec{p}_1,\vec{p}_2,y_1 - y_2\right)
   \varphi \left(\vec{p}_B,\vec{k}_B, y_2 - Y_B\right) .
  \end{align}
  In order to follow a MRK setup  
  we demand, as we did in Section~\ref{sec:4j-partonic},  
  that the rapidities of the produced particles 
  obey $Y_A > y_1 > y_2 > Y_B$, while $k_1^2$ 
  and $k_2^2$ are well
  above the resolution scale of the detectors.
  $x_{J_{A,B}}$ are the longitudinal momentum fractions
  of the two external jets, connected to the respective rapidities 
  $Y_{A,B}$ by the relation 
  $x_{J_{A,B}} = k_{A,B} \, e^{\, \pm \, Y_{A,B}} / \sqrt{s}$.

  Our goal is to define new observables for which
  the BFKL dynamics would surface in a distinct form. 
  Moreover, we request that our observables should be
  rather insensitive to possible higher-order corrections. 
  The related experimental observable we propose corresponds to
  the mean value (with $M,N,L$ being positive integers)
  \begin{align}
  \label{Cmnl}
   &{\cal C}_{MNL} \, = \,
   \langle \cos(M \,\phi_1) \cos(N \,\phi_2) \cos (L \,\phi_3 ) \rangle 
   \\ \nonumber =  
   & \frac{\int_0^{2 \pi} d \vartheta_A \int_0^{2 \pi} 
   d \vartheta_B \int_0^{2 \pi} d \vartheta_1 
   \int_0^{2 \pi} d \vartheta_2 
   \cos(M \phi_1) \cos(N \phi_2)  \cos (L \phi_3 )
   \; d \sigma^{4-{\rm jet}} }{\int_0^{2 \pi} d \vartheta_A 
   d \vartheta_B d \vartheta_1 d \vartheta_2
   \; d \sigma^{4-{\rm jet}} } \; ,
  \end{align}
  with $\phi_1$, $\phi_2$ and $\phi_3$ defined in Eq.~(\ref{phi123}).  
  The numerator in Eq.~(\ref{Cmnl}) actually reads
  \begin{align}
  \label{CmnlNUM}
  & \int_0^{2 \pi} d \vartheta_A \int_0^{2 \pi} 
  d \vartheta_B \int_0^{2 \pi} d \vartheta_1 
   \int_0^{2 \pi} d \vartheta_2 \, 
   \cos(M \phi_1) \cos(N \phi_2) \cos (L \phi_3 ) 
    \\ &\times \, 
     \frac{d\sigma^{4-{\rm jet}}}
        {dk_A \, dY_A \, d\vartheta_A \, 
         dk_B \, dY_B \, d\vartheta_B \, 
         dk_1 \, dy_1 d\vartheta_1 \, 
         dk_2 \, dy_2 d\vartheta_2} =  
         \nonumber \\ &\times \, 
   \frac{16 \pi^4 \, C_F \, \asb^4}{N_C^3} \, 
   \frac{x_{J_A} \, x_{J_B}}{k_A \, k_B \, k_1\, k_2} \,
   \int d^2 \vec{p}_A \int d^2 \vec{p}_B  
   \int d^2 \vec{p}_1 \int d^2 \vec{p}_2 \,
   \nonumber \\  & \times \, 
   \delta^{(2)} \left(\vec{p}_A + \vec{k}_1- \vec{p}_1\right) 
   \delta^{(2)} \left(\vec{p}_B - \vec{k}_2- \vec{p}_2\right) \,
   \nonumber \\ & \times \, 
   \left(\frac{N_C}{C_F}f_g(x_{J_A},\mu_F)
   +\sum_{r=q,\bar q}f_r(x_{J_A},\mu_F)\right) \,
   \nonumber \\ & \times \, 
   \left(\frac{N_C}{C_F}f_g(x_{J_B},\mu_F)
   +\sum_{s=q,\bar q}f_s(x_{J_B},\mu_F)\right)
   \nonumber \\ & \times \, 
   \left( 
   \tilde{\Omega}_{M,N,L} 
   + \tilde{\Omega}_{M,N,-L} 
   + \tilde{\Omega}_{M,-N,L} 
   + \tilde{\Omega}_{M,-N,-L} 
   \right.  
    \nonumber \\ \nonumber & 
    \left.
   + \; \tilde{\Omega}_{-M,N,L} 
   + \tilde{\Omega}_{-M,N,-L} 
   + \tilde{\Omega}_{-M,-N,L} 
   + \tilde{\Omega}_{-M,-N,-L} 
   \right) .
   \end{align}
   The quantity $ \tilde{\Omega}_{m,n,l}$ 
   is simply a convolution of BFKL gluon
   Green's functions, given in Eq.~(\ref{omega_mnl_tilde}).

  \subsection{Integration over the final-state phase space} 
   \label{sub:4j-phase-space}
   
  As anticipated, we would like to consider quantities 
  that are easily measured
  experimentally and, moreover, we want to eliminate 
  as much as possible any dependence
  on higher-order corrections. 
  Thus, we need to consider
  ratios similar to Eq.~(\ref{Rmnlpqr}), 
  which are defined on a partonic level though.
  Therefore, in order to provide testable theoretical predictions 
  against any current and forthcoming experimental data, 
  we proceed in two steps. Firstly, 
  we impose LHC kinematical cuts by integrating 
  ${\cal C}_{M N L}$ over the momenta of the tagged jets.
  More precisely,
  \begin{align}
  \label{Cmnl_int}
   & 
   C_{MNL} =
   \int_{Y_A^{\rm min}}^{Y_A^{\rm max}} \hspace{-0.25cm} dY_A
   \int_{Y_B^{\rm min}}^{Y_B^{\rm max}} \hspace{-0.25cm} dY_B
   \int_{k_A^{\rm min}}^{k_A^{\rm max}} \hspace{-0.25cm} dk_A
   \int_{k_B^{\rm min}}^{k_B^{\rm max}} \hspace{-0.25cm} dk_B
   \int_{k_1^{\rm min}}^{k_1^{\rm max}} \hspace{-0.25cm} dk_1
   \int_{k_2^{\rm min}}^{k_2^{\rm max}} \hspace{-0.25cm} dk_2
   \\ \nonumber & \times \,  
   \delta\left(Y_A - Y_B - Y\right) {\cal C}_{MNL} \; ,
  \end{align}
  where the rapidity $Y_A$ of the most forward jet $k_A$ 
  is restricted to $0 < Y_A < 4.7$
  and the rapidity $Y_B$ of the most backward jet $k_B$ 
  is restricted to $-4.7 < Y_B < 0$
  while their difference $Y = Y_A - Y_B$ 
  is kept fixed at definite values 
  within the range $6.5 < Y < 9$. Obviously, 
  the last condition on the allowed values
  of $Y$ makes both the integration ranges 
  over $Y_A$ and $Y_B$ smaller than 4.7 units of rapidity.
  Secondly, we remove the zeroth conformal spin 
  contribution responsible for 
  any collinear contamination 
  (contributions that originate at $\varphi_0$)
  and we minimise possible higher-order effects 
  by introducing the ratios
  \begin{eqnarray}
  \label{RPQRMNL}
  R_{PQR}^{MNL} \, = \, \frac{C_{MNL}}{C_{PQR}}
  \label{RmnlqprNew} \; ,
  \end{eqnarray}
  where $M, N, L, P, Q, R$ are positive definite integers.

  \subsection{Numerical analysis} 
   \label{sub:4j-numerical}
   
  In this Section the results for the ratios 
  $R_{PQR}^{MNL}$ in Eq.~(\ref{RmnlqprNew}) 
  are presented as functions of the 
  rapidity difference $Y$ between the outermost jets 
  for different momenta configurations and for two
  center-of-mass energies: $\sqrt s = 7$ and $\sqrt s = 13$ TeV. 
  For the transverse momenta $k_A$, $k_B$, $k_1$ and $k_2$ 
  the following cuts are imposed:
  \begin{enumerate}
  \item 
  \begin{align}
  \label{cut1}
  &k_A^{\rm min} = 35\, \text{GeV} \; , \,\,\, 
   k_A^{\rm max} = 60\, \text{GeV} \; , \\ 
  &k_B^{\rm min} = 45\, \text{GeV} \; , \,\,\, 
   k_B^{\rm max} = 60\, \text{GeV} \; , \nonumber\\
  &k_1^{\rm min} = 20\, \text{GeV} \; , \,\,\,
   k_1^{\rm max} = 35\, \text{GeV} \; , \nonumber\\\nonumber 
  &k_2^{\rm min} = 60\, \text{GeV} \; , \,\,\, 
   k_2^{\rm max} = 90\, \text{GeV} \; ;
  \end{align}
  \item 
  \begin{align}
  \label{cut2}
  &k_A^{\rm min} = 35\, \text{GeV} \; , \,\,\,
   k_A^{\rm max} = 60\, \text{GeV} \; , \\
  &k_B^{\rm min} = 45\, \text{GeV} \; , \,\,\,
   k_B^{\rm max} = 60\, \text{GeV} \; , \nonumber\\
  &k_1^{\rm min} = 25\, \text{GeV} \; , \,\,\,
   k_1^{\rm max} = 50\, \text{GeV} \; , \nonumber\\\nonumber 
  &k_2^{\rm min} = 60\, \text{GeV} \; , \,\,\,
   k_2^{\rm max} = 90\, \text{GeV} \; .
  \end{align}
  \end{enumerate}
  
  To keep things simple, in both cuts, 
  $k_2$ has been set larger
  than all the other three-jet momenta,  
  while only the range of $k_1$ is changed.
  In the cut defined in Eq.~(\ref{cut1}), 
  $k_1$ is smaller that all the other
  three-jet momenta whereas in the cut defined 
  in Eq.~(\ref{cut2}), the allowed
  $k_1$ values overlap
  with the ranges of $k_A$ and $k_B$.

   \subsubsection{Results and discussion}
   \label{ssub:4j-results}
   
\begin{figure}[H]
   \begin{center}
   
      \includegraphics[scale=.38]{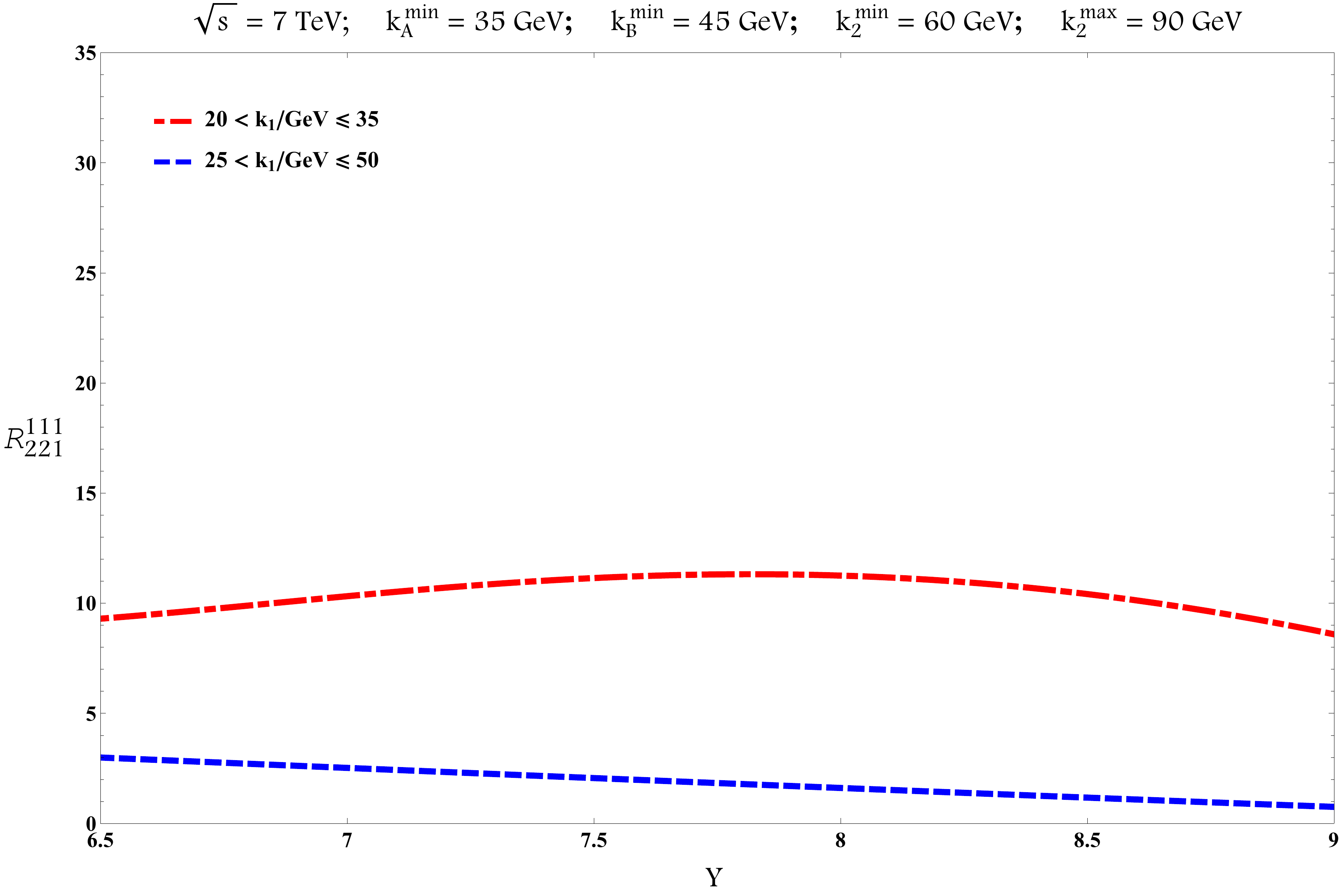}
   
      \vspace{.5cm}
   
      \includegraphics[scale=.38]{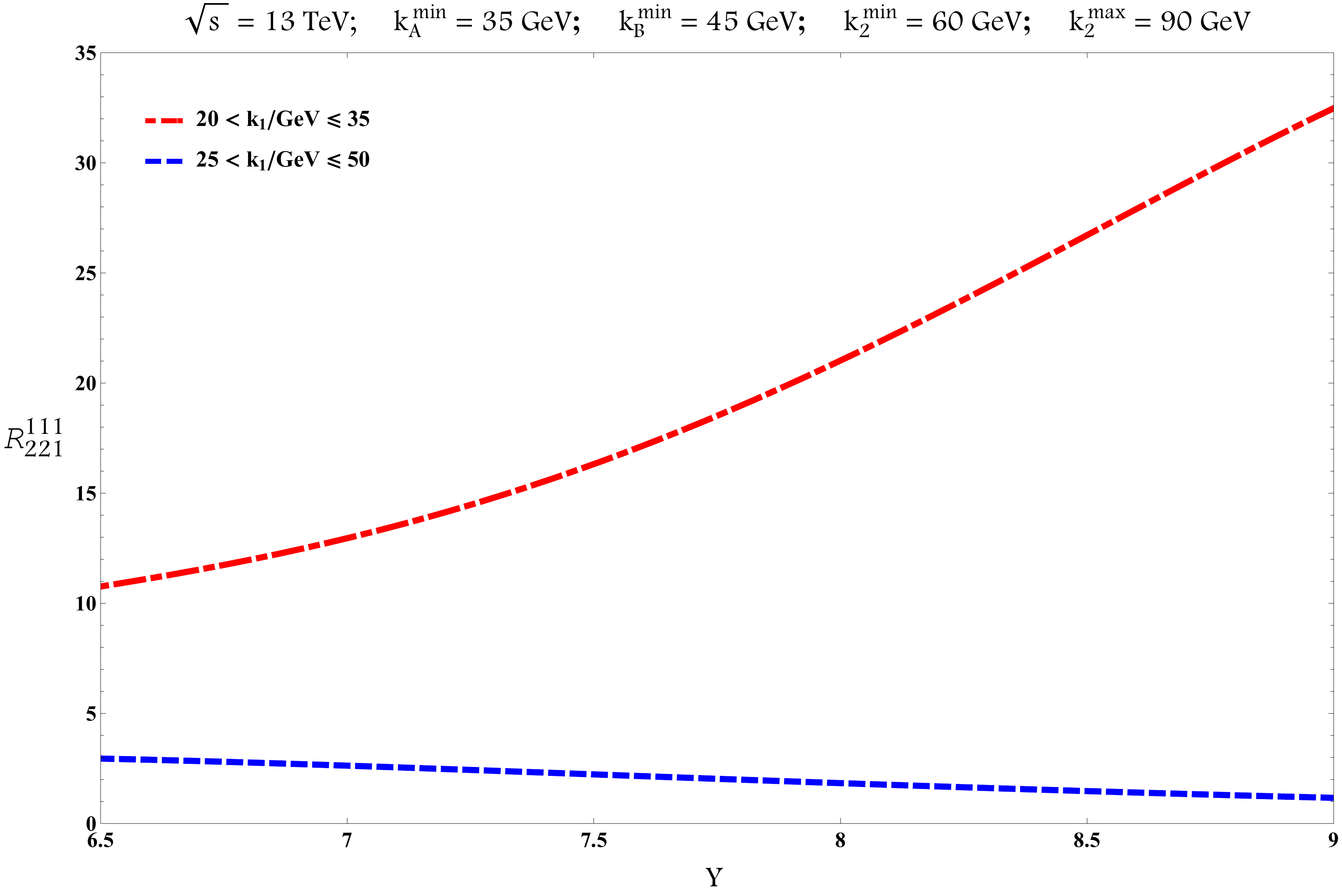}
   
   \caption[$Y$-dependence of the $R^{111}_{221}$ four-jet ratios]
   {$Y$-dependence of $R^{111}_{221}$
   for $\sqrt s = 7$ TeV (top) and for $\sqrt s = 13$ TeV (bottom).} 
   \label{fig:3}
   \end{center}
   \end{figure}

\begin{figure}[H]
   \begin{center}
   
      \includegraphics[scale=0.38]{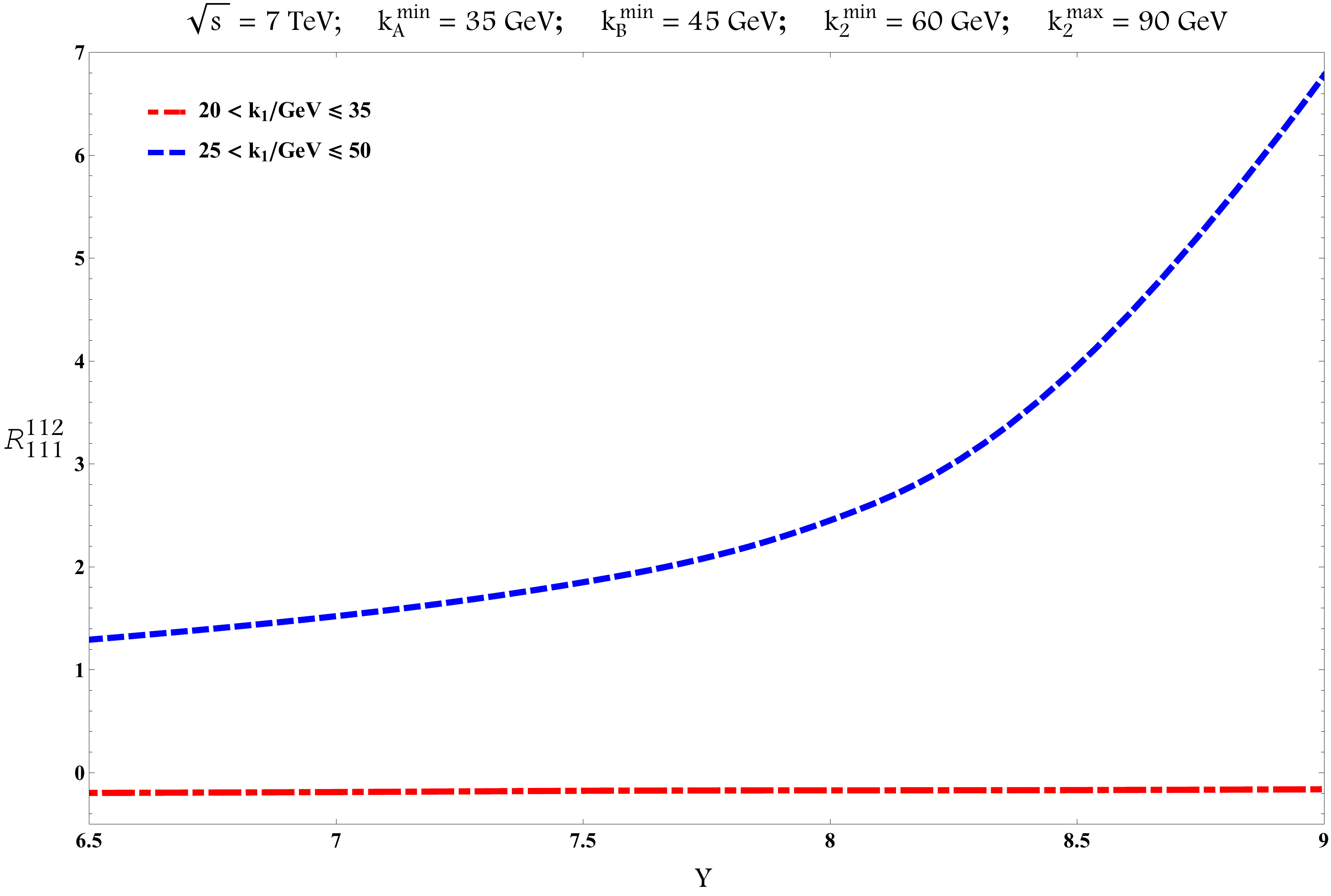}
   
      \vspace{.5cm}
   
      \includegraphics[scale=0.38]{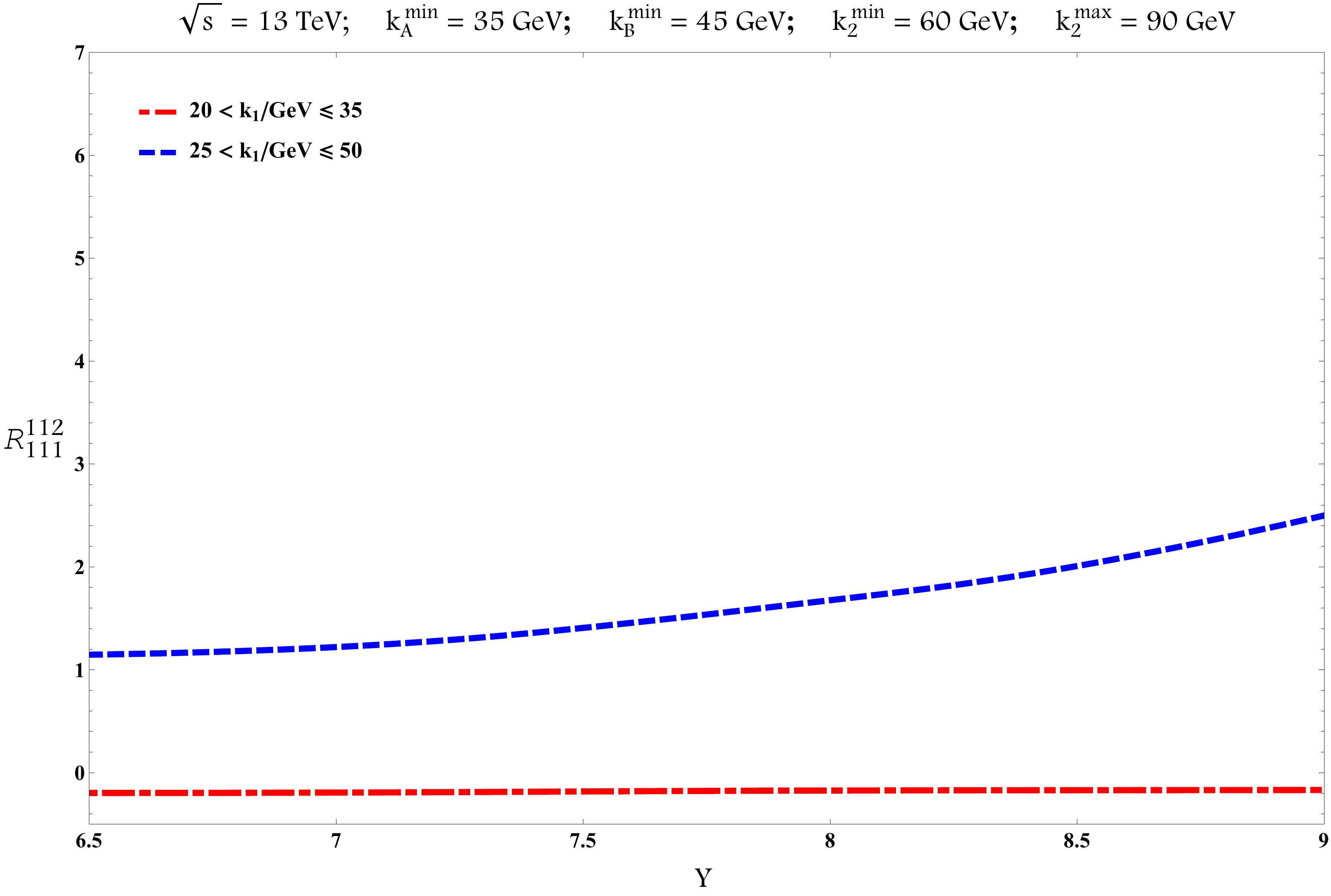}
   
   \caption[$Y$-dependence of the $R^{112}_{111}$ four-jet ratios]
   {$Y$-dependence of $R^{112}_{111}$
   for $\sqrt s = 7$ TeV (top) and for $\sqrt s = 13$ TeV (bottom).} 
   \label{fig:4}
   \end{center}
   \end{figure}

\begin{figure}[H]
   \begin{center}
   
      \includegraphics[scale=0.38]{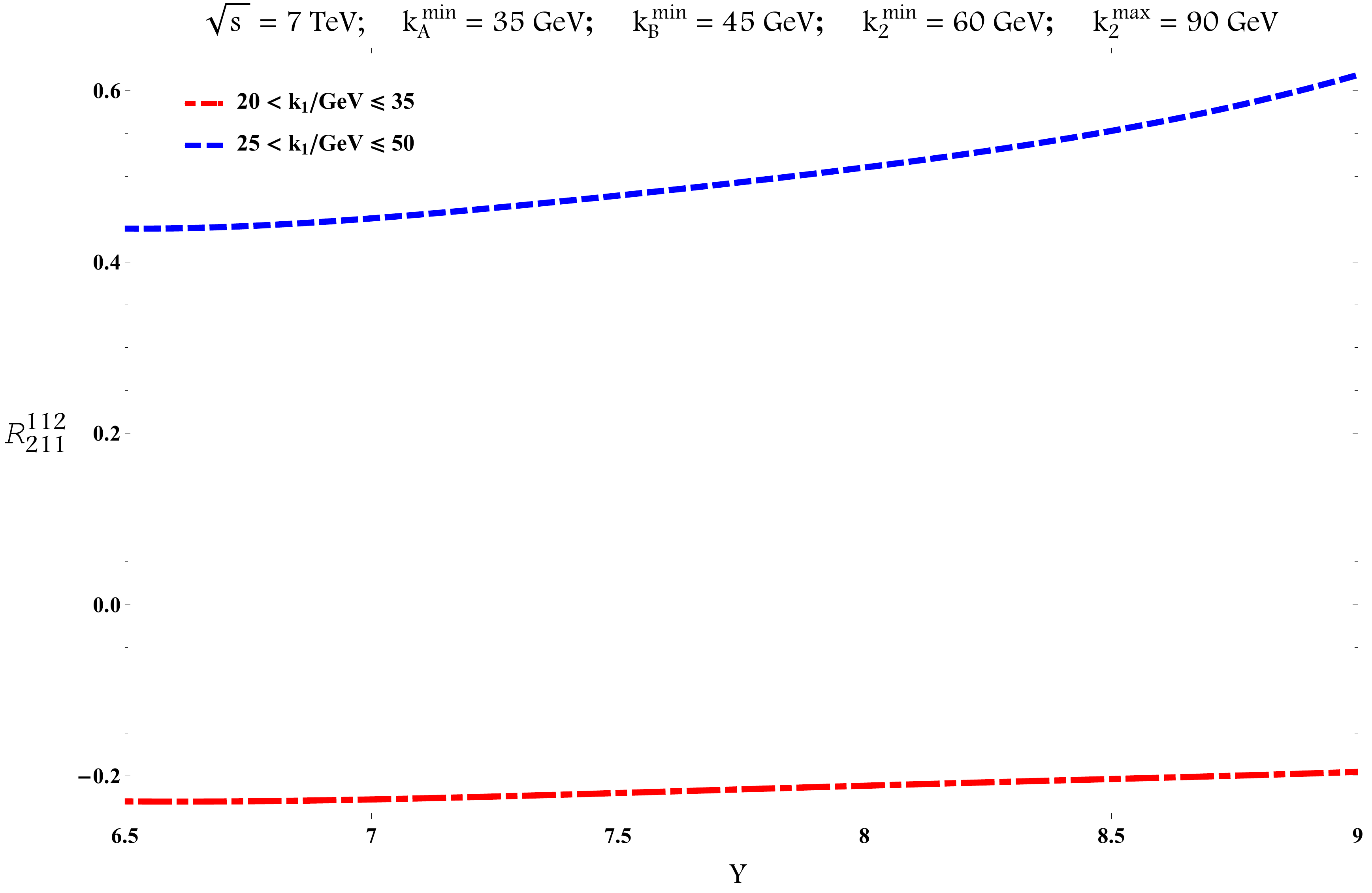}
   
      \vspace{.5cm}
   
      \includegraphics[scale=0.38]{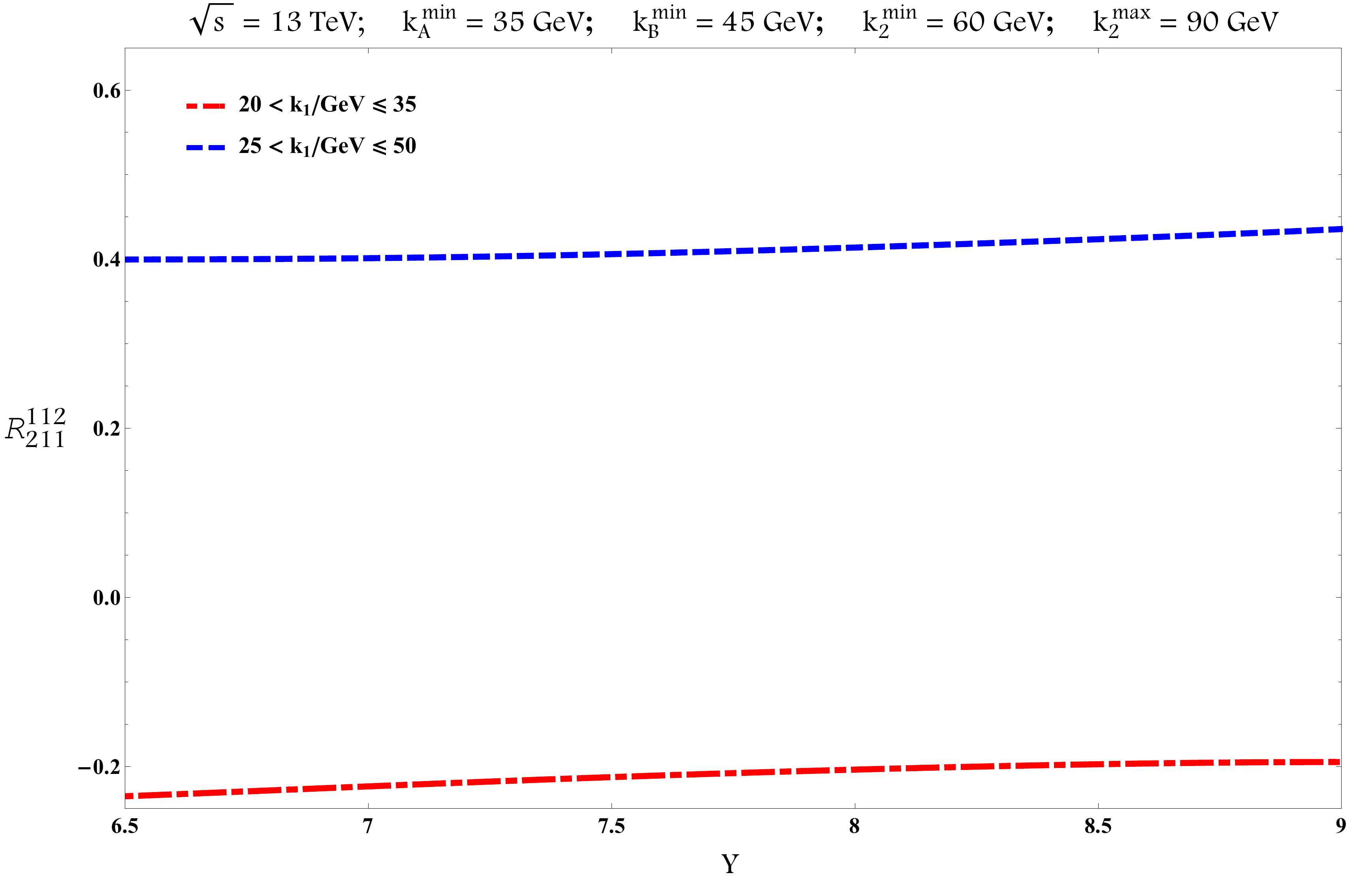}
   
   \caption[$Y$-dependence of the $R^{112}_{211}$ four-jet ratios]
   {$Y$-dependence of $R^{112}_{211}$
   for $\sqrt s = 7$ TeV (top) and for $\sqrt s = 13$ TeV (bottom).} 
   \label{fig:5}
   \end{center}
   \end{figure}

\begin{figure}[H]
   \begin{center}
   
      \includegraphics[scale=0.38]{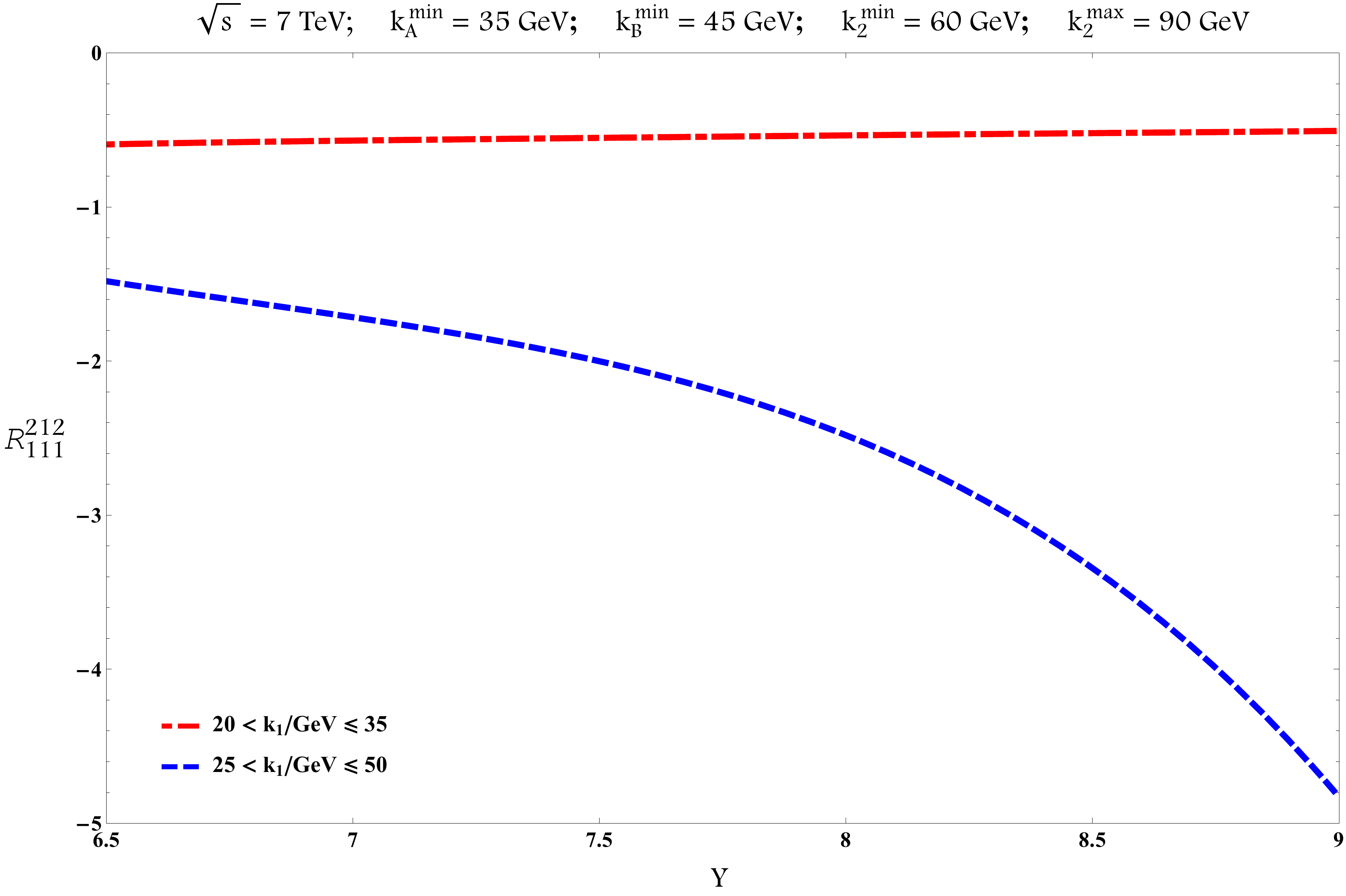}
   
      \vspace{.5cm}
   
      \includegraphics[scale=0.38]{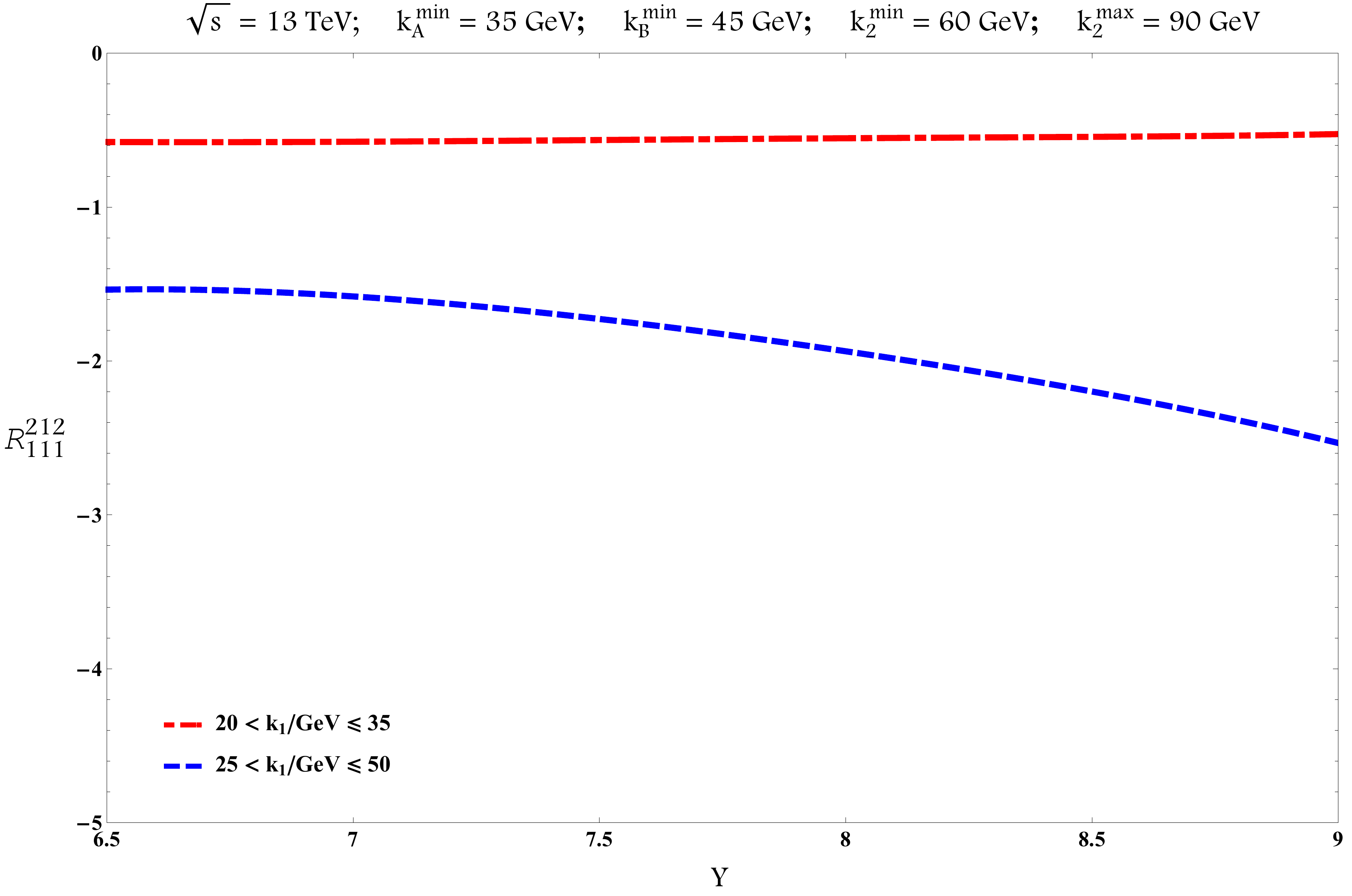}
   
   \caption[$Y$-dependence of the $R^{212}_{111}$ four-jet ratios]
   {$Y$-dependence of $R^{212}_{111}$
   for $\sqrt s = 7$ TeV (top) and for $\sqrt s = 13$ TeV (bottom).} 
   \label{fig:6}
   \end{center}
   \end{figure}

\begin{figure}[H]
   \begin{center}
   
      \includegraphics[scale=0.38]{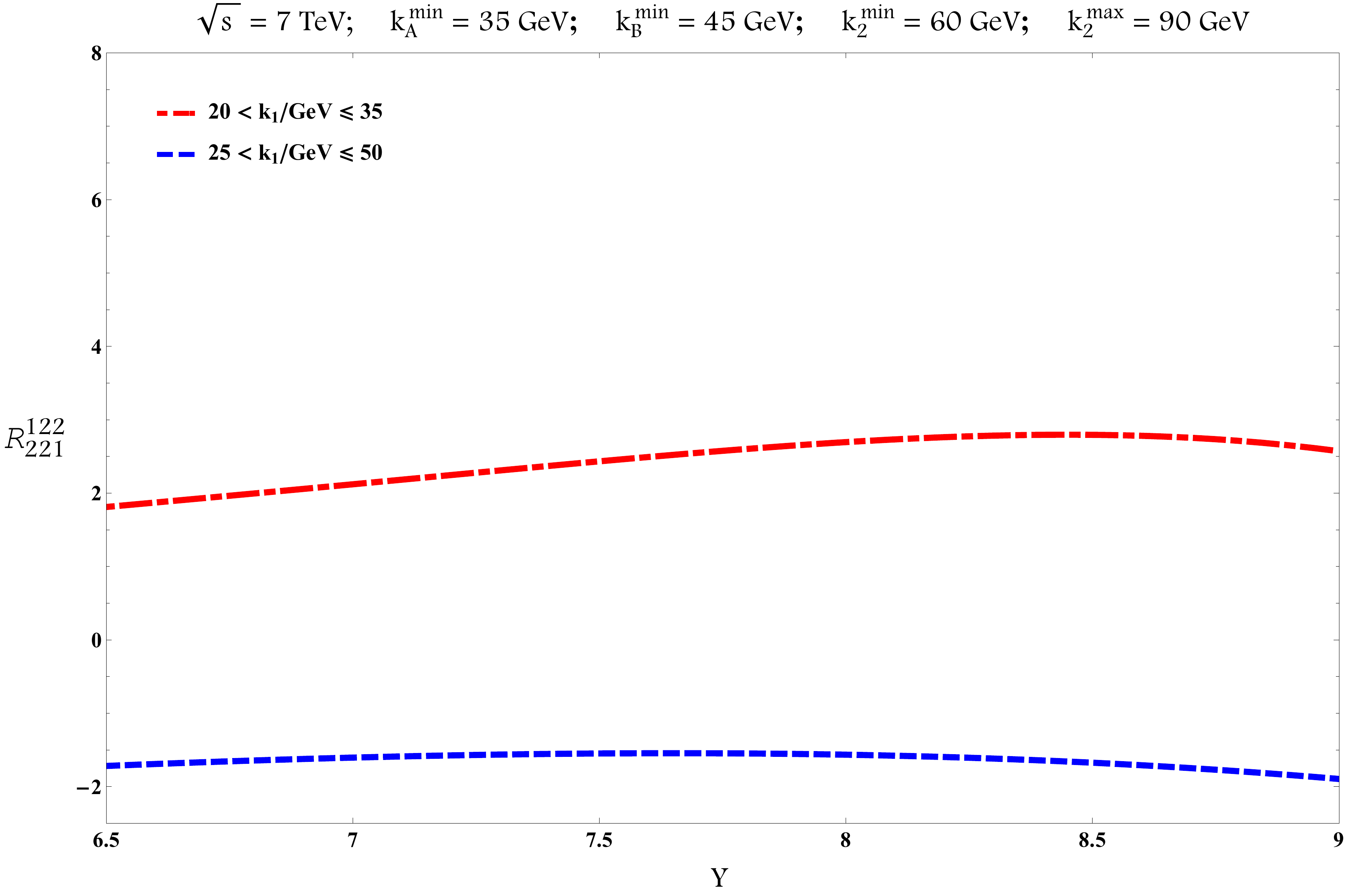}
   
      \vspace{.5cm}
   
      \includegraphics[scale=0.38]{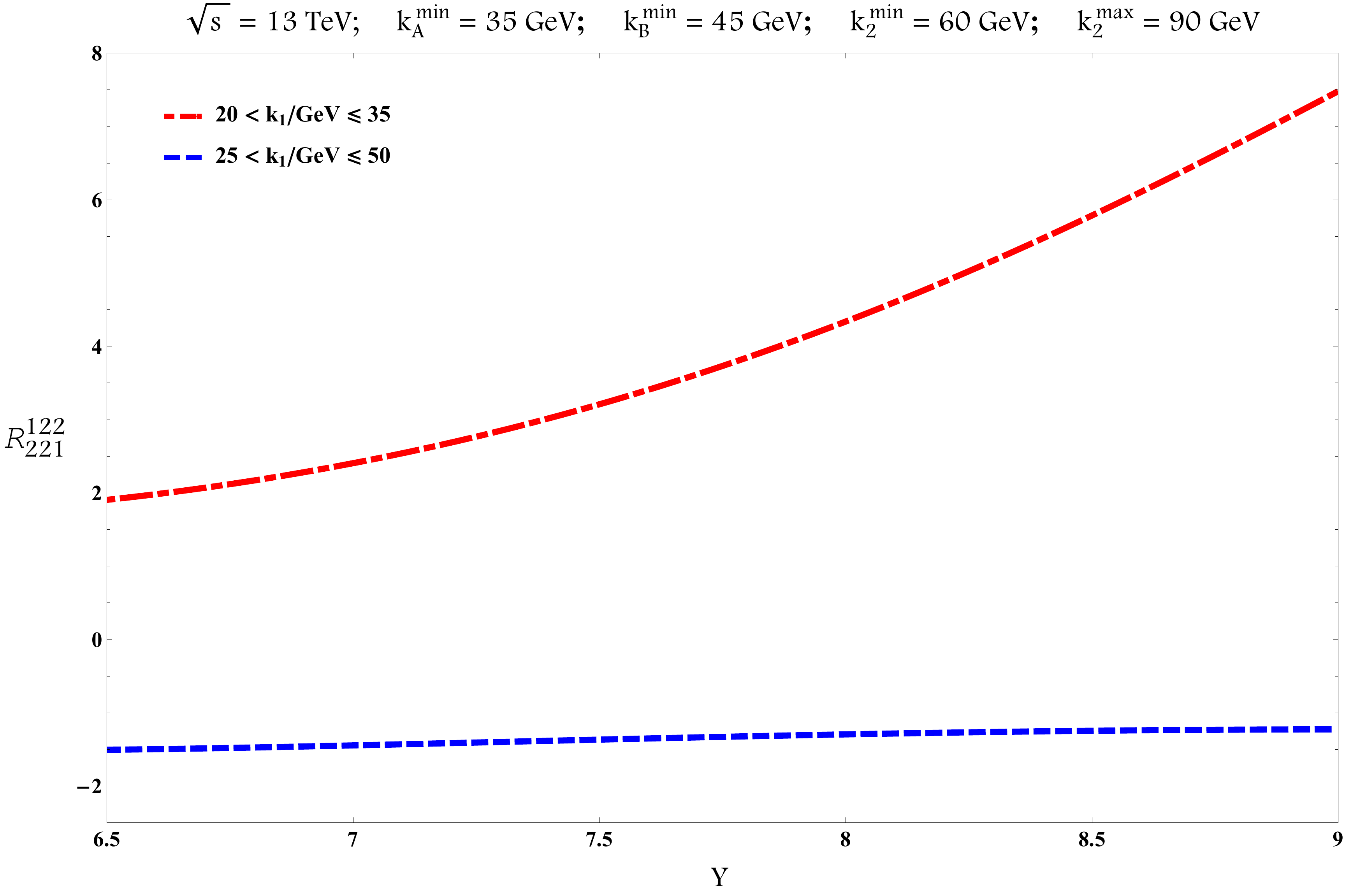}
   
   \caption[$Y$-dependence of the $R^{122}_{221}$ four-jet ratios]
   {$Y$-dependence of $R^{122}_{221}$
   for $\sqrt s = 7$ TeV (top) and for $\sqrt s = 13$ TeV (bottom).} 
   \label{fig:7}
   \end{center}
   \end{figure}

\begin{figure}[H]
   \begin{center}
   
      \includegraphics[scale=0.38]{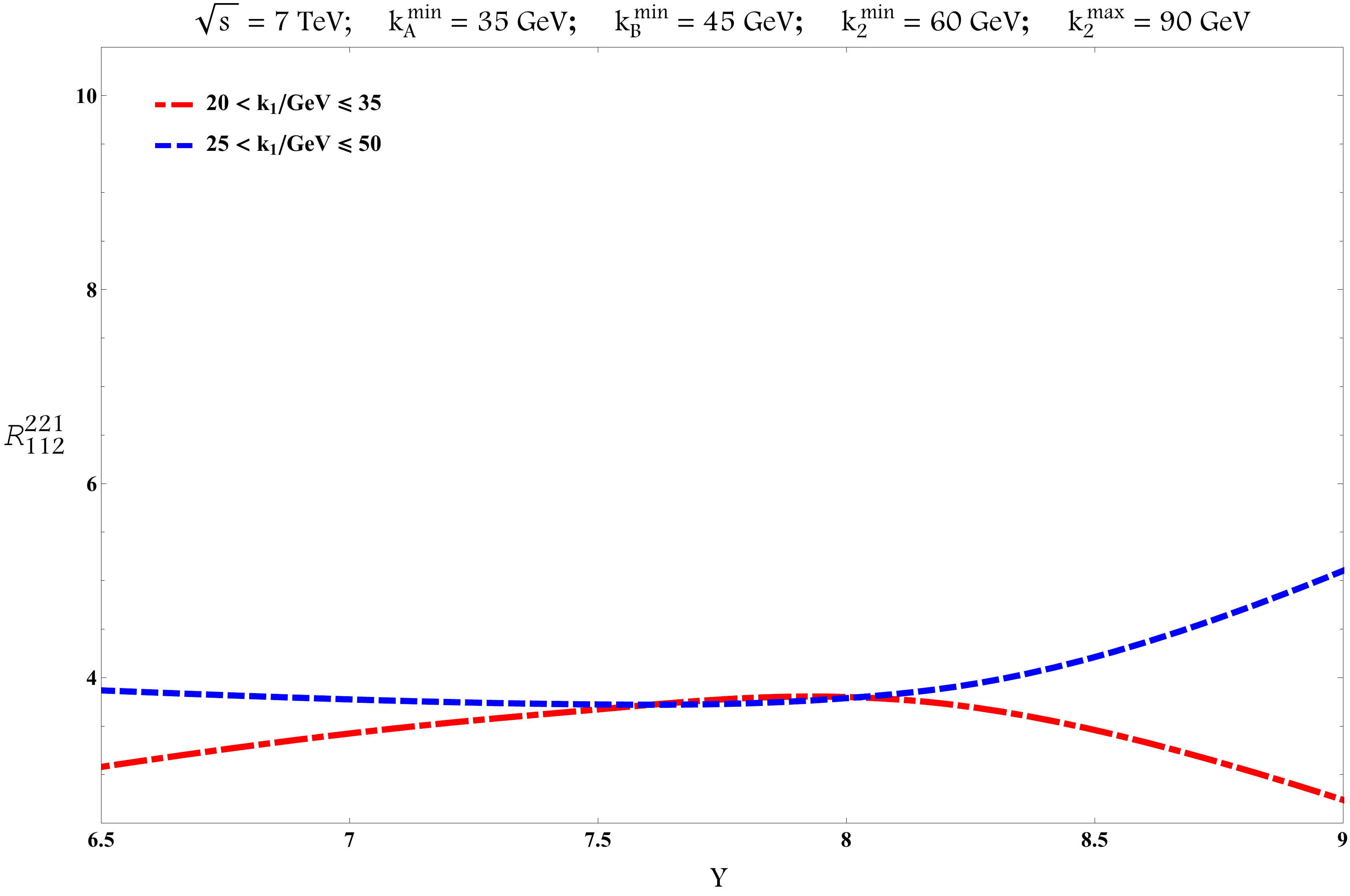}
   
      \vspace{.5cm}
   
      \includegraphics[scale=0.38]{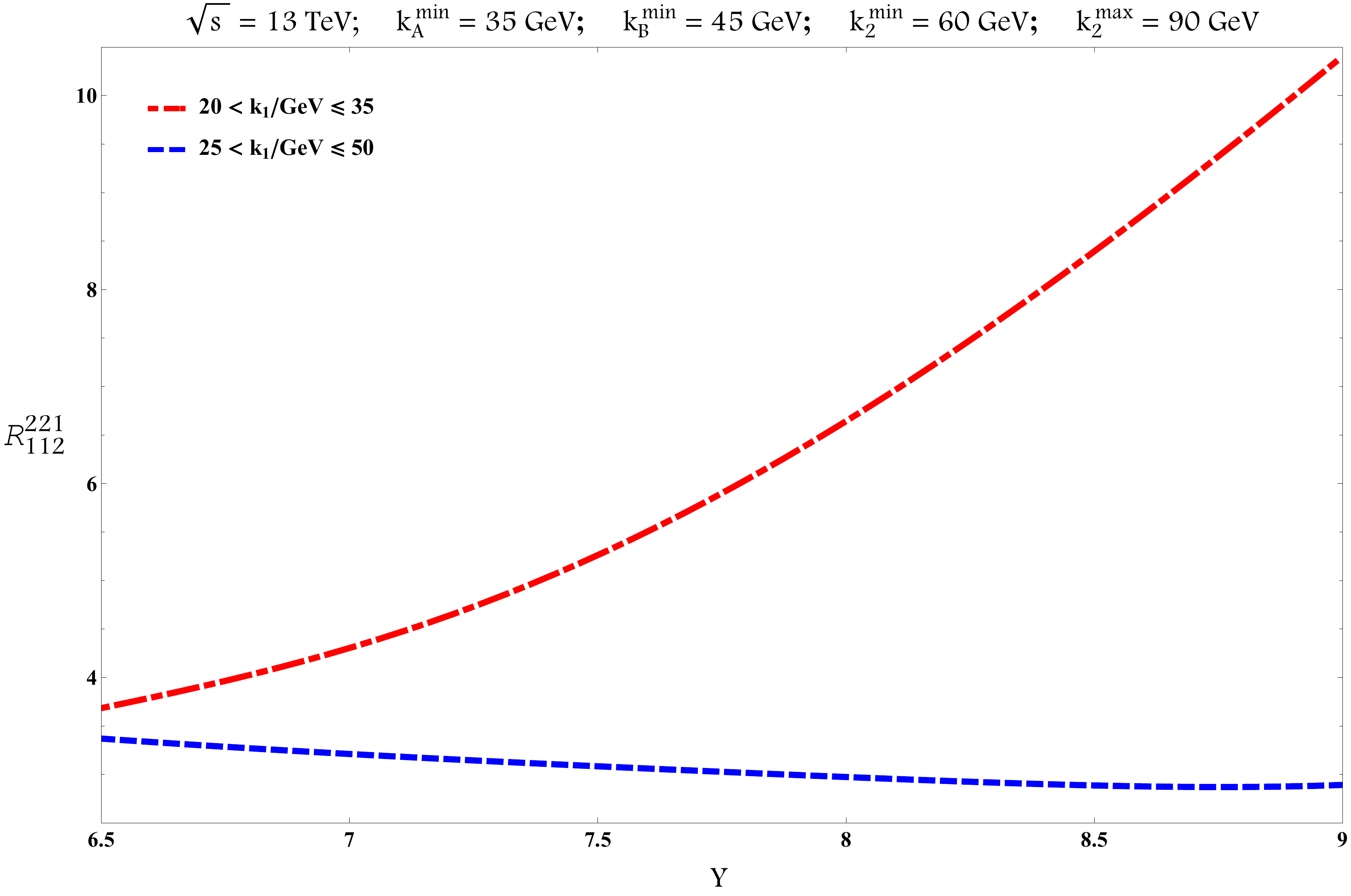}
   
   \caption[$Y$-dependence of the $R^{221}_{112}$ four-jet ratios]
   {$Y$-dependence of $R^{221}_{112}$
   for $\sqrt s = 7$ TeV (top) and for $\sqrt s = 13$ TeV (bottom).} 
   \label{fig:8}
   \end{center}
   \end{figure}

   The results for the
   ratios $R^{111}_{221}$, $R^{112}_{111}$, $R^{112}_{211}$, 
   $R^{212}_{111}$, $R^{122}_{221}$, $R^{221}_{112}$ 
   are shown in Figs.~\ref{fig:3}$-$\ref{fig:8}.
   We plot the ratios for the cut defined in Eq.~(\ref{cut1}) 
   with a red dot-dashed
   line and the ratios for the cut defined in Eq.~(\ref{cut2}) 
   with a blue dashed line.
   We place the $\sqrt s = 7$ TeV results
   on the top of each figure and the $\sqrt s = 13$ TeV results 
   at the bottom.
   
   The functional dependence of the ratios $R^{MNL}_{PQR}$ 
   on the rapidity difference between $k_A$ and $k_B$ is rather smooth. 
   We can further notice that there are ratios
   with an almost linear behaviour with $Y$ 
   and with a rather small slope.  
   To be specific, the ratios represented 
   by the blue curve in Fig.~\ref{fig:3} 
   and the red curve in Figs.~\ref{fig:4},~\ref{fig:5} and~\ref{fig:6} 
   demonstrate this linear behaviour in a striking fashion. 
   Furthermore, whenever a ratio exhibits a linear dependence on $Y$ 
   (for a certain kinematical cut of $k_1$) at colliding energy 7 TeV, 
   we observe that the ratio maintains almost 
   the exact same linear behaviour (with very similar actual values) 
   at 13 TeV as well.
   
   On the other hand, there are configurations 
   for which the functional dependence on Y 
   is much stronger and far from linear. 
   In Fig.~\ref{fig:4}, the blue curve on the top 
   rises from $\sim1.2$ at $Y=6.5$ to $\sim 6.8$ at $Y=9$, 
   whereas in Fig.~\ref{fig:6} on the top it drops 
   from $\sim (-1.5)$ to $\sim (-4.8)$ for the same variation in $Y$. 
   Generally, if for some ratio there is a strong functional dependence 
   on $Y$ for a $k_1$ of intermediate size (blue curve), 
   this dependence is `softened' at higher colliding energy 
   (see plots in Figs.~\ref{fig:4},~\ref{fig:5},~\ref{fig:6} 
   and~\ref{fig:8}). However, for a $k_1$ of smaller size 
   (red curve), we see that the functional dependence on $Y$ 
   gets stronger at 13 TeV (Figs.~\ref{fig:3},~\ref{fig:7} 
   and~\ref{fig:8}), unless of course it exhibits a linear behaviour 
   as was discussed in the previous paragraph.
   
   In all plots presented in Figs.~\ref{fig:3}$-$\ref{fig:8}, 
   there is no red or blue curve that changes sign 
   in the interval $6.5 < Y < 9$.
   Moreover, if a ratio $R^{MNL}_{PQR}$ 
   is positive (negative) at 7 TeV, 
   it will continue being positive (negative) at 13 TeV, 
   disregarding the specific functional behaviour on $Y$.
   
   In contrast to our main observation in Chapter~\ref{chap:3j} 
   where in general, 
   for most of the three-jet observables $R^{MN}_{PQ}$ 
   there were no significant changes after increasing 
   the colliding energy from 7 to 13 TeV, here we notice that, 
   depending on the kinematical cut, 
   an increase in the colliding energy may lead 
   to a noticeable change to the shape of the functional Y dependence, 
   {\it e.g.} red curve in Fig.~\ref{fig:3}, 
   blue and red curve in Fig.~\ref{fig:8}. 
   This is a very interesting point for the following reason. 
   If a BFKL-based analysis for an observable dictates that 
   the latter does not change much when the energy increases, 
   this fact actually indicates that a kind of asymptotia 
   has been reached, {\it e.g.} the slope of the Green's function 
   plotted as a function of the rapidity for very large rapidities. 
   In asymptotia, the dynamics is driven by pure BFKL effects 
   whereas pre-asymptotic effects are negligible. 
   In the present study, we have a mixed picture. 
   We have ratios that do not really change when the energy increases 
   and other ratios for which a higher colliding energy changes 
   their functional dependence on Y. A crucial point that allows us 
   to speak about pre-asymptotic effects, which in itself infers 
   that BFKL is still the relevant dynamics, 
   was outlined previously in this Section: 
   despite the fact that for some cases 
   we see a different functional dependence on Y 
   after raising the colliding energy, 
   it is important to note that we observe no change of sign 
   for any ratio $R^{MNL}_{PQR}$. 
   Therefore, the four-jet ratio observables we are studying here 
   are more sensitive to pre-asymptotic effects 
   than the related three-jet ratio observables 
   studied in Chapter~\ref{chap:3j}. 
   Nevertheless, by imposing different kinematical cuts 
   one can change the degree of importance of these effects. 
   To conclude with, carefully combined choice of cuts for 
   the $R^{MNL}_{PQR}$ observables and a detailed confrontation 
   between theoretical predictions and data may turn out 
   to be an excellent way to probe deeper into the BFKL dynamics.
    
   \subsubsection{Used tools}
   \label{ssub:4j-tools}
   
   The numerical computation of all the observables 
   shown in this Chapter was done
   in \textsc{Fortran}. \textsc{Mathematica}
   was used for various cross-checks.
   We used the NLO MSTW 2008 PDF sets~\cite{Martin:2009iq} whereas
   regarding the strong coupling a two-loop running coupling setup 
   with $\alpha_s\left(M_Z\right)=0.11707$ and five quark flavours was used.
   \cod{Vegas}~\cite{VegasLepage:1978} 
   as implemented in the \cod{Cuba} 
   library~\cite{Cuba:2005,ConcCuba:2015}
   was our main integration routine. 
   We also made use of a modified version 
   of the \cod{Psi}~\cite{RpsiCody:1973} routine and
   the library \cod{Quadpack}~\cite{Quadpack:book:1983}.

 \section{Summary} 
 \label{sec:4j-summary}
 
 New observables were proposed to study four-jet production at hadron colliders in terms of its azimuthal-angle dependences. These correspond to the ratios of correlation functions of products of cosines of azimuthal-angle differences among the tagged jets. A single BFKL ladder approach was used, with inclusive production of two forward/backward and two further, more central, tagged jets. The dependence on the transverse momenta and rapidities of the two central jets is a distinct signal of BFKL dynamics.
 
 The interesting patterns, similar to oscillation modes of a two-dimensional membrane, that the $\mathcal{R}^{MNL}_{PQR}$ exhibit, are the result of our first analysis at parton level and for final-state fixed kinematics done in Section~\ref{sec:4j-partonic}.  
 Then (Section~\ref{sec:4j-hadronic}), a full phenomenological study of LHC inclusive
 four-jet production was presented making use of the BFKL resummation framework.
 Our study was focused on azimuthal-angle dependent observables, investigating at hadronic level the $R^{MNL}_{PQR}$ ratios  
 at two different center-of-mass energies, $\sqrt s = 7, 13$ TeV. 

 An \emph{asymmetric} kinematical cut 
 with respect to the transverse momentum
 of the most forward ($k_A$) and most backward ($k_B$) jets, which is arguably 
 a more interesting kinematical configuration that a \emph{symmetric} cut, was chosen.
 The asymmetry was realised by imposing different lower limits
 to $k_A$ and $k_B$ ($k_A^{min} = 35$ GeV and $k_B^{min} = 45$ GeV).
 Additionally, we demanded for $k_2$ to be larger than both $k_A$ and $k_B$
 whereas the value of the transverse momentum $k_1$ was allowed to be 
 either smaller
 than both $k_A$ and $k_B$ or overlapping the $k_A$ and $k_B$ ranges.
 We presented the dependence of several $R^{MNL}_{PQR}$ on the rapidity
 interval $Y$ between $k_A$ and $k_B$.
 A smooth functional dependence of the ratios on $Y$ seems to be the rule. 
 The ratios we presented show in some cases considerable changes when 
 the colliding energy increases from 7 to 13 TeV which tells us that
 pre-asymptotic effects do play a role for the azimuthal ratios in inclusive
 four-jet production.

\renewcommand{\theequation}
             {\arabic{chapter}.\arabic{equation}}
\chapter{Conclusions and Outlook}
\label{chap:conclusions}

 \section{Conclusions} 
 \label{sec:conclusions}
 
 We brought exhaustive examples of testable predictions to probe QCD in the high-energy limit through the study of distinct inclusive hadronic processes. 
 
 The first reaction (Chapter~\ref{chap:mn-jets}) we investigated is the inclusive production of two jets featuring large transverse momenta and well separated in rapidity, known as Mueller--Navelet jets. We gave predictions~\cite{Caporale:2014gpa} with full NLA BFKL accuracy for the jet azimuthal correlations in kinematical ranges already covered by LCH data~\cite{Khachatryan:2016udy}, showing how a fair agreement between theory and experiment is reached when the $\mu_R$ and $\mu_F$ scales are optimised according to the BLM procedure (Section~\ref{sec:bfkl_blm}). 
 In spite of this, there are still other issues which deserve some care and have not been taken into account both in theoretical and experimental analyses so far. 
 
 One one side, the comparison of BFKL-inspired calculations with data needs to be extended to kinematical regimes where the two jets are emitted with \emph{asymmetric} transverse momenta. 
 In this way the Born contribution, which essentially comes from the production of back-to-back jets, is suppressed and the effects of the additional undetected hard gluon radiation is enhanced, thus giving us the chance to magnify and definitely figure out the size of the BFKL resummation, with respect to descriptions based on the fixed-order DGLAP approach.
 As a first step in this direction, we compared~\cite{Celiberto:2015yba} full NLA BFKL predictions with NLO fixed-order DGLAP calculations in the \emph{high-energy} limit, considering \emph{asymmetric} momentum configurations. 
 
 On the other side, for a given value of the jet rapidity separation, the rapidity of one of the two jets could be so small, that this jet is actually produced in the central region, rather than in the fragmentation region of the parent proton. Central jets originate from small-$x$ partons, and the collinear approach for the description of the Mueller--Navelet jet vertices may not hold at small $x$. Therefore we proposed to return back to the original Mueller--Navelet idea, to study the inclusive production of two forward jets separated by a large rapidity gap, removing from the analysis those regions where jets are produced at central rapidities. This allowed us to give the first phenomenological predictions~\cite{Celiberto:2016ygs} for Mueller--Navelet jet at the center-of-mass energy of 13~TeV, currently active at the LHC.
 
 The second reaction (Chapter~\ref{chap:dihadron}) we investigated is the inclusive dihadron production. This process has much in common with the well known Mueller--Navelet jet process. Hadrons can, however, be detected at
 much smaller values of the transverse momentum than jets, thus allowing us to explore an additional
 kinematical range, supplementary to the one studied with Mueller--Navelet jets. Furthermore, it has given us the opportunity to constrain not only the PDFs for the initial proton, but also
 the parton FFs describing the detected hadron in the final state. In a first phenomenological analysis~\cite{Celiberto:2016hae}, we have shown how the discrepancy between predictions with partial NLA BFKL accuracy 
 and full LLA BFKL calculations is significantly reduced via the use of the BLM scale setting. Then~\cite{Celiberto:2017ptm}, we gave the first predictions for hadrons' azimuthal correlations at~7 and~13 TeV in the full NLA BFKL approach, considering the effect of choosing different values for the factorisation scale $\mu_F$. We also gauged the uncertainty coming from the use of different PDF and FF parameterisations. Inclusive dihadron production represents so a new suitable channel to get a better understanding of the QCD dynamics in the high-energy limit. 
 
 We extended our analysis to the study of more exclusive processes, where one (Chapter~\ref{chap:3j}) or two (Chapter~\ref{chap:4j}) 
 jets are always tagged in the final state in more central 
 regions of the detectors, together with other two forward/backward ones. By demanding a strong ordering in rapidity 
 among the jets, according to MRK, we generalised our formalism to account for high-energy resummation effects. This allowed us to define new, suitable BFKL observables, sensitive to the azimuthal configurations of the tagged extra particles. We started from the partonic level, by giving predictions~\cite{Caporale:2015vya,Caporale:2015int} for azimuthal quantities averaged on the hard cross section. 
 Then, we presented the first phenomenological analyses at hadronic level~\cite{Caporale:2016soq,Caporale:2016xku} and for different final-state kinematical ranges, showing the weak dependence of our observables on the rapidity interval between the two outermost jets.
 Finally~\cite{Caporale:2016zkc}, we studied the effect of higher-order BFKL corrections, using the BLM method to optimise the renormalisation scale $\mu_R$ and considering three distinct 
 setups for the final-state phase space. The general outcome is that the NLA corrections are moderate and our proposed
 observables exhibit a very good perturbative stability.
 
 In view of all these considerations, we encourage experimental collaborations to consider \emph{asymmetric} configurations in their next Mueller--Navelet jet study, as well as to include inclusive dihadron production and inclusive multi-jet production processes in the program of future analyses 
 at the LHC, making use of new effective paths to improve our knowledge about the dynamics of strong interactions in the Regge limit.

 \section{Outlook} 
 \label{sec:outlook}
 
 The study of semi-hard processes is a wide research field, its wealthy phenomenology offering us a faultless chance to test perturbative QCD in the high-energy limit. 
 An ample range of new ide{\ae} can be guessed to extend our understanding of the BFKL dynamics in this kinematical regime. We mention and propose here some possible next studies, which are strictly related to the analysis presented in this thesis.
 
 For all the considered processes, it would be very important to compare our results with fixed-order perturbative calculations based on the DGLAP factorisation, paying particular attention to the Mueller--Navelet and the inclusive dihadron production reactions, where BFKL predictions with full NLA BFKL accuracy have been already provided. Furthermore, we plan to extend our analysis by investigating the effect of using \emph{asymmetric} cuts for the transverse momenta even in the case of hadrons,
 as well as studying less inclusive reactions where at least 
 one charged light hadron is always tagged in the final state. 
 If, together with the hadron, a forward jet is also emitted, 
 we will have the opportunity to study \emph{hadron-jet} correlations, which clearly enrich the exclusiveness of the process. On one side, the hadron tagging introduces dependence on FFs; on the other side, the larger rapidity values for which jets can be detected with respect to hadrons permit to consider final-state kinematics \emph{asymmetric} also in rapidity.
 
 A further way to disentangle the applicability border of our approach is to make comparisons with some other theoretical predictions which include higher-twist effects. For the last point, one can consider an alternative, higher-twist production 
 mechanism, related to \emph{multi-parton} interactions in QCD~\cite{Diehl:2011yj,Maciula:2014pla,Jung:2011yt,Baranov:2015nma}. The \emph{double-parton scattering} contribution to the Mueller--Navelet jet production was considered in Refs.~\cite{Ducloue:2015jba,Maciula:2014pla}, using different approaches. It would be very interesting to estimate the effect of the multi-particle interactions also in the other processes we proposed, \emph{i.a.} in the inclusive four-jet production~\cite{Maciula:2015vza,Kutak:2016mik,Kutak:2016ukc}. 
 
 As for the inclusive multi-jet production processes, full NLA BFKL studies are needed~\cite{3jet:nla-wip}, as well as comparisons with predictions from the BFKL-inspired Monte Carlo \cod{BFKLex}~\cite{Chachamis:2011rw,Chachamis:2011nz,Chachamis:2012fk,
 Chachamis:2012qw,Caporale:2013bva,Chachamis:2015zzp,Chachamis:2015ico,Chachamis:2016ejm}. 
 Results from general-purpose Monte Carlos tools should also be pursued.
 
 Finally, the inclusion of other resummation effects should be accounted for, such as the threshold-log resummation~\cite{Jager:2004jh,Kidonakis:1998bk,deFlorian:2007fv,Catani:1996yz} and, in the specific case of Mueller--Navelet jets, the resummation to all orders of logarithms in the jet-cone radius $R$, which arise when the correspondence between the jet momentum and the original parton's momentum is strongly affected by radiation at angles larger than $R$ (\emph{micro-jets})~\cite{Catani:2013oma,Dasgupta:2014yra}.


\end{document}